\documentclass[11pt,a4paper,twoside]{mydiss}
\usepackage{hyperref}
\usepackage{makeidx}
\usepackage{mathrsfs}

\makeindex

\usepackage{graphicx}
\usepackage{psfig}

\usepackage{fancyhdr}
\renewcommand{\chaptermark}[1]{\markboth{#1}{}}
\renewcommand{\sectionmark}[1]{\markright{\thesection\ #1}}
\usepackage{amsmath,amssymb}
\usepackage{bbm}

\def\slasha#1{\setbox0=\hbox{$#1$}#1\hskip-\wd0\hbox to\wd0{\hss\sl/\/\hss}}

\def\periodb#1{\setbox0=\hbox{$#1$}#1\hskip-\wd0\hbox to\wd0{-}}
\def\contra#1#2{\,{\buildrel \,\hbox{$
    \vrule height 3pt width 1pt depth 0pt
    \vrule height 3pt width #1pt depth -2pt
    \vrule height 3pt width 1pt depth 0pt $}
    \over {#2} }\,}
\newcommand{\zero}[1]{{\stackrel{\circ}{#1}}{}}     
\newcommand{\lsc}{\{\hspace{-0.1cm}[}
\newcommand{\rsc}{]\hspace{-0.1cm}\}}
\newcommand{\unit}{\mathbbm{1}}   
\newcommand{\mub}{{\bar{\mu}}}     
\newcommand{\id}{\mathrm{id}}   
\newcommand{\Ric}{\mathrm{Ric}}   
\newcommand{\sigmab}{{\bar{\sigma}}}     
\newcommand{\CA}{\mathcal{A}}    
\newcommand{\CCA}{\mathscr{A}}    
\newcommand{\CAh}{{\hat{\mathcal{A}}}}    
\newcommand{\CAt}{\tilde{\mathcal{A}}}    
\newcommand{\CB}{\mathcal{B}}    
\newcommand{\CC}{\mathcal{C}}    
\newcommand{\CCC}{\mathscr{C}}    
\newcommand{\CD}{\mathcal{D}}    
\newcommand{\CCD}{\mathscr{D}}    
\newcommand{\CCDb}{{\bar{\mathscr{D}}}}    
\newcommand{\CF}{\mathcal{F}}    
\newcommand{\CCF}{\mathscr{F}}    
\newcommand{\CG}{\mathcal{G}}    
\newcommand{\CCG}{\mathscr{G}}    
\newcommand{\CH}{\mathcal{H}}    
\newcommand{\CI}{\mathcal{I}}    
\newcommand{\CK}{\mathcal{K}}    
\newcommand{\CCK}{\mathscr{K}}    
\newcommand{\CL}{\mathcal{L}}    
\newcommand{\CCL}{\mathscr{L}}    
\newcommand{\CM}{\mathcal{M}}    
\newcommand{\CN}{\mathcal{N}}    
\newcommand{\CCO}{\mathscr{O}}    
\newcommand{\CO}{\mathcal{O}}    
\newcommand{\COb}{\bar{\mathcal{O}}}    
\newcommand{\CP}{\mathcal{P}}    
\newcommand{\CCP}{\mathscr{P}}    
\newcommand{\CQ}{\mathcal{Q}}    
\newcommand{\sts}{\CP^{3|4}}

\newcommand{\CCR}{\mathscr{R}}
\newcommand{\CS}{\mathcal{S}}    
\newcommand{\CT}{\mathcal{T}}    
\newcommand{\CCT}{\mathscr{T}}    
\newcommand{\CU}{\mathcal{U}}    
\newcommand{\CUt}{\tilde{\mathcal{U}}}    
\newcommand{\CV}{\mathcal{V}}    
\newcommand{\CCV}{\mathscr{V}}    
\newcommand{\CW}{\mathcal{W}}    
\newcommand{\CX}{\mathcal{X}}    
\newcommand{\CCX}{\mathscr{X}}    
\newcommand{\CZ}{\mathcal{Z}}    
\newcommand{\CE}{\mathcal{E}}    
\newcommand{\fra}{\mathfrak{a}}    
\newcommand{\frg}{\mathfrak{g}}    
\newcommand{\frA}{\mathfrak{A}}    
\newcommand{\frE}{\mathfrak{E}}    
\newcommand{\frG}{\mathfrak{G}}    
\newcommand{\frH}{\mathfrak{H}}    
\newcommand{\frh}{\mathfrak{h}}    
\newcommand{\frM}{\mathfrak{M}}    
\newcommand{\frS}{\mathfrak{S}}    
\newcommand{\frU}{\mathfrak{U}}    
\newcommand{\frX}{\mathfrak{X}}    
\newcommand{\cb}{\,,\,}             
\newcommand{\FK}{\mathbbm{K}}     
\newcommand{\FR}{\mathbbm{R}}     
\newcommand{\FC}{\mathbbm{C}}     
\newcommand{\FH}{\mathbbm{H}}     
\newcommand{\CPP}{{\mathbbm{C}P}}    
\newcommand{\RPS}{{\mathbbm{R}P}}    
\newcommand{\NN}{\mathbbm{N}}     
\newcommand{\TT}{\mathbbm{T}}     
\newcommand{\PP}{\mathbbm{P}}     
\newcommand{\RZ}{\mathbbm{Z}}     
\newcommand{\dd}{\mathrm{d}}     
\newcommand{\dpar}{\partial}     
\newcommand{\dparb}{{\bar{\partial}}}     
\newcommand{\derr}[2]{\frac{\dpar #1}{\dpar #2}}   
\newcommand{\embd}{{\hookrightarrow}}     
\newcommand{\diag}{{\mathrm{diag}}}     
\newcommand{\rk}{\mathrm{rk}}    
\newcommand{\de}{\mathrm{e}}     
\newcommand{\di}{\mathrm{i}}     
\newcommand{\bi}{{\bar{\imath}}}     
\newcommand{\Be}{{\mathbf{e}}}     
\newcommand{\bj}{{\bar{\jmath}}}     
\newcommand{\etab}{{\bar{\eta}}}     
\newcommand{\thetab}{{\bar{\theta}}}     
\newcommand{\bz}{{\bar{z}}}     
\newcommand{\by}{{\bar{y}}}     
\newcommand{\bpsi}{{\bar{\psi}}}     
\newcommand{\btheta}{{\bar{\theta}}}     
\newcommand{\bchi}{{\bar{\chi}}}     
\newcommand{\bl}{{\bar{\lambda}}}     
\newcommand{\hl}{{\hat{\lambda}}}     
\newcommand{\hpsi}{{\hat{\psi}}}     
\newcommand{\bZ}{{\bar{Z}}}     
\newcommand{\bV}{{\bar{V}}}     
\newcommand{\bv}{{\bar{v}}}     
\newcommand{\bW}{{\bar{W}}}     
\newcommand{\bE}{{\bar{E}}}     
\newcommand{\bTheta}{{\bar{\Theta}}}     
\newcommand{\rone}{\mathbf{1}}    
\newcommand{\rthree}{\mathbf{3}}    
\newcommand{\rfour}{\mathbf{4}}    
\newcommand{\bw}{{\bar{w}}}     
\newcommand{\ald}{{\dot{\alpha}}}     
\newcommand{\ed}{{\dot{1}}}     
\newcommand{\zd}{{\dot{2}}}     
\newcommand{\bed}{{\dot{\beta}}}     
\newcommand{\gad}{{\dot{\gamma}}}     
\newcommand{\ded}{{\dot{\delta}}}     
\newcommand{\eps}{{\varepsilon}}     
\newcommand{\eand}{{~~~\mbox{and}~~~}}     
\newcommand{\ewith}{{~~~\mbox{with}~~~}}     
\newcommand{\kernel}{{\mathrm{ker}}}     
\newcommand{\spn}{{\mathrm{span}}}     
\newcommand{\etr}{{\mathrm{etr}}}     
\newcommand{\edet}{{\mathrm{edet}}}     
\newcommand{\str}{{\mathrm{str}}}     
\newcommand{\sdet}{{\mathrm{sdet}}}     
\newcommand{\rep}[1]{\mathbf{#1}}   
\newcommand{\der}[1]{\frac{\dpar}{\dpar #1}}   
\newcommand{\dder}[1]{\frac{\dd}{\dd #1}}   
\newcommand{\Der}[1]{\frac{\delta}{\delta #1}}   
\newcommand{\tr}{\,\mathrm{tr}\,}     
\newcommand{\tra}[1]{\,\mathrm{tr}_{#1}\,}     

\newcommand{\dual}{{^\vee}}     
\newcommand{\sU}{\mathsf{U}}     
\newcommand{\sSU}{\mathsf{SU}}     
\newcommand{\sMat}{\mathsf{Mat}}     
\newcommand{\sSL}{\mathsf{SL}}     
\newcommand{\sSp}{\mathsf{Sp}}     
\newcommand{\sGL}{\mathsf{GL}}     
\newcommand{\agl}{\mathfrak{gl}}     
\newcommand{\asu}{\mathfrak{su}}     
\newcommand{\sSO}{\mathsf{SO}}     
\newcommand{\sEnd}{\mathsf{End}}     
\newcommand{\sO}{\mathsf{O}}     
\newcommand{\sSpin}{\mathsf{Spin}}     
\newcommand{\ssp}{\FR^{4|4\CN}}     
\newcommand{\sspdef}{\FR^{4|4\CN}_\hbar}     
\newcommand{\remark}[1]{}     
\newcommand{\z}[1]{{\stackrel{\circ}{#1}}{}}     
\newcommand{\drd}{{\dot{3}}}    

\newcommand{\SA}{\CP^{3\oplus 2|0}}     
\newcommand{\CSA}{\CCP^{3\oplus 2|0}}     
\newcommand{\SB}{\CP^{3\oplus 1|0}}     
\newcommand{\CSB}{\CCP^{3\oplus 1|0}}     

\usepackage{picins}

\begin{document}

\begin{titlepage}
\vspace*{2.1cm}
\centerline{\includegraphics[width=15cm,height=20.0cm]{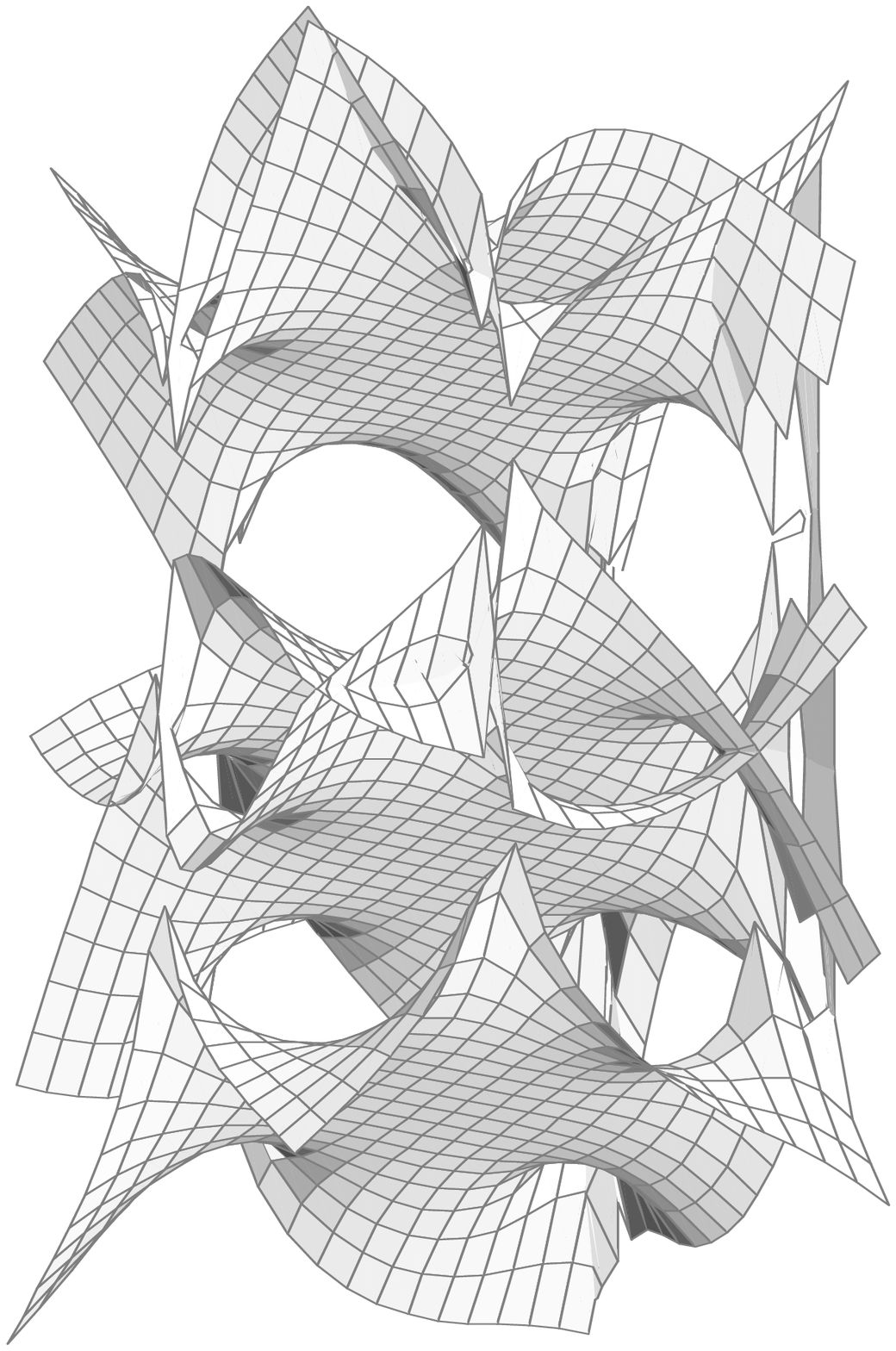}}
\vspace{-23.4cm}
\begin{center}
\Huge Aspects of Twistor Geometry\\\index{twistor}
and Supersymmetric Field Theories\\
   within Superstring Theory\index{string theory}
\end{center}
\vspace{17.4cm}
\begin{center}
\large
Von der Fakult\"{a}t f\"{u}r Mathematik und Physik der Universit\"{a}t Hannover\\
zur Erlangung des Grades\\
Doktor der Naturwissenschaften\\
Dr.\ rer.\ nat.\\
genehmigte Dissertation von\\[0.2cm]
{\Large Christian S\"{a}mann}\\[0.2cm]
geboren am 23.\ April 1977 in Fulda
\end{center}

\end{titlepage}


\newpage
\pagestyle{empty}{} \vspace*{1cm}


\newpage

\pagestyle{empty} {} \vspace*{5cm} \noindent {\em For there is
nothing hidden, except that it should be made
known;\\
neither was anything made secret, but that it should come to
light.}\\\hspace*{6.8cm}Mark 4,22\\
\vspace*{4cm}\\
\hspace*{7cm}{\em Wir m\"{u}ssen wissen, wir werden wissen.}\\
\hspace*{11cm}David Hilbert
\newpage
\pagestyle{empty}{} \vspace*{1cm}


\newpage

\pagestyle{empty} {} \vspace*{5cm}
\begin{center}
{\em To those who taught me}
\end{center}


\newpage
\pagestyle{empty}\vspace*{18.5cm}

Betreuer: Prof. Dr. Olaf Lechtenfeld und Dr. Alexander D.~Popov

Referent: Prof. Dr. Olaf Lechtenfeld

Korreferent: Prof. Dr. Holger Frahm

\vspace{0.5cm}

Tag der Promotion: 30.01.2006

\vspace{0.5cm}

Schlagworte:\, Nichtantikommutative Feldtheorie,\,
Twistorgeometrie,\, Stringtheorie\index{twistor}

String Theory\index{string theory}

\vspace{0.5cm}

ITP-UH-26/05


\newpage

\pagestyle{empty}

\begin{center}\LARGE\bfseries{\sc Zusammenfassung}\end{center}

Die Resultate, die in dieser Arbeit vorgestellt werden, lassen
sich im Wesentlichen zwei Forschungsrichtungen in der
Stringtheorie zuordnen: Nichtantikommutative Feldtheorie sowie
Twistorstringtheorie.\index{twistor}

Nichtantikommutative Deformationen von Superr\"{a}umen entstehen auf
nat\"{u}rliche Wei\-se bei Typ II Superstringtheorie in einem
nichttrivialen Graviphoton-Hinter\-grund, und solchen\index{graviphoton}
Deformationen wurde in den letzten zwei Jahren viel Beachtung
geschenkt. Zu\-n\"{a}chst konzentrieren wir uns auf die Definition der
nichtantikommutativen Deformation von $\CN=4$ super
Yang-Mills-Theorie. Da es f\"{u}r die Wirkung dieser Theorie keine
Superraumformulierung gibt, weichen wir statt dessen auf die
\"{a}quivalenten {\em constraint equations} aus. W\"{a}hrend der\index{constraint equations}
Herleitung der deformierten Feldgleichungen schlagen wir ein
nichtantikommutatives Analogon zu der Seiberg-Witten-Abbildung
vor.

Eine nachteilige Eigenschaft nichantikommutativer Deformationen
ist, dass sie Supersymmetrie teilweise brechen (in den einfachsten
F\"{a}llen halbieren sie die Zahl der erhaltenen Superladungen). Wir
stellen in dieser Arbeit eine sog.\ Drinfeld-Twist-Technik vor,
mit deren Hilfe man supersymmetrische Feldtheorien derart
reformulieren kann, dass die gebrochenen Supersymmetrien wieder
manifest werden, wenn auch in einem {\em getwisteten} Sinn. Diese
Reformulierung erm\"{o}glicht es, bestimmte chirale Ringe zu
definieren und ergibt supersymmetrische
Ward-Takahashi-Identit\"{a}ten, welche von gew\"{o}hnlichen
supersymmetrischen Feldtheorien bekannt sind. Wenn man Seibergs
{\em naturalness argument}, welches die Symmetrien von
Niederenergie-Wirkungen betrifft, auch im nichtantikommutativen
Fall zustimmt, so erh\"{a}lt man Nichtrenormierungstheoreme selbst f\"{u}r
nichtantikommutative Feldtheorien.

Im zweiten und umfassenderen Teil dieser Arbeit untersuchen wir
detailliert geome\-trische Aspekte von Supertwistorr\"{a}umen, die\index{twistor}
gleichzeitig Calabi-Yau-Super\-man\-nig\-fal\-tigkeiten sind und\index{Calabi-Yau}
dadurch als {\em target space} f\"{u}r topologische String\-theorien\index{target space}
geeignet sind. Zun\"{a}chst stellen wir die Geometrie des bekanntesten
Beispiels f\"{u}r einen solchen Supertwistorraum, $\CPP^{3|4}$, vor\index{twistor}
und f\"{u}hren die Penrose-Ward-Transformation, die bestimmte
holomorphe Vektorb\"{u}ndel \"{u}ber dem Supertwistorraum mit L\"{o}sungen zu\index{twistor}
den $\CN=4$ supersymmetrischen selbstdualen Yang-Mills-Gleichungen
verbindet, explizit aus. Anschlie{\ss}end diskutieren wir mehrere
dimensionale Reduktionen des Supertwistorraumes $\CPP^{3|4}$ und\index{twistor}
die implizierten Ver\"{a}nderungen an der Penrose-Ward-Transformation.

Fermionische dimensionale Reduktionen bringen uns dazu, exotische
Supermannigfaltigkeiten, d.h. Supermannigfaltigkeiten mit
zus\"{a}tzlichen (bosonischen) nilpotenten Dimensionen, zu studieren.
Einige dieser R\"{a}ume k\"{o}nnen als {\em target space} f\"{u}r topologische\index{target space}
Strings dienen und zumindest bez\"{u}glich des Satzes von Yau f\"{u}gen
diese sich gut in das Bild der
Calabi-Yau-Super\-man\-nig\-fal\-tigkeiten ein.\index{Calabi-Yau}

Bosonische dimensionale Reduktionen ergeben die
Bogomolny-Gleichungen sowie Matrixmodelle, die in Zusammenhang mit
den ADHM- und Nahm-Gleichungen stehen. (Tats\"{a}chlich betrachten wir
die Supererweiterungen dieser Gleichungen.) Indem wir bestimmte
Terme zu der Wirkung dieser Matrixmodelle hinzuf\"{u}gen, k\"{o}nnen wir
eine komplette \"{A}quivalenz zu den ADHM- und Nahm-Gleichungen
erreichen. Schlie{\ss}lich kann die nat\"{u}rliche Interpretation dieser
zwei Arten von BPS-Gleichungen als spezielle
D-Branekonfigurationen in Typ IIB Superstringtheorie vollst\"{a}ndig\index{D-brane}
auf die Seite der topo\-logischen Stringtheorie \"{u}bertragen werden.
Dies f\"{u}hrt zu einer Korrespondenz zwischen topologischen und
physikalischen D-Branesystemen und er\"{o}ffnet die interessante\index{D-brane}
Perspektive, Resultate von beiden Seiten auf die jeweils andere
\"{u}bertragen zu k\"{o}nnen.


\newpage
\pagestyle{empty}{} \vspace*{1cm}


\newpage

\pagestyle{empty}

\begin{center}\LARGE\bfseries{\sc Abstract}\end{center}

There are two major topics within string theory to which the\index{string theory}
results presented in this thesis are related: non-anticommutative
field theory on the one hand and twistor string theory on the\index{string theory}\index{twistor}\index{twistor!string theory}
other hand.

Non-anticommutative deformations of superspaces arise naturally in\index{super!space}
type II superstring theory in a non-trivial graviphoton background\index{graviphoton}\index{string theory}
and they have received much attention over the last two years.
First, we focus on the definition of a non-anticommutative
deformation of $\CN=4$ super Yang-Mills theory. Since there is no\index{N=4 super Yang-Mills theory@$\CN=4$ super Yang-Mills theory}\index{Yang-Mills theory}
superspace formulation of the action of this theory, we have to\index{super!space}
resort to a set of constraint equations defined on the superspace\index{constraint equations}
$\FR^{4|16}_\hbar$, which are equivalent to the $\CN=4$ super
Yang-Mills equations. In deriving the deformed field equations, we
propose a non-anticommutative analogue of the Seiberg-Witten map.\index{Seiberg-Witten map}

A mischievous property of non-anticommutative deformations is that
they partially break supersymmetry (in the simplest case, they\index{super!symmetry}
halve the number of preserved supercharges). In this thesis, we
present a so-called Drinfeld-twisting technique, which allows for
a reformulation of supersymmetric field theories on
non-anticommutative superspaces in such a way that the broken\index{non-anticommutative superspace}\index{super!space}
supersymmetries become manifest even though in some sense twisted.
This reformulation enables us to define certain chiral rings and\index{chiral!ring}
it yields supersymmetric Ward-Takahashi-identities, well-known
from ordinary supersymmetric field theories. If one agrees with
Seiberg's naturalness arguments concerning symmetries of
low-energy effective actions also in the non-anticommutative
situation, one even arrives at non-renormalization theorems for\index{non-renormalization theorems}\index{Theorem!non-renormalization}
non-anticommutative field theories.\index{non-anticommutative field theories}

In the second and major part of this thesis, we study in detail
geometric aspects of supertwistor spaces which are simultaneously\index{twistor}\index{twistor!space}
Calabi-Yau supermanifolds and which are thus suited as target\index{Calabi-Yau}\index{Calabi-Yau supermanifold}\index{super!manifold}
spaces for topological string theories. We first present the\index{topological!string}
geometry of the most prominent example of such a supertwistor\index{twistor}
space, $\CPP^{3|4}$, and make explicit the Penrose-Ward transform\index{Penrose-Ward transform}
which relates certain holomorphic vector bundles over the\index{holomorphic!vector bundle}
supertwistor space to solutions to the $\CN=4$ supersymmetric\index{twistor}\index{twistor!space}
self-dual Yang-Mills equations. Subsequently, we discuss several
dimensional reductions of the supertwistor space $\CPP^{3|4}$ and\index{dimensional reduction}\index{twistor}\index{twistor!space}
the implied modifications to the Penrose-Ward transform.\index{Penrose-Ward transform}

Fermionic dimensional reductions lead us to study exotic\index{dimensional reduction}
supermanifolds, which are supermanifolds with additional even\index{super!manifold}
(bosonic) nilpotent dimensions. Certain such spaces can be used as
target spaces for topological strings, and at least with respect\index{target space}\index{topological!string}
to Yau's theorem, they fit nicely into the picture of Calabi-Yau\index{Calabi-Yau}\index{Theorem!Yau}
supermanifolds.\index{super!manifold}

Bosonic dimensional reductions yield the Bogomolny equations\index{Bogomolny equations}\index{dimensional reduction}
describing static mo\-nopole configurations as well as matrix
models related to the ADHM- and the Nahm equations. (In fact, we\index{Nahm equations}\index{matrix model}
describe the superextensions of these equations.) By adding
certain terms to the action of these matrix models, we can render\index{matrix model}
them completely equivalent to the ADHM and the Nahm equations.\index{Nahm equations}
Eventually, the natural interpretation of these two kinds of BPS
equations by certain systems of D-branes within type IIB\index{D-brane}
superstring theory can completely be carried over to the\index{string theory}
topological string side via a Penrose-Ward transform. This leads\index{Penrose-Ward transform}\index{topological!string}
to a correspondence between topological and physical D-brane\index{D-brane}
systems and opens interesting perspectives for carrying over
results from either sides to the respective other one.


\newpage
\pagestyle{empty}{} \vspace*{1cm}


 \tableofcontents

\chapter{Introduction}

\section{High-energy physics and string theory}\index{string theory}

Today, there are essentially two well-established approaches to
describing fundamental physics, both operating in different
regimes: Einstein's theory of General Relativity\footnote{or more
appropriately: General Theory of Relativity}, which governs the
dynamics of gravitational effects on a large scale from a few
millimeters to cosmological distances and the framework called
quantum field theory, which incorporates the theory of special
relativity into quantum mechanics and captures phenomena at scales
from a fraction of a millimeter to $10^{-19}$m. In particular,
there is the quantum field theory called the standard model of
elementary particles, which is a quantum gauge theory with gauge
group $\sSU(3)\times \sSU(2)\times \sU(1)$ and describes the
electromagnetic, the weak and the strong interactions on equal
footing. Although this theory has already been developed between
1970 and 1973, it still proves to be overwhelmingly consistent
with the available experimental data today.

Unfortunately, a fundamental difference between these two
approaches is disturbing the beauty of the picture. While General
Relativity is a classical description of spacetime dynamics in
terms of the differential geometry of smooth manifolds, the\index{manifold}
standard model has all the features of a quantum theory as e.g.\
uncertainty and probabilistic predictions. One might therefore
wonder whether it is possible or even necessary to quantize
gravity.

The first question for the possibility of quantizing gravity is
already not easy to answer. Although promoting supersymmetry to a\index{super!symmetry}
local symmetry almost immediately yields a classical theory
containing gravity, the corresponding quantum field theory is
non-renormalizable. That is, an infinite number of renormalization
conditions is needed at the very high energies near the Planck
scale and the theory thus looses all its predictive
power\footnote{It is an amusing thought to imagine that
supergravity was indeed the correct theory and therefore nature
was in principle unpredictable.}. Two remedies to this problem are
conceivable: either to assume that there are additional degrees of
freedom between the standard model energy scale and the Planck
scale or to assume some underlying dependence of the infinite
number of renormalization conditions on a finite
subset\footnote{See also the discussion in {\tt
http://golem.ph.utexas.edu/$\sim$distler/blog/archives/000639.html}.}.

Today, there are essentially two major approaches to quantizing
gravity, which are believed to overcome the above mentioned
shortcoming: string theory, which trades the infinite number of\index{string theory}
renormalization conditions for an infinite tower of higher-spin
gauge symmetries, and the so-called loop quantum gravity approach
\cite{Rovelli:LR}. As of now, it is not even clear whether these
two approaches are competitors or merely two aspects of the same
underlying theory. Furthermore, there is no help to be expected
from experimental input since on the one hand, neither string
theory nor loop quantum gravity have yielded any truly verifiable\index{string theory}
(or better: falsifiable) results so far and on the other hand
there is simply no quantitative experimental data for any kind of
quantum gravity effect up to now.

The second question of the need for quantum gravity is often
directly answered positively, due to the argument given in
\cite{Eppley:1977} which amounts to a violation of uncertainty if
a classical gravitational field is combined with quantum
fields\footnote{It is argued that if measurement by a
gravitational wave causes a quantum mechanical wave function to
collapse then the uncertainty relation can only be preserved if
momentum conservation is violated. On the other hand, if there is
no collapse of the wave function, one could transmit signals
faster than with light.}. This line of reasoning has, however,
been challenged until today, see e.g.\ \cite{Callender:2001hx},
and it seems to be much less powerful than generally believed.

There is another reason for quantizing gravity, which is, however,
of purely aesthetical value: A quantization of gravity would most\index{quantization}
likely allow for the unification of all the known forces within
one underlying principle. This idea of unification of forces dates
back to the electro-magnetic unification by James Clerk Maxwell,
was strongly supported by Hermann Weyl and Albert Einstein and
found its present climax in the electroweak unification by Abdus
Salam and Steven Weinberg. Furthermore, there is a strong argument
which suggest that quantizing gravity makes unification or at
least simultaneous quantization of all other interactions\index{quantization}
unavoidable from a phenomenological point of view: Because of the
weakness of gravity compared to the other forces there is simply
no decoupling regime which is dominated by pure quantum gravity
effects and in which all other particle interactions are
negligible.

Unification of General Relativity and the standard model is
difficult due to the fundamental difference in the ways both
theories describe the world. In General Relativity, gravitational
interactions deform spacetime, and reciprocally originate from
such deformations. In the standard model, interactions arise from
the exchange of messenger particles. It is furthermore evident
that in order to quantize gravity, we have to substitute spacetime
by something more fundamental, which still seems to be completely
unknown.

Although the critical superstring theories, which are currently
the only candidate for a unified description of nature including a
quantum theory of gravity, still do not lead to verifiable
results, they may nevertheless be seen as a guiding principle for
studying General Relativity and quantum field theories. For this
purpose, it is important to find string/gauge field theory
dualities, of which the most prominent example is certainly the
AdS/CFT correspondence \cite{Maldacena:1997re}. These dualities\index{AdS/CFT correspondence}
provide a dictionary between certain pairs of string theories and
gauge theories, which allows to perform field theoretic
calculations in the mathematically often more powerful framework
of string theory.\index{string theory}

The recently proposed twistor string theory \cite{Witten:2003nn}\index{string theory}\index{twistor}\index{twistor!string theory}
gives rise to a second important example of such a duality. It has
been in its context that string theoretical methods have led for
the first time\footnote{Another string inspired prediction of
real-world physics has arisen from the computation of shear
viscosity via AdS/CFT-inspired methods in
\cite{Policastro:2001yc}.} to field theoretic predictions, which
would have been almost impossible to make with state-of-the-art
quantum field theoretical\footnote{One might actually wonder about
the perfect timing of the progress in high energy physics: These
calculations are needed for the interpretation of the results at
the new particle accelerator at CERN, which will start collecting
data in 2007.} technology.

As a large part of this thesis will be devoted to studying certain
aspects of this twistor string theory, let us present this theory\index{string theory}\index{twistor}\index{twistor!string theory}
in more detail. Twistor string theory was introduced in 2003 by
Edward Witten \cite{Witten:2003nn} and is essentially founded on
the marriage of Calabi-Yau and twistor geometry in the\index{Calabi-Yau}\index{twistor}
supertwistor space $\CPP^{3|4}$. Both of these geometries will\index{twistor!space}
therefore accompany most of our discussion.

Calabi-Yau manifolds are complex manifolds which have a trivial\index{Calabi-Yau}\index{complex!manifold}\index{manifold}
first Chern class. They are Ricci-flat and come with a holomorphic\index{Chern class}\index{Ricci-flat}\index{first Chern class}
volume element. The latter property allows to define a
Chern-Simons action on these spaces, which will play a crucial
r{\^o}le throughout this thesis. Calabi-Yau manifolds naturally emerge\index{Calabi-Yau}\index{manifold}
in string theory as candidates for internal compactification\index{string theory}
spaces. In particular, topological strings of B-type -- a\index{topological!string}
subsector of the superstrings in type IIB superstring theory --\index{string theory}
can be consistently defined on spaces with vanishing first Chern
number only and their dynamics is then governed by the\index{Chern number}\index{first Chern number}
above-mentioned Chern-Simons theory.\index{Chern-Simons theory}

Twistor geometry, on the other hand, is a novel description of\index{twistor}
spacetime, which was introduced in 1967 by Roger Penrose
\cite{Penrose:1967wn}. Although this approach has found many
applications in both General Relativity and quantum theory, it is
still rather unknown in the mathematical and physical communities
and it has only been recently that new interest was sparked among
string theorists by Witten's seminal paper \cite{Witten:2003nn}.
Interestingly, twistor geometry was originally designed as a\index{twistor}
unified framework for quantum theory and gravity, but so far, it
has not yielded significant progress in this direction. Its value
in describing various aspects of field theories, however, keeps
growing.

Originally, Witten showed that the topological B-model on the\index{topological!B-model}
supertwistor space $\CPP^{3|4}$ in the presence of $n$ ``almost\index{twistor}\index{twistor!space}
space-filling\footnote{a restriction on the fermionic worldvolume
directions of the D-branes}'' D5-superbranes is equivalent to\index{D-brane}
$\CN=4$ self-dual Yang-Mills theory. By adding D1-instantons, one\index{Yang-Mills theory}\index{instanton}\index{self-dual Yang-Mills theory}
can furthermore complete the self-dual sector to the full $\CN=4$
super Yang-Mills theory. Following Witten's paper, various further\index{N=4 super Yang-Mills theory@$\CN=4$ super Yang-Mills theory}\index{Yang-Mills theory}
target spaces for twistor string theory have been considered as\index{string theory}\index{target space}\index{twistor}\index{twistor!string theory}
well \cite{Popov:2004nk,Ahn:2004xs,Saemann:2004tt,Park:2004bw,
Giombi:2004xv,Wolf:2004hp,Chiou:2005jn,Popov:2005uv,Chiou:2005pu},
which lead, e.g., to certain dimensional reductions of the\index{dimensional reduction}
supersymmetric self-dual Yang-Mills equations. There has been a
vast number of publications dedicated to apply twistor string\index{twistor}
theory to determining scattering amplitudes in ordinary and
supersymmetric gauge theories (see e.g.\ \cite{TwistorWorkshop}\index{twistor}
and \cite{TwistorWorkshop2} for an overview), but only half a year
after Witten's original paper, disappointing results appeared. In
\cite{Berkovits:2004jj}, it was discovered that it seems hopeless
to decouple conformal supergravity from the part relevant for the
description of super Yang-Mills theory in twistor string theory\index{Yang-Mills theory}\index{string theory}\index{twistor}\index{twistor!string theory}
already at one-loop level. Therefore, the results for gauge theory
loop amplitudes are mostly obtained today by ``gluing together''
tree level amplitudes.

Nevertheless, research on twistor string theory continued with a\index{string theory}\index{twistor}\index{twistor!string theory}
more mathematically based interest. As an important example, the
usefulness of Calabi-Yau supermanifolds in twistor string theory\index{Calabi-Yau}\index{Calabi-Yau supermanifold}\index{string theory}\index{super!manifold}\index{twistor}\index{twistor!string theory}
suggests an extension of the famous mirror conjecture to
supergeometry. This conjecture states that Calabi-Yau manifolds\index{Calabi-Yau}\index{super!geometry}\index{manifold}
come in pairs of families, which are related by a mirror map.
There is, however, a class of such manifolds, the so-called rigid\index{rigid}\index{manifold}
Calabi-Yau manifolds, which cannot allow for an ordinary mirror. A\index{Calabi-Yau}
resolution to this conundrum had been proposed in
\cite{Sethi:1994ch}, where the mirror of a certain rigid\index{rigid}
Calabi-Yau manifold was conjectured to be a supermanifold. Several\index{Calabi-Yau}\index{super!manifold}\index{manifold}
publications in this direction have appeared since, see
\cite{Kumar:2004dj,Ahn:2004xs,Belhaj:2004ts,Ricci:2005cp,AhlLaamara:2006rk}
and references therein.

Returning now to the endeavor of quantizing gravity, we recall
that it is still not known what ordinary spacetime should exactly
be replaced with. The two most important extensions of spacetime
discussed today are certainly supersymmetry and noncommutativity.\index{super!symmetry}
The former extension is a way to avoid a severe restriction in
constructing quantum field theories: An ordinary bosonic symmetry
group, which is nontrivially combined with the Poincar{\'e} group of
spacetime transformations renders all interactions trivial. Since
supersymmetry is a fermionic symmetry, this restriction does not\index{super!symmetry}
apply and we can extend the set of interesting theories by some
particularly beautiful ones. Furthermore, supersymmetry seems to\index{super!symmetry}
be {\em the} ingredient to make string theory well-defined.\index{string theory}
Although, supersymmetry preserves the smooth underlying structure\index{super!symmetry}
of spacetime and can be nicely incorporated into the quantum field
theoretic framework, there is a strong hint that this extension is
a first step towards combining quantum field theory with gravity:
As stated above, we naturally obtain a theory describing gravity
by promoting supersymmetry to a local symmetry. Besides being in\index{super!symmetry}
some cases the low-energy limit of certain string theories, it is
believed that this so-called supergravity is the only consistent
theory of an interacting spin $\frac{3}{2}$-particle, the
superpartner of the spin 2 graviton.

Nevertheless, everything we know today about a possible quantum
theory of gravity seems to tell us that a smooth structure of
spacetime described by classical manifolds can not persist to\index{manifold}
arbitrarily small scales. One rather expects a deformation of the
coordinate algebra which should be given by relations like
\begin{equation*}
{}[\hat{x}^\mu,\hat{x}^\nu]\sim \Theta^{\mu\nu}\eand
\{\hat{\theta}^\alpha,\hat{\theta}^\beta\}\sim C^{\alpha\beta}~
\end{equation*}
for the bosonic and fermionic coordinates of spacetime. The idea
of bosonic deformations of spacetime coordinates can in fact be
traced back to work by H.~S.~Snyder in 1947 \cite{Snyder:1946qz}.
In the case of fermionic coordinates, a first model using a
deformed coordinate algebra appeared in \cite{Schwarz:1982pf}.
Later on, it was found that both deformations naturally arise in
various settings in string theory.\index{string theory}

So far, mostly the simplest possible deformations of ordinary
(super)spaces have been considered, i.e.\ those obtained by
constant deformation parameters $\Theta^{\mu\nu}$ and
$C^{\alpha\beta}$ on flat spacetimes. The non-(anti)commutative
field theories defined on these deformed spaces revealed many
interesting features, which are not common to ordinary field
theories. Further hopes, as e.g.\ that noncommutativity could tame
field theoretic singularities have been shattered with the
discovery of UV/IR mixing in amplitudes within noncommutative\index{UV/IR mixing}
field theories.

The fact that such deformations are unavoidable for studying
nontrivial string backgrounds have kept the interest in this field
alive and deformations have been applied to a variety of theories.
For $\CN=4$ super Yang-Mills theory, the straightforward\index{N=4 super Yang-Mills theory@$\CN=4$ super Yang-Mills theory}\index{Yang-Mills theory}
superspace approach broke down, but by considering so-called\index{super!space}
constraint equations, which live on an easily deformable\index{constraint equations}
superspace, also this theory can be rendered non-anticommutative,\index{super!space}
and we will discuss this procedure in this thesis.

Among the most prominent recent discoveries\footnote{or better:
``recently recalled discoveries''} in noncommutative geometry is
certainly the fact that via a so-called Drinfeld twist, one can in\index{Drinfeld twist}
some sense undo the deformation. More explicitly, Lorentz
invariance is broken to some subgroup by introducing a nontrivial\index{Lorentz!invariance}
deformation tensor $\Theta^{\mu\nu}$. The Drinfeld twist, however,\index{Drinfeld twist}
allows for a recovering of a twisted Lorentz symmetry. This
regained symmetry is important for discussing fundamental aspects
of noncommutative field theory as e.g.\ its particle content and
formal questions like the validity of Haag's theorem. In this
thesis, we will present the application of a similar twist in the\index{twist}
non-anticommutative situation and regain a twisted form of the
supersymmetry, which had been broken by non-anticommutativity.\index{super!symmetry}
This allows us to carry over several useful aspects of
supersymmetric field theories to non-anticommutative ones.

\section{Epistemological remarks}

String theory is certainly the physical theory which evokes the\index{string theory}
strongest emotions among professional scientists. On the one hand,
there are the advocates of string theory, never tired of stressing\index{string theory}
its incredible inherent beauty and the deep mathematical results
arising from it. On the other hand, there are strong critics, who
point out that so far, string theory had not made any useful\index{string theory}
predictions\footnote{It is doubtful that these critics would
accept the exception of twistor string theory, which led to new\index{string theory}\index{twistor}\index{twistor!string theory}
ways of calculating certain gauge theory amplitudes.} and that the
whole endeavor had essentially been a waste of money and brain
power, which had better been spent on more down-to-earth
questions. For this reason, let us briefly comment on string
theory from an epistemological point of view.\index{string theory}

The epistemological model used implicitly by today's physics
community is a mixture of rationalism and empiricism as both
doctrines by themselves have proven to be insufficient in the
history of natural sciences. The most popular version of such a
mixture is certainly Popper's {\em critical rationalism}
\cite{Popper:1979}, which is based on the observation that no
finite number of experiments can verify a scientific theory but a
single negative outcome can falsify it. For the following
discussion we will adopt this point of view.

Thus, we assume that there is a certain pool of theories, which
are in an evolutionary competition with each other. A theory is
permanently excluded from the pool if one of its predictions
contradicts an experimental result. Theories can be added to this
pool if they have an equal or better predictive power as any other
member of this pool. Note that the way these models are created is
-- contrary to many other authors -- of no interest to Popper.
However, we have to restrict the set of possible theories, which
we are admitting in the pool: only those, which can be
experimentally falsified are empirical and thus of direct
scientific value; all other theories are
metaphysical\footnote{Contrary to the logical positivism, Popper
attributes some meaning to such theories in the process of
developing new theories.}. One can therefore state that when Pauli
postulated the existence of the neutrino which he thought to be
undetectable, he introduced a metaphysical theory to the pool of
competitors and he was aware that this was a rather inappropriate
thing to do. Luckily, the postulate of the existence of the
neutrino became an empirical statement with the discovery of
further elementary forces and the particle was finally discovered
in 1956. Here, we have therefore the interesting example of a
metaphysical theory, which became an empirical one with improved
experimental capabilities.

In Popper's epistemological model, there is furthermore the class
of self-immunizing theories. These are theories, which constantly
modify themselves to fit new experimental results and therefore
come with a mechanism for avoiding being falsified. According to
Popper, these theories have to be discarded altogether. He applied
this reasoning in particular to dogmatic political concepts like
e.g.\ Marxism and Plato's idea of the perfect state. At first
sight, one might count supersymmetry to such self-immunizing\index{super!symmetry}
theories: so far, all predictions for the masses of the
superpartners of the particles in the standard model were
falsified which resulted in successive shifts of the postulated
supersymmetry breaking scales out of the reach of the then\index{super!symmetry}
up-to-date experiments. Besides self-immunizing, the theory even
becomes ``temporarily metaphysical'' in this way. However, one has
to take into account that it is not supersymmetry per se which is\index{super!symmetry}
falsified, but the symmetry breaking mechanisms it can come with.
The variety of such imaginable breaking mechanisms remains,
however, a serious problem.

When trying to put string theory in the context of the above\index{string theory}
discussed framework, there is clearly the observation that so far,
string theory has not made any predictions which would allow for a\index{string theory}
falsification. At the moment, it is therefore at most a
``temporarily metaphysical'' theory. Although it is reasonable to
expect that with growing knowledge of cosmology and string theory\index{string theory}
itself, many predictions of string theory will eventually become
empirical, we cannot compare its status to the one of the neutrino
at the time of its postulation by Pauli, simply for the reason
that string theory is not an actually fully developed theory. So\index{string theory}
far, it appears more or less as a huge collection of related and
interwoven ideas\footnote{For convenience sake, we will label this
collection of ideas by {\em string theory}, even though this\index{string theory}
nomenclature is clearly sloppy.} which contain strong hints of
being capable of explaining both the standard model and General
Relativity on equal footing. But without any doubt, there are many
pieces still missing for giving a coherent picture; a background
independent formulation -- the favorite point brought regularly
forth by advocates of loop quantum gravity -- is only one of the
most prominent ones.

The situation string theory is in can therefore be summarized in\index{string theory}
two points. First, we are clearly just in the process of
developing the theory; it should not yet be officially added to
our competitive pool of theories. For the development of string
theory, it is both necessary and scientifically sound to use\index{string theory}
metaphysical guidelines as e.g.\ beauty, consistency, mathematical
fertility and effectiveness in describing the physics of the
standard model and General Relativity. Second, it is desirable to
make string theory vulnerable to falsification by finding\index{string theory}
essential features of all reasonable string theories.
Epistemologically, this is certainly the most important task and,
if successful, would finally turn string theory into something\index{string theory}
worthy of being called a fully physical theory.

Let us end these considerations with an extraordinarily optimistic
thought: It could also be possible that there is only one unique
theory, which is consistent with all we know so far about the
world. If this were true, we could immediately abandon most of the
epistemological considerations made so far and turn to a purely
rationalistic point of view based on our preliminary results about
nature so far. That is, theories in our pool would no longer be
excluded from the pool by experimental falsification but by
proving their mathematical or logical inconsistency with the need
of describing the standard model and General Relativity in certain
limits. This point of view is certainly very appealing. However,
even if our unreasonably optimistic assumption was true, we might
not be able to make any progress without the help of further
experimental input.

Moreover, a strong opposition is forming against this idea, which
includes surprisingly many well-known senior scientists as e.g.\
Leonard Susskind \cite{Susskind:2003kw} and Steven Weinberg
\cite{Weinberg:2005fh}. In their approach towards the fundamental
principles of physics, which is known as {\em the landscape}, the
universe is divided into a statistical ensemble of sub-universes,
each with its own set of string compactification parameters and
thus its own low-energy effective field theory. Together with the
{\em anthropic principle}\/\footnote{Observers exist only in
universes which are suitable for creating and sustaining them.},
this might explain why our universe actually is as it is. Clearly,
the danger of such a concept is that questions which might in fact
be answerable by physical principles can easily be discarded as
irrelevant due to anthropic reasoning.

\section{Outline}

In this thesis, the material is presented in groups of subjects,
and it has been mostly ordered in such a way that technical terms
are not used before a definition is given. This, however, will
sometimes lead to a considerable amount of material placed between
the introduction of a concept and its first use. By adding as many
cross-references as possible, an attempt is made to compensate for
this fact.

Definitions and conventions which are not introduced in the body\index{body}
of the text, but might nevertheless prove to be helpful, are
collected in appendix A.

The thesis starts with an overview of the necessary concepts in
complex geometry. Besides the various examples of certain complex
manifolds as e.g.\ flag manifolds and Calabi-Yau spaces, in\index{Calabi-Yau}\index{complex!manifold}\index{flag manifold}\index{manifold}
particular the discussion of holomorphic vector bundles and their\index{holomorphic!vector bundle}
description in terms of Dolbeault and \v{C}ech cohomologies is
important.

It follows a discussion about basic issues in supergeometry. After\index{super!geometry}
briefly reviewing supersymmetry, which is roughly speaking the\index{super!symmetry}
physicist's name for a $\RZ_2$-grading, an overview of the various\index{Z2-grading@$\RZ_2$-grading}
approaches to superspaces is given. Moreover, the new results\index{super!space}
obtained in \cite{Saemann:2004tt} on exotic supermanifolds are\index{exotic!supermanifold}\index{super!manifold}
presented here. These spaces are supermanifolds endowed with
additional even nilpotent directions. We review the existing
approaches for describing such manifolds and introduce an\index{manifold}
integration operation on a certain class of them, the so-called
thickened and fattened complex manifolds. We furthermore examine\index{complex!manifold}\index{fattened complex manifold}\index{manifold}
the validity of Yau's theorem for such exotic Calabi-Yau\index{Calabi-Yau}\index{Theorem!Yau}
supermanifolds, and we find, after introducing the necessary\index{super!manifold}
tools, that the results fit nicely into the picture of ordinary
Calabi-Yau supermanifolds which was presented in\index{Calabi-Yau}\index{Calabi-Yau supermanifold}\index{super!manifold}
\cite{Rocek:2004bi}. We close the chapter with a discussion of
spinors in arbitrary dimensions during which we also fix all the\index{Spinor}
necessary reality conditions used throughout this thesis.

The next chapter deals with the various field theories which are
vital for the further discussion. It starts by recalling
elementary facts on supersymmetric field theories, in particular
their quantum aspects as e.g.\ non-renormalization theorems. It\index{non-renormalization theorems}\index{Theorem!non-renormalization}
follows a discussion of super Yang-Mills theories in various
dimensions and their related theories as chiral or self-dual
subsectors and dimensional reductions thereof. The second group of\index{dimensional reduction}
field theories that will appear in the later discussion are
Chern-Simons-type theories (holomorphic Chern-Simons theory and\index{Chern-Simons theory}
holomorphic BF-theories), which are introduced as well.
Eventually, a few remarks are made about certain aspects of
conformal field theories which will prove useful in what follows.

The aspects of string theory entering into this thesis are\index{string theory}
introduced in the following chapter. We give a short review on
string theory basics and superstring theories before elaborating\index{string theory}
on topological string theories. One of the latter, the topological\index{topological!string}
B-model, will receive much attention later due to its intimate
connection with holomorphic Chern-Simons theory. We will\index{Chern-Simons theory}\index{connection}
furthermore need some background information on the various types
of D-branes which will appear naturally in the models on which we\index{D-brane}
will focus our attention. We close this chapter with a few rather
general remarks on several topics in string theory.\index{string theory}

Noncommutative deformations of spacetime and the properties of
field theories defined on these spaces is the topic of the next
chapter. After a short introduction, we present the result of
\cite{Saemann:2004cf}, i.e.\ the non-anticommutative deformation
of $\CN=4$ super Yang-Mills equations using an equivalent set of
constraint equations on the superspace $\FR^{4|16}$. The second\index{constraint equations}\index{super!space}
half of this chapter is based on the publication
\cite{Ihl:2005zd}, in which the analysis of
\cite{Chaichian:2004za} on a Lorentz invariant interpretation of
noncommutative spacetime was extended to the non-anticommutative\index{noncommutative spacetime}
situation. This Drinfeld-twisted supersymmetry allows for carrying\index{Twisted supersymmetry}\index{super!symmetry}
over various quantum aspects of supersymmetric field theories to
the non-anticommutative situation.

The following chapter on twistor geometry constitutes the main\index{twistor}
part of this thesis. After a detailed introduction to twistor
geometry, integrability and the Penrose-Ward transform, we present\index{Penrose-Ward transform}\index{integrability}\index{twistor}
in four sections the results of the publications
\cite{Popov:2004rb,Saemann:2004tt,Popov:2005uv,Saemann:2005ji}.

First, the Penrose-Ward transform using supertwistor spaces is\index{Penrose-Ward transform}\index{twistor}\index{twistor!space}
discussed in complete detail, which gives rise to an equivalence
between the topological B-model and thus holomorphic Chern-Simons\index{topological!B-model}
theory on the supertwistor space $\CPP^{3|4}$ and $\CN=4$\index{twistor}\index{twistor!space}
self-dual Yang-Mills theory. While Witten \cite{Witten:2003nn} has\index{Yang-Mills theory}\index{self-dual Yang-Mills theory}
motivated this equivalence by looking at the field equations of
these two theories on the linearized level, the publication
\cite{Popov:2004rb} analyzes the complete situation to all orders
in the fields. We furthermore scrutinize the effects of the
different reality conditions which can be imposed on the
supertwistor spaces.\index{twistor}\index{twistor!space}

This discussion is then carried over to certain exotic
supermanifolds, which are simultaneously Calabi-Yau\index{Calabi-Yau}\index{exotic!supermanifold}\index{super!manifold}
supermanifolds. We report here on the results of
\cite{Saemann:2004tt}, where the possibility of using exotic
supermanifolds as a target space for the topological B-model was\index{exotic!supermanifold}\index{super!manifold}\index{target space}\index{topological!B-model}
examined. After restricting the structure sheaf of $\CPP^{3|4}$ by\index{sheaf}\index{structure sheaf}
combining an even number of Gra{\ss}mann-odd coordinates into
Gra{\ss}mann-even but nilpotent ones, we arrive at Calabi-Yau\index{Calabi-Yau}
supermanifolds, which allow for a twistor correspondence with\index{super!manifold}\index{twistor}\index{twistor!correspondence}
further spaces having $\FR^4$ as their bodies. Also a Penrose-Ward
transform is found, which relates holomorphic vector bundles over\index{Penrose-Ward transform}\index{holomorphic!vector bundle}
the exotic Calabi-Yau supermanifolds to solutions of bosonic\index{Calabi-Yau}\index{Calabi-Yau supermanifold}\index{exotic!Calabi-Yau supermanifold}\index{super!manifold}
subsectors of $\CN=4$ self-dual Yang-Mills theory.\index{Yang-Mills theory}\index{self-dual Yang-Mills theory}

Subsequently, the twistor correspondence as well as the\index{twistor}\index{twistor!correspondence}
Penrose-Ward transform are presented for the case of the\index{Penrose-Ward transform}
mini-supertwistor space, a dimensional reduction of the $\CN=4$\index{dimensional reduction}\index{mini-supertwistor space}\index{twistor}\index{twistor!space}
supertwistor space discussed previously.  This variant of the
supertwistor space $\CPP^{3|4}$ has been introduced in
\cite{Chiou:2005jn}, where it has been shown that twistor string
theory with the mini-supertwistor space as a target space is\index{mini-supertwistor space}\index{string theory}\index{target space}\index{twistor!space}\index{twistor!string theory}
equivalent to $\CN=8$ super Yang-Mills theory in three dimensions.\index{Yang-Mills theory}
Following Witten \cite{Witten:2003nn}, D1-instantons were added\index{instanton}
here to the topological B-model in order to complete the arising\index{topological!B-model}
BPS equations to the full super Yang-Mills theory. Here, we\index{Yang-Mills theory}
consider the geometric and field theoretic aspects of the same
situation {\em without} the D1-branes as done in
\cite{Popov:2005uv}. We identify the arising dimensional reduction\index{dimensional reduction}
of holomorphic Chern-Simons theory with a holomorphic BF-type\index{Chern-Simons theory}
theory and describe a twistor correspondence between the\index{twistor}\index{twistor!correspondence}
mini-supertwistor space and its moduli space of sections.\index{mini-supertwistor space}\index{moduli space}\index{twistor!space}
Furthermore, we establish a Penrose-Ward transform between this\index{Penrose-Ward transform}
holomorphic BF-theory and a super Bogomolny model on $\FR^3$. The\index{Bogomolny model}\index{holomorphic BF-theory}\index{super!Bogomolny model}
connecting link in this correspondence is a partially holomorphic
Chern-Simons theory on a Cauchy-Riemann supermanifold which is a\index{Chern-Simons theory}\index{super!manifold}
real one-dimensional fibration over the mini-supertwistor space.\index{fibration}\index{mini-supertwistor space}\index{twistor}\index{twistor!space}

While the supertwistor spaces examined so far naturally yield\index{twistor}\index{twistor!space}
Penrose-Ward transforms for certain self-dual subsectors of super\index{Penrose-Ward transform}
Yang-Mills theories, the superambitwistor space $\CL^{5|6}$\index{ambitwistor space}\index{twistor}\index{twistor!ambitwistor}\index{twistor!space}
introduced in the following section as a quadric in\index{quadric}
$\CPP^{3|3}\times \CPP^{3|3}$ yields an analogue equivalence
between holomorphic Chern-Simons theory on $\CL^{5|6}$ and full\index{Chern-Simons theory}
$\CN=4$ super Yang-Mills theory. After developing this picture to\index{N=4 super Yang-Mills theory@$\CN=4$ super Yang-Mills theory}\index{Yang-Mills theory}
its full extend as given in \cite{Popov:2004rb}, we moreover
discuss in detail the geometry of the corresponding dimensional
reduction yielding the mini-superambitwistor space $\CL^{4|6}$.\index{ambitwistor space}\index{dimensional reduction}\index{mini-superambitwistor space}\index{twistor}\index{twistor!ambitwistor}\index{twistor!space}

The Penrose-Ward transform built upon the space $\CL^{4|6}$ yields\index{Penrose-Ward transform}
solutions to the $\CN=8$ super Yang-Mills equations in three
dimensions as was shown in \cite{Saemann:2005ji}. We review the
construction of this new supertwistor space by dimensional\index{twistor}\index{twistor!space}
reduction of the superambitwistor space $\CL^{5|6}$ and note that\index{ambitwistor space}\index{twistor!ambitwistor}
the geometry of the mini-superambitwistor space comes with some\index{mini-superambitwistor space}
surprises. First, this space is not a manifold, but only a\index{manifold}
fibration. Nevertheless, it satisfies an analogue to the\index{fibration}
Calabi-Yau condition and therefore might be suited as a target\index{Calabi-Yau}
space for the topological B-model. We conjecture that this space\index{topological!B-model}
is the mirror to a certain mini-supertwistor space. Despite the\index{mini-supertwistor space}\index{twistor}\index{twistor!space}
strange geometry of the mini-superambitwistor space, one can\index{ambitwistor space}\index{mini-superambitwistor space}\index{twistor!ambitwistor}
translate all ingredients of the Penrose-Ward transform to this\index{Penrose-Ward transform}
situation and establish a one-to-one correspondence between
generalized holomorphic bundles over the mini-superambitwistor\index{twistor}\index{twistor!ambitwistor}
space and solutions to the $\CN=8$ super Yang-Mills equations in
three dimensions. Also the truncation to the Yang-Mills-Higgs
subsector can be conveniently described by generalized holomorphic
bundles over formal {\em sub-neighborhoods} of the
mini-ambitwistor space.\index{ambitwistor space}\index{twistor}\index{twistor!ambitwistor}\index{twistor!space}

We close this chapter with a presentation of the ADHM and the Nahm
constructions, which are intimately related to twistor geometry\index{twistor}
and which will allow us to identify certain field theories with
D-brane configurations in the following.\index{D-brane}

The next to last chapter is devoted to matrix models. We briefly\index{matrix model}
recall basic aspects of the most prominent matrix models and
introduce the new models, which were studied in
\cite{Lechtenfeld:2005xi}. In this paper, we construct two matrix
models from twistor string theory: one by dimensional reduction\index{dimensional reduction}\index{matrix model}\index{string theory}\index{twistor}\index{twistor!string theory}
onto a rational curve and another one by introducing
noncommutative coordinates on the fibres of the supertwistor space\index{twistor}\index{twistor!space}
$\CP^{3|4}\rightarrow \CPP^1$. Examining the resulting actions, we
note that we can relate our matrix models to a recently proposed\index{matrix model}
string field theory. Furthermore, we comment on their physical\index{string field theory}
interpretation in terms of D-branes of type IIB, critical $\CN=2$\index{D-brane}
and topological string theory. By extending one of the models, we\index{string theory}\index{topological!string}
can carry over all the ingredients of the super ADHM construction\index{ADHM construction}
to a D-brane configuration in the supertwistor space $\CP^{3|4}$\index{D-brane}\index{twistor}\index{twistor!space}
and establish a correspondence between a D-brane system in ten
dimensional string theory and a topological D-brane system. The\index{string theory}
analogous correspondence for the Nahm construction is also
established.

After concluding in the last chapter, we elaborate on the
remaining open questions raised by the results presented in this
thesis and mention several directions for future research.

\section{Publications}

During my PhD-studies, I was involved in the following
publications:

\begin{enumerate}
\item[1.] C.~S{\"a}mann and M.~Wolf,
{\em Constraint and super Yang-Mills equations on the deformed
superspace $\FR_\hbar^{(4|16)}$,} JHEP {\bf 0403} (2004) 048\index{super!space}
[hep-th/0401147].
\item[2.] A.~D.~Popov and C.~S{\"a}mann, {\em On supertwistors, the\index{twistor}
Penrose-Ward transform and N = 4 super Yang-Mills theory,}\index{Penrose-Ward transform}\index{Yang-Mills theory} Adv.\
Theor.\ Math.\ Phys.\  {\bf 9} (2005) 931 [hep-th/0405123].
\item[3.] C.~S\"{a}mann,
{\em The topological B-model on fattened complex manifolds and\index{complex!manifold}\index{fattened complex manifold}\index{topological!B-model}\index{manifold}
subsectors of $\CN=4$ self-dual Yang-Mills theory,} JHEP {\bf0501}\index{Yang-Mills theory}\index{self-dual Yang-Mills theory}
(2005) 042 [hep-th/0410292].
\item[4.] A.~D.~Popov, C.~S{\"a}mann and M.~Wolf,
{\em The topological B-model on a mini-supertwis\-tor space and\index{topological!B-model}
supersymmetric Bogomolny monopole equations,} JHEP {\bf 0510}
(2005) 058 [hep-th/0505161].
\item[5.] M.~Ihl and C.~S{\"a}mann, {\em Drinfeld-twisted supersymmetry and\index{Twisted supersymmetry}\index{super!symmetry}
non-anticommutative superspace,} JHEP {\bf 0601} (2006) 065\index{non-anticommutative superspace}\index{super!space}
[hep-th/0506057].
\item[6.] C.~S{\"a}mann,
{\em On the mini-superambitwistor space and $\CN=8$ super\index{ambitwistor space}\index{mini-superambitwistor space}\index{twistor}\index{twistor!ambitwistor}\index{twistor!space}
Yang-Mills theory,} hep-th/0508137.\index{Yang-Mills theory}
\item[7.] O.~Lechtenfeld and C.~S{\"a}mann, {\em Matrix
models and D-branes in twistor string theory}, JHEP {\bf 0603}\index{D-brane}\index{matrix model}\index{string theory}\index{twistor}\index{twistor!string theory}
(2006) 002 [hep-th/0511130].
\end{enumerate}

\chapter{Complex Geometry}

In this chapter, we review the basic notions of complex geometry,
which will be heavily used throughout this thesis due to the
intimate connection of this subject with supersymmetry and the\index{connection}\index{super!symmetry}
topological B-model. The following literature has proven to be\index{topological!B-model}
useful for studying this subject: \cite{Nakahara,Hwang} (complex
geometry), \cite{Joyce:2001,Greene:1996cy,Scheidegger} (Calabi-Yau\index{Calabi-Yau}
geometry), \cite{Popov:1998fb,Ivanova:2000xr} (Dolbeault- and
\v{C}ech-description of holomorphic vector bundles),\index{holomorphic!vector bundle}
\cite{Burns,Manetti} (deformation theory),\index{deformation theory}
\cite{GriffithsHarris,Hartshorne} (algebraic geometry).

\section{Complex manifolds}\label{sComplexManifolds}\index{complex!manifold}\index{manifold}

\subsection{Manifolds}\label{ssManifolds}

Similarly to the structural richness one gains when turning from
real analysis to complex analysis, there are many new features
arising when turning from real (and smooth) to complex manifolds.\index{complex!manifold}\index{manifold}
For this, the requirement of having smooth transition functions\index{transition function}
between patches will have to be replaced by demanding that the
transition functions are holomorphic.\index{transition function}

\paragraph{Holomorphic maps.} A map $f:\FC^m\rightarrow\index{holomorphic!map}
\FC^n:(z^1,\ldots ,z^m)\mapsto(w^1,\ldots ,w^n)$ is called {\em
holomorphic} if all the $w^i$ are holomorphic in each of the
coordinates $z^j$, where $1\leq i \leq n$ and $1\leq j \leq m$.

\paragraph{Complex manifolds.}\label{pComplexManifolds} Let $M$ be a topological space with\index{complex!manifold}\index{manifold}
an open covering $\frU$. Then $M$ is called a {\em complex
manifold} of {\em dimension $n$} if for every $U\in\frU$ there is\index{complex!manifold}\index{manifold}
a homeomorphism\footnote{i.e.\ $\phi_U$ is bijective and $\phi_U$
and $\phi^{-1}_U$ are continuous} $\phi_U: U\rightarrow \FC^n$
such that for each $U\cap V\neq \varnothing$ the {\em transition
function} $\phi_{UV}:=\phi_U\phi^{-1}_V$, which maps open subsets\index{transition function}
of $\FC^n$ to $\FC^n$, is holomorphic. A pair $(U,\phi_U)$ is
called a chart and the collection of all charts form a {\em
holomorphic structure}.\index{holomorphic!structure}

\paragraph{Gra{\ss}mannian manifolds.}\label{pGrassmannManifolds} An\index{manifold}
ubiquitous example of complex manifolds are {\em Gra{\ss}mannian\index{complex!manifold}
manifolds}. Such manifolds $G_{k,n}(\FC)$ are defined as the space
of $k$-dimensional vector subspaces in $\FC^n$. The most common
example is $G_{1,n}$ which is the {\em complex projective space}\index{complex!projective space}
$\CPP^n$. This space is globally described by {\em homogeneous
coordinates}\index{homogeneous coordinates}
$(\omega^1,\ldots ,\omega^{n+1})\in\FC^n\backslash\{(0,\ldots ,0)\}$
together with the identification
$(\omega^1,\ldots ,\omega^{n+1})\sim(t\omega^1,\ldots ,t\omega^{n+1})$ for
all $t\in \FC^\times$. An open covering of $\CPP^n$ is given by
the collection of open patches $U_j$ for which $\omega^j\neq 0$.
On such a patch $U_j$, we can introduce $n$ {\em inhomogeneous
coordinates} $(z^1,\ldots ,\hat{z}^j,\ldots ,z^{n+1})$ with\index{homogeneous coordinates}\index{inhomogeneous coordinates}
$z^i=\frac{\omega^i}{\omega^j}$, where the hat indicates an
omission. For convenience, we will always shift the indices on the
right of the omission to fill the gap, i.e.\ $z^i\rightarrow
z^{i-1}$ for $i>j$.

\paragraph{Theorem.} (Chow) Since we will often use complex projective\index{Theorem!Chow}
spaces and their subspaces, let us recall the following theorem by
Chow: Any submanifold of $\CPP^m$ can be defined by the zero locus\index{Theorem!Chow}
of a finite number of homogeneous polynomials. Note that the zero
locus of a set of polynomials is in general not a manifold (due to\index{manifold}
singularities), but an {\em algebraic variety}.\index{algebraic variety}

\paragraph{Flag manifolds.}\label{pFlagManifolds} Complex flag manifolds\index{flag manifold}\index{manifold}
are a major tool in the context of twistor geometry and the\index{twistor}
Penrose-Ward correspondence, cf. chapter \ref{chTwistorGeometry}.
They can be considered as generalizations of projective spaces and
Gra{\ss}mann manifolds. An $r$-tuple $(L_1,\ldots ,L_r)$ of vector spaces\index{manifold}
of dimensions $\dim_\FC L_i=d_i$ with $L_1\subset\ldots \subset
L_r\subset\FC^n$ and $0<d_1<\ldots <d_r<n$ is called a {\em flag} in
$\FC^n$. The {\em (complex) flag manifold} $F_{d_1\ldots d_r,n}$ is\index{flag manifold}\index{manifold}
the compact space
\begin{equation}
F_{d_1\ldots d_r,n}\ :=\ \{\mbox{all flags }(L_1,\ldots ,L_r)\mbox{ with
}\dim_\FC L_i\ =\ d_i,~i\ =\ 1,\ldots ,r\,\}~.
\end{equation}
Simple examples of flag manifolds are $F_{1,n}=\CPP^{n-1}$ and\index{flag manifold}\index{manifold}
$F_{k,n}=G_{k,n}(\FC)$. The flag manifold $F_{d_1\ldots d_r,n}$ can
also be written as the coset space
\begin{equation}
F_{d_1\ldots d_r,n}\ =\ \frac{\sU(n)}{\sU(n-d_r)\times\ldots \times
\sU(d_2-d_1)\times\sU(d_1)}~,
\end{equation}
and therefore its dimension is
\begin{equation}
\dim_\FC
F_{d_1\ldots d_r,n}\ =\ d_1(n-d_1)+(d_2-d_1)(n-d_2)+\ldots +(d_r-d_{r-1})(n-d_r)~.
\end{equation}

\paragraph{Weighted projective spaces.} A further generalization of\index{weighted projective spaces}
complex projective spaces are spaces which are obtained from\index{complex!projective space}
$(\FC^{m+1})\backslash\{0\}$ with coordinates $(z^i)$ by the
identification $(z^1,z^2,\ldots ,z^{m+1})\sim(t^{q_1}
z^1,t^{q_2}z^2,\ldots ,t^{q_{m+1}}z^{m+1})$ with $t\in\FC^\times$.
These spaces are called {\em weighted projective spaces} and\index{weighted projective spaces}
denoted by $W\CPP^{m}(q_1,\ldots ,q_{m+1})$. Note that
$W\CPP^m(1,\ldots ,1)=\CPP^m$.

A subtlety when working with weighted projective spaces is the\index{weighted projective spaces}
fact that they may not be smooth but can have non-trivial fixed
points under the coordinate identification, which lead to
singularities. Therefore, these spaces are mostly used as
embedding spaces for smooth manifolds.\index{manifold}

\paragraph{Stein manifolds.} A complex manifold that can be\index{Stein manifold}\index{complex!manifold}\index{manifold}
embedded as a closed submanifold into a complex Euclidean space is
called a {\em Stein manifold}. Such manifolds play an important\index{Stein manifold}\index{manifold}
r{\^o}le in making \v{C}ech cohomology sets on a manifold independent\index{Cech cohomology@\v{C}ech cohomology}
of the covering, see section \ref{subseccohomology},
\ref{pCechcohomologysets}.

\paragraph{Equivalence of manifolds.} Two complex manifolds\index{complex!manifold}\index{manifold}
$M$ and $N$ are {\em biholomorphic} if there is a biholomorphic\index{biholomorphic}
map\footnote{a holomorphic map with a holomorphic inverse}\index{holomorphic!map}
$m:M\rightarrow N$. This is equivalent to the fact that there is
an identical cover $\frU$ of $M$ and $N$ and that there are
biholomorphic functions $h_a$ on each patch $U_a\in\frU$ such that\index{biholomorphic}
we have the following relation between the transition functions:\index{transition function}
$f^M_{ab}=h_a^{-1}\circ f^N_{ab}\circ h_b$ on $U_a\cap U_b\neq
\varnothing$. Two complex manifolds are called {\em diffeomorphic}\index{complex!manifold}\index{diffeomorphic}\index{manifold}
if their underlying smooth manifolds are diffeomorphic. The
transition functions of two diffeomorphic manifolds on an\index{transition function}
identical cover $\frU$ are related by $f^M_{ab}=s_a^{-1}\circ
f^N_{ab}\circ s_b$ on nonempty intersections $U_a\cap U_b\neq
\varnothing$, where the $s_a$ are smooth functions on the patches
$U_a$.

We call complex manifolds {\em smoothly equivalent} if they are\index{complex!manifold}\index{smoothly equivalent}\index{manifold}
diffeomorphic and {\em holomorphically equivalent} if they are\index{diffeomorphic}\index{holomorphically equivalent}
biholomorphic. In one dimension, holomorphic equivalence implies\index{biholomorphic}
conformal equivalence, cf.\ section \ref{ssCFTbasics}.

\paragraph{Functions on manifolds.} Given a manifold $M$, we will\index{manifold}
denote the set of functions $\{f:M\rightarrow \FC\}$ on $M$ by
$\CCF(M)$. Smooth functions will be denoted by $C^\infty(M)$ and
holomorphic functions by $\CO(M)$.

\subsection{Complex structures}\index{complex!structure}

It is quite obvious that many real manifolds of even dimension\index{manifold}
might also be considered as complex manifolds after a change of\index{complex!manifold}
variables. The tool for making this statement exact is a complex
structure.\index{complex!structure}

\paragraph{Modules and vector spaces.}\index{modules}
A {\em left module} over a ring $\Lambda$ (or an
$\Lambda$-left-module) is an Abelian group $G$ together with an
operation $(\lambda\in\Lambda,a\in G)\mapsto\lambda a\in G$, which
is linear in both components. Furthermore, we demand that this
operation is associative, i.e.\ $(\lambda\mu)a=\lambda(\mu a)$ and
normalized according to $\unit_\Lambda a=a$.

Analogously, one defines a {\em right module} with right
multiplication and that of a {\em bimodule} with simultaneously
defined, commuting left and right multiplication.

A {\em vector space} is a module over a field and in particular, a
complex vector space is a module over $\FC$. Later on, we will
encounter supervector spaces which are modules over $\RZ_2$-graded\index{modules}\index{super!vector space}
rings, cf.\ \ref{ssSuperspaces}, \ref{pRmn}.

\paragraph{Complex structures.} Given a real vector space $V$, a\index{complex!structure}
{\em complex structure} on $V$ is a map $I:V\rightarrow V$ with
$I^2=-\unit_V$. This requires the vector space to have even
dimensions and is furthermore to be seen as a generalization of
$\di^2=-1$. After defining the scalar multiplication of a complex
number $(a+\di b)\in \FC$ with a vector $v\in V$ as $(a+\di b)v:=a
v+b Iv$, $V$ is a complex vector space. On the other hand, each
complex vector space has a complex structure given by $Iv=\di v$.\index{complex!structure}

\paragraph{Canonical complex structure.} The obvious\index{canonical complex structure}\index{complex!structure}
identification of $\FC^n$ with $\FR^{2n}$ is obtained by equating
$z^i=x^i+\di y^i$, which induces the {\em canonical complex
structure}\index{canonical complex structure}\index{complex!structure}
\begin{equation}
\begin{aligned}
I(x^1,\ldots ,x^n,y^1,\ldots ,y^n)&\ =\ (-y^1,\ldots ,-y^n,x^1,\ldots ,x^n)~,\\\mbox{and
thus }~~~I&\ =\ \left(\begin{array}{cc} 0 & -\unit_n \\ \unit_n & 0
\end{array}\right)~.
\end{aligned}
\end{equation}

\paragraph{Almost complex structure.} Given a real differentiable\index{almost complex structure}\index{complex!structure}
manifold $M$ of dimension $2n$, an {\em almost complex structure}\index{manifold}
is a smooth tensor field $I$ of type (1,1) on each patch of $M$,
such that at each point $x\in M$, $I_x$ is a complex structure on\index{complex!structure}
$T_xM$. The pair $(M,I)$ is called an {\em almost complex
manifold}. Note that each real manifold with even dimension\index{almost complex manifold}\index{complex!manifold}\index{manifold}
locally admits such a tensor field, and the equations
$I_b{}^a\der{x^a} f=\di \der{x^b} f$ are just the Cauchy-Riemann
equations. Thus, holomorphic maps $f:\FC^n\supset U\rightarrow\index{holomorphic!map}
\FC^m$ are exactly those which preserve the almost complex
structure.\index{almost complex structure}\index{complex!structure}

\paragraph{Complexification.} Given a real space $S$ with a real\index{complexification}
scalar multiplication $\cdot:\FR\times S\rightarrow S$, we define
its {\em complexification} as the tensor product $S^c=S\otimes_\FR\index{complexification}
\FC$. We will encounter an example in the following paragraph.

\paragraph{Holomorphic vector fields.}\label{pholomorphicvectorfields}\index{holomorphic!vector fields}
Consider the complexification of the tangent space\index{complexification}
$TM^c=TM\otimes_\FR\FC$. This space decomposes at each point $x$
into the direct sum of eigenvectors of $I$ with eigenvalues $+\di$
and $-\di$, which we denote by $T^{1,0}_xM$ and $T^{0,1}_xM$,
respectively, and therefore we have $TM^c=T^{1,0}M\oplus
T^{0,1}M$. Sections of $T^{1,0}M$ and $T^{0,1}M$ are called {\em
vector fields of type $(1,0)$} and $(0,1)$, respectively. Vector
fields of type $(1,0)$ whose action on arbitrary functions will be
holomorphic will be called {\em holomorphic vector fields} and\index{holomorphic!vector fields}
{\em antiholomorphic vector fields} are defined analogously. This
means in particular that a vector field $X$ given locally by
$X=\xi^i\der{z^i}$, where $(\der{z^1},\ldots ,\der{z^n})$ is a local
basis of $T^{1,0} M$, is a holomorphic vector field if the $\xi^i$
are holomorphic functions. We will denote the space of vector
fields on $M$ by $\CCX(M)$. The above basis is complemented by the
basis $(\der{\bz^1},\ldots ,\der{\bz^n})$ of $T^{0,1}M$ to a full
local basis of $TM^c$.

\paragraph{Integrable complex structures.} If an almost complex\index{complex!structure}\index{integrable}\index{integrable complex structure}
structure is induced from a holomorphic structure,\index{holomorphic!structure} cf.\
\ref{pComplexManifolds}, one calls this almost complex structure\index{almost complex structure}\index{complex!structure}
{\em integrable}. Thus, an almost complex manifold with an\index{almost complex manifold}\index{complex!manifold}\index{integrable}\index{manifold}
integrable complex structure is a complex manifold.\index{complex!structure}\index{integrable complex structure}

\paragraph{Newlander-Nirenberg theorem.} Let $(M,I)$ be an almost\index{Theorem!Newlander-Nirenberg}
complex manifold. Then the following statements are equivalent:\index{complex!manifold}\index{manifold}
\begin{enumerate}
\item The almost complex structure $I$ is integrable.\index{almost complex structure}\index{complex!structure}\index{integrable}
\item The {\em Nijenhuis tensor}\index{Nijenhuis tensor}
$N(X,Y)=\tfrac{1}{4}([X,Y]+I[X,IY]+I[IX,Y]-[IX,IY])$ (the torsion)
vanishes for arbitrary vector fields $X,Y\in \CCX(M)$.
\item The Lie bracket $[X,Y]$ closes in $T^{1,0}M$, i.e.\ for
$X,Y\in T^{1,0}M$, $[X,Y]\in T^{1,0}M$.
\end{enumerate}

\paragraph{Complex differential forms.} Analogously to complex\index{complex!differential forms}
tangent spaces, we introduce the space of complex differential
forms on a complex manifold $M$ as the complexification of the\index{complex!differential forms}\index{complex!manifold}\index{complexification}\index{manifold}
space of real differential forms:
$\Omega^{q}(M)^c:=\Omega^q(M)\otimes_\FR \FC$. Consider now a
$q$-form $\omega\in\Omega^{q}(M)^c$. If $\omega(V_1,\ldots ,V_q)=0$
unless $r$ of the $V_i$ are elements of $T^{1,0}M$ and $s=q-r$ of
them are elements of $T^{0,1}M$, we call $\omega$ a form of {\em
bidegree $(r,s)$}. We will denote the space of forms of bidegree\index{bidegree}
$(r,s)$ on $M$ by $\Omega^{r,s}(M)$. It is now quite obvious that
$\Omega^{q}$ (uniquely) splits into
$\bigoplus_{r+s=q}\Omega^{r,s}(M)$.

Clearly, elements of $\Omega^{1,0}$ and $\Omega^{0,1}$ are dual to
elements of $T^{1,0}M$ and $T^{0,1}M$, respectively. Local bases
for $\Omega^{1,0}$ and $\Omega^{0,1}$ dual to the ones given in
\ref{pholomorphicvectorfields} are then given by $(\dd z^1,\ldots ,\dd
z^n)$ and $(\dd \bz^1,\ldots ,\dd \bz^n)$ and satisfy the
orthogonality relations $\langle \dd
z^i,\der{z^j}\rangle=\delta^i_j$, $\langle \dd
\bz^\bi,\der{z^j}\rangle=\langle \dd z^i,\der{\bz^\bj}\rangle=0$
and $\langle \dd \bz^\bi,\der{\bz^\bj}\rangle=\delta^\bi_\bj$.

\paragraph{The exterior derivative.}\index{exterior derivative}
The {\em exterior derivative} $\dd$ maps a form of bidegree\index{bidegree}
$(r,s)$ to a form which is the sum of an $(r+1,s)$-form and an
$(r,s+1)$-form: Given an $(r,s)$ form $\omega$ on a complex
manifold $M$ by\index{complex!manifold}\index{manifold}
\begin{equation}
\omega\ =\ \frac{1}{r!s!}\,\omega_{i_1\ldots i_r\bi_{r+1}\ldots \bi_{r+s}} \dd
z^{i_1}\wedge\ldots \dd z^{i_r}\wedge\ldots \dd \bz^{i_{r+s}}~,
\end{equation}
we define
\begin{equation}\nonumber
\dd \omega\ =\ \frac{1}{r!s!}\left(\dpar_k\omega_{i_1\ldots \bi_{r+s}} \dd
z^k\wedge\dd z^{i_1}\wedge\ldots \dd
\bz^{i_{r+s}}+\dparb_{\bar{k}}\omega_{i_1\ldots \bi_{r+s}} \dd
z^{i_1}\wedge\ldots \dd \bz^{\bar{k}}\wedge\ldots \dd
\bz^{\bi_{r+s}}\right)~,
\end{equation}
which agrees with the definition of $\dd$ on $M$ interpreted as a
real manifolds. One therefore splits $\dd=\dpar+\dparb$, where\index{manifold}
$\dpar:\Omega^{r,s}(M)\rightarrow\Omega^{r+1,s}(M)$ and
$\dparb:\Omega^{r,s}(M)\rightarrow\Omega^{r,s+1}(M)$. The
operators $\dpar$ and $\dparb$ are called the {\em Dolbeault
operators}. A holomorphic $r$-form is given by an\index{Dolbeault operators}
$\omega\in\Omega^{r,0}(M)$ satisfying $\bar{\dpar}\omega=0$ and
holomorphic $0$-forms are holomorphic functions. The Dolbeault
operators are nilpotent, i.e.\ $\dpar^2=\dparb^2=0$, and therefore\index{Dolbeault operators}
one can construct the Dolbeault cohomology groups, see section\index{Dolbeault cohomology}
\ref{subseccohomology}.

\paragraph{Real structure.} A {\em real structure} $\tau$ on a\index{real structure}
complex vector space $V$ is an antilinear involution\index{antilinear involution}\index{involution}
$\tau:V\rightarrow V$. This implies that $\tau^2(v)=v$ and
$\tau(\lambda v)=\bl v$ for all $\lambda \in \FC$ and $v\in V$.
Therefore, a real structure maps a complex structure $I$ to $-I$.\index{complex!structure}\index{real structure}
One can use such a real structure to reduce a complex vector space
to a real vector subspace. A real structure on a complex manifold\index{complex!manifold}\index{manifold}
is a complex manifold with a real structure on its tangent spaces.
For an example, see the discussion in section
\ref{ssTwistorSpace}, \ref{pRealStructureP3}.

\subsection{Hermitian structures}\index{Hermitian structure}

\paragraph{Hermitian inner product.}\label{phermiteaninnerproduct}\index{Hermitian inner product}
Given a complex vector space $(V,I)$, a {\em Hermitian inner
product} is an inner product $g$ satisfying $g(X,Y)=g(I X,I Y)$\index{Hermitian inner product}
for all vectors $X,Y\in V$ (``$I$ is $g$-orthogonal''). Note that
every inner product $\tilde{g}$ can be turned into a Hermitian one
by defining $g=\frac{1}{2}(\tilde{g}(X,Y)+\tilde{g}(IX,IY))$. To
have an {\em almost Hermitian inner product} on an almost complex\index{Hermitian inner product}\index{almost Hermitian inner product}
manifold $M$, one smoothly defines a $g_x$ on $T_xM$ for every\index{manifold}
$x\in M$.

\paragraph{Hermitian structure.} Every Hermitian inner product $g$\index{Hermitian inner product}\index{Hermitian structure}
can be uniquely extended to a {\em Hermitian structure} $h$, which
is a map $h:V\times V\rightarrow \FC$ satisfying
\begin{enumerate}
\item $h(u,v)$ is $\FC$-linear in $v$ for every $u\in V$
\item $\overline{h(u,v)}=h(v,u)$ for all vectors $u,v\in V$.
\item $h(u,u)\geq 0$ for all vectors $u\in V$ and
$h(u,u)=0\Leftrightarrow u=0$.
\end{enumerate}
For Hermitian structures on an almost complex manifold $M$, we\index{Hermitian structure}\index{almost complex manifold}\index{complex!manifold}\index{manifold}
demand additionally that $h$ understood as a map
$h:\Gamma(TM)\times \Gamma(TM)\rightarrow \CCF(M)$ maps every pair
of smooth sections to smooth functions on $M$.

\paragraph{Hermitian metric.} When interpreting a smooth manifold\index{manifold}\index{metric}
$M$ as a complex manifold via an integrable almost complex\index{complex!manifold}\index{integrable}
structure, one can extend the Riemannian metric $g$ to a map\index{metric}
$\tilde{g}_x:T_xM^c\times T_xM^c\rightarrow \FC$ by
\begin{equation}
\tilde{g}_x:(X+\di Y,U+iV)\ \mapsto\ 
g_x(X,U)-g_x(Y,V)+\di(g_x(X,V)+g_x(Y,U))~.
\end{equation}
A metric obtained in this way and satisfying\index{metric}
$\tilde{g}_x(I_xX,I_xY)=\tilde{g}_x(X,Y)$ is called a {\em
Hermitian metric}. Given again bases $(\der{z^i})$ and\index{metric}
$(\der{\bz^i})$ spanning locally $T^{1,0}M$ and $T^{0,1}M$,
respectively, we have
\begin{equation}
g_{ij}\ =\ g_{\bi\bj}\ =\ 0~~~\mbox{and}~~~ g\ =\ g_{i\bj}\dd z^i\otimes\dd
\bz^\bj+g_{\bi j}\dd\bz^\bi\otimes\dd z^j
\end{equation}
for a Hermitian metric $g$. A complex manifold with a Hermitian\index{complex!manifold}\index{manifold}\index{metric}
metric is called a {\em Hermitian manifold}.\index{Hermitian manifold}

\paragraph{Theorem.} A complex manifold always admits a Hermitian\index{complex!manifold}\index{manifold}
metric. Given a Riemannian metric on a complex manifold, one\index{metric}
obtains a Hermitian metric e.g.\ by the construction described in
\ref{phermiteaninnerproduct}.

\paragraph{K\"{a}hler form.} Given a Hermitian manifold $(M,g)$, we define\index{Hermitian manifold}\index{K\"{a}hler!form}\index{manifold}
a tensor field $J$ of type $(1,1)$ by $J(X,Y)=g(IX,Y)$ for every
pair of sections $(X,Y)$ of $TM$. As
$J(X,Y)=g(IX,Y)=g(IIX,IY)=-g(IY,X)=-J(Y,X)$, the tensor field is
antisymmetric and defines a two-form, the {\em K\"{a}hler form} of the\index{K\"{a}hler!form}
Hermitian metric $g$. As easily seen, $J$ is invariant under the\index{metric}
action of $I$. Let $m$ be the complex dimension of $M$. One can
show that $\wedge^m J$ is a nowhere vanishing, real $2m$-form,
which can serve as a volume element and thus every Hermitian
manifold (and so also every complex manifold) is orientable.\index{Hermitian manifold}\index{complex!manifold}\index{manifold}

\paragraph{K\"{a}hler manifold.} A {\em K\"{a}hler manifold} is a\index{K\"{a}hler!manifold}\index{manifold}
Hermitian manifold $(M,g)$ on which one of the following three\index{Hermitian manifold}
equivalent conditions holds:
\begin{enumerate}
\item The K\"{a}hler form $J$ of $g$ satisfies $\dd J=0$.\index{K\"{a}hler!form}
\item The K\"{a}hler form $J$ of $g$ satisfies $\nabla J=0$.
\item The almost complex structure satisfies $\nabla I=0$,\index{almost complex structure}\index{complex!structure}
\end{enumerate}
where $\nabla$ is the Levi-Civita connection of $g$. The metric\index{Levi-Civita connection}\index{connection}\index{metric}
$g$ of a K\"{a}hler manifold is called a {\em K\"{a}hler metric}.\index{K\"{a}hler!manifold}\index{K\"{a}hler!metric}\index{manifold}

\paragraph{K\"{a}hler potential.} Given a K\"{a}hler manifold\index{K\"{a}hler!manifold}\index{K\"{a}hler!potential}\index{manifold}
$(M,g)$ with K\"{a}hler form $J$, it follows from $\dd\index{K\"{a}hler!form}
J=(\dpar+\dparb)\di g_{i\bj}\dd z^i\wedge \dd \bz^\bj=0$ that
\begin{equation}\label{potident}
\derr{g_{i\bj}}{z^l}\ =\ \derr{g_{l\bj}}{z^i}\eand
\derr{g_{i\bj}}{\bz^l}\ =\ \derr{g_{l\bj}}{\bz^i}~.
\end{equation}
Thus, we can define a local real function $\CCK$ such that
$g=\dpar\dparb \CCK$ and $J=\di \dpar\dparb \CCK$. This function
is called the {\em K\"{a}hler potential} of $g$. Conversely, if a\index{K\"{a}hler!potential}
metric is derived from a K\"{a}hler potential, it automatically\index{metric}
satisfies \eqref{potident}.

\paragraph{Examples.} A simple example is the K\"{a}hler metric on\index{K\"{a}hler!metric}\index{metric}
$\FC^m$ obtained from the K\"{a}hler potential $\CCK=\frac{1}{2}\sum\index{K\"{a}hler!potential}
z^i\bz^\bi$, which is the complex analog of $(\FR^{2m},\delta)$.
Also easily seen is the fact that any orientable complex manifold\index{complex!manifold}\index{manifold}
$M$ with $\dim_\FC M=1$ is K\"{a}hler: since $J$ is a real two-form,
$\dd J$ has to vanish on $M$. These manifolds are called {\em\index{manifold}
Riemann surfaces}. Furthermore, any complex submanifold of a\index{Riemann surface}
K\"{a}hler manifold is K\"{a}hler.\index{K\"{a}hler!manifold}\index{manifold}

An important example is the complex projective space $\CPP^n$,\index{complex!projective space}
which is also a K\"{a}hler manifold. In homogeneous coordinates\index{K\"{a}hler!manifold}\index{homogeneous coordinates}\index{manifold}
$(\omega^i)$ and inhomogeneous coordinates $(z^i)$ (see\index{inhomogeneous coordinates}
\ref{pGrassmannManifolds}), one can introduce a positive definite
function
\begin{equation}
\CCK_i\ =\ \sum_{j=1}^{n+1}\left|\frac{\omega^j}{\omega^i}\right|^2\ =\ \sum_{j=1}^n|z^j_i|^2+1
\end{equation}
on the patch $U_i$, which globally defines a closed two-form $J$
by $J:=\di \dpar\dparb\ln \CCK_j$, as one easily checks. From this
form, we obtain a metric by $g(X,Y):=J(X,IY)$, the {\em\index{metric}
Fubini-Study metric} of $\CPP^n$. In components, it reads on the\index{Fubini-Study metric}
patch $U_i$
\begin{equation}
g_i(X,Y)\ =\ 2\sum_{j,\bj}
\frac{\delta_{j\bj}\CCK_i-z_i^j\bz_i^\bj}{\CCK_i^2}X^j\bar{Y}^\bj~.
\end{equation}
Note that $S^2\cong \CPP^1$ is the only sphere which admits a
complex structure. Above we saw that it is also a K\"{a}hler manifold.\index{K\"{a}hler!manifold}\index{complex!structure}\index{manifold}

\paragraph{K\"{a}hler differential geometry.} On a K\"{a}hler manifold\index{K\"{a}hler!manifold}\index{manifold}
$(M,g)$ with K\"{a}hler potential $\CCK$, the components of the\index{K\"{a}hler!potential}
Levi-Civita connection simplify considerably. We introduce the\index{Levi-Civita connection}\index{connection}
Christoffel symbols as in Riemannian geometry by
\begin{equation}
\Gamma^i_{jk}\ =\ \tfrac{1}{2}g^{il}\left(\derr{g_{lk}}{x^j}+\derr{g_{lj}}{x^k}
-\derr{g_{jk}}{x^l}\right)~.
\end{equation}
Upon turning to complex coordinates and using the identity
\eqref{potident}, we see that
\begin{equation}
\Gamma^l_{jk}\ =\ g^{l\bar{s}}\derr{g_{k\bar{s}}}{z^j}\eand
\Gamma^{\bar{l}}_{\bj\bar{k}}\ =\ g^{\bar{l}s}\derr{g_{\bar{k}s}}{z^{\bj}}~,
\end{equation}
and all other components vanish. Connections of this type, which\index{connection}
are metric compatible ($\nabla_k\index{metric compatible}\index{metric}
g_{i\bj}=\nabla_{\bar{k}}g_{i\bj}=0$) are called {\em Hermitian
connections}.\index{Hermitian connections}\index{connection}

The torsion and curvature tensors are again defined by\index{curvature}
\begin{eqnarray}
T(X,Y)&=&\nabla_X Y-\nabla_Y X-[X,Y]~,\\
R(X,Y)Z&=&\nabla_X\nabla_YZ-\nabla_Y\nabla_XZ-\nabla_{[X,Y]} Z~,
\end{eqnarray}
and the only non-vanishing components of the Riemann tensor and
the Ricci tensor are\index{Ricci tensor}
\begin{equation}
R_{i\bj
k\bar{l}}\ =\ g_{i\bar{s}}\derr{\Gamma^{\bar{s}}_{\bj\bar{l}}}{z^k}\eand
\Ric_{\bi j}\ :=\ R^{\bar{k}}_{~\bi\bar{k}
j}\ =\ -\derr{\Gamma^{\bar{k}}_{\bi\bar{k}}}{z^j}~,
\end{equation}
respectively.

\paragraph{The Ricci form.} Given the Ricci tensor $\Ric$ on a\index{Ricci form}\index{Ricci tensor}
K\"{a}hler manifold $M$, we define the {\em Ricci form} $\CCR$ by\index{K\"{a}hler!manifold}\index{manifold}
\begin{equation}
\CCR(X,Y)\ :=\ \Ric(IX,Y).
\end{equation}
Thus we have in components $\CCR=\di R_{i\bj}\dd z^i\wedge \dd
\bz^\bj$. Note that on a K\"{a}hler manifold with metric\index{K\"{a}hler!manifold}\index{manifold}\index{metric}
$g_{\mu\nu}$, the Ricci form is closed and can locally be\index{Ricci form}
expressed as $\CCR=\di \dpar\dparb\ln G$, where
$G=\det(g_{\mu\nu})=\sqrt{g}$. Furthermore, its cohomology class
is (up to a real multiple) equal to the Chern class of the\index{Chern class}
canonical bundle on $M$.\index{canonical bundle}

A manifold with vanishing Ricci form is called {\em Ricci-flat}.\index{Ricci form}\index{Ricci-flat}\index{manifold}
K\"{a}hler manifolds with this property are called {\em Calabi-Yau\index{Calabi-Yau}\index{K\"{a}hler!manifold}
manifolds} and will be discussed in section \ref{secCalabiYau}.

\paragraph{Monge-Amp{\`e}re equation.} A differential equation of the\index{Monge-Amp{\`e}re equation}
type
\begin{equation}
(\dpar_x\dpar_x u)(\dpar_y\dpar_y u)-(\dpar_x\dpar_y
u)^2\ =\ f(x,y,u,\dpar_x u,\dpar_y u)
\end{equation}
is called a {\em Monge-Amp{\`e}re equation}. The condition of\index{Monge-Amp{\`e}re equation}
vanishing Ricci form obviously yields such an equation. We will\index{Ricci form}
explicitly discuss a related example in section \ref{ssExoticCYs}.

\paragraph{Hyper-K\"{a}hler manifold.} A {\em hyper-K\"{a}hler manifold}\index{K\"{a}hler!manifold}\index{hyper-K\"{a}hler manifold}\index{manifold}
is a Riemannian manifold with three K\"{a}hler structures $I,J$ and
$K$ which satisfy $IJK=-1$. Equivalently, one can define a
hyper-K\"{a}hler manifold as a Riemannian manifold with holonomy group\index{K\"{a}hler!manifold}\index{holonomy group}\index{hyper-K\"{a}hler manifold}\index{manifold}
contained in $\sSp(m)$, which is the group of $m\times m$
quaternionic unitary matrices with $m$ being half the complex
dimension of the manifold.\index{manifold}

\paragraph{'t Hooft tensors.} The {\em 't Hooft tensors} (or eta-symbols) are given by\index{t Hooft tensors@'t Hooft tensors}
\begin{equation}
\eta^{i(\pm)}_{\mu\nu}\ :=\ \eps_{i\mu\nu
4}\pm\delta_{i\mu}\delta_{\nu4}\mp\delta_{i\nu}\delta_{\mu 4}
\end{equation}
and satisfy the relation
$\eta^{i(\pm)}_{\mu\nu}=\pm*\eta^{i(\pm)}_{\mu\nu}$, where $*$ is
the Hodge star operator. They form three K\"{a}hler structures,
which give rise to a hyper-K\"{a}hler structure on the Euclidean
spacetime $\FR^4$. Note furthermore that any space of the form
$\FR^{4m}$ with $m\in \NN$ is evidently a hyper-K\"{a}hler
manifold.\index{K\"{a}hler!manifold}\index{hyper-K\"{a}hler manifold}\index{manifold}

\section{Vector bundles and sheaves}

\subsection{Vector bundles}\label{ssvectorbundles}

\paragraph{Homotopy lifting property.} Let $E$, $B$, and $X$ be\index{homotopy lifting property}
topological spaces. A map $\pi:E\rightarrow B$ is said to have the
{\em homotopy lifting property} with respect to the space $X$ if,\index{homotopy lifting property}
given the commutative diagram
\begin{equation}
\begin{aligned}
\begin{picture}(50,50)
\put(0.0,40.0){\makebox(0,0)[c]{$X\times \{0\}$}}
\put(0.0,0.0){\makebox(0,0)[c]{$X\times [0,1]$}}
\put(65.0,40.0){\makebox(0,0)[c]{$E$}}
\put(65.0,0.0){\makebox(0,0)[c]{$B$}}
\put(27.5,0.0){\vector(1,0){25}}
\put(27.5,40.0){\vector(1,0){25}}
\put(0.0,32.0){\vector(0,-1){20}}
\put(65.0,32.0){\vector(0,-1){20}}
\put(73.0,23.0){\makebox(0,0)[c]{$\pi$}}
\put(10.0,23.0){\makebox(0,0)[c]{$p$}}
\put(38,8.0){\makebox(0,0)[c]{$h_t$}}
\put(38,48.0){\makebox(0,0)[c]{$h$}}
\end{picture}
\end{aligned}
\end{equation}
there is a map $G:X\times [0,1]\rightarrow E$, which gives rise to
two commutative triangles. That is, $G(x,0)=h(x)$ and $\pi\circ
G(x,t)=h_t(x)$. Note that we assumed that all the maps are
continuous.

\paragraph{Fibration.} A {\em fibration} is a continuous map\index{fibration}
$\pi:E\rightarrow B$ between topological spaces $E$ and $B$, which
satisfies the homotopy lifting property for all topological spaces\index{homotopy lifting property}
$X$.

\paragraph{Complex vector bundles.} A {\em complex vector bundle}\index{complex!vector bundle}
$E$ over a complex manifold $M$ is a vector bundle\index{complex!manifold}\index{manifold}
$\pi:E\rightarrow M$, and for each $x\in M$, $\pi^{-1}(x)$ is a
complex vector space. As we will see in the following paragraph,
holomorphic vector bundles are complex vector bundles which allow\index{complex!vector bundle}\index{holomorphic!vector bundle}
for a trivialization with holomorphic transition functions.\index{transition function}\index{trivialization}

Every vector bundle is furthermore a fibration. The prove for this\index{fibration}
can be found e.g.\ in \cite{Hatcher:2002}.

In the following, we will denote the space of smooth sections of
the vector bundle $\pi:E\rightarrow M$ by $\Gamma(M,E)$.

\paragraph{Holomorphic vector bundle.} A {\em holomorphic vector\index{holomorphic!vector bundle}
bundle} $E$ of rank $k$ over a manifold $M$ with $\dim_\FC M=n$ is\index{manifold}
a $(k+n)$-dimensional complex manifold $E$ endowed with a\index{complex!manifold}
holomorphic projection $\pi:E\rightarrow M$ satisfying the
conditions
\begin{enumerate}
\item $\pi^{-1}(p)$ is a $k$-dimensional complex vector space for all $p\in
M$.
\item For each point $p\in M$, there is a neighborhood $U$ and a
biholomorphism $\phi_U:\pi^{-1}(U)\rightarrow U\times \FC^k$. The
maps $\phi_U$ are called {\em local trivializations}.\index{local trivialization}\index{trivialization}
\item The {\em transition functions} $f_{UV}$\index{transition function}
are holomorphic maps $U\cap V\rightarrow \sGL(k,\FC)$.\index{holomorphic!map}
\end{enumerate}
Holomorphic vector bundles of dimension $k=1$ are called {\em line\index{holomorphic!vector bundle}
bundles}.

\paragraph{Examples.} Let $M$ be a complex manifold of dimension\index{complex!manifold}\index{manifold}
$m$. The holomorphic tangent bundle $T^{1,0}M$, its dual, the
holomorphic cotangent bundle $T^{1,0}\dual M$, in fact all the
bundles $\Lambda^{p,0}M$ with $0\leq p\leq m$ are holomorphic
vector bundles. The complex line bundle $K_M:=\Lambda^{m,0}M$ is\index{holomorphic!vector bundle}
called the {\em canonical bundle}; $K_M$ is also a holomorphic\index{canonical bundle}
bundle.

On the spaces $\CPP^m$, one defines the {\em tautological line
bundle} as: $\FC^{m+1}\rightarrow \CPP^m$. One can proof that the\index{tautological line bundle}
canonical bundle over $\CPP^m$ is isomorphic to the ($m+1$)th\index{canonical bundle}
exterior power of the tautological line bundle. For more details\index{tautological line bundle}
on these line bundles, see also the remarks in \ref{plinebundles}.

\paragraph{Holomorphic structures.}\label{pholomorphicstructure}\index{holomorphic!structure}
Given a complex vector bundle $E$ over $M$, we define the bundle\index{complex!vector bundle}
of $E$-valued forms on $M$ by
$\Lambda^{p,q}E:=\Lambda^{p,q}M\otimes E$. An operator $\dparb:
\Gamma(M,\Lambda^{p,q}E)\rightarrow \Gamma(M,\Lambda^{p,q+1}E)$ is
called a {\em holomorphic structure} if it satisfies $\dparb^2=0$.\index{holomorphic!structure}
It is obvious that the action of $\dparb$ is independent of the
chosen trivialization, as the transition functions are holomorphic\index{transition function}\index{trivialization}
and $\dparb$ does not act on them. Note furthermore that the
operator $\dparb$ satisfies a graded Leibniz rule when acting on
the wedge product of a $(p,q)$-form $\omega$ and an arbitrary form
$\sigma$:
\begin{equation}
\dparb(\omega\wedge\sigma)\ =\ (\dparb\omega)\wedge\sigma+(-1)^{p+q}
\omega\wedge(\dparb\sigma)~.
\end{equation}

\paragraph{Theorem.}\label{holvecbundleholstructure} A complex
vector bundle $E$ is holomorphic if and only if there exists a\index{complex!vector bundle}
holomorphic structure $\dparb$ on $E$. For more details on this\index{holomorphic!structure}
statement, see section \ref{subseccohomology}.

\paragraph{Connections and curvature.} Given a complex vector bundle\index{complex!vector bundle}\index{connection}\index{curvature}
$E\rightarrow M$, a {\em connection} is a $\FC$-linear map
$\nabla:\Gamma(M,E)\rightarrow \Gamma(M,\Lambda^1 E)$ which
satisfies the Leibniz rule
\begin{equation}
\nabla(f \sigma)\ =\ \dd f\otimes \sigma+f\nabla \sigma~,
\end{equation}
where $f\in C^\infty(M)$ and $\sigma\in \Gamma(M,E)$. A connection\index{connection}
gives a means of transporting frames of $E$ along a path in $M$.
Given a smooth path $\gamma:[0,1]\rightarrow M$ and a frame
$\Be_0$ over $\gamma(0)$, there is a unique frame $\Be_t$
consisting of sections of $\gamma^* E$ such that
\begin{equation}
\nabla_{\dot{\gamma}(t)}\Be_t\ =\ 0
\end{equation}
for all $t\in[0,1]$. This frame is called the {\em parallel
transport} of $\Be_0$ along $\gamma$. As we can parallel transport\index{parallel transport}
frames, we can certainly do the same with vector fields.

The {\em curvature} associated to $\nabla$ is defined as the\index{curvature}
two-form $\CF_\nabla=\nabla^2$. Given locally constant sections
$(\sigma_1,\ldots ,\sigma_k)$ over $U$ defining a basis for each fibre
over $U$, we can represent a connection by a collection of\index{connection}
one-forms $\omega_{ij}\in\Gamma(U,\Lambda^1 U)$:
$\nabla\sigma_i=\omega_{ij}\otimes\sigma_j$. The components of the
corresponding curvature\index{curvature}
$\CF_\nabla\sigma_i=\CF_{ij}\otimes\sigma_j$ are easily calculated
to be $\CF_{ij}=\dd \omega_{ij}+\omega_{ik}\wedge\omega_{kj}$.
Roughly speaking, the curvature measures the difference between\index{curvature}
the parallel transport along a loop and the identity.\index{parallel transport}

Identifying $\nabla^{0,1}$ with the holomorphic structure\index{holomorphic!structure}
$\dparb$, one immediately sees from the theorem
\ref{holvecbundleholstructure} that the $(0,2)$-part of the
curvature of a holomorphic vector bundle has to vanish.\index{curvature}\index{holomorphic!vector bundle}

\paragraph{Chern connection.} Conversely, given a Hermitian structure on a\index{Chern connection}\index{Hermitian structure}\index{connection}
holomorphic vector bundle with holomorphic structure $\dparb$,\index{holomorphic!structure}\index{holomorphic!vector bundle}
there is a unique connection $\nabla$, the {\em Chern connection},\index{Chern connection}\index{connection}
for which $\nabla^{0,1}=\dparb$.

\paragraph{Connections on Hermitian manifolds.} On a Hermitian\index{Hermitian manifold}\index{connection}\index{manifold}
manifold $(M,I,h)$, there are two natural connections: the
Levi-Civita connection and the Chern connection. They both\index{Chern connection}\index{Levi-Civita connection}
coincide if and only if $h$ is K\"{a}hler.

\paragraph{Holonomy groups.} Let $M$ be a manifold of dimension $d$ endowed\index{holonomy group}\index{manifold}
with a connection $\nabla$. A vector $V\in T_p M$ will be\index{connection}
transformed to another vector $V'\in T_pM$ when parallel
transported along a closed curve through $p$. The group of all\index{parallel transport}
such transformations is called the {\em holonomy group} of the\index{holonomy group}
manifold $M$. Using the Levi-Civita connection which will not\index{Levi-Civita connection}\index{connection}\index{manifold}
affect the length of the vector $V$ during the parallel transport,\index{parallel transport}
the holonomy group will be a subgroup of $\sSO(d)$ on real\index{holonomy group}
manifolds and a subgroup of $\sU(d)$ for K\"{a}hler manifolds. Flat\index{K\"{a}hler!manifold}\index{manifold}
manifolds will clearly have the trivial group consisting only of
the element $\unit$ as their holonomy groups. Complex manifolds,\index{complex!manifold}\index{holonomy group}\index{manifold}
whose holonomy groups are $\sSU(d)$ are called {\em Calabi-Yau\index{Calabi-Yau}
manifolds} and will be discussed in section \ref{secCalabiYau}.\index{manifold}

\paragraph{Characteristic classes.} {\em Characteristic classes}\index{characteristic class}
are subsets of cohomology classes and are used to characterize
topological properties of manifolds and bundles. Usually they are\index{manifold}
defined by polynomials in the curvature two-form $\CF_\nabla$.\index{curvature}
Therefore, every trivial bundle has a trivial characteristic
class, and thus these classes indicate the nontriviality of a\index{characteristic class}
bundle. In the following, we will restrict our discussion mainly
to Chern classes, as they play a key r{\^o}le in the definition of\index{Chern class}
Calabi-Yau manifolds.\index{Calabi-Yau}\index{manifold}

\paragraph{Chern class.} Given a complex vector bundle\index{Chern class}\index{complex!vector bundle}
$E\rightarrow M$ with fibres $\FC^k$ endowed with a connection\index{connection}
$\omega$ defining a field strength $\CF$, we define the {\em total\index{field strength}
Chern class}\footnote{named after Shiing-shen Chern, who\index{Chern class}
introduced it in the 1940s} by
\begin{equation}
c(\CF)\ :=\ \det\left(\unit+\frac{\di \CF}{2\pi}\right)~.
\end{equation}
One can split $c(\CF)$ into the direct sums of forms of even
degrees:
\begin{equation}
c(\CF)\ =\ 1+c_1(\CF)+c_2(\CF)+\ldots ~.
\end{equation}
The $2j$-form $c_j(\CF)$ is called the {\em j-th Chern class}.\index{Chern class}
Note that when talking about the Chern class of a manifold, one\index{manifold}
means the Chern class of its tangent bundle calculated from the
curvature of the Levi-Civita connection.\index{Levi-Civita connection}\index{connection}\index{curvature}

\paragraph{Chern number.}\label{pChernnumber} If $M$ is compact\index{Chern number}
and of real dimension $2d$, one can pair any product of Chern
classes of total degree $2d$ with oriented homology classes of $M$\index{Chern class}
which results in integers called the {\em Chern numbers} of $E$.\index{Chern number}
As a special example, consider the possible first Chern classes of\index{Chern class}\index{first Chern class}
a line bundle $L$ over the Riemann sphere $\CPP^1\cong S^2$. It is\index{Riemann sphere}
$H^2(S^2)\cong \RZ$ and the number corresponding to the first
Chern class of the line bundle $L$ is called the {\em first Chern\index{Chern class}\index{first Chern class}
number}.

\paragraph{Properties of Chern classes.} The zeroth Chern class is\index{Chern class}
always equal to 1. For a manifold $M$ with dimension $d$, we\index{manifold}
clearly have $c_n(\CF)=0$ for $n>d$.

\paragraph{Calculating Chern classes.} A simple method for\index{Chern class}
calculating Chern classes is available if one can diagonalize
$\CF$ by an element $g\in\sGL(k,\FC)$ such that $g^{-1}\CF g=$
$\diag(x_1,\ldots ,x_n)=:D$. One then easily derives that
\begin{equation}
\begin{aligned}
\det(\unit+D)&\ =\ \det(\diag(1+x_1,\ldots ,1+x_n))\\&\ =\ 1+\tr
D+\tfrac{1}{2}((\tr D)^2-\tr D^2)+\ldots +\det D~.
\end{aligned}
\end{equation}

\paragraph{Theorem.} Consider two complex vector bundles\index{complex!vector bundle}
$E\rightarrow M$ and $F\rightarrow M$ with total Chern classes\index{Chern class}\index{total Chern class}
$c(E)$ and $c(F)$. Then the total Chern class of a Whitney sum
bundle\footnote{A {\em Whitney sum} of two vector bundles over a\index{Whitney sum bundle}
manifold $M$ yields the vector bundle whose fibres are the direct\index{manifold}
sums of the fibres of the original two bundles.} $(E\oplus
F)\rightarrow M$ is given by $c(E\oplus F)=c(E)\wedge c(F)$. In
particular, the first Chern classes add: $c_1(E\oplus\index{Chern class}\index{first Chern class}
F)=c_1(E)+c_1(F)$.

\paragraph{Whitney product formula.} Given a short exact sequence\index{Whitney product formula}\index{short exact sequence}
of vector bundles $A$, $B$ and $C$ as
\begin{equation}
0\ \rightarrow\  A\ \rightarrow\  B\ \rightarrow\  C\ \rightarrow\  0~,
\end{equation}
we have a splitting $B=A\oplus C$ and together with the above
theorem, we obtain the formula $c(A)\wedge c(C)=c(B)$. This
formula will be particularly useful for calculating the Chern
classes of the superambitwistor space $\CL^{5|6}$, see the short\index{Chern class}\index{ambitwistor space}\index{twistor}\index{twistor!ambitwistor}\index{twistor!space}
exact sequence \eqref{ambiexactseq}.

\paragraph{Further rules for calculations.} Given two vector bundles $E$ and $F$ over
a complex manifold $M$, we have the following formul\ae{}:\index{complex!manifold}\index{manifold}
\begin{align}
c_1(E\otimes F)&\ =\ \rk(F)c_1(E)+\rk(E)c_1(F)~,\\
c_1\left(E^{\dual}\right)&\ =\ -c_1(E)~.
\end{align}

\paragraph{Chern classes from degeneracy loci.}\label{pdegeneracy}\index{Chern class}
Chern classes essentially make statements about the degeneracy of
sets of sections of vector bundles via a Gau{\ss}-Bonnet formula. More
precisely, given a vector bundle $E$ of rank $e$ over $M$, the
$c_{e+1-i}$th Chern class is Poincar{\'e}-dual to the degeneracy cycle\index{Chern class}
of $i$ generic global sections. This degeneracy locus is obtained
by arranging the $i$ generic sections in an $e\times
i$-dimensional matrix $C$ and calculating the locus in $M$, where
$C$ has rank less than $i$. We will present an example in
paragraph \ref{plinebundles}. For more details, see e.g.\
\cite{GriffithsHarris}.

\paragraph{Chern character.}\label{pCherncharacter} Let us also briefly\index{Chern character}
introduce the characteristic classes called Chern characters,\index{characteristic class}
which play an important r{\^o}le in the Atiyah-Singer index theorem.
We will need them for instanton configurations, in which the\index{instanton}
number of instantons is given by an integral over the second Chern
character. One defines the {\em total Chern character} of a\index{Chern character}\index{total Chern character}
curvature two-form $\CF$ as\index{curvature}
\begin{equation}
\mathrm{ch}(\CF)\ =\ \tr \exp\left(\frac{\di\CF}{2\pi}\right)
\end{equation}
and the {\em $j$-th Chern character} as a part of the\index{Chern character}
corresponding Taylor expansion
\begin{equation}
\mathrm{ch}_j(\CF)\ =\ \frac{1}{j!}\tr\left(\frac{\di\CF}{2\pi}\right)^j~.
\end{equation}
Note that $\mathrm{ch}(\CF)$ is a polynomial of finite order on a
finite-dimensional manifold. Furthermore, one can express Chern\index{manifold}
characters in terms of Chern classes,\index{Chern class} e.g.\
\begin{equation}
\mathrm{ch}_1(\CF)\ =\ c_1(\CF)\eand\mathrm{ch}_2(\CF)\ =\ \tfrac{1}{2}(c_1(\CF)^2-2c_2(\CF))~.
\end{equation}
The zeroth Chern character $\mathrm{ch}_0(\CF)$ is simply the\index{Chern character}
dimension of the vector bundle associated to the curvature\index{curvature}
two-form $\CF$.

\subsection{Sheaves and line bundles}\label{ssSheaves}

\paragraph{Sheaf.} A {\em presheaf} $\frS$ on a topological space $X$\index{presheaf}\index{sheaf}
is an association of a group\footnote{Usually, the definition of a
sheaf involves only Abelian groups, but extensions to non-Abelian\index{sheaf}
groups are possible, see e.g.\ the discussion in
\cite{Popov:1998pc}.} $\frS(U)$ to every open set $U\subset X$
together with a restriction map $\rho_{UV}:\frS(V)\rightarrow
\frS(U)$ for $U\subset V\subset X$, which satisfies
$\rho_{UW}=\rho_{UV}\circ\rho_{VW}$ for $U\subset V\subset
W\subset X$ and $\rho_{WW}=\id$. A presheaf becomes a {\em sheaf}\index{presheaf}\index{sheaf}
under two additional conditions:
\begin{enumerate}
\item Sections are determined by local data: Given two sections
$\sigma, \tau\in\frS(V)$ with $\rho_{UV}(\sigma)=\rho_{UV}(\tau)$
for every open set $U\subset V$, we demand that $\sigma=\tau$ on
$V$.
\item Compatible local data can be patched together: If
$\sigma\in\frS(U)$ and $\tau\in\frS(V)$ such that $\rho_{(U\cap
V)U}(\sigma)=\rho_{(U\cap V)V}(\tau)$ then there exists an
$\chi\in\frS(U\cup V)$ such that $\rho_{U(U\cup V)}(\chi)=\sigma$
and $\rho_{V(U\cup V)}(\chi)=\tau$.
\end{enumerate}

\paragraph{Turning a presheaf into a sheaf.} One can associate a\index{presheaf}\index{sheaf}
sheaf $\frS$ to a presheaf $\frS_0$ on a topological space $X$ by
the following construction: Consider two local sections $s$ and
$s'\in\frS_0(U)$ for an open set $U\subset X$. We call $s$ and
$s'$ equivalent at the point $x\in X$ if there is a neighborhood
$V_x\subset U$, such that $\rho_{V_xU}(s)=\rho_{V_xU}(s')$. The
corresponding equivalence classes are called {\em germs} of\index{germs}
sections in the point $x$ and the space of germs at $x$ is denoted
by $\frS_x$. We can now define the sheaf $\frS$ as the union of\index{sheaf}
the spaces of germs $\frS:=\bigcup_{x\in X} \frS_x$, as this union\index{germs}
clearly has the required properties.

\paragraph{Subsheaf.} A {\em subsheaf} of a sheaf $\frS$\index{sheaf}\index{subsheaf}
over a topological space $X$ is a sheaf $\frS'$ over $X$ such that
$\frS'(U)$ is a subgroup of $\frS(U)$ for any open set $U\subset
X$. The restriction maps on $\frS'$ are inherited from the ones on
$\frS$.

\paragraph{Examples.} Examples for sheaves are the sheaf of\index{sheaf}
holomorphic functions $\CCO(U)$, the sheaves of continuous and
smooth functions\footnote{Note that $C^0(U,\frS)$ will denote the
set of \v{C}ech 0-cochains taking values in the sheaf $\frS$.}\index{sheaf}
$C^0(U)$ and $C^\infty(U)$ and the sheaves of smooth $(r,s)$-forms
$\varOmega^{r,s}(U)$, where $U$ is a topological space (a complex
manifold in the latter example).\index{complex!manifold}\index{manifold}

\paragraph{Structure sheaf.}\label{pStructureSheaf} One can interpret\index{sheaf}\index{structure sheaf}
a manifold $M$ as a {\em locally ringed space}, which\footnote{A\index{manifold}
special case of locally ringed spaces are the better-known {\em
schemes}.} is a topological space $M$ together with a sheaf $F$ of\index{sheaf}
commutative rings on $M$. This sheaf $F$ is called the structure
sheaf of the locally ringed space and one usually denotes it by\index{structure sheaf}
$\CCO_M$. In the case that $(M,\CCO_M)$ is a complex manifold, $F$\index{complex!manifold}\index{manifold}
is the sheaf of holomorphic functions on $M$.\index{sheaf}

\paragraph{Locally free sheaf.} A sheaf $\frE$ is {\em locally free}\index{locally free sheaf}\index{sheaf}
and {\em of rank $r$} if there is an open covering $\{U_j\}$ such
that $\frE|_{U_j}\cong \CCO^{\oplus r}_{U_j}$. One can show that
(isomorphism classes of) locally free sheaves of rank $r$ over a\index{morphisms!isomorphism}
manifold $M$ are in one-to-one correspondence with (isomorphism\index{manifold}
classes of) vector bundles of rank $r$ over $M$. The sheaf $\frE$\index{sheaf}
corresponding to a certain vector bundle $E$ is given by the sheaf
dual to the sheaf of sections of $E$. For this reason, the terms\index{sheaf}
vector bundle and (locally free) sheaf are often used sloppily for
the same object.

We will denote by $\CCO(U)$ the sheaf of holomorphic functions,\index{sheaf}
and the holomorphic vector bundle over $U$, whose sections\index{holomorphic!vector bundle}
correspond to elements of $\CCO(U)$, by $\CO(U)$.

\paragraph{Holomorphic line bundles.}\label{plinebundles} A\index{holomorphic!line bundle}
{\em holomorphic line bundle} is a holomorphic vector bundle of\index{holomorphic!vector bundle}
rank 1. Over the Riemann sphere $\CPP^1\cong S^2$, these line\index{Riemann sphere}
bundles can be completely characterized by an integer $d\in\RZ$,
cf.\ \ref{pChernnumber}.

Given the standard patches $U_+$ and $U_-$ on the Riemann sphere\index{Riemann sphere}
$\CPP^1$ with the inhomogeneous coordinates $\lambda_\pm$ glued\index{homogeneous coordinates}\index{inhomogeneous coordinates}
via $\lambda_\pm=1/\lambda_\mp$ on the intersection $U_+\cap U_-$
of the patches, the {\em holomorphic line bundle} $\CO(d)$ is\index{holomorphic!line bundle}
defined by its transition function $f_{+-}=\lambda^d_+$ and thus\index{transition function}
we have $z_+=\lambda_+^d z_-$, where $z_\pm$ are complex
coordinates on the fibres over $U_\pm$.

For $d\geq 0$, global sections of the bundle $\CO(d)$ are
polynomials of degree $d$ in the inhomogeneous coordinates\index{homogeneous coordinates}\index{inhomogeneous coordinates}
$\lambda_\pm$ and homogeneous polynomials of degree $d$ in
homogeneous coordinates. The $\CO(d)$ line bundle has first Chern\index{homogeneous coordinates}
number $d$, since --  according to the Gau{\ss}-Bonnet formula of
paragraph \ref{pdegeneracy} -- the first Chern class is Poincar{\'e}\index{Chern class}\index{first Chern class}
dual to the degeneracy loci of one generic global section. These
loci are exactly the $d$ points given by the zeros of a degree $d$
polynomial. Furthermore, the first Chern class is indeed\index{Chern class}\index{first Chern class}
sufficient to characterize a complex line bundle up to topological
(smooth) equivalence, and therefore it also suffices to
characterize a holomorphic line bundle up to holomorphic\index{holomorphic!line bundle}
equivalence.

The complex conjugate bundle to $\CO(d)$ is denoted by
$\bar{\CO}(d)$. Its sections have transition functions\index{transition function}
$\bar{\lambda}_+^d$: $\bar{z}_+=\bar{\lambda}_+^d \bar{z}_-$.

This construction can be generalized to higher-dimensional complex
projective spaces $\CPP^n$. Recall that these spaces are covered\index{complex!projective space}
by $n+1$ patches. In terms of the homogeneous coordinates\index{homogeneous coordinates}
$\lambda_i$, $i=0,\ldots,n$, the line bundle $\CO(d)\rightarrow
\CPP^n$ is defined by the transition function\index{transition function}
$f_{ij}=(\lambda_j/\lambda_i)^d$.

We will sometimes use the notation $\CO_{\CPP^n}(d)$, to label the
line bundle of degree $d$ over $\CPP^n$. Furthermore,
$\CO_{\CPP^n}$ denotes the trivial line bundle over $\CPP^n$, and
$\CO^k(d)$ is defined as the direct sum of $k$ line bundles of
rank $d$.

Note that bases of the (1,0)-parts of the tangent and the
cotangent bundles of the Riemann sphere $\CPP^1$ are sections of\index{Riemann sphere}
$\CO(2)$ and $\CO(-2)$, respectively. Furthermore, the canonical
bundle of $\CPP^n$ is $\CO(-n-1)$ and its tautological line bundle\index{canonical bundle}\index{tautological line bundle}
is $\CO(-1)$.

\paragraph{Theorem.} (Grothendieck) Any holomorphic bundle $E$ over\index{Theorem!Grothendieck}
$\CPP^1$ can be decomposed into a direct sum of holomorphic line
bundles. This decomposition is unique up to permutations of\index{holomorphic!line bundle}
holomorphically equivalent line bundles. The Chern numbers of the\index{Chern number}\index{holomorphically equivalent}
line bundles are holomorphic invariants of $E$, but only their sum
is also a topological invariant.

\subsection{Dolbeault and \v{C}ech cohomology}\label{subseccohomology}\index{Cech cohomology@\v{C}ech cohomology}

There are two convenient descriptions of holomorphic vector
bundles: the Dolbeault and the \v{C}ech description. Since the\index{holomorphic!vector bundle}
Penrose-Ward transform (see chapter \ref{chTwistorGeometry})\index{Penrose-Ward transform}
heavily relies on both of them, we recollect here the main aspects
of these descriptions and comment on their equivalence.

\paragraph{Dolbeault cohomology groups.} As the Dolbeault operator\index{Dolbeault cohomology}
$\dparb$ is nilpotent, one can introduce the Dolbeault complex\index{Dolbeault complex}
\begin{eqnarray}
\Omega^{r,0}(M)\ \stackrel{\bar{\dpar}}{\longrightarrow}\
\Omega^{r,1}(M)\ \stackrel{\bar{\dpar}}{\longrightarrow}\ \ldots  \
\stackrel{\bar{\dpar}}{\longrightarrow}\ \Omega^{r,m}(M)
\end{eqnarray}
on a complex manifold $M$ together with the $(r,s)$th {\em\index{complex!manifold}\index{manifold}
$\dparb$-cohomology group}\index{d-cohomology group@$\dparb$-cohomology group}
\begin{equation}
H^{r,s}_{\bar{\dpar}}(M)\ =\ \frac{\mathrm{cocycles}}{\mathrm{coboundaries}}\ =\ \frac{Z^{r,s}_{\bar{\dpar}}(M)}{B^{r,s}_{\bar{\dpar}}(M)}~.\index{coboundaries}\index{cocycles}
\end{equation}
Here, the {\em cocycles} $Z^{r,s}_\dparb(M)$ are the elements\index{cocycles}
$\omega$ of $\Omega^{r,s}(M)$ which are closed, i.e.\
$\dparb\omega=0$ and the {\em coboundaries} are those elements\index{coboundaries}
$\omega$ which are exact, i.e.\ for $s>0$ there is a form
$\tau\in\Omega^{r,s-1}(M)$ such that $\dparb\tau=\omega$.

The {\em Hodge number} $h^{r,s}$ is the complex dimension of\index{Hodge number}
$H^{r,s}_{\bar{\dpar}}(M)$. The corresponding {\em Betti number}\index{Betti number}
of the de Rham cohomology of the underlying real manifold is given\index{manifold}
by $b_k=\sum_{p=0}^k h^{p,k-p}$ and the {\em Euler number} of a\index{Euler number}
$d$-dimensional real manifold is defined as $\chi=\sum_{p=0}^d\index{manifold}
(-1)^p b_p$.

The Poincar{\'e} lemma can be directly translated to the complex\index{Poincar{\'e} lemma}
situation and thus every $\dparb$-closed form is locally
$\dparb$-exact.

\paragraph{Holomorphic vector bundles and Dolbeault cohomology.}\index{Dolbeault cohomology}\index{holomorphic!vector bundle}
Assume that $G$ is a group having a representation in terms of\index{representation}
$n\times n$ matrices. We will denote by $\frS$ the sheaf of smooth\index{sheaf}
$G$-valued functions on a complex manifold $M$ and by $\frA$ the\index{complex!manifold}\index{manifold}
sheaf of flat (0,1)-connections on a principal $G$-bundle\index{connection}\index{sheaf}
$P\rightarrow M$, i.e.\ germs of solutions to\index{germs}
\begin{equation}\label{eomdol}
\dparb \CA^{0,1}+\CA^{0,1}\wedge \CA^{0,1}\ =\ 0~.
\end{equation}

Note that elements $\CA^{0,1}$ of $\Gamma(M,\frA)$ define a
holomorphic structure $\dparb_\CA=\dparb+\CA^{0,1}$ on a trivial\index{holomorphic!structure}
rank $n$ complex vector bundle over $M$. The moduli space $\CM$ of\index{complex!vector bundle}\index{moduli space}
such holomorphic structures is obtained by factorizing\index{holomorphic!structure}
$\Gamma(M,\frA)$ by the group of gauge transformations, which is\index{gauge transformations}\index{gauge!transformation}
the set of elements $g$ of $\Gamma(M,\frS)$ acting on elements
$\CA^{0,1}$ of $\Gamma(M,\frA)$ as
\begin{equation}
\CA^{0,1}\ \mapsto\  g\CA^{0,1}g^{-1}+g\dparb g^{-1}~.
\end{equation}
Thus, we have $\CM\cong \Gamma(M,\frA)/\Gamma(M,\frS)$ and this is
the description of holomorphic vector bundles in terms of\index{holomorphic!vector bundle}
Dolbeault cohomology.\index{Dolbeault cohomology}

\paragraph{\v{C}ech cohomology sets.}\label{pCechcohomologysets} Consider a trivial\index{Cech cohomology@\v{C}ech cohomology}
principal $G$-bundle $P$ over a complex manifold $M$ covered by a\index{complex!manifold}\index{manifold}
collection of patches $\frU=\{ U_a\}$ and let $G$ have a
representation in terms of $n\times n$ matrices. Let $\frG$ be an\index{representation}
arbitrary sheaf of $G$-valued functions on $M$. The set of {\em\index{sheaf}
\v{C}ech $q$-cochains} $C^q(\frU,\frG)$ is the collection
$\psi=\{\psi_{a_0\ldots a_q}\}$ of sections of $\frG$ defined on
nonempty intersections $ U_{a_0}\cap\ldots \cap  U_{a_q}$.
Furthermore, we define the sets of \v{C}ech 0- and 1-cocycles by\index{cocycles}
\begin{align}
& Z^0(\frU,\frG)\ :=\ \{\,\psi\in
C^0(\frU,\frG)~|~\psi_a\ =\ \psi_b~\mbox{on}~ U_a\cap
 U_b\ \neq\  \varnothing\}\ =\ \Gamma(\frU,\frG)~,\\\nonumber
& Z^1(\frU,\frS)\ :=\ \{\,\chi\in
C^1(\frU,\frG)~|~\chi_{ab}\ =\ \chi^{-1}_{ba}~\mbox{on}~
 U_a\cap  U_b\ \neq\ 
\varnothing,~\\&\hspace{4.9cm}\chi_{ab}\chi_{bc}\chi_{ca}\ =\ \unit~\mbox{on}~
 U_a\cap  U_b\cap  U_c\ \neq\  \varnothing\}~.
\end{align}
This definition implies that the \v{C}ech 0-cocycles are\index{cocycles}
independent of the covering: it is $Z^0(\frU,\frG)= Z^0(M,\frG)$,
and we define the {\em zeroth \v{C}ech cohomology set} by\index{Cech cohomology@\v{C}ech cohomology}
$\check{H}^0(M,\frG):= Z^0(M,\frG)$. Two 1-cocycles $\chi$ and\index{cocycles}
$\tilde{\chi}$ are called {\em equivalent} if there is a 0-cochain
$\psi\in C^0(\frU,\frG)$ such that
$\tilde{\chi}_{ab}=\psi_a\chi_{ab}\psi_b^{-1}$ on all $ U_a\cap
U_b\neq \varnothing$. Factorizing $Z^1(\frU,\frG)$ by this
equivalence relation gives the {\em first \v{C}ech cohomology set}\index{Cech cohomology@\v{C}ech cohomology}
$\check{H}^1(\frU,\frG)\cong
 Z^1(\frU,\frG)/C^0(\frU,\frG)$.

If the patches $ U_a$ of the covering $\frU$ are Stein manifolds,\index{Stein manifold}\index{manifold}
one can show that the first \v{C}ech cohomology sets are\index{Cech cohomology@\v{C}ech cohomology}
independent of the covering and depend only on the manifold $M$,\index{manifold}
e.g.\ $\check{H}^1(\frU,\frS)=\check{H}^1(M,\frS)$. This is well
known to be the case in the situations we will consider later on,
i.e.\ for purely bosonic twistor spaces. Let us therefore imply\index{twistor}\index{twistor!space}
that all the coverings in the following have patches which are
Stein manifolds unless otherwise stated.\index{Stein manifold}\index{manifold}

Note that in the terms introduced above, we have $\CM\cong
\check{H}^0(M,\frA)/\check{H}^0(M,\frS)$.

\paragraph{Abelian \v{C}ech cohomology.}\label{pAbeliancech} If\index{Abelian \v{C}ech cohomology}\index{Cech cohomology@\v{C}ech cohomology}
the structure group $G$ of the bundle $P$ defined in the previous
paragraph is Abelian, one usually replaces in the notation of the
group action the multiplication by addition to stress
commutativity. Furthermore, one can then define a full {\em
Abelian \v{C}ech complex} from the operator\footnote{The\index{Abelian \v{C}ech complex}
corresponding picture in the non-Abelian situation has still not
been constructed in a satisfactory manner.}
$\check{\dd}:C^q(M,\frS)\rightarrow C^{q+1}(M,\frS)$ whose action
on \v{C}ech $q$-cochains $\psi$ is given by
\begin{equation}\label{eqdcheck}
(\check{\dd}\psi)_{a_0,a_1,\ldots ,a_{q+1}}\ :=\ 
\sum_{\nu=0}^{q+1}(-1)^\nu
\psi_{a_0,a_1,\ldots ,\hat{a}_\nu,\ldots ,a_{q+1}}~,
\end{equation}
where the hat $\hat{\cdot}$ denotes an omission. The nilpotency of
$\check{\dd}$ is easily verified, and the {\em Abelian \v{C}ech
cohomology} $\check{H}^q(M,\frS)$ is the cohomology of the\index{Abelian \v{C}ech cohomology}\index{Cech cohomology@\v{C}ech cohomology}
\v{C}ech complex.

More explicitly, we will encounter the following three Abelian
\v{C}ech cohomology groups: $\check{H}^0(M,\frS)$, which is the\index{Abelian \v{C}ech cohomology}\index{Cech cohomology@\v{C}ech cohomology}
space of global sections of $\frS$ on $M$, $\check{H}^1(M,\frS)$,
for which the cocycle and coboundary conditions read
\begin{equation}
\chi_{ac}\ =\ \chi_{ab}+\chi_{bc}\eand \chi_{ab}\ =\ \psi_a-\psi_b~,
\end{equation}
respectively, where $\chi\in C^1(M,\frS)$ and $\psi\in
C^0(M,\frS)$, and $\check{H}^2(M,\frS)$, for which the cocycle and
coboundary conditions read
\begin{equation}
\varphi_{abc}-\varphi_{bcd}+\varphi_{cda}-\varphi_{dab}\ =\ 0\eand
\varphi_{abc}\ =\ \chi_{ab}-\chi_{ac}+\chi_{bc}~,
\end{equation}
where $\varphi\in C^2(M,\frS)$, as one easily derives from
\eqref{eqdcheck}.

\paragraph{Holomorphic vector bundles and \v{C}ech\index{holomorphic!vector bundle}
cohomology.}\label{pholvecinCech} Given a complex manifold $M$,\index{complex!manifold}\index{manifold}
let us again denote the sheaf of smooth $G$-valued functions on\index{sheaf}
$M$ by $\frS$. We introduce additionally its subsheaf of\index{subsheaf}
holomorphic functions and denote it by $\frH$.

Contrary to the connections used in the Dolbeault description, the\index{connection}
\v{C}ech description of holomorphic vector bundles uses transition\index{holomorphic!vector bundle}
functions to define vector bundles. Clearly, such a collection of
transition functions has to belong to the first \v{C}ech cocycle\index{transition function}
set of a suitable sheaf $\frG$. Furthermore, we want to call two\index{sheaf}
vector bundles equivalent if there exists an element $h$ of
$C^0(M,\frG)$ such that
\begin{equation}
f_{ab}\ =\ h^{-1}_a\tilde{f}_{ab}h_b~~~\mbox{on all}~~ U_a\cap
U_b\ \neq\ \varnothing~.
\end{equation}
Thus, we observe that holomorphic and smooth vector bundles have
transition functions which are elements of the \v{C}ech cohomology\index{Cech cohomology@\v{C}ech cohomology}\index{transition function}
sets $\check{H}^1(M,\frH)$ and $\check{H}^1(M,\frS)$,
respectively.

\paragraph{Equivalence of the Dolbeault and \v{C}ech
descriptions.} For simplicity, let us restrict our considerations
to topologically trivial bundles, which will prove to be
sufficient. To connect both descriptions, let us first introduce
the subset $\frX$ of $C^0(M,\frS)$ given by a collection of
$G$-valued functions $\psi=\{\psi_a\}$, which fulfill
\begin{equation}
\psi_a\dparb\psi_a^{-1}\ =\ \psi_b\dparb\psi_b^{-1}
\end{equation}
on any two arbitrary patches $U_a$, $U_b$ from the covering $\frU$
of $M$. Due to \eqref{eomdol}, elements of $\check{H}^0(M,\frA)$
can be written as $\psi\dparb\psi^{-1}$ with $\psi\in\frX$. Thus,
for every $\CA^{0,1}\in \check{H}^0(M,\frA)$ we have corresponding
elements $\psi\in\frX$. One of these $\psi$ can now be used to
define the transition functions of a topologically trivial rank\index{transition function}
$n$ holomorphic vector bundle $E$ over $M$ by the formula\index{holomorphic!vector bundle}
\begin{equation}
f_{ab}\ =\ \psi^{-1}_a\psi_b~~~\mbox{on}~~ U_a\cap
U_b\ \neq\ \varnothing~.
\end{equation}
It is easily checked that the $f_{ab}$ constructed in this way are
holomorphic. Furthermore, they define holomorphic vector bundles\index{holomorphic!vector bundle}
which are topologically trivial, but not holomorphically trivial.
Thus, they belong to the kernel of a map
$\rho:\check{H}^1(M,\frH)\rightarrow \check{H}^1(M,\frS)$.

Conversely, given a transition function $f_{ab}$ of a\index{transition function}
topologically trivial vector bundle on the intersection $ U_a\cap
U_b$, we have
\begin{equation}
0=\dparb f_{ab}\ =\ \dparb
(\psi_a^{-1}\psi_b)\ =\ \psi_a^{-a}(\psi_a\dparb\psi_a^{-1}-\psi_b\dparb\psi_b^{-1})\psi_b\ =\ 
\psi_a^{-1}(\CA_a-\CA_b)\psi_b~.
\end{equation}
Hence on $ U_a\cap U_b$, we have $\CA_a=\CA_b$ and we have defined
a global $(0,1)$-form $\CA^{0,1}:=\psi\dparb\psi^{-1}$.

The bijection between the moduli spaces of both descriptions is\index{moduli space}
easily found. We have the short exact sequence\index{short exact sequence}
\begin{equation}
0\rightarrow\frH\ \stackrel{i}{\longrightarrow}\ \frS\
\stackrel{\delta^0}{\longrightarrow}\ \frA \
\stackrel{\delta^1}{\longrightarrow}\ 0~,
\end{equation}
where $i$ denotes the embedding of $\frH$ in $\frS$, $\delta^0$ is
the map $\frS\ni\psi\mapsto \psi\dparb\psi^{-1}\in\frA$ and
$\delta^1$ is the map $\frA\ni\CA^{0,1}\mapsto \dparb
\CA^{0,1}+\CA^{0,1}\wedge\CA^{0,1}$. This short exact sequence\index{short exact sequence}
induces a long exact sequence of cohomology groups
\begin{equation*}
0\ \rightarrow\  \check{H}^0(M,\frH)\ \stackrel{i_*}{\rightarrow}\
\check{H}^0(M,\frS)\ \stackrel{\delta^0_*}{\rightarrow}\
\check{H}^0(M,\frA)\ \stackrel{\delta^1_*}{\rightarrow}\
\check{H}^1(M,\frH)\ \stackrel{\rho}{\rightarrow}\
\check{H}^1(M,\frS)\ \rightarrow\  \ldots ~,
\end{equation*}
and from this we see that $\kernel\,\rho\cong
\check{H}^0(M,\frA)/\check{H}^0(M,\frS)\cong \CM$. Thus, the
moduli spaces of both descriptions are bijective and we have the\index{moduli space}
equivalence
\begin{equation}
(E,f_{+-}\!= \unit_n,A^{0,1})\ \sim\
(\tilde{E},\tilde{f}_{+-},\tilde{A}^{0,1}= 0)~.
\end{equation}
This fact is at the heart of the Penrose-Ward transform, see\index{Penrose-Ward transform}
chapter \ref{chTwistorGeometry}.

\paragraph{Remark concerning supermanifolds.} In the later\index{super!manifold}
discussion, we will need to extend these results to supermanifolds
and exotic supermanifolds, see chapter \ref{chsupergeometry}. Note\index{exotic!supermanifold}
that this is not a problem, as our above discussion was
sufficiently abstract. Furthermore, we can assume that the patches
of a supermanifold are Stein manifolds if and only if the patches\index{Stein manifold}\index{super!manifold}\index{manifold}
of the corresponding body are Stein manifolds since infinitesimal\index{body}
neighborhoods cannot be covered partially. Recall that having
patches which are Stein manifolds render the \v{C}ech cohomology\index{Cech cohomology@\v{C}ech cohomology}\index{Stein manifold}\index{manifold}
sets independent of the covering.

\subsection{Integrable distributions and Cauchy-Riemann\index{integrable}\index{integrable distribution}
structures}\label{ssCRstructures}

Cauchy-Riemann structures are a generalization of the concept of\index{Cauchy-Riemann structure}
complex structures to real manifolds of arbitrary dimension, which\index{complex!structure}\index{manifold}
we will need in discussing aspects of the mini-twistor geometry in\index{twistor}
section \ref{sPWMini}.

\paragraph{Integrable distribution.} Let $M$ be a smooth\index{integrable}\index{integrable distribution}
manifold of real dimension $d$ and $T_\FC M$ its complexified\index{manifold}
tangent bundle. A subbundle $\CT\subset T_\FC M$ is said to be
{\em integrable} if\index{integrable}
\begin{enumerate}
\item $\CT\cap\bar{\CT}$ has constant rank $k$,
\item $\CT$ and\footnote{We use the same letter for the bundle
$\CT$ and a distribution generated by its sections.}
$\CT\cap\bar{\CT}$ are closed under the Lie bracket.
\end{enumerate}
Given an integrable distribution $\CT$, we can choose local\index{integrable}\index{integrable distribution}
coordinates $u^1,\ldots,u^l,v^1,\ldots,v^k$,
$x^1,\ldots,x^m,y^1,\ldots,y^m$ on any patch $U$ of the covering
of $M$ such that $\CT$ is locally spanned by the vector fields
\begin{equation}
\der{v^1},\ldots,\der{v^k},\der{\bw^1},\ldots,\der{\bw^m}~,
\end{equation}
where $\bw^1=x^1-\di y^1,\ldots,\bw^m=x^m-\di y^m$
\cite{Nirenberg}.

\paragraph{$\CT$-differential.} For any smooth function $f$\index{T-differential@$\CT$-differential}
on $M$, let $\dd_\CT f$ denote the restriction of $\dd f$ to
$\CT$, i.e.\ $\dd_\CT$ is the composition
\begin{equation}
C^\infty(M)\ \stackrel{\dd}{\longrightarrow}\ \Omega^1(M)\
\longrightarrow\ \Gamma(M,\CT^*)~,
\end{equation}
where $\Omega^1(M):=\Gamma(M,T^* M)$ and $\CT^*$ denotes the sheaf\index{sheaf}
of (smooth) one-forms dual to $\CT$ \cite{Rawnsley}. The operator
$\dd_\CT$ can be extended to act on relative $q$-forms from the
space $\Omega^q_\CT(M):=\Gamma(M,\Lambda^q\CT^*)$,
\begin{equation}
\dd_\CT\ :\ \Omega^q_\CT(M)\ \rightarrow\
\Omega_\CT^{q+1}(M)~,~~~\mbox{for}~~~q\ \geq\ 0~.
\end{equation}
This operator is called a {\em $\CT$-differential}.\index{T-differential@$\CT$-differential}

\paragraph{$\CT$-connection.} Let $E$ be\index{T-connection@$\CT$-connection}\index{connection}
a smooth complex vector bundle over $M$. A covariant differential\index{complex!vector bundle}
(or connection) on $E$ along the distribution $\CT$ -- a {\em\index{connection}
$\CT$-connection} \cite{Rawnsley} -- is a $\FC$-linear mapping\index{T-connection@$\CT$-connection}
\begin{equation}
\nabla_\CT\ :\ \Gamma(M,E)\ \rightarrow\ \Gamma(M,\CT^*\otimes E)
\end{equation}
satisfying the Leibniz formula
\begin{equation}
\nabla_\CT(f\sigma)\ =\ f\nabla_\CT \sigma+\dd_\CT f\otimes
\sigma~,
\end{equation}
for a local section $\sigma\in\Gamma(M,E)$ and a local smooth
function $f$. This $\CT$-connection extends to a map\index{T-connection@$\CT$-connection}\index{connection}
\begin{equation}
\nabla_\CT\ :\ \Omega^q_\CT(M,E)\ \rightarrow\
\Omega_\CT^{q+1}(M,E)~,
\end{equation}
where $\Omega^q_\CT(M,E):=\Gamma(M,\Lambda^q\CT^*\otimes E)$.
Locally, $\nabla_\CT$ has the form
\begin{equation}
\nabla_\CT\ =\ \dd_\CT+A_\CT~,
\end{equation}
where the standard $\sEnd E$-valued $\CT$-connection one-form\index{T-connection@$\CT$-connection}\index{connection}
$A_\CT$ has components only along the distribution $\CT$.

\paragraph{$\CT$-flat vector bundles.} As usual, $\nabla^2_\CT$\index{T-flat vector bundle@$\CT$-flat vector bundle}
naturally induces a relative 2-form
\begin{equation}
\CF_\CT\in\Gamma(M,\Lambda^2\CT^*\otimes \sEnd E)
\end{equation}
which is the curvature of $A_\CT$. We say that $\nabla_\CT$ (or\index{curvature}
$A_\CT$) is flat if $\CF_\CT=0$. For a flat $\nabla_\CT$, the pair
$(E,\nabla_\CT)$ is called a $\CT$-flat vector bundle\index{T-flat vector bundle@$\CT$-flat vector bundle}
\cite{Rawnsley}.

Note that the complete machinery of Dolbeault and \v{C}ech
descriptions of vector bundles naturally generalizes to $\CT$-flat
vector bundles. Consider a manifold $M$ covered by the patches\index{T-flat vector bundle@$\CT$-flat vector bundle}\index{manifold}
$\frU:=\{U_{(a)}\}$ and a topologically trivial vector bundle
$(E,f_{+-}=\unit,\nabla_\CT)$ over $M$, with an expression
\begin{equation}\label{eq:4.28}
A_\CT|_{U_{(a)}}\ =\ \psi_a\dd_\CT\psi_a^{-1}
\end{equation}
of the flat $\CT$-connection, where the $\psi_a$ are smooth\index{T-connection@$\CT$-connection}\index{connection}
$\sGL(n,\FC)$-valued superfunctions on every patch $U_{(a)}$, we
deduce from the triviality of $E$ that $\psi_a\dd_\CT\psi_a^{-1}
=\psi_b\dd_\CT\psi_b^{-1}$ on the intersections $U_{(a)}\cap
U_{(b)}$. Therefore, it is $\dd_\CT(\psi_+^{-1}\psi_-)= 0$ and we
can define a $\CT$-flat complex vector bundle $\tilde{E}$ with the\index{complex!vector bundle}
canonical flat $\CT$-connection $\dd_\CT$ and the transition\index{T-connection@$\CT$-connection}\index{connection}
function $ \tilde{f}_{ab}\ :=\  \psi_a^{-1}\psi_b$. The bundles
$E$ and $\tilde{E}$ are equivalent as smooth bundles but not as
$\CT$-flat bundles. However, we have an equivalence of the
following data:
\begin{equation}
(E,f_{+-}\!= \unit_n,A_\CT)\ \sim\
(\tilde{E},\tilde{f}_{+-},\tilde{A}_\CT= 0)~,
\end{equation}
similarly to the holomorphic vector bundles discussed in the\index{holomorphic!vector bundle}
previous section.

\paragraph{Cauchy-Riemann structures.}\label{pCRmanifolds} A\index{Cauchy-Riemann structure}
{\em Cauchy-Riemann structure} on a smooth manifold $M$ of real\index{manifold}
dimension $d$ is an integrable distribution, which is a complex\index{integrable}\index{integrable distribution}
subbundle $\CCDb$ of rank $m$ of the complexified tangent bundle
$T_\FC M$. The pair $(M,\CCDb)$ is called a {\em Cauchy-Riemann
manifold of dimension $d=\dim_\FR M$, of rank $m=\dim_\FC \CCDb$\index{Cauchy-Riemann manifold}\index{manifold}
and of codimension $d-2m$}. In particular, a Cauchy-Riemann
structure on $M$ in the special case $d=2m$ is a complex structure\index{Cauchy-Riemann structure}\index{complex!structure}
on $M$. Thus, the notion of Cauchy-Riemann manifolds generalizes\index{Cauchy-Riemann manifold}\index{manifold}
the one of complex manifolds. Furthermore, given a vector bundle\index{complex!manifold}
$E$ over $M$, the pair $(E,\nabla_\CCDb)$, where $\nabla_\CCDb$ is
a $\CCDb$-connection, is a {\em Cauchy-Riemann vector bundle}.\index{Cauchy-Riemann vector bundle}\index{connection}

\section{Calabi-Yau manifolds}\label{secCalabiYau}\index{Calabi-Yau}\index{manifold}

Calabi-Yau manifolds are compact $d$-dimensional K\"{a}hler manifolds\index{Calabi-Yau}\index{K\"{a}hler!manifold}\index{manifold}
with holonomy group $\sSU(d)$. E.~Calabi conjectured in 1954 that\index{holonomy group}
such manifolds should admit a Ricci-flat metric in every K\"{a}hler\index{Ricci-flat}\index{manifold}\index{metric}
class. In 1971, this conjecture was proven by S.~T.~Yau.

\subsection{Definition and Yau's theorem}\label{subYautheorem}\index{Theorem!Yau}

\paragraph{Calabi-Yau manifolds.} A {\em local Calabi-Yau manifold} is a\index{Calabi-Yau}\index{manifold}
complex K\"{a}hler manifold with vanishing first Chern class. A {\em\index{Chern class}\index{K\"{a}hler!manifold}\index{first Chern class}
Calabi-Yau manifold} is a compact local Calabi-Yau manifold.\index{Calabi-Yau}

The notion of a local Calabi-Yau manifold stems from physicists\index{Calabi-Yau}\index{manifold}
and using it has essentially two advantages: First, one can
consider sources of fluxes on these spaces without worrying about
the corresponding ``drains''. Second, one can easily write down
metrics on many local Calabi-Yau manifolds, as e.g.\ on the\index{Calabi-Yau}\index{manifold}\index{metric}
conifold \cite{PandoZayas:2000sq}. We will sometimes drop the word\index{conifold}
``local'' if the context determines the situation.

\paragraph{Theorem.} (Yau) Yau has proven that for every complex\index{Theorem!Yau}
K\"{a}hler manifold $M$ with vanishing first Chern class $c_1=0$ and\index{Chern class}\index{K\"{a}hler!manifold}\index{first Chern class}\index{manifold}
K\"{a}hler form $J$, there exists a unique Ricci-flat metric on $M$ in\index{K\"{a}hler!form}\index{Ricci-flat}\index{metric}
the same K\"{a}hler class as $J$.

This theorem is particularly useful, as it links the relatively
easily accessible first Chern class to the existence of a\index{Chern class}\index{first Chern class}
Ricci-flat metric. The latter property is hard to check explicitly\index{Ricci-flat}\index{metric}
in most cases, in particular, because no Ricci-flat metric is
known on any (compact) Calabi-Yau manifold. Contrary to that, the\index{Calabi-Yau}\index{manifold}
first Chern class is easily calculated, and we will check the\index{Chern class}\index{first Chern class}
Calabi-Yau property of our manifolds in this way.\index{Calabi-Yau}\index{manifold}

\paragraph{Holonomy of a Calabi-Yau manifold.} Ricci-flatness of a\index{Calabi-Yau}\index{Ricci-flat}\index{manifold}
$d$-dimensional complex manifold $M$ implies the vanishing of the\index{complex!manifold}
trace part of the Levi-Civita connection, which in turn restricts\index{Levi-Civita connection}\index{connection}
the holonomy group of $M$ to $\sSU(d)$. In fact, having holonomy\index{holonomy group}
group $\sSU(d)$ is equivalent for a $d$-dimensional compact
complex manifold to being Calabi-Yau. For such manifolds with\index{Calabi-Yau}\index{complex!manifold}\index{manifold}
holonomy group $\sSU(d)$, it can furthermore be shown that\index{holonomy group}
$h^{0,d}=h^{d,0}=1$ and $h^{0,i}=h^{i,0}=0$ for $1<i<d$. The
nontrivial element of $H^{d,0}(M)$ defines the {\em holomorphic
volume form} $\Omega^{d,0}$, one of the key properties of a\index{holomorphic!volume form}
Calabi-Yau manifold, which we will exploit to define the action of\index{Calabi-Yau}\index{manifold}
holomorphic Chern-Simons theory, see section \ref{sshCStheory}.\index{Chern-Simons theory}
Arranging the Hodge numbers similarly as in Pascal's triangle, one\index{Hodge number}
obtains the {\em Hodge diamond}, which looks, e.g.\ for $d=2$ as\index{Hodge diamond}
\begin{equation}
\begin{array}{ccccc} & & h^{0,0} & &\\
& h^{1,0}& & h^{0,1} &\\
h^{2,0} & & h^{1,1} & & h^{0,2}\\
& h^{2,1}& & h^{1,2} &\\
& & h^{2,2} & &
\end{array}~~~\ =\ ~~~\begin{array}{ccccc} & & 1 & &\\
& 0& & 0 &\\
1 & & 20 & & 1\\
& 0 & & 0 &\\
& & 1 & &
\end{array}~.
\end{equation}

\paragraph{Equivalent definitions of Calabi-Yau manifolds.} Let us\index{Calabi-Yau}\index{manifold}
summarize all equivalent conditions on a compact complex manifold\index{complex!manifold}
$M$ of dimension $n$ for being a Calabi-Yau manifold:\index{Calabi-Yau}
\begin{itemize}
\item $M$ is a K\"{a}hler manifold\index{K\"{a}hler!manifold}\index{manifold}
with vanishing first Chern class.\index{Chern class}\index{first Chern class}
\item $M$ admits a Levi-Civita connection with $\sSU(n)$ holonomy.\index{Levi-Civita connection}\index{connection}
\item $M$ admits a nowhere vanishing holomorphic $(n,0)$-form
$\Omega^{n,0}$.
\item $M$ admits a Ricci-flat K\"{a}hler metric.\index{K\"{a}hler!metric}\index{Ricci-flat}\index{metric}
\item $M$ has a trivial canonical bundle.\index{canonical bundle}
\end{itemize}

\paragraph{Deformations of Calabi-Yau manifolds.} Let us briefly\index{Calabi-Yau}\index{manifold}
comment on the moduli space parameterizing deformations of\index{moduli space}
Calabi-Yau manifolds, which preserve Ricci flatness. For a more\index{Calabi-Yau}\index{manifold}
general discussion of deformation theory, see section\index{deformation theory}
\ref{secdeformationtheory}.

Consider a Calabi-Yau manifold $M$ with Ricci-flat metric $g$ in\index{Calabi-Yau}\index{Ricci-flat}\index{manifold}\index{metric}
the K\"{a}hler class $J$. Deformations of the metric consist of
pure index type ones and such of mixed type ones $\delta g=\delta
g_{ij}\dd z^i\dd z^j+\delta g_{i\bj}\dd z^i\dd z^\bj+ c.c.$. The
deformation of mixed type are given by elements of $H^{1,1}(M)$
and are associated to deformations of the K\"{a}hler class $J$
which -- roughly speaking -- determines the size of the Calabi-Yau\index{Calabi-Yau}
manifold. Deformations of pure type are associated with elements\index{manifold}
of $H^{1,2}(M)$ and demand a redefinition of coordinates to yield
a Hermitian metric. Therefore the complex structure is deformed,\index{complex!structure}\index{metric}
which determines the shape of the Calabi-Yau manifold.\index{Calabi-Yau}\index{manifold}

\paragraph{Comments on the moduli spaces.} Above we saw that\index{moduli space}
the moduli space of deformations of a Calabi-Yau manifold\index{Calabi-Yau}\index{manifold}
apparently decomposes into a K\"{a}hler moduli space and a complex
structure moduli space. In fact, the situation is unfortunately\index{complex!structure}
more subtle, and we want to briefly comment on this.

Given a K\"{a}hler form $J=\di g_{i\bj}\dd z^i\wedge\dd \bz^\bj$\index{K\"{a}hler!form}
on a Calabi-Yau manifold $M$, there is the positivity constraint\index{Calabi-Yau}\index{manifold}
for volumes $\int_{S_r}J^{\wedge r}>0$ for submanifolds
$S_r\subset M$ with complex dimension $r$. For any allowed
K\"{a}hler structure $J$, $sJ$ is also allowed for $s\in\FR^{>0}$.
Thus, the moduli space of K\"{a}hler forms is a cone.\index{K\"{a}hler!form}\index{moduli space}
Nevertheless, it is well known from string-theoretic arguments
that all elements of $H^{2}(M,\FR)$ should be admitted. The
solution is, to allow neighboring K\"{a}hler cones to exist,
sharing a common wall and interpreting them as belonging to a
Calabi-Yau manifold with different topology, which solves the\index{Calabi-Yau}\index{manifold}
positivity problem. Passing through the wall of a K\"{a}hler cone
changes the topology but preserves the Hodge numbers.\index{Hodge number}

The complex structure moduli space has similar singularities: If\index{complex!structure}\index{moduli space}
the Calabi-Yau manifold is defined by a homogeneous polynomial $P$\index{Calabi-Yau}\index{manifold}
in a projective space, points $z_0$ where
$P(z_0)=\dpar_{z^i}P(z)|_{z=z_0}=0$ are called the discriminant
locus in the moduli space. The Calabi-Yau manifold fails to be a\index{Calabi-Yau}\index{discriminant locus}\index{moduli space}\index{manifold}
complex manifold there, as the tangent space is not well defined,\index{complex!manifold}
but collapses to a point. Note, however, that in string theory,\index{string theory}
such geometric transitions do not cause any problems.

\paragraph{K3 manifolds.} A {\em K3 manifold} is a complex\index{K3 manifold}\index{manifold}
K\"{a}hler manifold $M$ of complex dimension 2 with $\sSU(2)$ holonomy\index{K\"{a}hler!manifold}
and thus it is a Calabi-Yau manifold. All K3 manifolds can be\index{Calabi-Yau}\index{K3 manifold}
shown to be smoothly equivalent. They have Euler number\index{Euler number}\index{smoothly equivalent}
$\chi(M)=24$ and Pontryagin classes $p_q(M)=1$. Their only
nontrivial Hodge number (i.e.\ the Hodge number not fixed in the\index{Hodge number}
Hodge diamond by the Calabi-Yau property) is $h^{1,1}=20$. K3\index{Calabi-Yau}\index{Hodge diamond}
manifolds play an important r{\^o}le in string theory\index{string theory}\index{manifold}
compactifications. The K3 manifold's name stems from the three\index{K3 manifold}\index{manifold}
mathematicians Kummer, K\"{a}hler and Kodaira who named it in the
1950s shortly after the K2 mountain was climbed for the first
time.

\paragraph{Rigid Calabi-Yau manifolds.} There is a class of\index{Calabi-Yau}\index{rigid}\index{rigid Calabi-Yau manifold}\index{manifold}
so-called {\em rigid Calabi-Yau manifolds}, which do not allow for
deformations of the complex structure. This fact causes problems\index{complex!structure}
for the mirror conjecture, see section \ref{ssMirror}, as it
follows that the mirrors of these rigid Calabi-Yau manifolds have\index{Calabi-Yau}\index{rigid}\index{rigid Calabi-Yau manifold}\index{manifold}
no K\"{a}hler moduli, which is inconsistent with them being K\"{a}hler
manifolds.\index{K\"{a}hler!manifold}\index{manifold}

\subsection{Calabi-Yau 3-folds}\label{ssCY3folds}\index{Calabi-Yau}

Calabi-Yau 3-folds play a central r{\^o}le in the context of string\index{Calabi-Yau}
compactification, see section \ref{typetwosuperstrings},
\ref{pCompactification}. A ten-dimensional string theory is\index{string theory}
usually split into a four-dimensional theory and a six-dimensional
$\CN=2$ superconformal theory. For a theory to preserve $\CN=2$
supersymmetry, the manifold has to be K\"{a}hler, conformal\index{super!symmetry}\index{manifold}
invariance demands Ricci-flatness. Altogether, the six-dimensional\index{Ricci-flat}
theory has to live on a Calabi-Yau 3-fold.\index{Calabi-Yau}

\paragraph{Triple intersection form.} On a Calabi-Yau 3-fold $M$, one\index{Calabi-Yau}
can define a topological invariant called the {\em triple
intersection form}:
\begin{equation}
I^{1,1}:\left(H_{\bar{\dpar}}^{1,1}(M)\right)^{\wedge
  3}\ \rightarrow\ \FR,~~~
I^{1,1}(A,B,C)\ :=\ \int_M A\wedge B\wedge C~.
\end{equation}

\paragraph{Calabi-Yau manifolds in weighted projective spaces.}\index{Calabi-Yau}\index{weighted projective spaces}\index{manifold}
Calabi-Yau manifolds can be described by the zero locus of
polynomials in weighted projective spaces, which is the foundation\index{weighted projective spaces}
of toric geometry. For example, a well-known group of Calabi-Yau\index{Calabi-Yau}
manifolds are the {\em quintics} in $\CPP^4$ defined by a\index{quintic}\index{manifold}
homogeneous\footnote{Homogeneity follows from the fact that
$p(\lambda z_0,\ldots ,\lambda z_5)$ has to vanish for all $\lambda$.}
quintic polynomial $q(z_0,\ldots z_4)$:\index{quintic}
\begin{equation}
M_q\ =\ \{(z_0,\ldots ,z_4)\in\CPP^4:q(z_0,\ldots ,z_4)=0\}~.
\end{equation}
Another example is the complete intersection of two cubics in
$\CPP^5$:
\begin{equation}
M_c\ =\ \{(z_0,\ldots ,z_4)\in\CPP^5:c_1(z_0,\ldots ,z_5)=c_2(z_0,\ldots ,z_5)=0\}~,
\end{equation}
where $c_1$ and $c_2$ are homogeneous cubic polynomials.

\paragraph{Calabi-Yau manifolds from vector bundles over\index{Calabi-Yau}\index{manifold}
$\CPP^1$.}\label{pCYoverCP1} A very prominent class of local
Calabi-Yau manifolds can be obtained from the vector bundles\index{Calabi-Yau}\index{manifold}
$\CO(a)\oplus\CO(b)\rightarrow \CPP^1$, where the Calabi-Yau
condition of vanishing first Chern class amounts to $a+b=-2$.\index{Calabi-Yau}\index{Chern class}\index{first Chern class}

To describe these bundles, we will always choose the standard
inhomogeneous coordinates $\lambda_\pm$ on the patches $U_\pm$\index{homogeneous coordinates}\index{inhomogeneous coordinates}
covering the base $\CPP^1$, together with the coordinates
$z^1_\pm$ and $z^2_\pm$ in the fibres over the patches $U_\pm$.
The transition functions on the overlap are implicitly given by\index{transition function}
\begin{equation}
z^1_+\ =\ \lambda_+^az^1_-~,~~~z^2_+\ =\ \lambda_+^bz^2_-~,~~~\lambda_+\ =\ \frac{1}{\lambda_-}~.
\end{equation}
The holomorphic volume form on these spaces, whose existence is\index{holomorphic!volume form}
granted by vanishing of the first Chern class, can be defined to\index{Chern class}\index{first Chern class}
be $\Omega^{3,0}_\pm=\pm\dd z_\pm^1\wedge\dd z_\pm^2\wedge\dd
\lambda_\pm$.

In more physical terms, this setup corresponds to a
$(\beta,\gamma)$-system of weight $a/2$ (and $b/2$), where the two\index{bg-system@$(\beta,\gamma)$-system}
bosonic fields describe the sections of the $\CO(a)\oplus\CO(b)$
vector bundle over $\CPP^1$.

One of the most common examples is the bundle
$\CO(0)\oplus\CO(2)\rightarrow \CPP^1$, which is, e.g., the
starting point in the discussion of Dijkgraaf and Vafa relating
matrix model computations to effective superpotential terms in\index{matrix model}
supersymmetric gauge theories \cite{Dijkgraaf:2002fc}. Switching
to the coordinates $x=z^1_+$, $u=2z^2_-$, $v=2z^2_+$, $y=2\lambda
z^2_-$, we can describe the above Calabi-Yau as $\FC\times A_1$,\index{Calabi-Yau}
where the $A_1$ singularity is given by\footnote{In general, $A_k$
is the space $\FC^2/\RZ_{k+1}$.} $uv-y^2=0$. Note that $A_1$ is a
local K3 manifold.\index{K3 manifold}\index{manifold}

Another example is the resolved conifold\index{conifold}\index{resolved conifold}
$\CO(-1)\oplus\CO(-1)\rightarrow \CPP^1$, which we will discuss in
the following section. Note that the (projective) twistor space of\index{twistor}\index{twistor!space}
$\FC^4$, $\CO(1)\oplus\CO(1)\rightarrow \CPP^1$, is not a
Calabi-Yau manifold, however, it can be extended by fermionic\index{Calabi-Yau}\index{manifold}
coordinates to a Calabi-Yau supermanifold, see section\index{Calabi-Yau supermanifold}\index{super!manifold}
\ref{ssupertwistorspaces}.

\subsection{The conifold}\label{ssConifold}\index{conifold}

\paragraph{The conifold.} The {\em conifold} is the algebraic\index{conifold}
variety $\CCC$ defined by the equation
\begin{equation}\label{conifolddefinition}
f(w)\ =\ w_1^2+w_2^2+w_3^2+w_4^2\ =\ 0
\end{equation}
in $\FC^4$ as a complex three-dimensional subspace of codimension
1. One immediately notes that in the point $\vec{w}=(w_i)=0$, the
tangent space of $\CCC$ collapses to a point which is indicated by
a simultaneous vanishing of the defining equation $f(w)$ and all
of its derivatives. Such points on a algebraic variety are called\index{algebraic variety}
{\em double points} and are points at which varieties fail to be\index{double points}
smooth. Thus, $\CCC$ is only a manifold for $w\neq 0$. As we will\index{manifold}
see later on, there are two possible ways of repairing this
singularity: a resolution and a deformation (or smoothing). Away
from the origin, the shape is most efficiently determined by
intersecting $\CCC$ with a seven sphere in $\FR^8$:
\begin{equation}
|w_1|^2+|w_2|^2+|w_3|^2+|w_4|^2\ =\ r^2~.
\end{equation}
Splitting $w_i$ in real and imaginary components $x_i$ and $y_i$,
one obtains the equations
$\vec{x}^2-2\di\vec{x}\cdot\vec{y}-\vec{y}\,^2=0$ for the conifold\index{conifold}
and obviously $\vec{x}^2+\vec{y}\,^2=r^2$ for the sphere. The
intersection is given by the three equations
\begin{equation}
\vec{x}\cdot\vec{y}\ =\ 0,~~~\vec{x}^2\ =\ \vec{y}\,^2\ =\ \frac{r^2}{2}~.
\end{equation}
The last two equations define 3-spheres, the first one reduces one
3-sphere to a 2-sphere, since fixing $\vec{x}$ requires $\vec{y}$
to be orthogonal, leaving an $S^2$. The radius of the spheres is
$r/\sqrt{2}$, so that the conifold is indeed a cone over the base\index{conifold}
$B=S^2\times S^3$. This base space is also known as the space
$T^{1,1}$,
\begin{equation}
B\ =\ T^{1,1}\ =\ \sSU(2)\times \frac{\sSU(2)}{\sU(1)}\ \cong\ 
\frac{\sSO(4)}{\sU(1)}~.
\end{equation}

By changing coordinates to $z_{1,3}=w_3\pm \di w_4$ and
$z_{2,4}=\di w_1\mp w_2$, one obtains another defining equation
for $\CCC$:
\begin{equation}
z_1z_3-z_2z_4\ =\ 0~.
\end{equation}
This equation leads to a definition using the determinant of a
matrix, which will be quite useful later on:
\begin{equation}
\mathcal{W}\ =\ \left(\begin{array}{cc}z_1 & z_2 \\ z_4 & z_3
\end{array}\right),~~~\det \mathcal{W}\ =\ 0~.
\end{equation}
From this matrix, one can introduce the radial coordinate of the
conifold by\index{conifold}
\begin{equation}
r=\tr (\mathcal{W}^{\dagger} \mathcal{W})\in \FR
\end{equation}
which parameterizes the distance from the origin in $\FC^4$:
$\vec{x}^2+\vec{y}\hspace{0.02cm}{}^2=r^2$. We see that the
geometry is invariant under $r\rightarrow \lambda r$, so that
$\dpar_r$ is a Killing vector.\index{Killing vector}

Let the angles of the $S^3$ be denoted by $(\theta_1,\phi_1,\psi)$
and the ones of the $S^2$ by $(\theta_2,\phi_2)$. Then by taking
$\psi$ and combining it with the radius, to $v=r\de^{\di\psi}$,
one gets a complex cone over $\CPP^1\times \CPP^1$.

\begin{figure}[h]
\centerline{\includegraphics[width=15cm,totalheight=4.7cm]
{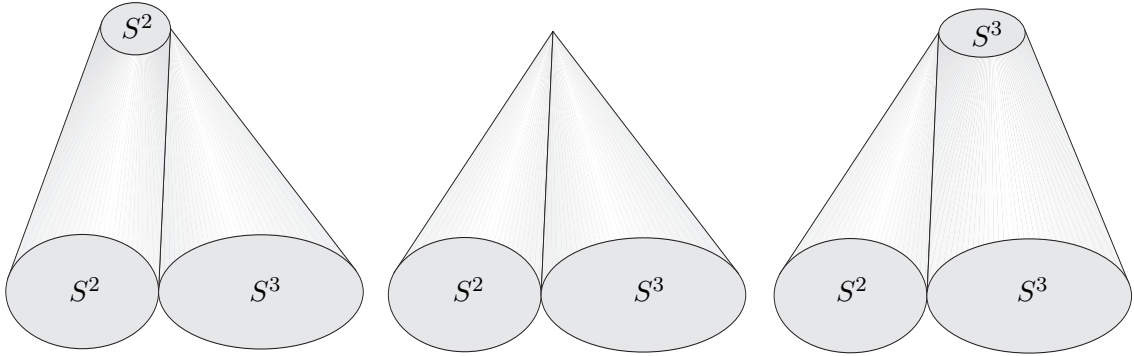}} \setlength{\unitlength}{1cm}
\begin{picture}(0.0,0.0)(+0.4,-0.1)
\put(2.1,4.8){\makebox(0,0)[c]{$S^2$}}
\put(13.4,4.7){\makebox(0,0)[c]{$S^3$}}
\put(1.4,1.3){\makebox(0,0)[c]{$S^2$}}
\put(3.8,1.3){\makebox(0,0)[c]{$S^3$}}
\put(6.5,1.3){\makebox(0,0)[c]{$S^2$}}
\put(8.9,1.3){\makebox(0,0)[c]{$S^3$}}
\put(11.6,1.3){\makebox(0,0)[c]{$S^2$}}
\put(14.0,1.3){\makebox(0,0)[c]{$S^3$}}
\end{picture}
\caption[]{The resolved, the singular and the deformed conifolds.}\index{conifold}\index{deformed conifold}
\label{fig:conifoldfig}
\end{figure}

\paragraph{The deformed conifold.} The deformation $\CCC_{\mathrm{def}}$\index{conifold}\index{deformed conifold}
of the conifold is obtained by deforming the defining equation
\eqref{conifolddefinition} to
\begin{equation}
f\ =\ w_1^2+w_2^2+w_3^2+w_4^2\ =\ z_1z_3-z_2z_4\ =\ \eps~.
\end{equation}
Due to $\vec{x}^2-\vec{y}^2=\eps$, the range of the radial
coordinate $r^2=\vec{x}^2+\vec{y}^2$ is $\eps \leq r<\infty$.
Thus, the tip of the conifold was pushed away from the origin to\index{conifold}
the point $r^2=\eps^2$, which corresponds to $\vec{x}^2=\eps^2$,
$\vec{y}^2=0$. The deformed conifold can be identified with\index{conifold}\index{deformed conifold}
$T^*S^3$. In the case of the singular conifold, the base\index{singular conifold}
$S^3\times S^2$ completely shrank to a point. Here, we note that
the $S^3$ at the tip, given by $\vec{x}^2=\eps$, has finite radius
$r=\eps$ and only the $S^2$ given by $\vec{x}\cdot\vec{y}=0$,
$\vec{y}^2=0$ shrinks to a point. This is depicted in figure
\ref{fig:conifoldfig}.

\paragraph{The resolved conifold.} The {\em resolved conifold}~~$\CCC_{\mathrm{res}}$ is\index{conifold}\index{resolved conifold}
defined by replacing the defining equation of the conifold $\det
\CW=0$ with
\begin{equation}\label{eqdefresconf}
\CW\left(\begin{array}{c} \lambda_\ed\\ \lambda_\zd
\end{array}\right)\ =\ \left(\begin{array}{cc}z_1 & z_2 \\ z_4 & z_3
\end{array}\right)\left(\begin{array}{c} \lambda_\ed\\ \lambda_\zd
\end{array}\right)\ =\ 
\left(\begin{array}{c} 0\\0
\end{array}\right)~~~\mbox{for}~~(\lambda_\ald)\ \neq\ \left(\begin{array}{c} 0\\0
\end{array}\right)~.
\end{equation}
Here, $(\lambda_\ald)\neq 0$ is a homogeneous coordinate on the
Riemann sphere $\CPP^1\cong S^2$. Switching to the inhomogeneous\index{Riemann sphere}
coordinates $\lambda_+:=\frac{\lambda_\zd}{\lambda_\ed}$ and
$\lambda_-:=\frac{\lambda_\ed}{\lambda_\zd}$, we note that
solutions to \eqref{eqdefresconf} are of the form
\begin{equation}
\CW\ =\ \left(\begin{array}{cc} -z_2 \lambda_+ & z_2 \\
-z_3\lambda_+ & z_3
\end{array}\right)\ =\ \left(\begin{array}{cc} z_1 & z_1\lambda_-\\
z_4 & z_4\lambda_-
\end{array}\right)~,
\end{equation}
and thus the coordinates $(z_2,z_3,\lambda_+)$ and
$(z_1,z_4,\lambda_-)$ describe $\CCC_{\mathrm{res}}$ on two
patches $U_\pm$ with transition functions $z_1=-\lambda_+ z_2$ and\index{transition function}
$z_4=-\lambda_+ z_3$. Up to a sign, which can easily be absorbed
by a redefinition of the coordinates, this is the rank two vector
bundle $\CO(-1)\oplus\CO(-1)\rightarrow \CPP^1$. Contrary to the
case of the deformed conifold, the $S^3$ at the tip vanishes while\index{conifold}\index{deformed conifold}
the $S^2$ keeps its finite size.

\paragraph{Metric on the conifold.} Recall that the metric for a\index{conifold}\index{metric}
{\em real} $d$-dimensional cone takes the form
\begin{equation}
g_{mn}\dd x^m\dd x^n\ =\ \dd \rho^2+\rho^2 h_{ij}\dd x^i\dd x^j,
\end{equation}
where $h_{ij}$ is the metric on the $(d-1)$-dimensional base\index{metric}
space. If this base is not the space $S^{d-1}$, there is a
singularity at $\rho=0$. In the case of the conifold, the base\index{conifold}
manifold is $S^2\times S^3$, and thus the singularity at $\rho=0$\index{manifold}
is the one already present in the discussion above. A detailed
discussion of the explicit form of the natural metric on the\index{metric}
conifold is found in \cite{PandoZayas:2000sq} and the references\index{conifold}
therein.

\paragraph{The conifold transition.}\label{pconifoldtrans} The transition from a\index{conifold}\index{conifold transition}
deformed conifold through a singular conifold to a resolved\index{deformed conifold}\index{singular conifold}
conifold is an allowed process in string theory which amounts to a\index{string theory}
topology change. An application of this transition is found in the
famous large $N$ duality in \cite{Vafa:2000wi}: In type IIA string
theory compactified on the deformed conifold, i.e.\ on $T^*S^3$,\index{conifold}\index{deformed conifold}\index{string theory}
wrapping $N$ $D6$-branes around the $S^3$ produces $U(N)$
Yang-Mills theory in the remaining four noncompact directions\index{Yang-Mills theory}
filling spacetime. In the large $N$ limit, this is equivalent to
type IIA string theory on the small resolution, i.e. on\index{string theory}
$\CO(-1)\oplus \CO(-1)\rightarrow\CPP^1$. The inverse process is
found in the mirror picture of this situation: $N$ $D5$-branes
wrapped around the sphere of the small resolution give rise to
$U(N)$ Yang-Mills theory, the large $N$ limit corresponds to type\index{Yang-Mills theory}
IIB string theory compactified on the deformed conifold $T^*S^3$.\index{conifold}\index{deformed conifold}\index{string theory}

\section{Deformation theory}\label{secdeformationtheory}\index{deformation theory}

Deformation theory is an important tool in twistor theory as well\index{deformation theory}\index{twistor}
as in the Kodaira-Spencer theory of gravity\index{Kodaira-Spencer theory}
\cite{Bershadsky:1993cx}, the closed string theory corresponding\index{closed string}\index{string theory}
to the topological B-model. In the former theory, one considers\index{topological!B-model}
deformations of a $\CPP^1$ which is holomorphically embedded in an
open subspace of $\CPP^3$. These deformations are called {\em
relative deformations}. The Kodaira-Spencer theory of gravity, on\index{Kodaira-Spencer theory}\index{relative deformation}
the other hand, is a theory which describes the deformation of the
total complex structure of a Calabi-Yau manifold as a result of\index{Calabi-Yau}\index{complex!structure}\index{manifold}
closed string interactions.\index{closed string}

\subsection{Deformation of compact complex
manifolds}\label{ssDeformationCCM}\index{complex!manifold}\index{manifold}

\paragraph{Deformation of complex structures.} Consider a complex\index{complex!structure}
manifold $M$ covered by patches $U_a$ on which there are\index{manifold}
coordinates $z_a=(z_a^i)$ together with transition functions\index{transition function}
$f_{ab}$ on nonempty intersections $U_a\cap U_b\neq \varnothing$
satisfying the compatibility condition (cf. section
\ref{subseccohomology}, \ref{pholvecinCech}) $f_{ac}=f_{ab}\circ
f_{bc}$ on all $U_a\cap U_b\cap U_c\neq \varnothing$. A {\em
deformation of the complex structure} is obtained by making the\index{complex!structure}\index{deformation of the complex structure}
coordinates $z_a$ and the transition functions $f_{ab}$ depend on\index{transition function}
an additional set of parameters $t=(t^1,t^2,\ldots )$ such that
\begin{equation}\label{deformation}
z_a(t)\ =\ f_{ab}(t,z_b(t))\eand
f_{ac}(t,z_c(t))\ =\ f_{ab}(t,f_{bc}(t,z_c(t)))~,
\end{equation}
and $z_a(0)$ and $f_{ab}(0)$ are the coordinates and transition
functions we started from.\index{transition function}

\paragraph{Infinitesimal deformations.} One linearizes the second
equation in \eqref{deformation} by differentiating it with respect
to the parameter $t$ and considering
$Z_{ac}:=\derr{f_{ac}}{t}|_{t=0}$. This leads to the {\em
linearized cocycle condition}
\begin{equation}\label{lincocylcecond}
Z_{ac}\ =\ Z_{ab}+Z_{bc}~,
\end{equation}
and thus the vector field $Z_{ac}$ is an element of the \v{C}ech
1-cocycles on $M$ with values in the sheaf of germs of holomorphic\index{cocycles}\index{germs}\index{sheaf}
vector fields $\frh$ (see section \ref{subseccohomology},
\ref{pAbeliancech}). Trivial deformations, on the other hand, are
those satisfying $f_{ab}(t,h_b(t,z_b(t)))=h_a(t,z_a(t))$, where
the $h_a$ are holomorphic functions for fixed $t$, as then the
manifolds for two arbitrary parameters of $t$ are biholomorphic\index{biholomorphic}\index{manifold}
and thus equivalent. Infinitesimally, this amounts to
$Z_{ab}=Z_a-Z_b$, where $Z_a=\derr{h_a}{t}$. The latter equation
is the Abelian coboundary condition, and thus we
conclude\footnote{being slightly sloppy} that nontrivial
infinitesimal deformations of the complex structure of a complex\index{complex!structure}
manifold $M$ are given by the first \v{C}ech cohomology group\index{Cech cohomology@\v{C}ech cohomology}\index{manifold}
$\check{H}(M,\frh)$.

These considerations motivate the following theorem:

\paragraph{Theorem.} (Kodaira-Spencer-Nirenberg) If\index{Theorem!Kodaira}
$\check{H}^1(M,\frh)=0$, any small deformation of $M$ is trivial.
If $\check{H}^1(M,\frh)\neq 0$ and $\check{H}^2(M,\frh)=0$ then
there exists a complex manifold $\CM$ parameterizing a family of\index{complex!manifold}\index{manifold}
complex structures on $M$ such that the tangent space to $\CM$ is\index{complex!structure}
isomorphic to $\check{H}^1(M,\frh)$.

Thus, the dimension of the \v{C}ech cohomology group\index{Cech cohomology@\v{C}ech cohomology}
$\check{H}^1(M,\frh)$ gives the number of parameters of
inequivalent complex structures on $M$, while\index{complex!structure}
$\check{H}^2(M,\frh)$ gives the {\em obstructions} to the
construction of deformations.

Note that this theorem can also be adapted to be suited for
deformations of complex vector bundles over a fixed manifold $M$.\index{complex!vector bundle}\index{manifold}

\paragraph{Beltrami differential.}\label{pBeltrami} Given a complex\index{Beltrami differential}
manifold $M$ with a Dolbeault operator $\dparb$, one can describe\index{manifold}
perturbations of the complex structure by adding a\index{complex!structure}
$T^{1,0}M$-valued (0,1)-form $A$, which would read in local
coordinates $(z^i)$ as
\begin{equation}
\tilde{\dparb}\ :=\ \dd\bz^\bi\der{\bz^\bi}+\dd\bz^\bi
A^j_\bi\der{z^j}~.
\end{equation}
For such an operator $\tilde{\dparb}$ to define a complex
structure, it has to satisfy $\tilde{\dparb}^2=0$. One can show\index{complex!structure}
that this integrability condition amounts to demanding that\index{integrability}
$\check{H}^2(M,\frh)=0$.

\paragraph{Rigid manifolds.} A complex manifold $M$ with vanishing\index{complex!manifold}\index{rigid}\index{manifold}
first \v{C}ech cohomology group is called {\em rigid}.\index{Cech cohomology@\v{C}ech cohomology}

\paragraph{Example.} Consider the complex line bundles $\CO(n)$\index{O(n)@$\CO(n)$}
over the Riemann sphere $\CPP^1$. The \v{C}ech cohomology group\index{Cech cohomology@\v{C}ech cohomology}\index{Riemann sphere}
$\check{H}^0(\CPP^1,\CO(n))$ vanishes for $n<0$ and amounts to
global sections of $\CO(n)$ otherwise. As $\frh=\CO(2)$, we\index{O(n)@$\CO(n)$}
conclude by using Serre duality\footnote{The spaces
$\check{H}^i(M,\CO(E))$ and $\check{H}^{n-i}(M,\CO(E\dual\otimes
\Lambda^{n,0}))$ are dual, where $M$ is a compact complex manifold\index{complex!manifold}\index{manifold}
of dimension $n$, $E$ a holomorphic vector bundle and\index{holomorphic!vector bundle}
$\Lambda^{n,0}$ a $(n,0)$-form. One can then pair elements of
these spaces and integrate over $M$.} that $\dim
\check{H}^1(\CPP^1,\CO(2))=\dim \check{H}^0(\CPP^1,\CO(-4))=0$ and
thus $\CO_{\CPP^1}(n)$ is rigid.\index{rigid}

\subsection{Relative deformation theory}\index{deformation theory}\index{relative deformation}

\paragraph{Normal bundle.}\label{pnormalbundle} Given a manifold $X$\index{normal bundle}\index{manifold}
and a submanifold $Y\subset X$, we define the {\em normal bundle}
$\CN$ to $Y$ in $X$ via the short exact sequence\index{short exact sequence}
\begin{equation}
1\ \rightarrow\  TY\ \rightarrow\  TX|_Y\ \rightarrow\  \CN\ \rightarrow\  1~.
\end{equation}
Therefore, $\CN=\frac{TX|_Y}{TY}$. This space can roughly be seen
as the local orthogonal complement to $Y$ in $X$. For complex
manifolds, it is understood that one considers the holomorphic\index{complex!manifold}\index{manifold}
tangent spaces $T^{1,0}X$ and $T^{1,0}Y$.

\paragraph{Infinitesimal motions.} Deformations of $Y$ in $X$
will obviously be described by elements of
$\check{H}^0(Y,\CO(\CN))$ at the infinitesimal level, but let us
be more explicit.

Assume that $Y$ is covered by patches $U_a$ with coordinates
$z_a(t)$ and transition functions $f_{ab}(t,z_b(t))$ satisfying\index{transition function}
the linearized cocycle condition \eqref{lincocylcecond}.
Furthermore, let $h_a(t,z_a(t))$ be a family of embeddings of
$U_a$ into $X$, which is holomorphic for each $t$ in the
coordinates $z_a(t)$ and satisfies the conditions
\begin{equation}\label{reldefconds}
h_a(0,z_a(0))\ =\ \id\eand
h_b(t,z_b(t))\ =\ h_a(t,f_{ab}(t,z_b(t)))~~~\mbox{on}~~~U_a\cap U_b~.
\end{equation}
One can again linearize the latter condition and consider the
vectors only modulo tangent vectors to $Y$ (which would correspond
to moving $Y$ tangent to itself in $X$, leaving $Y$ invariant).
One obtains $\derr{h_a}{t}=\derr{h_b}{t}$ and therefore these
vector fields define global sections of the normal bundle.\index{normal bundle}

Obstructions to these deformations can be analyzed by considering
the second order expansion of \eqref{reldefconds}, which leads to
the condition that the first \v{C}ech cohomology group\index{Cech cohomology@\v{C}ech cohomology}
$\check{H}^1(Y,\CO(\CN))$ must be trivial.

Altogether, we can state that

\paragraph{Theorem.} (Kodaira) If $\check{H}^1(Y,\CO(\CN))=0$\index{Theorem!Kodaira}
then there exists a $d=\dim_\FC \check{H}^0(Y,\CO(\CN))$ parameter
family of deformations of $Y$ inside $X$.

\paragraph{Examples.} Consider a projective line $Y=\CPP^1$ embedded in
the complex projective space $\CPP^3$ and let $X$ be a\index{complex!projective space}
neighborhood of $Y$ in $\CPP^3$. The normal bundle is just\index{normal bundle}
$\CN=\CO(1)\oplus\CO(1)$ and we have furthermore
\begin{equation}
\check{H}^0(Y,\CO(\CN))\ \cong\ \check{H}^0(Y,\CO(1))\oplus\check{H}^0(Y,\CO(1))\ \cong\ \FC^4\eand
\check{H}^1(Y,\CO(\CN))\ =\ 0~.
\end{equation}
We will make extensive use of this example later in the context of
the twistor correspondence.\index{twistor}\index{twistor!correspondence}

As another example, consider the resolved conifold\index{conifold}\index{resolved conifold}
$X=\CO(-1)\oplus\CO(-1)\rightarrow \CPP^1$, which we discussed 
above. There are no deformations of the base space $Y=\CPP^1$
inside this vector bundle, as
$\check{H}^0(Y,\CO(\CN))=\varnothing$, and thus $Y$ is rigid in\index{rigid}
$X$. For example, if one would wrap D-branes around this $\CPP^1$,\index{D-brane}
they are fixed and cannot fluctuate.

\chapter{Supergeometry}\label{chsupergeometry}

The intention of this chapter is to give a concise review of the
geometric constructions motivated by supersymmetry and fix the\index{super!symmetry}
relevant conventions. Furthermore, we discuss so-called exotic
supermanifolds, which are supermanifolds with additional even\index{exotic!supermanifold}\index{super!manifold}
nilpotent directions, reporting some novel results.

Physicists use the prefix ``super'' to denote objects which come
with a $\RZ_2$-grading. With this grading, each superobject can be\index{Z2-grading@$\RZ_2$-grading}
decomposed into an {\em even} or {\em bosonic} part and an {\em
odd} or {\em fermionic} part, the latter being nilquadratic. It
can thus capture the properties of the two fundamental species of
elementary particles: bosons (e.g.\ photons) and fermions (e.g.\
electrons).

The relevant material to this chapter is found in the following
references: \cite{Argyres,Bilal:2001nv,Lykken:1996xt,Gates:1983nr}
(general supersymmetry),\index{super!symmetry}
\cite{DeWitt:1992cy,Berezin:1966nc,Maninbook,Cartier:2002zp}
(superspaces and supermanifolds),\index{super!manifold}\index{super!space}
\cite{Eastwood:1986,Eastwood:1992,Saemann:2004tt} (exotic
supermanifolds and thickenings).\index{exotic!supermanifold}\index{super!manifold}\index{thickening}

\section{Supersymmetry}\index{super!symmetry}

\paragraph{Need for supersymmetry.} The Coleman-Mandula theorem\index{super!symmetry}\index{Theorem!Coleman-Mandula}
\cite{Coleman:1967ad} states that if the S-matrix of a quantum
field theory in more than 1+1 dimensions possesses a symmetry
which is not a direct product with the Poincar{\'e} group, the
S-matrix is trivial. The only loophole to this theorem
\cite{Haag:1974qh} is to consider an additional $\RZ_2$-graded
symmetry which we call {\em supersymmetry} (SUSY). Although SUSY\index{super!symmetry}
was introduced in the early 1970s and led to a number of
aesthetically highly valuable theories, it is still unknown if it
actually plays any r{\^o}le in nature. The reason for this is mainly
that supersymmetry -- as we have not detected any superpartners to\index{super!symmetry}
the particle spectrum of the standard model -- is broken by some
yet unknown mechanism. However, there are some phenomenological
hints for the existence of supersymmetry from problems in the\index{super!symmetry}
current non-supersymmetric standard model of elementary particles.
Among those are the following:
\begin{itemize}
\item The gauge couplings of the standard model seem to unify at
$M_U\sim 2\cdot 10^{16}$GeV in the minimal supersymmetric standard
model (MSSM), while there is no unification in non-SUSY theories.\index{minimal supersymmetric standard model}
\item The {\em hierarchy problem}, i.e.\ the mystery of the\index{hierarchy problem}
unnaturally big ratio of the Planck mass to the energy scale of
electroweak symmetry breaking ($\sim 300$GeV), which comes with
problematic radiative corrections of the Higgs mass. In the MSSM,
these corrections are absent.
\item Dark matter paradox: the {\em neutralino}, one of the extra particles in the\index{dark matter paradox}\index{neutralino}
supersymmetric standard model, might help to explain the missing
dark matter in the universe. This dark matter is not observed but
needed for correctly explaining the dynamics in our galaxy and
accounts for $25\%$ of the total matter\footnote{another $70\%$
stem from so-called {\em dark energy}} in our universe.
\end{itemize}
Other nice features of supersymmetry are seriously less radiative\index{super!symmetry}
corrections and the emergence of gravity if supersymmetry is
promoted to a local symmetry. Furthermore, all reasonable string
theories appear to be supersymmetric. Since gravity has eventually
to be reconciled with the standard model (and -- as mentioned in
the introduction -- string theories are candidates with very few
competitors) this may also be considered as a hint.

There might even be a mathematical reason for considering at least
supergeometry: mirror symmetry (see section \ref{ssMirror})\index{mirror symmetry}\index{super!geometry}
essentially postulates that every family of Calabi-Yau manifolds\index{Calabi-Yau}\index{manifold}
comes with a mirror family, which has a rotated Hodge diamond. For\index{Hodge diamond}
some so-called ``rigid'' Calabi-Yau manifolds this is impossible,\index{Calabi-Yau}\index{rigid}\index{manifold}
but there are proposals that the corresponding mirror partners
might be Calabi-Yau supermanifolds \cite{Sethi:1994ch}.\index{Calabi-Yau}\index{Calabi-Yau supermanifold}\index{super!manifold}

Be it as it may, we will soon know more about the phenomenological
value of supersymmetry from the experimental results that will be\index{super!symmetry}
found at the new large hadron collider (LHC) at CERN.

\subsection{The supersymmetry\index{super!symmetry}
algebra}\label{ssSupersymmetryAlgebra}

\paragraph{The supersymmetry algebra.} The supersymmetric extension of the\index{super!symmetry}
Poincar{\'e} algebra on a pseudo-Euclidean four-dimensional space is\index{Poincar{\'e} algebra}
given by
\begin{equation}\label{susyalgebra}
\begin{aligned}
&[P_\rho,M_{\mu\nu}]\ =\ \di(\eta_{\mu\rho}P_\nu-\eta_{\nu\rho}P_\mu)~,\\
&[M_{\mu\nu},M_{\rho\sigma}]\ =\ -\di(\eta_{\mu\rho}M_{\nu\sigma}-\eta_{\mu\sigma}M_{\nu\rho}-
\eta_{\nu\rho}M_{\mu\sigma}+\eta_{\nu\sigma}M_{\mu\rho})~,\\
&[P_\mu,Q_{\alpha i}]\ =\ 0~,~~~[P_\mu,\bar{Q}_\ald^i]\ =\ 0~,~~~\\
&[M_{\mu\nu},Q_{i \alpha}]\ =\ \di (\sigma_{\mu\nu})_\alpha{}^\beta
Q_{i\beta}~,~~~[M_{\mu\nu},\bar{Q}^{i\ald}]\ =\ \di
(\bar{\sigma}_{\mu\nu})^\ald{}_\bed
\bar{Q}^{i\bed}~.\\
&\{Q_{\alpha i},\bar{Q}^j_\bed\}\ =\ 2\sigma_{\alpha\bed}^\mu P_\mu
\delta_i^j~,~~~\{Q_{\alpha i},Q_{\beta
j}\}\ =\ \eps_{\alpha\beta}Z_{ij}~,~~~
\{\bar{Q}^i_{\ald},\bar{Q}^j_\bed\}\ =\ \eps_{\ald\bed}\bar{Z}^{ij}~,
\end{aligned}
\end{equation}
where $\eta_{\mu\nu}$ is either the Euclidean metric, the\index{metric}
Minkowskian one or the Kleinian one with
$(\eta_{\mu\nu})=\diag(-1,-1,+1,+1)$. The generators $P$, $M$, $Q$
and $\bar{Q}$ correspond to translations, Lorentz transformations
and supersymmetry transformations (translations in chiral\index{super!symmetry}
directions in superspace), respectively. The terms\index{super!space}
$Z^{ij}=-Z^{ji}$ are allowed central extensions of the algebra,
i.e.\ $[Z^{ij},\cdot]=0$. We will almost always put them to zero
in the following.

The indices $i,j$ run from $1$ up to the number of
supersymmetries, usually denoted by $\CN$. In four dimensions, the
indices $\alpha$ and $\ald$ take values $1,2$. In particular, we
have therefore $4\CN$ supercharges $Q_{\alpha i}$ and
$\bar{Q}_\ald^i$.

\paragraph{Sigma matrix convention.}\label{pSigmaMatrixConvention}
On four-dimensional Minkowski space, we use
\begin{equation}
\sigma_0\ :=\  \left(\begin{array}{cc}
1 & 0 \\
0 & 1 \\
\end{array}\right)
~~~~~ \sigma_1\ :=\  \left(\begin{array}{cc}
0 & 1 \\
1 & 0 \\
\end{array}\right)
~~~~~ \sigma_2\ :=\  \left(\begin{array}{cc}
0 & -\mathrm{i} \\
\mathrm{i} & 0 \\
\end{array}\right)
~~~~~ \sigma_3\ :=\  \left(\begin{array}{cc}
1 & 0 \\
0 & -1 \\
\end{array}\right)\nonumber
\end{equation}
and thus $\sigma_\mu=(\unit,\sigma_i)$ together with the
definition $\bar{\sigma}_\mu=(\unit,-\sigma_i)$, where $\sigma_i$
denotes the three Pauli matrices. On Euclidean spacetime, we\index{Pauli matrices}
define
\begin{equation}
\sigma_1\ :=\  \left(\begin{array}{cc}
0 & 1 \\
1 & 0 \\
\end{array}\right)
~~~~~ \sigma_2\ :=\  \left(\begin{array}{cc}
0 & -\mathrm{i} \\
\mathrm{i} & 0 \\
\end{array}\right)
~~~~~ \sigma_3\ :=\  \left(\begin{array}{cc}
1 & 0 \\
0 & -1 \\
\end{array}\right)
~~~~~ \sigma_4\ :=\  \left(\begin{array}{cc}
\di & 0 \\
0 & \di \\
\end{array}\right)
\nonumber
\end{equation}
with $\sigma_\mu=(\sigma_i,\di\unit)$,
$\bar{\sigma}_\mu=(-\sigma_i,-\di\unit)$ and on Kleinian space
$\FR^{2,2}$ we choose
\begin{equation}
\sigma_1\ :=\  \left(\begin{array}{cc}
0 & 1 \\
-1 & 0 \\
\end{array}\right)
~~~~~ \sigma_2\ :=\  \left(\begin{array}{cc}
0 & -\di \\
-\di & 0 \\
\end{array}\right)
~~~~~ \sigma_3\ :=\  \left(\begin{array}{cc}
1 & 0 \\
0 & -1 \\
\end{array}\right)
~~~~~ \sigma_4\ :=\  \left(\begin{array}{cc}
-\di & 0 \\
0 & -\di \\
\end{array}\right)~.\nonumber
\end{equation}
In the above supersymmetry algebra, we also made use of the\index{super!symmetry}
symbols
\begin{equation}
\sigma^{\nu\mu}{}_\alpha{}^\beta\ :=\ \tfrac{1}{4}(
\sigma^\nu_{\alpha\ald}\bar{\sigma}^{\mu\ald\beta}-
\sigma^\mu_{\alpha\ald}\bar{\sigma}^{\nu\ald\beta})\eand
\bar{\sigma}^{\nu\mu}{}^\ald{}_\bed\ :=\ \tfrac{1}{4}(
\bar{\sigma}^{\nu\ald\alpha}\sigma^{\mu}_{\alpha\bed}-
\bar{\sigma}^{\mu\ald\alpha}\sigma^{\nu}_{\alpha\bed})~.
\end{equation}

\paragraph{Immediate consequences of supersymmetry.} While every\index{super!symmetry}
irreducible representation of the Poincar{\'e} algebra corresponds to\index{Poincar{\'e} algebra}\index{irreducible representation}\index{representation}
a particle, every irreducible representation of the supersymmetry\index{super!symmetry}
algebra corresponds to several particles, which form a
supermultiplet. As $P^2$ commutes with all generators of the\index{supermultiplet}
supersymmetry algebra, all particles in a supermultiplet have the\index{super!symmetry}
same mass. Furthermore, the energy $P_0$ of any state is always
positive, as
\begin{equation}\label{EnergyPositive}
2\sigma^\mu_{\alpha\ald}\langle \psi|P_\mu|\psi \rangle\ =\ \langle
\psi|\{Q_{\alpha i},\bar{Q}^i_\ald\}|\psi \rangle\ =\ ||Q_{\alpha
i}|\psi\rangle||^2+||\bar{Q}^i_\ald|\psi\rangle||^2\ \geq\  0~.
\end{equation}
And finally, we can deduce that a supermultiplet contains an equal\index{supermultiplet}
number of bosonic and fermionic degrees of freedom (i.e.\ physical
states with positive norm). To see this, introduce a parity\index{parity}
operator $\CP$ which gives $-1$ and $1$ on a bosonic and fermionic
state, respectively. Consider then
\begin{equation}
2 \sigma^\mu_{\alpha\ald}\tr(\CP P_\mu)\ =\ \tr(\CP\{Q_{\alpha
i},\bar{Q}^i_\ald\})\ =\ \tr(-Q_{\alpha i}\CP
\bar{Q}^i_\ald+\CP\bar{Q}^i_\ald Q_{\alpha i})\ =\ 0~,
\end{equation}
where we used the facts that $\CP$ anticommutes with $Q_{\alpha
i}$ and that the trace is cyclic. Any non-vanishing $P_\mu$ then
proves the above statement.

\subsection{Representations of the supersymmetry\index{representation}\index{super!symmetry}
algebra}\label{sssusyrep}

\paragraph{Casimir operators of the supersymmetry algebra.}\index{Casimir}\index{super!symmetry}
To characterize all irreducible representations of the\index{irreducible representation}\index{representation}
supersymmetry algebra, we need to know its Casimir operators.\index{Casimir}\index{super!symmetry}
Recall that the Casimirs of the Poincar{\'e} algebra are the mass\index{Poincar{\'e} algebra}
operator $P^2=P_\mu P^\mu$ with eigenvalues $m^2$ together with
the square of the Pauli-Ljubanski vector\index{Pauli-Ljubanski vector}
$W_\mu=\frac{1}{2}\eps_{\mu\nu\rho\sigma}P^\nu W^{\rho\sigma}$
with eigenvalues $-m^2s(s+1)$ for massive and $W_\mu=\lambda
P_\mu$ for massless states, where $s$ and $\lambda$ are the spin
and the helicity, respectively.\index{helicity}

In the super Poincar{\'e} algebra, $P^2$ is still a Casimir, while\index{Casimir}\index{Poincar{\'e} algebra}\index{super!Poincar{\'e} algebra}
$W^2$ has to be replaced by the {\em superspin operator} given by\index{super!spin}
$C^2=C_{\mu\nu}C^{\mu\nu}$ with
\begin{equation}
C_{\mu\nu}\ =\ (W_\mu-\tfrac{1}{4}\bar{Q}^i_\ald\bar{\sigma}^{\ald\beta}_\mu
Q_{i\beta})P_\nu-(W_\nu-\tfrac{1}{4}\bar{Q}^i_\ald\bar{\sigma}^{\ald\beta}_\nu
Q_{i\beta})P_\mu~.
\end{equation}
Thus $C^2=P^2 W^2-\frac{1}{4}(P\cdot W)^2$, and in the massive
case, this operator has eigenvalues $-m^4s(s+1)$, where $s$ is
called the {\em superspin}.\footnote{When having additional\index{super!spin}
conformal invariance, one can define the analog of a {\em
superhelicity}.}\index{helicity}

\paragraph{Massless representations.}\label{pmasslessrep}\index{representation}
First, let us consider massless representations in a frame with
$P_\mu=(E,0,0,E)$ which leads to
\begin{equation}
\sigma^\mu P_\mu\ =\ \left(\begin{array}{cc} 0 & 0 \\ 0 & 2E
\end{array}\right)~.
\end{equation}
From the relation $\{Q_{\alpha
i},\bar{Q}^j_\bed\}=2\sigma_{\alpha\bed}^\mu P_\mu$ one deduces by
a similar argument to \eqref{EnergyPositive} that
$Q_{i1}=\bar{Q}^i_{\dot{1}}=0$. Together with the supersymmetry\index{super!symmetry}
algebra, this also implies that the central charges $Z^{ij}$ have\index{central charge}
to vanish. The remaining supercharges $Q_{i2}$ and
$\bar{Q}^i_{\dot{2}}$ are proportional to fermionic annihilation
and creation operators, respectively. From the commutation
relations of these operators with the generator of rotations
$J_3=M_{12}$, one sees that they indeed lower respectively rise
the helicity of a state by $\frac{1}{2}$. Choosing a highest\index{helicity}
weight state $|h\rangle$ annihilated by all the $Q_{i2}$s, we can
built a supermultiplet by acting with the $\bar{Q}^i_{\dot{2}}$s\index{supermultiplet}
on it. Altogether, one obtains $2^{\CN}$ states due to the
nilpotency of the $\bar{Q}^i_\zd$s.

\paragraph{Massless supermultiplets.} For $\CN=1$, a\index{supermultiplet}
supermultiplet consists of fields of helicity\index{helicity}
$(\lambda_0,\lambda_0+\frac{1}{2})$. Since we do not want to
exceed $\lambda=1$ for physical\footnote{Otherwise, our
supermultiplet will necessarily contain gravitini and gravitons.}\index{supermultiplet}
reasons, we thus find the chiral (scalar) multiplet with
helicities $(0,\frac{1}{2})$ together with its CPT conjugate
$(-\frac{1}{2},0)$ and the vector (gauge) multiplet with
helicities $(\frac{1}{2},1)$ and $(-1,-\frac{1}{2})$.

For $\CN=2$, a supermultiplet consists of fields with helicities\index{supermultiplet}
$(\lambda_0,\lambda_0+\frac{1}{2},\lambda_0+\frac{1}{2},\lambda_0+1)$,
and we therefore find the vector multiplet\index{vector multiplet}
$(0,\frac{1}{2},\frac{1}{2},1)$ and its corresponding CPT
conjugate $(-1,-\frac{1}{2},-\frac{1}{2},0)$. Note that this
multiplet amounts essentially to the sum of a chiral and a vector
multiplet in $\CN=1$ language. Furthermore, there is the\index{vector multiplet}
hypermultiplet, consisting of fields with helicities\index{hypermultiplet}
$(-\frac{1}{2},0,0,\frac{1}{2})$, which can but does not
necessarily have to be its own CPT adjoint.

For $\CN=4$, there is a single multiplet $(-1,4\times
-\frac{1}{2},6\times 0,4\times \frac{1}{2},1)$ with helicities not
larger than one: a gauge potential (helicities $\pm 1$), four Weyl
fermions and their conjugates with helicities $\pm\frac{1}{2}$ and
three complex scalars with helicity $0$.\index{helicity}

\paragraph{Massive representations.}\index{representation}
For massive representations, we choose the frame
$P_\mu=(m,0,0,0)$. By an appropriate $\sU(\CN)$ rotation of the
generators, we can bring the matrix of central charges $Z^{ij}$ to\index{central charge}
a block diagonal form $(Z^{ij})=\diag(z^k)$, where the $z^k$ are
antisymmetric $2\times 2$ matrices. Here, we assumed that $\CN$
was even. If $\CN$ was odd there would be an additional zero
eigenvalue of the matrix $(Z^{ij})$. The supercharges can be
rearranged to fermionic creation and annihilation operators
according to
\begin{equation}
a_\alpha^r\ :=\ \tfrac{1}{\sqrt{2}}\left(Q_\alpha^{(2r-2)+1}+\eps_{\alpha\beta}(Q^{2r}_\beta)^\dagger\right)~,~~~
b_\alpha^r\ :=\ \tfrac{1}{\sqrt{2}}\left(Q_\alpha^{(2r-2)+1}-\eps_{\alpha\beta}(Q^{2r}_\beta)^\dagger\right)
\end{equation}
with $r=1,\ldots ,\frac{\CN}{2}$, for which the only non-vanishing
anticommutators are\index{commutators}
\begin{equation}
\{a_\alpha^r,(a^s_\beta)^\dagger\}\ =\ (2m-q_r)\delta_{rs}\delta_{\alpha\beta}~,~~~
\{b_\alpha^r,(b^s_\beta)^\dagger\}\ =\ (2m+q_r)\delta_{rs}\delta_{\alpha\beta}~,
\end{equation}
where $q_r$ is the upper right entry of $z^r$. The positivity of
the Hilbert space requires $2m\geq |q_r|$ for all $r$. For values
of $q_r$ saturating the boundary, the corresponding operators
$a_\alpha^r$ and $b_\alpha^r$ have to be put to zero.

Thus, we obtain $2\CN-2k$ fermionic oscillators amounting to
$2^{2(\CN-k)}$ states, where $k$ is the number of $q_r$ for which
$2m=|q_r|$. The multiplets for $k>0$ are called {\em short
multiplets} or {\em BPS multiplets}, in the case $k=\frac{\CN}{2}$\index{BPS multiplet}\index{short multiplet}
one calls them {\em ultrashort multiplets}.\index{ultrashort multiplet}

\section{Supermanifolds}\label{secsupermanifolds}\index{super!manifold}

For supersymmetric quantum field theories, a representation of the\index{representation}
super Poincar{\'e} algebra on fields is needed. Such representations\index{Poincar{\'e} algebra}\index{super!Poincar{\'e} algebra}
can be defined by using functions which depend on both commuting
and anticommuting coordinates. Note that such a $\RZ_2$-grading of\index{Z2-grading@$\RZ_2$-grading}
the coordinates comes with a $\RZ_2$-grading of several other
objects, as e.g.\ derivatives, integral forms, vector fields etc.

There are basically three approaches to $\RZ_2$-graded coordinates
on spaces:
\begin{itemize}
\item The first one just introduces a set of Gra{\ss}mann variables,\index{Gra{\ss}mann variable}
which serve as formal parameters in the calculation and take the
r{\^o}le of the anticommuting coordinates. This setup is the one most
commonly used in physics. Deeper formalizations can be found, and
we briefly present the sheaf-theoretic approach, in which a\index{sheaf}
supermanifold is interpreted as an ordinary manifold with a\index{super!manifold}\index{manifold}
structure sheaf enlarged to a supercommutative ring, cf. the\index{sheaf}\index{structure sheaf}
definition of a locally ringed space in \ref{ssSheaves},
\ref{pStructureSheaf}.
\item The second one, pioneered by A.~Rogers and B.~S.~DeWitt,
allows the coordinates to take values in a Gra{\ss}mann algebra. This\index{Gra{\ss}mann algebra}
approach, though mathematically in many ways more appealing than
the first one, has serious drawbacks, as physics seems to be
described in an unnatural manner.
\item A unifying approach has been proposed by A.~Schwarz
\cite{Schwarz:1984} by defining all objects of supermathematics in
a categorial language. This approach, however, also comes with
some problematic aspects, which we will discuss later.
\end{itemize}

\subsection{Supergeneralities}\label{ssSupergeneralities}

\paragraph{$\RZ_2$-grading.} A set $S$ is said to posses a\index{Z2-grading@$\RZ_2$-grading}
{\em $\RZ_2$-grading} if one can associate to each element $s\in
S$ a number $\tilde{s}\in\{0,1\}$, its {\em parity}. If there is a\index{parity}
product structure defined on $S$, the product has furthermore to
respect the grading, i.e.\
\begin{equation}
s_1\cdot
s_2\ =\ s_3~~~\Rightarrow~~~\tilde{s}_3\equiv\tilde{s}_1+\tilde{s}_2\mod
2~.
\end{equation}
In the following, we will sometimes use a tilde over an index to
refer to the grading or parity of the object naturally associated\index{parity}
to that index. Objects with parity 0 are called {\em even}, those
with parity 1 are called {\em odd}.

\paragraph{Supervector space.} A {\em supervector space} is a\index{super!vector space}
$\RZ_2$-graded vector space. In some cases, one considers a {\em
supervector space} as a module over a ring with nilpotent\index{super!vector space}
elements. Here, the multiplication with elements of the ring has
to respect the grading. A supervector space of dimension $m|n$ is\index{super!vector space}
the span of a basis with $m$ even elements and $n$ odd elements.

\paragraph{Sign rule.} A heuristic sign rule which can be used\index{sign rule}
as a guideline for operating with $\RZ_2$-graded objects is the
following: If in a calculation in an ordinary algebra one has to
interchange two terms $a$ and $b$ in a monomial then in the
corresponding superalgebra, one has to insert a factor of\index{super!algebra}
$(-1)^{\tilde{a}\tilde{b}}$.

\paragraph{Supercommutator.} The {\em supercommutator} is the\index{super!commutator}
natural generalization of the commutator for $\RZ_2$-graded rings
reflecting the above sign rule. Depending on the grading of the\index{sign rule}
involved objects, it behaves as a commutator or an anticommutator:
\begin{equation}
\lsc a,b\rsc\ :=\ a\cdot b-(-1)^{\tilde{a}\tilde{b}}b\cdot a~.
\end{equation}
From this definition, we immediately conclude that
\begin{equation}
\lsc a,b\rsc\ =\ -(-1)^{\tilde{a}\tilde{b}}\lsc b,a\rsc~.
\end{equation}
Note that instead of explicitly writing commutators and\index{commutators}
anticommutators in the supersymmetry algebra \eqref{susyalgebra},\index{super!symmetry}
we could also have used supercommutators everywhere.\index{super!commutator}

\paragraph{Super Jacobi identity.} In an associative $\RZ_2$-graded ring $A$, the\index{super!Jacobi identity}
supercommutator satisfies the following {\em super Jacobi\index{super!commutator}
identity}:
\begin{equation}\label{superJacobi}
\lsc a,\lsc b,c \rsc\rsc+(-1)^{\tilde{a}(\tilde{b}+\tilde{c})}\lsc
b,\lsc c,a \rsc\rsc+(-1)^{\tilde{c}(\tilde{a}+\tilde{b})}\lsc
c,\lsc a,b \rsc\rsc\ =\ 0
\end{equation}
for $a,b,c\in A$, as one easily verifies by direct calculation.

\paragraph{Supercommutative rings.} A $\RZ_2$-graded ring $A$ is\index{super!commutative rings}
called {\em supercommutative} if the supercommutator $\lsc a,b\index{super!commutator}
\rsc$ vanishes for all elements $a,b\in A$.

\paragraph{Superalgebra.} One can lift a supervector space $V$ to\index{super!algebra}\index{super!vector space}
a {\em superalgebra} by endowing it with an associative
multiplication respecting the grading (i.e.\
$\widetilde{ab}\equiv\tilde{a}+\tilde{b}\mod 2$) and a unit
$\unit$ with $\tilde{\unit}=0$. If we have an additional bracket
on $V$ which satisfies the super Jacobi identity\index{super!Jacobi identity}
\eqref{superJacobi}, we obtain a corresponding super Lie algebra\index{Lie algebra}\index{super!Lie algebra}
structure.

\paragraph{Super Poisson structure.}\label{psuperPoisson} A\index{super!Poisson structure}
{\em super Poisson structure} is a super Lie algebra structure\index{Lie algebra}\index{super!Lie algebra}
which satisfies the super Jacobi identity and the equations\index{super!Jacobi identity}
\begin{equation}
\lsc f,gh\rsc\ =\ \lsc f,g\rsc h+(-1)^{\tilde{f}\tilde{g}}g \lsc
f,h\rsc\eand\lsc fg,h\rsc\ =\ f \lsc g,h\rsc
+(-1)^{\tilde{g}\tilde{h}}\lsc f,h\rsc g~.
\end{equation}

\paragraph{Supermatrices.} Linear transformations on a supervector
space are described by {\em supermatrices}. Given a supervector\index{super!vector space}
space $V$ of dimension $m|n$, the basis $e$ is a tuple of $m$ even
and $n$ odd elements of $V$, and we will always assume this order
of basis vectors in the following. Even supermatrices are those,
which preserve the parity of the basis vectors and thus have the\index{parity}
block structure
\begin{equation}\label{standardsmatrix}
K=\left(\begin{array}{cc} A & B\\ C & D
\end{array}\right)~,
\end{equation}
with the elements $A$ and $D$ being even and the elements $B$ and
$C$ being odd. The blocks of odd supermatrices have inverse
parities. Note that there are furthermore supermatrices which are
not $\RZ_2$-graded in the above scheme. They are said to have {\em
mixed parity}. Due to their existence, the supermatrices do not\index{parity}
form a supervector space.\index{super!vector space}

\paragraph{Supertrace.} The {\em supertrace} of  a standard\index{super!trace}
supermatrix which has the form \eqref{standardsmatrix} is defined
by
\begin{equation}
\str(K)\ =\ \tr A-\tr D~.
\end{equation}
This definition ensures that $\str(KL)=\str(LK)$, and -- after a
suitable definition -- invariance under transposition of the
matrix $K$. Furthermore, we have
$\str(\unit_{m|n})=\tr(\unit_m)-\tr(\unit_n)=m-n$.

\paragraph{Superdeterminant.} A {\em superdeterminant} is easily defined\index{super!determinant}
by integrating the classical variational law
\begin{equation}
\delta \ln \det K\ =\ \tr(K^{-1}\delta K)
\end{equation}
together with the boundary condition $\sdet(\unit_{m|n})=1$. This
definition yields for a matrix $K$ of the form
\eqref{standardsmatrix}
\begin{equation}
\sdet(K)\ =\ \frac{\det (A-D B^{-1} C)}{\det
(B)}\ =\ \frac{\det(A)}{\det(B)\det(\unit_m-C A^{-1}DB^{-1})}~.
\end{equation}
We will present the derivation of this result in an analogous case
in \ref{pedet} of section \ref{ssExoticCYs}.

Our definition preserves in particular the product rule for
ordinary determinants, i.e.\ we have $\sdet(KL)=\sdet(K)\sdet(L)$.
The superdeterminant $\sdet(\cdot)$ is also called the {\em\index{super!determinant}
Berezinian}.\index{Berezinian}

\paragraph{Almost nilpotent algebra.} An {\em almost nilpotent algebra} is\index{almost nilpotent algebra}
an associative, finite-dimensional, unital, $\RZ_2$-graded
supercommutative algebra in which the ideal of nilpotent elements
has codimension 1.

\subsection{Gra{\ss}mann variables}\index{Gra{\ss}mann variable}

\paragraph{Gra{\ss}mann variables.} Define a set of formal variables\index{Gra{\ss}mann variable}
$\lambda:=\{\theta^i\}$ which satisfy the algebra
\begin{equation}
 \{\theta^i,\theta^j\}\ =\ \theta^i\theta^j+\theta^j\theta^i\ :=\ 0~.
\end{equation}
The elements of this set are called {\em Gra{\ss}mann variables}.\index{Gra{\ss}mann variable}
Trivial consequences of the algebra are their anticommutativity:
$\theta^i\theta^j=-\theta^j\theta^i$ and their nilquadraticy:
$(\theta^i)^2=0$. The parity of a Gra{\ss}mann variables is odd:\index{Gra{\ss}mann variable}\index{parity}
$\tilde{\theta^i}=1$.

\paragraph{Gra{\ss}mann algebras.}\label{pGrassmannalgebras} The algebra generated by a set of\index{Gra{\ss}mann algebra}
$N\in\NN\cup\{\infty\}$ Gra{\ss}mann variables over $\FC$ or $\FR$ is\index{Gra{\ss}mann variable}
called the {\em Gra{\ss}mann algebra} $\Lambda_N$. A Gra{\ss}mann algebra\index{Gra{\ss}mann algebra}
is a $\RZ_2$-graded algebra, and thus every element
$z\in\Lambda_N$ can be decomposed into an even part
$z_0\in\Lambda_0$ with $\tilde{z}_0=0$ and an odd part
$z_1\in\Lambda_1$ with $\tilde{z}_1=1$ as well as into a {\em
body} $z_B\in \Lambda_B:=\Lambda_N\cap \FC$ and a {\em soul}\index{body}\index{soul}
$z_S\in\Lambda_S:=\Lambda_N\backslash \FC$. Note that the soul is
nilpotent, and an element $z$ of the Gra{\ss}mann algebra $\Lambda_N$\index{Gra{\ss}mann algebra}
has the multiplicative inverse
$z^{-1}=\frac{1}{z_B}\sum_{i=0}^N(-\frac{z_S}{z_B})^i$ if and only
if the body is non-vanishing. Elements of a Gra{\ss}mann algebra are\index{Gra{\ss}mann algebra}\index{body}
also called {\em supernumbers}.\index{super!number}

Note that a Gra{\ss}mann algebra is an almost nilpotent algebra.\index{Gra{\ss}mann algebra}\index{almost nilpotent algebra}

\paragraph{Derivatives with respect to Gra{\ss}mann variables.} Recall\index{Gra{\ss}mann variable}
that a derivative is a linear map which annihilates constants and
satisfies a Leibniz rule. For Gra{\ss}mann variables, one easily finds\index{Gra{\ss}mann variable}
that the most appropriate definition of a derivative is
\begin{equation}
\der{\theta^i}(a+\theta^i b)\ :=\ b~,
\end{equation}
where $a$ and $b$ are arbitrary constants in $\theta^i$. Due to
the nilpotency of Gra{\ss}mann variables, this definition fixes the\index{Gra{\ss}mann variable}
derivative completely, and it gives rise to the following super
Leibniz rule:\index{super!Leibniz rule}
\begin{equation}
\der{\theta^i}(ab)\ =\ \left(\der{\theta^i}a\right)b+
(-1)^{\tilde{a}}a\left(\der{\theta^i}b\right)~.
\end{equation}
Note that in our conventions, all the derivatives with respect to
Gra{\ss}mann variables act from the left.\index{Gra{\ss}mann variable}

\paragraph{Integration over Gra{\ss}mann variables.}\label{pGrassmannIntegral}\index{Gra{\ss}mann variable}
The corresponding rule for an integration $\int\dd \theta^i$ is
fixed by demanding that $\int\dd \theta^i$ is a linear functional
and that\footnote{no sum over $i$ implied} $\der{\theta^i}\int
\dd\theta^i f=\int \dd \theta^i \der{\theta^i} f=0$, where $f$ is
an arbitrary function of $\theta^i$. The latter condition is the
foundation of integration by parts and Stokes' formula. Thus we
have to define
\begin{equation}
\int \dd \theta^i~(a+\theta^i b)\ :=\ b~,
\end{equation}
and integration over a Gra{\ss}mann variable is equivalent to\index{Gra{\ss}mann variable}
differentiating with respect to it. This integration prescription
was first introduced by F.~A.~Berezin, one of the pioneers of
Gra{\ss}mann calculus, and is therefore called {\em Berezin
integration}.\index{Berezin integration}

When performing a change of coordinates, the Jacobian is replaced
by the Berezinian, i.e.\ the usual determinant is replaced by the\index{Berezinian}
superdeterminant, and we will encounter several examples for this\index{super!determinant}
later on.

\paragraph{Complex conjugation of Gra{\ss}mann variables.}\index{Gra{\ss}mann variable}
After defining a complex conjugation on
$*:\lambda\rightarrow\lambda$, we call the elements of $\lambda$
{\em complex} Gra{\ss}mann variables and those elements $\xi\in\index{Gra{\ss}mann variable}
\mathrm{span}(\lambda)=\Lambda$ for which $\xi^*=\xi$ {\em real}.

We will have to introduce different explicit antilinear
involutions defining reality conditions for Gra{\ss}mann variables in\index{Gra{\ss}mann variable}\index{antilinear involution}\index{involution}
our discussion later on. However, we can already fix two
conventions: First, our reality conditions will always be
compatible with
\begin{equation}
\overline{\der{\theta}}\ =\ \der{\bar{\theta}}~.
\end{equation}
Furthermore, we adopt the following convention for the conjugation
of products of Gra{\ss}mann variables and supernumbers in general:\index{Gra{\ss}mann variable}\index{super!number}
\begin{equation}
\tau(\theta^1\theta^2)\ =\ \tau(\theta^2)\tau(\theta^1)~~~\mbox{and}~~~\tau(z^1
z^2)\ =\ \tau(z^2)\tau(z^1)~.
\end{equation}
This choice is almost dictated by the fact that we need the
relation $(AB)^\dagger=B^\dagger A^\dagger$ for matrix-valued
superfunctions. A slight drawback here is that the product of two
real objects will be imaginary. This is furthermore the most
common convention used for supersymmetry in Minkowski space, and\index{super!symmetry}
the difference to the convention
$\tau(\theta^1\theta^2)=\tau(\theta^1)\tau(\theta^2)$ is just a
factor of $\di$ in the Gra{\ss}mann generators. A more detailed
discussion can be found in \cite{Cartier:2002zp}.

\subsection{Superspaces}\label{ssSuperspaces}\index{super!space}

\paragraph{Superspace from an enlarged structure sheaf.} A\index{sheaf}\index{structure sheaf}\index{super!space}
{\em superspace} is a pair $(M,\CO_M)$, where $M$ is a topological
space and $\CO_M$ is a sheaf of supercommutative rings such that\index{sheaf}\index{super!commutative rings}
the stalk $\CO_{M,x}$ at any point $x \in M$ is a local ring.
Thus, $(M,\CO_M)$ is a locally ringed space with a structure sheaf\index{sheaf}\index{structure sheaf}
which is supercommutative.

\paragraph{Superspace according to A.~Schwarz.} We will describe the\index{super!space}
categorial approach of A.~Schwarz in more detail in section
\ref{sspartiallformal}.

\paragraph{The space $\FR^{m|n}$.}\label{pRmn} Given a set of $n$
Gra{\ss}mann variables $\{\theta^i\}$, the space $\FR^{0|n}$ is the\index{Gra{\ss}mann variable}
set of points denoted by the formal coordinates $\theta^i$. The
space $\FR^{m|n}$ is the cartesian product $\FR^m\times
\FR^{0|n}$, and we say that $\FR^{m|n}$ is of dimension $m|n$.
This construction straightforwardly generalizes to the complex
case $\FC^{m|n}$. Besides being the simplest superspace,\index{super!space}
$\FR^{m|n}$ will serve as a local model (i.e.\ a patch) for
supermanifolds. In the formulation of Manin, we can put\index{super!manifold}
$\FR^{m|n}=(\FR^m,\Lambda_n)$.

\paragraph{Maps on $\FR^{m|n}$.} A function $f:\FR^{m|n}\rightarrow
\FR\otimes\Lambda_n$ is an element of
$\CCF(\FR^m)\otimes\Lambda_n$ and we will denote this set by
$\CCF(\FR^{m|n})$. Smooth functions will correspondingly be
denoted by $C^\infty(\FR^{m|n})$. Choosing coordinates
$(x^i,\theta^j)$, where $1\leq i\leq m$ and $1\leq j \leq n$, we
can write $f$ as
\begin{equation}
f(x,\theta)\ =\ f_0(x)+f_j(x)\theta^j+f_{k_1k_2}(x)\theta^{k_1}\theta^{k_2}+\ldots 
+f_{l_1\ldots l_n}(x)\theta^{l_1}\ldots \theta^{l_n}~.
\end{equation}
We will often also use the notation $f(x,\theta)=f_I(x)\theta^I$,
where $I$ is a multiindex. Note that functions on $\FR^{0|n}$ are
just the supernumbers defined in \ref{pGrassmannalgebras}, and if\index{super!number}
$f_0$ is nowhere vanishing then there is a function $f^{-1}(x)$,
which is given by the inverse of the supernumber $f(x)$, such that\index{super!number}
$f(x)f^{-1}(x)$ is the constant function with value $1$.

The formula for the inverse of a supernumber can be generalized to\index{super!number}
matrix valued supernumbers (and therefore to matrix valued
superfunctions) $\psi\in\sGL(n,\FR)\otimes\Lambda_n$:
\begin{equation*}
\psi^{-1}\ =\ \psi_B^{-1}-\psi_B^{-1}\psi_S\psi_B^{-1}+
\psi_B^{-1}\psi_S\psi_B^{-1}\psi_S\psi_B^{-1}-
\psi_B^{-1}\psi_S\psi_B^{-1}\psi_S\psi_B^{-1}\psi_S\psi_B^{-1}+\ldots ~,
\end{equation*}
where $\psi=\psi_B+\psi_S$ is the usual decomposition into body\index{body}
and soul.\index{soul}

\paragraph{Superspace for $\CN$-extended supersymmetry.}\label{pNSuperspace}\index{N-extended supersymmetry@$\CN$-extended supersymmetry}\index{super!space}\index{super!symmetry}
The {\em superspace for $\CN$-extended supersymmetry} in four
dimensions is the space $\FR^{4|4\CN}$ (or $\FC^{4|4\CN}$ as the
complex analogue), i.e.\ a real four-dimensional space $\FR^{4}$
with arbitrary signature endowed additionally with $4\CN$ Gra{\ss}mann
coordinates. These coordinates are grouped into Weyl spinors\index{Spinor}
$\theta^{\alpha i}$ and $\btheta^i_\ald$, where $\alpha, \ald=1,2$
and $i=1,\ldots ,\CN$. The spinor indices are raised and lowered with\index{Spinor}
the antisymmetric $\eps$-symbol defined by
$\eps_{12}=\eps_{\dot{1}\dot{2}}=-\eps^{12}=-\eps^{\dot{1}\dot{2}}=-1$.

For simplicity, let us introduce the following shorthand notation:
we can drop undotted contracted spinor indices if the left one is\index{Spinor}
the upper index and dotted contracted spinor indices if the right
one is the upper one, i.e.\
$\theta\theta=\theta^\alpha\theta_\alpha$ and
$\btheta\btheta=\btheta_\ald\btheta^\ald$.

\paragraph{Spinorial notation.}\label{pSpinorialNotation} It will often be convenient\index{Spinor}
to rewrite the spacetime coordinates $x^\mu$ in spinor notation,
using (local) isomorphisms as e.g.\ $\sSO(4)\simeq\index{morphisms!isomorphism}
\sSU(2)\times\sSU(2)$ for a Euclidean spacetime\footnote{Similar
isomorphisms also exist for Minkowski and Kleinian signature.} by\index{Kleinian signature}\index{morphisms!isomorphism}
$x^{\alpha\ald}=-\di\sigma_\mu^{\alpha\ald}x^\mu$. The sigma
matrices are determined by the signature of the metric under\index{metric}
consideration (see section \ref{ssSupersymmetryAlgebra},
\ref{pSigmaMatrixConvention}). This will simplify considerably the
discussion at many points later on, however, it requires some care
when comparing results from different sources.

As shorthand notations for the derivatives, we will use in the
following
\begin{equation}
\dpar_{\alpha\ald}\ :=\ \der{x^{\alpha\ald}}~,~~~\dpar_{\alpha
i}\ :=\ \der{\theta^{\alpha i}} \eand
\dparb^i_\ald\ :=\ \der{\bar{\theta}_i^\ald}~.
\end{equation}

\paragraph{Representation of the supersymmetry algebra.} On the\index{representation}\index{super!symmetry}
superspace $\FR^{4|4\CN}$ described by the coordinates\index{super!space}
$(x^\mu,\theta^{\alpha i},\bar{\theta}^i_\ald)$, one can define a
representation of the supersymmetry algebra by introducing the\index{representation}\index{super!symmetry}
action of the superderivatives on functions\footnote{One should
stress that the convention presented here is suited for Minkowski
space and differs from one used later when discussing supertwistor\index{twistor}
spaces for Euclidean superspaces.} $f\in\CCF(\FR^{4|4\CN})$\index{super!space}
\begin{equation}
D_{\alpha i}f\ :=\ \dpar_{\alpha i}
f+\bar{\theta}^\ald_i\dpar_{\alpha\ald} f\eand
\bar{D}^i_{\ald}f\ :=\ -\bar{\dpar}^i_{\ald} f-\theta^{\alpha
i}\dpar_{\alpha\ald} f~,
\end{equation}
as well as the action of the supercharges
\begin{equation}
Q_{\alpha i}f\ :=\ \dpar_{\alpha i}
f-\bar{\theta}^\ald_i\dpar_{\alpha\ald} f\eand
\bar{Q}^i_{\ald}f\ :=\ -\bar{\dpar}^i_{\ald} f+\theta^{\alpha
i}\dpar_{\alpha\ald} f~.
\end{equation}
The corresponding transformations induced by the supercharges
$Q_{\alpha i}$ and $\bar{Q}^i_\ald$ in superspace read\index{super!space}
\begin{equation}
\begin{aligned}
\delta x^{\alpha\ald}&\ =\ \theta^{\alpha i}\xi^\ald_i~,~~~&\delta
\theta^{\alpha
i}&\ =\ 0~,~~~&\delta \thetab^\ald_i&\ =\ \xi^\ald_i~,\\
\delta x^{\alpha\ald}&\ =\ \xi^{\alpha i}\thetab^\ald_i~,~~~&\delta
\theta^{\alpha i}&\ =\ \xi^{\alpha i}~,~~~&\delta \thetab^\ald_i&\ =\ 0~,
\end{aligned}
\end{equation}
respectively, where $(\xi^{\alpha i},\xi^\ald_i)$ are odd
parameters.

\paragraph{Chiral superspaces and chiral coordinates.} The\index{chiral!coordinates}\index{chiral!superspace}\index{super!space}
superspace $\FR^{4|4\CN}$ splits into the two {\em chiral
superspaces} $\FR^{4|2\CN}_L$  and $\FR^{4|2\CN}_R$ where the\index{chiral!superspace}
subscripts $L$ and $R$ stand for left-handed (chiral) and
right-handed (anti-chiral). The theories under consideration often
simplify significantly when choosing the appropriate coordinate
system for the chiral superspaces. In the left-handed case, we\index{chiral!superspace}\index{super!space}
choose
\begin{equation}
(y^{\alpha\ald}_L:=x^{\alpha\ald}+ \theta^{\alpha
i}\btheta^\ald_i,~ \theta^{\alpha i},~\btheta^\ald_i)~.
\end{equation}
The representations of the superderivatives and the supercharges\index{representation}
read in these chiral coordinates as\index{chiral!coordinates}
\begin{equation}
\begin{aligned}
D_{\alpha i}f&\ =\ \dpar_{\alpha i}
f+2\btheta^\ald_i\dpar^L_{\alpha\ald}
f~,&\bar{D}^i_{\ald}f&\ =\ -\dparb^i_{\ald}f~,\\
Q_{\alpha i}f&\ =\ \dpar_{\alpha i} f~,&\bar{Q}^i_{\ald}f&\ =\
-\dparb^i_{\ald}f+2\theta^{\alpha i}\dpar^L_{\alpha\ald} f~,
\end{aligned}
\end{equation}
where $\dpar^L_{\alpha\ald}$ denotes a derivative with respect to
$y^{\alpha\ald}$. Due to
$\dpar^L_{\alpha\ald}=\dpar_{\alpha\ald}$, we can safely drop the
superscript ``$L$'' in the following.

One defines the anti-chiral coordinates accordingly as\index{chiral!coordinates}
\begin{equation}
(y_R^{\alpha\ald}:=x^{\alpha\ald}- \theta^{\alpha
i}\btheta^\ald_i,~ \theta^{\alpha i},~\btheta^\ald_i)~.
\end{equation}

Note that we will also work with superspaces of Euclidean\index{super!space}
signature and the complexified superspaces, in which $\theta$ and
$\thetab$ are not related via complex conjugation. In these cases,
we will often denote $\thetab$ by $\eta$.

\subsection{Supermanifolds}\index{super!manifold}

\paragraph{Supermanifolds.} Roughly speaking, a\index{super!manifold}
{\em supermanifold} is defined to be a topological space which is
locally diffeomorphic to $\FR^{m|n}$ or $\FC^{m|n}$. In general, a\index{diffeomorphic}
supermanifold contains a purely bosonic part, the {\em body},\index{body}\index{super!manifold}
which is parameterized in terms of the supermanifold's bosonic
coordinates. The body of a supermanifold is a real or complex\index{body}
manifold by itself. The $\RZ_2$-grading of the superspace used for\index{Z2-grading@$\RZ_2$-grading}\index{super!space}\index{manifold}
parameterizing the supermanifold induces a grading on the ring of\index{super!manifold}
functions on the supermanifold. For objects like subspaces, forms
etc.\ which come with a dimension, a degree etc., we use the
notation $i|j$, where $i$ and $j$ denote the bosonic and fermionic
part, respectively.

Due to the different approaches to supergeometry, we recall the\index{super!geometry}
most basic definitions used in the literature. For a more
extensive discussion of supermanifolds, see \cite{Cartier:2002zp}\index{super!manifold}
and references therein as well as the works
\cite{Berezin:1987wh,Kostant:1975qe,Leites:1980}.

\paragraph{The parity inverting operator $\Pi$.} Given a vector\index{parity}\index{parity inverting operator}
bundle $E\rightarrow M$, the parity inverting operator $\Pi$ acts
by reversing the parity of the fiber coordinates.

\paragraph{Examples of simple supermanifolds.}\label{psmexamples}\index{super!manifold}
Consider the tangent bundle $TM$ over a manifold $M$ of dimension\index{manifold}
$n$. The dimension of $TM$ is $2n$, and a point in $TM$ can be
locally described by $n$ coordinates on the base space $(x^i)$ and
$n$ coordinates in the fibres $(y^i)$. The parity inverted tangent\index{parity}
bundle $\Pi TM$ is of dimension $n|n$ and locally described by the
$n$ coordinates $(x^i)$ on the base space together with the $n$
Gra{\ss}mann coordinates $(\theta^i)$ in the fibres. More explicitly,
we have e.g.\ $\Pi T\FR^4=\FR^{4|4}$, the superspace for $\CN=1$\index{super!space}
supersymmetry.\index{super!symmetry}

Another example which we will often encounter is the space $\Pi
\CO(n)\rightarrow\CPP^1$ which is described by complex variables
$\lambda_\pm$ and Gra{\ss}mann variables $\theta_\pm$ on the two\index{Gra{\ss}mann variable}
standard patches $U_\pm$ of $\CPP^1$ with
$\theta_+=\lambda_+^n\theta_-$ on $U_+\cap U_-$. This bundle has
first Chern number $-n$, as in fermionic integration, the Jacobian\index{Chern number}\index{first Chern number}
is replaced by an inverse of the Jacobian (the Berezinian).\index{Berezinian}

\paragraph{Supermanifolds in the sheaf-theoretic approach.} We do not want\index{sheaf}\index{super!manifold}
to repeat the formal discussion of \cite{Maninbook} at this point,
but merely make some remarks. It is clear that a supermanifold\index{super!manifold}
will be a superspace as defined above with some additional\index{super!space}
restrictions. These restrictions basically state that it is
possible to decompose a supermanifold globally into its\index{super!manifold}
{\em{}body}, which is a (in some sense maximal) ordinary real or\index{body}
complex manifold, and into its {\em soul}, which is the\index{complex!manifold}\index{soul}\index{manifold}
``infinitesimal cloud'' surrounding the body and complementing it\index{body}
to the full supermanifold.\index{super!manifold}

Let us consider as an example the chiral superspace $\FR^{4|4\CN}$\index{chiral!superspace}\index{super!space}
and the complex projective superspace $\CPP^{3|4}$. Their bodies
are the spaces $\FR^{4|0}=\FR^4$ and $\CPP^{3|0}=\CPP^3$,
respectively.

\paragraph{Supermanifolds according to B.~S.~DeWitt.} This\index{super!manifold}
construction of a supermanifold will not be used in this thesis
and is only given for completeness sake.

First, we define the {\em superdomain} $\FR^m_c\times \FR^n_a$ to\index{super!domain}
be an open superspace described by $m+n$ real coordinates $u^i\in\index{super!space}
\Lambda_0$ and $v^j\in \Lambda_1$, with $i=1,\ldots m$,
$j=1,\ldots n$. Note that $\FR^m_c\times \FR^n_a$ is {\em not} a
supervector space in this approach to supermathematics.\index{super!vector space}

Furthermore, a topology on this space can be obtained from the
topology of the embedded real space $\FR^m$ via the canonical
projection
\begin{equation}
\pi:\FR^m_c\times \FR^n_a\ \rightarrow\  \FR^m~.
\end{equation}
That is, a subset $Y\subset\FR^m_c\times \FR^n_a$ is open if its
projection $\pi(Y)$ onto $\FR^m$ is open. Therefore, a superdomain\index{super!domain}
is not Hausdorff, but only projectively Hausdorff.

A {\em supermanifold} of dimension $m|n$ is then a topological\index{super!manifold}
space which is locally diffeomorphic to $\FR^m_c\times \FR^n_a$.\index{diffeomorphic}

The definition of the body of such a supermanifold is a little\index{body}\index{super!manifold}
more subtle, as one expects the body to be invariant under
coordinate transformations. This implies that we introduce
equivalence classes of points on such supermanifolds, and only\index{super!manifold}
then we can define a body as the real manifold which consists of\index{body}\index{manifold}
all these equivalence classes. For further details, see
\cite{DeWitt:1992cy} or \cite{Cartier:2002zp}.

\subsection{Calabi-Yau supermanifolds and Yau's\index{Calabi-Yau}\index{Calabi-Yau supermanifold}\index{super!manifold}
theorem}\label{ssCYsupermanifolds}

\paragraph{Calabi-Yau supermanifolds.}\label{pCYsupermanifolds}\index{Calabi-Yau}\index{Calabi-Yau supermanifold}\index{super!manifold}
A {\em Calabi-Yau supermanifold} is a supermanifold which has
vanishing first Chern class. Thus, Calabi-Yau supermanifolds come\index{Chern class}\index{first Chern class}
with a nowhere vanishing holomorphic measure $\Omega$. Note,
however, that $\Omega$ is not a differential form in the Gra{\ss}mann
coordinates, since Gra{\ss}mann differential forms are dual to
Gra{\ss}mann vector fields and thus transform contragrediently to
them. Berezin integration, however is equivalent to\index{Berezin integration}
differentiation, and thus a volume element has to transform as a
product of Gra{\ss}mann vector fields, i.e.\ with the inverse of the
Jacobi determinant. Such forms are called integral forms and for
short, we will call $\Omega$ a {\em holomorphic volume form},\index{holomorphic!volume form}
similarly to the usual nomenclature for Calabi-Yau manifolds.\index{Calabi-Yau}\index{manifold}

\paragraph{Comments on the definition.} This definition has become
common usage, even if not all such spaces admit a Ricci-flat\index{Ricci-flat}
metric. Counterexamples to Yau's theorem for Calabi-Yau\index{Calabi-Yau}\index{metric}\index{Theorem!Yau}
supermanifolds can be found in \cite{Rocek:2004bi}.\index{super!manifold}

Nevertheless, one should remark that vanishing of the first Chern
class -- and not Ricci-flatness -- is necessary for a consistent\index{Chern class}\index{Ricci-flat}\index{first Chern class}
definition of the topological B-model on a manifold (see section\index{topological!B-model}\index{manifold}
\ref{ssTopologicalBModel}). And, from another viewpoint, it is
only with the help of a holomorphic volume form that one can give\index{holomorphic!volume form}
an action for holomorphic Chern-Simons theory (see section\index{Chern-Simons theory}
\ref{sshCStheory}). Thus, the nomenclature is justified from a
physicist's point of view.

\paragraph{Examples.}\label{psCYex} The most important example discussed in
recent publications is certainly the space $\CPP^{3|4}$ and its
open subset
\begin{equation}
\CP^{3|4}\ =\ \FC^2\otimes \CO(1)\oplus\FC^4\otimes\Pi\CO(1)
\ \rightarrow\  \CPP^1~.
\end{equation}
The latter space is clearly a Calabi-Yau supermanifold, since its\index{Calabi-Yau}\index{Calabi-Yau supermanifold}\index{super!manifold}
first Chern class is trivial.\footnote{Recall that $\Pi\CO(1)$\index{Chern class}\index{PO(n)@$\Pi\CO(1)$}\index{first Chern class}
contributes $-1$ to the total first Chern number, see\index{Chern number}\index{first Chern number}
\ref{psmexamples}.} The space $\CP^{3|4}$ is covered by two
patches $\CU_\pm$, on which its holomorphic volume form is given\index{holomorphic!volume form}
by
\begin{equation}
\hat\Omega^{3,0|4,0}_\pm\ =\ \pm\dd z_\pm^1\wedge\dd z_\pm^2\wedge\dd
\lambda_\pm\,\dd\eta_1^\pm\dd\eta_2^\pm\dd\eta_3^\pm\dd\eta_4^\pm~,
\end{equation}
where $\lambda_\pm$ is the coordinate on the base space, while
$z^\alpha_\pm$ and $\eta_i^\pm$ are coordinates of the bosonic and
fermionic line bundles, respectively. Note that the body of a\index{body}
Calabi-Yau supermanifold is not a Calabi-Yau manifold, in general,\index{Calabi-Yau}\index{Calabi-Yau supermanifold}\index{super!manifold}\index{manifold}
as also in the case of the above example: the body of $\CP^{3|4}$\index{body}
is $\CO(1)\oplus\CO(1)\rightarrow \CPP^1$ which is not a
Calabi-Yau manifold.\index{Calabi-Yau}\index{manifold}

A further class of examples for superspaces with the Calabi-Yau\index{Calabi-Yau}\index{super!space}
property is given by the weighted projective spaces\index{weighted projective spaces}
$W\CPP^{3|2}(1,1,1,1|p,q)$ with $p+q=4$, which were proposed as
target spaces for the topological B-model in \cite{Witten:2003nn}\index{target space}\index{topological!B-model}
and studied in detail in \cite{Popov:2004nk}.

To complete the list, we will also encounter the superambitwistor\index{twistor}\index{twistor!ambitwistor}
space $\CL^{5|6}$, which is a quadric in the product of two\index{quadric}
supertwistor spaces, and the mini-supertwistor space\index{mini-supertwistor space}\index{twistor}\index{twistor!space}
$\CP^{2|4}:=\CO(2)\oplus\Pi\CO(1)\otimes\FC^4$. The corresponding
mini-superambitwistor space $\CL^{4|6}$ is not a supermanifold,\index{ambitwistor space}\index{mini-superambitwistor space}\index{super!manifold}\index{twistor}\index{twistor!ambitwistor}\index{twistor!space}
see section \ref{ssminisuperambi1}.

\paragraph{Yau's theorem on supermanifolds.} In\index{super!manifold}\index{Theorem!Yau}
\cite{Rocek:2004bi}, it was shown that Yau's theorem is not valid
for all supermanifolds. That is, even if the first Chern class is\index{Chern class}\index{first Chern class}\index{super!manifold}
vanishing on a supermanifold with K\"{a}hler form $J$, this does\index{K\"{a}hler!form}
not imply that this supermanifold admits a super Ricci-flat metric\index{Ricci-flat}\index{metric}
in the same K\"{a}hler class as $J$. To construct a
counterexample, one can start from a K\"{a}hler manifold with\index{K\"{a}hler!manifold}\index{manifold}
vanishing first Chern class and one fermionic and an arbitrary\index{Chern class}\index{first Chern class}
number of bosonic dimensions. One finds that such a supermanifold\index{super!manifold}
admits a Ricci-flat metric if and only if its scalar curvature is\index{Ricci-flat}\index{curvature}\index{metric}
vanishing \cite{Rocek:2004bi}. As the weighted projective spaces\index{weighted projective spaces}
$W\CPP^{m,1}(1,\ldots,1|m)$ provide examples, for which this
condition is not met, we find that the na\"ive form of Yau's
theorem is not valid for supermanifolds.\index{super!manifold}

In a following paper \cite{Zhou:2004su}, it was conjectured that
this was an artifact of supermanifolds with one fermionic\index{super!manifold}
dimensions, but in the paper \cite{Rocek:2004ha} published only
shortly afterwards, counterexamples to the na\"ive form of Yau's
theorem with two fermionic dimensions were presented.

\section{Exotic supermanifolds}\label{sExoticSupermanifolds}\index{exotic!supermanifold}\index{super!manifold}

In this section, we want to give a brief review of the existing
extensions or generalizations of supermanifolds, having additional\index{super!manifold}
dimensions described by even nilpotent coordinates. Furthermore,
we will present a discussion of Yau's theorem on exotic\index{Theorem!Yau}
supermanifolds. In the following, we shall call every (in a\index{super!manifold}
well-defined way generalized) manifold which is locally described\index{manifold}
by $k$ even, $l$ even and nilpotent and $q$ odd and nilpotent
coordinates an {\em exotic supermanifold} of dimension $(k\oplus\index{exotic!supermanifold}\index{super!manifold}
l|q)$. In section \ref{spwexotic}, some of the exotic
supermanifolds defined in the following will serve as target\index{exotic!supermanifold}\index{super!manifold}
spaces for a topological B-model.\index{topological!B-model}

\subsection{Partially formal supermanifolds}\index{partially formal supermanifold}\index{super!manifold}
\label{sspartiallformal}

\paragraph{Supermathematics via functors.}
The objects of supermathematics, as e.g.\ supermanifolds or\index{super!manifold}
supergroups, are naturally described as covariant functors from
the category of Gra{\ss}mann algebras to corresponding categories of\index{Gra{\ss}mann algebra}
ordinary mathematical objects, as manifolds or groups,\index{manifold}
\cite{Schwarz:1984}. A generalization of this setting is to
consider covariant functors with the category of almost nilpotent
(AN) algebras as domain \cite{Konechny:1997hr,Konechny:1997hrb}.
Recall that an AN algebra $\Xi$ can be decomposed into an even
part $\Xi_0$ and an odd part $\Xi_1$ as well as in the canonically
embedded ground field (i.e.\ $\FR$ or $\FC$), $\Xi_B$, and the
nilpotent part $\Xi_S$. The parts of elements $\xi\in\Xi$
belonging to $\Xi_B$ and $\Xi_S$ are called the {\em body} and the\index{body}
{\em soul} of $\xi$, respectively.\index{soul}

\paragraph{Superspaces and superdomains.}\index{super!domain}\index{super!space}
A {\em superspace} is a covariant functor from the category of AN
algebras to the category of sets. Furthermore, a {\em topological
superspace} is a functor from the category of AN algebras to the\index{super!space}\index{topological!superspace}
category of topological spaces.

Consider now a tuple
$(x^1,\ldots ,x^k,y^1,\ldots ,y^l,\zeta^1,\ldots ,\zeta^q)$ of $k$ even, $l$
even and nilpotent and $q$ odd and nilpotent elements of an AN
algebra $\Xi$, i.e.\ $x^i\in\Xi_0$, $y^i\in\Xi_0\cap\Xi_S$ and
$\zeta^i\in\Xi_1$. The functor from the category of AN algebras to
such tuples is a superspace denoted by $\FR^{k\oplus l|q}$. An\index{super!space}
open subset $U^{k\oplus l|q}$ of $\FR^{k\oplus l|q}$, which is
obtained by restricting the fixed ground field $\Xi_B$ of the
category of AN algebras to an open subset, is called a {\em
superdomain} of dimension $(k\oplus l|q)$. After defining a graded\index{super!domain}
basis $(e_1,\ldots ,e_k,f_1,\ldots ,f_{l},\eps_1,\ldots ,\eps_q)$ consisting
of $k+l$ even and $q$ odd vectors, one can consider the set of
linear combinations
$\left\{x^ie_i+y^jf_j+\zeta^\alpha\eps_\alpha\right\}$ which forms
a {\em supervector space} \cite{Konechny:1997hr,Konechny:1997hrb}.\index{super!vector space}

Roughly speaking, one defines a partially formal
supermanifold\footnote{This term was introduced in\index{partially formal supermanifold}\index{super!manifold}
\cite{Kontsevich:1997}.} of dimensions $(k\oplus l|q)$ as a
topological superspace smoothly glued together from superdomains\index{super!domain}\index{super!space}\index{topological!superspace}
$U^{k\oplus l|q}$. Although we will not need the exact definition
in the subsequent discussion, we will nevertheless give it here
for completeness sake.

\paragraph{Maps between superspaces.}\index{super!space}
We define a {\em map between two superspaces} as a natural
transformation of functors. More explicitly, consider two
superspaces $\CM$ and $\CN$. Then a map $F:\CM\rightarrow\CN$ is a\index{super!space}
map between superspaces if $F$ is compatible with the morphisms of
AN algebras $\alpha:\Xi\rightarrow\Xi'$. We call a smooth map
$\kappa:\FR^{k\oplus l|q}_\Xi\rightarrow\FR^{k'\oplus l'|q'}_\Xi$
between two superdomains $\Xi_0$-{\em smooth} if for every\index{super!domain}
$x\in\FR^{k\oplus l|q}_\Xi$ the tangent map
$(\kappa_\Xi)_*:T_x\rightarrow T_{\kappa_\Xi(x)}$ is a
homomorphism of $\Xi_0$-modules. Furthermore, we call a map\index{modules}
$\kappa:\FR^{k\oplus l|q}\rightarrow\FR^{k'\oplus l'|q'}$ {\em
smooth} if for all AN algebras $\Xi$ the maps $\kappa_\Xi$ are
$\Xi_0$-smooth.

\paragraph{Partially formal supermanifolds.}\index{partially formal supermanifold}\index{super!manifold}
Now we can be more precise: A {\em partially formal supermanifold
of dimension} $(k\oplus l|q)$ is a superspace locally equivalent\index{partially formal supermanifold}\index{super!manifold}\index{super!space}
to superdomains of dimension $(k\oplus l|q)$ with smooth\index{super!domain}
transition functions on the overlaps. Thus, a partially formal\index{transition function}
supermanifold is also an exotic supermanifold.\index{exotic!supermanifold}\index{super!manifold}

However, not every exotic supermanifold is partially formal. We\index{exotic!supermanifold}\index{super!manifold}
will shortly encounter examples of such cases: exotic
supermanifolds, which are constructed using a particular AN\index{exotic!supermanifold}\index{super!manifold}
algebra instead of working with the category of AN algebras.

The definitions used in this section stem from
\cite{Konechny:1997hr,Konechny:1997hrb}, where one also finds
examples of applications.

Unfortunately, it is not clear how to define a general integration
over the even nilpotent part of such spaces; even the existence of
such an integral is questionable. We will comment on this point
later on. This renders partially formal supermanifolds useless as\index{partially formal supermanifold}\index{super!manifold}
target spaces for a topological string theory, as we need an\index{string theory}\index{target space}\index{topological!string}
integration to define an action. Therefore, we have to turn to
other generalizations.

\subsection{Thick complex manifolds}\index{complex!manifold}\index{thick complex manifold}\index{manifold}

\paragraph{Formal neighborhoods.}\index{formal neighborhood}
Extensions to $m$-th formal neighborhoods of a submanifold $X$ in
a manifold $Y\supset X$ and the more general thickening procedure\index{thickening}\index{manifold}
have been proposed and considered long ago\footnote{In fact, the
study of infinitesimal neighborhoods goes back to \cite{Grauert}\index{infinitesimal neighborhood}
and \cite{Griffiths}. For a recent review, see
\cite{Camacho:2002}.} in the context of twistor theory, in\index{twistor}
particular for ambitwistor spaces, e.g.\ in\index{ambitwistor space}\index{twistor!ambitwistor}\index{twistor!space}
\cite{Witten:1978xx,Eastwood:1987,LeBrun:1986,Eastwood:1986}. We
will ignore this motivation and only recollect the definitions
needed for our discussion in chapter \ref{chTwistorGeometry}.

\paragraph{Thickening of complex manifolds.} Given a complex manifold\index{complex!manifold}\index{thickening}\index{manifold}
$X$ with structure sheaf $\CO_X$, we consider a sheaf of\index{sheaf}\index{structure sheaf}
$\FC$-algebras $\CO_{(m)}$ on $X$ with a homomorphism
$\alpha:\CO_{(m)}\rightarrow\CO_X$, such that locally $\CO_{(m)}$
is isomorphic to $\CO[y]/(y^{m+1})$ where $y$ is a formal
(complex) variable and $\alpha$ is the obvious projection. The
resulting ringed space or scheme $X_{(m)}:=(X,\CO_{(m)})$ is
called a {\em thick complex manifold}. Similarly to the\index{complex!manifold}\index{thick complex manifold}\index{manifold}
nomenclature of supermanifolds, we call the complex manifold $X$\index{super!manifold}
the {\em body} of $X_{(m)}$.\index{body}

\paragraph{Example.}
As a simple example, let $X$ be a closed submanifold of the
complex manifold $Y$ with codimension one. Let $\CI$ be the ideal\index{complex!manifold}\index{manifold}
of functions vanishing on $X$. Then $\CO_{(m)}=\CO_Y/\CI^{m+1}$ is
called an {\em infinitesimal neighborhood} or the $m$-th {\em\index{infinitesimal neighborhood}
formal neighborhood} of $X$. This is a special case of a thick\index{formal neighborhood}
complex manifold. Assuming that $X$ has complex dimension $n$,\index{complex!manifold}\index{manifold}
$\CO_{(m)}$ is also an exotic supermanifold of dimension $(n\oplus\index{exotic!supermanifold}\index{super!manifold}
1|0)$. More explicitly, let $(x^1,\ldots ,x^n)$ be local coordinates
on $X$ and $(x^1,\ldots ,x^n,y)$ local coordinates on $Y$. Then the
ideal $\CI$ is generated by $y$ and $\CO_{(m)}$ is locally a
formal polynomial in $y$ with coefficients in $\CO_X$ together
with the identification $y^{m+1}\sim 0$. Furthermore, one has
$\CO_{(0)}=\CO_X$.

Returning to the local description as a formal polynomial in $y$,
we note that there is no object $y^{-1}$ as it would violate
associativity by an argument like $0=y^{-1} y^{m+1}=y^{-1}y
y^m=y^m$. However, the inverse of a formal polynomial in $y$ is
defined if (and only if) the zeroth order monomial has an inverse.
Suppose that $p=a+\sum_{i=1}^mf_i y^i=a+b$. Then we have
$p^{-1}=\frac{1}{a}\sum_{i=0}^m(-\frac{b}{a})^i$, analogously to
the inverse of a supernumber.\index{super!number}

\paragraph{Vector bundles.}
A {\em holomorphic vector bundle} on $(X,\CO_{(m)})$ is a locally\index{holomorphic!vector bundle}
free sheaf of $\CO_{(m)}$-modules.\index{modules}\index{sheaf}

The {\em tangent space} of a thick complex manifold is the sheaf\index{complex!manifold}\index{sheaf}\index{thick complex manifold}\index{manifold}
of derivations $D:\CO_{(m)}\rightarrow\CO_{(m)}$. Let us consider
again our above example $X_{(m)}=(X,\CO_{(m)})$. Locally, an
element of $T X_{(m)}$ will take the form $D=f
\der{y}+\sum_jg^j\der{x^j}$ together with the differentiation
rules
\begin{equation}
\der{y}\,y\ =\ 1~,~~~\der{y}\,x^i\ =\ \der{x^i}\,y\ =\ 0~,~~~\der{x^i}\,x^j\ =\ \delta_i^j~.
\end{equation}

All this and the introduction of cotangent spaces for thick
complex manifolds is found in \cite{Eastwood:1986}.\index{complex!manifold}\index{thick complex manifold}\index{manifold}

\paragraph{Integration on thick complex manifolds.}\index{complex!manifold}\index{thick complex manifold}\index{manifold}
In defining a (definite) integral over the nilpotent formal
variable $y$, which is needed for formulating hCS theory by giving
an action, one faces the same difficulty as in the case of Berezin
integration: the integral should not be taken over a specific\index{Berezin integration}
range as we integrate over an infinitesimal neighborhood which\index{infinitesimal neighborhood}
would give rise to infinitesimal intervals. Furthermore, this
neighborhood is purely formal and so has to be the integration.
Recall that a suitable integration $I$ should satisfy the
rule\footnote{This rule can also be used to fix Berezin
integration, cf. section \ref{secsupermanifolds},\index{Berezin integration}
\ref{pGrassmannIntegral}.} $DI=ID=0$, where $D$ is a derivative
with respect to a variable over which $I$ integrates. The first
requirement $DI=0$ states that the result of definite integration
does not depend on the variables integrated over. The requirement
$ID=0$ for integration domains with vanishing boundary (or
functions vanishing on the boundary) is the foundation of Stokes'
formula and integration by parts. It is easy to see that the
condition $DI=ID=0$ demands that
\begin{equation}
I\ =\ c\cdot\frac{\dpar^m}{\dpar y^m}~,
\end{equation}
where $y$ is the local formal variable from the definition of
$X_{(m)}$ and $c$ is an arbitrary normalization constant, e.g.\
$c=1/m!$ would be most convenient. Thus, we define
\begin{equation}
\int \dd y~ f\ :=\ \frac{1}{m!}\,\frac{\dpar^m}{\dpar y^m}\, f~.
\end{equation}
This definition only relies on an already well-defined operation
and thus is well-defined itself.\footnote{From this definition, we
see the problem arising for partially formal supermanifolds: The\index{partially formal supermanifold}\index{super!manifold}
integration process on thick complex manifolds returns the\index{complex!manifold}\index{thick complex manifold}\index{manifold}
coefficient of the monomial with highest possible power in $y$.
For partially formal supermanifolds, where one works with the\index{partially formal supermanifold}\index{super!manifold}
category of AN algebras, such a highest power does not exist as it
is different for each individual AN algebra.} Additionally, it
also agrees with the intuitive picture that the integral of a
constant over an infinitesimal neighborhood should vanish.\index{infinitesimal neighborhood}
Integration over a thick complex manifold is an\index{complex!manifold}\index{thick complex manifold}\index{manifold}
integro-differential operation.

\paragraph{Change of coordinates.}
Consider now a change of coordinates $(x^1,\ldots ,x^n,y)\rightarrow
(\tilde{x}^1,\ldots ,\tilde{x}^n,\tilde{y})$ which leaves invariant
the structure of the thick complex manifold. That is,\index{complex!manifold}\index{thick complex manifold}\index{manifold}
$\tilde{x}^i$ is independent of $y$, and $\tilde{y}$ is a
polynomial only in $y$ with vanishing zeroth order coefficient and
non-vanishing first order coefficient. Because of
$\dpar_{\tilde{y}}=\frac{\dpar y}{\dpar \tilde{y}}\,\dpar_y$, we
have the following transformation of a volume element under such a
coordinate change:
\begin{equation}
\dd \tilde{x}^1\ldots \dd \tilde{x}^n\dd
\tilde{y}\ =\ \det\left(\frac{\dpar \tilde{x}^i}{\dpar x^j}\right)\dd
x^1\ldots \dd x^n\left(\frac{\dpar y}{\dpar \tilde{y}}\right)^m\dd y~.
\end{equation}

The theorems in \cite{Eastwood:1986} concerning obstructions to
finding $X_{(m+1)}$ given $X_{(m)}$ will not be needed in the
following, as we will mainly work with order one thickenings (or\index{thickening}
fattenings) and in the remaining cases, the existence directly
follows by construction.

\subsection{Fattened complex manifolds}\index{complex!manifold}\index{fattened complex manifold}\index{manifold}

\paragraph{Fattening of complex manifolds.}\index{complex!manifold}\index{manifold}
Fattened complex manifolds \cite{Eastwood:1992} are\index{fattened complex manifold}
straightforward generalizations of thick complex manifolds.\index{thick complex manifold}
Consider again a complex manifold $X$ with structure sheaf\index{sheaf}\index{structure sheaf}
$\CO_X$. The {\em $m$-th order fattening with codimension $k$} of
$X$ is the ringed space $X_{(m,k)}=(X,\CO_{(m,k)})$ where
$\CO_{(m,k)}$ is locally isomorphic to
\begin{equation}
\CO[y^1,\ldots ,y^k]/(y^1,\ldots ,y^k)^{m+1}~.
\end{equation}
Here, the $y^i$ are again formal complex variables. We also demand
the existence of the (obvious) homomorphism
$\alpha:\CO_{(m,k)}\rightarrow\CO_X$. It follows immediately that
a fattening with codimension 1 is a thickening. Furthermore, an\index{thickening}
$(m,k)$-fattening of an $n$-dimensional complex manifold $X$ is an\index{complex!manifold}\index{manifold}
exotic supermanifold of dimension $(n\oplus k|0)$ and we call $X$\index{exotic!supermanifold}\index{super!manifold}
the {\em body} of $X_{(m,k)}$.\index{body}

As in the case of thick complex manifolds, there are no inverses\index{complex!manifold}\index{thick complex manifold}\index{manifold}
for the $y^i$, but the inverse of a formal polynomial $p$ in the
$y^i$ decomposed into $p=a+b$, where $b$ is the nilpotent part of
$p$, exists again if and only if $a\neq 0$ and it is then given by
$p^{-1}=\frac{1}{a}\sum_{i=0}^m(-\frac{b}{a})^i$. A {\em
holomorphic vector bundle} on $\CO_{(m,k)}$ is a locally free\index{holomorphic!vector bundle}
sheaf of $\CO_{(m,k)}$-modules. The {\em tangent space} of a thick\index{modules}\index{sheaf}
complex manifold is also generalized in an obvious manner.\index{complex!manifold}\index{manifold}

\paragraph{Integration on fattened complex manifolds.}\index{complex!manifold}\index{fattened complex manifold}\index{manifold}
We define the integral analogously to thick complex manifolds as\index{thick complex manifold}
\begin{equation}
\int \dd y^1\ldots \dd y^k~f\ :=\ \frac{1}{m!}\,\frac{\dpar^m}{\dpar
(y^1)^m}~\ldots ~\frac{1}{m!}\,\frac{\dpar^m}{\dpar (y^k)^m}\, f~.
\end{equation}
A change of coordinates
$(x^1,\ldots ,x^n,y^1,\ldots ,y^k)\rightarrow(\tilde{x}^1,\ldots ,\tilde{x}^n,\tilde{y}^1,\ldots ,\tilde{y}^k)$
must again preserve the structure of the fat complex manifold:\index{complex!manifold}\index{manifold}
$\tilde{x}^i$ is independent of the $y^i$ and the $\tilde{y}^i$
are nilpotent polynomials in the $y^i$ with vanishing monomial of
order 0 and at least one non-vanishing monomial of order 1.
Evidently, all the $\tilde{y}^i$ have to be linearly independent.
Such a coordinate transformation results in a more complicated
transformation law for the volume element:
\begin{equation}
\dd \tilde{x}^1\ldots \dd \tilde{x}^n\dd
\tilde{y}^1\ldots \dd\tilde{y}^k\ =\ \det\left(\frac{\dpar
\tilde{x}^i}{\dpar x^j}\right)\dd x^1\ldots \dd x^n\left(\frac{\dpar
y^{i_1}}{\dpar \tilde{y}^1}\ldots \frac{\dpar y^{i_k}}{\dpar
\tilde{y}^k}\right)^m\dd y^{i_1}\ldots \dd y^{i_k}~,
\end{equation}
where a sum over the indices $(i_1,\ldots ,i_k)$ is implied. In this
case, the coefficient for the transformation of the nilpotent
formal variables cannot be simplified. Recall that in the case of
ordinary differential forms, the wedge product provides the
antisymmetry needed to form the determinant of the Jacobi matrix.
In the case of Berezin integration, the anticommutativity of the\index{Berezin integration}
derivatives with respect to Gra{\ss}mann variables does the same for\index{Gra{\ss}mann variable}
the inverse of the Jacobi matrix. Here, we have neither of these
and therefore no determinant appears.

\paragraph{Thick and fattened supermanifolds.}\index{super!manifold}
After thickening or fattening a complex manifold, one can readily\index{complex!manifold}\index{thickening}\index{manifold}
add fermionic dimensions. Given a thickening of an
$n$-dimen\-sional complex manifold of order $m$, the simplest\index{complex!manifold}\index{manifold}
example is possibly $\Pi T X_{(m)}$, an $(n\oplus 1|n+1)$
dimensional exotic supermanifold. However, we will not study such\index{exotic!supermanifold}\index{super!manifold}
objects in the following.

\subsection{Exotic Calabi-Yau supermanifolds and Yau's\index{Calabi-Yau}\index{Calabi-Yau supermanifold}\index{exotic!Calabi-Yau supermanifold}\index{super!manifold}
theorem}\label{ssExoticCYs}

\paragraph{Exotic Calabi-Yau supermanifolds.}\index{Calabi-Yau}\index{Calabi-Yau supermanifold}\index{exotic!Calabi-Yau supermanifold}\index{super!manifold}
Following the convention for supermanifolds (cf. section
\ref{secsupermanifolds}, \ref{pCYsupermanifolds}), we shall call
an exotic supermanifold Calabi-Yau if its first Chern class\index{Calabi-Yau}\index{Chern class}\index{exotic!supermanifold}\index{first Chern class}
vanishes and it therefore comes with a holomorphic volume form.\index{holomorphic!volume form}
For exotic supermanifolds, too, the Calabi-Yau property is not\index{Calabi-Yau}\index{exotic!supermanifold}\index{super!manifold}
sufficient for the existence of a Ricci-flat metric, as we will\index{Ricci-flat}\index{metric}
derive in the following.

\paragraph{Exotic trace and exotic determinant.}\label{pedet}\index{exotic!determinant}\index{exotic!trace}
We start from a $(k\oplus l|q)$-dimensional exotic supermanifold\index{exotic!supermanifold}\index{super!manifold}
with local coordinate vector
$(x^1,\ldots ,x^k,y^1,\ldots ,y^l,\zeta^1,\ldots ,\zeta^q)^T$. An element of
the tangent space is described by a vector
$(X^1,\ldots ,X^k,Y^1,\ldots ,Y^l,Z^1,\ldots ,Z^q)^T$. Both the metric and\index{metric}
linear coordinate transformations on this space are defined by
nonsingular matrices
\begin{equation}
K\ =\  \left(\begin{array}{ccc} A & B & C\\ D & E & F\\ G & H & J
\end{array}\right)~,
\end{equation}
where the elements $A,B,D,E,J$ are of even and $G,H,C,F$ are of
odd parity. As a definition for the {\em extended supertrace} of\index{parity}\index{super!trace}
such matrices, we choose
\begin{equation}
\etr(K)\ :=\ \tr(A)+\tr(E)-\tr(J)~,
\end{equation}
which is closely related to the supertrace and which is the\index{super!trace}
appropriate choice to preserve cyclicity: $\etr(KM)=\etr(MK)$.
Similarly to \cite{DeWitt:1992cy}, we define the extended
superdeterminant by\index{super!determinant}
\begin{equation}
\delta\ln\edet(K)\ :=\ \etr(K^{-1}\delta K)~~~\mbox{together
with}~~~\edet(\unit)\ :=\ 1~,
\end{equation}
which guarantees $\edet(KM)=\edet(K)\edet(M)$. Proceeding
analogously to \cite{DeWitt:1992cy}, one decomposes $K$ into the
product of a lower triangular matrix, a block diagonal matrix and
an upper diagonal matrix. The triangular matrices can be chosen to
have only 1 as diagonal entries and thus do not contribute to the
total determinant. The block diagonal matrix is of the form
\begin{equation}
K'\ =\ \left(\begin{array}{ccc} A & 0 & 0\\
0 & E-DA^{-1}B & 0 \\ 0 & 0 & R
\end{array}\right)~,
\end{equation}
with $R=J-GA^{-1}C-(H-GA^{-1}B)(E-DA^{-1}B)^{-1}(F-DA^{-1}C)$. The
determinant of a block diagonal matrix is easily calculated and in
this case we obtain
\begin{equation}
\edet(K)\ =\ \edet(K')\ =\ \frac{\det(A)\det(E-DA^{-1}B)}{\det(R)}~.
\end{equation}
Note that for the special case of no even nilpotent dimensions,
for which one should formally set $B=D=F=H=0$, one recovers the
formul\ae{} for the supertrace ($E=0$) and the superdeterminant\index{super!determinant}\index{super!trace}
($E=\unit$ to drop the additional determinant).

\paragraph{Yau's theorem on exotic supermanifolds.}\index{exotic!supermanifold}\index{super!manifold}\index{Theorem!Yau}
In \cite{Rocek:2004bi}, the authors found that K\"{a}hler
supermanifolds with one fermionic dimension admit Ricci-flat\index{Ricci-flat}\index{super!manifold}
supermetrics if and only if the body of the K\"{a}hler\index{body}
supermanifold admits a metric with vanishing scalar\index{super!manifold}\index{metric}
curvature,\footnote{For related work, see\index{curvature}
\cite{Zhou:2004su,Rocek:2004ha,Lindstrom:2005uh}.} and thus Yau's
theorem (see section \ref{subYautheorem}) is only valid under
additional assumptions. Let us investigate the same issue for the
case of a $(p\oplus 1|0)$-dimensional exotic supermanifold $Y$\index{exotic!supermanifold}\index{super!manifold}
with one even nilpotent coordinate $y$. We denote the ordinary
$p$-dimensional complex manifold embedded in $Y$ by $X$. The\index{complex!manifold}\index{manifold}
extended K\"{a}hler potential on $Y$ is given by a real-valued\index{K\"{a}hler!potential}
function $\CCK=f^0+f^1y\bar{y}$, such that the metric takes the\index{metric}
form
\begin{equation}
g:=\left(\dpar_i\dparb_{\bar{\jmath}}\CCK\right)\ =\ \left(\begin{array}{cc}
f^0_{,i\bar{\jmath}}+f^1_{,i\bar{\jmath}}\,y\bar{y} & f^1_{,i}\,y \\
f^1_{,\bar{\jmath}}\,\bar{y} & f^1
\end{array}\right)~.
\end{equation}
For the extended Ricci-tensor to vanish, the extended K\"{a}hler
potential has to satisfy the Monge-Amp{\`e}re equation\index{K\"{a}hler!potential}\index{Monge-Amp{\`e}re equation}
$\edet(g):=\edet(\dpar_i\dparb_{\bar{\jmath}}\CCK)=1$. In fact, we
find
\begin{align*}
\edet(g)\ =\ &\det\left(f^0_{,i\bar{\jmath}}+f^1_{,i\bar{\jmath}}y\bar{y}\right)\left(f^1-
f^1_{\bar{m}}g^{\bar{m}n}f^1_{,n}y\bar{y}\right)\\
\ =\ &\det\left[\left(f^0_{,i\bar{\jmath}}+f^1_{,i\bar{\jmath}}y\bar{y}\right)\left(\sqrt[p]{f^1}-
\frac{f^1_{\bar{m}}g^{\bar{m}n}f^1_{,n}y\bar{y}}{p
(f^1)^{\frac{p-1}{p}}}\right)\right]\\
\ =\ &\det\left[f^0_{,i\bar{\jmath}}\sqrt[p]{f^1}+\left(f^1_{,i\bar{\jmath}}\sqrt[p]{f^1}-
f^0_{,i\bar{\jmath}}\frac{f^1_{\bar{m}}g^{\bar{m}n}f^1_{,n}}{p
(f^1)^{\frac{p-1}{p}}}\right)y\bar{y}
\right]\\
\ =\ &\det\left[f^0_{,l\bar{\jmath}}\sqrt[p]{f^1}\right]
\det\left[\delta_i^k+\left(g^{\bar{m}k}f^1_{,i\bar{m}}-
\delta_i^k\frac{f^1_{\bar{m}}g^{\bar{m}n}f^1_{,n}}{p
f^1}\right)y\bar{y} \right]~,
\end{align*}
where $g^{\bar{m}n}$ is the inverse of $f^0_{,n\bar{m}}$. Using
the relation $\ln\det(A)=\tr \ln(A)$, we obtain
\begin{equation}
\edet(g)\ =\ \det\left[f^0_{,l\bar{\jmath}}\sqrt[p]{f^1}\right]\left(1+\left(g^{\bar{m}i}f^1_{,i\bar{m}}-
\frac{f^1_{\bar{m}}g^{\bar{m}n}f^1_{,n}}{f^1}\right)y\bar{y}
\right)~.
\end{equation}
From demanding extended Ricci-flatness, it follows that\index{Ricci-flat}
\begin{equation}\label{cond0}
f^1\ =\ \frac{1}{\det\left(f^0_{,l\bar{\jmath}}\right)}\eand
\left(g^{\bar{\jmath}i}f^1_{,i\bar{\jmath}}-
\frac{f^1_{\bar{\jmath}}g^{\bar{\jmath}i}f^1_{,i}}{
f^1}\right)\ =\ 0~.
\end{equation}
The second equation can be simplified to
\begin{equation}
g^{\bar{\jmath}i}\left(f^1_{,i\bar{\jmath}}-
\frac{f^1_{\bar{\jmath}}f^1_{,i}}{ f^1}\right)\ =\  f^1
g^{\bar{\jmath}i}\left(\ln(f^1)\right)_{,i\bar{\jmath}}\ =\ 0~,
\end{equation}
and together with the first equation in \eqref{cond0}, it yields
\begin{equation}
g^{\bar{\jmath}i}\left(\ln
\frac{1}{\det\left(f^0_{,k\bar{\jmath}}\right)}\right)_{,i\bar{\jmath}}\ =\ 
-g^{\bar{\jmath}i}\left(\ln\det\left(f^0_{,k\bar{\jmath}}\right)\right)_{,i\bar{\jmath}}\ =\ 
-g^{\bar{\jmath}i}R_{,i\bar{\jmath}}\ =\ 0~.
\end{equation}
This equation states that an exotic supermanifold $Y$ of dimension\index{exotic!supermanifold}\index{super!manifold}
$(p\oplus 1|0)$ admits an extended Ricci-flat metric if and only\index{Ricci-flat}\index{metric}
if the embedded ordinary manifold $X$ has vanishing scalar\index{manifold}
curvature. A class of examples for which this additional condition\index{curvature}
is not satisfied are the weighted projective spaces\index{weighted projective spaces}
$W\CPP^{m-1\oplus1|0}(1,\ldots ,1\oplus m|\cdot)$, which have
vanishing first Chern class but do not admit a K\"{a}hler metric\index{Chern class}\index{K\"{a}hler!metric}\index{first Chern class}\index{metric}
with vanishing Ricci scalar.

Thus, we obtained exactly the same result as in
\cite{Rocek:2004bi}, which is somewhat surprising as the
definition of the extended determinant involved in our calculation
strongly differs from the definition of the superdeterminant.\index{super!determinant}
However, this agreement might be an indication that fattened
complex manifolds -- together with the definitions made above --\index{complex!manifold}\index{fattened complex manifold}\index{manifold}
fit nicely in the whole picture of extended Calabi-Yau spaces.\index{Calabi-Yau}

\section{Spinors in arbitrary dimensions}\index{Spinor}

The main references for this section are
\cite{Berg:2000aa,Weinberg:2000cr} and appendix B of
\cite{Polchinski:1998rr}.

\subsection{Spin groups and Clifford algebras}\index{Clifford algebra}\index{spin group}

\paragraph{Spin group.} The {\em spin group} $\sSpin(p,q)$ is the\index{spin group}
{\em double cover} (or {\em universal cover}) of the Lorentz group\index{Lorentz!group}\index{double cover}\index{universal cover}
$\sSO(p,q)$. Explicitly, it is defined by the short exact sequence\index{short exact sequence}
\begin{equation}
1\ \rightarrow\  \RZ_2\ \rightarrow\  \sSpin(p,q)\ \rightarrow\ 
\sSO(p,q)\ \rightarrow\  1~.
\end{equation}

\paragraph{Clifford algebra.} Let $V$ be a $(p+q)$-dimensional vector\index{Clifford algebra}
space $V$ with a pseudo-Euclidean scalar product $g_{AB}$
invariant under the group $\sO(p,q)$. Consider furthermore $p+q$
symbols $\gamma_A$ with a product satisfying
\begin{equation}
\gamma_A\gamma_B+\gamma_B\gamma_A\ =\ -2g_{AB}\unit~.
\end{equation}
The {\em Clifford algebra} $\CCC(p,q)$ is then a $2^{p+q}$\index{Clifford algebra}
dimensional vector space spanned by the basis
\begin{equation}
(\unit,\gamma_A,\gamma_A\gamma_B,\ldots ,\gamma_1\ldots \gamma_{p+q})~.
\end{equation}
Note that this is a $\RZ_2$-graded algebra,
$\CCC(p,q)=\CCC_+(p,q)\oplus\CCC_-(p,q)$, where $\CCC_+(p,q)$ and
$\CCC_-(p,q)$ denote the elements consisting of an even and odd
number of symbols $\gamma_A$, respectively.

\paragraph{Representation of the Clifford algebra.} A\index{Clifford algebra}\index{representation}
faithful representation of the Clifford algebra for $d=2k+2$ can\index{faithful representation}
be found by recombining the generators $\gamma_A$ as follows:
\begin{equation}
\begin{aligned}
\gamma_{0\pm}\ =\ \tfrac{1}{2}(\pm\gamma_0+\gamma_1)\eand
\gamma_{a\pm}\ =\ \tfrac{1}{2}(\gamma_{2a}\pm\di\gamma_{2a+1})~~~\mbox{for}~~~a\ \neq\ 
0~.
\end{aligned}
\end{equation}
This yields the fermionic oscillator algebra
\begin{equation}
\{\gamma_{a+},\gamma_{b-}\}\ =\ \delta^{ab}~,~~~
\{\gamma_{a+},\gamma_{b+}\}\ =\ \{\gamma_{a-},\gamma_{b-}\}\ =\ 0~,
\end{equation}
and by the usual highest weight construction, one obtains a
$2^{k+1}$-dimensional representation. That is, starting from a\index{representation}
state $|h\rangle$ with $\gamma_{a-}|h\rangle=0$ for all $a$, we
obtain all the states by acting with arbitrary combinations of the
$\gamma_{a+}$ on $|h\rangle$. As every $\gamma_{a+}$ can appear at
most once, this leads to $2^{k+1}$ states, which can be
constructed iteratively, see \cite{Polchinski:1998rr}. Given such
a representation for $d=2k+2$, one can construct a representation\index{representation}
for $d=2k+3$ by adding the generator
$\gamma_d=\di^{-k}\gamma_0\ldots \gamma_{d-2}$. Thus, the faithful
representations of the Clifford algebra on a space with dimension\index{Clifford algebra}\index{faithful representation}\index{representation}
$d$ are $2^{\left[\frac{d}{2}\right]}$-dimensional.

\paragraph{Embedding a Spin group in a Clifford algebra.} Given a\index{Clifford algebra}\index{spin group}
Clifford algebra $\CCC(p,q)$, the generators
\begin{equation}
\Sigma_{AB}\ =\ -\tfrac{\di}{4}[\gamma_A,\gamma_B]
\end{equation}
form a representation of the Lie algebra of $\sSpin(p,q)$. This is\index{Lie algebra}\index{representation}
the {\em Dirac representation}, which is a reducible\index{Dirac representation}
representation of the underlying Lorentz algebra. As examples,\index{Lorentz!algebra}
consider in four dimensions the decomposition of the Dirac
representation into two Weyl representations\index{Dirac representation}\index{Weyl representation}\index{representation}
$\rep{4}_{\mathrm{Dirac}}=\rep{2}+\rep{2}'$ as well as the similar
decomposition in ten dimensions:
$\rep{32}_{\mathrm{Dirac}}=\rep{16}+\rep{16}'$.

\paragraph{Examples.} The following table contains those examples of
spin groups which are most frequently encountered. For further\index{spin group}
examples and more details, see \cite{Bryant:2000}.
\begin{align*}
\sSpin(2)&\ \cong\ \sU(1)&\sSpin(3)&\ \cong\ \sSU(2)&\sSpin(4)&\ \cong\ \sSU(2)\times\sSU(2)\\
\sSpin(1,1)&\ \cong\ \FR^\times&\sSpin(2,1)&\ \cong\ \sSL(2,\FR)&\sSpin(3,1)&\ \cong\ \sSL(2,\FC)\\
\sSpin(5)&\ \cong\ \mathsf{Sp}(2)&\sSpin(6)&\ \cong\ \sSU(4)&\sSpin(2,2)&\ \cong\ \sSL(2,\FR)\times\sSL(2,\FR)\\
\sSpin(4,1)&\ \cong\ \mathsf{Sp}(1,1)&\sSpin(5,1)&\ \cong\ \sSL(2,\FH)
\end{align*}

\subsection{Spinors}\label{ssSpinors}\index{Spinor}

\paragraph{Spinors.} A {\em spinor} on a spacetime with\index{Spinor}
Lorentz group $\sSO(p,q)$ is an element of the representation\index{Lorentz!group}\index{representation}
space of the group $\sSpin(p,q)$. Generically, a (Dirac) spinor is\index{Spinor}
thus of complex dimension $2^{[(p+q)/2]}$.

\paragraph{Minkowski space.} On $d$-dimensional Minkowski space, the
$2^{[\frac{d}{2}]}$-dimensional Dirac representation splits into\index{Dirac representation}\index{representation}
two {\em Weyl representations}, which are the two sets of\index{Weyl representation}
eigenstates of the chirality operator
\begin{equation}
\gamma\ =\ \di^{-k}\gamma_0\gamma_1\ldots\gamma_{d-1}~,
\end{equation}
where $\gamma$ has eigenvalues $\pm 1$. This operator can be used
to define a projector onto the two Weyl representation:\index{Weyl representation}\index{representation}
\begin{equation}
P_\pm\ :=\ \frac{\unit\pm \gamma}{2}~.
\end{equation}

In dimensions $d=0,1,2,3,4 \mod 8$, one can also impose a {\em
Majorana condition} on a Dirac spinor, which demands that a\index{Majorana condition}\index{Spinor}
so-called Majorana spinor $\psi$ is its own charge conjugate:
\begin{equation}
\psi^c\ =\ \psi~~~\mbox{with}~~~\psi^c\ :=\ C\gamma_0\psi^*~.
\end{equation}
Here, $C$ is the charge conjugation operator, satisfying\index{charge conjugation operator}
\begin{equation}
C\gamma_\mu C^{-1}\ =\ -\gamma_\mu^T\eand
C\gamma_0(C\gamma_0)^*\ =\ \unit~.
\end{equation}
The latter equation implies $(\psi^c)^c=\psi$. (Note that in the
remaining cases $d=5,6,7 \mod 8$, one can group the spinors into\index{Spinor}
doublets and impose a symplectic Majorana condition. We will\index{Majorana condition}\index{symplectic Majorana condition}
encounter such a condition in the case of Gra{\ss}mann variables on\index{Gra{\ss}mann variable}
Euclidean spacetime in \ref{peuclideancase}.)

The Majorana condition is essentially equivalent to the Weyl\index{Majorana condition}
condition in dimensions $d=0,4 \mod 8$. In dimensions $d=2 \mod
8$, one can impose both the Weyl and the Majorana conditions\index{Majorana condition}
simultaneously, which yields {\em Majorana-Weyl spinors}. The\index{Majorana-Weyl spinor}\index{Spinor}
latter will appear when discussing ten-dimensional super
Yang-Mills theory in section \ref{ssmaxsusy}.\index{Yang-Mills theory}

Two-spinors, in particular the commuting ones needed in twistor\index{Spinor}\index{twistor}\index{two-spinors}
theory, will be discussed in \ref{ptwospinors} of section
\ref{ssmotivation} in more detail.

\paragraph{Euclidean space.} The discussion of Euclidean spinors\index{Spinor}
is quite parallel, and one basically identifies the properties of
representations of $\sSpin(p)$ with those of $\sSpin(p+1,1)$. The\index{representation}
Dirac representation decomposes again into two Weyl\index{Dirac representation}
representations, and one can impose a Majorana condition for\index{Majorana condition}
$d=0,1,2,6,7 \mod 8$. In the cases $d=3,4,5 \mod 8$, one has to
switch to a pseudoreal representation.\index{representation}

\paragraph{Vectors from spinors.} The generators of the Clifford\index{Spinor}
algebra can be interpreted as linear maps on the spinor space.
Thus they (and their reduced versions) can be used to convert
vector indices into two spinor indices and vice versa. We already\index{Spinor}
used this fact in introducing the notation
$x^{\alpha\ald}:=-\di\sigma_\mu^{\alpha\ald}x^\mu$. In particular,
this example together with conventions for commuting two-spinors\index{Spinor}\index{two-spinors}
are given in section \ref{ssmotivation}, \ref{ptwospinors}. For
more details in general dimensions, see \cite{Penrosebooks}.

\paragraph{Reality conditions.} A real structure is an antilinear\index{real structure}
involution $\tau$, which gives rise to a reality condition by\index{involution}
demanding that $\tau(\cdot)=\cdot$. The real structures which we\index{real structure}
will define live on superspaces with four- or three-dimensional\index{super!space}
bodies. In the four-dimensional case, there are two such
involutions for Kleinian signature\footnote{i.e.\ signature (2,2)}\index{Kleinian signature}\index{involution}
on the body, and each one for bodies with Euclidean and Minkowski\index{body}
signature. In the three-dimensional case, there is evidently just
a Euclidean and a Minkowski signature possible on the body. We\index{body}
want to stress in advance that contrary to the Minkowskian
signature $(3,1)$, the variables $\theta^{\alpha i}$ and
$\eta^{\ald}_i=\thetab^{\ald}_i$ are independent for both
signatures (4,0) and (2,2).

In the following, we will consider the superspaces $\FR^{4|4\CN}$\index{super!space}
and $\FR^{3|4\CN}$ with coordinates
$(x^{\alpha\ald},\eta^\ald_i,\theta^{\alpha i})$ and
$(y^{\ald\bed},\eta^\ald_i,\theta^{\alpha i})$, respectively. The
latter coordinates are obtained from dimensional reduction via the\index{dimensional reduction}
formula $y^{\ald\bed}:=-\di x^{(\ald\bed)}$, see section
\ref{ssRelatedTheories}, \ref{pdimreduction} for more details.

\paragraph{Kleinian case.} For this case, we introduce two
real structures $\tau_1$ and $\tau_0$, which act on the bosonic\index{real structure}
coordinates of our superspace as\index{super!space}
\begin{equation}
\begin{aligned}
\tau_1(x^{2\zd})\ :=\ \bar{x}^{1\ed}~,~~~~~
\tau_1(x^{2\ed})\ :=\ \bar{x}^{1\zd}~,\\
\tau_0(x^{\alpha\ald})\ :=\ \bar{x}^{\alpha\ald}~.\hspace{1.8cm}
\end{aligned}
\end{equation}
For $\tau_1$, we can thus extract the real coordinates $x^\mu\in
\FR^{2,2}$, $\mu=1,\ldots ,4$ by
\begin{equation}
x^{2\zd}\ =\ \bar{x}^{1\ed}\ =\ -(x^4+\di x^3)\eand
x^{2\ed}\ =\ \bar{x}^{1\zd}\ =\ -(x^2-\di x^1)~.
\end{equation}
and the real coordinates $x^a\in \FR^{2,1}$, $a=1,2,3$ by
\begin{equation}
y^{\ed\ed}\ =\ -\bar{y}^{\zd\zd}\ =\ (x^1+\di x^2)\ =:\
y~,~~~y^{\ed\zd}\ =\ \bar{y}^{\ed\zd}\ =\ -x^3~.
\end{equation}
For the fermionic coordinates, the actions of the two real
structures read as\index{real structure}
\begin{equation}\label{tau1}
\tau_1\left(\begin{array}{c}\theta^{1i} \\ \theta^{2i}
\end{array}\right)\ =\ 
\left(\begin{array}{c}\bar{\theta}^{2i} \\ \bar{\theta}^{1i}
\end{array}\right)~,~~~
\tau_1\left(\begin{array}{c}\eta^{\dot{1}}_{i} \\[0.8mm] \eta^{\dot{2}}_{i}
\end{array}\right)\ =\ 
\left(\begin{array}{c}\bar{\eta}^{\dot{2}}_{i} \\[0.8mm]
\bar{\eta}^{\dot{1}}_{i}
\end{array}\right)
\end{equation}
and
\begin{equation}\label{tau2}
\tau_0(\theta^{\alpha i})\ =\ \bar{\theta}^{\alpha i}~~~\mbox{and}~~~
\tau_0(\eta^{\ald}_i)\ =\ \bar{\eta}^{\ald}_i~,
\end{equation}
matching the definition for commuting spinors. The resulting\index{Spinor}
Majorana-condition is then
\begin{align}\label{majorana1}
\tau_1(\theta^{\alpha i})\ =\ \theta^{\alpha i}~~~\mbox{and}~~~
\tau_1(\eta^\ald_i)\ =\ \eta^\ald_i~~&\Leftrightarrow~~
\theta^{2i}\ =\ \bar{\theta}^{1i}~~~\mbox{and}~~~
\eta_i^{\dot{2}}\ =\ \bar{\eta}_i^{\dot{1}}~,\\
\label{majorana1.5} \tau_0(\theta^{\alpha i})\ =\ \theta^{\alpha
i}~~~\mbox{and}~~~
\tau_0(\eta^\ald_i)\ =\ \eta^\ald_i~~&\Leftrightarrow~~ \theta^{\alpha
i}\ =\ \bar{\theta}^{\alpha i}~~~\mbox{and}~~~
\eta_i^{\dot{\alpha}}\ =\ \bar{\eta}_i^{\dot{\alpha}}~.
\end{align}

\paragraph{Minkowski case.} Here, we define a real structure $\tau_M$ by\index{real structure}
the equations
\begin{equation}
\tau_M(x^{\alpha\bed})\ =\ -\overline{x^{\beta\ald}}\eand
\tau_M(\eta_i^\ald)\ =\ \overline{\theta^{\alpha i}}~,
\end{equation}
where the indices $\alpha=\ald$ and $\beta=\bed$ denote the same
number.

\paragraph{Euclidean case.}\label{peuclideancase} In the Euclidean
case, the real structure acts on the bosonic coordinates according\index{real structure}
to
\begin{equation}
\tau_{-1}(x^{2\zd})\ :=\ \bar{x}^{1\ed}~,~~~
\tau_{-1}(x^{2\ed})\ :=\ -\bar{x}^{1\zd}~
\end{equation}
and the prescription for a change to real coordinates
$x^\mu\in\FR^4$ reads as
\begin{equation}
x^{2\zd}\ =\ \bar{x}^{1\ed}\ =\ -(-x^4+\di x^3)\eand
x^{2\ed}\ =\ -\bar{x}^{1\zd}\ =\ (x^2-\di x^1)
\end{equation}
in four bosonic dimensions. In the three-dimensional case, we have
\begin{equation}
y^{\ed\ed}\ =\ -\bar{y}^{\zd\zd}\ =\ (x^1+\di x^2)\ =:\
y~,~~~y^{\ed\zd}\ =\ \bar{y}^{\ed\zd}\ =\ -x^3~.
\end{equation}
Here, we can only fix a real structure on the fermionic\index{real structure}
coordinates if the number of supersymmetries $\CN$ is even (see
e.g.\ \cite{Kotrla:1984ky, Lukierski:1986jw}). In these cases, one
groups together the fermionic coordinates in pairs and defines
matrices
\begin{equation*}
(\eps_r{}^s)\ :=\ \left(\begin{array}{cc} 0 & -1 \\ 1 & 0
\end{array}\right)~,~~r,s\ =\ 1,2~~~\mbox{and}~~~
(T_i{}^j)\ :=\ \left(\begin{array}{ccc} \eps & 0 \\
0 & \eps
\end{array}\right)~,~~i,j\ =\ 1,\ldots ,4~.
\end{equation*}
 The action of $\tau_{-1}$ is then given by
\begin{equation}\label{tau3}
\tau_{-1}\left(\begin{array}{cc}\theta^{11} & \theta^{1 2} \\
\theta^{2 1} & \theta^{2 2}
\end{array}\right)\ =\ \left(\begin{array}{cc}
0 & -1 \\ 1 & 0
\end{array}\right)
\left(\begin{array}{cc}\bar{\theta}^{1 1} & \bar{\theta}^{1 2} \\
\bar{\theta}^{2 1} & \bar{\theta}^{2 2}
\end{array}\right)\left(\begin{array}{cc}
0 & -1 \\ 1 & 0
\end{array}\right)
\end{equation}
for $\CN{=}2$ and by
\begin{equation*}
\tau_{-1}\left(\begin{array}{ccc}\theta^{11} & \cdots &\theta^{1 4} \\
\theta^{2 1} & \cdots & \theta^{2 4}
\end{array}\right)\ =\ \left(\begin{array}{cc}
0 & -1 \\ 1 & 0
\end{array}\right)
\left(\begin{array}{ccc}\bar{\theta}^{1 1} & \cdots &\bar{\theta}^{1 4} \\
\bar{\theta}^{2 1} & \cdots & \bar{\theta}^{2 4}
\end{array}\right)\left(\begin{array}{cccc}
0 & -1 & 0 & 0 \\ 1 & 0 & 0 & 0 \\
0 & 0 & 0 & -1\\
0 & 0 & 1 & 0
\end{array}\right)
\end{equation*}
for $\CN{=}4$. The last equation can also be written in components
as
\begin{subequations}
\begin{equation}
\tau_{-1}(\theta^{\alpha i})\ =\ -\eps^{\alpha\beta}
T_j{}^i\bar{\theta}^{\beta j}~,
\end{equation}
where there is a summation over $\beta$ and $j$. The same
definition applies to $\eta^\ald_i$:
\begin{equation}\label{tau5}
\tau_{-1}(\eta_i^\ald)\ =\ \eps^{\ald\bed}
T_i{}^j\bar{\eta}_{j}^\bed~.
\end{equation}
\end{subequations}
The reality conditions here are symplectic Majorana
conditions, which read explicitly\index{Majorana condition}\index{symplectic Majorana condition}
\begin{equation}\label{realcond}
\tau_{-1}(\theta^{\alpha i})\ =\ \theta^{\alpha i}~~~\mbox{and}~~~
\tau_{-1}(\eta_i^\ald)\ =\ \eta_i^\ald~.
\end{equation}
We have for instance for $\CN{=}4$
\begin{equation}\label{realcnd}
\tau\left(\begin{array}{cccc} \eta_1^\ed & \eta_2^\ed &\eta_3^\ed
&\eta_4^\ed \\\eta_1^\zd &\eta_2^\zd &\eta_3^\zd &\eta_4^\zd
\end{array}\right)\ =\ \left(\begin{array}{cccc}
-\bar{\eta}^{\dot{2}}_2 &\bar{\eta}^{\dot{2}}_1
&-\bar{\eta}^{\dot{2}}_4 &\bar{\eta}^{\dot{2}}_3 \\
\bar{\eta}^{\dot{1}}_2 &-\bar{\eta}^{\dot{1}}_1
&\bar{\eta}^{\dot{1}}_4 &-\bar{\eta}^{\dot{1}}_3 \\
\end{array}\right)~.
\end{equation}

\chapter{Field Theories}

The purpose of this chapter is to give an overview of the field
theories we will encounter in this thesis. Basic facts together
with the necessary well-known results are recalled for convenience
and in order to fix our notation.

\section{Supersymmetric field theories}

First, let us briefly recall some elementary facts on
supersymmetric field theories which will become useful in the
subsequent discussion. In particular, we will discuss the $\CN=1$
superfield formalism and present some features of supersymmetric
quantum field theories as supersymmetric Ward-Takahashi identities\index{Ward-Takahashi identities}
and non-renormalization theorems. The relevant references for this\index{non-renormalization theorems}\index{Theorem!non-renormalization}
section are
\cite{Wess:1992cp,Buchbinder:1998qv,Lykken:1996xt,Argyres,Bilal:2001nv}.

\subsection{The $\CN=1$ superspace formalism}\index{super!space}

When discussing the massless representations of the $\CN=1$\index{representation}
supersymmetry algebra in \ref{sssusyrep}, \ref{pmasslessrep}, we\index{super!symmetry}
encountered two multiplets: the chiral multiplet with fields\index{chiral!multiplet}
having helicities $(0,\frac{1}{2})$ and the vector multiplet\index{vector multiplet}
consisting of fields with helicities $(\frac{1}{2},1)$. There is a
nice way of representing both multiplets as superfunctions on the
superspace $\FR^{4|4}$, which allows us to easily write down\index{super!space}
supersymmetric actions and often simplifies further examinations
of supersymmetric theories significantly. Throughout this section,
we will assume a superspace with Minkowski signature.\index{super!space}

\paragraph{General superfield.} A general superfield on the $\CN=1$
superspace $\FR^{4|4}$ with coordinates $(x^\mu,\theta^{\alpha\index{super!space}
i},\thetab^i_\ald)$ can be expanded as a power series in the
Gra{\ss}mann variables with highest monomial $\theta^2\btheta^2$.\index{Gra{\ss}mann variable}
However, this representation of the supersymmetry algebra is\index{representation}\index{super!symmetry}
reducible and by applying different constraints onto the general
superfield, we will obtain two irreducible representations: the\index{irreducible representation}\index{representation}
chiral superfield and the vector superfield.\index{vector superfield}

\paragraph{Chiral superfield.} Chiral and anti-chiral
superfields $\Phi$ and $\bar{\Phi}$ are defined via the condition\index{chiral!superfields}
\begin{equation}
\bar{D}_\ald \Phi\ =\ 0\eand D_\alpha \bar{\Phi}\ =\ 0~,
\end{equation}
respectively. These conditions are most generally solved by
restricting the functions $\Phi$ and $\bar{\Phi}$ to the chiral
and anti-chiral subspaces $\FR^{4|2}_L$ and $\FR^{4|2}_R$ of the
superspace $\FR^{4|4}$:\index{super!space}
\begin{equation}
\Phi\ =\ \Phi(y_L^{\alpha\ald},\theta^\alpha)\eand
\bar{\Phi}\ =\ \bar{\Phi}(y_R^{\alpha\ald},\btheta^\ald)~.
\end{equation}
Let us now focus on the chiral superfields, the anti-chiral ones\index{chiral!superfields}
are obtained by complex conjugation. Their component expansion
reads as\footnote{Recall our convention for spinor bilinears,\index{Spinor}
e.g.\ $\theta\theta=\theta^\alpha\theta_\alpha$ and
$\btheta\btheta=\btheta_\ald\btheta^\ald$.}
\begin{equation}
\Phi(y,\theta)\ =\ \phi(y)+\sqrt{2}\theta\psi(y)-\theta\theta F(y)~,
\end{equation}
where $\phi$ is a complex scalar with helicity 0, $\psi_\alpha$ is\index{helicity}
a Weyl spinor with helicity $\frac{1}{2}$ and $F$ is an auxiliary\index{Spinor}
field which causes the supersymmetry algebra to close off-shell.\index{super!symmetry}
The field $\Phi$ now contains the complete chiral multiplet and\index{chiral!multiplet}
the supersymmetry transformations are easily read off to be\index{super!symmetry}
\begin{equation}
\delta \phi\ =\ \sqrt{2}\eps
\psi~,~~~\delta\psi_\alpha\ =\ \sqrt{2}\dpar_{\alpha\ald}\bar{\eps}^\ald-\sqrt{2}F\eps_\alpha\eand
\delta F\ =\ \sqrt{2}\dpar_{\alpha\ald}\psi^\alpha\bar{\eps}^\ald~.
\end{equation}
Note that this superfield contains $4$ real bosonic and $4$ real
fermionic degrees of freedom off-shell. On-shell, the component
$F$ becomes an auxiliary field and we are left with $2$ real
bosonic and $2$ real fermionic degrees of freedom.

Correspondingly, the complex conjugate field $\bar{\Phi}$ is an
anti-chiral superfield containing the anti-chiral multiplet with\index{chiral!multiplet}
fields of helicity $0$ and $-\frac{1}{2}$.\index{helicity}

\paragraph{Vector superfield.} To represent the vector multiplet\index{vector multiplet}\index{vector superfield}
containing fields of helicity $\pm\frac{1}{2}$ and $\pm1$, it is\index{helicity}
clear that we will need both left- and right-handed Gra{\ss}mann
variables, and the vector superfield will be a function on the\index{Gra{\ss}mann variable}\index{vector superfield}
full $\CN=1$ superspace $\FR^{4|4}$. Na\"ively, this gives rise to\index{super!space}
$16$ components in the superfield expansion. However, by imposing
the so-called {\em Wess-Zumino gauge}, one can reduce the\index{Wess-Zumino gauge}
components and obtain the following field expansion
\begin{equation}\label{expVWZ}
V_{\mathrm{WZ}}\ =\ \di\theta\btheta\sigma^\mu A_\mu(x)+\di
\theta\theta\btheta \bar{\lambda}(x)-\di \btheta\btheta\theta
\lambda(x)+\tfrac{1}{2}\theta\theta\btheta\btheta D(x)~,
\end{equation}
giving rise to the {\em real} Lie algebra valued vector superfield\index{Lie algebra}\index{vector superfield}
$V_{\mathrm{WZ}}=-V_{\mathrm{WZ}}^\dagger$, where we chose the
generators of the gauge group to be anti-Hermitian. A disadvantage
of this gauge is that it is not invariant under supersymmetry\index{super!symmetry}
transformations, i.e.\ any supersymmetry transformation will cause
additional terms in the expansion \eqref{expVWZ} to appear, which,
however, can subsequently be gauged away.

A gauge transformation is now generated by a Lie algebra valued\index{Lie algebra}\index{gauge!transformation}
chiral superfield $\Lambda$ and acts on a vector superfield $V$ by\index{vector superfield}
\begin{equation}\label{gaugetrafoV}
\de^V\ \rightarrow\  \de^{i\Lambda^\dagger}\de^V\de^{-\di \Lambda}~.
\end{equation}
There are two corresponding field strengths defined by\index{field strength}
\begin{equation}
W_\alpha\ =\ -\tfrac{1}{4}\bar{D}\bar{D}\left(\de^{-2V}D_\alpha\de^{2V}\right)\eand
\bar{W}_\ald\ =\ -\tfrac{1}{4}DD\left(\de^{2V}\bar{D}_\ald\de^{-2V}\right)~,
\end{equation}
the first of which is chiral (since $\bar{D}^3=0$), the second
anti-chiral. Both field strengths transform covariantly under the\index{field strength}
gauge transformations \eqref{gaugetrafoV}:\index{gauge transformations}\index{gauge!transformation}
\begin{equation}
W_\alpha\ \rightarrow\  \de^{\di \Lambda} W_\alpha
\de^{-\di\Lambda}\eand \bar{W}_\ald\ \rightarrow\  \de^{\di
\Lambda^\dagger} \bar{W}_\ald \de^{-\di\Lambda^\dagger}~.
\end{equation}
Eventually, let us stress that all the above formul\ae{} were
given for a non-Abelian gauge group and simplify considerably for
Abelian gauge groups.

\paragraph{SUSY invariant actions.} Actions which are invariant
under supersymmetry are now easily constructed by considering\index{super!symmetry}
polynomials in superfields and integrating over the appropriate
superspace. When constructing such actions, one has however to\index{super!space}
guarantee that the action is Hermitian and that additional
symmetries, as e.g.\ gauge invariance are manifest. The former is
easily achieved by adding complex conjugated terms for chiral
expressions. An example for such a gauge invariant action is
\begin{equation}\label{SUSYactionN1YM}
S\sim\int\dd^4 x~\tr\left(\int \dd^2 \theta~WW+\int \dd^2
\btheta~\bar{W}\bar{W}\right)~,
\end{equation}
corresponding to $\CN=1$ super Yang-Mills theory. Note that both\index{Yang-Mills theory}
terms in \eqref{SUSYactionN1YM} are real and equal. Furthermore, a
coupling to chiral matter is achieved via an additional term $\sim
\int \dd^4x\dd^4\theta\,\, \bar{\Phi}\de^{2V}\Phi$ in the action.
In the latter case, gauge transformations act on the chiral\index{gauge transformations}\index{gauge!transformation}
superfields according to $\Phi\rightarrow\de^{\di\Lambda}\Phi$ and
$\bar{\Phi}\rightarrow \bar{\Phi}\de^{-\di\Lambda^\dagger}$.

\paragraph{On-shell massive representations for superspin $0$.} By\index{representation}\index{super!spin}
combining two complex massless chiral superfields via the\index{chiral!superfields}
equations
\begin{equation}
-\tfrac{1}{4}\bar{D}^2\bar{\Phi}+m\Phi\ =\ 0\eand
-\tfrac{1}{4}D^2\Phi+m\bar{\Phi}\ =\ 0~,
\end{equation}
one can find real on-shell representations in terms of $\CN=1$\index{representation}
superfields also for a massive multiplet of superspin 0. This fact\index{super!spin}
is essential in constructing the Wess-Zumino model.\index{Wess-Zumino model}

\subsection{The Wess-Zumino model}\label{WessZumino}\index{Wess-Zumino model}

One of the most popular supersymmetric field theories is the
Wess-Zumino model. It is well-suited as a toy model to demonstrate\index{Wess-Zumino model}
features of supersymmetric field theories arising due to their
supersymmetry, as e.g.\ non-renormalization theorems.\index{non-renormalization theorems}\index{super!symmetry}\index{Theorem!non-renormalization}

\paragraph{Action.} This model was proposed by J.~Wess and
B.~Zumino in \cite{Wess:1974tw} and is given by the action
\begin{equation}\label{actionWZM}
S_{\mathrm{WZM}}\ :=\ \int \dd^4 x\dd^4\theta~\bar{\Phi}\Phi+\int
\dd^4x\dd^2\theta~\CCL_c(\Phi)+\int
\dd^4x\dd^2\btheta~\bar{\CCL_c}(\bar{\Phi})~,
\end{equation}
where $\CCL_c$ is a holomorphic function of a complex field, the
{\em chiral superpotential}. The kinetic term arises from\index{chiral!superpotential}
$\bar{\Phi}\Phi$ after Taylor-expanding the chiral superfields\index{chiral!superfields}
around the non-chiral coordinate $x^{\alpha\ald}$. While $\CCL_c$
is classically unrestricted, renormalizability demands that it is
at most a third-order polynomial, and we will adapt the common
notation
\begin{equation}
\CCL_c\ :=\ \frac{m}{2}\Phi^2+\frac{\lambda}{3!}\Phi^3~.
\end{equation}
We have dropped the monomial of order 1, as it simply amounts to a
constant shift in the superfield $\Phi$.

\paragraph{Equations of motion.} The corresponding equations of
motion are easily derived to be
\begin{equation}
-\tfrac{1}{4}\bar{D}^2\bar{\Phi}+\CCL_c'(\Phi)\ =\ 0\eand
-\tfrac{1}{4}D^2\Phi+\bar{\CCL}_c'(\bar{\Phi})\ =\ 0~.
\end{equation}

\paragraph{The Landau-Ginzburg model.} \label{LandauGinzburg} A possibility\index{Landau-Ginzburg model}
of generalizing the action \eqref{actionWZM} is to allow for
several chiral superfields. Such a model with $n$ massless chiral\index{chiral!superfields}
superfields $\Phi_a$ and a polynomial interaction is called a {\em
Landau-Ginzburg model} and its action reads as\index{Landau-Ginzburg model}
\begin{equation}\label{LGaction}
S=\int \dd^4 x\left(\int\dd^4\theta~
\CCK(\Phi_a,\bar{\Phi}_a)+\frac{1}{2}\int\dd^2\theta~
\CCL_c(\Phi_a) +\frac{1}{2}\int\dd^2\theta~
\bar{\CCL_c}(\bar{\Phi}_a)\right)~,
\end{equation}
where $\CCL_c(\Phi_a)$ is again called the (chiral)
superpotential. The vacua of the theory are the critical points of
$\CCL_c(\Phi_a)$.

The function $\CCK(\Phi_a,\bar{\Phi}_a)$ can be considered as a
{\em K\"{a}hler potential} and defines the {\em K\"{a}hler metric}\index{K\"{a}hler!metric}\index{K\"{a}hler!potential}\index{metric}
$g_{i\bj}:=\dpar_i\dpar_\bj \CCK(\Phi_a,\bar{\Phi}_a)$. Note that
the component fields in the action \eqref{LGaction} couple via the
K\"{a}hler metric $g_{i\bj}$ and higher derivatives of the K\"{a}hler\index{K\"{a}hler!metric}\index{metric}
potential. For vanishing $\CCL_c$, the Landau-Ginzburg is a
nonlinear sigma model (cf.\ section \ref{ssnonlinsigmamodel}),\index{nonlinear sigma model}\index{sigma model}
which defines a K\"{a}hler geometry via $\CCK$. One can also obtain a
supersymmetric nonlinear sigma model from a K\"{a}hler geometry.\index{nonlinear sigma model}\index{sigma model}

It is known that the Landau-Ginzburg models with a single chiral\index{Landau-Ginzburg model}
superfield $\Phi$ and polynomial interaction
$\CCL_c(\Phi)=\Phi^{P+2}$ has central charge $c_P=\frac{3P}{P+2}$\index{central charge}
at its infrared fixed point and can be shown to be essentially the
$P$-th minimal model.\index{minimal model}

\subsection{Quantum aspects}\label{ssquantumaspects}

The heavy constraints imposed by supersymmetry on a quantum field\index{super!symmetry}
theory become manifest at quantum level: the additional symmetry
leads to a cancellation of contributions from certain Feynman
diagrams and the vacuum energy does not receive any quantum
corrections. Furthermore, there is additional structure found in
the correlation functions, the so-called chiral rings, which we\index{chiral!ring}
will discuss momentarily. Certain properties of these rings lead
quite directly to non-renormalization theorems, which strongly\index{non-renormalization theorems}\index{Theorem!non-renormalization}
constrain the allowed quantum corrections and simplify
considerably the study of a supersymmetric quantum field theory.

\paragraph{Quantization.} Consider a quantum field theory with a set\index{quantization}
of fields $\varphi$ and an action functional $S[\varphi]$ which
splits into a free and an interaction part
$S[\varphi]=S_0[\varphi]+S_{\mathrm{int}}[\varphi]$. The
generating functional is given by
\begin{equation}
\CZ[J]\ :=\ \int \CCD \varphi \de^{\di S[\varphi]+\int \dd^4 x~\varphi
J}
\end{equation}
from which the $n$-point correlation functions are defined by
\begin{equation}
G_n(x_1,\ldots ,x_n)\ :=\ \left.\frac{1}{\CZ[J]} \frac{\delta^n
\CZ[J]}{\delta J(x_1)\ldots \delta J(x_n)}\right|_{J=0}~.
\end{equation}
Perturbation theory is done in terms of the power expansion of the
following reformulation of the generating functional:
\begin{equation}
\CZ[J]\ =\ \de^{\di S_{\mathrm{int}}\left[\frac{\delta}{\delta
J}\right]}\CZ_0[J]\ewith \CZ_0[J]\ :=\ \int \CCD \varphi \de^{\di
S_0[\varphi]+\int \dd^4 x~\varphi J}~,
\end{equation}
which will yield a power series in the coupling constants
contained in $S_{\mathrm{int}}$.

\paragraph{Low energy effective action.} Consider a quantum field\index{low energy effective action}
theory with fields $\phi$ and action $S$. We choose some cutoff
$\Lambda$ and decompose the fields into high- and low-frequency
parts:
\begin{equation}
\phi\ =\ \phi_H+\phi_L\ewith\phi_H:
\omega\ >\ \Lambda~~~\phi_L:\omega\ <\ \Lambda~.
\end{equation}
In the path integral, we then perform the integration of all
high-frequency fields $\phi_H$ and arrive at
\begin{equation}
\int \CCD \phi_L\CCD\phi_H\de^{\di S[\phi_L,\phi_H]}\ =\ \int \CCD
\phi_L \de^{\di S_\Lambda[\phi_L]}~,
\end{equation}
where $S_\Lambda[\phi_L]$ is the so-called\footnote{This effective
action is not to be confused with the 1PI effective action related
to the one particle irreducible diagrams and calculated from the
standard generating functional via a Legendre transform.} {\em low
energy effective action} or {\em Wilsonian effective action}. For\index{low energy effective action}
more details, see e.g.\ \cite{Polchinski:1992ed}.

\paragraph{Chiral rings and correlation functions.} The chiral\index{chiral!ring}
rings of operators in supersymmetric quantum field theories are
cohomology rings of the supercharges $Q_{i \alpha}$ and
$\bar{Q}^i_\ald$. Correlation functions which are built out of
elements of a single such chiral ring have peculiar properties.\index{chiral!ring}

Recall that the vacuum is annihilated by both the supersymmetry\index{super!symmetry}
operators $Q_{\alpha i}$ and $\bar{Q}^i_\ald$. Using this fact, we
see that $Q_{\alpha i}$- and $\bar{Q}^i_\ald$-exact operators
cause a correlation function built of $Q_{\alpha i}$ or
$\bar{Q}^i_\ald$-closed operators to vanish, e.g.\
\begin{equation}
\begin{aligned}
\langle \lsc Q,A \rsc
 \bar{\CO}_1 \ldots \bar{\CO}_n\rangle\ =\ & \langle
\lsc Q,A \bar{\CO}_1 \ldots \bar{\CO}_n\rsc \rangle\pm\langle A
\lsc Q,\bar{\CO}_1 \rsc
 \ldots \bar{\CO}_n\rangle\\&\pm\ldots\pm\langle A
\bar{\CO}_1 \ldots \lsc Q,\bar{\CO}_n\rsc  \rangle\\
\ =\ &\langle Q A \bar{\CO}_1 \ldots \bar{\CO}_n\rangle\pm\langle
A \bar{\CO}_1  \ldots \bar{\CO}_n  Q\rangle\ =\ 0~.
\end{aligned}
\end{equation}
The resulting two cohomology rings are called the {\em chiral} and
{\em anti-chiral ring}.\index{anti-chiral ring}\index{chiral!ring}

\paragraph{Supersymmetric Ward-Takahashi identities.}\label{psusyWTI} The\index{Ward-Takahashi identities}
existence of chiral rings in our theory leads to supersymmetric\index{chiral!ring}
Ward-Takahashi identities. Since one can write any derivative with\index{Ward-Takahashi identities}
respect to bosonic coordinates as an anticommutator of the
supercharges due to the supersymmetry algebra, any such derivative\index{super!symmetry}
of correlation functions built purely out of chiral or anti-chiral
operators will vanish:
\begin{equation}
\der{x^\mu}\langle \CO_1 \ldots \CO_n\rangle\ =\  \der{x^\mu}\langle
\bar{\CO}_1 \ldots \bar{\CO}_n\rangle \ =\ 0~.
\end{equation}
The correlation functions do not depend on the bosonic coordinates
of the operators, and hence one can move them to a far distance of
each other, which causes the correlation function to
factorize\footnote{This observation has first been made in
\cite{Novikov:1983ee}.}:
\begin{equation}
\langle\bar{\CO}_1(x_1)\ldots\bar{\CO}_n(x_n)\rangle\ =\
\langle\bar{\CO}_1(x^\infty_1)\rangle\ldots\langle\bar{\CO}_n(x^\infty_n)\rangle~.
\end{equation}
Such a correlation function therefore does not contain any contact
terms. This phenomenon is called {\em clustering} in the\index{clustering}
literature.

Another direct consequence of the existence of chiral rings is the\index{chiral!ring}
holomorphic dependence of the chiral correlation functions on the
coupling constants, i.e.
\begin{equation}
\der{\bl}\langle\CO_1\ldots\CO_n\rangle\ =\ 0~.
\end{equation}
As an illustrative example for this, consider the case of a
$\CN=1$ superpotential ``interaction'' term added to the
Lagrangian,
\begin{equation}
\CL_{\mathrm{W}}\ =\ \int\dd^2\theta~ \lambda \Phi+
\int\dd^2\bar{\theta}~\bl\bar{\Phi}~,
\end{equation}
where
$\Phi=\phi(y)+\sqrt{2}\theta^\alpha\psi_\alpha(y)-\theta^2F(y)$ is
a chiral superfield and one of the supersymmetry transformations\index{super!symmetry}
is given by $\{Q_\alpha,\psi_\beta\}\sim\eps_{\alpha\beta}F$. Then
we have
\begin{equation}\label{WardIdentity3}
\begin{aligned}
\der{\bl}\langle\CO_1\ldots\CO_n\rangle&\ =\ \int\dd^4
y\dd^2\bar{\theta}\langle\CO_1\ldots\CO_n\bar{\Phi}\rangle\ \sim\
\int\dd^4 y\langle\CO_1\ldots\CO_n F\rangle\\ &\ \sim\ \int\dd^4
y\langle\CO_1\ldots\CO_n\{\bar{Q}_\ald,\bar{\psi}^\ald\}\rangle\
=\ 0~.
\end{aligned}
\end{equation}

\paragraph{Non-renormalization theorems.} It is in the\index{non-renormalization theorems}\index{Theorem!non-renormalization}
non-renormalization theorems\footnote{For more details and a
summary of non-renormalization theorems, see
\cite{Buchbinder:2004bu}.} that supersymmetric field theories
reveal their full power as quantum field theories.

\begin{itemize}
\item Every term in the effective action of an $\CN=1$
supersymmetric quantum field theory can be written as an integral
over the full superspace.\index{super!space}
\item The general structure of the effective action of the
Wess-Zumino model is given by\index{Wess-Zumino model}
\begin{equation*}
\Gamma[\Phi,\bar{\Phi}]\ =\ \sum_n\int \dd^4 x_1\ldots \dd^4x_n\int
\dd^4\theta~f(x_1,\ldots,x_n)F_1(x_1,\theta)\ldots
F_n(x_n,\theta)~,
\end{equation*}
where the $F_i$ are local functions of the fields $\Phi$,$
\bar{\Phi}$ and their covariant derivatives.\index{covariant derivative}
\item The superpotential of the Wess-Zumino model is not\index{Wess-Zumino model}
renormalized at all. For more details on this point, see section
\ref{ssapplications}, \ref{pnonren}.
\item This is also true for $\CN=1$ super Yang-Mills theories.
The renormalization of the kinetic term only happens through the
gauge coupling and here not beyond one-loop order.
\item All vacuum diagrams sum up to zero and thus, consistently with
our analysis of the supersymmetry algebra, the vacuum energy is\index{super!symmetry}
indeed zero.
\item The action of $\CN=2$ supersymmetric theories can always be
written as
\begin{equation}
\tfrac{1}{16\pi}\mathrm{Im}\int \dd^4
x\dd^2\theta^1\dd^2\theta^2~\CCF(\Psi)~,
\end{equation}
where $\CCF$ is a holomorphic function of $\Psi$ called the {\em
prepotential}. The field $\Psi$ is the $\CN=2$ chiral superfield
composed of a $\CN=1$ chiral superfield $\Phi$ and the super field
strength $W_\alpha$ according to\index{field strength}
\begin{equation}
\Psi\ =\ \Phi(y,\theta^1)+\sqrt{2}\theta^{2\alpha}W_\alpha(y,\theta^1)+
\theta^{2\alpha}\theta^2_\alpha G(y,\theta^1)~.
\end{equation}
For $\CN=2$ super Yang-Mills theory, the prepotential is $\CCF\sim\index{Yang-Mills theory}
\tr(\Psi^2)$.
\item The $\beta$-function for $\CN=4$ super Yang-Mills theory\index{N=4 super Yang-Mills theory@$\CN=4$ super Yang-Mills theory}\index{Yang-Mills theory}
vanishes and hence the coupling constant does not run.
\end{itemize}

\section{Super Yang-Mills theories}\label{sSYMtheories}

In the following section, we describe the maximally supersymmetric
Yang-Mills theories one obtains from $\CN=1$ in ten dimensions by
dimensional reduction. The key references for our discussion are\index{dimensional reduction}
\cite{Brink:1976bc,Harnad:1985bc,Harnad:1984vk} (super Yang-Mills)
and \cite{Dorey:2002ik,Tong:2005un} (instantons and monopoles).\index{instanton}
Further references are found in the respective sections.

\subsection{Maximally supersymmetric Yang-Mills
theories}\label{ssmaxsusy}

\paragraph{Preliminaries.} We start from
$d$-dimensional Minkowski space $\FR^{1,d-1}$ with Minkowski
metric $\eta_{\mu\nu}=\diag(+1,-1,\ldots ,-1)$. On this space,\index{metric}
consider a vector bundle with a connection, i.e.\ a one-form\index{connection}
$A_\mu$ taking values in the Lie algebra of a chosen gauge group\index{Lie algebra}
$G$. We will always assume that the corresponding generators are
anti-Hermitian. The associated {\em field strength} is defined by\index{field strength}
\begin{equation}
F_{\mu\nu}\ :=\ [\nabla_\mu,\nabla_\nu]\ =\ \dpar_\mu A_\nu-\dpar_\nu
A_\mu+[A_\mu,A_\nu]~.
\end{equation}
Consider furthermore a spinor $\lambda$ transforming in the double\index{Spinor}
cover $\sSpin(1,d-1)$ of $\sSO(1,d-1)$ and in the adjoint
representation of the gauge group $G$. Its {\em covariant\index{representation}
derivative} is defined by $\nabla_\mu\lambda:=\dpar_\mu
\lambda+[A_\mu,\lambda]$.

{\em Gauge transformations}, which are parameterized by smooth\index{gauge transformations}\index{gauge!transformation}
sections $g$ of the trivial bundle $G\times \FR^{1,d-1}$, will act
on the above fields according to
\begin{equation}
A_\mu \ \mapsto\  g^{-1} A_\mu g+g^{-1}\dpar_\mu g~,~~~
F_{\mu\nu}\ \mapsto\  g^{-1} F_{\mu\nu} g~,~~~\lambda\ \mapsto\  g^{-1}
\lambda g~,
\end{equation}
and thus the terms
\begin{equation}
\tr\left(-\tfrac{1}{4} F_{\mu\nu}F^{\mu\nu}\right)\eand
\tr\left(\di\bl \Gamma^\mu \nabla_\mu \lambda\right)
\end{equation}
are gauge invariant quantities. Note that $\Gamma^\mu$ is a
generator of the Clifford algebra $\CCC(1,d-1)$.\index{Clifford algebra}

\paragraph{$\CN=1$ super Yang-Mills theory.} Recall that a gauge\index{Yang-Mills theory}
potential in $d$ dimensions has $d-2$ degrees of freedom, while
the counting for a Dirac spinor yields\index{Spinor}
$2^{\lfloor\frac{d}{2}\rfloor}$. By additionally imposing a
Majorana or a Weyl condition, we can further halve the degrees of
freedom of the spinor. Thus the action\index{Spinor}
\begin{equation}
S=\int \dd^d x~ \tr\left(-\tfrac{1}{4} F_{\mu\nu}F^{\mu\nu}+\di\bl
\Gamma^\mu \nabla_\mu \lambda\right)
\end{equation}
can only posses a linear supersymmetry in dimensions four, six and\index{super!symmetry}
ten. More explicitly, supersymmetry is possible in $d=10$ with
both the Majorana and the Weyl condition imposed on the spinor\index{Spinor}
$\lambda$, in $d=6$ with the Weyl condition imposed on $\lambda$
and in $d=4$ with either the Majorana or the Weyl
condition\footnote{Here, actually both are equivalent.} imposed on
$\lambda$. These theories will then have $\CN=1$ supersymmetry.\index{super!symmetry}

In the following, we will always be interested in maximally
supersymmetric Yang-Mills theories and thus start from $\CN=1$ in
$d=10$ with 16 supercharges. Higher numbers of supersymmetries
will lead to a graviton appearing in the supermultiplet, and in\index{supermultiplet}
supergravity \cite{VanNieuwenhuizen:1981ae}, one in fact considers
theories with 32 supercharges. On the other hand, as we saw by the
above considerations of degrees of freedom, we cannot construct
$\CN=1$ supersymmetric field theories in higher dimensions.
Further theories will then be obtained by dimensional reduction.\index{dimensional reduction}

\paragraph{$\CN=1$ SYM theory in $d=10$.} This theory is defined
by the action\footnote{In this section, we will always denote the
fields of the ten-dimensional theory by a hat.}
\cite{Brink:1976bc}
\begin{equation}
S=\int \dd^{10}
x~\tr\left(-\tfrac{1}{4}\hat{F}_{MN}\hat{F}^{MN}+\tfrac{\di}{2}\hat{\bl}
\Gamma^M \hat{\nabla}_M \hat{\lambda}\right)~,
\end{equation}
where $\hat{\lambda}$ is a 16-dimensional Majorana-Weyl spinor and\index{Majorana-Weyl spinor}\index{Spinor}
therefore satisfies
\begin{equation}
\hat{\lambda}\ =\  C\hat{\bl}^T\eand \hat{\lambda}\ =\ + \Gamma
\hat{\lambda}~.
\end{equation}
Here, $C$ is the charge conjugation operator and\index{charge conjugation operator}
$\Gamma=\di\Gamma_0\ldots \Gamma_{9}$. The supersymmetry\index{super!symmetry}
transformations are given by
\begin{equation}
\delta \hat{A}_M\ =\ \di \bar{\alpha}\Gamma_M \hat{\lambda} \eand
\delta\hat{\lambda}\ =\ \Sigma_{MN} \hat{F}^{MN}\alpha~.
\end{equation}

\paragraph{Constraint equations.} Instead of deriving the\index{constraint equations}
equations of motion of ten-dimen\-sional SYM theory from an
action, one can also use so-called {\em constraint equations},\index{constraint equations}
which are the compatibility conditions of a linear system and thus\index{compatibility conditions}\index{linear system}
fit naturally in the setting of integrable systems. These\index{integrable}
constraint equations are defined on the superspace $\FR^{10|16}$\index{constraint equations}\index{super!space}
with Minkowski signature on the body. They read\index{body}
\begin{equation}
\{ \hat{\nabla}_A,\hat{\nabla}_B\}\ =\ 2\Gamma^M_{AB}\hat{\nabla}_M~,
\end{equation}
where $\hat{\nabla}_M=\dpar_M+\hat{\omega}_M$ is the covariant
derivative in ten dimensions and\index{covariant derivative}
\begin{equation}
\hat{\nabla}_A\ :=\ D_A+\hat{\omega}_A\ :=\ \der{\theta^A}+\Gamma^M_{AB}\theta^B\der{x^M}+\hat{\omega}_A
\end{equation}
is the covariant superderivative. Note that both the fields
$\hat{\omega}_M$ and $\hat{\omega}_A$ are superfields. From these
potentials, we construct the spinor superfield and the bosonic\index{Spinor}
curvature\index{curvature}
\begin{equation}
\hat{\lambda}^B\ :=\ \tfrac{1}{10}\Gamma^{M
AB}[\hat{\nabla}_M,\hat{\nabla}_A]\eand
\hat{F}_{MN}\ :=\ [\hat{\nabla}_M,\hat{\nabla}_N]~.
\end{equation}
Using Bianchi identities and identities for the Dirac matrices in
ten dimensions, one obtains the superfield equations
\begin{equation}
\Gamma^M_{AB}\hat{\nabla}_M\hat{\lambda}^B\ =\ 0\eand\hat{\nabla}^M
\hat{F}_{MN}+\tfrac{1}{2}\Gamma_{N
AB}\{\hat{\lambda}^A,\hat{\lambda}^B\}\ =\ 0~.
\end{equation}
One can show that these equations are satisfied if and only if
they are satisfied to zeroth order in their $\theta$-expansion
\cite{Harnad:1985bc}. We will present this derivation in more
detail for the case of $\CN=4$ SYM theory in four dimensions in
section \ref{subsecN4}.

\paragraph{Dimensional reduction.} A {\em dimensional reduction} of\index{dimensional reduction}
a theory  from $\FR^d$ to $\FR^{d-q}$ is essentially a
Kaluza-Klein compactification on the $q$-torus $T^q$, cf.\ also\index{Kaluza-Klein}
section \ref{ssTduality}. The fields along the compact directions
can be expanded as a discrete Fourier series, where the radii of
the cycles spanning $T^q$ appear as inverse masses of higher
Fourier modes. Upon taking the size of the cycles to zero, the
higher Fourier modes become infinitely massive and thus decouple.
In this way, the resulting fields become independent of the
compactified directions. The Lorentz group on $\FR^d$ splits\index{Lorentz!group}
during this process into the remaining Lorentz group on
$\FR^{d-q}$ and an internal symmetry group $\sSO(q)$. When
dimensionally reducing a supersymmetric gauge theory, the latter
group will be essentially the R-symmetry of the theory and the
number of supercharges will remain the same, see also
\cite{Seiberg:1997ax} for more details.

Let us now exemplify this discussion with the dimensional
reduction of ten-dimen\-sional $\CN=1$ SYM theory to $\CN=4$ SYM\index{dimensional reduction}
theory in four dimensions.

\paragraph{$\CN=4$ SYM theory in $d=4$.} The dimensional reduction\index{dimensional reduction}
from $d=10$ to $d=4$ is easiest understood by replacing each
spacetime index $M$ by $(\mu,ij)$. This reflects the underlying
splittings of $\sSO(9,1)\rightarrow \sSO(3,1)\times\sSO(6)$ and
$\sSpin(9,1)\rightarrow \sSpin(3,1)\times\sSpin(6)$, where
$\sSpin(6)\cong\sSU(4)$. The new index $\mu$ belongs to the
four-dimensional vector representation of $\sSO(3,1)$, while the\index{representation}
indices $ij$ label the representation of $\sSpin(6)$ by
antisymmetric tensors of $\sSU(4)$. Accordingly, the gauge
potential $\hat{A}_M$ is split into $(A_\mu,\phi_{ij})$ as
\begin{equation}
\begin{aligned}
\hat{A}_\mu\ =\ A_\mu\eand\phi_{i4}\ =\ \frac{\hat{A}_{i+3}+\di
\hat{A}_{i+6}}{\sqrt{2}}\ewith
\phi^{ij}\ :=\ \tfrac{1}{2}\eps^{ijkl}\phi_{kl}~.
\end{aligned}
\end{equation}
The gamma matrices decompose as
\begin{equation}
\Gamma^\mu\ =\ \gamma^\mu\otimes \unit\eand
\Gamma^{ij}\ =\ \gamma_5\otimes \left(\begin{array}{cc}0 & \rho^{ij}\\
\rho_{ij} & 0
\end{array}\right)~,
\end{equation}
where $\Gamma^{ij}$ is antisymmetric in $ij$ and $\rho^{ij}$ is a
$4\times 4$ matrix given by
\begin{equation}
\left(\rho_{ij}\right)_{kl}\ :=\ \eps_{ijkl}\eand
\left(\rho^{ij}\right)_{kl}\ :=\ \frac{1}{2}\eps^{ijmn}\eps_{mnkl}~.
\end{equation}
With these matrices, one finds that
\begin{equation}
\Gamma\ =\ \Gamma_0\cdots\Gamma_9\ =\ \gamma_5\otimes \unit_8\eand
C_{10}\ =\ C\otimes \left(\begin{array}{cc} 0 & \unit_4\\ \unit_4 & 0
\end{array}\right)~,
\end{equation}
where $C$ is again the charge conjugation operator. For the spinor\index{Spinor}\index{charge conjugation operator}
$\lambda$, we have
\begin{equation}
\lambda\ =\ \left(\begin{array}{c}L \chi^i\\R \tilde{\chi}_i
\end{array}\right)\ewith \tilde{\chi}_i\ =\  C(\bar{\chi}^i)^T~,~~~
L=\frac{\unit+\gamma_5}{2}~,~~~R\ =\ \frac{\unit-\gamma_5}{2}~.
\end{equation}
The resulting action and further details on the theory are found
in section \ref{subsecN4}.

\paragraph{Remark on the Euclidean case.} Instead of compactifying
the ten-dimensional theory on the torus $T^6$, one can also
compactify this theory on the {\em Minkowski torus} $T^{5,1}$,
using an appropriate decomposition of the Clifford algebra\index{Clifford algebra}
\cite{Belitsky:2000ws} and adjusted reality conditions on the
fields. This derivation, however, leads to a non-compact
R-symmetry group and one of the scalars having a negative kinetic
term. As it is not possible to start from an $\CN=1$ SYM action in
ten Euclidean dimensions containing Majorana-Weyl spinors, it is\index{Majorana-Weyl spinor}\index{Spinor}
better to adjust the $\CN=4$ SYM action on four-dimensional
Minkowski space ``by hand'' for obtaining the corresponding
Euclidean action by using a Wick rotation. This is consistent with
the procedure we will use later on: to consider all fields and
symmetry groups in the complex domain and apply the desired
reality conditions only later on.

\paragraph{Further dimensional reductions.} There are further\index{dimensional reduction}
dimensional reductions which are in a similar spirit to the above
discussed reduction from ten-dimensional $\CN=1$ SYM theory to
$\CN=4$ SYM theory in four dimensions. Starting from $\CN=1$ SYM
theory in ten dimensions (six dimensions), one obtains $\CN=2$ SYM
theory in six dimensions (four dimensions). Equally well one can
continue the reduction of $\CN=2$ SYM theory in six dimensions to
$\CN=4$ SYM theory in four dimensions. The reduction of $\CN=4$
SYM theory in four dimensions to three dimensions leads to a
$\CN=8$ SYM theory, where $\CN=8$ supersymmetry arises from\index{N=8 SYM theory@$\CN=8$ SYM theory}\index{super!symmetry}
splitting the complex $\sSpin(3,1)\cong\sSL(2,\FC)$ supercharges
in four dimensions into real $\sSpin(2,1)\cong\sSL(2,\FR)$ ones in
three dimensions. We will discuss this case in more detail in
section \ref{ssRelatedTheories}.

\subsection{$\CN=4$ SYM theory in four dimensions}\label{subsecN4}

The maximally supersymmetric Yang-Mills theory in four dimensions\index{Yang-Mills theory}\index{maximally supersymmetric Yang-Mills theory}\index{super!symmetric Yang-Mills theory}
is the one with $\CN=4$ supersymmetry and thus with 16\index{super!symmetry}
supercharges. This theory received much attention, as it is a
conformal theory even at quantum level and therefore its
$\beta$-function vanishes. In fact, both perturbative
contributions and instanton corrections are finite, and it is\index{instanton}
believed that $\CN=4$ SYM theory is {\em finite at quantum level}.
In the recent development of string theory, this theory played an\index{string theory}
important r{\^o}le in the context of the AdS/CFT correspondence\index{AdS/CFT correspondence}
\cite{Maldacena:1997re} and twistor string theory\index{string theory}\index{twistor}\index{twistor!string theory}
\cite{Witten:2003nn}.

\paragraph{Action and supersymmetry transformations.} The field\index{super!symmetry}
content of four-dimen\-sion\-al $\CN=4$ SYM theory obtained by
dimensional reduction as presented above consists of a gauge\index{dimensional reduction}
potential $A_\mu$, four chiral and anti-chiral spinors\index{Spinor}
$\chi^i_\alpha$ and $\chi_i^\ald$ and three complex scalars
arranged in the antisymmetric matrix $\phi_{ij}$. These fields are
combined in the action
\begin{equation}
\begin{aligned}
S=\int\dd^4 x~\tr&\left\{-\tfrac{1}{4} F_{\mu\nu}F^{\mu\nu}
+\tfrac{1}{2}\nabla_\mu\phi_{ij}\nabla^\mu\phi^{ij}-\tfrac{1}{4}\left[\phi_{ij},\phi_{kl}\right][\phi^{ij},\phi^{kl}]\right.\\
&\left.+\di\bar{\chi}\gamma^\mu\nabla_\mu L\chi-\tfrac{\di}{2}
\left(\bar{\tilde{\chi}}^i[L\chi^j,\phi_{ij}] -\bar{\chi}_i[R
\tilde{\chi}_j,\phi^{ij}]\right) \right\}~,
\end{aligned}
\end{equation}
where we introduced the shorthand notation
$\phi^{ij}:=\frac{1}{2}\eps^{ijkl}\phi_{kl}$ which also implies
$\phi_{ij}=\frac{1}{2}\eps_{ijkl}\phi^{kl}$. The corresponding
supersymmetry transformations are parameterized by four complex\index{super!symmetry}
spinors $\alpha^i$ which satisfy the same Majorana condition as\index{Majorana condition}\index{Spinor}
$\chi^i$. We have
\begin{equation}\label{SUSYN4}
\begin{aligned}
\delta A_\mu &\ =\  \di\left(\bar{\alpha}_i\gamma_\mu
L\chi^i-\bar{\chi}_i\gamma_\mu L\alpha^i\right)~,\\
\delta \phi_{ij} &\ =\ \di\left(\bar{\alpha}_j R\tilde{\chi}_i-
\bar{\alpha}_i R\tilde{\chi}_j+\eps_{ijkl}\bar{\tilde{\alpha}}^k L
\chi^l\right)~,\\
\delta L\chi^i&\ =\ \sigma_{\mu\nu}F^{\mu\nu}L\alpha^i-\gamma^\mu
\nabla_\mu
\phi^{ij}R\tilde{\alpha}_j+\tfrac{1}{2}[\phi^{ik},\phi_{kj}]
L \alpha^j~,\\
\delta R\tilde{\chi}_i&\ =\ \sigma_{\mu\nu}F^{\mu\nu}
R\tilde{\alpha}_i+\gamma^\mu\nabla_\mu\phi_{ij}L\alpha^j+
\tfrac{1}{2}[\phi_{ik},\phi^{kj}]R\tilde{\alpha}_j~.
\end{aligned}
\end{equation}
One should stress that contrary to the Yang-Mills supermultiplets\index{supermultiplet}
for $\CN\leq 3$, the $\CN=4$ supermultiplet is irreducible.

\paragraph{Underlying symmetry groups.} Besides the supersymmetry\index{super!symmetry}
already discussed in the upper paragraph, the theory is invariant
under the Lorentz group $\sSO(3,1)$ and the R-symmetry group\index{Lorentz!group}
$\sSpin(6)\cong \sSU(4)$. Note, however, that the true
automorphism group of the $\CN=4$ supersymmetry algebra is the\index{morphisms!automorphism}\index{super!symmetry}
group $\sU(4)$. Due to its adjoint action on the fields, the sign
of the determinant is not seen and therefore only the subgroup
$\sSU(4)$ of $\sU(4)$ is realized.

As already mentioned, this theory is furthermore conformally
invariant and thus we have the conformal symmetry group
$\sSO(4,2)\cong\sSU(2,2)$, which also gives rise to conformal
supersymmetry additionally to Poincar{\'e} supersymmetry.\index{conformal supersymmetry}\index{super!symmetry}

Altogether, the underlying symmetry group is the supergroup
$\sSU(2,2|4)$. As this is also the symmetry of the space
$AdS_5\times S^5$, $\CN=4$ SYM theory is one of the major
ingredients of the AdS/CFT correspondence.\index{AdS/CFT correspondence}

\paragraph{$\CN=3$ and $\CN=4$ SYM theories.} The automorphism\index{morphisms!automorphism}
group of the $\CN=3$ supersymmetry algebra is $\sU(3)$. However,\index{super!symmetry}
the R-symmetry group of $\CN=3$ SYM theory is only $\sSU(3)$. With
respect to the field content and its corresponding action and
equations of motion, $\CN=3$ and $\CN=4$ SYM theory are completely
equivalent. When considering the complexified theories, one has to
impose an additional condition in the case $\CN=4$, which reads
\cite{Witten:1978xx}
$\phi_{ij}=\frac{1}{2}\eps_{ijkl}\bar{\phi}^{kl}$ and makes the
fourth supersymmetry linear.\index{super!symmetry}

\paragraph{Spinorial notation.} Let us switch now to spinorial notation (see also\index{Spinor}
\ref{ssSuperspaces}, \ref{pSpinorialNotation}), which will be much
more appropriate for our purposes. Furthermore, we will choose a
different normalization for our fields to match the conventions in
the publications we will report on.

In spinorial notation, we essentially substitute indices $\mu$ by\index{Spinor}
pairs $\alpha\ald$, i.e.\ we use the spinor representation\index{representation}
$(\mathbf{\frac{1}{2},\bar{\frac{1}{2}}})$ equivalent to the
$\mathbf{4}$ of $\sSO(3,1)$. The Yang-Mills field strength thus\index{field strength}
reads as\footnote{The following equations include the
decomposition of the field strength into self-dual and\index{field strength}
anti-self-dual parts, see the next section.}\index{anti-self-dual}
\begin{equation}
[\nabla_{\alpha\ald},\nabla_{\beta\bed}]\ =:\ 
F_{\alpha\ald,\beta\bed}\ =\  \eps_{\ald\bed}
f_{\alpha\beta}+\eps_{\alpha\beta} f_{\ald\bed}~,
\end{equation}
and the action takes the form
\begin{align}\label{spinoraction}
S\ =\ &\int\dd^4 x~\tr\left\{ f^{\ald\bed}f_{\ald\bed} +
f^{\alpha\beta}f_{\alpha\beta}+\nabla^{\alpha\ald}
 \phi^{ij}\nabla_{\alpha\ald}
\phi_{ij}+\tfrac{1}{8}[ \phi^{ij}, \phi^{kl}][ \phi_{ij},
\phi_{kl}]+\right.
\\&\left.+\eps^{\alpha\beta}\eps^{\bed\dot{\gamma}}
(\chi^i_\alpha (\nabla_{\beta\bed}
\bar{\chi}_{i\dot{\gamma}})-(\nabla_{\beta\bed}\chi^i_\alpha)
\bar{\chi}_{i\dot{\gamma}})-\eps^{\alpha\beta}\chi_\alpha^k[\chi_\beta^l,
\phi_{kl}]
-\eps^{\ald\bed}\bar{\chi}_{i\ald}[\bar{\chi}_{j\bed},\phi^{ij}]
\right\}~.\nonumber
\end{align}
In this notation, we can order the field content according to
helicity. The fields $( f_{\alpha\beta},\chi^i_\alpha,$\index{helicity}
$\phi_{ij}, \bar{\chi}_{i\ald},f_{\ald\bed})$ are of helicities
$(+1,+\frac{1}{2},0,-\frac{1}{2},-1)$, respectively.

\paragraph{Equations of motion.} The equations of motion of
$\CN=4$ SYM theory are easily obtained by varying
\eqref{spinoraction} with respect to the different fields. For the
spinors, we obtain the equations
\begin{subequations}\label{EOMN4}
\begin{align}
\eps^{\alpha\beta}\nabla_{\alpha\ald}\chi^i_\beta+[\phi^{ij},\bar{\chi}_{j\ald}] & \ =\  0~,\\
\eps^{\ald\bed}\nabla_{\alpha\ald}\bar{\chi}_{i\bed}+[
\phi_{ij},\chi^j_\alpha]& \ =\  0~,
\end{align}
and the bosonic fields are governed by the equations
\begin{align}
\eps^{\ald\bed}\nabla_{\gamma\ald} f_{\bed\gad}+\eps^{\alpha\beta}
\nabla_{\alpha\gad} f_{\beta\gamma}&\ =\ {\tfrac{1}{2}}
[\nabla_{\gamma\gad} \phi_{ij}, \phi^{ij}] +
\{\chi^i_\gamma,\bar{\chi}_{i\gad}\}~,\\
\eps^{\alpha\beta}\eps^{\ald\bed}\nabla_{\alpha\ald}
\nabla_{\beta\bed} \phi_{ij} -{\tfrac{1}{2}}[ \phi^{kl},[
\phi_{kl}, \phi_{ij}]]
&\ =\ {\tfrac{1}{2}}\eps_{ijkl}\eps^{\alpha\beta}
\{\chi^k_\alpha,\chi^l_\beta\}+\eps^{\ald\bed}\{\bar{\chi}_{i\ald},
\bar{\chi}_{j\bed}\}~.
\end{align}
\end{subequations}

\paragraph{Remark on superspace formulation.} There is a\index{super!space}
formulation of $\CN=4$ SYM theory in the $\CN=1$ superfield
formalism. The field content is reconstructed from three chiral
superfields plus a vector superfield and the corresponding action\index{chiral!superfields}\index{vector superfield}
reads \cite{Gates:1983nr,Kovacs:1999fx}
\begin{equation}
\begin{aligned}
S=\tr\int\dd^4 x&\left(\int \dd^4\theta
~\de^{-V}\Phi_I^\dagger\de^V\Phi^I+\tfrac{1}{4}\left(\int \dd^2
\theta~ \tfrac{1}{4} W^\alpha W_\alpha + c.c.\right)+\right.\\
&+\left.~\di \tfrac{\sqrt{2}}{3!}\left(\int \dd^2\theta
~\eps_{IJK}\Phi^I[\Phi^J,\Phi^K]+\int
\dd^2\btheta~\eps^{IJK}\Phi_I^\dagger[\Phi_J^\dagger,\Phi^\dagger_K]\right)\right)~,
\end{aligned}
\end{equation}
where $I,J,K$ run from $1$ to $3$. A few remarks are in order
here: Only a $\sSU(3)\times \sU(1)$ subgroup of the R-symmetry
group $\sSU(4)$ is manifest. This is the R-symmetry group of
$\CN=3$ SYM theory, and this theory is essentially equivalent to
$\CN=4$ SYM theory as mentioned above. Furthermore, one should
stress that this is an $\CN=1$ formalism only, and far from a
manifestly off-shell supersymmetric formulation of the theory. In
fact, such a formulation would require an infinite number of
auxiliary fields. For further details, see \cite{Kovacs:1999fx}.

\paragraph{Constraint equations for $\CN=4$ SYM theory.} Similarly\index{constraint equations}
to the ten-dimensional SYM theory, one can derive the equations of
motion of $\CN=4$ SYM theory in four dimensions from a set of
constraint equations on $\FR^{4|16}$. They read as\index{constraint equations}
\begin{equation}\label{constraintN4SYM}
\begin{aligned}
\{\nabla_{\alpha i},\nabla_{\beta
j}\}\ =\ -2\eps_{\alpha\beta}\phi_{ij}~,~~~
\{\bar{\nabla}^i_{\ald},\bar{\nabla}^j_{\bed}\}\ =\ -2\eps_{\ald\bed}\phi^{ij}~,\\
\{\nabla_{\alpha i},\bar{\nabla}^j_{\bed},\}\ =\ -2\delta^j_i
\nabla_{\alpha\bed}~,\hspace{2.8cm}
\end{aligned}
\end{equation}
where we introduced the covariant derivatives\index{covariant derivative}
\begin{equation}
\nabla_{\alpha i}\ =\ D_{\alpha i}+\lsc\omega_{\alpha
i},\cdot\rsc~,~~~
\nabla^i_{\ald}\ =\ \bar{D}^i_\ald-\lsc\bar{\omega}^i_{\ald},\cdot\rsc~,~~~
\nabla_{\alpha\ald}\ =\ \dpar_{\alpha\ald}+\lsc
A_{\alpha\ald},\cdot\rsc~.
\end{equation}
We can now define the superfields whose components will be formed
by the field content of $\CN=4$ SYM theory. As such, we have the
bosonic curvature\index{curvature}
\begin{equation}
[\nabla_{\alpha\ald},\nabla_{\beta\bed}]\ =:\ F_{\alpha\ald\beta\bed}\ =\ 
\eps_{\ald\bed}{f}_{\alpha\beta}+\eps_{\alpha\beta}{f}_{\ald\bed}
\end{equation}
and the two superspinor fields\index{Spinor}\index{super!spin}
\begin{equation}
[\nabla_{\alpha i},\nabla_{\beta\bed}]\ =:\ 
\eps_{\alpha\beta}\bchi_{i\bed}\eand
[\bar{\nabla}^i_\ald,\nabla_{\beta\bed}]\ =:\ 
\eps_{\ald\bed}\chi^i_\beta~.
\end{equation}
Using the graded Bianchi identities (cf.\ \eqref{superJacobi}) for\index{graded Bianchi identities}
all possible combinations of covariant derivatives introduced\index{covariant derivative}
above, we obtain the equations of motion of $\CN=4$ SYM theory
\eqref{EOMN4} with all the fields being superfields.

\paragraph{Superfield expansions.}\label{N4superfieldexp} It can be shown that the
equations \eqref{EOMN4} with all fields being superfields are
satisfied if and only if they are satisfied to zeroth order in the
superfield expansion. To prove this, we need to calculate
explicitly the shape of this expansion, which can always be done
following a standard procedure: We first impose a transverse gauge
condition
\begin{equation}
\theta^{\alpha i}\omega_{\alpha
i}-\bar{\theta}^\ald_i\bar{\omega}^i_\ald\ =\ 0~,
\end{equation}
which allows us to introduce the fermionic Euler operator\index{Euler operator}
\begin{equation}
\CD\ =\ \theta\nabla+\bar{\theta}\bar{\nabla}\ =\ \theta
D+\bar{\theta}\bar{D}~.
\end{equation}
Together with the constraint equations \eqref{constraintN4SYM}, we\index{constraint equations}
then easily obtain
\begin{subequations}
\begin{align}
(1+\CD)\omega_{\alpha i}&\ =\ 2\btheta^\ald_i
A_{\alpha\ald}-2\eps_{\alpha\beta} \theta^{\beta j}\phi_{ij}~,\\
(1+\CD)\bar{\omega}^i_\ald&\ =\  2\theta^{\alpha
i}A_{\alpha\ald}-\eps_{\ald\bed}
\eps^{ijkl}\btheta^\bed_j\phi_{kl}~,
\end{align}
and with the graded Bianchi-identities we can calculate
\begin{align}
\CD A_{\alpha\ald}
&\ =\ -\eps_{\alpha\beta}\theta^{i\beta}\bar{\chi}_{i\ald}+
\eps_{\ald\bed}\bar{\theta}^\bed_i\chi^i_\alpha~,\\
\CD \phi_{ij} &\ =\  \eps_{ijkl}\theta^{k\alpha}\chi^l_\alpha
-\bar{\theta}^\ald_i\bar{\chi}_{j\ald}+\bar{\theta}^\ald_j\bar{\chi}_{i\ald}~,\\
\CD\chi^i_\alpha &\ =\  -2\theta^{i\beta}
f_{\alpha\beta}+{\tfrac{1}{2}}
\eps_{\alpha\beta}\eps^{iklm}\theta^{\beta j}[ \phi_{lm},
\phi_{jk}]- \eps^{ijkl}\bar{\theta}^\ald_j\nabla_{\alpha\ald}
\phi_{kl}~,\\
\CD\bar{\chi}_{i\ald}& \ =\  2\theta^{j\alpha}\nabla_{\alpha\ald}
\phi_{ij}+ 2\bar{\theta}^\bed_i
f_{\ald\bed}+{\tfrac{1}{2}}\eps_{\ald\bed}
\eps^{jklm}\bar{\theta}^\bed_j[ \phi_{lm}, \phi_{ik}]~.
\end{align}
\end{subequations}
From the above equations, one can recursively reconstruct the
exact field expansion of the superfields whose zeroth order
components form the $\CN=4$ supermultiplet. In the following, we\index{supermultiplet}
will only need a detailed expansion in $\theta$ which, up to
quadratic order in the $\theta$s, is given by
\begin{subequations}
\begin{align}
A_{\alpha\ald}\ =\ &
\zero{A}_{\alpha\ald}+\eps_{\alpha\beta}\zero{\bar{\chi}}_{i\ald}\theta^{i\beta}
-\eps_{\alpha\beta}\zero{\nabla}_{\alpha\ald}\zero{\phi}_{ij}\theta^{i\beta}\theta^{j\gamma}
\ +\ \cdots~, \\
\phi_{ij}\ =\ &
\zero{\phi}_{ij}-\eps_{ijkl}\zero{\chi}^l_\alpha\theta^{k\alpha}-
\eps_{ijkl}(\delta^l_m\zero{f}_{\beta\alpha}+{\tfrac{1}{4}}
\eps_{\beta\alpha}\eps^{lnpq}[\zero{\phi}_{pq},\zero{\phi}_{mn}])\theta^{k\alpha}
\theta^{m\beta}\ +\ \cdots~,\\\nonumber \chi^i_\alpha\ =\ &
\zero{\chi}^i_\alpha-(2\delta^i_j\zero{f}_{\beta\alpha}+{\tfrac{1}{2}}
\eps_{\beta\alpha}\eps^{iklm}[\zero{\phi}_{lm},\zero{\phi}_{jk}])\theta^{\beta
j}\ +&\\\nonumber &\left\{{\tfrac{1}{2}}
\delta^i_j\eps^{\ald\bed}(\eps_{\gamma\alpha}
\zero{\nabla}_{\beta\ald}\zero{\bar{\chi}}_{k\bed}+\eps_{\gamma\beta}\zero{\nabla}_{\alpha\ald}\zero{\bar{\chi}}_{k\bed})\
-\right.\\
&\left.~{\tfrac{1}{4}}\eps_{\alpha\beta}\eps^{ipmn}(\eps_{jkpq}
[\zero{\phi}_{mn},\zero{\chi}^q_\gamma]+\eps_{mnkq}[\zero{\phi}_{jp},\zero{\chi}^q_\gamma])
\right\}\theta^{\beta j}\theta^{k\gamma}\ +\
\cdots~,\\
\bar{\chi}_{i\ald}\ =\ &
\zero{\bar{\chi}}_{i\ald}+2\zero{\nabla}_{\alpha\ald}\zero{\phi}_{ij}\theta^{j\alpha}+(\eps_{ijkl}
\zero{\nabla}_{\alpha\ald}\zero{\chi}^l_\beta+\eps_{\alpha\beta}[\zero{\phi}_{ij},\zero{\bar{\chi}}_{k\ald}])
\theta^{j\alpha}\theta^{k\beta}\ +\ \cdots~.
\end{align}
\end{subequations}
Therefore, the equations \eqref{EOMN4} with all the fields being
superfields are indeed equivalent to the $\CN=4$ SYM equations.

\subsection{Supersymmetric self-dual Yang-Mills theories}\label{sssSDYM}

In the following, we will always restrict ourselves to four
dimensional spacetimes with Euclidean $(\eps=-1)$ or Kleinian
$(\eps=+1)$ signature. Furthermore, we will label the Gra{\ss}mann
variables on $\FR^{4|2\CN}_R$ by $\eta^\ald_i$, and since the Weyl\index{Gra{\ss}mann variable}
spinors $\chi$ and $\bar{\chi}$ are no longer related via complex\index{Spinor}
conjugation we redenote $\bar{\chi}$ by $\tilde{\chi}$.

\paragraph{Self-dual Yang-Mills theory.} Self-dual\index{Yang-Mills theory}\index{self-dual Yang-Mills theory}
Yang-Mills (SDYM) fields on $\FR^{4,0}$ and $\FR^{2,2}$ are
solutions to the self-duality equations
\begin{equation}\label{SDcond}
F_{\mu\nu}\ =\ \tfrac{1}{2}\eps_{\mu\nu\rho\sigma}F^{\rho\sigma}~~~
\mbox{or}~~~F\ =\ \ast F
\end{equation}
which are equivalently written in spinor notation as\index{Spinor}
\begin{equation}
f_{\ald\bed}\ :=\ -\frac{1}{2}\eps^{\alpha\beta}(\dpar_{\alpha\ald}A_{\beta\bed}-
\dpar_{\beta\bed}A_{\alpha\ald}
+[A_{\alpha\ald},A_{\beta\bed}])\ =\ 0~.
\end{equation}
Solutions to these equations form a subset of the solution space
of Yang-Mills theory. If such a solution is of finite energy, it\index{Yang-Mills theory}
is called an {\em instanton}. Recall that an arbitrary Yang-Mills\index{instanton}
field strength decomposes into a self-dual $f_{\alpha\beta}$ and\index{field strength}
an anti-self-dual part $f_{\ald\bed}$:\index{anti-self-dual}
\begin{equation}
F_{\alpha\ald\beta\bed}\ =\ \eps_{\ald\bed}f_{\alpha\beta}+\eps_{\alpha\beta}f_{\ald\bed}~,
\end{equation}
where the former is of helicity $+1$ and has equal electric and\index{helicity}
magnetic components and the latter is of helicity $-1$ and has
magnetic and electric components of opposite signs. Furthermore,
it is known that the only local symmetries of the self-dual
Yang-Mills equations (for a semisimple gauge group $G$) are the
conformal group\footnote{The conformal group on $\FR^{p,q}$ is
given by $\sSO(p+1,q+1)$.} and the gauge symmetry
\cite{Popov:1998pc}.

\paragraph{Supersymmetric extensions of the SDYM equations.} As the
self-dual Yang-Mills equations form a subsector of the full
Yang-Mills theory, a possible supersymmetric extension of the\index{Yang-Mills theory}
self-duality equations can be obtained by taking the full set of
SYM field equations and imposing certain constraints on them.
These constraints have to include \eqref{SDcond} and keep the
resulting set of equations invariant under supersymmetry\index{super!symmetry}
transformations. This works for SYM theories with $\CN\leq 3$, and
the field content of the full $\CN$-extended SYM theory splits
into a self-dual supermultiplet and an anti-self-dual\index{anti-self-dual}\index{supermultiplet}
supermultiplet:
\begin{equation}\label{tablesupermultiplets}
\begin{array}{c|ccccc}
 & h\ =\ 1 & h\ =\ \tfrac{1}{2} & h\ =\ 0 & h\ =\ -\tfrac{1}{2} & h\ =\ -1\\\hline
\CN\ =\ 0 & f_{\alpha\beta} & & & & f_{\ald\bed}\\
\CN\ =\ 1 & f_{\alpha\beta} & \lambda_\alpha & & \lambda_\ald &
f_{\ald\bed}\\
\CN\ =\ 2 & f_{\alpha\beta} & \lambda^i_\alpha & \phi^{[12]}~~
\phi_{[12]}&
\lambda_{\ald i} & f_{\ald\bed}\\
\CN\ =\ 3 & f_{\alpha\beta} & \lambda_\alpha ~~ \chi^i_\alpha &
\phi^{[ij]}~~ \phi_{[ij]}&
\chi_{\ald i} ~~ \lambda_\ald& f_{\ald\bed}\\
\CN\ =\ 4 & f_{\alpha\beta} & \chi^i_\alpha &
\phi^{[ij]}\ =\ \tfrac{1}{2}\eps^{ijkl} \phi_{[kl]}& \chi_{\ald i}&
f_{\ald\bed}
\end{array}
\end{equation}
where each column consists of fields with a certain helicity and\index{helicity}
each row contains a supermultiplet for a certain value of $\CN$.\index{supermultiplet}
The indices $i,j,\ldots$ always run from $1$ to $\CN$. From the
table \eqref{tablesupermultiplets}, we see that for $\CN{=}4$, the
situation is more complicated, as the SYM multiplet
$(f_{\alpha\beta},\chi^{\alpha i},\phi^{ij},\tilde{\chi}_{\ald
i},f_{\ald\bed})$, where the fields have the helicities
$(+1,+\frac{1}{2},0,-\frac{1}{2},-1)$, is irreducible. By
introducing an additional field $G_{\ald\bed}$ with helicity -1,\index{helicity}
which takes in some sense the place of $f_{\ald\bed}$, one can
circumvent this problem (see e.g.\
\cite{Siegel:1992za,Devchand:1996gv}). The set of physical fields
for $\CN{=}4$ SYM theory consists of the self-dual and the
anti-self-dual field strengths of a gauge potential\index{anti-self-dual}\index{field strength}
$\CA_{\alpha\ald}$, four spinors $\chi^i_\alpha$ together with\index{Spinor}
four spinors $\tilde{\chi}_{\ald
i}\sim\eps_{ijkl}\tilde{\chi}_\ald^{jkl}$ of opposite chirality
and six real (or three complex) scalars $\phi^{ij}=\phi^{[ij]}$.
For $\CN{=}4$ super SDYM theory, the multiplet is joined by an
additional spin-one field
$G_{\ald\bed}\sim\eps_{ijkl}G^{ijkl}_{\ald\bed}$ with helicity\index{helicity}
$-1$ and the multiplet is -- after neglecting the vanishing
anti-self-dual field strength $f_{\ald\bed}$ -- identified with\index{anti-self-dual}\index{field strength}
the one of $\CN=4$ SYM theory.

\paragraph{Equations of motion.} Using the above mentioned
auxiliary field $G_{\ald\bed}$, we arrive at the following
equations of motion:
\begin{equation}\label{SDYMeom}
\begin{aligned}
f_{\ald\bed}&\ =\ 0~,\\
\nabla_{\alpha\ald}\chi^{\alpha i}&\ =\ 0~,\\
\Box\phi^{ij}&\ =\ -\tfrac{\eps}{2}\{\chi^{\alpha i},\chi^j_{\alpha}\}~,\\
\nabla_{\alpha\ald}\tilde{\chi}^{\ald ijk}&\ =\ +2
\eps\,[\phi^{[ij},\chi^{k]}_{\alpha}]~,\\
\eps^{\ald\dot{\gamma}}\nabla_{\alpha\ald}G^{ijkl}_{\dot{\gamma}
\dot{\delta}}&\ =\ +\eps\{\chi^{[i}_{\alpha},\tilde{\chi}^{jkl]}_{\dot{\delta}}\}
-\eps\,[\phi^{[ij},\nabla_{\alpha\dot{\delta}}\phi^{kl]}]~,
\end{aligned}
\end{equation}
where we introduced the shorthand notations
$\square:=\frac{1}{2}\nabla_{\alpha\ald}\nabla^{\alpha\ald}$ and
$\eps=\pm 1$ distinguishes between Kleinian and Euclidean
signature on the spacetime under consideration.

\paragraph{Action for $\CN=4$ SDYM theory.} The action reproducing
the above equations of motion was first given in
\cite{Siegel:1992za} and reads with our scaling of fields as
\begin{equation}\label{LagrangianSDYM}
S\ =\ \int
\dd^4x~\tr\left(G^{\ald\bed}f_{\ald\bed}+\tfrac{\eps}{2}\eps_{ijkl}\tilde{\chi}^{\ald
ijk} \nabla_{\alpha\ald}\chi^{\alpha
l}+\tfrac{\eps}{2}\eps_{ijkl}\phi^{ij}\square\phi^{kl}+\eps_{ijkl}\phi^{ij}\chi^{\alpha
k}\chi_\alpha^l\right)~,
\end{equation}
where $G_{\ald\bed}:=\frac{1}{4!}\eps_{ijkl}G_{\ald\bed}^{ijkl}$.
Note that although the field content appearing in this action is
given by the multiplet $(f_{\alpha\beta},\chi^{\alpha
i},\phi^{ij},\tilde{\chi}_{\ald i},f_{\ald\bed},G_{\ald\bed})$,
$f_{\ald\bed}$ vanishes due to the SDYM equations of motion and
the supermultiplet of non-trivial fields is\index{supermultiplet}
$(f_{\alpha\beta},\chi^{\alpha i},\phi^{ij},\tilde{\chi}_{\ald
i},G_{\ald\bed})$. These degrees of freedom match exactly those of
the full $\CN{=}4$ SYM theory and often it is stated that they are
the same. Following this line, one can even consider the full
$\CN{=}4$ SYM theory and $\CN{=}4$ SDYM theory as the same
theories on linearized level, which are only distinguished by
different interactions.

\paragraph{Constraint equations.}\label{pconstraintSDYM} Similarly to the case of the\index{constraint equations}
full $\CN=4$ SYM theory, one can obtain the equations of motion
\eqref{SDYMeom} also from a set of constraint equations. These\index{constraint equations}
constraint equations live on the chiral superspace\index{chiral!superspace}\index{super!space}
$\FR^{4|2\CN}_R$ with coordinates $(x^{\alpha\ald},\eta^\ald_i)$
and read explicitly as
\begin{equation}\label{constraintN4SDYM}
\begin{aligned}
{}[\nabla_{\alpha\ald},\nabla_{\beta\bed}]+[\nabla_{\alpha\bed},&
\nabla_{\beta\ald}]\ =\ 0~,~~~
[\nabla_{\ald}^i,\nabla_{\beta\bed}]+[\nabla_{\bed}^i,\nabla_{\beta\ald}]\ =\ 0~,\\
&\{\nabla_\ald^i,\nabla_\bed^j\}+\{\nabla_\bed^i,\nabla_\ald^j\}\ =\ 0~,
\end{aligned}
\end{equation}
where we have introduced covariant derivatives\index{covariant derivative}
\begin{equation}
\nabla_{\alpha\ald}\ :=\ \der{x^{\alpha\ald}}
+\CA_{\alpha\ald}~~~\mbox{and}~~~
\nabla_\ald^i\ :=\ \der{\eta^\ald_i}+\CA_\ald^i~.
\end{equation}
Note that the gauge potentials $\CA_{\alpha\ald}$ and $\CA^i_\ald$
are functions on the chiral superspace $\FR^{4|2\CN}_R$. Equations\index{chiral!superspace}\index{super!space}
\eqref{constraintN4SDYM} suggest the introduction of the following
self-dual super gauge field strengths:\index{field strength}
\begin{equation}
\begin{aligned}
{}[\nabla_{\alpha\ald},\nabla_{\beta\bed}]\ =\ \eps_{\ald\bed}
f_{\alpha\beta}(x,\eta)~,~~~
[\nabla^i_{\ald},\nabla_{\beta\bed}]\ =\ \eps_{\ald\bed}
f^i_{\beta}(x,\eta)~,\\
\{\nabla^i_{\ald},\nabla^j_{\bed}\}\ =\ \eps_{\ald\bed}
f^{ij}(x,\eta)~,\hspace{2.8cm}
\end{aligned}
\end{equation}
and by demanding that $f^{ij}$ is antisymmetric and
$f_{\alpha\beta}$ is symmetric, these equations are equivalent to
\eqref{constraintN4SDYM}. The lowest components of
$f_{\alpha\beta}$, $f_\alpha^i$ and $f^{ij}$ will be the SDYM
field strength, the spinor field $\chi_\alpha^i$ and the scalars\index{Spinor}\index{field strength}
$\phi^{ij}$, respectively. By using Bianchi identities for the
self-dual super gauge field strengths, one can show that these\index{field strength}
definitions yield superfield equations which agree in zeroth order
with the component equations of motion \eqref{SDYMeom}
\cite{Devchand:1996gv}.

To show the actual equivalence of the superfield equations with
the equations \eqref{SDYMeom}, one proceeds quite similarly to the
full SYM case, cf.\ \ref{N4superfieldexp}. We impose the
transverse gauge condition $\eta^\ald_i\CA_\ald^i=0$ and introduce
an Euler operator\index{Euler operator}
\begin{equation}
\CD\ :=\ \eta^\ald_i\nabla^i_\ald\ =\ \eta^\ald_i\dpar^i_\ald~,
\end{equation}
which yields the following relations:
\begin{equation}
\begin{aligned}
 \CD f_{\alpha\beta}&\ =\ \tfrac{1}{2}\ \eta^{\ald}_i \nabla_{(\alpha\ald} \chi^i_{\beta)}\\
 \CD \chi^j_{\alpha}&\ =\  2\ \eta^{\ald}_i\nabla_{\alpha\ald} \phi^{ij}\\
 \CD \phi^{jk} &\ =\  \eta^{\ald}_i \tilde{\chi}^{ijk}_{\ald}\\
 \CD \tilde{\chi}^{jkl}_\bed
 &\ =\  \eta^{\ald}_i ( G^{ijkl}_{\ald\bed}  + \eps_{\ald\bed} [ \phi^{i[j}, \phi^{kl]}])\\
 \CD G^{jklm}_{\bed\gad} &\ =\  -\eta_{i(\bed}\left( \tfrac{2}{3} [ \phi^{i[j} , \tilde{\chi}^{klm]}_{\gad)} ]
  - \tfrac{1}{3} [ \phi^{[jk} , \tilde{\chi}^{lm]i}_{\gad)} ]  \right)
\end{aligned}
\end{equation}
as well as
\begin{equation}
\begin{aligned}
(1+\CD)\CA^i_\ald&\ =\ 2\eps_{\ald\bed}\eta^\bed_j \phi^{ij}~,\\
\CD\CA_{\alpha\ald}&\ =\ -\eps_{\ald\bed}\eta^\bed_i\chi^i_\alpha~,
\end{aligned}
\end{equation}
from which the field expansion can be reconstructed explicitly.
For our purposes, it will always be sufficient to know that
\begin{equation}\label{fieldexpansiondevchand}
\begin{aligned}
\CA_{\alpha\ald}&\ =\ A_{\alpha\ald}-
\eps_{\ald\bed}\eta^\bed_i\chi^i_\alpha+\ldots -\tfrac{1}{12}
\eps_{\ald\bed}\eta^\bed_i\eta^\gad_j\eta^\ded_k\eta^{\dot{\eps}}_l\nabla_{\alpha\gad}
G_{\ded\dot{\eps}}^{ijkl}~,\\\CA^i_\ald&\ =\
\eps_{\ald\bed}\eta^\bed_j\phi^{ik}+\tfrac{2}{3}\eps_{\ald\bed}\eta^\bed_j\eta^\gad_k\tilde{\chi}^{ijk}_\gad
+\tfrac{1}{4}\eps_{\ald\bed}\eta^\bed_j\eta^\gad_k\eta^\ded_l
\left(G^{ijkl}_{\gad\ded}+\eps_{\gad\ded}\ldots \right)~,
\end{aligned}
\end{equation}
as this already determines the field content completely.

\paragraph{From SYM theory to super SDYM theory.} Up to $\CN=3$, solutions to the
supersymmetric SDYM equations form a subset of the corresponding
full SYM equations. By demanding an additional condition, one can
restrict the constraint equations of the latter to the ones of the\index{constraint equations}
former \cite{Semikhatov:1982ig,Volovich:1983ii}. For $\CN=1$, the
condition to impose is $[\nabla_{i\alpha},\nabla_{\beta\bed}]=0$,
while for $\CN=2$ and $\CN=3$, one has to demand that
$\{\nabla_{i\alpha},\nabla_{j\beta}\}=\eps_{\alpha\beta}\phi_{ij}=0$.
For $\CN=4$, one can use the same condition as for $\CN=3$, but
one has to drop the usual reality condition
$\phi_{ij}=\frac{1}{2}\eps_{ijkl}\bar{\phi}^{kl}$, which renders
the fourth supersymmetry nonlinear. Alternatively, one can follow\index{super!symmetry}
the discussion in \cite{Witten:2003nn}, where $\CN=4$ SYM and
$\CN=4$ SDYM theories are considered as different weak coupling\index{weak coupling}
limits of an underlying field theory including an auxiliary field.

\subsection{Instantons}\index{instanton}

\paragraph{Meaning of instantons.} The dominant contribution to the\index{instanton}
partition function
\begin{equation}
\CZ\ :=\ \int \CCD \varphi \de^{-S_E[\varphi,\dpar_\mu \varphi]}
\end{equation}
of a quantum field theory defined by a (Euclidean) action $S_E$
stems from the minima of the action functional
$S_E[\varphi,\dpar_\mu \varphi]$. In non-Abelian gauge theories,
one calls the local minima, which exist besides the global one,
{\em instantons}. Instantons therefore cannot be studied\index{instanton}
perturbatively, but they are non-perturbative effects.

Although they did not give rise to an explanation of quark
confinement, instantons found various other applications in QCD\index{confinement}\index{instanton}
and supersymmetric gauge theories. In mathematics, they are
related to certain topological invariants on four-manifolds.

\paragraph{Instantons in Yang-Mills theory.} Consider now such a\index{Yang-Mills theory}\index{instanton}
non-Abelian gauge theory on Euclidean spacetime $\FR^4$, which
describes the dynamics of a gauge potential $A$, a Lie algebra\index{Lie algebra}
valued connection one-form on a bundle $E$, and its field strength\index{connection}\index{field strength}
$F=\dd A+A\wedge A$. The corresponding Yang-Mills energy is given
by the functional $-\frac{1}{2}\int_{\FR^4} \tr(F\wedge \ast F)$.

We will restrict our considerations to those gauge configurations
with finite energy, i.e.\ the gauge potential has to approach pure
gauge at infinity. This essentially amounts to considering the\index{gauge!pure gauge}
theory on $S^4$ instead of $\FR^4$. We can then define the
topological invariant $-\frac{1}{2(2\pi)^2}\int_{\FR^4}
\tr(F\wedge F)$, which is the {\em instanton number} and counts\index{instanton}\index{instanton number}
instantons contained in the considered configuration. Note that
this invariant corresponds to a nontrivial second Chern character\index{Chern character}
of the curvature $F$. Recall that one can write this second Chern\index{curvature}
character in terms of first and second Chern classes, see section\index{Chern class}
\ref{ssvectorbundles}, \ref{pCherncharacter}.

We can decompose the energy functional into
\begin{equation}
0\ \leq\  \tfrac{1}{2}\int_{\FR^4} \tr((\ast F+\de^{-\di \theta}
F)\wedge(F+\de^{i\theta}\ast F))\ =\ \int_{\FR^4}\tr(\ast F\wedge
F+2\cos \theta F\wedge F)
\end{equation}
for all real $\theta$. Therefore we have
\begin{equation}
\tfrac{1}{2}\int_{\FR^4}\tr(\ast F\wedge F)\ \leq\ 
\tfrac{1}{2}\left|\int_{\FR^4}\tr (F\wedge F)\right|
\end{equation}
The configurations satisfying this bound are called BPS, cf.\
\ref{pBPSmonopoles}, and they form minima of the energy
functional. For such configurations, either the self-dual or the
anti-self-dual Yang-Mills equation holds:\index{anti-self-dual}
\begin{equation}
F=\pm\ast F~.
\end{equation}
The name instantons stems from the fact that these configurations\index{instanton}
are localized at spacetime points.

In our conventions, an instanton is a self-dual gauge field\index{instanton}
configuration with positive topological charge $k$.
Anti-instantons have negative such charge and satisfy the\index{anti-instanton}\index{instanton}
anti-self-duality equations.\index{anti-self-dual}

\paragraph{Abelian instantons.} From the above definition of the\index{instanton}
instanton number, it is clear that in the Abelian case, where\index{instanton number}
$F=\dd A$, no instanton solutions can exist:
\begin{equation}
\begin{aligned}
-\frac{1}{2(2\pi)^2}\int_{\FR^4} \tr (F\wedge
F)&\ =\ -\frac{1}{2(2\pi)^2}\int_{\FR^4} \tr \dd(A\wedge
F)\\&\ =\ -\frac{1}{2(2\pi)^2}\int_{S^3} \tr (A\wedge F)\ =\ 0~,
\end{aligned}
\end{equation}
where $S^3$ is the sphere at spatial infinity, on which the
curvature $F$ vanishes. Note, however, that the situation is\index{curvature}
different on noncommutative spacetime, where Abelian instantons do\index{instanton}\index{noncommutative spacetime}
exist.

\paragraph{Moduli space of instantons.} On a generic four-dimensional\index{instanton}\index{moduli space}
Riemann manifold $M$, the moduli space of instantons is the space\index{manifold}
of self-dual gauge configurations modulo gauge transformations. It\index{gauge transformations}\index{gauge!transformation}
is noncompact and for $k$ $\sU(N)$ instantons of dimension\index{instanton}
\begin{equation}
4 N k-\frac{N^2-1}{2}(\chi+\sigma)~,
\end{equation}
where $\chi$ and $\sigma$ are the Euler characteristics and the
signature of $M$, respectively.

\paragraph{Construction of instantons.} There is a number of\index{instanton}
methods for constructing instantons, which are almost all inspired
by twistor geometry. We will discuss them in detail in section\index{twistor}
\ref{sSolutionGT}. There, one finds in particular a discussion of
the well-known {\em ADHM construction} of instantons.\index{ADHM construction}\index{instanton}

\paragraph{Supersymmetric instantons.} Note furthermore that in\index{instanton}
$\CN=1$ supersymmetric gauge theories, instanton configurations
break half of the supersymmetries, as they appear on the
right-hand side of the supersymmetry transformations,\index{super!symmetry} cf.\
\eqref{SUSYN4}. From \eqref{SUSYN4} we also see that this holds
for supersymmetric instanton configurations up to $\CN=3$.\index{instanton}

\subsection{Related field theories}\label{ssRelatedTheories}

In this section, we want to briefly discuss two related field
theories which will become important in the later discussion:
$\CN=8$ SYM theory in three dimensions and the super Bogomolny\index{N=8 SYM theory@$\CN=8$ SYM theory}
model. These theories are obtained by reduction of $\CN=4$ SYM
theory and $\CN$-extended SDYM theory from four to three
dimensions.

\paragraph{Dimensional reduction $\FR^4\rightarrow\index{dimensional reduction}
\FR^3$.}\label{pdimreduction}Recall that the rotation group
$\sSO(4)$ of $(\FR^4,\delta_{\mu\nu})$ is locally isomorphic to
$\sSU(2)_L\times \sSU(2)_R\cong\sSpin(4)$. The rotation group
$\sSO(3)$ of $(\FR^3,\delta_{ab})$ with $a,b=1,2,3$ is locally
$\sSU(2)\cong \sSpin(3)$, which can be interpreted as the diagonal
group $\diag(\sSU(2)_L\times \sSU(2)_R)$ upon dimensional
reduction to three dimensions. Therefore, the distinction between\index{dimensional reduction}
undotted, i.e.\ $\sSU(2)_L$, and dotted, i.e.\ $\sSU(2)_R$,
indices disappears.

Explicitly, the dimensional reduction $\FR^4\rightarrow \FR^3$ is\index{dimensional reduction}
now performed by introducing the new coordinates\footnote{The fact
that we dimensionally reduce by the coordinate $x^2$ is related to
our sigma matrix convention.}
\begin{equation}\label{dimredcoordinates}
\begin{aligned}
y^{\ald\bed}\ :=\  -\di x^{(\ald\bed)}\eand x^{[\ald\bed]}\ =\
-\eps^{\ald\bed}x^2\hspace{2cm}\\~~\mbox{with}~~y^{\ed\ed}\ =\
-\bar{y}^{\zd\zd}\ =\ (-\di x^4-x^3)\ =:\ y\eand y^{\ed\zd}\ =\
\bar{y}^{\ed\zd}\ =\ - x^1~
\end{aligned}
\end{equation}
together with the derivatives
\begin{equation}\label{dimredderivatives}
\dpar_{\ald\ald}\ :=\ \der{y^{\ald\ald}}\eand \dpar_{\ed\zd}\ :=\
\frac{1}{2}\der{y^{\ed\zd}}~.
\end{equation}
More abstractly, this splitting corresponds to the decomposition
$\rfour=\rthree\oplus\rone$ of the irreducible real vector
representation $\rfour$ of the group $\sSU(2)_L\times \sSU(2)_R$\index{representation}
into two irreducible real representations $\rthree$ and $\rone$ of
the group $\sSU(2)$.

The four-dimensional gauge potential $A_{\alpha\ald}$ is split
into a three-dimensional gauge potential $A_{(\ald\bed)}$ and a
Higgs field $\Phi$
\begin{equation}
B_{\ald\bed}\ =\ A_{\ald\bed}-\tfrac{\di}{2}\eps_{\ald\bed}\Phi~,
\end{equation}
which motivates the introduction of the following differential
operator and covariant derivative:\index{covariant derivative}
\begin{equation}
\nabla_{\ald\bed}\ :=\ \dpar_{\ald\bed}+B_{\ald\bed}\eand
D_{\ald\bed}\ :=\ \nabla_{(\ald\bed)}\ =\ \dpar_{\ald\bed}+A_{\ald\bed}~.
\end{equation}

\paragraph{Yang-Mills-Higgs theory.}\label{pYMHiggs}\index{Yang-Mills-Higgs theory}
Yang-Mills-Higgs theory is defined in $d$ dimensions by the action
\begin{equation}
S=\int \dd^dx~
\tr\left(-\tfrac{1}{4}F_{\mu\nu}F^{\mu\nu}+\nabla_\mu\phi\nabla^\mu\phi-\tfrac{\gamma}{4}(\phi\phi^*-1)^2\right)~,
\end{equation}
where $F$ is as usually the field strength of a gauge potential\index{field strength}
and $\phi$ is a complex scalar. The potential term can in
principle be chosen arbitrarily, but renormalizability restricts
it severely. The equations of motion of this theory read
\begin{equation}
\nabla_\mu F^{\mu\nu}\ =\ -[\nabla^\nu\phi,\phi]\eand
\nabla^\mu\nabla_\mu\phi\ =\ \gamma\phi(\phi\phi^*-1)~.
\end{equation}
In our considerations, we will only be interested in a
three-dimensional version of this theory with vanishing potential
term $\gamma=0$, which can be obtained from four-dimensional
Yang-Mills theory via a dimensional reduction.\index{Yang-Mills theory}\index{dimensional reduction}

\paragraph{$\CN=8$ SYM theory in three dimensions.} This theory is\index{N=8 SYM theory@$\CN=8$ SYM theory}
obtained by dimensionally reducing $\CN=1$ SYM theory in ten
dimensions to three dimensions, or, equivalently, by dimensionally
reducing four-dimensional $\CN=4$ SYM theory to three dimensions.
As a result, the 16 real supercharges are re-arranged in the
latter case from four spinors transforming as a $\mathbf{2}_\FC$\index{Spinor}
of $\sSpin(3,1)\cong\sSL(2,\FC)$ into eight spinors transforming
as a $\mathbf{2}$ of $\sSpin(2,1)\cong \sSL(2,\FR)$.

The automorphism group of the supersymmetry algebra is\index{morphisms!automorphism}\index{super!symmetry}
$\sSpin(8)$, and the little group of the remaining Lorentz group\index{Lorentz!group}
$\sSO(2,1)$ is trivial. As massless particle content, we therefore
expect bosons transforming in the $\mathbf{8}_v$ and fermions
transforming in the $\mathbf{8}_c$ of $\sSpin(8)$. One of the
bosons will, however, appear as a dual gauge potential on $\FR^3$
after dimensional reduction, and therefore only a $\sSpin(7)$\index{dimensional reduction}
R-symmetry group is manifest in the action and the equations of
motion. Altogether, we have a gauge potential $A_a$ with
$a=1,\ldots,3$, seven real scalars $\phi^i$ with $i=1,\ldots,7$
and eight spinors $\chi^j_\ald$ with $j=1,\ldots,8$.\index{Spinor}

Moreover, recall that in four dimensions, $\CN=3$ and $\CN=4$
super Yang-Mills theories are equivalent on the level of field
content and corresponding equations of motion. The only
difference\footnote{In the complexified case, one has an
additional condition which takes the shape
$\phi_{ij}=\frac{1}{2}\eps_{ijkl}\phi^{kl}$, cf.\
\cite{Witten:1978xx}.} is found in the manifest R-symmetry groups
which are $\sSU(3)\times \sU(1)$ and $\sSU(4)$, respectively. This
equivalence obviously carries over to the three-dimensional
situation: $\CN=6$ and $\CN=8$ super Yang-Mills theories are
equivalent regarding their field content and the equations of
motion.

\paragraph{The super Bogomolny model.}\label{psB} We start from the\index{Bogomolny model}\index{super!Bogomolny model}
$\CN$-extended supersymmetric SDYM equations on $\FR^4$, i.e. the
first $\CN$ equations of \eqref{SDYMeom} in which the R-symmetry
indices $i,j,\ldots$ are restricted to $1,\ldots,\CN$. After
performing the dimensional reduction as presented in\index{dimensional reduction}
\ref{pdimreduction}, one arrives at the field content
\begin{equation}
A_{\ald\bed},\chi^i_\ald,\Phi,\phi^{ij},\tilde{\chi}_{i\ald},
G_{\ald\bed}
\end{equation}
with helicities $(1,\frac{1}{2},0,0,-\frac{1}{2},-1)$, where we
used the shorthand notations
\begin{equation}\label{shorthand-sB}
\tilde{\chi}_{i\ald}\ :=\
\tfrac{1}{3!}\eps_{ijkl}\tilde{\chi}_\ald^{jkl}\eand G_{\ald\bed}\
:=\  \tfrac{1}{4!}\eps_{ijkl}G^{ijkl}_{\ald\bed}~.
\end{equation}
The supersymmetric extension of the Bogomolny equations now read\index{Bogomolny equations}
\begin{equation}\label{eom-sB}
\begin{aligned}
f_{\ald\bed}&\ =\ -\tfrac{\di}{2}D_{\ald\bed}\Phi~,
\\\eps^{\bed\gad}D_{\ald\bed}\chi^i_\gad&\ =\
-\tfrac{\di}{2}[\Phi,\chi^i_\ald]~,
\\\triangle\phi^{ij}&\ =\ -\tfrac{1}{4}[\Phi,[\phi^{ij},\Phi]]+
\eps^{\ald\bed}\{\chi^i_\ald,\chi^j_\bed\}~,
\\\eps^{\bed\gad}D_{\ald\bed}\tilde{\chi}_{i\gad}&\ =\
-\tfrac{\di}{2}[\tilde{\chi}_{i\ald},\Phi]+
2\di[\phi_{ij},\chi^j_\ald]~,\\
\eps^{\bed\gad}D_{\ald\bed}G_{\gad\ded}&\ =\
-\tfrac{\di}{2}[G_{\ald\ded},\Phi]+
\di\{\chi^i_\ald,\tilde{\chi}_{i\ded}\}
-\tfrac{1}{2}[\phi_{ij},D_{\ald\ded}\phi^{ij}]+
\tfrac{\di}{4}\eps_{\ald\ded}[\phi_{ij},[\Phi,\phi^{ij}]]~.
\end{aligned}
\end{equation}
Here, we have used the fact that we have a decomposition of the
field strength in three dimensions according to\index{field strength}
\begin{equation}
F_{\ald\bed\gad\ded}\ =\ [D_{\ald\bed},D_{\gad\ded}]\ =:\
\eps_{\bed\ded}f_{\ald\gad}+\eps_{\ald\gad}f_{\bed\ded}
\end{equation}
with $f_{\ald\bed}=f_{\bed\ald}$. We have also introduced the
abbreviation
$\triangle:=\frac{1}{2}\eps^{\ald\bed}\eps^{\gad\ded}D_{\ald\gad}
D_{\bed\ded}$.

For $\CN=8$, one can write down the following action functional
leading to the equations \eqref{eom-sB}:
\begin{equation}\label{action-sB}
\begin{aligned}
S_{\mathrm{sB}}\ =\ &\int\dd^3 x~\tr\Big\{\Big.G^{\ald\bed}
\left(f_{\ald\bed}+\tfrac{\di}{2}D_{\ald\bed}\Phi\right)+
\di\eps^{\ald\ded}\eps^{\bed\gad}\chi^i_\ald
D_{\ded\bed}\tilde{\chi}_{i\gad}+\\&+
\tfrac{1}{2}\phi_{ij}\triangle\phi^{ij}-\tfrac{1}{2}
\eps^{\ald\ded}\chi^i_\ald[\tilde{\chi}_{i\ded},\Phi]-
\eps^{\ald\gad}\phi_{ij}\{\chi^i_\ald,\chi^j_\gad\}+\tfrac{1}{8}
[\phi_{ij},\Phi][\phi^{ij},\Phi]\Big.\Big\}~.
\end{aligned}
\end{equation}
In this expression, we have again used the shorthand notation
$\phi_{ij}:=\frac{1}{2!}\eps_{ijkl}\phi^{kl}$.

\paragraph{Constraint equations.}\label{psBconstraints} Similarly to the SYM\index{constraint equations}
and the (super-)SDYM equations, one can give a set of constraint
equations on $\FR^{3|2\CN}$, which are equivalent to the super\index{constraint equations}
Bogomolny equations on $\FR^3$. For this, we introduce the\index{Bogomolny equations}
first-order differential operators
$\nabla_{\ald\bed}:=\dpar_{(\ald\bed)}+\CB_{\ald\bed}$ and
$D_\ald^i=\der{\eta_i^\ald}+\CA_\ald^i=:\dpar_\ald^i+\CA_\ald^i$,
where
$\CB_{\ald\bed}:=\CA_{\ald\bed}-\frac{\di}{2}\eps_{\ald\bed}\Phi$.
Then the appropriate constraint equations read as\index{constraint equations}
\begin{equation}\label{mpconstraint}
       [\nabla_{\ald\gad},\nabla_{\bed\ded}]\ =:\ \eps_{\gad\ded}
       \Sigma_{\ald\bed}~,
       ~~~
       {[D_\ald^i,\nabla_{\bed\gad}]}\ =:\ \di\eps_{\ald\gad}
       \Sigma^i_\bed
       \eand
       \{D^i_\ald,D^j_\bed\}\ =:\ \eps_{\ald\bed}\Sigma^{ij}~,
\end{equation}
where $\Sigma_{\ald\bed}=\Sigma_{\bed\ald}$ and
$\Sigma^{ij}=-\Sigma^{ji}$. Note that the first equation in
\eqref{mpconstraint} immediately shows that
$f_{\ald\bed}=-\tfrac{\di}{2}D_{\ald\bed}\Phi$ and thus the
contraction of the first equation of \eqref{mpconstraint} with
$\eps^{\gad\ded}$ gives
$\Sigma_{\ald\bed}=f_{\ald\bed}-\tfrac{\di}{2}D_{\ald\bed}\Phi=
2f_{\ald\bed}$. The graded Bianchi identities for the differential\index{graded Bianchi identities}
operators $\nabla_{\ald\bed}$ and $D^i_\ald$ yield in a
straightforward manner further field equations, which allow us to
identify the superfields $\Sigma^i_\ald$ and $\Sigma^{ij}$ with
the spinors $\chi^i_\ald$ and the scalars $\phi^{ij}$,\index{Spinor}
respectively. Moreover, $\tilde{\chi}_{i\ald}$ is given by
$\tilde{\chi}_{i\ald}:=\frac{1}{3}\eps_{ijkl}D^j_\ald\phi^{kl}$
and $G_{\ald\bed}$ is defined by
$G_{\ald\bed}:=-\frac{1}{4}D^i_{(\ald}\tilde{\chi}_{i\bed)}$.
Collecting the above information, one obtains the superfield
equations for $\CA_{\ald\bed}$, $\chi^i_\ald$, $\Phi$,
$\phi^{ij}$, $\tilde{\chi}_{i\ald}$ and $G_{\ald\bed}$ which take
the same form as \eqref{eom-sB} but with all the fields now being
superfields. Thus, the projection of the superfields onto the
zeroth order components of their $\eta$-expansions gives
\eqref{eom-sB}.

Similarly to all the previous constraint equations, one can turn\index{constraint equations}
to transverse gauge and introduce the Euler operator $\CD:=\index{Euler operator}
\eta^\ald_iD^i_\ald$ to recover the component fields in the
superfield expansion of the superconnection $\CA$. The explicit\index{connection}
result is obtained straightforwardly to be
\begin{subequations}\label{superexp4}
\begin{align}\nonumber
\CB_{\ald\bed}\ =\ &
\z{\CB}_{\ald\bed}-\di\eps_{\bed\gad_1}\eta_{j_1}^{\gad_1}
\z{\chi}^{j_1}_\ald+
\tfrac{1}{2!}\eps_{\bed\gad_1}\eta_{j_1}^{\gad_1}
\eta_{j_2}^{\gad_2}
\nabla_{\ald\gad_2}\z{\phi}^{j_1j_2}-\tfrac{1}{2\cdot3!}
\eps_{\bed\gad_1}\eta_{j_1}^{\gad_1}\eta_{j_2}^{\gad_2}
\eta_{j_3}^{\gad_3}\eps^{j_1j_2j_3k}\nabla_{\ald\gad_2}
\z{\tilde{\chi}}_{k\gad_3}~-\\&
-\tfrac{1}{4!}\eps_{\bed\gad_1}\eta_{j_1}^{\gad_1}
\eta_{j_2}^{\gad_2}\eta_{j_3}^{\gad_3}\eta_{j_4}^{\gad_4}
\eps^{j_1j_2j_3j_4}\nabla_{\ald\gad_2}\z{G}_{\gad_3\gad_4}+
\cdots\\\nonumber
 \CA^i_\ald\ =\
&\tfrac{1}{2!}\eps_{\ald\gad_1}\eta_{j_1}^{\gad_1}
\z{\phi}^{ij_1}- \tfrac{1}{3!}\eps_{\ald\gad_1}\eta_{j_1}^{\gad_1}
\eta_{j_2}^{\gad_2}\eps^{ij_1j_2k}\z{\tilde{\chi}}_{k\gad_2}
+\\&+\tfrac{3}{2\cdot4!}\eps_{\ald\gad_1}\eta_{j_1}^{\gad_1}
\eta_{j_2}^{\gad_2}\eta_{j_3}^{\gad_3}
\eps^{ij_1j_2j_3}\z{G}_{\gad_2\gad_3}+\cdots
\end{align}
\end{subequations}
The equations \eqref{mpconstraint} are satisfied for these
expansions if the supersymmetric Bogomolny equations\index{Bogomolny equations}
\eqref{eom-sB} hold for the physical fields appearing in the above
expansions and vice versa.

\paragraph{BPS monopoles.}\label{pBPSmonopoles} The Bogomolny\index{BPS monopole}
equations appear also as the defining equation for
Bogomolny-Prasad-Sommerfield (BPS) monopole configurations
\cite{Bogomolny:1975de,Prasad:1975kr}, see also
\cite{Harvey:1996ur}. We start from the Yang-Mills-Higgs
Lagrangian\footnote{For convenience, we will switch again to
vector indices $a,b,\ldots$ ranging from $1$ to $3$ in the
following.} given in \ref{pYMHiggs}, and note that its energy
functional for static configurations $(A_a,\phi)$ is given by
\begin{equation}
E=\tfrac{1}{4}\int \dd^3 x~ \tr (F_{ab}F_{ab}+2D_a\phi D_a\phi)~.
\end{equation}
To guarantee finite energy, we have to demand that
\begin{equation}
\lim_{|r|\rightarrow \infty}\tr(F_{ab}F_{ab})\ =\ 0\eand
\lim_{|r|\rightarrow \infty}\tr(D_a\phi D_a\phi)\ =\ 0
\end{equation}
sufficiently rapidly. The energy functional has a lower bound,
which can be calculated to be
\begin{equation}
\begin{aligned}\label{BPSbound}
E&\ =\ -\tfrac{1}{4}\int \dd^3 x~
\tr(F_{ab}\mp\eps_{abc}D_c\phi)(F_{ab}\mp\eps_{abd}D_d\phi)\mp\int
\dd^3 x~\tr\left(\tfrac{1}{2}\eps_{abc} F_{bc}D_c\phi\right)\\
&\ =\ -\tfrac{1}{4}\int \dd^3 x~
\tr(F_{ab}\mp\eps_{abc}D_c\phi)(F_{ab}\mp\eps_{abd}D_d\phi)\pm
4\pi Q\ \geq\  4\pi|Q|~.
\end{aligned}
\end{equation}
Here, we could choose the absolute value as one of the bounds
always becomes trivial. We found the magnetic charge $Q$ which can
also be understood as the magnetic flux through a sphere around
the origin with infinite radius:
\begin{equation}
Q\ =\ -\frac{1}{8\pi}\int \dd^3
x~\tr(\eps_{abc}F_{ab}D_c\phi)\ =\ -\frac{1}{4\pi}\int_{S^2_\infty}\dd
s_a~ \tr \left(\tfrac{1}{2}\eps_{abc}F_{bc}\phi\right)~.
\end{equation}
The configurations $(A_a,\phi)$, which satisfy the bound
\eqref{BPSbound} are called {\em BPS monopoles} and necessarily\index{BPS monopole}
fulfill the (first order) Bogomolny equations\index{Bogomolny equations}
\begin{equation}\label{BPSequation}
F_{ab}\ =\ \eps_{abc}D_c \phi~.
\end{equation}
Inversely, those finite energy configurations $(A_a,\phi)$ which
satisfy the Bogomolny equations \eqref{BPSequation} are BPS\index{Bogomolny equations}
monopoles.

\paragraph{Monopole solutions.} A twistor-inspired solution\index{twistor}
generating technique, the Nahm construction, is presented in
section \ref{sSolutionGT}.

\section{Chern-Simons theory and its relatives}\index{Chern-Simons theory}

In this section, we briefly review basic and relevant facts on
Chern-Simons theory. A broader discussion can be found in\index{Chern-Simons theory}
\cite{Freed:1992vw} and \cite{Dunne:1998qy}. Subsequently, we
present some related models, which we will encounter later on. In
particular, we will present a holomorphic Chern-Simons theory\index{Chern-Simons theory}
\cite{Witten:1992fb}, which will play a vital r{\^o}le in chapter
\ref{chTwistorGeometry}.

\subsection{Basics}

\paragraph{Motivation.} Chern-Simons theory is a completely new\index{Chern-Simons theory}
type of gauge theories, which was accidently discovered by
Shiing-Shen Chern and James Harris Simons when studying Pontryagin
densities of 3-manifolds \cite{Chern:1974ft}. It is crucial in
3-manifold topology and knot theory and its partition function
defines the Witten-Reshetikhin-Turaev invariant, a topological
invariant of 3-manifolds. Furthermore, perturbation theory gives
rise to an infinite number of other topological invariants.

Chern-Simons theories are deeply connected to anyons, particles
living in two dimension which have magnetic flux tied to their
electric charge and -- considering a large wavelength limit -- to
a description of the Landau problem of charged particles moving in
a plane under the influence of a magnetic field perpendicular to
the plane.

\paragraph{Abelian Chern-Simons and Maxwell theory.} The difference\index{Maxwell theory}
between Chern-Simons theory and ordinary Maxwell theory is easiest\index{Chern-Simons theory}
seen comparing the Lagrangians and the equations of motion:
\begin{subequations}
\begin{align}
\CL_{\mathrm{M}}&\ =\ -\tfrac{1}{4}F^{\mu\nu}F_{\mu\nu}-A_\mu J^\mu~,&
\dpar_\mu F^{\mu\nu}&\ =\ J^\nu~,\\
\CL_{\mathrm{CS}}&\ =\ \tfrac{\kappa}{2}\eps^{\mu\nu\rho}A_\mu
\dpar_\nu A_\rho-A_\mu J^\mu~, &
\tfrac{\kappa}{2}\eps^{\mu\nu\rho}F_{\nu\rho}&\ =\ J^\mu~,
\end{align}
\end{subequations}
Gauge invariance is not obvious, as the Lagrangian is not
exclusively defined in terms of the invariant field strength\index{field strength}
$F_{\mu\nu}$. Nevertheless, one easily checks that gauge
transforming $\mathcal{L}_{\mathrm{CS}}$ leads to a total
derivative, which vanishes for manifolds without boundary.\index{manifold}

\paragraph{Solutions.} The solutions to the Chern-Simons field
equations $F_{\mu\nu}=\frac{1}{\kappa}\eps_{\mu\nu\rho}J^\rho$ are
trivial for vanishing source. To get nontrivial solution, there
are several possibilities: One can consider couplings to matter
fields and to a Maxwell term (the latter provides a new mass
generation formalism for gauge fields besides the Higgs
mechanism), nontrivial topology and boundaries of the
configuration space or generalize the action to non-Abelian gauge
fields and incorporating gravity.

\paragraph{Non-Abelian Chern-Simons theory.} Consider a vector\index{Chern-Simons theory}
bundle $E$ over a three-dimensional real manifold $M$ with a\index{manifold}
connection one-form $A$. Non-Abelian Chern-Simons theory is then\index{Chern-Simons theory}\index{connection}
defined by the action
\begin{equation}
\mathcal{L}_{CS}\ =\ \kappa\eps^{\mu\nu\rho}\tr\left(A_\mu\dpar_\nu
A_\rho+\tfrac{2}{3}A_\mu A_\nu A_\rho\right)
\end{equation}

Under a transformation $\delta A_\mu$, the Lagrangian changes
according to
\begin{equation}
\delta\mathcal{L}_{\mathrm{CS}}\ =\ \kappa\eps^{\mu\nu\rho}\tr(\delta
A_\mu F_{\nu\rho})
\end{equation}
with the standard non-Abelian field strength $F_{\mu\nu}=\dpar_\mu\index{field strength}
A_\nu-\dpar_\nu A_\mu+[A_\mu,A_\nu]$. The equations of motion take
the same form as in the Abelian case
\begin{equation}
\kappa\eps^{\mu\nu\rho}F_{\nu\rho}\ =\ J^\mu~.
\end{equation}
Under the non-Abelian gauge transformation, the Lagrangian\index{gauge!transformation}
transforms into
\begin{equation}
\mathcal{L}'_{\mathrm{CS}}\ =\ \mathcal{L}_{\mathrm{CS}}-(\mbox{tot.
derivative})-w(g)
\end{equation}
where $w(g)$ describes a winding number\index{winding number}
\begin{equation}
\int_M w(g)\ =\ 8\pi^2\kappa N~,~~N\in\RZ~.
\end{equation}
This gives rise to a quantization condition for $\kappa$ if we\index{quantization}
demand that the partition function $\de^{\di S_{CS}}$ is invariant
under gauge transformations\index{gauge transformations}\index{gauge!transformation}

\paragraph{Topological invariance.} Note that the energy-momentum
tensor of Chern-Simons theory vanishes:\index{Chern-Simons theory}\index{energy-momentum tensor}
\begin{equation}
T^{\mu\nu}\ =\ \frac{2}{\sqrt{\det g}}\frac{\delta S_{CS}}{\delta
g_{\mu\nu}}\ =\ 0~,
\end{equation}
which is due to the fact that $\mathcal{L}_{\mathrm{CS}}$ is
independent of the metric. Therefore, Chern-Simons theory is a\index{Chern-Simons theory}\index{metric}
{\em topological field theory}.\index{topological!field theory}

\paragraph{Quantization.} Canonical quantization of the system is\index{quantization}
straightforward as the components of the gauge fields are
canonically conjugate to each other:
\begin{equation}
[A_i(\vec{x}),A_j(\vec{y})]\ =\ \frac{\di}{\kappa}\eps_{ij}
\delta(\vec{x}-\vec{y})~,
\end{equation}
where $i,j=1,2$.

\subsection{Holomorphic Chern-Simons theory}\label{sshCStheory}\index{Chern-Simons theory}

Holomorphic Chern-Simons theory is besides super Yang-Mills theory\index{Chern-Simons theory}\index{Yang-Mills theory}
the most important field theory we will consider. Its omnipresence
is simply due to the fact that the open topological B-model on a\index{topological!B-model}
Calabi-Yau threefold containing $n$ space-filling D5-branes is\index{Calabi-Yau}
equivalent to holomorphic Chern-Simons theory on the same\index{Chern-Simons theory}
Calabi-Yau manifold with gauge group $\sGL(n,\FC)$, as we will see\index{Calabi-Yau}\index{manifold}
in section \ref{ssequivalencehCS}.

\paragraph{Setup.} We start from a complex $d$-dimensional
manifold $M$ over which we consider a holomorphic principal\index{manifold}
$G$-bundle $P$, where $G$ is a semisimple Lie (matrix) group with
Lie algebra $\frg$. Consider furthermore a connection one-form\index{Lie algebra}\index{connection}
(i.e.\ a Lie algebra valued one-form) $A$ on $P$, which is carried
over to the associated holomorphic vector bundle $E\rightarrow M$\index{holomorphic!vector bundle}
of $P$. We define the corresponding field strength by $F=\dd\index{field strength}
A+A\wedge A$, and denote by $A^{0,1}$ and $F^{0,2}$ the
$(0,1)$-part and the $(0,2)$-part of $A$ and $F$, respectively.
Note that $F^{0,2}=\dparb A^{0,1}+A^{0,1}\wedge A^{0,1}$.

\paragraph{Equations of motion.} Analogously to Chern-Simons
theory without sources, the equations of motion of holomorphic\index{Chern-Simons theory}
Chern-Simons theory simply read
\begin{equation}\label{hCSeom}
F^{0,2}\ =\ \dparb A^{0,1}+A^{0,1}\wedge A^{0,1}\ =\ 0~,
\end{equation}
and thus $\dparb_A=\dparb+A^{0,1}$ defines a holomorphic structure\index{holomorphic!structure}
on $E$, see \ref{pholomorphicstructure} in section
\ref{ssvectorbundles}. One can state that the Dolbeault
description of holomorphic vector bundles is in fact a description\index{holomorphic!vector bundle}
via holomorphic Chern-Simons theory.\index{Chern-Simons theory}

\paragraph{Action.} If $M$ is a Calabi-Yau threefold and thus\index{Calabi-Yau}
comes with a holomorphic $(3,0)$-form $\Omega^{3,0}$, one can
write down an action of holomorphic Chern-Simons theory which\index{Chern-Simons theory}
reproduces the equation \eqref{hCSeom}:
\begin{equation}
S_{\mathrm{hCS}}\ =\ \tfrac{1}{2}\int_M\Omega^{3,0}\wedge
\tr\left(A^{0,1}\wedge \dparb A^{0,1}+\tfrac{2}{3}A\wedge A\wedge
A\right)~.
\end{equation}
This action has been introduced in \cite{Witten:1992fb}.

\paragraph{Remarks.} In his paper \cite{Witten:1992fb}, Witten
remarks that hCS theory is superficially non-renormalizable by
power counting but that its symmetries suggest that it should be
finite at quantum level. This conclusion is in agreement with
holomorphic Chern-Simons theory being equivalent to a string\index{Chern-Simons theory}
theory.

\subsection{Related field theories}\label{ssrelFieldTheories}

\paragraph{Topological BF-theory.} This theory\index{topological!BF-theory}
\cite{Blau:1989bq,Horowitz:1989ng} is an extension of Chern-Simons
theory to manifolds with arbitrary dimension. Consider a\index{Chern-Simons theory}\index{manifold}
semisimple Lie matrix group $G$ with Lie algebra $\frg$.\index{Lie algebra}
Furthermore, let $M$ be a real manifold of dimension $d$, $P$ a\index{manifold}
principal $G$-bundle over $M$ and $A$ a connection one-form on\index{connection}
$P$. The associated curvature is -- as usual -- given by $F=\dd\index{curvature}
A+A\wedge A$. Then the action of topological BF-theory is given by\index{topological!BF-theory}
\begin{equation}\label{actiontBF}
S_{\mathrm{BF}}\ =\ \int_M \tr(B\wedge F)~,
\end{equation}
where $B$ is a $(d-2)$-form in the adjoint representation of the\index{representation}
gauge group $G$. That is, a gauge transformation $g\in\Gamma(P)$\index{gauge!transformation}
act on the fields $A$ and $B$ according to
\begin{equation}
A\ \mapsto\  g^{-1}Ag+g^{-1}\dd g\eand B\ \mapsto\  g^{-1}Bg~.
\end{equation}
The equations of motion of \eqref{actiontBF} read as
\begin{equation}
F\ =\ 0\eand \dd B+ A\wedge B-(-1)^d B\wedge A\ =\ 0~,
\end{equation}
and thus BF-theory describes flat connections and $\dd_A$-closed\index{connection}
$(d-2)$-forms on an $d$-dimensional real manifold.\index{manifold}

\paragraph{Holomorphic BF-theory.}\label{pholBF} Holomorphic BF-theory\index{holomorphic BF-theory}
\cite{Popov:1999cq,Ivanova:2000xr,Ivanova:2000af} is an extension
of topological BF-theory to the complex situation. As such, it can\index{topological!BF-theory}
also be considered as an extension of holomorphic Chern-Simons
theory to complex manifolds of complex dimensions different from\index{Chern-Simons theory}\index{complex!manifold}\index{manifold}
three. Let us consider the same setup as above, but now with $M$ a
complex manifold of (complex) dimension $d$. Then the\index{complex!manifold}\index{manifold}
corresponding action reads
\begin{equation}
S_{\mathrm{hBF}}\ =\ \int_M \tr(B\wedge F^{0,2})~,
\end{equation}
where $B$ is here a $(d,d-2)$-form on $M$ in the adjoint
representation of the gauge group $G$ and $F^{0,2}$ is the\index{representation}
$(0,2)$-part of the curvature $F$. If $M$ is a Calabi-Yau\index{Calabi-Yau}\index{curvature}
manifold, there is a natural holomorphic volume form\index{holomorphic!volume form}\index{manifold}
$\Omega^{d,0}$ on $M$, and one can alternatively introduce the
action
\begin{equation}
S_{\mathrm{hBF}}\ =\ \int_M \Omega^{d,0}\wedge\tr(B\wedge F^{0,2})~,
\end{equation}
where $B$ is now a $(0,d-2)$-form on $M$. The equations of motion
in the latter case read as
\begin{equation}
\dparb A^{0,1}+A^{0,1}\wedge A^{0,1}\ =\ 0\eand \dparb B+A^{0,1}\wedge
B-(-1)^d B\wedge A^{0,1}\ =\ 0~.
\end{equation}
This theory is sometimes called holomorphic $\theta$BF-theory,
where $\theta=\Omega^{d,0}$.

We will encounter an example for such a holomorphic BF-theory when\index{holomorphic BF-theory}
discussing the topological B-model on the mini-supertwistor space\index{mini-supertwistor space}\index{topological!B-model}\index{twistor}\index{twistor!space}
in section \ref{sPWMini}.

\section{Conformal field theories}

A {\em conformal field theory} is a (quantum) field theory, which\index{conformal field theory}
is invariant under (local) conformal, i.e.\ angle-preserving,
coordinate transformations. Such field theories naturally arise in
string theory, quantum field theory, statistical mechanics and\index{string theory}
condensed matter physics. Usually, conformal field theories are
considered in two dimensions, but e.g.\ also $\CN=4$ super
Yang-Mills theory in four dimensions is conformal, even at quantum\index{N=4 super Yang-Mills theory@$\CN=4$ super Yang-Mills theory}\index{Yang-Mills theory}
level. Among the many available introductions to conformal field
theory, very useful ones are\index{conformal field theory} e.g.\
\cite{Ginsparg:1988ui,Schellekens:1996tg,DiFrancesco:1997nk}. A
very concise introduction can moreover be found in the first
chapter of \cite{Polchinski:1994mb}.

\subsection{CFT basics}\label{ssCFTbasics}

\paragraph{The conformal group.} Infinitesimal conformal
transformations $x^\mu\rightarrow x^\mu+\eps^\mu$ have to preserve
the square of the line element up to a local factor $\Omega(x)$,
and from
\begin{equation}
\dd s^2\ \rightarrow\  \dd
s^2+(\dpar_\mu\eps_\nu+\dpar_\nu\eps_\mu)\dd x^\mu\dd x^\nu~,
\end{equation}
we therefore conclude that
$(\dpar_\mu\eps_\nu+\dpar_\nu\eps_\mu)\sim \eta_{\mu\nu}$. On the
two-dimensional plane with complex coordinates $z=x^1+\di x^2$,
these equations are simply the Cauchy-Riemann equations
\begin{equation}
\dpar_1\eps_1\ =\ \dpar_2\eps_2\eand \dpar_1\eps_2\ =\ -\dpar_2\eps_1~.
\end{equation}
Two-dimensional, local conformal transformations are thus given by
holomorphic functions and these transformations are generated by
\begin{equation}
\ell_n\ =\ -z^{n+1}\dpar_z\eand \bar{\ell}_n\ =\ -\bz^{n+1}\dpar_\bz~,
\end{equation}
which are the generators of the {\em Witt-algebra}\/\footnote{the
algebra of Killing vector fields on the Riemann sphere.}\index{Killing vector}\index{Riemann sphere}
\begin{equation}\label{WittAlgebra}
[\ell_m,\ell_n]\ =\ (m-n)\ell_{m+n}~,~~~
[\bar{\ell}_m,\bar{\ell}_n]\ =\ (m-n)\bar{\ell}_{m+n}~,~~~
[\ell_m,\bar{\ell}_n]\ =\ 0~.
\end{equation}
On the compactification $\CPP^1$ of the complex plane $\FC$, the
{\em global} conformal transformations are the so-called {\em
M\"{o}bius transformations}, which are maps\index{M\"{o}bius transformations}
$z\mapsto\frac{az+b}{cz+d}$ with $ad-bc=1$. Note that these maps
form a group $\cong \sSL(2,\FC)/\RZ_2\cong\sSO(3,1)$ and map
circles on the sphere onto circles. They are generated by
$\ell_{-1},\ell_0,\ell_1$ and their complex conjugates.

\paragraph{Exemplary theory.} To briefly discuss relevant properties of conformal
field theories, we will use an exemplary theory, which is
introduced in this paragraph. Consider the two-dimensional field
theory given by the action
\begin{equation}\label{exempmodel}
S=\tfrac{1}{4\pi}\int \dd^2z~ \dpar X\dparb X~,
\end{equation}
where $X=X(z,\bz)$ is a function\footnote{The notation $X(z,\bz)$
here merely implies that $X$ is a priori a general, not
necessarily holomorphic function. Sometimes, however, it is also
helpful to consider a complexified situation, in which $z$ and
$\bz$ are independent, complex variables.} on $\FC$ and $\dpar$
and $\dparb$ denote derivatives with respect to $z$ and $\bz$.
Furthermore, the normalization of the measure $\dd^2z$ is chosen
such that $\int \dd^2 z\delta^2(z,\bz)=1$. The equation of motion
following from this action simply reads $\dpar\dparb X(z,\bz)=0$
and the solutions to these equations are harmonic functions
$X(z,\bz)$.

\paragraph{Operator equation.} On the quantum level, the above\index{operator equation}
mentioned equation of motion is only true up to contact terms, as
one easily derives
\begin{equation}\label{opeqn}
\begin{aligned}
0&\ =\ \int \CCD X \Der{X(z,\bz)}\de^{-S}X(z',\bz')\\&\ =\  \langle
\delta^2(z-z',\bz-\bz')\rangle+\tfrac{1}{2\pi}\dpar_z\dparb_\bz\langle
X(z,\bz)X(z',\bz')\rangle~.
\end{aligned}
\end{equation}
Such an equation is called an {\em operator equation}. By\index{operator equation}
introducing {\em normal ordering}\index{normal ordering}
\begin{equation}
:\CO(X):\,\ =\ \exp\left(\tfrac{1}{2}\int \dd^2 z\dd^2 \bz'~
\ln|z-z'|^2 \Der{X(z,\bz)}\Der{X(z',\bz')}\right)\CO(X)~,
\end{equation}
we can cast the operator equation \eqref{opeqn} into the classical\index{operator equation}
form
\begin{equation}
\dpar_z\dpar_\bz:X(z,\bz)X(z',\bz'):\,\ =\ 0~,
\end{equation}
where
\begin{equation}
:X(z,\bz)X(z',\bz'):\,\ =\ X(z,\bz)X(z',\bz')+\ln|z-z'|^2~.
\end{equation}
Taylor expanding the above equation, we obtain an example of an
{\em operator product expansion}:\index{operator product expansion}
\begin{equation*}
X(z,\bz)X(0,0)\ =\ -\ln|z|^2+:X^2(0,0):+z:X\dpar X(0,0):+\bz:X\dparb
X(0,0):+\ldots~.
\end{equation*}

\paragraph{Energy-momentum tensor.}\label{penergymomentum}\index{energy-momentum tensor}
The {\em energy-momentum tensor} naturally appears as Noether
current for conformal transformations. Consider an infinitesimal
such transformation $z'=z+\eps g(z)$, which leads to a field
transformation $\delta X=-g(z)\dpar X-\bar{g}(\bz)\dparb X$. The
Noether currents are $j(z)=\di g(z)T(z)$ and $\bj(\bz)=\di
\bar{g}(\bz)\tilde{T}(\bz)$, where we have in our exemplary theory
\begin{equation}\label{energymomentumtensor}
T(z)\ =\ -\tfrac{1}{2}:\dpar X\dpar X:\eand
\tilde{T}(\bz)\ =\ -\tfrac{1}{2}:\dparb X\dparb X:~.
\end{equation}
From the condition that in the divergence $\dparb j-\dpar \bj$ of
$j$, each term has to vanish separately\footnote{$g$ and $\bar{g}$
are ``linearly independent''}, and the fact that the
energy-momentum tensor is the Noether current for rigid\index{energy-momentum tensor}\index{rigid}
translations, one derives that the only nontrivial components of
the tensor $T$ are $T_{zz}=T(z)$, $T_{\bz\bz}=\tilde{T}(\bar{z})$.
(Back in real coordinates $x=\mathrm{Re}(z),y=\mathrm{Im}(z)$,
this is equivalent to the energy-momentum tensor having vanishing\index{energy-momentum tensor}
trace, and one can also take this property as a definition for a
conformal field theory.) One can derive furthermore that in any\index{conformal field theory}
given conformal field theory, the operator product expansion of\index{operator product expansion}
the {\em energy-momentum tensor} $T(z)$ is given by\index{energy-momentum tensor}
\begin{equation}
T(z)T(w)\ =\ \frac{c/2}{(z-w)^4}+\frac{2T(w)}{(z-w)^2}+\frac{\dpar_w
T(w)}{z-w}+\ldots ~,
\end{equation}
where $c$ is called the {\em central charge} of the theory.\index{central charge}

\paragraph{Radial quantization.} Let us take a short glimpse at\index{quantization}
quantum aspects of conformal field theories. For this, we
compactify the complex plane along the $x$-axis to an infinitely
long cylinder and map it via $z\mapsto \de^{z}$ to the annular
region $\FC^\times$. Time now runs radially and equal time lines
are circles having the origin as their center. The equal time
commutators of operators can here be easily calculated via certain\index{commutators}
contour integrals. Take e.g.\ charges $Q_i[C]=\oint_C\frac{\dd
z}{2\pi\di} j_i$ and three circles $C_1,C_2,C_3$ with constant
times $t_1>t_2>t_3$. Then the expression
\begin{equation}
Q_1[C_1]Q_2[C_2]-Q_1[C_3]Q_2[C_2]~,
\end{equation}
which vanishes classically, will turn into the commutator
\begin{equation}
[Q_1,Q_2][C_2]\ =\ \oint_{C_2}\frac{\dd
z_2}{2\pi\di}\mathrm{Res}_{z\rightarrow z_2} j_1(z)j_2(z_2)
\end{equation}
when considered as an expectation value, i.e.\ when inserted into
the path integral. The residue arises by deforming $C_1-C_3$ to a
contour around $z_2$, which is possible as there are no further
poles present. The operator order yielding the commutator is due
to the fact that any product of operators inserted into the path
integral will be automatically time-ordered, which corresponds to
a {\em radial ordering} in our situation.

\paragraph{Virasoro algebra.}\label{pVirasoro} Upon radial quantization, the mode\index{Virasoro algebra}\index{quantization}
expansion of the energy-momentum tensor $T(z)=\sum_n L_n z^{-n-2}$\index{energy-momentum tensor}
and $\tilde{T}(\bar{z})=\sum_n \bar{L}_n \bar{z}^{-n-2}$ together
with the inverse relations
\begin{equation}
L_m\ =\ \oint_C\frac{\dd z}{2\pi\di}z^{m+1}T(z)\eand
\bar{L}_m\ =\ -\oint_C\frac{\dd
\bar{z}}{2\pi\di}\bar{z}^{m+1}\tilde{T}(\bar{z})
\end{equation}
then lead immediately to the {\em Virasoro algebra}\index{Virasoro algebra}
\begin{equation}\label{VirasoroAlgebra}
[L_n,L_m]\ =\ (n-m)L_{n+m}+\frac{c}{12}n(n^2-1)\delta_{m+n,0}~.
\end{equation}
This algebra is the central extension of the Witt algebra\index{Witt algebra}
\eqref{WittAlgebra}.

\paragraph{Canonical quantization.} To canonically quantize our\index{quantization}
exemplary model \eqref{exempmodel}, we can use the fact that any
harmonic field\footnote{i.e.\ $\dpar\dparb X=0$} $X$ can be
(locally) expanded as the sum of a holomorphic and an
antiholomorphic function. That is, we expand $\dpar X$ as a
Laurent series in $z$ with coefficients $\alpha_m$ and $\dparb X$
in $\bz$ with coefficients $\tilde{\alpha}_m$. Integration then
yields
\begin{equation}\label{LaurentX}
X\ =\ x-\di\tfrac{\alpha'}{2}p\ln|z|^2+\di\sqrt{\tfrac{2}{\alpha'}}
\sum_{\stackrel{m=-\infty}{m\neq 0}}^\infty\frac{1}{m}\left(
\frac{\alpha_m}{z^m}+\frac{\tilde{\alpha}_m}{\bz^m}\right)~,
\end{equation}
where we singled out the zeroth order in both series and
identified the log-terms arising from $\dpar X$ and $\dparb X$
with translations and thus momentum. The radial quantization\index{quantization}
procedure then yields the relations
\begin{equation}
[\alpha_m,\alpha_n]\ =\ [\tilde{\alpha}_m,\tilde{\alpha}_n]\ =\ m\delta_{m+n}\eand
[x,p]\ =\ \di~.
\end{equation}
We will examine the spectrum of this theory for $26$ fields
$X^\mu$ in section \ref{ssSTQuantization}.

\paragraph{Primary fields.} A {\em tensor} or {\em primary field}\index{primary field}
$\phi(w)$ in a conformal field theory transforms under general\index{conformal field theory}
conformal transformations as
\begin{equation}
\phi'(z',\bz')\ =\ (\dpar_zz')^{-h_\phi}(\dpar_\bz\bz')^{-\bar{h}_\phi}\phi(z,\bz)~,
\end{equation}
where $h_\phi$ and $\bar{h}_\phi$ are the {\em conformal weights}\index{conformal weight}
of the field $\phi(w)$. Furthermore, $h_\phi+\bar{h}_\phi$
determine its {\em scaling dimension}, i.e.\ its behavior under\index{scaling dimension}
scaling, and $h_\phi-\bar{h}_\phi$ is the field's spin. With the
energy-momentum tensor $T(z)$, such a field $\phi$ has the\index{energy-momentum tensor}
following operator product expansion:\index{operator product expansion}
\begin{equation}
T(z)\phi(w)\ =\ \frac{h_\phi}{(z-w)^2}+\frac{\dpar_w
\phi(w)}{z-w}+\ldots ~.
\end{equation}
For the modes appearing in the expansion $\phi(z)=\sum_n \phi_n
z^{-n-h_\phi}$, we thus have the algebra
\begin{equation}
[L_n,\phi_m]\ =\ (n(h_\phi-1)-m)\phi_{m+n}~.
\end{equation}

\paragraph{Current algebras.} Currents in a conformal field theory\index{conformal field theory}\index{current algebra}
are (1,0)-tensor $j^a(z)$ with the operator product expansion\index{operator product expansion}
\begin{equation}
j^a(z)j^b(0)\sim \frac{k^{ab}}{z^2}+\di \frac{f^{ab}{}_c}{z}
j^c(0)~.
\end{equation}
The Laurent expansion $j^a(z)=\sum_{m=-\infty}^\infty
\frac{j_m^a}{z^{m+1}}$ then leads to the {\em current algebra} or\index{current algebra}
{\em Kac-Moody algebra}\index{Kac-Moody algebra}
\begin{equation}
[j_m^a,j_n^b]\ =\ mk^{ab}\delta_{m+n,0}+\di f^{ab}{}_cj^c_{m+n}~.
\end{equation}

\paragraph{Further theories.} The exemplary theory
\eqref{exempmodel} is certainly one of the most important
conformal field theories. Further examples are given by the $bc$-
and the $\beta\gamma$-systems
\begin{equation}
S_{bc}\ =\ \int \dd^2z~ b\dparb c\eand S_{\beta\gamma}\ =\ \int \dd^2z~
\beta\dparb \gamma~,
\end{equation}
which serve e.g.\ as Faddeev-Popov ghosts for the Polyakov string
and the superstring, see also section \ref{ssN1ST}. In the former
theory, the fields $b$ and $c$ are anticommuting fields and
tensors of weight $(\lambda,0)$ and $(1-\lambda,0)$. This theory
is purely holomorphic and in the operator product expansion of the\index{operator product expansion}
energy-momentum tensor, an additional contribution to the central\index{energy-momentum tensor}
charge of $c=-3(2\lambda-1)^2+1$ and $\tilde{c}=0$ appears for
each copy of the $bc$-system. The $\beta\gamma$-system has
analogous properties, but the fields $\beta$ and $\gamma$ are here
commuting and the central charge contribution has an opposite\index{central charge}
sign. Recall the relation between $\beta\gamma$-systems and local
Calabi-Yau manifolds of type $\CO(a)\oplus\CO(-2-a)\rightarrow\index{Calabi-Yau}\index{manifold}
\CPP^1$ discussed in section \ref{ssCY3folds}, \ref{pCYoverCP1}.
The case of a $bc$-system with equal weights $h_b=h_c=\frac{1}{2}$
will be important when discussing the superstring. Here, one
usually relabels $b\rightarrow \psi$ and $c\rightarrow \bpsi$

\subsection{The $\CN=2$ superconformal algebra}\index{super!conformal algebra}

\paragraph{Constituents.} The $\CN=2$ superconformal algebra (SCA) is\index{super!conformal algebra}
generated by the energy-momentum tensor $T(z)$, two supercurrents\index{energy-momentum tensor}
$G^+(z)$ and $G^-(z)$, which are primary fields of the Virasoro\index{primary field}
algebra with weight $\frac{3}{2}$ and a $\sU(1)$ current $J(z)$,
which is a primary field of weight $1$. The supercurrents\index{primary field}
$G^\pm(z)$ have $\sU(1)$ charges $\pm1$. This mostly fixes the
operator product expansion of the involved generators to be\index{operator product expansion}
\begin{equation}
\begin{aligned}
T(z)T(w) &\ =\  \frac{c/2}{(z-w)^4} + \frac{2 T(w)}{(z-w)^2} +
\frac{\dpar_w T(w)}{z-w} + \cdots ~,\\
T(z)G^{\pm}(w) &\ =\  \frac{3/2}{(z-w)^2}\, G^\pm(w) +
\frac{\dpar_w G^{\pm}(w)}{z-w} + \cdots ~,\\
T(z)J(w) &\ =\  \frac{J(w)}{(z-w)^2} + \frac{\dpar_w J(w)}{z-w} + \cdots~,\\
G^{+}(z)G^{-}(w) &\ =\  \frac{2\,c/3}{(z-w)^3} + \frac{2J(w)}{(z-w)^2}
+\frac{2T(w) + \dpar_w J(w)}{z-w} + \cdots~,\\
J(z)G^{\pm}(w) &\ =\  \pm\, \frac{G^{\pm}(w)}{z-w} + \cdots~,\\
J(z)J(w) &\ =\   \frac{c/3}{(z-w)^2} + \cdots~,
\end{aligned}
\end{equation}
where the dots stand for regular terms. Additionally to the mode
expansion of the energy-momentum tensor given in \ref{pVirasoro},\index{energy-momentum tensor}
we have the mode expansions for the two supercurrents and the
$\sU(1)$ current
\begin{equation}
G^\pm(z)\ =\ \sum_{n=-\infty}^\infty
G^\pm_{n\pm\eta\pm\frac{1}{2}}z^{-(n\pm\eta\pm\frac{1}{2})-\frac{3}{2}}\eand
J(z)\ =\ \sum_{n=-\infty}^\infty J_n z^{-n-1}~,
\end{equation}
where $\eta\in[-\frac{1}{2},\frac{1}{2})$. The latter parameter is
responsible for the boundary conditions of the supercurrents, and
by substituting $z\rightarrow \de^{2\pi\di} z$, we obtain
\begin{equation}
G^\pm(\de^{2\pi\di}z)\ =\ -\de^{\mp
2\pi\di(\eta+\frac{1}{2})}G^\pm(z)~,
\end{equation}
and therefore in superstring theory, $\eta=-\frac{1}{2}$ will\index{string theory}
correspond to the Neveu-Schwarz (NS) sector, while $\eta=0$ is\index{Neveu-Schwarz}
related with the Ramond (R) sector, cf.\ section\index{Ramond}
\ref{typetwosuperstrings}.

\paragraph{The algebra.} The operator product expansion\index{operator product expansion}
essentially fixes the algebra in terms of the modes introduced for
the generators. First, we have the Virasoro algebra\index{Virasoro algebra}
\begin{equation}
[L_n,L_m]\ =\ (n-m)L_{n+m}+\frac{c}{12}n(n^2-1)\delta_{m+n,0}~.
\end{equation}
Second, there is the algebra for the $\sU(1)$ current and its
commutation relation with the Virasoro generators
\begin{equation}
\begin{aligned}
\left[ J_m, J_n\right] & \ =\  \tfrac{c}{3} m\, \delta_{m+n,0}~, \\
\left[ L_n, J_m\right] & \ =\  -m\, J_{m+n}~.
\end{aligned}
\end{equation}
Eventually, there are the relations involving the two
supercurrents $G^\pm$:
\begin{equation}
\begin{aligned}
 \left[ L_n, G^{\pm}_{m \pm a}\right] & \ =\ 
          \left( \tfrac{n}{ 2} - (m \pm a)\right) \, G^{\pm}_{m + n\pm a}~, \\
 \left[ J_n, G^{\pm}_{m \pm a}\right] & \ =\ \pm\, G^{\pm}_{m + n \pm a}~,\\
\left\{G^+_{n+a}, G^-_{m-a}\right\} &\ =\  2\, L_{m+n} +
  (n-m+2a)\, J_{n+m}\tfrac{c}{ 3} \left( (n+a)^2 - \tfrac{1 }{ 4} \right)
  \delta_{m+n,0}~,
\end{aligned}
\end{equation}
where we used the shorthand notation $a=\eta+\frac{1}{2}$.

\paragraph{The $\CN=(2,2)$ SCA.} This algebra is obtained by
adding a second, right-moving $\CN=2$ SCA algebra with generators
$\tilde{T}(\bar{z})$, $\tilde{G}^\pm(\bar{z})$ and
$\tilde{J}(\bar{z})$.

\paragraph{Representations of the $\CN=(2,2)$ SCA.} There are three\index{representation}
well-established representations of the $\CN=2$ SCA. Most
prominently, one can define a supersymmetric nonlinear sigma model\index{nonlinear sigma model}\index{sigma model}
in two dimensions, which possesses $\CN=2$ superconformal
symmetry. In section \ref{ssnonlinsigmamodel}, we will discuss
such models in more detail. Furthermore, there are the
Landau-Ginzburg theories discussed in \ref{LandauGinzburg},
section \ref{WessZumino} and the so-called minimal models. For\index{minimal model}
more details, see \cite{Greene:1996cy,Scheidegger}.

\chapter{String Theory}

String theory is certainly the most promising and aesthetically\index{string theory}
satisfying candidate for a unification of the concepts of quantum
field theory and general relativity. Although there is still no
realistic string theory describing accurately all the measured\index{string theory}
features of the (known) elementary particles, ``existence proofs''
for standard-model-like theories arising from string theories have
been completed, see e.g.\ \cite{Lust:2004ks}. One of the most
important current problems is the selection of the correct
background in which string theory should be discussed; the\index{string theory}
achievement of moduli stabilization (see \cite{DeWolfe:2005uu} and
references therein) show that there is progress in this area.
Among the clearly less appealing approaches is the
``landscape''-concept discussed in \cite{Susskind:2003kw}.

The relevant literature to this chapter is
\cite{Green:1987mn,Green:1987sp,Polchinski:1998rq,Polchinski:1998rr,
Polchinski:1994mb,Schwarz:2000ew,Szabo:2002ca} (general and
$\CN=1$ string theory),\index{string theory}
\cite{Marcus:1992wi,Martinec:1997cw,Lechtenfeld:1999gd} ($\CN=2$
string theory), \cite{Polchinski:1996na,Sen:1998rg,Johnson:2000ch}\index{N=2 string theory@$\CN=2$ string theory}\index{string theory}
(D-branes),\index{D-brane}
\cite{Witten:1991zz,Greene:1996cy,Hori:Trieste,Mirrorbook}
(topological string theory and mirror symmetry).\index{mirror symmetry}\index{string theory}\index{topological!string}

\section{String theory basics}\index{string theory}

In this section, we will briefly recall the elementary facts on
the bosonic string. This theory can be regarded as a toy model to
study features which will also appear in the later discussion of
the superstring.

\subsection{The classical string}

\paragraph{Historical remarks.} Strings were originally
introduced in the late 1960s to describe confinement in a quantum\index{confinement}
field theory of the strong interaction, but during the next years,
QCD proved to be the much more appropriate theory. Soon thereafter
it was realized that the spectrum of an oscillating string
contains a spin-2 particle which behaves as a graviton and
therefore string theory should be used for unification instead of\index{string theory}
a model of hadrons. After the first ``superstring revolution'' in\index{superstring revolution}
1984/1985, string theory had become an established branch of\index{string theory}
theoretical physics and the five consistent superstring theories
had been discovered. In the second superstring revolution around\index{superstring revolution}
1995, dualities relating these five string theories were found,
giving a first taste of non-perturbative string theory.\index{string theory}
Furthermore, one of the most important objects of study in string
theory today, the concept of the so-called D-branes, had been\index{D-brane}\index{string theory}
introduced.

\paragraph{Bosonic string actions.} Consider a two-dimensional
(pseudo-)Riemannian manifold $\Sigma$ described locally by\index{manifold}
coordinates $\sigma^0$ and $\sigma^1$ and a metric\index{metric}
$\gamma_{\alpha\beta}$ of Minkowski signature $(-1,+1)$. This
space is called the {\em worldsheet} of the string and is the\index{worldsheet}
extended analogue of the worldline of a particle. Given a further
Riemannian manifold $M$ and a map $X:\Sigma\rightarrow M$ smoothly\index{manifold}
embedding the worldsheet $\Sigma$ of the string into the {\em\index{worldsheet}
target space} $M$, we can write down a string action (the {\em\index{target space}
Polyakov action})\index{Polyakov action}
\begin{equation}\label{PolyakovAction}
S\ =\ -\frac{T}{2}\int \dd^2\sigma~
\sqrt{\gamma}\gamma^{\alpha\beta}\dpar_\alpha X^\mu\dpar_\beta
X_\mu~,
\end{equation}
where $T$ is the tension of the string. This meaning becomes even
clearer, when we recast this action into the form of the {\em
Nambu-Goto action}\index{Nambu-Goto}\index{Nambu-Goto action}
\begin{equation}
S\ =\ -T\int \dd^2\sigma~ \sqrt{-\det \dpar_\alpha X^\mu\dpar_\beta
X_\mu}~,
\end{equation}
which is equal to $-T$ times the area of the worldsheet of the\index{worldsheet}
string. Note that more frequently, one encounters the constants
$l_s=\sqrt{\frac{1}{2\pi T}}$, which is the {\em string length}\index{string length}
and the {\em Regge slope}  $\alpha'=\frac{1}{2\pi T}=l_s^2$.\index{Regge slope}

In general, we have $\sigma^0$ run in an arbitrary interval of
``time'' and $\sigma^1$ run between $0$ and $\pi$ if the ``spatial
part'' of the worldsheet is noncompact and between $0$ and $2\pi$\index{worldsheet}
else.

\paragraph{Equations of motion.} The equations of motion obtained
by varying \eqref{PolyakovAction} with respect to the worldsheet\index{worldsheet}
metric read as\index{metric}
\begin{equation}
\dpar_\alpha X^\mu\dpar_\beta
X_\mu\ =\ \tfrac{1}{2}\gamma_{\alpha\beta}\gamma^{\gamma\delta}\dpar_\gamma
X^\mu\dpar_\delta X_\mu~,
\end{equation}
which implies that the induces metric\index{metric}
$h_{\alpha\beta}:=\dpar_\alpha X^\mu\dpar_\beta X_\mu$ is
proportional to the worldsheet metric.\index{worldsheet}\index{metric}

\paragraph{Closed and Neumann boundary conditions.} To determine\index{Neumann boundary conditions}
the variation with respect to $X$, we have to impose boundary
conditions on the worldsheet. The simplest case is the one of\index{worldsheet}
periodic boundary conditions, in which the spatial part of the
worldsheet becomes compact:\index{worldsheet}
\begin{equation}
\begin{aligned}
X^\mu(x,2\pi)\ =\ X^\mu(x,0)~,~~~\dpar^\mu X^\nu(x,2\pi)\ =\ \dpar^\mu
X^\nu(x,0)~,\\
\gamma_{\alpha\beta}(x,2\pi)\ =\ \gamma_{\alpha\beta}(x,0)~.\hspace{2.5cm}
\end{aligned}
\end{equation}
This describes a closed string, where all boundary terms clearly\index{closed string}
vanish. The same is true if we demand that
\begin{equation}\label{nbc}
n^\alpha\dpar_\alpha X^\mu\ =\ 0~~~\mbox{on}~~~\dpar \Sigma~,
\end{equation}
where $n^\alpha$ is normal to $\dpar\Sigma$, as the boundary term
in the variation of the action with respect to $X$ is evidently
proportional to $\dpar_\alpha X^\mu$. Taking a flat, rectangular
worldsheet, \eqref{nbc} reduces to $\dpar_1 X^\mu=0$. These\index{worldsheet}
conditions are called {\em Neumann boundary conditions} and\index{Neumann boundary conditions}
describe an open string whose endpoints can move freely in the\index{open string}
target space. Both the closed and the Neumann boundary conditions\index{Neumann boundary conditions}\index{target space}
yield
\begin{equation}
\dpar^\alpha\dpar_\alpha X^\mu\ =\ 0
\end{equation}
as further equations of motion.

\paragraph{Dirichlet boundary conditions.} One can also impose\index{Dirichlet boundary conditions}
so-called {\em Dirichlet boundary conditions}, which state that
the endpoints of a string are fixed in the spatial direction:
\begin{equation}
\dpar_0 X^\mu\ =\ 0~.
\end{equation}
However, these boundary conditions by themselves have some
unpleasant features: Not only do they break Poincar{\'e} symmetry, but
they also have momentum flowing off the endpoints of the open
strings. The true picture is that open strings with\index{open string}
Dirichlet-boundary conditions end on subspaces of the target
space, so called {\em D-branes}, which we will study in section\index{D-brane}\index{target space}
\ref{sDBranes}. For this reason, we will restrict ourselves here
to open strings with Neumann boundary conditions.\index{Neumann boundary conditions}\index{open string}

\paragraph{Symmetries.} The Polyakov action has a remarkable set\index{Polyakov action}
of symmetries:
\begin{itemize}
\item Poincar{\'e} symmetry in the target space\index{target space}
\item Diffeomorphism invariance on the worldsheet\index{worldsheet}
\item Weyl-invariance on the worldsheet
\end{itemize}\index{worldsheet}
Weyl-invariance means that the action is invariant under a local
rescaling of the worldsheet metric and thus the worldsheet action\index{worldsheet}\index{metric}
is {\em conformally invariant}.

\paragraph{The energy-momentum tensor.} The variation of the\index{energy-momentum tensor}
action with respect to the worldsheet metric yields the\index{worldsheet}\index{metric}
{}{}{}{}{\em energy-momentum tensor}\index{energy-momentum tensor}
\begin{equation}
T^{\alpha\beta}\ =\ -4\pi\sqrt{-\det
\gamma}\Der{\gamma_{\alpha\beta}}S~.
\end{equation}
Since the worldsheet metric is a dynamical field, the\index{worldsheet}\index{metric}
energy-momentum tensor vanishes classically. Furthermore, the\index{energy-momentum tensor}
trace of this tensor has to vanish already due to Weyl-invariance:
\begin{equation}
T_{\alpha}{}^\alpha\sim\gamma_{\alpha\beta}\Der{\gamma_{\alpha\beta}}S\ =\ 0~.
\end{equation}

\subsection{Quantization}\label{ssSTQuantization}\index{quantization}

\paragraph{Canonical quantization.} To quantize classical string\index{quantization}
theory given by the Polyakov action \eqref{PolyakovAction}, we\index{Polyakov action}
first fix the gauge for the worldsheet metric. In {\em conformally\index{worldsheet}\index{metric}
flat gauge}, we have
$(\gamma^{\alpha\beta})=\de^{\phi(\sigma)}\diag(-1,+1)$ and the
action \eqref{PolyakovAction} reduces to $D$ copies of our
exemplary theory \eqref{exempmodel} from section
\ref{ssCFTbasics}, for which we already discussed the quantization\index{quantization}
procedure. Note, however, that the creation and annihilation
operators receive an additional index for the $D$ dimensions of
spacetime and their algebra is modified to
\begin{equation}
[\alpha^\mu_m,\alpha^\nu_n]\ =\ [\tilde{\alpha}^\mu_m,\tilde{\alpha}^\nu_n]\ =\ 
m\delta_{m+n}\eta_{(M)}^{\mu\nu}~,
\end{equation}
where $\eta^{\mu\nu}_{(M)}$ is the Minkowski metric on the target\index{metric}
space manifold $M$. We thus have a quantum mechanical system\index{manifold}
consisting of the tensor product of 2$D$ harmonic oscillators and
a free particle. We therefore derive the states in our theory from
vacua $|k,0\rangle$, which are eigenstates of the momentum
operators $p^\mu=\alpha_0^\mu=\tilde{\alpha}_0^\mu$. These vacuum
states are furthermore annihilated by the operators
$\alpha^\mu_m$, $\tilde{\alpha}^\mu_m$ with $m<0$. The remaining
operators with $m>0$ are the corresponding creation operators. Due
to the negative norm of the oscillator states in the time
direction on the worldsheet, canonical quantization by itself is\index{quantization}\index{worldsheet}
insufficient and one needs to impose the further constraints
\begin{equation}\label{physicalstatecond}
(L_0-a)|\psi\rangle\ =\ (\tilde{L}_0-a)|\psi\rangle\ =\ 0\eand
L_n|\psi\rangle\ =\ \tilde{L}_n|\psi\rangle\ =\ 0~~~\mbox{for}~~~ n\ >\ 0~.
\end{equation}
This is a consequence of the above applied na\"ive gauge fixing
procedure and to be seen analogously to the Gupta-Bleuler
quantization prescription in quantum electrodynamics.\index{quantization}

Note that in the case of closed strings, one has an additional\index{closed string}
independent copy of the above Fock-space.

\paragraph{BRST quantization.} A more modern approach to\index{quantization}
quantizing the bosonic string is the BRST approach, from which the
above Virasoro constrains follow quite naturally: The physical
states belong here to the cohomology of the BRST operator. We will
not discuss this procedure but refer to the review material on
string theory, in particular to \cite{Polchinski:1998rq}.\index{string theory}

\paragraph{Virasoro generators and creation/annihilation
operators.} To expand the Virasoro generators in terms of the
creation and annihilation operators used above, we insert the
Laurent expansion \eqref{LaurentX} for $X^\mu$ into the
energy-momentum tensor \eqref{energymomentumtensor} and read of\index{energy-momentum tensor}
the coefficients of the total Laurent expansion. We find
\begin{equation}
L_0\ =\ \tfrac{1}{2}\alpha_0^2+\sum_{n=1}^\infty(\alpha_{-n}^\mu\alpha_{\mu
n})+\eps_0\ =\ \frac{\alpha'p^2}{4}+\sum_{n=1}^\infty(\alpha_{-n}^\mu\alpha_{\mu
n})+\eps_0~,
\end{equation}
where $\eps_0$ is a normal ordering constant, and\index{normal ordering}
\begin{equation}
L_m\ =\ \tfrac{1}{2}\sum_{n=-\infty}^\infty :
\alpha_{m-n}^\mu\alpha_{\mu n}:~,
\end{equation}
where $:\cdot:$ denotes creation-annihilation normal ordering. For\index{normal ordering}
the quantum operator $L_0$, the vacuum energy is formally
$\eps_0=\frac{d-2}{2}\zeta(-1)$, where $\zeta(-1)=\sum_n
n=-\frac{1}{12}$ after regularization.

\paragraph{Conformal anomaly.}\label{pConformalAnomaly} The\index{anomaly}
{\em conformal anomaly} or {\em Weyl anomaly} is the quantum
anomaly of local worldsheet symmetries. One can show that the\index{worldsheet}
anomaly related to Weyl invariance is proportional to the central
charge of the underlying conformal field theory. Since the\index{central charge}\index{conformal field theory}
appropriate ghost system for gauging the worldsheet symmetries is\index{worldsheet}
a $bc$-system with $\lambda=2$, the central charge is proportional\index{central charge}
to $D-26$, where $D$ is the number of bosons $(X^\mu)$. Thus, the
{\em critical dimension} of bosonic string theory, i.e.\ the\index{string theory}
dimension for which the total central charge vanishes, is $D=26$.\index{central charge}
From this, it also follows that $a=1$.

\paragraph{Open string spectrum.} The constraint $L_0=a$ in\index{open string}
\eqref{physicalstatecond} is essentially the mass-shell condition
\begin{equation}
m^2\ =\ -p_0^2\ =\ -\frac{1}{2\alpha'}\alpha_0^2\ =\ \frac{1}{\alpha'}(N-a)\ :=\ 
\frac{1}{\alpha'}\left(\sum_{n=1}^\infty
\alpha_{-n}\alpha_n\right)~.
\end{equation}
The ground state $|k,0\rangle$, for which $N=0$ has thus mass
$-k^2=m^2=-\frac{a}{\alpha'}<0$ and is in fact tachyonic.\index{tachyon}
Therefore, bosonic string theory is actually not a consistent\index{string theory}
quantum theory and should be regarded as a pedagogical toy
model\footnote{{\em Tachyon condensation} might be a remedy to\index{tachyon}\index{tachyon condensation}
this problem, but we will not go into details here.}.

The first excited level is given by $N=1$ and consists of
oscillator states of the form
$\alpha^\mu_{-1}|k;0\rangle=\zeta_\mu\alpha_{-1}|k;0\rangle$,
where $\zeta_\mu$ is some polarization vector. As mass, we obtain
$m^2=\frac{1}{\alpha'}(1-a)$ and from
\begin{equation}
L_1|k;\zeta\rangle\ =\ \sqrt{2\alpha'}(k\cdot\zeta)|k;0\rangle
\end{equation}
together with the physical state condition $L_1|k;\zeta\rangle=0$,
it follows that $k\cdot \zeta=0$. As mentioned in
\ref{pConformalAnomaly}, $a=1$ and therefore the first excited
level is massless. (Other values of $a$ would have led to further
tachyons and ghost states of negative norm.) Furthermore, due to\index{tachyon}
the polarization condition, we have $d-2=24$ independent
polarization states. These states therefore naturally correspond
to a massless spin 1 vector particle.

Higher excited states become significantly massive and are usually
discarded with the remark that they practically decouple.

\paragraph{Closed string spectrum.} In the case of closed strings, the\index{closed string}
physical state conditions \eqref{physicalstatecond} need to hold
for both copies $L_n$ and $\tilde{L}_n$ of the Virasoro
generators, corresponding to the right- and left-moving sectors.
Adding and subtracting the two conditions for $L_0$ and
$\tilde{L}_0$ yields equations for the action of the Hamiltonian
$H$ and the momentum $P$ on a physical state, which amount to
invariance under translations in space and time. These
considerations lead to the two conditions
\begin{equation}
m^2\ =\ \frac{4}{\alpha'}(N-1)\eand N\ =\ \tilde{N}~,
\end{equation}
where the first equation is the new mass-shell condition and the
second is the so-called level-matching condition.

The ground state $|k;0,0\rangle$ is evidently again a spin 0
tachyon and therefore unstable.\index{tachyon}

The first excited level is of the form
\begin{equation}
|k;\zeta\rangle\ =\ \zeta_{\mu\nu}(\alpha_{-1}^\mu|k;0\rangle\otimes\tilde{\alpha}^\nu_{-1}|k;0\rangle
\end{equation}
and describes massless states satisfying the polarization
condition $k^\mu\zeta_{\mu\nu}=0$. The polarization tensor
$\zeta_{\mu\nu}$ can be further decomposed into a symmetric, an
antisymmetric and a trace part according to
\begin{equation}
\zeta_{\mu\nu}\ =\ g_{\mu\nu}+B_{\mu\nu}+\eta_{\mu\nu}\Phi~.
\end{equation}
The symmetric part here corresponds to a spin 2 {\em graviton
field}, the antisymmetric part is called the {\em Neveu-Schwarz\index{Neveu-Schwarz}
$B$-field} and the scalar field $\Phi$ is the spin 0 {\em
dilaton}.

\paragraph{Chan-Paton factors.} We saw above that open strings\index{Chan-Paton factors}\index{open string}
contain excitations related to Abelian gauge bosons. To lift them
to non-Abelian states, one attaches non-dynamical degrees of
freedom to the endpoints of the open string, which are called {\em\index{open string}
Chan-Paton factors}. Here, one end will carry the fundamental\index{Chan-Paton factors}
representation and the other end the antifundamental\index{representation}
representation of the gauge group. Assigning Chan-Paton factors to\index{Chan-Paton factors}
both ends leads correspondingly to an adjoint representation. Note
that in the discussion of scattering amplitudes, one has to
appropriately take traces over the underlying matrices.

\section{Superstring theories}\label{sSST}

There are various superstring theories which have proven to be
interesting to study. Most conveniently, one can classify these
theories with the number of supersymmetries $(p,q)$ which square
to translations along the left- and right-handed light cone in the
1+1 dimensions of the worldsheet. The bosonic string considered\index{worldsheet}
above and living in 26 dimensions has supersymmetry $\CN=(0,0)$.\index{super!symmetry}
The type IIA and type IIB theories, which are of special interest
in this thesis, have supersymmetry $\CN=(1,1)$. The type I\index{super!symmetry}
theories are obtained from the type II ones by orbifolding with
respect to worldsheet parity and the heterotic string theories\index{parity}\index{worldsheet}
have supersymmetry $\CN=(0,1)$. Interestingly, it has been\index{super!symmetry}
possible to link all of the above supersymmetric string theories
to a master theory called M-theory \cite{Witten:1995ex}, on which\index{M-theory}
we do not want to comment further.

Besides the above theories with one supersymmetry, there are the\index{super!symmetry}
$\CN=2$ string theories with supersymmetry $\CN=(2,2)$ or
heterotic supersymmetry $\CN=(2,1)$. We will discuss the former
case at the end of this section. The latter has target space\index{target space}
$\FR^{2,2}$ for the right-handed sector and $\FR^{2,2}\times T^8$
for the left-handed sector. Also this theory has been conjectured
to be related to M-theory \cite{Martinec:1997cw}.\index{M-theory}

\subsection{$\CN=1$ superstring theories}\label{ssN1ST}

\paragraph{Preliminary remarks.} The motivation for turning to
superstring theories essentially consists of two points: First of
all, the bosonic spectrum contains a tachyon as we saw above and\index{tachyon}
therefore bosonic string theory is inconsistent as a quantum\index{string theory}
theory. Second, to describe reality, we will eventually need some
fermions in the spectrum and therefore bosonic string theory\index{string theory}
cannot be the ultimate answer. One might add a third reason for
turning to superstrings: The critical dimension of bosonic string
theory, 26, is much less aesthetical than the critical dimensions\index{string theory}
of $\CN=1$ superstring theory, 10, which includes the beautiful
mathematics of Calabi-Yau manifolds into the target space\index{Calabi-Yau}\index{target space}\index{manifold}
compactification process.

Note that there are several approaches to describe the
superstring, see also section \ref{ssFurtherDBranes},
\ref{pSDbranes}. Here, we will follow essentially the {\em
Ramond-Neveu-Schwarz} (RNS) formulation, which uses\index{Neveu-Schwarz}\index{Ramond}
two-dimensional worldsheet supersymmetry and the additional {\em\index{super!symmetry}\index{worldsheet}
Gliozzi-Scherck-Olive} (GSO) projection to ensure also target
space supersymmetry. Furthermore, the GSO projection guarantees a\index{GSO projection}\index{super!symmetry}\index{target space}
tachyon-free spectrum and modular invariance.\index{tachyon}

\paragraph{Superstring action.} A straightforward generalization of
the bosonic string action is given by
\begin{equation}
S=\tfrac{1}{4\pi}\int\dd^2z~ \left( \tfrac{2}{\alpha'} \dpar
X^\mu\bar{\dpar}X_\mu + \psi^\mu\bar{\dpar}\psi_\mu +
\tilde{\psi}^\mu\dpar\tilde{\psi}_\mu \right)~,
\end{equation}
where the two fermionic fields $\psi^\mu$ and $\tilde{\psi}^\mu$
are holomorphic and antiholomorphic fields, respectively. Recall
that we already discussed the conformal field theories for the
$\psi^\mu$ and the $\tilde{\psi}^\mu$ in section
\ref{ssCFTbasics}.

\paragraph{Boundary conditions.} From the equations of motion,
we get two possible boundary conditions leading to two sectors:
\begin{center}
\begin{tabular}{lll}
Ramond (R) & $\psi^\mu(0,\tau)=\tilde{\psi}^\mu(0,\tau)$ &\index{Ramond}
$\psi^\mu(\pi,\tau)=\tilde{\psi}^\mu(\pi,\tau)$~,
\\[0.2cm]
Neveu-Schwarz (NS) & $\psi^\mu(0,\tau)=-\tilde{\psi}^\mu(0,\tau)$\index{Neveu-Schwarz}
& $\psi^\mu(\pi,\tau)=\tilde{\psi}^\mu(\pi,\tau)$~.
\end{tabular}
\end{center}
It is useful to unify these boundary condition in the equations
\begin{equation}
\psi^\mu(z+2\pi)\ =\ \de^{2\pi\di\nu}\psi^\mu(z)\eand
\tilde{\psi}^\mu(\bz+2\pi)\ =\ \de^{-2\pi\di\tilde{\nu}}\tilde{\psi}^\mu(\bz)~,
\end{equation}
over the complex plane, with
$\nu,\tilde{\nu}\in\{0,\frac{1}{2}\}$.

\remark{By the doubling trick, we can combine $\psi$ and
$\tilde{\psi}$ in the following way:
\begin{equation}
\psi^\mu(\sigma,\tau)\ :=\ \left\{\begin{array}{ll}\psi^\mu(\sigma,\tau)
& 0\ <\ \sigma\ <\ \pi
\\ \tilde{\psi}^\mu(2\pi-\sigma,\tau) &
\pi\ <\ \sigma\ <\ 2\pi\end{array}\right.
\end{equation}}

\paragraph{Superconformal symmetry.} Recall that in the bosonic
case, the Virasoro generators appeared as Laurent coefficients in
the expansion of the energy-momentum tensor, which in turn is the\index{energy-momentum tensor}
Noether current for conformal transformations. For superconformal
transformations, we have the additional supercurrents
\begin{equation}
T_F(z)\ =\ \di\sqrt{\frac{2}{\alpha'}}\psi^\mu(z)\dpar X_\mu(z)\eand
T_F(\bz)\ =\ \di\sqrt{\frac{2}{\alpha'}}\tilde{\psi}^\mu(\bz)\dparb
X_\mu(\bz)~.
\end{equation}
Their Laurent expansions are given by
\begin{equation}
T_F(z)\ =\ \sum_{r\in\RZ+\nu}\frac{G_r}{z^{r+3/2}}\eand
\tilde{T}_F(\bz)\ =\ \sum_{r\in\RZ+\tilde{\nu}}\frac{\tilde{G}_r}{\bz^{r+3/2}}~,
\end{equation}
where $\nu$ and $\tilde{\nu}$ label again the applied boundary
conditions. The Laurent coefficients complete the Virasoro algebra\index{Virasoro algebra}
to its $\CN=(1,0)$ and $\CN=(0,1)$ supersymmetric extension. The
former reads
\begin{equation}
\begin{aligned}
{}[L_m,L_n]&\ =\ (m-n)L_{m+n}+\frac{c}{12}(n^3-n)\delta_{m+n,0}~,\\
\{G_r,G_s\}&\ =\ 2L_{r+s}+\frac{c}{12}(4r^2-1)\delta_{r+s,0}~,\\
[L_m,G_r]&\ =\ \frac{m-2r}{2} G_{m+r}~.
\end{aligned}
\end{equation}
This algebra is also called the {\em Ramond algebra} for $r,s$\index{Ramond}
integer and the {\em Neveu-Schwarz algebra} for $r,s$\index{Neveu-Schwarz}
half-integer.

\paragraph{Critical dimension.} Since in each of the
holomorphic (right-moving) and antiholomorphic (left-moving)
sectors, each boson contributes 1 and each fermion contributes
$\frac{1}{2}$ to the central charge, the total central charge is\index{central charge}
$c=\frac{3}{2}D$. This central charge has to compensate the
contribution from the superconformal ghosts, which is $-26+11$,
and thus the critical dimension of $\CN=1$ superstring theory is\index{string theory}
$10$.

\paragraph{Preliminary open superstring spectrum.} The mode expansions
of the right- and left-moving fermionic fields read
\begin{equation}
\psi^\mu(z)\ =\ \sum_{r\in\RZ+\nu}\frac{\psi_r^\mu}{z^{r+1/2}}\eand
\tilde{\psi}^\mu(\bz)\ =\ \sum_{r\in\RZ+\tilde{\nu}}\frac{\tilde{\psi}_r^\mu}{\bz^{r+1/2}}~,
\end{equation}
and after canonical quantization, one arrives at the algebra\index{quantization}
\begin{equation}
\{\psi_r^\mu,\psi_s^\nu\}\ =\ \{\tilde{\psi}_r^\mu,\tilde{\psi}_s^\nu\}\ =\ \eta^{\mu\nu}\delta_{r+s,0}~.
\end{equation}
The normal ordering constant $a$ appearing in\index{normal ordering}
\eqref{physicalstatecond} is found to be $a=0$ in the Ramond\index{Ramond}
sector and $a=\frac{1}{2}$ in the Neveu-Schwarz sector. The\index{Neveu-Schwarz}
physical state conditions \eqref{physicalstatecond} are extended
by the demand
\begin{equation}
G_r|\mathrm{phys}\rangle\ =\ 0~~~\mbox{for}~~~r\ >\ 0~,
\end{equation}
and the level number $N$ is modified to
\begin{equation}
N=\sum_{n=1}^\infty
\alpha_{-n}\cdot\alpha_n+\sum_{r>0}r\psi_{-r}\cdot\psi_r~.
\end{equation}

Equipped with these results, we see that the NS ground state is
again tachyonic, and has mass $m^2=-\frac{1}{2\alpha'}$. The first\index{tachyon}
excited levels consist of the massless states
$\psi^\mu_{-\frac{1}{2}}|k;0\rangle_{\mathrm{NS}}$, and can again
be related to spacetime gauge potentials.

The R sector contains (massless) zero modes $\psi_0^\mu$
satisfying the ten-dimensional Dirac algebra
$\{\psi_0^\mu,\psi_0^\nu\}=\eta^{\mu\nu}$ and all states in the R
sector are spacetime fermions. Note that already in the ground
state, we cannot expect spacetime supersymmetry between the R and\index{super!symmetry}
NS sectors due to the strong difference in the number of states.

\paragraph{Preliminary closed string spectrum.} There\index{closed string}
are evidently four different pairings of the fermion boundary
conditions for closed strings giving rise to the four sectors of\index{closed string}
the closed superstring. Spacetime bosons are contained in the
NS-NS and the R-R sectors, while spacetime fermions are in the
NS-R and the R-NS sectors.

While the NS-NS ground state contains again a tachyon, the\index{tachyon}
remaining states in the first levels form all expected states of
supergravity.

\paragraph{GSO projection.} To obtain (local) supersymmetry in\index{GSO projection}\index{super!symmetry}
the target space of the theory, we have to apply the so-called\index{target space}
{\em Gliozzi-Scherck-Olive} (GSO) projection, which eliminates
certain states from the na\"ive superstring spectrum.

Explicitly, the GSO projection acts on the NS sector by keeping\index{GSO projection}
states with an odd number of $\psi$ excitations, while removing
all other states. This clearly eliminates the tachyonic NS-vacuum,\index{tachyon}
and the ground states become massless. More formally, one can
apply the projection operator
$P_\mathrm{GSO}=\frac{1}{2}(1-(-1)^F)$, where $F$ is the fermion
number operator.

In the R sector, we apply the same projection operator, but
replace $(-1)^F$ by
\begin{equation}\label{FermionNumber}
(-1)^F\ \rightarrow\ \pm\Gamma(-1)^F~,
\end{equation}
where $\Gamma$ is the ten-dimensional chirality operator
$\Gamma=\Gamma^0\ldots \Gamma^9$. This projection reduces the zero
modes in the R ground state to the appropriate number to match the
new massless NS ground states. This is an indication for the fact
that the GSO projection indeed restores spacetime supersymmetry.\index{GSO projection}\index{super!symmetry}

Note that the massless ground states of the theory are
characterized by their representation of the little group\index{representation}
$\sSO(8)$ of the Lorentz group $\sSO(1,9)$. The NS sector is just\index{Lorentz!group}
$\mathbf{8}_v$, while there the two possibilities for the R
sector, depending on the choice of GSO projection: $\mathbf{8}_s$\index{GSO projection}
and $\mathbf{8}_c$.

\paragraph{Green-Schwarz action.} For completeness sake, let us\index{Green-Schwarz action}
give the covariant Green-Schwarz action of the type IIB
superstring in Nambu-Goto form, which will be needed in the\index{Nambu-Goto}
definition of the IKKT model in section \ref{ssIKKTMM}. The action
reads as
\begin{equation}
\begin{aligned}\label{stringactionIIBNG}
S_{\mathrm{GS}}\ =\ -T\int
\dd^2\sigma~\Big(\sqrt{-\tfrac{1}{2}\Sigma}+ \di \eps^{ab}\dpar_a
X^\mu(\thetab^1\Gamma_\mu\dpar_b
\theta^1&+\thetab^2\Gamma_\mu\dpar_b\theta^2)\\&+\eps^{ab}\thetab^1\Gamma^\mu
\dpar_a\theta^1\thetab^2\Gamma_\mu\dpar_b\theta^2\Big)~,
\end{aligned}
\end{equation}
where $\theta^1$ and $\theta^2$ are Majorana-Weyl spinors in\index{Majorana-Weyl spinor}\index{Spinor}
ten-dimensions and
\begin{equation}
\Sigma^{\mu\nu}\ =\ \eps^{ab}\Pi_a^\mu\Pi_b^\nu~~~\mbox{with}~~~
\Pi_a^\mu\ =\ \dpar_a X^\mu-\di \thetab^1\Gamma^\mu\dpar_a\theta^1+\di
\thetab^2\Gamma^\mu\dpar_a\theta^2~.
\end{equation}

\subsection{Type IIA and type IIB string theories}\label{typetwosuperstrings}

Recall that we had two different possibilities of defining the
fermion number operator in equation \eqref{FermionNumber}. For open
strings, both choices are in principle equivalent but for closed\index{open string}
strings, the relative sign between left- and right-moving sectors
are important.

\paragraph{Type IIA.} In this case, we choose the opposite GSO
projections for the left- and the right-moving sectors. The\index{GSO projection}
resulting theory is therefore non-chiral, and the field content
can be classified under the little group $\sSO(8)$ as
$(\mathbf{8}_v\oplus\mathbf{8}_s)\otimes
(\mathbf{8}_v\oplus\mathbf{8}_c)$

\paragraph{Type IIB.} Here, we choose the same GSO projection on\index{GSO projection}
both sectors, which will lead to a chiral theory with field
content $(\mathbf{8}_v\oplus\mathbf{8}_s)\otimes
(\mathbf{8}_v\oplus\mathbf{8}_s)$.

\paragraph{The R-R sectors.} Constructing vertex operators for
the R-R sector states leads to antisymmetric tensors $G$ of even rank
$n$ for type IIA and odd rank $n$ for type IIB, satisfying Maxwell
equations. Thus, we get the following potentials $C$ with $G=\dd\index{Maxwell equations}
C$:
\begin{center}
\begin{tabular}{cl}
type IIA & ~~~$C_{(1)}~~C_{(3)}~~C_{(5)}~~C_{(7)}$\\[0.1cm] type
IIB & $C_{(0)}~~C_{(2)}~~C_{(4)}~~C_{(6)}~~C_{(8)}$
\end{tabular}
\end{center}
Each potential of rank $k$ has a Hodge dual of rank $8-k$ via
\begin{equation}
*\dd C_{(k)}\ =\ \dd \tilde{C}_{(8-k)}~,
\end{equation}
since the target space has dimension 10.\index{target space}

\paragraph{Compactification.}\label{pCompactification} To obtain a
phenomenologically relevant theory, one evidently has to reduce
the number of large dimensions from ten to four. Since the
geometry of spacetime is determined dynamically, one can imagine
that there are certain solutions to the string theory under\index{string theory}
consideration, which correspond to a compactification of  the
theory on a six-dimensional manifold. Particularly nice such\index{manifold}
manifolds besides the six-dimensional torus are Calabi-Yau\index{Calabi-Yau}
manifolds and compactifying a ten-dimensional string theory on\index{string theory}\index{manifold}
such a space often yields standard model like physics, with many
parameters as masses, coupling constants, numbers of quark and
lepton families determined by the explicit geometry of the chosen
Calabi-Yau threefold.\index{Calabi-Yau}

\subsection{T-duality for type II superstrings}\label{ssTduality}\index{T-duality}

In this section, let us briefly describe the symmetry called
T-duality in string theory. This symmetry has no analogue in field\index{T-duality}\index{string theory}
theory and is therefore truly stringy.

\paragraph{T-duality for closed strings.} Assume we quantize one\index{T-duality}\index{closed string}
of the nine spatial dimensions on a circle as $X^9\sim X^9+2\pi
R$, where $R$ is the radius of the circle. It follows that the
momentum along this direction is quantized $p_0^9=\frac{n}{R}$
with $n\in\RZ$. Recall the expansion of the string embedding
function
\begin{equation}\label{xexp}
X^\mu(\tau,\sigma)\ =\ x_0^\mu+\tilde{x}_0^\mu+\sqrt{\frac{\alpha'}{2}}
(\alpha_0^\mu+\tilde{\alpha}_0^\mu)\tau+\sqrt{\frac{\alpha'}{2}}
(\alpha_0^\mu-\tilde{\alpha}_0^\mu)\sigma+\ldots~,
\end{equation}
where the dots denote oscillator terms. Moreover, the center of
mass spacetime momentum reads as
\begin{equation}\label{p1}
p_0^\mu\ =\ \sqrt{\frac{1}{2\alpha'}}\left(\alpha^\mu_0+\tilde{\alpha}^\mu_0\right)~.
\end{equation}
With the quantization condition, we thus obtain\index{quantization}
$\alpha_0^9+\tilde{\alpha}^9_0=\frac{2n}{R}\sqrt{\frac{\alpha'}{2}}$.
The compactification also constrains the coordinate in the $X^9$
direction according to
\begin{equation}
X^9(\tau,\sigma+2\pi)\ =\ X^9(\tau,\sigma)+2\pi w R~,
\end{equation}
where the integer $w$ is the {\em winding number} and describes,\index{winding number}
how often the closed string is wound around the compactified\index{closed string}
direction. Together with the expansion \eqref{xexp}, we derive the
relation
\begin{equation}\label{p2}
\alpha_0^9-\tilde{\alpha}_0^9\ =\ wR\sqrt{\frac{2}{\alpha'}}~.
\end{equation}
Putting \eqref{p1} and \eqref{p2} together, we obtain furthermore
that
\begin{equation}
\alpha_0^9\ =\ \sqrt{\alpha'}{2}\left(\frac{n}{R}+\frac{wR}{\alpha'}\right)\eand
\tilde{\alpha}_0^9\ =\ \sqrt{\alpha'}{2}\left(\frac{n}{R}-\frac{wR}{\alpha'}\right),
\end{equation}
and the mass formula for the spectrum gets modified to
\begin{equation}
\begin{aligned}
m^2\ =\ &-p_\mu
p^\mu\ =\ \frac{2}{\alpha'}(\alpha_0^9)^2+\frac{4}{\alpha'}(N-1)
=\frac{2}{\alpha'}(\tilde{\alpha}_0^9)^2+\frac{4}{\alpha'}(\tilde{N}-1)~\\
\ =\ &\frac{n^2}{R^2}+\frac{w^2R^2}{\alpha'{}^2}+\frac{2}{\alpha'}(N+\tilde{N}-2)~.
\end{aligned}
\end{equation}
We see that there are essentially two towers of states in the
game: the tower of Kaluza-Klein momentum states and the tower of\index{Kaluza-Klein}
winding states. Noncompact states are obtained for $n=w=0$. In the
limit of large radius $R\rightarrow \infty$, the winding states
become very massive and thus disappear, while the momentum states
form a continuum. In the opposite limit $R\rightarrow 0$, the
momentum states decouple and the winding states become continuous.

Note that all the formul\ae{} are symmetric under the interchange
\begin{equation}
n\ \leftrightarrow\  w\eand R\ \leftrightarrow\  \frac{\alpha'}{R}~,
\end{equation}
and this symmetry is called {\em T-duality}. In terms of\index{T-duality}
zero-modes, this symmetry corresponds to
\begin{equation}
\alpha_0^9\ \leftrightarrow\  \alpha_0^9\eand
\tilde{\alpha}_0^9\ \leftrightarrow\  -\tilde{\alpha}_0^9~.
\end{equation}
Note that T-duality therefore corresponds to a parity\index{T-duality}\index{parity}
transformation of the right-movers.

\paragraph{T-duality for open strings.} As open strings cannot\index{T-duality}\index{open string}
wrap around the compact dimensions, they are dimensionally reduced
by T-duality in the limit $R\rightarrow 0$. Although the interior\index{T-duality}
of the open strings still vibrate in all ten dimensions, the\index{open string}
endpoints are restricted to a nine-dimensional subspace. This is
also seen by adding the mode expansion of the open string with\index{open string}
reversed parity of the right-movers, which causes the momentum in\index{parity}
the T-dualized direction to vanish. The nine-dimensional subspace
is naturally explained in the language of D-branes, see section\index{D-brane}
\ref{sDBranes}.

\paragraph{T-duality for type II superstrings.} We saw that\index{T-duality}
T-duality corresponds to a parity change of the right-movers. By\index{parity}
target space supersymmetry, it must therefore also change the\index{super!symmetry}\index{target space}
parity of the right-moving fermion fields. This inverts the choice\index{parity}
of sign in the GSO projection and eventually turns the GSO\index{GSO projection}
projection for type IIA theory into the GSO projection of type IIB
theory. T-dualizing any odd number of target space dimensions thus\index{target space}
maps the two different type II superstring theories into each
other, while T-dualizing an even number of dimensions does not
modify the superstring theory's type.\index{string theory}

\subsection{String field theory}\label{sssft}\index{string field theory}

\paragraph{Motivation.} String field theory (SFT) is an attempt to\index{string field theory}
describe string theory in a background independent manner. All the\index{string theory}
excitations of the string are encoded in an infinite number of
fields, which in turn are recombined in a single {\em string
field} $\CA$. After quantizing this field, we have -- roughly
speaking -- an operator $\hat{\CA}$ for every string in the target
space. There are different SFTs, which describe the dynamics of\index{target space}
the string field. In the following, we will only be interested in
the Chern-Simons-like version formulated by Witten
\cite{Witten:1986qs}.

Although SFT found several successful applications, there are also
conceptual drawbacks. First of all, the close strings are still
missing or at least hidden in Witten's successful formulation.
Second, and most importantly, it contradicts the principle derived
from M-theory that branes and strings should be equally\index{M-theory}
fundamental.

\paragraph{Cubic SFT.} Take a $\RZ$-graded algebra
$\frA$ with an associative product $\star$ and a derivative $Q$
with $Q^2=0$ and $\widetilde{Q \CA}=\tilde{\CA}+1$ for any
$\CA\in\frA$. Assume furthermore a map $\int:\frA\rightarrow \FC$
which gives non-vanishing results only for elements of grading $3$
and respects the grading, i.e.\ $\int \CA\star
\CB=(-1)^{\tilde{\CA}\tilde{\CB}}\int \CB\star \CA$. The (formal)
action of cubic SFT is then
\begin{equation}
S=\tfrac{1}{2}\int \left(\CA\star
Q\CA+\tfrac{2}{3}\CA\star\CA\star\CA\right)~.
\end{equation}
This action is invariant under the gauge transformations $\delta\index{gauge transformations}\index{gauge!transformation}
\CA=Q\eps-\eps\star \CA+\CA\star \eps$. It can furthermore be
easily extended to allow for Chan-Paton factors by replacing\index{Chan-Paton factors}
$\frA$ by $\frA\otimes \agl(n,\FC)$ and $\int$ by $\int\otimes
\tr$.

\paragraph{Physical interpretation.} The physical interpretation
of the above construction is the following: $\CA$ is a {\em string
field} encoding all possible excitations of an open string. The\index{open string}
operator $\star$ glues the halves of two open strings together,
forming a third one and the operator $\int$ folds an open string
and glues its two halves together \cite{Witten:1986qs}.\index{open string}

\subsection{The $\CN=2$ string}\label{ssN2string}

\paragraph{Introduction.} Besides the bosonic string theory having a\index{string theory}
26-dimensional target space (and some consistency problems due to\index{target space}
a tachyon in the spectrum) and the super string theory with\index{string theory}\index{tachyon}
$\CN=1$ worldsheet supersymmetry having a 10-dimensional target\index{super!symmetry}\index{worldsheet}
space, the $\CN=2$ string living naturally in 4 dimensions
received much attention as a toy model. In our consideration, this
string will essentially serve as a model for some D-brane\index{D-brane}
configurations arising in the context of twistor geometry. For\index{twistor}
more details see
\cite{Marcus:1992wi,Marcus:1992xt,Lechtenfeld:1999gd,Gluck:2003pa}
and references therein.

\paragraph{Action.} The action of the $\CN=2$ string is given by a
two-dimensional $\CN=2$ supergravity model with chiral matter
coupled to it. The $\CN=2$ supergravity multiplet here consists of
a zweibein $e^a_\alpha$, a complex gravitino $(\chi_\alpha,
\chi^*_\alpha)$ and a $\sU(1)$ gauge potential $A_\alpha$. The
chiral matter is captured by the components of a $\CN=2$ chiral
superfield $X^i\sim x^i+\theta\psi^i$, where $i=1,\ldots ,d$ and $d$
is the target space dimension. The corresponding action reads as\index{target space}
\begin{equation}
\begin{aligned}
S=\int
\dd^2z~\sqrt{\eta}\big(&\tfrac{1}{2}\eta^{\alpha\beta}\dpar_\alpha
x_i\dpar_\beta \bar{x}^i+\di
\bar{\psi}^i\slasha{D}\psi_i+A_\alpha\bar{\psi}^i\gamma^\alpha\psi_i+\\
&+(\dpar_\alpha
\bar{x}^i+\bar{\psi}^i\chi_\alpha)\chi_\beta\gamma^\alpha\gamma^\beta\psi_i+\mathrm{c.c.}\big)~.
\end{aligned}
\end{equation}

\paragraph{Critical dimension.} As usual, the critical dimension
is calculated by adding all the contributions of the necessary
ghosts systems. Here, we have again one $bc$-ghost system for
worldsheet reparameterizations, a complex $\beta\gamma$-system for\index{worldsheet}
the supersymmetry and a $b'c'$-system with weights $(1,0)$ for the\index{super!symmetry}
$\sU(1)$-symmetry. Together with the matter fields, we have
$c=-2+D$ in total, where $D$ is the complex dimension. Thus, the
$\CN=2$ string has critical dimension $4$.

\paragraph{Spectrum and symmetries.} In the
following we will always assume the metric on the target space\index{target space}\index{metric}
$\FR^4$ of the $\CN=2$ string to be either
Euclidean\footnote{Considering a Euclidean target space, however,\index{target space}
yields no propagating degrees of freedom.} or Kleinian, i.e.\
$\eta_{\mu\nu}=\diag(+1,+1,-1,-1)$. The underlying worldsheet\index{worldsheet}
theory \cite{Brink:1976vg} is $\CN=2$ supergravity coupled to two
$\CN=2$ massless chiral multiplets, the latter forming the\index{chiral!multiplet}
ordinary sigma model describing a string. The corresponding action\index{sigma model}
is  $\CN=2$ supersymmetric and Weyl invariant on the worldsheet.\index{worldsheet}
Furthermore,there is a global $\sU(1,1)$ target space symmetry.\index{target space}

\paragraph{Amplitudes.} Upon quantization, one finds a single\index{quantization}
massless open string state $|k\rangle$ in the spectrum, which can\index{open string}
be endowed with Chan-Paton factors. On the interaction side, the\index{Chan-Paton factors}
structure of amplitudes is rather simple. All $n$-point functions
with $n>3$ vanish identically for both open and closed strings.\index{closed string}
The lower amplitudes give rise to the effective field theory.

\paragraph{Effective field theory.} It has been shown in
\cite{Ooguri:1990ww} that the $\CN=2$ open string is equivalent to\index{open string}
self-dual Yang-Mills theory in 2+2 dimensions. It was also proven\index{Yang-Mills theory}\index{self-dual Yang-Mills theory}
there that the $\CN=2$ closed string is equivalent to self-dual\index{closed string}
supergravity. In \cite{Siegel:1992za}, it was argued that the
appropriate field theory is rather a fully supersymmetrized
version, and thus the $\CN=2$ critical string should correspond to
$\CN=4$ supersymmetric self-dual Yang-Mills theory. Note that the\index{Yang-Mills theory}\index{self-dual Yang-Mills theory}
D-branes of $\CN=2$ string theory will be discussed in\index{D-brane}\index{N=2 string theory@$\CN=2$ string theory}\index{string theory}
\ref{ssFurtherDBranes}, \ref{pN2DBranes}.

\section{Topological string theories}\index{topological!string}

Topological string theories are obtained from the physical\index{topological!string}
description of strings moving on Calabi-Yau manifolds after\index{Calabi-Yau}\index{manifold}
twisting the field content which turns the usual supersymmetric
sigma model into a topological field theory. They describe\index{sigma model}\index{topological!field theory}
subsectors of the physical string, which are under control and
suited for extensive study. We will be mostly interested in the
so-called topological B-model, as it nicely reduces to holomorphic\index{topological!B-model}
Chern-Simons theory.\index{Chern-Simons theory}

Besides the topological field theories which we will obtain in the
following by twisting the field content of a nonlinear sigma
model, there are further field theories giving rise to a\index{nonlinear sigma model}\index{sigma model}
topological string representation of the $\CN=(2,2)$\index{representation}\index{topological!string}
superconformal algebra: the Landau-Ginzburg model and the\index{Landau-Ginzburg model}\index{super!conformal algebra}
so-called minimal models.\index{minimal model}

\subsection{The nonlinear sigma model and its\index{nonlinear sigma model}\index{sigma model}
twists}\label{ssnonlinsigmamodel}

\paragraph{Sigma models.} A theory which contains a scalar field\index{sigma model}
$\phi$ mapping some spacetime to some (usually Riemannian)
manifold $X$ is called a {\em sigma model}. The sigma model is\index{sigma model}\index{manifold}
called {\em linear} if the target manifold $X$ is a linear space,\index{linear space}\index{manifold}
otherwise it is called {\em nonlinear}.

\paragraph{Nonlinear sigma model.} The most convenient starting\index{nonlinear sigma model}\index{sigma model}
point of discussing topological strings is the standard nonlinear\index{topological!string}
sigma model in two dimensions which describes maps $\Phi$ from a\index{sigma model}
Riemann surface $\Sigma$ to a target manifold $X$ with Riemannian\index{Riemann surface}\index{manifold}
metric $(g_{IJ})$ and Riemann tensor $(R_{IJKL})$. This model is\index{metric}
defined by the action
\begin{equation}\label{nlsaction}
\begin{aligned}
S\ =\ 2t\int_\Sigma\dd^2z~\left(\tfrac{1}{2}g_{IJ}(\phi)\dpar_z\phi^I\dpar_{\bar{z}}\phi^J
+ \tfrac{\di}{2}g_{IJ}(\phi)\psi_-^ID_z\psi_-^J + \right.
\\ \left.\tfrac{\di}{2}g_{IJ}\psi_+^ID_{\bar{z}}\psi_+^J +
\tfrac{1}{4}R_{IJKL}\psi_+^I\psi_+^J\psi_-^K\psi_-^L\right)~,
\end{aligned}
\end{equation}
where $D_z$ is the pullback of the Levi-Civita connection\index{Levi-Civita connection}\index{connection}
$\Gamma^I_{JK}$ on $TX$ and the $\phi^I$ are coordinates on $X$.
If we denote the canonical and anticanonical line
bundles\footnote{i.e.\ bundles of one-forms of type $(1,0)$ and
$(0,1)$, respectively} over $\Sigma$ by $K$ and $\bar{K}$, the
fermions are sections of the following bundles:
\begin{equation}
\psi_+^I\in\Gamma(K^{1/2}\otimes\Phi^*(TX))\eand\psi_-^I\in\Gamma(\bar{K}^{1/2}\otimes\Phi^*(TX))~.
\end{equation}
The supersymmetry transformations leaving \eqref{nlsaction}\index{super!symmetry}
invariant are given by
\begin{equation}
\begin{aligned}
\delta \phi^I&\ =\ \di\eps_-\psi_+^I+\di\eps_+\psi_-^I~,\\
\delta
\psi_+^I&\ =\ -\eps_-\dpar_z\phi^I-\di\eps_+\psi_-^K\Gamma^I_{KM}\psi_+^M~,\\
\delta\psi_-^I&\ =\ -\eps_+\dpar_\bz\phi^I-\di\eps_-\psi_+^K\Gamma^I_{KM}\psi_-^M~.
\end{aligned}
\end{equation}

\paragraph{$\CN=2$ supersymmetry.} If $X$ is K\"{a}hler, we gain\index{super!symmetry}
additional $\CN=2$ supersymmetry: The indices $I,J,K,\ldots $ split
into holomorphic and antiholomorphic parts: $i,\bi,\ldots $ and we
have the following field content:
\begin{equation}
\begin{aligned}
\phi^i&\in T^{1,0}X~,&\psi^i_+&\in K^{1/2}\otimes\Phi^*(T^{1,0}
X)~,&\psi^i_-&\in \bar{K}^{1/2}\otimes\Phi^*(T^{1,0} X)~,\\
\phi^\bi&\in T^{0,1}X~,&\psi^\bi_+&\in
K^{1/2}\otimes\Phi^*(T^{0,1} X)~,&\psi^\bi_-&\in
\bar{K}^{1/2}\otimes\Phi^*(T^{0,1} X)~
\end{aligned}
\end{equation}
together with the action
\begin{equation*}
S\ =\ 2t\int_\Sigma\dd^2z~\left(\tfrac{1}{2}g_{IJ}(\phi)\dpar_z\phi^I\dpar_{\bar{z}}\phi^J
+ \di\psi_-^\bi D_z\psi_-^ig_{\bi i} + \di\psi_+^\bi D_z\psi_+^i
g_{\bi i} + R_{i\bi
j\bj}\psi_+^i\psi_+^\bi\psi_-^j\psi_-^\bj\right)~.
\end{equation*}
The supersymmetry transformations under which this action is\index{super!symmetry}
invariant are given by
\begin{equation}
\begin{aligned}
\delta \phi^i&\ =\ \di\alpha_-\psi^i_++\di \alpha_+\psi^i_-~,&\delta
\phi^\bi&\ =\ \di\tilde{\alpha}_-\psi^\bi_++\di
\tilde{\alpha}_+\psi^\bi_-~,\\
\delta\psi^i_+&\ =\ -\tilde{\alpha}_-\dpar_z\phi^i-\di\alpha_+\psi^j_-\Gamma^i_{jm}\psi_+^m
~,&\delta\psi^\bi_+&\ =\ -\alpha_-\dpar_z\phi^\bi-\di\tilde{\alpha}_+\psi^\bj_-
\Gamma^\bi_{\bar{j}\bar{m}}\psi_+^{\bar{m}}~,
\\
\delta\psi^i_-&\ =\ -\tilde{\alpha}_+\dpar_\bz\phi^i-\di\alpha_-\psi^j_+\Gamma^i_{jm}\psi_-^m
~,&\delta\psi^\bi_-&\ =\ -\alpha_+\dpar_\bz\phi^\bi-\di\tilde{\alpha}_-\psi^\bj_+
\Gamma^\bi_{\bar{j}\bar{m}}\psi_-^{\bar{m}}~,
\end{aligned}
\end{equation}
where the infinitesimal fermionic parameters
$\alpha_+,\tilde{\alpha}_+$ and $\alpha_-,\tilde{\alpha}_-$ are
holomorphic sections of $\bar{K}^{1/2}$ and $K^{1/2}$,
respectively.

\paragraph{Twist of the nonlinear sigma model.} The nonlinear sigma\index{nonlinear sigma model}\index{sigma model}\index{twist}
model defined in the previous paragraph can now be twisted in two
possible ways resulting in the topological A- and B-model. On each
pair of spinors $(\psi_+^i,\psi_+^\bi)$, we can apply the\index{Spinor}
following twists:
\begin{center}
\begin{tabular}{rll}
untwisted & $\psi_+^i\in K^{1/2}\otimes\Phi^*(T^{1,0}X)$ &
$\psi_+^{\bi}\in
K^{1/2}\otimes\Phi^*(T^{0,1}X)$ \\[0.2cm]
$+$ twist& $\psi_+^i\in \Phi^*(T^{1,0}X)$ & $\psi_+^{\bi}\in
K\otimes\Phi^*(T^{0,1}X)$  \\[0.2cm]
$-$ twist& $\psi_+^i\in K\otimes\Phi^*(T^{1,0}X)$ &
$\psi_+^{\bi}\in \Phi^*(T^{0,1}X)$
\end{tabular}
\end{center}
Analogous twists can be applied on the pairs
($\psi_-^i,\psi_-^{\bi}$) with $K$ replaced by $\bar{K}$.

Equally well, one can consider this as a modification of the
underlying energy-momen\-tum tensor by
\begin{align}
T(z)\ \rightarrow\  T_{\mathrm{top}}(z)\ =\ T(z)\pm\tfrac{1}{2} \dpar
J(z)~,\\
\tilde{T}(\bar{z})\ \rightarrow\ 
\tilde{T}_{\mathrm{top}}(\bar{z})\ =\ \tilde{T}(\bar{z})\pm\tfrac{1}{2}
\dparb \tilde{J}(\bar{z})~.
\end{align}

Combining the twists on the $(\psi_+,\psi_-)$ sectors, we arrive
again at two possible total twists: the {\em A-twist} $(-,+)$ and\index{A-twist}
the {\em B-twist} $(-,-)$. Here, only the relative sign of the\index{B-twist}
twists in the two sectors matters, as other combinations are
obtained by complex conjugation. Half-twisted models have not
aroused much attention.

\subsection{The topological A-model}\index{topological!A-model}

We will not consider the topological A-model in detail but just\index{topological!A-model}
give a rough outline for completeness sake only.

\paragraph{Field content.} Due to the properties of the Gra{\ss}mann
coordinates interpreted as sections of different bundles over
$\Sigma$, we follow \cite{Witten:1991zz} and rename the fields
according to
\begin{equation}
\chi^i\ =\ \psi_+^i~,~~~\chi^\bi\ =\ \psi_-^\bi~,~~~\psi^\bi_z\ =\ \psi^\bi_+~,~~~
\psi^i_\bz\ =\ \psi^i_-~.
\end{equation}
The action thus reads in the new coordinates as
\begin{equation}
\begin{aligned}
S\ =\ 2t\int_\Sigma \dd^2 z~
&\left(\tfrac{1}{2}g_{IJ}\dpar_z\phi^I\dpar_\bz\phi^J+\di\psi^\bi_zD_\bz\chi^ig_{\bi
i}+\di \psi^i_zD_z\chi^\bi g_{\bi i}-R_{i\bi
j\bj}\psi^i_\bz\psi^\bi_z\chi^j\chi^\bj\right)~.
\end{aligned}
\end{equation}

\paragraph{Supersymmetry transformations.} The supersymmetry\index{super!symmetry}
transformations of the nonlinear sigma model become topological\index{nonlinear sigma model}\index{sigma model}
transformation laws after performing the A-twist. They are easily\index{A-twist}
derived by setting $\alpha_+=\tilde{\alpha}_-=0$ and by
introducing a BRST operator $Q$ arising from the topological
transformation laws as $\delta(\cdot)=-\di \alpha\{Q,\cdot\}$, one
can rewrite the action as
\begin{equation}\label{actionAmodeltop}
S=\di t\int_\Sigma\dd^2 z~\{Q,V\}+t\int_\Sigma \Phi^*(J)
\end{equation}
with
\begin{equation}
\begin{aligned}
V\ =\ g_{i\bj}(\psi^\bi_z\dpar_\bz \phi^j+\dpar_z \phi^\bi\psi^j_\bz)
\int_\Sigma\Phi^*(J)\ =\ \int_\Sigma\dd^2
z~\left(\dpar_z\phi^i\dpar_\bz\phi^\bj
g_{i\bj}-\dpar_\bz\phi^i\dpar_z\phi^\bj g_{i\bj}\right)~.
\end{aligned}
\end{equation}
The latter expression is the pull-back of the K\"{a}hler form $J=-\di\index{K\"{a}hler!form}
g_{i\bj}\dd z^i\dd z^\bj$ and its integral depends only on the
cohomology class of $J$ and the homotopy class of $\Phi$. In
general, one considers normalizations such that this integral
equals $2\pi n$, where $n$ is an integer called the {\em instanton\index{instanton}
number} or the {\em degree}.

Note that the reformulation done in \eqref{actionAmodeltop}
actually shows that the topological A-model is indeed a\index{topological!A-model}
topological field theory.\index{topological!field theory}

\paragraph{Observables.} Given an $n$-form $W$, one can map it to
a corresponding local operator $\CO_W$ by replacing $\dd z^i$ and
$\dd \bz^\bi$ in the basis of one-forms by $\chi^i$ and
$\chi^\bi$, respectively. Furthermore, it is
$\{Q,\CO_W\}=-\CO_{\dd W}$, where $\dd$ is the de-Rham
differential. Thus, there is a consistent map from the
BRST-cohomology of local operators to the de Rham cohomology, and
(when restricting to local operators) we can represent observables
by elements of the de Rham cohomology. There is an additional
``physical state condition'', which reduces the de Rham cohomology
to its degree $(1,1)$-subset. This subset corresponds to
deformations of the K\"{a}hler form and the topological A-model\index{K\"{a}hler!form}\index{topological!A-model}
therefore describes deformations of the K\"{a}hler moduli of its
target space.\index{target space}

\subsection{The topological B-model}\label{ssTopologicalBModel}\index{topological!B-model}

The topological B-model and its open string equivalent,\index{open string}\index{topological!B-model}
holomorphic Chern-Simons theory will concern us mostly in the\index{Chern-Simons theory}
later discussion, so let us be more explicit at this point.

\paragraph{Reformulation and supersymmetry.} We follow again\index{super!symmetry}
\cite{Witten:1991zz} and define the following new coordinates:
\begin{align}
\eta^\bi&\ =\ \psi^\bi_++\psi^\bi_-~,&\theta_i&\ =\ g_{i\bi}(\psi^\bi_+-\psi^\bi_-)~,
&\rho^i_z&\ =\ \psi^i_+~,&\rho^i_\bz&\ =\ \psi^i_-~,
\end{align}
where $\rho^i_z$ and $\rho^i_\bz$ are now one-forms with values in
$\Phi^*(T^{1,0}X)$ and $\Phi^*(T^{0,1}X)$, respectively. After
this redefinition, the action becomes
\begin{equation}
\begin{aligned}
S\ =\ t\int_\Sigma \dd^2 z~
&\Big(g_{IJ}\dpar_z\phi^I\dpar_\bz\phi^J+\di\eta^\bi(D_z\rho^i_\bz+D_\bz\rho^i_z)g_{i\bi}
+\\&\left.+\di\theta_i(D_\bz\rho_z^i-D_z\rho_\bz^i)+R_{i\bi
j\bj}\rho^i_z\rho^j_\bz\eta^\bi\theta_kg^{k\bj}\right)~.
\end{aligned}
\end{equation}

The supersymmetry transformations are reduced by $\alpha_\pm=0$\index{super!symmetry}
and $\tilde{\alpha}_\pm=\alpha$ to
\begin{equation}\label{eqSUSYred}
\delta \phi^i\ =\ 0~,~~~\delta
\phi^\bi\ =\ \di\alpha\eta^\bi~,~~~\delta\eta^\bi\ =\ \delta\theta_i\ =\ 0~,~~~\delta\rho^i\ =\ -\alpha\dd
\phi^i~.
\end{equation}
One can define a BRST operator from
$-\di\alpha\{Q,\cdot\}=\delta(\cdot)$ satisfying $Q^2=0$ modulo
equations of motion. With its help, one can write
\begin{equation}\label{eqBactionredf}
S=\di t\int\{Q,V\}+tW
\end{equation}
with
\begin{equation}
V\ =\ g_{i\bj}\left(\rho^i_z\dpar_\bz\phi^\bj+\rho^i_\bz\dpar_z\phi^\bj\right)\eand
W=\int_\Sigma\left(-\theta_iD\rho^i-\tfrac{\di}{2}R_{i\bi
j\bj}\rho^i\wedge \rho^j\eta^\bi \theta_k g^{k\bj}\right)~.
\end{equation}
Since one can show that the B-model is independent of the complex
structure on $\Sigma$ as well as the K\"{a}hler metric on $X$,\index{K\"{a}hler!metric}\index{complex!structure}\index{metric}
this model is a topological field theory. Furthermore, the theory\index{topological!field theory}
is mostly independent of $t\in \FR^+$, as the first term in the
action \eqref{eqBactionredf} changes by a term $\{Q,\cdot\}$ and
the second term can be readjusted by a redefinition of
$\theta\rightarrow \theta/t$. Thus the only dependence of
correlation functions on $t$ stems from $\theta$-dependence of the
observables. As this dependence can be clearly factored out, one
can perform all calculations in the large $t$-limit, and this
renders the B-model much simpler than the A-model: In this weak
coupling limit, one can simply expand around the bosonic minima of\index{weak coupling}
the action, which are constant maps $\Phi:\Sigma\rightarrow X$,
and thus the path integral becomes an ordinary integral over $X$.

\paragraph{Anomalies.} One can show that if $X$ is not a
Calabi-Yau manifold, the topological B-model is anomalous. This\index{Calabi-Yau}\index{topological!B-model}\index{manifold}
condition is stricter than for the A-model, and interestingly
reduces our target spaces to the mathematically most appealing\index{target space}
ones.

\paragraph{Ghost number.} The B-model has a $\RZ$-grading from a\index{ghost number}
quantum number called the {\em ghost number}. Putting
$\tilde{Q}=1$ and $\tilde{\phi}=0$, we obtain from the BRST
algebra \eqref{eqSUSYred} that $\tilde{\eta}=1$ and
$\tilde{\theta}=-1$. One can show that for a Calabi-Yau manifold\index{Calabi-Yau}\index{manifold}
$X$ of complex dimension $d$, a correlation function vanishes for
genus $g$, unless its total ghost number equals $2d(1-g)$.\index{ghost number}

\paragraph{Observables.} In the A-model, we could take the de Rham
cohomology as a model for our local operators. In the case of the
topological B-model, the situation is slightly more difficult. We\index{topological!B-model}
have to consider forms in the Dolbeault cohomology which take\index{Dolbeault cohomology}
values in the exterior algebra of the tangent bundle of $X$.
Consider an element $V$ of $\Lambda^q T^{1,0}X\otimes\Omega^{0,p}$
given by
\begin{equation}
V\ =\ V_{\bi_1\ldots \bi_p}{}^{j_1\ldots j_q}\dd \bz^{\bi_1}\wedge\ldots \wedge
\dd\bz^{\bi_p}\der{z^{j_1}}\wedge\ldots \wedge\der{z^{j_q}}~.
\end{equation}
We can again map $V$ to a local operator $\CO_V$ by replacing the
one-forms $\dd z^\bi$ by $\eta^{\bi}$ and the vector fields
$\der{z_j}$ by $\psi_j$. One then finds that
\begin{equation}
\{Q,\CO_V\}\ =\ -\CO_{\dparb V}
\end{equation}
and thus, we can consider the sheaf cohomology\index{sheaf}
$\oplus_{p,q}H^{0,p}(X,\Lambda^q T^{1,0}X)$ as the space of local
operators in the topological B-model. The BRST operator $Q$ is\index{topological!B-model}
naturally mapped to the Dolbeault operator $\dparb$. As in the
topological A-model, the Dolbeault cohomology is reduced to a\index{Dolbeault cohomology}\index{topological!A-model}
subset by a physical state condition: the group of Beltrami
differentials, introduced in \ref{ssDeformationCCM},\index{Beltrami differential}
\ref{pBeltrami}. Thus, it describes deformations of the complex
structure moduli of $X$.\index{complex!structure}

\paragraph{Correlation functions.} Given a set of points $x_\alpha$ on
$\Sigma$, the correlation function
\begin{equation}
\langle\prod_\alpha \CO_{\CV_\alpha}(x_\alpha)\rangle~,
\end{equation}
vanishes, unless the wedge product of all $\CV_\alpha$ is an
element of $H^{0,d}(X,\Lambda^d T^{1,0}X)$, where $d$ is the
dimension of the Calabi-Yau manifold: Any such element can be\index{Calabi-Yau}\index{manifold}
transformed into a top form by multiplying with the square of the
holomorphic volume form $\Omega^{d,0}$. This top-form is then\index{holomorphic!volume form}
integrated over, since, as stated above, the path integral reduces
to an integral over the Calabi-Yau manifold $X$ in the case of the\index{Calabi-Yau}\index{manifold}
topological B-model.\index{topological!B-model}

\paragraph{Comparison of the topological models.} The topological
A-model suffers some drawbacks compared to the topological\index{topological!A-model}
B-model: The moduli space of consistent maps from the worldsheet\index{moduli space}\index{worldsheet}
to the target space does not reduce as nicely as in the case of\index{target space}
the topological B-model and thus the calculation of the path\index{topological!B-model}
integral is considerably more difficult. This fact is also related
to the additional instanton corrections the partition function of\index{instanton}
the topological A-model receives.\index{topological!A-model}

However, the A-model is not as strictly restricted to having a
Calabi-Yau manifold as its target space as the topological\index{Calabi-Yau}\index{target space}\index{manifold}
B-model. Furthermore, physical quantities are more easily
interpreted in the framework of the A-twist. Therefore, one often\index{A-twist}
starts from the A-model and uses mirror symmetry, the T-duality on\index{T-duality}\index{mirror symmetry}
the level of topological sigma models, to switch to the B-model\index{sigma model}
and perform the calculations there. A mirror transformation of the
results leads then back to the A-model.

\subsection{Equivalence to holomorphic Chern-Simons
theory}\label{ssequivalencehCS}\index{Chern-Simons theory}

In this section, we will briefly describe the arguments for the
equivalence of the open topological B-model with a Calabi-Yau\index{Calabi-Yau}\index{topological!B-model}
threefold as target space and holomorphic Chern-Simons theory on\index{Chern-Simons theory}\index{target space}
the Calabi-Yau threefold as presented in \cite{Witten:1992fb}.\index{Calabi-Yau}

\paragraph{Argumentation via coupling.} We start from a
worldsheet $\Sigma$ which has a disjoint union of circles $C_i$ as\index{worldsheet}
its boundary $\dpar \Sigma$. For our strings to end within the
Calabi-Yau manifold $M$, we assume the target space to be filled\index{Calabi-Yau}\index{target space}\index{manifold}
with a stack of $n$ D5-branes, that is, we have a rank $n$ vector
bundle $E$ over $M$ with a gauge potential $A$, see section
\ref{ssTypeIIBranes}, \ref{pDRR}. Let us examine the consistency
condition for coupling the open topological B-model to the gauge\index{topological!B-model}
potential. This coupling is accomplished by adding the following
term to the Feynman path integral:
\begin{equation}
\int \CCD \Phi_i\exp(-S[\Phi_i])\cdot \Pi_i \tr P \exp
\oint_{C_i}\phi^*(\tilde{A})~,
\end{equation}
where $\phi:\Sigma\rightarrow M$ and
$\tilde{A}=\phi^*(A)-\di\eta^\bi F_{\bi j}\rho^j$ is the adjusted
gauge potential. Preservation of the BRST symmetry then demands
that $F^{0,2}:=\dpar A^{0,1}+A^{0,1}\wedge A^{0,1}=0$, where
$A^{0,1}$ is the (0,1)-part of the gauge potential $A$, see also
the discussion in the section \ref{sDBranes}. Thus, we can only
couple the topological B-model consistently to a gauge potential\index{topological!B-model}
if its (0,2)-part of the curvature vanishes. Via topological\index{curvature}
arguments, one can furthermore show that the only degrees of
freedom contained in the open topological B-model is the gauge\index{topological!B-model}
potential, and we can reduce this model to the action of
holomorphic Chern-Simons theory with the equations of motion\index{Chern-Simons theory}
$F^{0,2}=0$.

\paragraph{Argumentation via SFT.} Considering the open string\index{open string}
field theory presented in section \ref{sssft}, one can also show
that the open topological B-model reduces to the holomorphic\index{topological!B-model}
Chern-Simons action \cite{Witten:1992fb}.

\paragraph{Summary.} Altogether, we can state that the open
topological B-model describes the dynamics of holomorphic\index{topological!B-model}
structures $\dparb_A=\dparb+A^{0,1}$ on its target space. Note\index{target space}
that it is possible to extend the equivalence between the open
topological B-model and holomorphic Chern-Simons theory to the\index{Chern-Simons theory}\index{topological!B-model}
case of target spaces which are Calabi-Yau supermanifolds\index{Calabi-Yau}\index{Calabi-Yau supermanifold}\index{super!manifold}\index{target space}
\cite{Witten:2003nn}.

\subsection{Mirror symmetry}\label{ssMirror}\index{mirror symmetry}

\paragraph{Mirror symmetry and T-duality.} One of the most\index{T-duality}\index{mirror symmetry}
important symmetries in string theory is T-duality, which inverts\index{string theory}
the radius of a compactified dimension and thus exchanges winding
and momentum modes in the corresponding direction, see section
\ref{ssTduality}. This symmetry links e.g.\ type IIA and type IIB
superstring theories. On the level of the embedded topological
string theories, this symmetry might translate into the\index{topological!string}
conjectured mirror symmetry. The target space Calabi-Yau manifolds\index{Calabi-Yau}\index{mirror symmetry}\index{target space}\index{manifold}
would then come in mirror pairs, and mirror symmetry would
exchange K\"{a}hler and complex structure deformations. The complete\index{complex!structure}
statement of the mirror conjecture is that A-type topological
string theory with a Calabi-Yau manifold $M$ as a target space is\index{Calabi-Yau}\index{string theory}\index{target space}\index{topological!string}\index{manifold}
fully equivalent to B-type topological string theory with a\index{topological!string}
Calabi-Yau manifold $W$ as a target space, where $M$ and $W$ are\index{Calabi-Yau}\index{target space}\index{manifold}
mirror pairs, i.e.\ they have Hodge numbers with\index{Hodge number}
$h^{1,1}_M=h^{2,1}_W$ and $h^{2,1}_M=h^{1,1}_W$ in the
three-dimensional case. Such mirror pairs of Calabi-Yau manifolds\index{Calabi-Yau}\index{manifold}
are usually constructed via orbifolding varieties in complex and
weighted projective spaces or using toric geometry, see\index{weighted projective spaces} e.g.\
\cite{Greene:1996cy} for examples.

\paragraph{Mirror CFTs.} We mentioned above that the topological
A- and B-models are independent on the complex structure and\index{complex!structure}
K\"{a}hler moduli, respectively, and that this independence is due to
the $Q$-exactness of the moduli in the respective theories. In
this sense, the two models are complementary, and it is indeed
possible to consider not only the mirror symmetry of Calabi-Yau\index{Calabi-Yau}\index{mirror symmetry}
manifolds, but mirror symmetry of the whole field theories. A\index{manifold}
number of examples for such mirror pairs of conformal field theory\index{conformal field theory}
has indeed been found. Mirror symmetry has furthermore been\index{mirror symmetry}
extended from the set of CFTs defined via a nonlinear sigma model\index{nonlinear sigma model}\index{sigma model}
action having a Calabi-Yau manifold as a target space to more\index{Calabi-Yau}\index{target space}\index{manifold}
general models. Among those are nonlinear sigma models with\index{nonlinear sigma model}\index{sigma model}
non-compact or local Calabi-Yau manifolds as target space,\index{Calabi-Yau}\index{target space}\index{manifold}
Landau-Ginzburg models and minimal models. This extension has in\index{Landau-Ginzburg model}\index{minimal model}
fact been necessary since a direct calculation of a mirror theory
of a nonlinear sigma model can, e.g., yield a Landau-Ginzburg\index{nonlinear sigma model}\index{sigma model}
theory.

\paragraph{Consequences of mirror symmetry.} Mirror symmetry might\index{mirror symmetry}
be of vast importance in string theory. First of all, one expects\index{string theory}
it to give rise to a number of new string dualities, similarly to
the new dualities found with the help of T-duality. Second, it is\index{T-duality}
already a major calculatory tool within topological string theory.\index{string theory}\index{topological!string}
As we saw above, the B-model often allows for a mathematically
more tractable description, while the A-model is often more
closely related to physically interesting quantities. One could
thus imagine to work essentially in the A-model and switch via
mirror symmetry to the B-model, whenever a calculation is to be\index{mirror symmetry}
performed. Eventually, the results can then be retranslated to the
A-model.

Mirror symmetry even found applications in mathematics, when it\index{mirror symmetry}
was used for finding all the numbers $n_d$ of rational degree $d$
curves lying in the quintic embedded in $\CPP^4$\index{quintic}
\cite{Candelas:1990rm}. This result obtained by physicists was
preceded by more than a century of efforts by mathematicians.
Mirror symmetry related here the complicated problem of\index{mirror symmetry}
enumerative geometry to a much simpler problem in complex
geometry.

\section{D-Branes}\label{sDBranes}\index{D-brane}

Certainly one of the turning points in the development of string
theory was the discovery that besides the fundamental string,\index{string theory}
there are further objects, the so-called D-branes, which\index{D-brane}
unavoidably arise in string theories, and that these D-branes are
sources in the Ramond-Ramond sector with a nearly arbitrary\index{Ramond}
worldvolume dimension \cite{Polchinski:1995mt}. Roughly speaking,
a D-brane is a hypersurface on which open strings with Dirichlet\index{D-brane}\index{open string}
boundary condition can end, and which absorb the momentum flowing
off the endpoints of the string. Note that in our conventions, a
D$p$-brane will denote a D-brane with a worldvolume of dimensions\index{D-brane}
$(1,p)$ and $(a,b)$ with $a+b=p$ in $\CN=1$ and $\CN=2$ critical
superstring theories, respectively.

\subsection{Branes in type II superstring
theory}\label{ssTypeIIBranes}\index{string theory}

\paragraph{The NS-five brane.} As we saw before, the
NS-NS-sector contains an antisymmetric tensor of rank two which
has a Hodge dual $B_{(6)}$ by $*\dd B_{(2)}=\dd B_{(6)}$. This
potential couples naturally to the world volume of a
five-dimensional object, the NS-five brane:
\begin{equation}
S\ =\ Q_{\mathrm{NS5}}\int_{\mathcal{M}_6}B_{(6)}.
\end{equation}
The NS5-brane exists in both type IIA and type IIB superstring\index{NS5-brane}
theories.

\paragraph{D-branes in the R-R sector.}\label{pDRR} Generally\index{D-brane}
speaking, there are two different points of views for these
D-branes. First, one can understand a D$p$-brane as a\index{D-brane}
$p$-dimensional hyperplane on which open strings end. Second, a\index{open string}
D$p$-brane is a brane-like soliton of type IIA or type IIB\index{soliton}
supergravity in ten dimensions.

Recall from section \ref{sSST} that there are higher-form
potentials in the R-R sector of type II string theory. It is only\index{string theory}
natural to introduce sources to which these potentials can couple
electrically. This gives rise to hypersurfaces, the D$p$-branes,
with a $(p+1)$-dimensional worldvolume $\CM_{\mathrm{D}p}$ which
couple to the potentials $C_{(i)}$ via \cite{Polchinski:1995mt}
\begin{equation}\label{Dbranecoupling}
\mu_p\int_{\CM_{\mathrm{D}p}}C_{(p+1)}~,
\end{equation}
where $\mu_p$ is the corresponding charge. Since the higher-form
potentials are of even and odd rank in type IIA and type IIB
string theory, respectively, this construction yields D$0,2,4$ and\index{string theory}
$6$-branes in type IIA and D$(-1),1,3,5$ and $7$-branes in type
IIB string theory.\index{string theory}

A stack of $n$ such D$p$-branes naturally comes with a rank $n$
vector bundle $E$ over their $p+1$-dimensional worldvolume
together with a connection one-form $A$. This field arises from\index{connection}
the Chan-Paton factors attached as usually to the ends of an open\index{Chan-Paton factors}
string.

In the following, we will mostly discuss D-branes within type IIB\index{D-brane}
superstring theory.\index{string theory}

\paragraph{D-brane dynamics.} The action for a D9-brane is the\index{D-brane}
{\em Born-Infeld action}\index{Born-Infeld action}
\begin{equation}
S_{\mathrm{BI}}\ =\ \frac{1}{(4\pi^2\alpha')^5g_s}\int
\dd^{10}x~\sqrt{-\det(\eta_{\mu\nu}+T^{-1}F_{\mu\nu})}~,
\end{equation}
where $T=\frac{1}{2\pi\alpha'}$ is the string tension and
$F_{\mu\nu}$ the field strength of the gauge potential $A$ living\index{field strength}
on the D-brane's worldvolume. The actions for lower-dimensional\index{D-brane}
D-branes are obtained by dimensional reduction, which converts the\index{dimensional reduction}
gauge potential components $A^\mu$ in the reduced dimensions to
Higgs-fields $X^i$. Expanding the determinant and taking the field
theory limit $\alpha'\rightarrow 0$ in which all massive string
modes decouple, yields the ten-dimensional Yang-Mills equations
(or a dimensional reduction thereof).\index{dimensional reduction}

The theory describing the dynamics in the worldvolume is therefore
$\CN=1$ super Yang-Mills theory reduced from ten dimensions to the\index{Yang-Mills theory}
worldvolume of the D-brane. The resulting Higgs fields describe\index{D-brane}
the motion of the D-brane in the directions of the target space\index{target space}
normal to the worldvolume of the D-brane.

Note that on curved spaces, one often has to consider twisted
supersymmetry as linear realizations may not be compatible with\index{Twisted supersymmetry}\index{super!symmetry}
the geometry \cite{Bershadsky:1995qy}. One therefore uses (a
supersymmetric extension of) the Hermitian Yang-Mills
equations\footnote{Depending on their explicit shape, these
equations are also called generalized Hitchin equations,
Donaldson-Uhlenbeck-Yau equations and Hermite-Einstein equations.}
\begin{equation}\label{hym}
F^{0,2}\ =\ F^{2,0}\ =\ 0\eand k^{d-1}\wedge F\ =\ \gamma k^d~,
\end{equation}
which are also reduced appropriately from ten to $p+1$ dimensions,
see e.g. \cite{Iqbal:2003ds}. Here, $k$ is the K\"{a}hler form of the\index{K\"{a}hler!form}
target space and $\gamma$ is the slope of $E$, i.e.\ a constant\index{target space}
encoding information about the first Chern class of the vector\index{Chern class}\index{first Chern class}
bundle $E$. These equations imply the (dimensionally reduced,
supersymmetric) Yang-Mills equations.

\subsection{Branes within branes}

\paragraph{Instanton configurations.} From comparing the\index{instanton}
amplitude of a closed string being exchanged between two parallel\index{closed string}
D-branes to the equivalent one-loop open string vacuum amplitude,\index{D-brane}\index{open string}
one derives for the coupling $\mu_p$ in \eqref{Dbranecoupling}
that $\mu_p=(2\pi)^{-p}\alpha'{}^{-\frac{1}{2}(p+1)}$.
Furthermore, the anomalous coupling of gauge brane fields with
bulk fields have to satisfy certain conditions which restrict them
to be given by
\begin{equation}\label{anomalouscoupling}
\mu_p\int_{\CM_{\mathrm{D}p}}\sum_i i^*C_{(i)}\wedge \tr\de^{2\pi
\alpha'F+B}\wedge\sqrt{\hat{A}(4\pi^2\alpha'R)}~,
\end{equation}
where $i$ is the embedding of $\CM_{\mathrm{D}p}$ into spacetime
and $\hat{A}$ is the A-roof genus\footnote{If the normal bundle\index{normal bundle}
$\CN$ of $\CM_{\mathrm{D}p}$ in spacetime has non-vanishing
curvature $R_\CN$, we additionally have to divide by\index{curvature}
$\sqrt{\hat{A}(4\pi^2\alpha' R_\CN)}$.} (Dirac genus), which is
equivalent to the Todd class if $\CM$ is a Calabi-Yau manifold. By\index{Calabi-Yau}\index{manifold}
expanding the exponent and the Dirac genus in
\eqref{anomalouscoupling}, one picks up a term
\begin{equation}
\mu_p\frac{(2\pi\alpha')^2}{2}\int C_{p-3}\wedge \tr F\wedge F~,
\end{equation}
and thus one learns that instanton configurations on $E$ give rise\index{instanton}
to D$(p-4)$-branes, where each instanton carries exactly one unit
of D$(p-4)$-brane charge. Similarly, the first term from the
expansion of the Dirac genus gives rise to D$(p-4)$-branes when
$\CM_{\mathrm{D}p}$ is wrapped on a surface with non-vanishing
first Pontryagin class, e.g.\ on a K3 surface.

A bound state of a stack of D$p$-branes with a D($p$-4)-brane can
therefore be described in two possible ways. On the one hand, we
can look at this state from the perspective of the
higher-dimensional D$p$-brane. Here, we find that the D($p$-4)
brane is described by a gauge field strength $F$ on the bundle $E$\index{field strength}
over the worldvolume of the D$p$-brane with a nontrivial second
Chern character $ch_2(E)$. The instanton number (the number of\index{Chern character}\index{instanton}\index{instanton number}
D($p$-4) branes) is given by the corresponding second Chern class.\index{Chern class}
In particular, the bound state of a stack of BPS D$3$-branes with
a D(-1)-brane is given by a self-dual field strength $F=\ast F$ on\index{field strength}
$E$ with $-\frac{1}{8\pi^2}\int F\wedge F=1$. On the other hand,
one can adapt the point of view of the D($p$-4)-brane inside the
D$p$-brane and consider the dimensional reduction of the $\CN=1$\index{dimensional reduction}
super Yang-Mills equations from ten dimensions to the worldvolume
of the D($p$-4)-branes. To complete the picture, one has to add
strings with one end on the D$p$-brane and the other one on the
D($p$-4)-branes. Furthermore, one has to take into account that
the presence of the D$p$-brane will halve the number of
supersymmetries once more, usually to a chiral subsector. In the
case of the above example of D$3$- and D(-1)-branes, this will
give rise to the ADHM equations discussed in section
\ref{sSolutionGT}.

\paragraph{Monopole configurations.} Similarly, one obtains
monopole configurations \cite{Callan:1997kz}, but here the D-brane\index{D-brane}
configuration, consisting of a bound state of
D$p$-D($p$-2)-branes, is slightly more involved. One can again
discuss this configuration from both the perspectives of the D3-
and the D1-branes. From the perspective of the D1-branes, the
bound state is described by the Nahm equations presented in\index{Nahm equations}
section \ref{ssNahm}.

\subsection{Physical B-branes}

\paragraph{Boundary conditions.} As stated above, a D-brane in type\index{D-brane}
II string theory is a Ramond-Ramond charged BPS state. When\index{Ramond}\index{string theory}
compactifying this theory on Calabi-Yau manifolds, one has to\index{Calabi-Yau}\index{manifold}
consider boundary conditions corresponding to BPS states in the
appropriate $\CN=2$ superconformal field theory (SCFT), and there\index{conformal field theory}
are precisely two possibilities: the so-called A-type boundary
condition and the B-type boundary condition \cite{Ooguri:1996ck}.
Therefore, D-branes on Calabi-Yau manifolds come in two kinds:\index{Calabi-Yau}\index{D-brane}\index{manifold}
A-branes and B-branes. We will only be concerned with the latter
ones.

Recall that the $\CN=(2,2)$ superconformal algebra is generated by\index{super!conformal algebra}
a holomorphic set of currents $T(z),G^\pm(z),J(z)$ and an
antiholomorphic one
$\tilde{T}(\bar{z}),\tilde{G}^\pm(\bar{z}),\tilde{J}^\pm(\bar{z})$.
The B-type boundary condition is then given by
\begin{equation}
G^\pm(z)\ =\ \tilde{G}^\pm(\bar{z})\eand J(z)\ =\ \tilde{J}(\bar{z})~,
\end{equation}
see e.g.\ \cite{Diaconescu:2001ze}.

\paragraph{Dynamics of a stack of D-branes.} Consider now a stack\index{D-brane}
of $n$ D$p$-branes\footnote{Here, a D$p$-brane has $p+1$ real
dimensions in the Calabi-Yau manifold.} which are B-branes in a\index{Calabi-Yau}\index{manifold}
$d$-dimensional Calabi-Yau manifold $M$ with K\"{a}hler form $k$. As\index{K\"{a}hler!form}
the open strings living on a brane come with Chan-Paton degrees of\index{open string}
freedom, our D$p$-branes come with a vector bundle $E$ of rank $n$
and a gauge theory determining the connection $A$ on $E$. Let us\index{connection}
denote the field strength corresponding to $A$ by $F$. The\index{field strength}
dynamics of $A$ is then governed by the generalized Hitchin
equations \cite{Harvey:1996gc} (cf.\ equations \eqref{hym})
\begin{subequations}
\begin{align}\label{HEhCS}
F^{0,2}\ =\ &F^{2,0}\ =\ 0~,\\\label{HEmustable} k^{d-1}&\wedge F\ =\ \gamma
k^d~,\\\label{HEdimred} \dparb_A X_i\ =\ 0~~~&\mbox{and}~~~
[X_i,X_j]\ =\ 0~,
\end{align}
\end{subequations}
where $\gamma$ is again a constant determined by the magnetic flux
of the gauge bundle. The fields $X_i$ represent the normal motions
of the B-brane in $M$.

\paragraph{The six-dimensional case.} Consider now the case $p+1=d=6$.
Then we are left with equations \eqref{HEhCS} and
\eqref{HEmustable}, which can also be obtained from the instanton\index{instanton}
equations of a twisted maximally supersymmetric Yang-Mills theory,\index{Yang-Mills theory}\index{maximally supersymmetric Yang-Mills theory}\index{super!symmetric Yang-Mills theory}
reduced from ten to six dimensions
\cite{Bershadsky:1995qy,Iqbal:2003ds,Nekrasov:2004js}. It is not
clear whether there is any difference to the holomorphic
Chern-Simons theory obtained by Witten in \cite{Witten:1992fb} as\index{Chern-Simons theory}
argued in \cite{Nekrasov:2004js}: equations \eqref{HEhCS} are
obviously the correct equations of motion and \eqref{HEmustable}
combines with $\sU(N)$ gauge symmetry to a $\sGL(N,\FC)$ gauge
symmetry.

Lower-dimensional branes, as e.g.\ D2- and D0-branes correspond to
gauge configurations with nontrivial second and third Chern
classes, respectively, and thus they are instantons of this\index{Chern class}\index{instanton}
maximally supersymmetric Yang-Mills theory \cite{Nekrasov:2004js}.\index{Yang-Mills theory}\index{maximally supersymmetric Yang-Mills theory}\index{super!symmetric Yang-Mills theory}

\paragraph{Remark concerning topological A-branes.} For quite some
time, only special Lagrangian submanifolds were thought to give\index{special Lagrangian submanifold}
rise to a topological A-brane. For a Calabi-Yau threefold\index{Calabi-Yau}
compactification, this would imply that those branes always have a
worldvolume of real dimension three. However, Kapustin and Orlov
have shown \cite{Kapustin:2003qq} that it is necessary to extend
this set to coisotropic Lagrangian submanifolds, which allow for
further odd-dimensional topological A-branes.

\subsection{Topological B-branes}

\paragraph{Holomorphic submanifolds.} Recall that the complex
structure $I$ of the Calabi-Yau manifold does not interchange\index{Calabi-Yau}\index{complex!structure}\index{manifold}
normal and tangent directions of a boundary consistently defined
in the topological B-model. Therefore, a D-brane should wrap a\index{D-brane}\index{topological!B-model}
holomorphically embedded submanifold $\CC$ of the ambient
Calabi-Yau manifold $M$ and this restriction will preserve the\index{Calabi-Yau}\index{manifold}
topological symmetry of our model
\cite{Becker:1995kb,Ooguri:1996ck}. Thus, there are topological
B-branes with worldvolumes of dimension $0,2,4$ and $6$.

\paragraph{Chan-Paton degrees.} Furthermore, the open topological
strings ending on a stack of B-branes will also carry Chan-Paton\index{topological!string}
degrees of freedom, which in turn will lead to a complex vector
bundle $E$ over $\CC$. However, one is forced to impose a boundary\index{complex!vector bundle}
condition \cite{Witten:1992fb,Hori:2000ck}: the vanishing of the
variation of the action from the boundary term. This directly
implies that the curvature $F$ of $E$ is a 2-form of type (1,1)\index{curvature}
and in particular $F^{0,2}$ vanishes. Therefore, the underlying
gauge potential $A^{0,1}$ defines a holomorphic structure and $E$\index{holomorphic!structure}
becomes a holomorphic vector bundle. The gauge theory describing\index{holomorphic!vector bundle}
the D-brane dynamics is holomorphic Chern-Simons theory, as shown\index{Chern-Simons theory}\index{D-brane}
in \cite{Witten:1992fb}. Note that the equations of motion
\begin{equation}
F^{0,2}\ =\ 0~.
\end{equation}
differ from the one of their BPS-cousins only by the second
equation in \eqref{hym}. This equation is a (BPS) stability
condition on the vector bundle $E$.

\paragraph{Lower-dimensional B-branes.} B-branes, which do not
fill the complete Calabi-Yau manifold $M$ are described by\index{Calabi-Yau}\index{manifold}
dimensional reductions of hCS theory\index{dimensional reduction}
\cite{Vafa:2001qf,Neitzke:2004ni} and we have again additional
(Higgs) fields $X_i$, which are holomorphic sections of the normal
bundle of the worldvolume $\CC$ in $M$ with values in\index{normal bundle}
$\sGL(n,\FC)$ satisfying $[X_i,X_j]=0$. They describe fluctuations
of the B-branes in the normal directions. Explicitly, the
equations governing the fields present due to the B-branes read as
\begin{equation}\label{hitchineq}
F^{0,2}\ =\ F^{2,0}\ =\ 0~,~~~\dparb_A X_i\ =\ 0~,~~~[X_i,X_j]\ =\ 0~.
\end{equation}
These equations are a subset of the generalized Hitchin equations
\eqref{HEhCS}-\eqref{HEdimred}. The missing equation
\eqref{HEmustable} completes \eqref{HEhCS} to the Hermitian
Yang-Mills equations. According to a theorem by Donaldson,
Uhlenbeck and Yau (see e.g.\ \cite{Aspinwall:2004jr} and\index{Theorem!Yau}
references therein), the existence of a Hermitian Yang-Mills
connection is equivalent to $E$ being $\mu$-stable, which in turn\index{connection}
is equivalent to the BPS condition at large radius.

For the latter remark, recall that the actually appropriate
description of B-branes is the derived category of coherent\index{derived category}
sheaves, see e.g.\ \cite{Aspinwall:2004jr} and references therein.
A topological B-brane is simultaneously a physical B-brane if it
satisfies some stability condition which is equivalent to the
B-type BPS condition. Thus, we saw above that the physical
B-branes are a subsector of the topological B-branes.

\paragraph{Topological and physical D-branes.} As one can nicely\index{D-brane}
embed the topological open string into the physical open string\index{open string}
(and therefore physical D-branes into topological ones), we expect\index{D-brane}
that lower-dimensional topological branes, which are bound states
in a topological D5-brane should appear as gauge configurations in
six-dimensional twisted super Yang-Mills theory with nontrivial\index{Yang-Mills theory}
Chern classes. In particular, a D2-brane should correspond to an\index{Chern class}
instanton and thus to a nontrivial second Chern class\index{instanton}
\cite{Neitzke:2004ni}. The term in the partition function
capturing this kind of instantons is $\exp(-\int_M k\wedge ch_2)$,\index{instanton}
where $k$ is the K\"{a}hler form of the ambient Calabi-Yau\index{Calabi-Yau}\index{K\"{a}hler!form}
manifold\footnote{Note at this point that while the A-model and\index{manifold}
the B-model depend on the K\"{a}hler structure moduli and the
complex structure moduli, respectively, the r{\^o}le is interchanged\index{complex!structure}
for D-branes: Lagrangian submanifolds couple naturally to the\index{D-brane}
holomorphic 3-form of a Calabi-Yau, while holomorphic submanifolds\index{Calabi-Yau}
naturally couple to the K\"{a}hler form \cite{Nekrasov:2004js}, as\index{K\"{a}hler!form}
seen in this example.} $M$.

Note that we will completely ignore closed strings interacting\index{closed string}
with the B-branes. Their vertex operators would give rise to
deformations of the complex structure described by a Beltrami\index{complex!structure}
differential, cf.\ section \ref{ssDeformationCCM},
\ref{pBeltrami}. The theory governing these deformations is the
Kodaira-Spencer theory of gravity \cite{Bershadsky:1993cx}.\index{Kodaira-Spencer theory}

\subsection{Further aspects of D-branes}\label{ssFurtherDBranes}\index{D-brane}

\paragraph{Non-BPS branes.} There are essentially two reasons for
extending the analysis of D-branes to non-BPS \cite{Sen:1999mg}\index{D-brane}
ones: First, stable non-BPS branes are part of the spectrum and
lead to non-trivial but calculable results in different limits of
the string coupling. Second, they give rise to worldvolume gauge
theories with broken supersymmetry and might therefore play an\index{super!symmetry}
important r{\^o}le in string compactifications yielding
phenomenologically relevant models.

As an example \cite{Sen:1999mg}, let us consider a
D$2p$-$\bar{\mathrm{D}}2p$-pair of BPS branes in type IIA
superstring theory. This configuration is invariant under\index{string theory}
orbifolding with respect to $(-1)^{F_L}$, where $F_L$ is the
spacetime fermion number of the left-movers. The bulk, however,
will be described by type IIB superstring theory after this\index{string theory}
orbifolding. This operation projects out modes in the open string\index{open string}
spectrum which would correspond to separating the two D-branes.\index{D-brane}
Thus, we arrive at a single object, a non-BPS D$2p$-brane in type
IIB superstring theory. However, the tachyonic mode, which is\index{string theory}\index{tachyon}
present from the very beginning for a D-brane-anti-D-brane pair,\index{D-brane}\index{anti-D-brane}
is not projected out.

Although the non-BPS D-brane considered above was unstable due to\index{D-brane}\index{non-BPS D-brane}
the existence of a tachyonic mode, there are certain\index{tachyon}
orbifold/orientifold compactifications in which the tachyonic
modes are projected out and therefore the non-BPS D-brane becomes\index{D-brane}\index{non-BPS D-brane}
stable.

\paragraph{D-branes in $\CN=2$ string theory.}\label{pN2DBranes}\index{D-brane}\index{N=2 string theory@$\CN=2$ string theory}\index{string theory}
Considering D-branes in critical $\CN=2$ string theory is not as
natural as in ten-dimensional superstring theories since the NS
sector is connected to the R sector via the $\CN=2$ spectral flow,
and it is therefore sufficient to consider the purely NS part of
the $\CN=2$ string. Nevertheless, one can confine the endpoints of
the open strings in this theory to certain subspaces and impose\index{open string}
Dirichlet boundary conditions to obtain objects which we will call\index{Dirichlet boundary conditions}
D-branes in $\CN=2$ string theory. Although the meaning of these\index{D-brane}\index{N=2 string theory@$\CN=2$ string theory}\index{string theory}
objects has not yet been completely established, there seem to be
a number of safe statements we can recollect. First of all, the
effective field theory of these D-branes is four-dimensional\index{D-brane}
(supersymmetric) SDYM theory reduced to the appropriate
worldvolume \cite{Martinec:1997cw,Gluck:2003pa}. The
four-dimensional SDYM equations are nothing but the Hermitian
Yang-Mills equations:
\begin{equation}
F^{2,0}\ =\ F^{0,2}\ =\ 0\eand k\wedge F\ =\ 0~,
\end{equation}
where $k$ is again the K\"{a}hler form of the background. The\index{K\"{a}hler!form}
Higgs-fields arising in the reduction process describe again
fluctuations of the D-branes in their normal directions.\index{D-brane}

As is familiar from the topological models yielding hCS theory, we
can introduce A- and B-type boundary conditions for the D-branes\index{D-brane}
in $\CN=2$ critical string theory. For the target space\index{string theory}\index{target space}
$\FR^{2,2}$, the A-type boundary conditions are compatible with
D-branes of worldvolume dimension (0,0), (0,2), (2,0) and (2,2)\index{D-brane}
only \cite{Junemann:2001sp,Gluck:2003pa}.

\paragraph{Super D-branes.}\label{pSDbranes} There are three\index{D-brane}
approaches of embedding worldvolumes into target spaces when\index{target space}
Gra{\ss}mann directions are involved. First, one has the
Ramond-Neveu-Schwarz (RNS) formulation\index{Neveu-Schwarz}\index{Ramond}
\cite{Neveu:1971rx,Ramond:1971gb}, which maps a super worldvolume
to a bosonic target space. This approach only works for a spinning\index{target space}
particle and a spinning string; no spinning branes have been
constructed so far. However, this formulation allows for a
covariant quantization. Second, there is the Green-Schwarz (GS)\index{quantization}
formulation \cite{Green:1983sg}, in which a bosonic worldvolume is
mapped to a target space which is a supermanifold. In this\index{super!manifold}\index{target space}
approach, the well-known $\kappa$-symmetry appears as a local
worldvolume fermionic symmetry. Third, there is the
doubly-supersymmetric formulation (see \cite{Sorokin:1999jx} and
references therein), which unifies in some sense both the RNS and
GS approaches. In this formulation, an additional superembedding
condition is imposed, which reduces the worldvolume supersymmetry\index{super!symmetry}
to the $\kappa$-symmetry of the GS approach.

In the following, we will often work implicitly with the doubly
supersymmetric approach.

\paragraph{Geometric engineering.} It is easily possible to\index{geometric engineering}
engineer certain D-brane configurations, which, when put in\index{D-brane}
certain Calabi-Yau compactifications, give rise to a vast variety\index{Calabi-Yau}
of field theories in four dimensions
\cite{Katz:1996fh,Mayr:1998tx}. Most prominently, one realizes
$\CN=2$ supersymmetric gauge theories from compactifications of
type II string theories. In particular, objects arising in field
theory, as e.g.\ the Seiberg-Witten torus, are easily interpreted
within such a compactification scheme.

Let us consider a popular example, which was developed in
\cite{Brunner:1999jq} and studied e.g.\ in \cite{Kachru:2000ih}
and \cite{Dijkgraaf:2002fc}. We start from the algebraic variety\index{algebraic variety}
\begin{equation}
xy\ =\ z^2-t^{2n}
\end{equation}
in $\FC^4$. For $n=1$, this is just the resolved conifold\index{conifold}\index{resolved conifold}
$\CO(-1)\oplus\CO(-1)\rightarrow \CPP^1$ with a rigid $\CPP^1$ at\index{rigid}
its tip, see section \ref{ssConifold}. For $n>1$, the geometry
also contains a $\CPP^1$ with normal bundle $\CO(0)\oplus\CO(-2)$\index{normal bundle}
but the deformation of the sphere inside this bundle is obstructed
at $n$-th order, which can be described by a superpotential
$W(\phi)$, which is a polynomial of $n+1$th order, where $t^2\sim
\phi$. We have therefore the coordinates
\begin{equation}
\lambda_+\ =\ \frac{1}{\lambda_-}~,~~~z_+^1\ =\ z_-^1~,~~~\lambda_+z_+^2\ =\ \lambda_-z_-^2+W'(z^1_+)
\end{equation}
on the two patches $\CU_\pm$ covering the deformation of
$\CO(0)\oplus\CO(-2)$ together with the identification
\begin{equation}
z^1_+\ =\ z^1_-\ =\ t~,~~~z^2_+\ =\ \tfrac{1}{2}x~,~~~z^2_-\ =\ \tfrac{1}{2}y
\eand z\ =\ (2\lambda_+z_+^2-W'(z_+^1))~.
\end{equation}
This geometry has rigid $\CPP^1$s at the critical points of the\index{rigid}
superpotential $W(z_+^1)$.

Without the deformation by $W(z^1_+)$, wrapping $n$ D5-branes
around the $\CPP^1$ of $\CO(0)\oplus\CO(-2)$ yields an $\CN=2$
$\sU(n)$ gauge theory on the remaining four dimensions of the
D5-branes, which are taken to extend in ordinary spacetime. The
deformation by $W(z^1_+)$ breaks $\CN=2$ supersymmetry down to\index{super!symmetry}
$\CN=1$ with vacua at the critical points of the superpotential.
One can now distribute the D5-branes among the $i$ critical points
of $W(z^1_+)$, each corresponding to a rigid $\CPP^1$, and thus\index{rigid}
break the gauge group according to $\sU(n)\rightarrow
\sU(n_1)\times\ldots\times\sU(n_i)$, with $n_1+\ldots+n_i=n$.

\subsection{Twistor string theory}\label{ssTST}\index{string theory}\index{twistor}\index{twistor!string theory}

A good source for further information and deeper review material
about the developments in twistor string theory is\index{string theory}\index{twistor}\index{twistor!string theory}
\cite{TwistorWorkshop} and \cite{TwistorWorkshop2}. The paper in
which twistor string theory was considered for the first time is\index{string theory}\index{twistor!string theory}
\cite{Witten:2003nn}.

\paragraph{Motivation.} Even after half a century of intense
research, we still do not completely understand quantum
chromodynamics. The most prominent point is probably the
phenomenon called {\em confinement}, i.e.\ the fact that quarks\index{confinement}
are permanently confined inside a bound state as the coupling
constant becomes large at low energies. To answer this and more
questions, string-gauge theory dualities are important. The most
prominent example is certainly the AdS/CFT correspondence\index{AdS/CFT correspondence}
\cite{Maldacena:1997re}.

Witten's motivation for the construction of twistor string theory\index{string theory}\index{twistor}\index{twistor!string theory}
was originally to find an alternative description of the string
theory side in the AdS/CFT correspondence, which is suited for\index{AdS/CFT correspondence}\index{string theory}
describing the small gauge coupling limit. The existence of a
radically different such description is in fact plausible, as many
theories change drastically their shape when considered in a
certain regime or after dualities have been applied. One aspect
should, however, remain conserved: the symmetry group
$P\sSU(2,2|4)$ or $P\sSU(4|4)$ of the target space $AdS_5\times\index{target space}
S^5$. The most natural space with this symmetry group is probably
the supertwistor space\footnote{We will consider twistor spaces in\index{twistor}\index{twistor!space}
more detail in chapter \ref{chTwistorGeometry}.} $\CPP^{3|4}$.
Since this space is in fact a Calabi-Yau supermanifold, one can\index{Calabi-Yau}\index{Calabi-Yau supermanifold}\index{super!manifold}
study the topological B-model having this space as a target space.\index{target space}\index{topological!B-model}

\paragraph{Twistor string theory.} Consider the supertwistor space\index{string theory}\index{twistor}\index{twistor!space}\index{twistor!string theory}
$\CPP^{3|4}$ with a stack of $n$ almost space-filling D5-branes.
Here, ``almost space-filling'' means that the fermionic
coordinates extend in the holomorphic directions only, while all
the antiholomorphic directions are completely ignored. Twistor\index{twistor}
string theory is now simply the topological B-model with\index{string theory}\index{topological!B-model}
$\CPP^{3|4}$ as its target space and the above given D-brane\index{D-brane}\index{target space}
configuration. This model can be shown to be equivalent to
holomorphic Chern-Simons theory on $\CPP^{3|4}$ describing\index{Chern-Simons theory}
holomorphic structures on a rank $n$ complex vector bundle. The\index{complex!vector bundle}\index{holomorphic!structure}
power of twistor string theory in describing gauge theories arises\index{string theory}\index{twistor}\index{twistor!string theory}
from the twistor correspondence and the Penrose-Ward transform,\index{Penrose-Ward transform}\index{twistor!correspondence}
see chapter \ref{chTwistorGeometry}.

\paragraph{Further twistor string theories.} Further topological\index{twistor}
string theories with a supertwistor space as target space have\index{target space}\index{twistor!space}
been considered. First, following the proposal in
\cite{Witten:2003nn}, the superambitwistor space has been\index{ambitwistor space}\index{twistor}\index{twistor!ambitwistor}\index{twistor!space}
considered in \cite{Neitzke:2004ni} and \cite{Aganagic:2004yh} as
a target space for the topological B-model. In particular, a\index{target space}\index{topological!B-model}
mirror conjecture was established between the superambitwistor\index{twistor}\index{twistor!ambitwistor}
space and the supertwistor space previously discussed by Witten.\index{twistor!space}
In \cite{Chiou:2005jn}, the discussion was extended to the
mini-supertwistor space, which will probably be the mirror of the\index{mini-supertwistor space}\index{twistor}\index{twistor!space}
mini-superambitwistor space introduced in \cite{Saemann:2005ji}.\index{ambitwistor space}\index{mini-superambitwistor space}\index{twistor!ambitwistor}
All of these spaces and their r{\^o}le in twistor geometry will be
extensively discussed in chapter \ref{chTwistorGeometry}.

\chapter{Non-(anti)commutative Field
Theories}\label{chNACFieldTheories}

In this chapter, we will be concerned with noncommutative
deformations of spacetime and non-anticommutative deformations of
superspace. Both noncommutativity\footnote{Note that it has become\index{super!space}
common usage to call a space noncommutative, while gauge groups
with the analogous property are called non-Abelian.} and
non-anticommutativity naturally arise in type II string theories
put in a constant NS-NS $B$-field background \cite{Connes:1997cr}
and a constant R-R graviphoton background \cite{deBoer:2003dn},\index{graviphoton}
respectively. Therefore these deformations seem to be unavoidable
when studying string theory in nontrivial backgrounds. Moreover,\index{string theory}
they can provide us with interesting toy models which are
well-suited for studying features of string theories (as e.g.\
non-locality) which do not appear within ordinary field theories.

\section{Comments on noncommutative field theories}\index{noncommutative field theories}

Over the last decade, there has been an immense effort by string
theorists to improve our understanding of string dynamics in
nontrivial backgrounds. Most prominently, Seiberg and Witten
\cite{Seiberg:1999vs} discovered that superstring theory in a\index{string theory}
constant Kalb-Ramond 2-form background can be formulated in terms\index{Kalb-Ramond}\index{Ramond}
of field theories on noncommutative spacetimes upon taking the\index{noncommutative spacetime}
so-called Seiberg-Witten zero slope limit. Subsequently, these
noncommutative variants of ordinary field theories were intensely
studied, revealing many interesting new aspects, such as UV/IR
mixing \cite{Minwalla:1999px}, the vastness of nontrivial\index{UV/IR mixing}
classical solutions to the field equations\footnote{See also the
discussion in sections \ref{sSolutionGT} and
\ref{ssClassicalSolsMM}.} and the nonsingular nature of the
noncommutative instanton moduli spaces, see\index{instanton}\index{moduli space} e.g.\
\cite{Nekrasov:2000zz}. It turned out that as low energy effective
field theories, noncommutative field theories exhibit many\index{noncommutative field theories}
manifestations of stringy features descending from the underlying
string theory. Therefore, these theories have proven to be an\index{string theory}
ideal toy model for studying string theoretic questions which
otherwise remain intractable, as e.g.\ tachyon condensation\index{tachyon}\index{tachyon condensation}
\cite{Sen:1998rg,David:2000um,Aganagic:2000mh,Kraus:2000wx,
Kajiura:2001pq} and further dynamical aspects of strings
\cite{Gross:2000ph} (for recent work, see e.g.\
\cite{Wimmer:2005bz, Popov:2005ik}). Noncommutativity has also
been used as a means to turn a field theory into a matrix model\index{matrix model}
\cite{Lechtenfeld:2005xi}. The results of this publication are
presented in section \ref{sTwistorMM}.

\subsection{Noncommutative deformations}

\paragraph{Deformation of the coordinate algebra.} In ordinary
quantum mechanics, the coordinate algebra on the phase space
$\FR^3\times \FR^3$ is deformed to the Heisenberg algebra\index{Heisenberg algebra}
\begin{equation}
[\hat{x}^i,\hat{p}_j]\ =\ \di \hbar
\delta^i_j~,~~~[\hat{x}^i,\hat{x}^j]\ =\ [\hat{p}_i,\hat{p}_j]\ =\ 0~.
\end{equation}
As these relations have not been verified to very low distances
(i.e.\ very high energies), a natural (relativistic)
generalization would look like
\begin{equation}\label{noncommdef}
[\hat{x}^\mu,\hat{x}^\nu]\ =\ \di \theta^{\mu\nu}~,
\end{equation}
where $\theta^{\mu\nu}$ is a constant of dimension $[L]^2$.
Clearly, by such a deformation, the Poincar{\'e} group is broken down
to the stabilizer subgroup of the deformation tensor $\theta$. The\index{stabilizer subgroup}
deformation of the space $\FR^4$ with coordinates satisfying the
algebra \eqref{noncommdef} will be denoted by $\FR^4_\theta$ and
called {\em noncommutative spacetime}.\index{noncommutative spacetime}

The first discussion of noncommutative spaces in a solid
mathematical framework has been presented by Alain Connes
\cite{Connes:1980ji}. Since then, noncommutative geometry has been
used in various areas of theoretical physics as e.g.\ in the
description of the quantum Hall effect in condensed matter physics
and in particular in string theory.\index{string theory}

For a review containing a rather formal introduction to
noncommutative geometry, see \cite{Varilly:1997qg}. Further useful
review papers are \cite{Douglas:2001ba} and \cite{Szabo:2001kg}.

\paragraph{Noncommutativity from string theory.} In 1997,\index{string theory}
noncommutative geometry was shown to arise in certain limits of
M-theory and string theory on tori\index{M-theory}\index{string theory}
\cite{Connes:1997cr,Douglas:1997fm}; several further appearances
have been discovered thereafter. Let us here briefly recall the
analysis of \cite{Seiberg:1999vs}.

Consider open strings in flat space and in the background of a\index{open string}
constant Neveu-Schwarz $B$-field on a D-brane with the action\index{D-brane}\index{Neveu-Schwarz}
\begin{equation}
S=\tfrac{1}{4\pi\alpha'}\int_\Sigma g_{MN}\dpar_\alpha
X^M\dpar^\alpha X^N-\tfrac{\di}{2}\int_{\dpar \Sigma}
B_{MN}X^M\dpar_T X^N~,
\end{equation}
where $\Sigma$ denotes the string worldsheet and $\dpar_T$ is a\index{worldsheet}
derivative tangential to the boundary $\dpar\Sigma$. We assume the
latter subspace to be mapped to the worldvolume of the D-brane.\index{D-brane}
For our purposes, it is enough to restrict ourselves to the case
where $\Sigma$ is the disc and map it conformally to the upper
half plane $H=\{z\in\FC\,|\,\mathrm{Im}(z)>0\}$. The resulting
equations of motion read
\begin{equation}
\left.g_{MN}(\dpar_z-\dparb_\bz)X^N+2\pi\alpha'B_{MN}(\dpar_z+\dparb_\bz)X^N\right|_{z=\bz}\ =\ 0~,
\end{equation}
from which we can calculate a propagator $\langle
X^M(z)X^N(z')\rangle$. At the boundary $\dpar\Sigma=\FR\subset\FC$
of $\Sigma$, where the open string vertex operators live we are\index{open string}
interested in, this propagator reads
\begin{equation}\label{propagator}
\begin{aligned}
\langle
X^M(\tau)X^N(\tau')\rangle&\ =\ -\alpha'G^{MN}\log(\tau-\tau')^2+\tfrac{\di}{2}\theta^{MN}\mathrm{sign}(\tau-\tau')\\&~~~\mbox{with}
~~~\theta^{MN}\ =\ 2\pi\alpha'\left(\frac{1}{g+2\pi\alpha'
B}\right)^{[MN]}~,
\end{aligned}
\end{equation}
where $\tau\in\FR$ parameterizes the boundary. Recall now that one
can calculate commutators of operators from looking at the short\index{commutators}
distance limit of operator products. From \eqref{propagator}, we
get
\begin{equation}
{}[X^M(\tau),X^N(\tau)]\ =\ T\left(X^M(\tau)X^N(\tau^-)-X^M(\tau)X^N(\tau^+)\right)\ =\ \di
\theta^{MN}~.
\end{equation}
Thus, the target space in our string configuration indeed proves\index{target space}
to carry a noncommutative coordinate algebra. Note, however, that
to be accurate, one has to carefully consider a zero slope limit
$\alpha'\rightarrow 0$ to decouple more complicated string
effects.

\paragraph{Two-oscillator Fock space.} In the following, let us
restrict ourselves to four dimensions and consider a self-dual
($\kappa=1$) or an anti-self-dual $(\kappa=-1)$ deformation tensor\index{anti-self-dual}
$\theta^{\mu\nu}$, which has components
\begin{equation}\label{thetacomponents}
\theta^{12}\ =\ -\theta^{21}\ =\ \kappa\theta^{34}\ =\
-\kappa\theta^{43}\ =\ \theta\ >\ 0~.
\end{equation}
After introducing the annihilation operators\footnote{The constant
$\eps=\pm1$ here distinguishes between a metric of Kleinian\index{metric}
signature $(2,2)$ for $\eps=+1$ and a metric of Euclidean
signature $(4,0)$ for $\eps=-1$ on $\FR^4$.}
\begin{equation}
a_1\ =\ x^1-\di \eps x^2\eand a_2\ =\ -\kappa\eps x^3+\di\eps x^4~,
\end{equation}
we find the appropriate representation space of the algebra\index{representation}
\eqref{noncommdef} to be the two-oscillator Fock space
$\CH=\mathrm{span}\{|n_1,n_2\rangle|n_1,n_2\in\NN\}$ with
\begin{equation}
|n_1,n_2\rangle\ =\ \frac{1}{\sqrt{n_1!n_2!}}\left(\hat{a}_1^\dagger\right)^{n_1}
\left(\hat{a}_2^\dagger\right)^{n_2}|0,0\rangle~.
\end{equation}
One can therefore picture functions on $\FR^4_\theta$ as (the
tensor product of two) infinite-dimen\-sional matrices
representing operators on $\CH$.

\paragraph{Derivatives and integrals.} The derivatives on noncommutative
spacetimes are given by inner derivations of the Heisenberg\index{noncommutative spacetime}
algebra \eqref{noncommdef}. We can define
\begin{equation}
\dpar_\mu\ \rightarrow\  \hat{\dpar}_\mu\ :=\ -\di
\theta_{\mu\nu}[\hat{x}^\nu,\cdot]~,
\end{equation}
where $\theta_{\mu\nu}$ is the inverse of $\theta^{\mu\nu}$. This
definition yields $\hat{\dpar}_\mu \hat{x}^\nu=\delta_\mu^\nu$,
analogously to the undeformed case. Furthermore, due to the
commutator in the action, the Leibniz rule holds as usual.

Integration is correspondingly defined by taking the trace over
the Fock space $\CH$ representing the noncommutative space
\begin{equation}
\int \dd^4 x \ \rightarrow\  (2\pi\theta)^2\tr_\CH~.
\end{equation}
The analogue to the fact that the integral over a total derivative
vanishes is here the vanishing of the trace of a commutator.
Equally well as the former does not hold for arbitrary functions,
the latter does not hold for arbitrary operators
\cite{Douglas:2001ba}.

\paragraph{Moyal-Weyl correspondence.} The Moyal-Weyl\index{Moyal-Weyl correspondence}
correspondence maps the operator formalism of noncommutative\index{operator formalism}
geometry to the star-product formalism,\index{star-product formalism} i.e.\
\begin{equation}
(\hat{f}(\hat{x}),\cdot)\longleftrightarrow(f(x),\star)~.
\end{equation}
This map can be performed by a double Fourier transform using the
formul\ae{}
\begin{equation}
\hat{f}(\hat{x})\ =\ \int \dd \alpha~
\de^{\di\alpha\hat{x}}\phi(\alpha)\eand \phi(\alpha)\ =\ \int\dd x~
\de^{-\di \alpha x}f(x)~.
\end{equation}
Consistency then requires the star product to be defined according
to
\begin{equation}
(f\star g)(x)\ :=\ f(x)
\exp\left(\tfrac{\di}{2}\overleftarrow{\dpar_\mu}
\theta^{\mu\nu}\overrightarrow{\dpar_\nu}\right)g(x)~,
\end{equation}
and the noncommutative deformation of spacetime is then written as
$[x^\mu\stackrel{\star}{,}x^\nu]=\di\theta^{\mu\nu}$. Note
furthermore that the star product is associative: $(f\star g)\star
h=f\star(g\star h)$ and behaves as one would expect under complex
conjugation: $(f\star g)^*=g^*\star f^*$. Under the integral, we
have the identities
\begin{equation}
\int \dd^4 x~(f\star g)(x)\ =\ \int \dd^4 x~(g\star f)(x)\ =\ \int \dd^4
x~ (f\cdot g)(x)~.
\end{equation}

\subsection{Features of noncommutative field
theories}\label{ssNCfeatures}\index{noncommutative field theories}

\paragraph{Noncommutative gauge theories.} As found above, a\index{noncommutative gauge theories}
derivative is mapped to a commutator on noncommutative spaces. We
can extend the arising commutator by a gauge potential
$\hat{A}_\mu$, which is a Lie algebra valued function on the\index{Lie algebra}
noncommutative space. We thus arrive at
\begin{equation}
\nabla_\mu \ \rightarrow\  [\hat{X}_\mu,\cdot]\ewith
\hat{X}_\mu\ =\ -\di\theta_{\mu\nu}\hat{x}^\nu\otimes
\unit_\CG+\hat{A}_\mu~,
\end{equation}
where $\unit_\CG$ is the unit of the gauge group $\CG$
corresponding to the Lie algebra under consideration. The field\index{Lie algebra}
strength is then given by
\begin{equation}
\hat{F}_{\mu\nu}\ =\ [\hat{X}_\mu,\hat{X}_\nu]+\di
\theta_{\mu\nu}\otimes \unit_\CG~,
\end{equation}
where the last term compensates the noncommutativity of the bare
derivatives. The Yang-Mills action becomes
\begin{equation}
S=\tr_\CH\otimes
\tr_\CG\left([\hat{X}_\mu,\hat{X}_\nu]+\di\theta_{\mu\nu}\otimes
\unit_\CG\right)^2~,
\end{equation}
which is the action of a matrix model with infinite-dimensional\index{matrix model}
matrices.

\paragraph{Gauge transformations.} The action of gauge transformations\index{gauge transformations}\index{gauge!transformation}
$\hat{g}$ is found by trivially translating their action from the
commutative case, i.e.\
\begin{equation}
\hat{A}_\mu\ \rightarrow\  \hat{g}^{-1} \hat{A}_\mu
\hat{g}+\hat{g}^{-1}\hat{\dpar}_\mu \hat{g}~.
\end{equation}
Let us now switch to the star product formalism and consider the
noncommutative analogue to infinitesimal Abelian gauge
transformations $\delta A_\mu=\dpar_\mu \lambda$, which reads\index{gauge transformations}\index{gauge!transformation}
\begin{equation}
\delta A_\mu\ =\ \dpar_\mu\lambda+\lambda\star
A_\mu-A_\mu\star\lambda~.
\end{equation}
We thus see that even in the case of an Abelian gauge group, the
group of gauge transformations is a non-Abelian one.\index{gauge transformations}\index{gauge!transformation}

It is important to stress that in noncommutative Yang-Mills
theory, not all gauge groups are admissible. This is due to the\index{Yang-Mills theory}
fact that the corresponding Lie algebras may no longer close under\index{Lie algebra}
star multiplication. As an example, consider the gauge group
$\sSU(2)$: The commutator $[x^\mu
\di\sigma^3\stackrel{\star}{,}x^\nu
\di\sigma^3]=-\di\theta^{\mu\nu}\unit_2$ is not an element of
$\asu(2)$.

\paragraph{Seiberg-Witten map.}\label{pSWmap} The last\index{Seiberg-Witten map}
observations seem intuitively to forbid the following statement:
There is a map, called {\em Seiberg-Witten map}\index{Seiberg-Witten map}
\cite{Seiberg:1999vs}, which links gauge equivalent configurations
in a commutative gauge theory to gauge equivalent configurations
in its noncommutative deformation, thus rendering both theories
equivalent via field redefinitions. The idea is to regularize the
low-energy effective theory of open strings in a $B$-field\index{open string}
background in two different ways, once using Pauli-Villars and
once with the point-splitting procedure. In the former case, we
obtain the ordinary Born-Infeld action yielding commutative\index{Born-Infeld action}
Yang-Mills theory as the effective theory. In the latter case,\index{Yang-Mills theory}
however, we obtain a noncommutative variant of the Born-Infeld
action, which gives rise to a noncommutative gauge theory. Since\index{Born-Infeld action}
the effective action should be independent of the regularization
process, both theories should be equivalent and connected via a
Seiberg-Witten map.\index{Seiberg-Witten map}

Consistency conditions imposed by the existence of a
Seiberg-Witten map like\index{Seiberg-Witten map}
\begin{equation}\label{eqSWmap}
\hat{A}(A+\delta_\lambda
A)\ =\ \hat{A}(A)+\delta_{\hat{\Lambda}}\hat{A}(A)~,
\end{equation}
where $\lambda$ and $\hat{\Lambda}$ describe infinitesimal
commutative and noncommutative gauge transformations,\index{gauge transformations}\index{gauge!transformation}
respectively, prove to be a helpful calculatory tool. We will make
use of a similar condition in a non-anticommutative deformed
situation in section \ref{ssNACSYM}, \ref{pNACSWmap}.

\paragraph{UV/IR mixing.} One of the hopes for noncommutative\index{UV/IR mixing}
field theories was that the divergencies which are ubiquitous in
ordinary quantum field theory would be tamed by the
noncommutativity of spacetime, since the latter implies
non-locality, which does the job in string theory. The situation,\index{string theory}
however, is even worse: besides some infinities inherited from the
commutative theory, certain ultraviolet singularities get mapped
to peculiar infrared singularities, even in massive scalar
theories. This phenomenon is known under the name or {\em UV/IR
mixing} and was first studied in \cite{Minwalla:1999px} and\index{UV/IR mixing}
\cite{VanRaamsdonk:2000rr}.

\paragraph{Noncommutative instantons.} Instantons in\index{instanton}
noncommutative gauge theories have some peculiar properties. First\index{noncommutative gauge theories}
of all, it is possible to have non-trivial instantons even for\index{instanton}
gauge group $\sU(1)$ as discussed in \cite{Nekrasov:1998ss}. This
is due to the above presented fact that even for an Abelian gauge
group, the group of gauge transformations is non-Abelian. In\index{gauge transformations}\index{gauge!transformation}
\cite{Nekrasov:1998ss}, it has moreover been shown that a suitable
deformation can resolve the singularities in the instanton moduli\index{instanton}
space.

\section{Non-anticommutative field theories}\index{non-anticommutative field theories}

Expanding essentially on the analysis of \cite{Ooguri:2003qp},
Seiberg \cite{Seiberg:2003yz} showed\footnote{For earlier work in
this area, see \cite{Schwarz:1982pf,Ferrara:2000mm,
Klemm:2001yu,deBoer:2003dn}.} that there is a deformation of
Euclidean $\CN=1$ superspace in four dimensions which leads to a\index{super!space}
consistent supersymmetric field theory with half of the
supersymmetries broken. The idea was to deform the algebra of the
anticommuting coordinates $\theta$ to the Clifford algebra\index{Clifford algebra}
\begin{equation}
\{\hat{\theta}^{A},\hat{\theta}^{B}\}\ =\ C^{A,B}~,
\end{equation}
which arises from considering string theory in a background with a\index{string theory}
constant graviphoton field strength. This discovery triggered many\index{field strength}\index{graviphoton}
publications, in particular, non-anticom\-mutativity for extended
supersymmetry was considered, as well\index{super!symmetry}
\cite{Ivanov:2003te,Ferrara:2003xk,Saemann:2004cf,DeCastro:2005vb}.

An alternative approach, which was followed in
\cite{Ferrara:2000mm}, manifestly preserves supersymmetry but\index{super!symmetry}
breaks chirality. This has many disadvantages, as without chiral
superfields, it is e.g.\ impossible to define super Yang-Mills\index{chiral!superfields}
theory in the standard superspace formalism.\index{super!space}

In section \ref{sDrinfeldTwistedSupersymmetry}, we will present an
approach in which supersymmetry {\em and} chirality are manifestly
and simultaneously preserved, albeit in a twisted form.

\subsection{Non-anticommutative deformations of superspaces}\index{super!space}

\paragraph{Associativity of the star product.} In the Minkowski case,
one can show that the deformations preserving the associativity of
the star product all satisfy
\begin{equation}
\{\hat{\theta}^A,\hat{\theta}^B\}\ =\ \{\hat{\btheta}^A,\hat{\btheta}^B\}\ =\ 
\{\hat{\theta}^A,\hat{\btheta}^B\}\ =\ 0~.
\end{equation}
These deformations are clearly too trivial, but one can circumvent
this problem by turning to Euclidean spacetime. Here, the most
general deformation compatible with associativity of the star
product reads
\begin{equation}
\{\hat{\theta}^A,\hat{\theta}^B\}\ \neq\ 
0\eand\{\hat{\btheta}^A,\hat{\btheta}^B\}\ =\ \{\hat{\theta}^A,\hat{\btheta}^B\}\ =\ 0~,
\end{equation}
which is possible, as $\theta$ and $\btheta$ are no longer related
by complex conjugation. For this reason, we will always consider
superspaces which have an Euclidean metric on their bodies in the\index{super!space}\index{metric}
following. To justify our use of the (Minkowski) superfield
formalism, we can assume to temporarily work with complexified
spacetime and field content and impose appropriate reality
conditions after all calculations have been performed.

\paragraph{The deformed superspace $\FR^{4|4\CN}_\hbar$.} The canonical\index{super!space}
deformation of $\ssp$ to $\sspdef$ amounts to putting
\begin{equation}\label{orddecnac}
\{\hat{\theta}^{\alpha i},\hat{\theta}^{\beta j}\}\ =\ \hbar
C^{\alpha i,\beta j}~,
\end{equation}
where the hats indicate again the deformed Gra{\ss}mann coordinates in
the operator representation.\index{representation}

As in the case of the noncommutative deformation, one can
equivalently deform the algebra of superfunctions $\CS$ on $\ssp$
to an algebra $\CS_\star$, in which the product is given by the
Moyal-type star product
\begin{equation}\label{candef}
f\star g\ =\ f\exp\left(-\tfrac{\hbar}{2}
\overleftarrow{Q}_{\alpha i} C^{\alpha i,\beta
j}\overrightarrow{Q}_{\beta j}\right) g~,
\end{equation}
where $\overleftarrow{Q}_{\alpha i}$ and
$\overrightarrow{Q}_{\beta j}$ are supercharges acting from the
right and the left, respectively. Recall that in our convention
for superderivatives, we have
\begin{equation}
\theta^{\alpha i}\overleftarrow{Q}_{\beta
j}\ =\ -\delta^i_j\delta_\beta^\alpha~.
\end{equation}

Contrary to the case of noncommutative deformations, an $\hbar$
was inserted into the definition of the deformation
\eqref{orddecnac} to indicate the different orders. Since the star
product \eqref{candef} is a finite sum due to the nilpotency of
the Gra{\ss}mann variables, power expansions in the deformation\index{Gra{\ss}mann variable}
parameter are even more important than in the noncommutative case.

All commutators involving this star multiplication will be denoted\index{commutators}
by a $\star$, e.g.\ the graded commutator will read as
\begin{equation}
\lsc f\stackrel{\star}{,}g\rsc\ :=\ f\star
g-(-1)^{\tilde{f}\tilde{g}}g\star f~.
\end{equation}
The new coordinate algebra obtained from this deformation reads as
\begin{equation}
\begin{aligned}
{}[x^{\alpha\ald}\stackrel{\star}{,}x^{\beta\bed}]\ =\ -\hbar
C^{\alpha i,j \beta}\bar{\theta}^\ald_i\bar{\theta}^\bed_j~,~~~
[x^{\alpha\ald}\stackrel{\star}{,}\theta^{\beta j}]\ =\ -\hbar
C^{\alpha i,j
\beta}\bar{\theta}^\ald_i~,\\
\{\theta^{\alpha i}\stackrel{\star}{,}\theta^{\beta j}\}\ =\ \hbar
C^{\alpha i,j \beta}~~\hspace{3cm}
\end{aligned}
\end{equation}
and all other supercommutators vanish, but after changing to the\index{commutators}\index{super!commutator}
chiral coordinates\index{chiral!coordinates}
\begin{equation}
(y^{\alpha\ald}:=x^{\alpha\ald}+\theta^{\alpha
i}\btheta^\ald_i,\theta^{\alpha i},\bar{\theta}^\ald_i)~,
\end{equation}
the coordinate algebra simplifies to
\begin{equation}
{}[y^{\alpha\ald}\stackrel{\star}{,}y^{\beta\bed}]\ =\ 0~,~~~
[y^{\alpha\ald}\stackrel{\star}{,}\theta^{\beta j}]\ =\ 0~,~~~
[\theta^{\alpha i}\stackrel{\star}{,}\theta^{\beta j}]\ =\ \hbar
C^{\alpha i,\beta j}~.
\end{equation}
This deformation has been shown to arise in string theory from\index{string theory}
open superstrings of type IIB in the background of a constant
graviphoton field strength\index{field strength}\index{graviphoton}
\cite{Ooguri:2003qp,Seiberg:2003yz,deBoer:2003dn}.

\paragraph{Deformed supersymmetry algebra.} The corresponding\index{super!symmetry}
deformed algebra of superderivatives and supercharges reads
as\footnote{For further reference, we present the algebra both in
spinor and vector notation.}\index{Spinor}
\begin{equation}\label{defalgebra}
\begin{aligned}
\{D_{\alpha i}\stackrel{\star}{,}D_{\beta j}\}&\ =\ 0~,~~~
\{\bar{D}^i_{\ald}\stackrel{\star}{,}\bar{D}^j_{\bed}\}\ =\ 0~,\\
\{D_{\alpha i}\stackrel{\star}{,}\bar{D}^j_{\bed}\}&\ =\
-2\delta_j^i\dpar_{\alpha\bed}\ =\ -2\di\delta_i^j\sigma^\mu_{\alpha\bed}\dpar_\mu~,\\
\{Q_{\alpha i}\stackrel{\star}{,}Q_{\beta j}\}&\ =\ 0~,\\
\{\bar{Q}^i_{\ald}\stackrel{\star}{,}\bar{Q}^j_{\bed}\}&\ =\
4\hbar C^{\alpha i,\beta j}\dpar_{\alpha\ald}\dpar_{\beta\bed}\ =\
-4\hbar
C^{\alpha i,\beta j}\sigma_{\alpha\ald}^\mu\sigma_{\beta\bed}^\nu\dpar_\mu\dpar_\nu~,\\
\{Q_{\alpha i}\stackrel{\star}{,}\bar{Q}^j_{\bed}\}&\ =\
2\delta_i^j\dpar_{\alpha\bed}\ =\
2\di\delta_i^j\sigma^\mu_{\alpha\bed}\dpar_\mu~.
\end{aligned}
\end{equation}
By inspection of this deformed algebra, it becomes clear that the
number of supersymmetries is reduced to $\CN/2$, since those
generated by $\bar{Q}^i_\ald$ are broken. On the other hand, it
still allows for the definition of chiral and anti-chiral
superfields as the algebra of the superderivatives $D_{\alpha i}$\index{chiral!superfields}
and $\bar{D}^i_\ald$ is undeformed. Because of this, graded
Bianchi identities are also retained,\index{graded Bianchi identities} e.g.\
\begin{equation*}
\lsc\nabla_a\stackrel{\star}{,}\lsc\nabla_b\stackrel{\star}{,}\nabla_c\rsc\rsc
+(-1)^{\tilde{a}(\tilde{b}+\tilde{c})}
\lsc\nabla_b\stackrel{\star}{,}\lsc\nabla_c\stackrel{\star}{,}\nabla_a\rsc\rsc+
(-1)^{\tilde{c}(\tilde{a}+\tilde{b})}
\lsc\nabla_c\stackrel{\star}{,}\lsc\nabla_a\stackrel{\star}{,}\nabla_b\rsc\rsc\ =\ 0~.
\end{equation*}

\paragraph{Consequence for field theories.} Field theories on
non-anticommutative superspaces are usually defined by replacing\index{non-anticommutative superspace}\index{super!space}
all ordinary products in the action written in the $\CN=1$
superfield formalism by star products. First of all, such theories
will evidently have non-Hermitian Lagrangians since -- roughly
speaking -- chiral parts of the action will get deformed, while
anti-chiral parts remain unchanged. This, however, allows for
renormalizable theories which have terms in their Lagrangian with
mass dimension larger than 4 \cite{Britto:2003kg}. Of particular
interest to our work is the question of renormalizability of
non-anticommutative field theories and here specifically of the\index{non-anticommutative field theories}
$\CN=\frac{1}{2}$ Wess-Zumino model, as discussed in\index{Wess-Zumino model}
\cite{Terashima:2003ri,Britto:2003aj,Britto:2003aj2,Grisaru:2003fd,Britto:2003kg,Lunin:2003bm,Berenstein:2003sr}.
For more recent work on the renormalizability of
non-anticommutative super Yang-Mills theory, see\index{Yang-Mills theory} e.g.\
\cite{Grisaru:2005we}.

\subsection{Non-anticommutative $\CN=4$ SYM
theory}\label{ssNACSYM}

\paragraph{Idea.} In the cases $\CN=1$ and $\CN=2$, one has
appropriate superspace formalisms at hand, which allow for a\index{super!space}
direct deformation of supersymmetric field theories by deforming
their actions in these formalisms. In the cases\footnote{These
cases are essentially equivalent, see section \ref{subsecN4}.}
$\CN=3$ and $\CN=4$, however, there is no such formalism. Instead,
one can use the constraint equations \eqref{constraintN4SYM} on\index{constraint equations}
$\FR^{4|16}$, which are equivalent to the $\CN=4$ SYM equations as
discussed in section \ref{subsecN4}. By considering these
constraint equations on the deformed space $\FR^{4|16}_\hbar$, one\index{constraint equations}
finds the equations of motion of the corresponding deformed
theory.

\paragraph{Deformed constraint equations.} We start from the constraint\index{constraint equations}
equations of $\CN=4$ SYM theory introduced in \ref{subsecN4} on
$\FR^{4|16}$ and follow the discussion of the undeformed case. On
the deformed space $\FR^{4|16}_\hbar$, they read as
\begin{equation}\label{defconstraint}
\begin{aligned}
\{\tilde{\nabla}_{\alpha
i}\stackrel{\star}{,}\tilde{\nabla}_{\beta
j}\}\ =\ -2\eps_{\alpha\beta}\tilde{\phi}_{ij}~,~~~
\{\tilde{\bar{\nabla}}^i_{\ald}\stackrel{\star}{,}\tilde{\bar{\nabla}}^j_{\bed}\}\ =\ -2\eps_{\ald\bed}\tilde{\phi}^{ij}~,\\
\{\tilde{\nabla}_{\alpha
i}\stackrel{\star}{,}\tilde{\bar{\nabla}}^j_{\bed},\}\ =\ -2\delta^j_i
\tilde{\nabla}_{\alpha\bed}~,\hspace{3cm}
\end{aligned}
\end{equation}
where we will use a tilde\footnote{Appearing over an exponent, the
tilde still denotes the corresponding parity.} to label fields\index{parity}
living on the deformed superspace $\FR^{4|16}_\hbar$. The\index{super!space}
covariant derivatives are obtained from super gauge potentials\index{covariant derivative}
\begin{equation}
\tilde{\nabla}_{\alpha i}\ =\ D_{\alpha i}+\lsc\tilde{\omega}_{\alpha
i}\stackrel{\star}{,}\cdot\rsc~,~~~
\tilde{\nabla}^i_{\ald}\ =\ \bar{D}^i_\ald-\lsc\tilde{\bar{\omega}}^i_{\ald}\stackrel{\star}{,}\cdot\rsc~,~~~
\tilde{\nabla}_{\alpha\ald}\ =\ \dpar_{\alpha\ald}+\lsc\tilde{A}_{\alpha\ald}\stackrel{\star}{,}\cdot\rsc~,
\end{equation}
and we define additionally the superspinor fields\index{Spinor}\index{super!spin}
\begin{equation}\label{defspinor}
[\tilde{\nabla}_{\alpha
i}\stackrel{\star}{,}\tilde{\nabla}_{\beta\bed}]\ =:\ \eps_{\alpha\beta}\tilde{\bar{\chi}}_{i\bed}\eand
[\tilde{\bar{\nabla}}^i_{\ald}\stackrel{\star}{,}\tilde{\nabla}_{\beta\bed}]\ =:\ \eps_{\ald\bed}\tilde{\chi}^i_{\beta}~.
\end{equation}
Proceeding further along the lines of the undeformed case, we
finally arrive at the $\CN=4$ SYM equations with all commutators\index{commutators}
replaced by star-commutators:
\begin{equation}\label{N4defeom}
\begin{aligned}
\tilde{\nabla}_{\alpha\ald}\tilde{\chi}^{i\beta}+{\tfrac{1}{2}}
\eps^{ijkl}[
\tilde{\phi}_{kl}\stackrel{\star}{,}\bar{\chi}_{j\ald}] & \ =\ 
0~,\\
\tilde{\nabla}_{\alpha\ald}\tilde{\bar{\chi}}^\bed_i+[
\tilde{\phi}_{ij}\stackrel{\star}{,}\tilde{\chi}^j_\alpha]& \ =\  0~,\\
\eps^{\ald\bed}\tilde{\nabla}_{\gamma\ald}
\tilde{f}_{\bed\gad}+\eps^{\alpha\beta}
\tilde{\nabla}_{\alpha\gad}
\tilde{f}_{\beta\gamma}&\ =\ {\tfrac{1}{4}}\eps^{ijkl}
[\tilde{\nabla}_{\gamma\gad} \tilde{\phi}_{ij}\stackrel{\star}{,}
\tilde{\phi}_{kl}] +
\{\tilde{\chi}^i_\gamma\stackrel{\star}{,}\tilde{\bar{\chi}}_{i\gad}\}~,\\
\tilde{\nabla}_{\alpha\ald}\tilde{\nabla}^{\alpha\ald}
\tilde{\phi}_{ij}
-{\tfrac{1}{4}}\eps^{klmn}[\tilde{\phi}_{mn}\stackrel{\star}{,}[\tilde{\phi}_{kl}\stackrel{\star}{,}
\tilde{\phi}_{ij}]] &\ =\ {\tfrac{1}{2}}\eps_{ijkl}
\{\tilde{\chi}^k_\alpha\stackrel{\star}{,}\tilde{\chi}^{l\beta}\}+\{\tilde{\bar{\chi}}_{i\ald}\stackrel{\star}{,}
\tilde{\bar{\chi}}^\ald_{j}\}~.
\end{aligned}
\end{equation}
Recall that all the fields appearing in the above equations are in
fact superfields on the deformed space $\FR^{4|16}_\hbar$, and we
still have to extract the zeroth order components and their
deformed equations of motion.

\paragraph{The Seiberg-Witten map.}\label{pNACSWmap} While the\index{Seiberg-Witten map}
derivation of the superfield expansion in the undeformed case was
quite simple by imposing transverse gauge and using the recursion
operator $\CD$, we face some difficulties in the deformed case.
Using again this Euler operator would lead to a highly nonlinear\index{Euler operator}
system of algebraic equations and the complete knowledge of the
superfield expansion, i.e.\ about $2^{16}$ terms for every field,
is needed to calculate corrections even to first order in $\hbar$.

Therefore we suggest an alternative approach based on a
generalization of the Seiberg-Witten map, cf.\ section\index{Seiberg-Witten map}
\ref{ssNCfeatures}, \ref{pSWmap}, which will yield the expansion
of the superfields order by order in $\hbar$. For this, let us
choose $\tilde{\omega}_{\alpha i}$ as the fundamental field of our
theory, i.e.\ all the other fields $\tilde{\bar{\omega}}^i_\ald$,
$\tilde{A}_{\alpha\ald}$, $\tilde{\phi}_{ij}$,
$\tilde{\chi}^i_\alpha$ and $\tilde{\bar{\chi}}_{i\ald}$ are fixed
for a certain $\tilde{\omega}_{\alpha i}$ by the constraint
equations \eqref{defconstraint} and the definitions\index{constraint equations}
\eqref{defspinor}.

First recall that infinitesimal gauge transformations of the\index{gauge transformations}\index{gauge!transformation}
undeformed and the deformed gauge potential are given by
\begin{equation}
\delta_\lambda\omega_{\alpha i} \ =\  D_{\alpha
i}\lambda+[\omega_{\alpha i},\lambda]\eand
\delta_{\tilde{\Lambda}}\tilde{\omega}_{\alpha i} \ =\  D_{\alpha
i}\tilde{\Lambda} +[\tilde{\omega}_{\alpha
i}\stackrel{\star}{,}\tilde{\Lambda}]~,
\end{equation}
respectively, where $\lambda$ and $\tilde{\Lambda}$ are even
superfields parameterizing the transformation. Analogously to the
noncommutative formula \eqref{eqSWmap}, the starting point is then
the equation
\begin{equation}\label{gaugetrafoSW}
\tilde{\omega}_{\alpha
i}(\omega+\delta_\lambda\omega,\bar{\omega}+\delta_\lambda\bar{\omega})
=\tilde{\omega}_{\alpha
i}(\omega,\bar{\omega})+\delta_{\tilde{\Lambda}}
\tilde{\omega}_{\alpha i}(\omega,\bar{\omega})~.
\end{equation}

\paragraph{Explicit solution.} To obtain the explicit form of the
Seiberg-Witten map, we can use the consistency condition that two\index{Seiberg-Witten map}
successive gauge transformations should, e.g.\ for a superfield\index{gauge transformations}\index{gauge!transformation}
$\tilde{\psi}$ in the fundamental representation, satisfy\index{fundamental representation}\index{representation}
\begin{equation}
[\delta_{\tilde{\Lambda}},\delta_{\tilde{\Sigma}}]\tilde{\psi}\ =\ -[\tilde{\Lambda}\stackrel{\star}{,}\tilde{\Sigma}]
\star\tilde{\psi}\ =\ \delta_{[\tilde{\Lambda}\stackrel{\star}{,}\tilde{\Sigma}]}\tilde{\psi}~.
\end{equation}
By the Seiberg-Witten map, gauge equivalent solutions get mapped\index{Seiberg-Witten map}
to deformed gauge equivalent solutions, and thus we can restrict
ourselves to gauge transformations of the type\index{gauge transformations}\index{gauge!transformation}
$\delta_\lambda\tilde{\psi}=-\tilde{\Lambda}_\lambda(\omega,\bar{\omega})\star\tilde{\psi}$.
Then one can simplify the above consistency condition to
\begin{equation}\label{conscond2}
\delta_\lambda\tilde{\Lambda}_\sigma-\delta_\sigma\tilde{\Lambda}_\lambda
+[\tilde{\Lambda}_\lambda\stackrel{\star}{,}\tilde{\Lambda}_\sigma]\ =\ \tilde{\Lambda}_{[\lambda,\sigma]}~.
\end{equation}

As for all the fields in our deformed theory, we assume that also
$\tilde{\Lambda}$ is a polynomial\footnote{In principle, it could
also be any power series.} in $\hbar$ and considering the first
order of $\hbar$ in \eqref{conscond2}, we arrive at
\begin{equation}\label{conscond3}
\delta_\lambda\tilde{\Lambda}^1_\sigma-\delta_\sigma\tilde{\Lambda}^1_\lambda
+[\lambda,\tilde{\Lambda}^1_\sigma]
+[\tilde{\Lambda}^1_\lambda,\sigma] -\tfrac{1}{2}C^{\alpha i,\beta
j}[\dpar_{\alpha i} \lambda,\dpar_{j \beta}
\sigma]\ =\ \tilde{\Lambda}^1_{[\lambda,\sigma]}~.
\end{equation}
Although it is not straightforward, it is possible to guess the
solution to this equation, which is given by
\begin{equation}
\begin{aligned}
\tilde{\Lambda}_\lambda&\ =\ \lambda-\tfrac{\hbar}{4}C^{\alpha i,\beta
j}
[\dpar_{\alpha i}\lambda,\Omega_{\beta j}]+\CO(\hbar^2)~,\\
\Omega_{\alpha i}&\ :=\ \omega_{\alpha
i}+\bar{\theta}^\bed_j\left(\bar{D}^j_\bed\omega_{\alpha i}
+D_{\alpha
i}\bar{\omega}^j_\bed+\{\bar{\omega}^j_\bed,\omega_{\alpha
i}\}\right)~.
\end{aligned}
\end{equation}
To verify this solution, note that infinitesimal gauge
transformations act on $\Omega_{\alpha i}$ as\index{gauge transformations}\index{gauge!transformation}
$\delta_\lambda\Omega_{\alpha i}=\dpar_{\alpha
i}\lambda+[\Omega_{\alpha i},\lambda]$, and therefore we have
$\tilde{\Lambda}_\lambda=\lambda-\tfrac{\hbar}{4}C^{\alpha i,\beta
j}[\dpar_{\alpha i}\lambda,\Omega_{\beta j}] +\CO(\hbar^2)$.

\paragraph{Field expansion.} Let us now consider the first order in $\hbar$ of the second
equation in \eqref{gaugetrafoSW}, which reads
\begin{equation}
\delta_\lambda\tilde{\omega}^1_{\alpha i}\ =\ D_{\alpha
i}\tilde{\Lambda}^1+[\tilde{\omega}^1_{\alpha i},\lambda]
+[\omega_{\alpha i},\tilde{\Lambda}^1]+{\tfrac{1}{2}}C^{\beta
j,k\gamma}\{\dpar_{\beta j}\omega_{\alpha i},
\dpar_{k\gamma}\lambda\}~.
\end{equation}
With our above result for $\tilde{\Lambda}_\lambda^1$, one finds
after some algebraic manipulations that
\begin{equation}
\tilde{\omega}_{\alpha i}^1\ =\ \tfrac{1}{4}C^{\beta j,k\gamma}
\{\Omega_{\beta j},\dpar_{k\gamma}\omega_{\alpha
i}+R_{k\gamma,\alpha i}\}
\end{equation}
with
\begin{equation}
R_{\alpha i,\beta j}\ :=\ \dpar_{\alpha i}\omega_{\beta j}+ D_{\beta
j}\Omega_{\alpha i}+\{\omega_{\beta j},\Omega_{\alpha i}\}~.
\end{equation}
Now that we have the definition of our fundamental field, we can
work through the constraint equations \eqref{defconstraint} and\index{constraint equations}
the definitions \eqref{defspinor} to obtain the first order in
$\hbar$ of the other fields. From the first constraint equation,
we immediately obtain
\begin{equation}
\tilde{\phi}^1_{ij}\ =\ {\tfrac{1}{2}}\eps^{\alpha\beta}\nabla_{(\alpha
i} \tilde{\omega}^1_{\beta
j)}+{\tfrac{1}{8}}\eps^{\alpha\beta}C^{m\delta,n\eps}
\{\dpar_{m\delta}\omega_{\alpha i},\dpar_{n\eps}\omega_{\beta
j}\}~,
\end{equation}
where the parentheses denote, as usual, symmetrization with
appropriate weight. From this solution, we can use the second
constraint equation to solve for the first order term in
$\tilde{\bar{\omega}}^j_\bed{}^1$. Together with the assumption
that $\bar{\nabla}^i_\ald \tilde{\bar{\omega}}^{j\, 1}_\bed=
\bar{\nabla}^j_\bed \tilde{\bar{\omega}}^{i\, 1}_\ald$, we find
the equation
\begin{equation}\label{eqnForOmegaBar}
\begin{aligned}
\bar{\nabla}^i_\ald \tilde{\bar{\omega}}^{j\, 1}_\bed \ &\ =\ \
{\tfrac{1}{2}}\eps_{\ald\bed}\eps^{ijkl}\tilde{\phi}^1_{kl}+
{\tfrac{1}{4}}C^{m\delta,n\eps}\{\dpar_{m\delta}
\bar{\omega}^i_\ald,\dpar_{n\eps}\bar{\omega}^j_\bed\}\\
\ &\ =\ \ \bar{D}^i_\ald\tilde{\bar{\omega}}^{j\, 1}_\bed
-\{\bar{\omega}^i_\ald, \tilde{\bar{\omega}}^{j\, 1}_\bed \}~.
\end{aligned}
\end{equation}
Recall that in the undeformed case, we used transverse gauge to
break super gauge invariance to ordinary gauge symmetry. Here, we
can impose a similar condition to simplify the situation:
\begin{equation}
\theta\tilde{\omega}-\btheta\tilde{\bar{\omega}}\ =\
\theta^{\alpha i}\tilde{\omega}_{\alpha
i}+\btheta^\ald_i\tilde{\bar{\omega}}^i_\ald\ =\ 0~,
\end{equation}
which is separately valid to all orders in $\hbar$. From this, we
obtain the further relation $\tilde{\bar{\omega}}^{i\,
1}_\ald=\bar{D}^i_\ald(\theta\tilde{\omega}^1)-
\btheta^\bed_j\bar{D}^i_\ald\tilde{\bar{\omega}}^{j\, 1}_\bed$,
which turns equation \eqref{eqnForOmegaBar} into
\begin{equation}\label{itOmegaBar}
\bar{D}^i_\ald\tilde{\bar{\omega}}^{j\,
1}_\bed-\btheta^\gad_l[\bar{\omega}^i_\ald,
\bar{D}^j_\bed\tilde{\bar{\omega}}^{l\,
1}_\gad]\ =\ K^{ij}_{\ald\bed}~,
\end{equation}
where we have abbreviated
\begin{equation}
K^{ij}_{\ald\bed}\ :=\ 
{\tfrac{1}{2}}\eps_{\ald\bed}\eps^{ijkl}\tilde{\phi}^1_{kl}+
{\tfrac{1}{4}}C^{m\delta,n\eps}\{\dpar_{m\delta}\bar{\omega}^i_\ald,
\dpar_{n\eps}\bar{\omega}^j_\bed\}+\{\bar{\omega}^i_\ald,\bar{D}^j_\bed
(\theta\tilde{\omega}^1)\}~.
\end{equation}
The expression for $\tilde{\bar{\omega}}^1$ is found by iterating
the equation \eqref{itOmegaBar}, which becomes a little technical.
We obtain
\begin{equation}\label{afterit}
\bar{D}_{\bar{A}}\tilde{\bar{\omega}}_{\bar{B}}^1\ =\ \sum_{|{\bar
I}|\ \leq\  8} (-1)^{\lfloor\frac{|{\bar
I}|}{2}\rfloor}\,\btheta^{{\bar I}}\,
\lsc\bar{\omega},K\rsc_{{\bar I},{\bar{A}}{\bar{B}}}~,
\end{equation}
where ``$\lfloor\ \rfloor$'' denotes the Gau{\ss} bracket, $|I|$ the
length of the multiindex $I$ and
\begin{equation}
\lsc\bar{\omega},K\rsc_{{\bar
I},{\bar{A}}{\bar{B}}}\ :=\ \lsc\bar{\omega}_{\bar{A}},\lsc\bar{\omega}_{\bar{B}},\lsc\bar{\omega}_{{\bar{A}}_1},
\cdots\lsc\bar{\omega}_{{\bar{A}}_{|{\bar
I}|-2}},K_{{\bar{A}}_{|{\bar I}|-1} {\bar{A}}_{|{\bar
I}|}}\rsc\cdots\rsc\rsc\rsc~.
\end{equation}
The sum in \eqref{afterit} is finite as the order of
$\bar{\theta}$ increases during the iteration. The first order
contribution is thus given by
\begin{equation}
\tilde{\bar{\omega}}_{\bar{A}}^1\ =\
\bar{D}_{\bar{A}}(\theta\tilde{\omega}^1)-\btheta^{\bar{B}}
\sum_{|{\bar I}|\ \leq\  8}(-1)^{\lfloor\frac{|{\bar I}|}{
2}\rfloor}\,\btheta^{{\bar I}}\, \lsc\bar{\omega},K\rsc_{{\bar
I},{\bar{A}}{\bar{B}}}~.
\end{equation}
From here on, it is easy to write down the first order deformation
of the remaining fields via the third constraint equation and the
definitions \eqref{defspinor}:
\begin{equation}
\begin{aligned}
{{\tilde{A}}}_{\alpha\bed}^1 &\ =\  {\tfrac{1}{8}}(\nabla_{\alpha i}
             {\tilde{\bar{\omega}}}^{i\,
1}_\bed-\bar{\nabla}^i_\bed{\tilde{\omega}}^1_{\alpha i}+
             {\tfrac{1}{2}}C^{m\delta,n\eps}\{\dpar_{m\delta}\omega_{\alpha i},
             \partial_{n\eps}\bar{\omega}^i_\bed\})\\
{\tilde{\bar{\chi}}}^1_{i\bed} &\ =\ 
-{\tfrac{1}{2}}\eps^{\alpha\beta}(
             \nabla_{\alpha i}{\tilde{A}}^1_{\beta\bed}-\nabla_{\beta\bed}{\tilde{\omega}}^1_{\alpha i}+
             {\tfrac{1}{2}}C^{m\delta,n\eps}\{\partial_{m\delta}\omega_{\alpha i},
             \partial_{n\eps} A_{\beta\bed}\})\\
{\tilde{\chi}}^{i\, 1}_\beta &\ =\  -{\tfrac{1}{2}}\eps^{\ald\bed}
             (\bar{\nabla}^i_\ald{\tilde{A}}^1_{\beta\bed}+\nabla_{\beta\bed}
             {\tilde{\bar{\omega}}}^{i\, 1}_\ald-{\tfrac{1}{2}}C^{m\delta,n\eps}
             \{\partial_{m\delta}\bar{\omega}^i_\ald,\partial_{n\eps} A_{\beta\bed}\})
\end{aligned}
\end{equation}

\paragraph{Deformed field equations.} So far, we computed the
first order deformations in $\hbar$ of the superfields and by
restricting to their zeroth order components, we obtained the
deformations of the $\CN=4$ SYM multiplet. It remains, however, to
calculate the zeroth order components of the superfield equations
\eqref{N4defeom}. For this, we need to know the explicit zeroth
order form of products $\theta^I\star \theta^J$ with $I,J$ being
multiindices. By induction, one can easily prove that
\begin{equation}\label{Wick}
\begin{aligned}
\theta^{A_1}\star\ldots \star\theta^{A_n}&\ =\ \theta^{A_1}\ldots \theta^{A_n}+\sum
\mbox{contractions}\\
&=\theta^{A_1}\ldots \theta^{A_n}+\sum_{i<j}
\theta^{A_1}\ldots \contra{29}{\theta^{A_i}\cdots\theta}^{A_j}\ldots \theta^{A_n}+\ldots ~,
\end{aligned}
\end{equation}
which resembles a fermionic Wick theorem and where a contraction
is defined as
\begin{equation}
\contra{15}{\theta^{A_i}\theta}^{A_j}\ :=\ \tfrac{\hbar}{2}C^{A_i,A_j}~.
\end{equation}
Note that signs appearing from the grading have to be taken into
account. For $n=2$, \eqref{Wick} is obvious, and for $n>2$ one can
show that
\begin{equation}
(\theta^{A_1}\cdots\theta^{A_n})\star\theta^{A_{n+1}}\ =\
\theta^{A_1}\cdots\theta^{A_n}\theta^{A_{n+1}}\ +\ \sum_{i=1}^n\
\theta^{A_1}\cdots\contra{33}{\theta^{A_i}\cdots\
\theta}^{A_{n+1}}~,
\end{equation}
which proves \eqref{Wick} by induction. Since we are interested
only in the zeroth order terms in \eqref{Wick}, let us define the
projection operator $\pi_\circ$, which extracts these terms. Then
we have
\begin{equation}
\begin{aligned}
\pi_\circ(\theta^I\star\theta^J)\ &\ =\ \
\pi_\circ((\theta^{A_1}\cdots\theta^{A_n})\star
(\theta^{B_1}\cdots\theta^{B_m}))\\ &\ =\ \
\delta_{nm}\frac{(-1)^{\frac{n}{2}(n-1)} \hbar^n}{2^n\,
n!}\sum_{\{i,j\}} \eps_{i_1\cdots i_n}\eps_{j_1\cdots j_n}
C^{A_{i_1},B_{j_1}}\cdots C^{A_{i_n},B_{j_n}}~,
\end{aligned}
\end{equation}
which is rather obvious, and we have
$\pi_\circ(\theta^I\star\theta^J)=\pi_\circ(\theta^J\star\theta^I)$
as a corollary.

To compute all the commutators appearing in the equations of\index{commutators}
motion \eqref{N4defeom}, let us expand every superfield as
\begin{equation}
\tilde{f}\ =\ \zero{\tilde{f}}+\sum_I\tilde{f}_I\theta^I+ \mbox{terms
containing $\bar{\theta}$}~.
\end{equation}
Given two superfunctions $\tilde{f}$ and $\tilde{g}$, we obtain
for the three possible cases of gradings $(0,0),$ $(1,1),(0,1)$
for the pair $(\tilde{f},\tilde{g})$ the following results:
\begin{equation}
\begin{aligned}
\pi_\circ([\tilde{f}\stackrel{\star}{,}\tilde{g}])\ \ =\  & \
[\zero{\tilde{f}},\zero{\tilde{g}}]+\sum_{|I|=|J|}(-1)^{\tilde{I}}
[\tilde{f}_I,\tilde{g}_J]\
\pi_\circ(\theta^I\star\theta^J)~,\\
\pi_\circ(\{\tilde{f}\stackrel{\star}{,}\tilde{g}\})\ \ =\  & \
\{\zero{\tilde{f}},\zero{\tilde{g}}\}+\sum_{|I|=|J|}
\{\tilde{f}_I,\tilde{g}_J\}\
\pi_\circ(\theta^I\star\theta^J)~,\\
\pi_\circ([\tilde{f}\stackrel{\star}{,}\tilde{g}])\ \ =\  & \
[\zero{\tilde{f}},\zero{\tilde{g}}]+\sum_{|I|=|J|}
(\tilde{f}_I\tilde{g}_J-(-1)^{\tilde{I}}\tilde{g}_J\tilde{f}_I)\
\pi_\circ(\theta^I\star\theta^J)~,
\end{aligned}
\end{equation}

Now we have all the necessary ingredients to derive the field
equations of $\CN=4$ SYM theory on $\FR^{4|16}_\hbar$. The
equations of motion for the eight Weyl spinors read\index{Spinor}
\begin{equation*}
\begin{aligned}
{\eps^{\alpha\beta}}\z{\tilde{\nabla}}_{\,\alpha\ald}\z{\tilde{\chi}}^i_\beta
+{\tfrac{1}{2}}
\eps^{ijkl}[\z{\tilde{\phi}}_{kl},\z{\tilde{\bar{\chi}}}_{j\ald}]\ =\ &
-\eps^{\alpha\beta}\sum_{|I|=|J|}
({\tilde{A}}_{\alpha\ald|I}{\tilde{\chi}}_{\beta|J}^i-(-1)^{\tilde{I}}
{\tilde{\chi}}^i_{\beta|J}{\tilde{A}}_{\alpha\ald|I})\,T^{IJ}\\
&-{\tfrac{1}{2}}\eps^{ijkl} \sum_{|I|=|J|} ({\tilde
{W}}_{kl|I}{\tilde{\bar{\chi}}}_{j\ald|J}-(-1)^{\tilde{I}}
{\tilde{\bar{\chi}}}_{j\ald|J}{\tilde{\phi}}_{kl|I})\,T^{IJ}~,\\
\eps^{\ald\bed}\zero{\tilde{\nabla}}_{\,\alpha\ald}\zero{\tilde{\bar{\chi}}}_{i\bed}+[\zero{\tilde{\phi}}_{ij},
\zero{\tilde{\chi}}^j_\alpha] \ =\ &-\eps^{\ald\bed}\sum_{|I|=|J|}
({\tilde{A}}_{\alpha\ald|I}{\tilde{\bchi}}_{i\bed|J}-(-1)^{\tilde{I}}
{\tilde{\bchi}}_{i\bed|J}{\tilde{A}}_{\alpha\ald|I})\,T^{IJ}\\
&-\sum_{|I|=|J|}(\tilde{\phi}_{ij|I}{\tilde{\chi}}^j_{\alpha|J}-(-1)^{\tilde{I}}{\tilde{\chi}}^j_{\alpha|J}\tilde{\phi}_{ij|I})\,T^{IJ}~,
\end{aligned}
\end{equation*}
where we introduced $T^{IJ}:=\pi_\circ(\theta^I\star\theta^J)$ for
brevity. For the bosonic fields, the equations of motion read as
\begin{equation*}
\begin{aligned}
\eps^{\ald\bed}\zero{\tilde{\nabla}}_{\gamma\ald}\zero{\tilde{f}}_{\bed\gad}+\eps^{\alpha\beta}
  \zero{\tilde{\nabla}}_{\alpha\gad}\zero{\tilde{f}}_{\beta\gamma}-{\tfrac{1}{4}}\eps^{ijkl}
  &
  [\zero{\tilde{\nabla}}_{\gamma\gad}\zero{\tilde{\phi}}_{ij},\zero{\tilde{\phi}}_{kl}] -\{\zero{\tilde{\chi}}^i_\gamma,
  \zero{\tilde{\bar{\chi}}}_{i\gad}\}\ =\ \\
& \kern-4cm-\sum_{|I|=|J|}(-1)^{\tilde{I}}\left\{\eps^{\ald\bed}
  [{\tilde{A}}_{\gamma\ald|I},\tilde{f}_{\bed\gad|J}]+\eps^{\alpha\beta}
  [{\tilde{A}}_{\alpha\gad|I},\tilde{f}_{\beta\gamma|J}]\right\}\,T^{IJ}\\
& \kern-3cm+\sum_{|I|=|J|}\left\{(-1)^{\tilde{I}}{\tfrac{1}{4}}
  \eps^{ijkl}[(\tilde{\nabla}_{\gamma\gad}\tilde{\phi}_{ij})_I,\tilde{\phi}_{kl|J}]+
  \{{\tilde{\chi}}^i_{\gamma|I},{\tilde{\bchi}}_{i\gad|J}\}\right\}\,
  T^{IJ}~,\\
  \eps^{\alpha\beta}\eps^{\ald\bed}\zero{\tilde{\nabla}}_{\alpha\ald}
  \zero{\tilde{\nabla}}_{\beta\bed}\zero{\tilde{\phi}}_{ij}
  -{\tfrac{1}{4}}\eps^{klmn}[\zero{\tilde{\phi}}_{mn},&[\zero{\tilde{\phi}}_{kl}~,
  \zero{\tilde{\phi}}_{ij}]]
  \ =\ {\tfrac{1}{2}}\eps_{ijkl}\eps^{\alpha\beta}
  \{\zero{\tilde{\chi}}^k_\alpha,\zero{\tilde{\chi}}^l_\beta\}+\eps^{\ald\bed}\{\zero{\tilde{\bar{\chi}}}_{i\ald},
  \zero{\tilde{\bar{\chi}}}_{j\bed}\}\\
&
\kern-5.2cm-\sum_{|I|=|J|}(-1)^{\tilde{I}}\left\{\eps^{\alpha\beta}
  \eps^{\ald\bed}[{\tilde{A}}_{\alpha\ald|I},(\tilde{\nabla}_{\beta\bed}
  \tilde{\phi}_{ij})_J]
  -{\tfrac{1}{4}}\eps^{klmn}[\tilde{\phi}_{mn|I}~,
  [\tilde{\phi}_{kl},\tilde{\phi}_{ij}]_J]
  \right\}\,T^{IJ}\\
& \kern-2.7cm+\sum_{|I|=|J|}\left\{{\tfrac{1}{2}}\eps_{ijkl}
  \eps^{\alpha\beta}\{{\tilde{\chi}}^k_{\alpha|I},{\tilde{\chi}}\
  ^l_{\beta|J}\}+\eps^{\ald\bed}
  \{{\tilde{\bchi}}_{i\ald|I},{\tilde{\bchi}}_{j\bed|J}\}\right\}\,T^{IJ}~.
\end{aligned}
\end{equation*}

\paragraph{Deformed $\CN=4$ multiplet.} It now remains to calculate
the bodies of the deformed superfields. This is a rather lengthy
but straightforward calculation which yields the following
results:
\begin{equation}
\begin{aligned}
\zero{\tilde{\phi}}_{ij}\ &\ =\ \
\zero{\phi}_{ij}+{\tfrac{\hbar}{2}}\,C^{m\delta,n\eps}\,
\eps_{\delta\eps}\{\zero{\phi}_{mi},\zero{\phi}_{jn}\}+ \CO(\hbar^2)~,\\
\zero{\tilde{A}}_{\alpha\bed}\ &\ =\ \
\zero{A}_{\alpha\bed}+{\tfrac{\hbar}{4}}\,C^{m\delta,n\eps}\,
\eps_{\alpha\delta}\{\zero{\phi}_{mn},\zero{A}_{\eps\bed}\}
+\CO(\hbar^2)~,\\
\zero{\tilde{\bar{\chi}}}_{i\bed}\ &\ =\ \
\zero{\bar{\chi}}_{i\bed}+{\tfrac{\hbar}{96}}\,
C^{m\delta,n\eps}\,[
11\eps_{\delta\eps}(\{\zero{\phi}_{mn},\zero{\bar{\chi}}_{i\bed}\}-
2\{\zero{\phi}_{in},\zero{\bar{\chi}}_{m\bed}\})\\
&\kern1.5cm-5(\eps_{mnij}\{\zero{A}_{\delta\bed},\zero{\chi}^j_\eps\})
]+\CO(\hbar^2)~,\\
\zero{\tilde{\chi}}^i_\beta\ &\ =\ \
\zero{\chi}^i_\beta+{\tfrac{\hbar}{16}}\,
C^{m\delta,n\eps}\,[\{\zero{\phi}_{mn},{\tfrac{4}{3}}\eps_{\eps\delta}
\zero{\chi}^i_\beta-{\tfrac{11}{3}}\eps_{\delta\beta} \zero{\chi}^i_\eps\}\\
&\kern1.5cm-\delta^i_m\{\zero{\phi}_{ln},{\tfrac{4}{3}}\eps_{\eps\delta}
\zero{\chi}^l_\beta+{\tfrac{7}{3}}\eps_{\eps\beta}
\zero{\chi}^l_\delta-{\tfrac{2}{3}}\eps_{\delta\beta}\zero{\chi}^l_\eps\}\\
&\kern1.5cm-\eps_{\beta\eps}\eps^{\ald\bed}
\{\zero{A}_{\delta\ald},12\delta^i_m\zero{\bar{\chi}}_{n\bed}-
{\tfrac{1}{2}}\delta^i_n\zero{\bar{\chi}}_{m\bed}\}]+\CO(\hbar^2)~.
\end{aligned}
\end{equation}
To obtain the final equations of motion for the $\CN=4$ SYM
multiplet, one has to substitute these expressions into the
deformed field equations. All the remaining superfield components
can be replaced with the corresponding undeformed components, as
we are only interested in terms of first order in $\hbar$ and
$T^{IJ}$ is at least of this order. This will eventually give rise
to equations of the type
\begin{equation}
\eps^{\alpha\beta}\zero{\nabla}_{\alpha\ald}\zero{\chi}^i_\beta+{\tfrac{1}{2}}
\eps^{ijkl}[\zero{W}_{kl},\zero{\bar{\chi}}_{j\ald}]  \ =\ 
\CO(\hbar)~,
\end{equation}
but actually performing this task leads to both unenlightening and
complicated looking expressions, so we refrain from writing them
down. To proceed in a realistic manner, one can constrain the
deformation parameters to obtain manageable equations of motion.

For instance, in order to compare the deformed equations of motion
with Seiberg's deformed $\CN=1$ equations\footnote{or similarly in
the case of the deformed $\CN=2$ equations in $\CN=1$ superspace\index{super!space}
language \cite{Araki:2003se}} \cite{Seiberg:2003yz}, one would
have to restrict the deformation matrix $C^{\alpha i,\beta j}$
properly and to put some of the fields, e.g., $\tilde{\phi}_{ij}$,
to zero.

\paragraph{Remarks on the Seiberg-Witten map.} Generalizing the\index{Seiberg-Witten map}
string theory side of the derivation of Seiberg-Witten maps seems\index{string theory}
to be nontrivial. The graviphoton used to deform the fermionic\index{graviphoton}
coordinates belongs to the R-R sector, while the gauge field
strength causing the deformation in the bosonic case sits in the\index{field strength}
NS-NS sector. This implies that the field strengths appear on
different footing in the vertex operators of the appropriate
string theory (type II with $\CN=2$, $d=4$). A first step might be\index{string theory}
to consider a ``pure gauge'' configuration in which the gluino and\index{gauge!pure gauge}
gluon field strengths vanish. The corresponding vertex operator in\index{field strength}
Berkovits' hybrid formalism on the boundary of the worldsheet of\index{worldsheet}
an open string contains the terms\index{open string}
\begin{equation}
V\ =\ {\tfrac{1}{2\alpha'}}\int \dd\tau~
(\dot{\theta}^\alpha\omega_\alpha+\dot{X}^\mu A_\mu
-\di\sigma^\mu_{\alpha\ald}\dot{\theta}^\alpha\btheta^\ald
A_\mu)~,
\end{equation}
with the formal (classical) gauge transformations\index{gauge transformations}\index{gauge!transformation}
$\delta_\lambda\omega_\alpha=D_\alpha\lambda$ and $\delta_\lambda
A_\mu=\dpar_\mu\lambda$. From here, one may proceed exactly as in
\cite{Seiberg:1996vs} using the deformation of
\cite{Seiberg:2003yz}: regularization of the action by
Pauli-Villars\footnote{Pauli-Villars was applied to supergravity,
e.g., in \cite{Gaillard:1994mn}.} and point-splitting procedures
lead to an undeformed and a deformed gauge invariance,
respectively. Although on flat Euclidean space, pure gauge is\index{gauge!pure gauge}
trivial, the two different gauge transformations obtained are not.\index{gauge transformations}\index{gauge!transformation}

More general, a Seiberg-Witten map is a translation rule between\index{Seiberg-Witten map}
two physically equivalent field theories. The fact that our choice
of deformation generically breaks half of the supersymmetry is not\index{super!symmetry}
in contradiction with the existence of a Seiberg-Witten map, but\index{Seiberg-Witten map}
may be seen analogously to the purely bosonic case: in both the
commutative and noncommutative theories, particle Lorentz
invariance is broken which is due to the background field\index{Lorentz!invariance}
($B$-field).

\section{Drinfeld-twisted supersymmetry}\label{sDrinfeldTwistedSupersymmetry}\index{Twisted supersymmetry}\index{super!symmetry}

Another development which attracted much attention recently began
with the realization that noncommutative field theories, although\index{noncommutative field theories}
manifestly breaking Poincar{\'e} symmetry, can be recast into a form
which is invariant under a twist-deformed action of the Poincar{\'e}
algebra \cite{Oeckl:2000eg,Chaichian:2004za,Chaichian:2004yh}. In\index{Poincar{\'e} algebra}
this framework, the commutation relation $[x^{\mu},x^{\nu}]=\di
\Theta^{\mu \nu}$ is understood as a result of the non-{\it
co}commutativity of the coproduct of a twisted Hopf Poincar{\'e}\index{coproduct}
algebra acting on the coordinates. This result can be used to show
that the representation content of Moyal-Weyl-deformed theories is\index{representation}
identical to that of their undeformed Lorentz invariant
counterparts. Furthermore, theorems in quantum field theory which
require Lorentz invariance for their proof can now be carried over\index{Lorentz!invariance}
to the Moyal-Weyl-deformed case using twisted Lorentz invariance.
For related works, see also
\cite{Banerjee:2004ev,Koch:2004ud,Dimitrijevic:2004rf,Matlock:2005zn,
Gonera:2005hg,Aschieri:2005yw,Calmet:2005qm,Lukierski:2005fc,Aschieri:2005zs,Chaichian:2005yp}.

The following section is based on the paper \cite{Ihl:2005zd} and
presents an extension of the analysis of
\cite{Chaichian:2004za,Chaichian:2004yh} to supersymmetric field
theories on non-anticommutative superspaces. We will use\index{non-anticommutative superspace}\index{super!space}
Drinfeld-twisted supersymmetry to translate properties of these\index{Twisted supersymmetry}\index{super!symmetry}
field theories into the non-anticommutative situation, where half
of the supersymmetries are broken.

Note that Drinfeld-twisted supersymmetry was already considered in\index{Twisted supersymmetry}\index{super!symmetry}
the earlier publication \cite{Kobayashi:2004ep} and there is some
overlap with our discussion in the case $\CN=1$. The analysis of
extended supersymmetries presented in this reference differs from
the one we will propose here. Furthermore, our discussion will
include several new applications of the re-gained twisted
supersymmetry. In the paper \cite{Zupnik:2005ut}, which appeared\index{Twisted supersymmetry}\index{super!symmetry}
simultaneously with \cite{Ihl:2005zd}, Drinfeld-twisted
$\CN=(1,1)$ supersymmetry has been discussed. More recent work in\index{super!symmetry}
this area is, e.g.,
\cite{Zheltukhin:2005wq,Banerjee:2005ig,Zheltukhin:2005yy,Zupnik:2006qa,Qureshi:2006qg}.

\subsection{Preliminary remarks}

\paragraph{Hopf algebra.} A {\em Hopf algebra} is\index{Hopf algebra}
an algebra $H$ over a field $\FK$ together with a {\em product}
$m$, a {\em unit} $\unit$, a {\em coproduct} $\Delta:H\rightarrow\index{coproduct}
H\otimes H$ satisfying $(\Delta\otimes\id)\Delta=(\id\otimes
\Delta)\Delta$, a {\em counit} $\eps:H\rightarrow \FK$ satisfying
$(\eps\otimes\id)\Delta=\id$ and $(\id\otimes\eps)\Delta=\id$ and
an {\em antipode} $S:H\rightarrow H$ satisfying
$m(S\otimes\id)\Delta=\eps\unit$ and $m(\id\otimes S)\Delta=\eps
\unit$. The maps $\Delta$, $\eps$ and $S$ are unital maps, that is
$\Delta(\unit)=\unit\otimes \unit$, $\eps(\unit)=1$ and
$S(\unit)=\unit$.

\paragraph{Hopf superalgebra.} Recall from section\index{Hopf superalgebra}\index{super!algebra}
\ref{ssSupergeneralities} that a {\em superalgebra} is a
supervector space endowed with {\em i)} an associative\index{super!vector space}
multiplication respecting the grading and {\em ii)} the graded
commutator $\lsc a,b\rsc=ab-(-1)^{\tilde{a}\tilde{b}}ba$. We fix
the following rule for the interplay between the multiplication
and the tensor product $\otimes$ in a superalgebra:\index{super!algebra}
\begin{equation}
(a_1\otimes a_2)(b_1\otimes b_2)\ =\
(-1)^{\tilde{a}_2\tilde{b}_1}(a_1b_1\otimes a_2b_2)~.
\end{equation}

A superalgebra is called a {\em Hopf superalgebra} if it is\index{Hopf superalgebra}\index{super!algebra}
endowed with a graded coproduct\footnote{In Sweedler's notation\index{coproduct}
with $\Delta(a)=\sum a_{(1)}\otimes a_{(2)}$, this amounts to
$\tilde{a}\equiv\tilde{a}_{(1)}+\tilde{a}_{(2)} \mod 2$.} $\Delta$
and a counit $\eps$, both of which are graded algebra morphisms,
i.e.\
\begin{equation}
\Delta(ab)\ =\ \sum
(-1)^{\tilde{a}_{(2)}\tilde{b}_{(1)}}a_{(1)}b_{(1)}\otimes
a_{(2)}b_{(2)}\eand \eps(ab)\ =\ \eps(a)\eps(b)~,
\end{equation}
and an antipode $S$ which is a graded algebra anti-morphism, i.e.\
\begin{equation}
S(ab)\ =\ (-1)^{\tilde{a}\tilde{b}}S(b)S(a)~.
\end{equation}
As usual, one furthermore demands that $\Delta$, $\eps$ and $S$
are unital maps, that $\Delta$ is coassociative and that $\eps$
and $S$ are counital. For more details, see \cite{Brouder:2004}
and references therein.

\paragraph{An extended graded Baker-Campbell-Hausdorff formula}\index{Baker-Campbell-Hausdorff formula}
First, note that $\de^{A\otimes B}\de^{-A\otimes B}$ is indeed
equal to $\unit\otimes\unit$ for any two elements $A,B$ of a
superalgebra. This is clear for $\tilde{A}=0$ or $\tilde{B}=0$.\index{super!algebra}
For $\tilde{A}=\tilde{B}=1$ it is most instructively gleaned from
\begin{equation*}
\left(\unit\otimes\unit+A\otimes B-\tfrac{1}{2}A^2\otimes
B^2+\ldots\right)\left(\unit\otimes\unit-A\otimes
B-\tfrac{1}{2}A^2\otimes B^2-\ldots\right)\ =\ \unit\otimes\unit~.
\end{equation*}

Now, for elements $A_I,B_J,D$ of a graded algebra, where the
parities of the elements $A_I$ and $B_J$ are all equal
$\tilde{A}=\tilde{A}_I=\tilde{B}_J$ and $\lsc A_I,A_J\rsc=\lsc
B_I,B_J\rsc=0$, we have the relation
\begin{align}\label{bchformula}
&\de^{C^{IJ}A_I\otimes B_J} \left(D\otimes \unit\right)
\de^{-C^{KL}A_K\otimes B_L}\\\nonumber&\hspace{1cm} \ =\
\sum_{n=0}^\infty
\frac{(-1)^{n\tilde{A}\tilde{D}+\frac{n(n-1)}{2}\tilde{A}}}{n!}C^{I_1J_1}\ldots
C^{I_nJ_n}\lsc A_{I_1},\lsc\ldots\lsc A_{I_n},D\rsc\rsc\rsc
\otimes B_{J_1}\ldots B_{J_n}~.
\end{align}
{\em Proof:} To verify this relation, one can simply adapt the
well-known iterative proof via a differential equation. First note
that
\begin{equation}
\de^{\lambda C^{IJ}A_I\otimes B_J}(C^{KL}A_K \otimes B_L)\ =\
(C^{KL}A_K\otimes B_L)\de^{\lambda C^{IJ}A_I\otimes B_J}~.
\end{equation}
Then define the function
\begin{equation}
F(\lambda)\ :=\ \de^{\lambda C^{IJ}A_I\otimes B_J}(D\otimes
1)\de^{-\lambda C^{KL}A_K\otimes B_L}~,
\end{equation}
which has the derivative
\begin{align}
\dder{\lambda}F(\lambda)\ =\ (C^{MN}A_M\otimes B_N)&\de^{\lambda
C^{IJ}A_I\otimes B_J}(D\otimes 1)\de^{-\lambda C^{KL}A_K\otimes
B_L}\\\nonumber&-\de^{\lambda C^{IJ}A_I\otimes B_J}(D\otimes
1)\de^{-\lambda C^{KL}A_K\otimes B_L}(C^{MN}A_M\otimes B_N)~.
\end{align}
Thus, we have the identity $\dder{\lambda}
F(\lambda)=[(C^{MN}A_M\otimes B_N),F(\lambda)]$, which, when
applied recursively together with the Taylor formula, leads to
\begin{equation}
F(1)\ =\ \sum_{n=0}^\infty \frac{1}{n!}
\left[C^{I_1J_1}A_{I_1}\otimes
B_{J_1}\left[\ldots\left[C^{I_nJ_n}A_{I_n}\otimes
B_{J_n},D\otimes\unit\right]\ldots\right]\right]~.
\end{equation}
Also recursively, one easily checks that
\begin{align}
&\left[C^{I_1J_1}A_{I_1}\otimes
B_{J_1}\left[\ldots\left[C^{I_nJ_n}A_{I_n}\otimes
B_{J_n},D\otimes\unit\right]\ldots\right]\right]\\\nonumber&\hspace{2cm}\
=\ (-1)^{\tilde{A}\tilde{D}}(-1)^\kappa C^{I_1J_1}\ldots
C^{I_nJ_n}\lsc A_{I_1},\lsc\ldots\lsc A_{I_n},D\rsc\rsc\rsc
\otimes B_{J_1}\ldots B_{J_n}~,
\end{align}
where $\kappa$ is given by
$\kappa=(n-1)\tilde{A}+(n-2)\tilde{A}+\ldots+\tilde{A}$.
Furthermore, we have
\begin{equation}
(-1)^\kappa\ =\ (-1)^{n^2-\sum_{i=1}^n i}\ =\
(-1)^{n^2+\sum_{i=1}^n i}\ =\ (-1)^{\frac{n(n-1)}{2}}~,
\end{equation}
which, together with the results above, proves formula
\eqref{bchformula}. This extended graded Baker-Campbell-Hausdorff
formula also generalizes straightforwardly to the case when\index{Baker-Campbell-Hausdorff formula}
$D\otimes \unit$ is replaced by $\unit \otimes D$.

\subsection{Drinfeld twist of the Euclidean super Poincar{\'e}\index{Drinfeld twist}\index{twist}
algebra}\label{sec:drinfeld}

\paragraph{Euclidean super Poincar{\'e} algebra.} The starting point of\index{Euclidean super Poincar{\'e} algebra}\index{Poincar{\'e} algebra}\index{super!Poincar{\'e} algebra}
our discussion is the ordinary Euclidean super Poincar{\'e}
algebra\footnote{or inhomogeneous super Euclidean algebra} $\frg$\index{Euclidean super Poincar{\'e} algebra}\index{Poincar{\'e} algebra}\index{super!Poincar{\'e} algebra}
on $\ssp$ without central extensions, which generates the
isometries on the space $\ssp$. More explicitly, we have the
generators of translations $P_\mu$, the generators of
four-dimensional rotations $M_{\mu\nu}$ and the $4\CN$
supersymmetry generators $Q_{\alpha i}$ and $\bar{Q}_\ald^i$.\index{super!symmetry}
Recall from section \ref{ssSupersymmetryAlgebra} that they satisfy
the algebra
\begin{equation}\label{sepalgebra}
\begin{aligned}
\begin{aligned}
{}[P_\rho,M_{\mu\nu}]&\ =\
\di(\delta_{\mu\rho}P_\nu-\delta_{\nu\rho}P_\mu)~,\\
{}[M_{\mu\nu},M_{\rho\sigma}]&\ =\
-\di(\delta_{\mu\rho}M_{\nu\sigma}-\delta_{\mu\sigma}M_{\nu\rho}-
\delta_{\nu\rho}M_{\mu\sigma}+\delta_{\nu\sigma}M_{\mu\rho})~,
\end{aligned}\hspace{0.65cm}\\
\begin{aligned}
{}[P_\mu,Q_{\alpha i}]&\ =\ 0~,&[P_\mu,\bar{Q}_\ald^i]&\ =\ 0~,&\\
[M_{\mu\nu},Q_{i \alpha}]&\ =\ \di
(\sigma_{\mu\nu})_\alpha{}^\beta
Q_{i\beta}~,&[M_{\mu\nu},\bar{Q}^{i\ald}]&\ =\ \di
(\bar{\sigma}_{\mu\nu})^\ald{}_\bed
\bar{Q}^{i\bed}~,\\
\{Q_{\alpha i},\bar{Q}^j_\bed\}&\ =\
2\delta^i_j\sigma_{\alpha\bed}^\mu P_\mu ~,&\{Q_{\alpha
i},Q_{\beta j}\}&\ =\ \{\bar{Q}^i_{\ald},\bar{Q}^j_\bed\}\ =\ 0~.
\end{aligned}
\end{aligned}
\end{equation}
Recall furthermore that the Casimir operators of the Poincar{\'e}\index{Casimir}
algebra used for labelling representations are $P^2$ and $W^2$,\index{representation}
where the latter is the square of the Pauli-Ljubanski operator
\begin{equation}
W_{\mu}\ =\ -\tfrac{1}{2} \eps_{\mu \nu \rho\sigma} M^{\nu \rho}
P^{\sigma}~.
\end{equation}
This operator is, however, not a Casimir of the super Poincar{\'e}\index{Casimir}
algebra; instead, there is a supersymmetric variant: the
(superspin) operator $\widetilde{C}^2$ defined as the square of\index{super!spin}
\begin{equation}
\widetilde{C}_{\mu \nu}\ =\ \widetilde{W}_{\mu} P_{\nu} -
\widetilde{W}_{\nu} P_{\mu},
\end{equation}
where $\widetilde{W}_{\mu}:= W_{\mu}- \frac{1}{4} \bar{Q}^i_{\ald}
\bar{\sigma}^{\ald \alpha}_{\mu}Q_{i \alpha}$.

\paragraph{Universal enveloping algebra.} A {\em universal\index{universal enveloping algebra}
enveloping algebra}\, $\CU(\fra)$ of a Lie algebra $\fra$ is an\index{Lie algebra}
associative unital algebra together with a Lie algebra
homomorphism $h:\fra\rightarrow \CU(\fra)$, satisfying the\index{Lie algebra}
following universality property: For any further associative
algebra $A$ with homomorphism $\phi:\fra\rightarrow A$, there
exists a unique homomorphism $\psi:\CU(\fra)\rightarrow A$ of
associative algebras, such that $\phi=\psi\circ h$. Every Lie
algebra has an universal enveloping algebra, which is unique up to\index{Lie algebra}\index{universal enveloping algebra}
algebra isomorphisms.\index{morphisms!isomorphism}

\paragraph{The universal enveloping algebra of $\frg$.} The universal\index{universal enveloping algebra}
enveloping algebra $\CU(\frg)$ of the Euclidean super Poincar{\'e}
algebra $\frg$ is a cosupercommutative Hopf superalgebra with\index{Euclidean super Poincar{\'e} algebra}\index{Hopf superalgebra}\index{Poincar{\'e} algebra}\index{super!Poincar{\'e} algebra}\index{super!algebra}
counit and coproduct defined by $\eps(\unit)=1$ and $\eps(x)=0$\index{coproduct}
otherwise, $\Delta(\unit)=\unit\otimes \unit$ and
$\Delta(x)=\unit\otimes x+x\otimes \unit$ otherwise.

\paragraph{Drinfeld twist.} Given a Hopf algebra $H$ with coproduct\index{Drinfeld twist}\index{Hopf algebra}\index{coproduct}
$\Delta$, a counital 2-cocycle $\CF$ is a counital element of
$H\otimes H$, which has an inverse and satisfies
\begin{equation}\label{cocyclecond}
\CF_{12}(\Delta\otimes\id)\CF\ =\ \CF_{23}(\id\otimes\Delta)\CF~,
\end{equation}
where we used the common shorthand notation
$\CF_{12}=\CF\otimes\unit$, $\CF_{23}=\unit\otimes\CF$ etc. As
done in \cite{Chaichian:2004za}, such a counital 2-cocycle $\CF\in
H\otimes H$ can be used to define a twisted Hopf
algebra\footnote{This twisting amounts to constructing a\index{Hopf algebra}
quasitriangular Hopf algebra, as discussed, e.g., in
\cite{Chari:1994pz}.} $H^\CF$ with a new coproduct given by\index{coproduct}
\begin{equation}
\Delta^\CF(Y)\ :=\ \CF\Delta(Y)\CF^{-1}~.
\end{equation}
The element $\CF$ is called a {\em Drinfeld twist}; such a\index{Drinfeld twist}
construction was first considered in \cite{Drinfeld:1989st}.

\paragraph{The Drinfeld twist of $\CU(\frg)$.} For our purposes,\index{Drinfeld twist}\index{twist}
i.e.\ to recover the canonical algebra of non-anticommutative
coordinates \eqref{orddecnac}, we choose the Abelian twist
$\CF\in\CU(\frg)\otimes\CU(\frg)$ defined by\index{twist}
\begin{equation}
\CF\ =\ \exp\left(-\frac{\hbar}{2} C^{\alpha i,\beta j} Q_{\alpha
i}\otimes Q_{\beta j}\right)~.
\end{equation}
As one easily checks, $\CF$ is indeed a counital 2-cocycle: First,
it is invertible and its inverse is given by
$\CF^{-1}=\exp\left(\frac{\hbar}{2} C^{\alpha i,\beta j} Q_{\alpha
i}\otimes Q_{\beta j}\right)$. (Because the $Q_{\alpha i}$ are
nilpotent, $\CF$ and $\CF^{-1}$ are not formal series but rather
finite sums.) Second, $\CF$ is counital since it satisfies the
conditions
\begin{equation}
(\eps\otimes \id)\CF\ =\ \unit\eand(\id\otimes \eps)\CF\ =\
\unit~,
\end{equation}
as can be verified without difficulty. Also, the remaining cocycle
condition \eqref{cocyclecond} turns out to be fulfilled since
\begin{equation}
\begin{aligned}
\CF_{12}(\Delta\otimes\id)\CF&\ =\
\CF_{12}\exp\left(-\frac{\hbar}{2} C^{\alpha i,\beta j}(Q_{\alpha
i}\otimes \unit+\unit\otimes
Q_{\alpha i})\otimes Q_{\beta j}\right)~,\\
\CF_{23}(\id\otimes\Delta)\CF&\ =\
\CF_{23}\exp\left(-\frac{\hbar}{2} C^{\alpha i,\beta j}Q_{\alpha
i}\otimes(Q_{\beta j}\otimes \unit+\unit\otimes Q_{\beta
j})\right)
\end{aligned}
\end{equation}
yields, due to the (anti)commutativity of the $Q_{\alpha i}$,
\begin{equation}
\CF_{12}\CF_{13}\CF_{23}\ =\ \CF_{23}\CF_{12}\CF_{13}~,
\end{equation}
which is obviously true.

\paragraph{Twisted multiplication and coproduct.} Note that after\index{coproduct}
introducing this Drinfeld twist, the multiplication in $\CU(\frg)$\index{Drinfeld twist}
and the action of $\frg$ on the coordinates remain the same. In
particular, the representations of the twisted and the untwisted\index{representation}
algebras are identical. It is only the action of $\CU(\frg)$ on
the tensor product of the representation space, given by the\index{representation}
coproduct, which changes.\index{coproduct}

Let us be more explicit on this point: the coproduct of the\index{coproduct}
generator $P_\mu$ does not get deformed, as $P_\mu$ commutes with
$Q_{\beta j}$:
\begin{equation}\label{eq:P}
\Delta^\CF(P_\mu)\ =\ \Delta(P_\mu)~.
\end{equation}
For the other generators of the Euclidean super Poincar{\'e} algebra,\index{Euclidean super Poincar{\'e} algebra}\index{Poincar{\'e} algebra}\index{super!Poincar{\'e} algebra}
the situation is slightly more complicated. Due to the rule
$(a_1\otimes a_2)(b_1\otimes
b_2)=(-1)^{\tilde{a}_2\tilde{b}_1}(a_1b_1\otimes a_2 b_2)$, where
$\tilde{a}$ denotes the Gra{\ss}mann parity of $a$, we have the\index{parity}
relations\footnote{Here, $I_k$ and $J_k$ are multi-indices, e.g.\
$I_k=i_k \alpha_k$.} (cf.\ equation \eqref{bchformula})
\begin{align}\nonumber
&\CF \left(D\otimes \unit\right)\CF^{-1}\ =\
\\\nonumber&\hspace{1cm} \sum_{n=0}^\infty
\frac{(-1)^{n\tilde{D}+\frac{n(n-1)}{2}}}{n!}\left(-\frac{\hbar}{2}\right)^nC^{I_1J_1}\ldots
C^{I_nJ_n}\lsc Q_{I_1},\lsc\ldots\lsc Q_{I_n},D\rsc\rsc\rsc
\otimes Q_{J_1}\ldots Q_{J_n}~,\\ &\CF \left(\unit\otimes
D\right)\CF^{-1}\ =\ \\&\hspace{1cm} \sum_{n=0}^\infty
\frac{(-1)^{n\tilde{D}+\frac{n(n-1)}{2}}}{n!}\left(-\frac{\hbar}{2}\right)^nC^{I_1J_1}\ldots
C^{I_nJ_n} Q_{I_1}\ldots Q_{I_n}\otimes \lsc
Q_{J_1},\lsc\ldots\lsc Q_{J_n},D\rsc\rsc\rsc~,\nonumber
\end{align}
where $\lsc\cdot,\cdot\rsc$ denotes the graded commutator. {}From
this, we immediately obtain
\begin{equation}
\Delta^\CF(Q_{\alpha i})\ =\ \Delta(Q_{\alpha i})~.
\end{equation}
Furthermore, we can also derive the expressions for
$\Delta^\CF(M_{\mu\nu})$ and $\Delta^\CF(\bar{Q}_\gad^k)$, which
read
\begin{align}
\Delta^\CF(M_{\mu\nu}) &\ =\ \Delta(M_{\mu\nu})+\frac{\di
\hbar}{2} C^{\alpha i,\beta j} \left[
(\sigma_{\mu\nu})_\alpha{}^\gamma Q_{i\gamma} \otimes Q_{\beta
j}+Q_{\alpha i} \otimes (\sigma_{\mu\nu})_\beta{}^\gamma
 Q_{j\gamma}\right]~, \\
\Delta^\CF(\bar{Q}^k_{\gad})&\ =\ \Delta(\bar{Q}^k_{\gad}) + \hbar
C^{\alpha i,\beta j} \left[ \delta^k_i\sigma_{\alpha\gad}^\mu
P_\mu \otimes Q_{\beta j}+Q_{\alpha i} \otimes \delta^k_j
\sigma_{\beta\gad}^\mu P_\mu \right].
\end{align}
The twisted coproduct of the Pauli-Ljubanski operator $W_{\mu}$\index{coproduct}
becomes
\begin{equation}\label{eq:Wtwist}
\Delta^\CF(W_{\mu})\ =\ \Delta(W_{\mu}) -\frac{\di \hbar}{4}
C^{\alpha i , j \beta} \eps_{\mu \nu \rho\sigma} \left( Q_{i
\alpha} \otimes (\sigma^{\nu \rho})_{\beta}{}^{\gamma} Q_{j
\gamma} P^{\sigma} + (\sigma^{\nu \rho})_{\alpha}{}^{\gamma} Q_{i
\gamma} P^{\sigma} \otimes Q_{j \beta} \right)~,
\end{equation}
while for its supersymmetric variant $\widetilde{C}_{\mu\nu}$, we
have
\begin{equation}\label{eq:C2}
\begin{aligned}
\Delta^\CF(\widetilde{C}_{\mu\nu}) &\ =\
\Delta(\widetilde{C}_{\mu\nu}) -\frac{\hbar}{2} C^{\alpha i , j
\beta} \left[ Q_{\alpha i} \otimes Q_{j \beta},
\Delta(\widetilde{C}_{\mu\nu})\right] \\ &\ =\
\Delta(\widetilde{C}_{\mu\nu})- \frac{\hbar}{2} C^{\alpha i , j
\beta} \left( \left[Q_{\alpha i},\widetilde{C}_{\mu\nu}\right]
\otimes Q_{j \beta} + Q_{i \alpha} \otimes
\left[Q_{j \beta},\widetilde{C}_{\mu\nu}\right]\right)\\
&\ =\ \Delta(\widetilde{C}_{\mu\nu})~,
\end{aligned}
\end{equation}
since $[Q_{i \alpha},\widetilde{C}_{\mu\nu}]=0$ by construction.

\paragraph{Representation on the algebra of superfunctions.} Given\index{representation}
a representation of the Hopf algebra $\CU(\frg)$ in an associative\index{Hopf algebra}
algebra consistent with the coproduct $\Delta$, one needs to\index{coproduct}
adjust the multiplication law after introducing a Drinfeld twist.\index{Drinfeld twist}
If $\CF^{-1}$ is the inverse of the element
$\CF\in\CU(\frg)\otimes \CU(\frg)$ generating the twist, the new
product compatible with $\Delta^\CF$ reads
\begin{equation}\label{multiplication}
a\star b\ :=\ m^\CF (a\otimes b)\ :=\ m\circ \CF^{-1}(a\otimes
b)~,
\end{equation}
where $m$ denotes the ordinary product $m(a\otimes b)=ab$.

Let us now turn to the representation of the Hopf superalgebra\index{Hopf superalgebra}\index{representation}\index{super!algebra}
$\CU(\frg)$ on the algebra
$\CS:=C^{\infty}(\FR^4)\otimes\Lambda_{4\CN}$ of superfunctions on
$\FR^{4|4\CN}$. On $\CS$, we have the standard representation of\index{representation}
the super Poincar{\'e} algebra in chiral coordinates\index{Poincar{\'e} algebra}\index{chiral!coordinates}\index{super!Poincar{\'e} algebra}
$(y^\mu,\theta^{\alpha i},\btheta^\ald_i)$:
\begin{equation}
\begin{aligned}
P_\mu f&\ =\ \di\dpar_\mu f~,
&M_{\mu\nu}f&\ =\ \di(y_\mu\dpar_\nu-y_\nu\dpar_\mu)f~,\\
Q_{\alpha i}f&\ =\ \der{\theta^{\alpha i}}f~,& \bar{Q}^i_\ald f&\
=\ \left(-\der{\btheta^\ald_i}f+2\di\theta^{\alpha
i}\sigma^\mu_{\alpha\ald}\dpar_\mu\right) f~,
\end{aligned}
\end{equation}
where $f$ is an element of $\CS$. After the twist, the
multiplication $m$ becomes the twist-adapted multiplication
$m^\CF$ \eqref{multiplication}, which reproduces the coordinate
algebra of $\sspdef$, e.g.\ we have
\begin{equation}
\begin{aligned}
\{\theta^{\alpha i}\stackrel{\star}{,}\theta^{\beta j}\}&\ :=\
m^\CF(\theta^{\alpha i}\otimes\theta^{\beta
j})+m^\CF(\theta^{\beta j}\otimes\theta^{\alpha i})\\&\ =\
\theta^{\alpha i}\theta^{\beta j}+\frac{\hbar}{2} C^{\alpha
i,\beta j} +\theta^{\beta j}\theta^{\alpha i}+\frac{\hbar}{2}
C^{\beta j,\alpha i}\\&\ =\ \hbar C^{\alpha i,\beta j}~.
\end{aligned}
\end{equation}
Thus, we have constructed a representation of the Euclidean super\index{representation}
Poincar{\'e} algebra on $\sspdef$ by employing $\CS_\star$, thereby\index{Poincar{\'e} algebra}
making twisted supersymmetry manifest.\index{Twisted supersymmetry}\index{super!symmetry}

\subsection{Applications}\label{ssapplications}

We saw in the above construction of the twisted Euclidean super
Poincar{\'e} algebra that our description is equivalent to the\index{Euclidean super Poincar{\'e} algebra}\index{Poincar{\'e} algebra}\index{super!Poincar{\'e} algebra}
standard treatment of Moyal-Weyl-deformed superspace. We can\index{super!space}
therefore use it to define field theories via their Lagrangians,
substituting all products by star products, which then will be
invariant under twisted super Poincar{\'e} transformations. This can
be directly carried over to quantum field theories, replacing the
products between operators by star products. Therefore, twisted
super Poincar{\'e} invariance, in particular twisted supersymmetry,\index{Twisted supersymmetry}\index{super!symmetry}
will always be manifest.

As a consistency check, we want to show that the tensor $C^{\alpha
i,\beta j}:=\{\theta^{\alpha i},\theta^{\beta j}\}_{\star}$ is
invariant under twisted super Poincar{\'e} transformations before
tackling more advanced issues. Furthermore, we want to relate the
representation content of the deformed theory with that of the\index{representation}
undeformed one by scrutinizing the Casimir operators of this\index{Casimir}
superalgebra. Eventually, we will turn to supersymmetric\index{super!algebra}
Ward-Takahashi identities and their consequences for\index{Ward-Takahashi identities}
renormalizability.

\paragraph{Invariance of $C^{\alpha i,\beta j}$.} The action of the
twisted supersymmetry charge on $C^{\alpha i,\beta j}$ is given by\index{Twisted supersymmetry}\index{super!symmetry}
\begin{equation}
\begin{aligned}
\hbar Q^\CF_{k\gamma} C^{\alpha i, \beta j} &\ =\
Q^\CF_{k\gamma}\left(\{\theta^{\alpha i}\stackrel{\star}{,}\theta^{\beta j}\}\right)\\
&\ :=\ m^\CF\circ\left(\Delta^\CF(Q_{k\gamma})(\theta^{\alpha
i}\otimes\theta^{\beta j}+
\theta^{j \beta} \otimes \theta^{i \alpha})\right)\\
&\ =\ m^\CF\circ\left(\Delta(Q_{k\gamma})(\theta^{\alpha
i}\otimes\theta^{\beta j}+ \theta^{j \beta} \otimes \theta^{i
\alpha})\right)\\ &\ =\ m \circ \CF^{-1} ( \delta_k^i
\delta_{\gamma}^{\alpha} \otimes \theta^{j \beta} +\delta_k^j
\delta_{\gamma}^{\beta} \otimes \theta^{i \alpha} - \theta^{i
\alpha} \otimes \delta_k^j \delta_{\gamma}^{\beta} - \theta^{j
\beta} \otimes \delta_k^i \delta_{\gamma}^{\alpha})\\ &\ =\ m (
\delta_k^i \delta_{\gamma}^{\alpha} \otimes \theta^{j \beta}
+\delta_k^j \delta_{\gamma}^{\beta} \otimes \theta^{i \alpha} -
\theta^{i \alpha} \otimes \delta_k^j \delta_{\gamma}^{\beta}
- \theta^{j \beta} \otimes \delta_k^i \delta_{\gamma}^{\alpha})\\
&\ =\ 0~.
\end{aligned}
\end{equation}
Similarly, we have
\begin{equation}
\begin{aligned}
\hbar (\bar{Q}^k_{\gad})^\CF C^{\alpha i,\beta j}&\ =\
m^\CF\circ\left(\Delta^\CF(\bar{Q}^k_{\gad}) (\theta^{\alpha
i}\otimes\theta^{\beta j}+ \theta^{\beta j} \otimes \theta^{\alpha
i})\right)\\ &\ =\ m^\CF\circ\left(\Delta(\bar{Q}^k_{\gad})
(\theta^{\alpha i}\otimes\theta^{\beta j}+
\theta^{\beta j} \otimes \theta^{\alpha i})\right)\\
&\ =\ 0~,
\end{aligned}
\end{equation}
and
\begin{equation}
\hbar P^\CF_{\mu \nu} C^{\alpha i,\beta j}\ =\ m^\CF\circ
\left(\Delta(P_{\mu}) (\theta^{\alpha i}\otimes\theta^{\beta j}+
\theta^{\beta j} \otimes \theta^{\alpha i})\right)\ =\ 0~.
\end{equation}
For the action of the twisted rotations and boosts, we get
\begin{align}
\hbar M^\CF_{\mu\nu} C^{\alpha i,\beta j} &\ =\ m^\CF \circ
\left(\Delta^\CF(M_{\mu \nu})(\theta^{\alpha i}\otimes
\theta^{\beta j}+ \theta^{\beta j} \otimes \theta^{\alpha i})
\right)\nonumber\\
&\ =\ m\circ\CF^{-1}\CF\Delta(M_{\mu\nu})\CF^{-1}(\theta^{\alpha
i}\otimes \theta^{\beta j}+ \theta^{\beta j} \otimes
\theta^{\alpha i})\\\nonumber &\ =\ m(\unit\otimes
M_{\mu\nu}+M_{\mu\nu}\otimes\unit)\left((\theta^{\alpha i}\otimes
\theta^{\beta j}+ \theta^{\beta j}\otimes \theta^{\alpha i})
-\hbar C^{\alpha i,\beta j}\unit\otimes\unit\right)\\\nonumber &\
=\ 0~,
\end{align}
where we made use of
$M_{\mu\nu}=\di(y_\mu\dpar_\nu-y_\nu\dpar_\mu)$. Thus, $C^{\alpha
i,\beta j}$ is invariant under the twisted Euclidean super
Poincar{\'e} transformations, which is a crucial check of the validity
of our construction.

\paragraph{Representation content.} An important feature of\index{representation}
noncommutative field theories was demonstrated recently\index{noncommutative field theories}
\cite{Chaichian:2004za,Chaichian:2004yh}: they share the same
representation content as their commutative counterparts. Of\index{representation}
course, one would expect this to also hold for non-anticommutative
deformations, in particular since the superfields defined, e.g.,
in \cite{Seiberg:2003yz} on a deformed superspace have the same\index{super!space}
set of components as the undeformed ones.

To decide whether the representation content in our case is the\index{representation}
same as in the commutative theory necessitates checking whether
the twisted action of the Casimir operators $P^2= P_{\mu} \star\index{Casimir}
P^{\mu}$ and $\widetilde{C}^2= \widetilde{C}_{\mu\nu} \star
\widetilde{C}^{\mu\nu}$ on elements of $\CU(\frg)\otimes\CU(\frg)
$ is altered with respect to the untwisted case. But since we have
already shown in (\ref{eq:P}) and (\ref{eq:C2}) that the actions
of the operators  $P_{\mu}$ and $\widetilde{C}_{\mu \nu}$ remain
unaffected by the twist, it follows immediately that the operators
$P^2$ and $\widetilde{C}^2$ are still Casimir operators in the\index{Casimir}
twisted case. Together with the fact that the representation space\index{representation}
considered as a module is not changed, this proves that the
representation content is indeed the same.\index{representation}

\paragraph{Chiral rings and correlation functions.} As discussed in\index{chiral!ring}
section \ref{ssquantumaspects}, the chiral rings of operators in
supersymmetric quantum field theories are cohomology rings of the
supercharges $Q_{i \alpha}$ and $\bar{Q}^i_\ald$ and correlation
functions which are built out of elements of a single such chiral
ring have peculiar properties.\index{chiral!ring}

In \cite{Seiberg:2003yz}, the anti-chiral ring was defined and\index{anti-chiral ring}\index{chiral!ring}
discussed for non-anticommutative field theories. The chiral ring,\index{non-anticommutative field theories}
however, lost its meaning: the supersymmetries generated by
$\bar{Q}^i_\ald$ are broken, cf.\ \eqref{defalgebra}, and
therefore the vacuum is expected to be no longer invariant under
this generator. Thus, the $\bar{Q}$-cohomology is not relevant for
correlation functions of chiral operators.

In our approach to non-anticommutative field theory, {\em twisted}
supersymmetry is manifest and therefore the chiral ring can be\index{chiral!ring}\index{super!symmetry}
treated similarly to the untwisted case as we want to discuss in
the following.

Let us assume that the Hilbert space $\CH$ of our quantum field
theory carries a representation of the Euclidean super Poincar{\'e}\index{representation}
algebra $\frg$, and that there is a unique, $\frg$ invariant
vacuum state $|0\rangle$. Although the operators $Q_{\alpha i}$
and $\bar{Q}^i_\ald$ are not related via Hermitian conjugation
when considering supersymmetry on Euclidean spacetime, it is still\index{super!symmetry}
natural to assume that the vacuum is annihilated by both
supercharges. The reasoning for this is basically the same as the
one employed in \cite{Seiberg:2003yz} to justify the use of
Minkowski superfields on Euclidean spacetime: one can obtain a
complexified supersymmetry algebra on Euclidean space from a\index{super!symmetry}
complexified supersymmetry algebra on Minkowski
space.\footnote{One can then perform all superspace calculations\index{super!space}
and impose suitable reality conditions on the component fields in
the end.} Furthermore, it has been shown that in the
non-anticommutative situation, just as in the ordinary undeformed
case, the vacuum energy of the Wess-Zumino model is not\index{Wess-Zumino model}
renormalized \cite{Britto:2003aj}.

We can now define the ring of chiral and anti-chiral operators by
the relations
\begin{equation}
\lsc \bar{Q}\stackrel{\star}{,}\CO \rsc\ =\ 0 \eand \lsc
Q\stackrel{\star}{,}\bar{\CO}\rsc\ =\ 0~,
\end{equation}
respectively. In a correlation function built from chiral
operators, $\bar{Q}$-exact terms, i.e.\ terms of the form $\lsc
\bar{Q}\stackrel{\star}{,}A\rsc$, do not contribute as is easily
seen from
\begin{align}\nonumber
\langle \lsc \bar{Q}\stackrel{\star}{,}A
\rsc\star{\CO}_1\star\ldots\star{\CO}_n\rangle\ =\ & \langle \lsc
\bar{Q}\stackrel{\star}{,}A\star{\CO}_1\star\ldots\star{\CO}_n\rsc\rangle\pm\langle
A\star\lsc \bar{Q}\stackrel{\star}{,}{\CO}_1
\rsc\star\ldots\star{\CO}_n\rangle\\&\pm\ldots\pm\langle\nonumber
A\star {\CO}_1\star\ldots\star\lsc \bar{Q}\stackrel{\star}{,}{\CO}_n\rsc \rangle\\
\ =\ &\langle \bar{Q}\nonumber
A\star{\CO}_1\star\ldots\star{\CO}_n\rangle\pm\langle
A\star{\CO}_1\star \ldots\star{\CO}_n \star\bar{Q}\rangle\\ \ =\ &
0~,\label{WardIdentity1}
\end{align}
where we used that $\bar{Q}$ annihilates both $\langle 0|$ and
$|0\rangle$, completely analogously to the case of untwisted
supersymmetry. Therefore, the relevant operators in the chiral\index{Twisted supersymmetry}\index{super!symmetry}
ring consist of the $\bar{Q}$-closed modulo the $\bar{Q}$-exact
operators. The same argument holds for the anti-chiral ring after\index{anti-chiral ring}\index{chiral!ring}
replacing $\bar{Q}$ with $Q$, namely
\begin{align}\nonumber
\langle \lsc Q\stackrel{\star}{,}A
\rsc\star\bar{\CO}_1\star\ldots\star\bar{\CO}_n\rangle\ =\ &
\langle \lsc
Q\stackrel{\star}{,}A\star\bar{\CO}_1\star\ldots\star\bar{\CO}_n\rsc\rangle\pm\langle
A\star\lsc Q\stackrel{\star}{,}\bar{\CO}_1
\rsc\star\ldots\star\bar{\CO}_n\rangle\\\nonumber&\pm\ldots\pm\langle
A\star \bar{\CO}_1\star\ldots\star\lsc
Q\stackrel{\star}{,}\bar{\CO}_n\rsc \rangle\\\nonumber \ =\
&\langle Q
A\star\bar{\CO}_1\star\ldots\star\bar{\CO}_n\rangle\pm\langle
A\star\bar{\CO}_1\star \ldots\star\bar{\CO}_n\star Q\rangle\\ \ =\
& 0~.\label{WardIdentity2}
\end{align}

\paragraph{Twisted supersymmetric Ward-Takahashi identities.} The\index{Ward-Takahashi identities}
above considered properties of correlation functions are
particularly useful since they imply a twisted supersymmetric
Ward-Takahashi identity: any derivative with respect to the
bosonic coordinates of an anti-chiral operator annihilates a
purely chiral or anti-chiral correlation function, cf.\ section
\ref{ssquantumaspects}, \ref{psusyWTI}. Recall that this is due to
the fact that $\dpar\sim \{Q,\bar{Q}\}$ and therefore any
derivative gives rise to a $Q$-exact term, which causes an
anti-chiral correlation function to vanish. Analogously, the
bosonic derivatives of chiral correlation functions vanish. Thus,
the correlation functions are independent of the bosonic
coordinates, and we can move the operators to a far distance of
each other, also in the twisted supersymmetric case:
\begin{equation}\label{clustering}
\langle\bar{\CO}_1(x_1)\star\ldots\star\bar{\CO}_n(x_n)\rangle\ =\
\langle\bar{\CO}_1(x^\infty_1)\rangle\star\ldots\star\langle\bar{\CO}_n(x^\infty_n)\rangle~.
\end{equation}
and we discover again that these correlation functions {\em
clusterize}.

Another direct consequence of \eqref{WardIdentity1} is the
holomorphic dependence of the chiral correlation functions on the
coupling constants, i.e.
\begin{equation}
\der{\bl}\langle\CO_1\star\ldots\star\CO_n\rangle\ =\ 0~.
\end{equation}
This follows in a completely analogous way to the ordinary
supersymmetric case, and for an example, see again
\ref{ssquantumaspects}, \ref{psusyWTI}.

\paragraph{Comments on non-renormalization theorems.}\label{pnonren}\index{non-renormalization theorems}\index{Theorem!non-renormalization}
A standard perturbative non-renormalization theorem for $\CN=1$
supersymmetric field theory states that every term in the
effective action can be written as an integral over
$\dd^2\theta\dd^2\bar{\theta}$. It has been shown in
\cite{Britto:2003aj} that this theorem also holds in the
non-anticommutative case. The same is then obviously true in our
case of twisted and therefore unbroken supersymmetry, and the\index{super!symmetry}
proof carries through exactly as in the ordinary case.

Furthermore, in a supersymmetric nonlinear sigma model, the\index{nonlinear sigma model}\index{sigma model}
superpotential is not renormalized. A nice argument for this fact
was given in \cite{Seiberg:1993vc}. Instead of utilizing Feynman
diagrams and supergraph techniques, one makes certain naturalness
assumptions about the effective superpotential. These assumptions
turn out to be strong enough to enforce a non-perturbative
non-renormalization theorem.\index{Theorem!non-renormalization}

In the following, let us demonstrate this argument in a simple
case, following closely \cite{Argyres}. Take a nonlinear sigma
model with superpotential\index{nonlinear sigma model}\index{sigma model}
\begin{equation}
\CW\ =\ \tfrac{1}{2}m\Phi^2+\tfrac{1}{3}\lambda \Phi^3~,
\end{equation}
where $\Phi=\phi+\sqrt{2}\theta\psi+\theta\theta F$ is an ordinary
chiral superfield. The assumptions we impose on the effective
action are the following:
\begin{itemize}
\setlength{\itemsep}{-1mm}
\item[$\triangleright$] Supersymmetry is also a symmetry of the\index{super!symmetry}
effective superpotential.
\item[$\triangleright$] The effective superpotential is holomorphic
in the coupling constants.
\item[$\triangleright$] Physics is smooth and regular under the possible
weak-coupling limits.
\item[$\triangleright$] The effective superpotential preserves the
$\sU(1)\times \sU(1)_R$ symmetry of the original superpotential
with charge assignments $\Phi:(1,1)$, $m:(-2,0)$,
$\lambda:(-3,-1)$ and $\dd^2\theta:(0,-2)$.
\end{itemize}
It follows that the effective superpotential must be of the form
\begin{equation}\label{eq:Weff}
\CW_{\mathrm{eff}}\ =\ m\Phi
W\left(\frac{\lambda\Phi}{m}\right)\ =\ \sum_i a_i \lambda^i
m^{1-i}\Phi^{i+2}~,
\end{equation}
where $W$ is an arbitrary holomorphic function of its argument.
Regularity of physics in the two weak-coupling limits
$\lambda\rightarrow 0$ and $m\rightarrow 0$ then implies that
$\CW_{\mathrm{eff}}=\CW$.

To obtain an analogous non-renormalization theorem in the\index{Theorem!non-renormalization}
non-anticommutative setting, we make similar assumptions about the
effective superpotential as above. We start from
\begin{equation}
\CW_{\star}\ =\ \tfrac{1}{2}m\Phi\star\Phi+\tfrac{1}{3}\lambda
\Phi\star\Phi\star\Phi~,
\end{equation}
and assume the following:
\begin{itemize}
\setlength{\itemsep}{-1mm}
\item[$\triangleright$] {\em Twisted} supersymmetry is a symmetry of the\index{super!symmetry}
effective superpotential. Note that this assumption is new
compared to the discussion in \cite{Britto:2003aj}. Furthermore,
arguments substantiating that the effective action can always be
written in terms of star products have been given in
\cite{Britto:2003aj2}.
\item[$\triangleright$] The effective superpotential is holomorphic
in the coupling constants. (This assumption is equally natural as
in the supersymmetric case, since it essentially relies on the
existence of chiral and anti-chiral rings, which we proved above\index{anti-chiral ring}\index{chiral!ring}
for our setting.)
\item[$\triangleright$] Physics is smooth and regular under the possible
weak-coupling limits.
\item[$\triangleright$] The effective superpotential preserves the
$\sU(1)\times \sU(1)_R$ symmetry of the original superpotential
with charge assignments $\Phi:(1,1)$, $m:(-2,0)$,
$\lambda:(-3,-1)$, $\dd^2\theta:(0,-2)$ and, additionally,
$C^{\alpha i,\beta j}:(0,2)$, $|C|\sim C^{\alpha i,\beta j}
C_{\alpha i,\beta j}: (0,4)$.
\end{itemize}
At first glance, it seems that one can now construct more
$\sU(1)\times \sU(1)_R$-symmetric terms in the effective
superpotential due to the new coupling constant $C$; however, this
is not true. Taking the $C\rightarrow 0$ limit, one immediately
realizes that $C$ can never appear in the denominator of any term.
Furthermore, it is not possible to construct a term containing $C$
in the nominator, which does not violate the regularity condition
in at least one of the other weak-coupling limits. Altogether, we
arrive at an expression similar to (\ref{eq:Weff})
\begin{equation}
\CW_{\mathrm{eff},\star}\ =\ \sum_i a_i \lambda^i m^{1-i}\Phi^{\star
i+2}~,
\end{equation}
and find that $\CW_{\mathrm{eff},\star}=\CW_\star$.

To compare this result with the literature, first note that, in a
number of papers, it has been shown that quantum field theories in
four dimensions with $\CN=\frac{1}{2}$ supersymmetry are\index{super!symmetry}
renormalizable to all orders in perturbation theory
\cite{Terashima:2003ri}-\cite{Berenstein:2003sr}. This even
remains true for generic $\CN=\frac{1}{2}$ gauge theories with
arbitrary coefficients, which do not arise as a
$\star$-deformation of $\CN =1$ theories. However, the authors of
\cite{Britto:2003aj,Grisaru:2003fd}, considering the
non-anticommutative Wess-Zumino model we discussed above, add\index{Wess-Zumino model}
certain terms to the action by hand, which seem to be necessary
for the model to be renormalizable. This would clearly contradict
our result $\CW_{\mathrm{eff},\star}=\CW_\star$. We conjecture
that this contradiction is merely a seeming one and that it is
resolved by a resummation of all the terms in the perturbative
expansion. A similar situation was encountered in
\cite{Britto:2003aj2}, where it was found that one could not write
certain terms of the effective superpotential using star products,
as long as they were considered separately. This obstruction,
however, vanished after a resummation of the complete perturbative
expansion and the star product was found to be sufficient to write
down the complete effective superpotential.

Clearly, the above result is stricter than the result obtained in
\cite{Britto:2003aj}, where less constraint terms in the effective
superpotential were assumed. However, we should stress that it is
still unclear to what extend the above assumptions on
$\CW_{\mathrm{eff},\star}$ are really natural. This question
certainly deserves further and deeper study, which we prefer to
leave to future work.

\chapter{Twistor Geometry}\label{chTwistorGeometry}

The main reason we will be interested in twistor geometry is its\index{twistor}
use in describing the solutions to certain Yang-Mills equations by
holomorphic vector bundles on a corresponding twistor space, which\index{holomorphic!vector bundle}\index{twistor}\index{twistor!space}
allows us to make contact with holomorphic Chern-Simons theory. We\index{Chern-Simons theory}
will completely ignore the gravity aspect of twistor theory.\index{twistor}

In this chapter, we will first deal with the twistor\index{twistor}
correspondence and its underlying geometrical structure. Then we
will discuss in detail the Penrose-Ward transform, which maps\index{Penrose-Ward transform}
solutions to certain gauge field equations to certain holomorphic
vector bundles over appropriate twistor spaces.\index{holomorphic!vector bundle}\index{twistor}\index{twistor!space}

The relevant literature to this chapter consists of
\cite{Penrosebooks,Wardbook,Masonbook,Woodhouse:id,TwistorPrimer,TwistorsAndGR,Popov:2004rb}.\index{twistor}

\section{Twistor basics}\index{twistor}

The twistor formalism was initially introduced by Penrose to give\index{twistor}
an appropriate framework for describing both general relativity
and quantum theory. For this, one introduces so-called twistors,\index{twistor}
which -- like the wave-function -- are intrinsically complex
objects but allow for enough algebraic structure to encode
spacetime geometry.

\subsection{Motivation}\label{ssmotivation}

\paragraph{Idea.} As mentioned above, the basic motivation of
twistor theory was to find a common framework for describing both\index{twistor}
general relativity and quantum mechanics. However, twistors found
a broad area of application beyond this, e.g.\ in differential
geometry.

In capturing both relativity and quantum mechanics, twistor theory\index{twistor}
demands some modifications of both. For example, it allows for the
introduction of nonlinear elements into quantum mechanics, which
are in agreement with some current interpretations of the
measurement process: The collapse of the wave-function contradicts
the principle of unitary time evolution, and it has been proposed
that this failure of unitarity is due to some overtaking nonlinear
gravitational effects.

The main two ingredients of twistor theory are non-locality in\index{twistor}
spacetime and analyticity (holomorphy) in an auxiliary complex
space, the {\em twistor space}. This auxiliary space can be\index{twistor}\index{twistor!space}
thought of as the space of light rays at each point in spacetime.\index{light ray}
Given an observer in a four-dimensional spacetime at a point $p$,
his {\em celestial sphere}, i.e.\ the image of planets, suns and\index{celestial sphere}
galaxies he sees around him, is the backward light cone at $p$
given by the 2-sphere
\begin{equation}
t\ =\ -1\eand x^2+y^2+z^2\ =\ 1~.
\end{equation}
From this, we learn that the twistor space of $\FR^4$ is\index{twistor}\index{twistor!space}
$\FR^4\times S^2$. On the other hand, this space will be
interpreted as the complex vector bundle $\CO(1)\oplus\CO(1)$ over\index{complex!vector bundle}
the Riemann sphere $\CPP^1$. The prescription for switching\index{Riemann sphere}
between the twistor space and spacetime is called the {\em\index{twistor}\index{twistor!space}
twistor} or {\em Klein correspondence}.\index{Klein correspondence}

Non-locality of the fields in a physical theory is achieved by
encoding the field information at a point in spacetime into
holomorphic functions on twistor space. By choosing an appropriate\index{twistor}\index{twistor!space}
description, one can cause the field equations to vanish on
twistor space, i.e.\ holomorphy of a function on twistor space\index{twistor}\index{twistor!space}
automatically guarantees that the corresponding field satisfies
its field equations.

\paragraph{Two-spinors.}\label{ptwospinors} Recall our convention\index{Spinor}\index{two-spinors}
for switching between vector and spinor indices:
\begin{equation}
x^{\alpha\ald}\ =\ -\di\sigma_\mu^{\alpha\ald}x^\mu\ =\ -\tfrac{\di}{2}\left(\begin{array}{cc}
-\di x^0-\di x^3 & -\di x^1- x^2\\ -\di x^1+ x^2 & -\di x^0+\di
x^3
\end{array}\right)~,
\end{equation}
where we used here the $\sigma$-matrices appropriate for signature
$($$-$$+$$+$$+$$)$. The inverse transformation is given by
$x^\mu=\frac{\di}{2}\tr(\sigma^\mu_{\alpha\ald}
(x^{\alpha\ald}))$.

The norm of such a vector is easily obtained from
$||x||^2=\eta_{\mu\nu}x^\mu
x^\nu=\det(x^{\alpha\ald})=\frac{1}{2}x_{\alpha\ald}x^{\alpha\ald}$.
From this formula, we learn that
\begin{equation}
\eta_{\mu\nu}\ =\ \tfrac{1}{2}\eps_{\alpha\beta}\eps_{\ald\bed}~,
\end{equation}
where $\eps_{\alpha\beta}$ is the antisymmetric tensor in two
dimensions. We choose again the convention
$\eps_{\ed\zd}=-\eps^{\ed\zd}=-1$ which implies that
$\eps_{\ald\bed}\eps^{\bed\gad}=\delta_\ald^\gad$, see also
section \ref{ssSuperspaces}, \ref{pNSuperspace}.

Recall now that any vector $x^{\mu}\sim x^{\alpha\ald}$ can be
decomposed into four (commuting) two-spinors according to\index{Spinor}\index{two-spinors}
\begin{equation}\label{spinordecomposition}
x^{\alpha\ald}\ =\ \tilde{\lambda}^\alpha\lambda^\ald+\kappa^\alpha\tilde{\kappa}^\ald~.
\end{equation}
If the vector $x^\mu$ is real then $\tilde{\lambda}$ and
$\tilde{\kappa}$ are related to $\lambda$ and $\kappa$ by complex
conjugation. If the real vector $x^\mu$ is a null and
future-pointing vector, its norm vanishes, and one can hence drop
one of the summands in \eqref{spinordecomposition}
\begin{equation}
x^{\alpha\ald}\ =\ \kappa^\alpha\bar{\kappa}^\ald~~~\mbox{with}~~~
\bar{\kappa}^\ald\ =\ \overline{\kappa^\alpha}~,
\end{equation}
where $\kappa^\alpha$ is an $\sSL(2,\FC)$-spinor, while\index{Spinor}
$\bar{\kappa}^\ald$ is an $\overline{\sSL(2,\FC)}$-spinor. Spinor
indices are raised and lowered with the $\eps$-tensor, i.e.\
\begin{equation}
\kappa_\alpha\ =\ \eps_{\alpha\beta}\kappa^\beta\eand
\bar{\kappa}^\ald\ =\ \eps^{\ald\bed}\bar{\kappa}_\bed~.
\end{equation}
As the spinors $\kappa^\alpha$ and $\bar{\kappa}^\ald$ are\index{Spinor}
commuting, i.e.\ they are {\em not} Gra{\ss}mann-valued, we have
\begin{equation}
\kappa_\alpha\kappa^\alpha\ =\ \bar{\kappa}^\ald\bar{\kappa}_\ald\ =\ 0~.
\end{equation}

\remark{
\begin{figure}[h]
\centerline{\includegraphics[width=4.5cm,totalheight=4.7cm]
{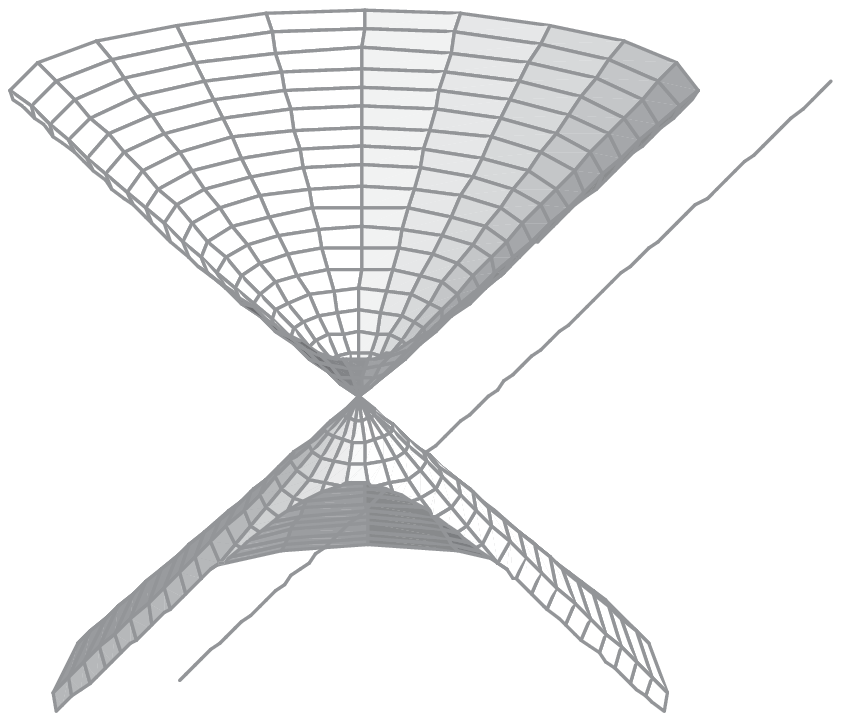}} \setlength{\unitlength}{1cm} \caption[]{The light
cone at the origin and a light ray passing through the cone.}\index{light ray}
\label{fig:lightcone}
\end{figure}}

\parpic[r]{\includegraphics[width=4.0cm,totalheight=4.0cm]
{lightcone.eps}}

\paragraph{Light rays.} A light ray in Minkowski space is\index{light ray}
parameterized by the equations $x^{\alpha\ald}=x_0^{\alpha\ald}+t
p^{\alpha\ald}$. For a general light ray, one can reparameterize\index{light ray}
this description such that $x_0^{\alpha\ald}$ is a null vector.
For light rays which intersect the light cone of the origin more\index{light ray}
than once, i.e.\ light rays which lie in a null hyperplane through
the origin, one chooses $x_0^{\alpha\ald}$ to be orthogonal to
$p^{\alpha\ald}$ with respect to the Minkowski metric. After\index{metric}
decomposing the vectors into spinors, we have\index{Spinor}
\begin{equation}
x^{\alpha\ald}\ =\ c \omega^\alpha
\bar{\omega}^\ald+t\lambda^\ald\bar{\lambda}^\alpha\eand
x^{\alpha\ald}\ =\ \zeta^\alpha\lambda^\ald+\bar{\zeta}^\ald\bar{\lambda}^\alpha+t\lambda^\ald\bar{\lambda}^\alpha
\end{equation}
in the general and special cases, respectively. We can reduce both
cases to the single equation
\begin{equation}\label{lightraysincidence}
\omega^\alpha\ =\ \di x^{\alpha\ald}\lambda_\ald
\end{equation}
by assuming $c=-\di (\bar{\omega}^\ald\lambda_\ald)^{-1}$ in the
general case and $\omega^\alpha=\di
\bar{\lambda}^\alpha(\bar{\zeta}^\bed\lambda_\bed)$ in the special
case. For all points $x^{\alpha\ald}$ on a light ray, equation\index{light ray}
\eqref{lightraysincidence}, the {\em incidence relations} hold.\index{incidence relation}

\paragraph{Twistors.} A {\em twistor} $Z^i$ is now defined as a\index{twistor}
pair of two-spinors $(\omega^\alpha,\lambda_\ald)$ which transform\index{Spinor}\index{two-spinors}
under a translation of the origin $0\rightarrow r^{\alpha\ald}$ as
\begin{equation}
(\omega^\alpha,\lambda_\ald)\ \rightarrow\  (\omega^\alpha-\di
r^{\alpha\ald}\lambda_\ald,\lambda_\ald)~.
\end{equation}
The spinors $\omega^\alpha$ and $\lambda_\ald$ are usually called\index{Spinor}
the {\em primary} and {\em secondary spinor parts} of the twistor\index{spinor parts}\index{twistor}
$Z^i$.

The twistor space $\TT$ is thus a four-dimensional complex vector\index{twistor}\index{twistor!space}
space, and we can introduce an anti-linear involution\index{involution}
$\tau:\TT\rightarrow \TT\dual$ by defining
\begin{equation}
Z^i\ =\ (\omega^\alpha,\lambda_\ald)\ \mapsto\ (\bar{\lambda}_\alpha,\bar{\omega}^\ald)\ =\ \bar{Z}_i~.
\end{equation}
Furthermore, we can define a Hermitian inner product $h(Z,U)$ for\index{Hermitian inner product}
two twistors $Z^i=(\omega^\alpha,\lambda_\ald)$ and\index{twistor}
$U^i=(\sigma^\ald,\mu_\alpha)$ via
\begin{equation}
h(Z,U)\ =\ Z^i\bar{U}_i\ =\ \omega^\alpha\bar{\mu}_\alpha+\lambda_\ald\bar{\sigma}^\ald~,
\end{equation}
which is not positive definite but of signature
$($$+$$+$$-$$-$$)$. This leads to the definition of {\em
null-twistors}, for which $Z^i \bar{Z}_i=0$. Since this constraint\index{null-twistors}\index{twistor}
is a real equation, the null-twistors form a real
seven-dimensional subspace $\TT_N$ in $\TT$. Furthermore, $\TT_N$
splits $\TT$ into two halves: $\TT^\pm$ with twistors of positive\index{twistor}
and negative norm.

As a relative phase between $\omega^\alpha$ and
$\bar{\lambda}^\alpha$ does not affect the underlying light ray,\index{light ray}
we can assume that $\omega^\alpha\bar{\lambda}_\alpha$ is purely
imaginary. Then the twistor underlying the light ray becomes a\index{light ray}\index{twistor}
null twistor, since $Z^i\bar{Z}_i=2 \mathrm{Re}
(\omega^\alpha\bar{\lambda}_\alpha)$.

There is a nice way of depicting twistors based on the so-called\index{twistor}
Robinson congruence. An example is printed on the back of this\index{Robinson congruence}
thesis. For more details, see appendix \ref{sQuinticRobinson}

\paragraph{Light rays and twistors.} We call a twistor {\em incident}\index{incident}\index{light ray}\index{twistor}
to a spacetime point $x^{\alpha\ald}$ if its spinor parts satisfy\index{Spinor}\index{spinor parts}
the incidence relations \eqref{lightraysincidence}.\index{incidence relation}

After we restrict $x^{\alpha\ald}$ to be real, there is clearly a
null twistor $Z^i$ incident to all points $x^{\alpha\ald}$ on a\index{incident}\index{twistor}
particular light ray, which is unique up to complex rescaling. On\index{light ray}
the other hand, every null twistor with non-vanishing secondary\index{twistor}
spinor ($\lambda$-) part corresponds to a light ray. The remaining\index{Spinor}\index{light ray}
twistors correspond to light rays through infinity and can be\index{twistor}
interpreted by switching to the conformal compactification of
Minkowski space. This is easily seen by considering the incidence
relations \eqref{lightraysincidence}, where -- roughly speaking --\index{incidence relation}
a vanishing secondary spinor part implies infinite values for\index{Spinor}
$x^{\alpha\ald}$ if the primary spinor part is finite.

Since the overall scale of the twistor is redundant, we rather\index{twistor}
switch to the {\em projective twistor space} $\PP\TT=\CPP^{3}$.\index{projective twistor space}\index{twistor!space}
Furthermore, non-vanishing of the secondary spinor part implies\index{Spinor}
that we take out a sphere and arrive at the space
$\CP^3=\CPP^3\backslash \CPP^1$. The restriction of $\CP^3$ to
null-twistors will be denoted by $\CP^3_N$. We can now state that\index{null-twistors}\index{twistor}
there is a one-to-one correspondence between light rays in\index{light ray}
(complexified) Minkowski space $M$ ($M^c$) and elements of
$\CP^3_N$ ($\CP^3$).

\paragraph{``Evaporation'' of equations of motion.} Let us make a
simple observation which will prove helpful when discussing
solutions to field equations in later sections. Consider a
massless particle with vanishing helicity. Its motion is\index{helicity}
completely described by a four momentum $p^{\alpha\ald}$ and an
initial starting point $x^{\alpha\ald}$. To each such motion,
there is a unique twistor $Z^i$. Thus, while one needs to solve\index{twistor}
equations to determine the motion of a particle in its phase
space, this is not so in twistor space: Here, the equations of\index{twistor}\index{twistor!space}
motion have\footnote{This terminology is due to
\cite{TwistorsAndGR}.} ``evaporated'' into the structure of the\index{twistor}
twistor space. We will encounter a similar phenomenon later, when\index{twistor!space}
discussing the Penrose and the Penrose-Ward transforms, which\index{Penrose-Ward transform}
encode solutions to certain field equations in holomorphic
functions and holomorphic vector bundles on the twistor space.\index{holomorphic!vector bundle}\index{twistor}\index{twistor!space}

\paragraph{Quantization.} The canonical commutation relations on\index{quantization}
Minkowski spacetime (the {\em Heisenberg algebra}) read\index{Heisenberg algebra}
\begin{equation}
[\hat{p}_\mu,\hat{p}_\nu]\ =\ [\hat{x}^\mu,\hat{x}^\nu]\ =\ 0\eand
[\hat{x}^\nu,\hat{p}_\mu]\ =\ \di \hbar \delta_\mu^\nu~,
\end{equation}
and induce canonical commutation relations for twistors:\index{twistor}
\begin{equation}
[\hat{Z}^i,\hat{Z}^j]\ =\ [\hat{\bZ}_i,\hat{\bZ}_j]\ =\ 0\eand[\hat{Z}^i,\hat{\bZ}_j]\ =\ \hbar
\delta_j^i~.
\end{equation}
Alternatively, one can also follow the ordinary canonical
quantization prescription\index{canonical quantization prescription}\index{quantization}
\begin{equation}
[\hat{f},\hat{g}]\ =\ \di \hbar \widehat{\{f,g\}}+\CO(\hbar^2)
\end{equation}
for the twistor variables and neglecting terms beyond linear order\index{twistor}
in $\hbar$, which yields the same result.

A representation of this algebra is easily found on the algebra of\index{representation}
functions on twistor space by identifying $\hat{Z}^i=Z^i$ and\index{twistor}\index{twistor!space}
$\hat{\bZ}_i=-\hbar\der{Z^i}$. This description is not quite
equivalent to the Bargmann representation, as there one introduces\index{Bargmann representation}\index{representation}
complex coordinates on the phase space, while here, the underlying
space is genuinely complex.

The helicity $2s=Z^i \bZ_i$ is augmented to an operator, which\index{helicity}
reads in symmetrized form
\begin{equation}
\hat{s}\ =\ \tfrac{1}{4}(\hat{Z}^i\hat{\bZ}_i+\hat{\bZ}_i
\hat{Z}^i)\ =\ -\frac{\hbar}{2}\left(Z^i\der{Z^i}+2\right)~,
\end{equation}
and it becomes clear that an eigenstate of the helicity operator\index{helicity}\index{helicity operator}
with eigenvalue $s\hbar$ must be a homogeneous twistor function\index{twistor}
$f(Z^i)$ of degree $-2s-2$. One might wonder, why this description
is asymmetric in the helicity, i.e.\ why e.g.\ eigenstates of\index{helicity}
helicity $\pm 2$ are described by homogeneous twistor functions of\index{twistor}
degree $-6$ and $+2$, respectively. This is due to the inherent
chirality of the twistor space. By switching to the dual twistor\index{twistor}\index{twistor!dual twistor}\index{twistor!space}
space $\PP\TT\dual$, one arrives at a description in terms of
homogeneous twistor functions of degree $2s+2$.\index{twistor}

\subsection{Klein (twistor-) correspondence}\index{twistor}

Interestingly, there is some work by Felix Klein
\cite{Klein:1,Klein:2} dating back to as early as 1870, in which
he discusses correspondences between points and subspaces of both
$\PP\TT$ and the compactification of $M^c$. In the following
section, this {\em Klein correspondence} or {\em twistor\index{Klein correspondence}\index{twistor}
correspondence} will be developed.

\paragraph{The correspondence $r\in M^c\leftrightarrow \CPP^1\subset
\PP\TT$.} The (projective) twistors satisfying the incidence\index{twistor}
relations \eqref{lightraysincidence} for a given fixed point
$r^{\alpha\ald}\in M^c$ are of the form $Z^i=(\di
r^{\alpha\ald}\lambda_\ald,\lambda_\ald)$ up to a scale. Thus, the
freedom we have is a projective two-spinor $\lambda_\ald$\index{Spinor} (i.e.\
in particular, at most one component of $\lambda_\ald$ can vanish)
and the components of this spinor are the homogeneous coordinates\index{Spinor}\index{homogeneous coordinates}
of the space $\CPP^1$. This is consistent with our previous
observations, as twistors incident to a certain point $r\in M^c$\index{incident}\index{twistor}
describe all light rays through $r$, and this space is the\index{light ray}
celestial sphere $S^2\cong \CPP^1$ at this point. Thus, the point\index{celestial sphere}
$r$ in $M$ corresponds to a projective line $\CPP^1$ in $\PP\TT$.

\paragraph{The correspondence $p\in\PP\TT\leftrightarrow \alpha
\mbox{-plane in}~M^c$.} Given a twistor $Z^i$ in $\PP\TT$, the\index{twistor}
points $x$ which are incident with $Z^i$ form a two-dimensional\index{incident}
subspace of $M^c$, a so-called {\em $\alpha$-plane}. These planes\index{a-plane@$\alpha$-plane}
are completely null, which means that two arbitrary points on such
a plane are always separated by a null distance. More explicitly,
an $\alpha$-plane corresponding to a twistor\index{a-plane@$\alpha$-plane}\index{twistor}
$Z^i=(\omega^\alpha,\lambda_\ald)$ is given by
\begin{equation}\label{alphaplanes}
x^{\alpha\ald}\ =\ x^{\alpha\ald}_0+\mu^\alpha\lambda^\ald~,
\end{equation}
where $x^{\alpha\ald}_0$ is an arbitrary solution to the incidence
relations \eqref{lightraysincidence} for the fixed twistor $Z^i$.\index{incidence relation}\index{twistor}
The two-spinor $\mu^\alpha$ then parameterizes the $\alpha$-plane.\index{Spinor}\index{a-plane@$\alpha$-plane}

\paragraph{Dual twistors.} The incidence relations for a {\em dual\index{incidence relation}\index{twistor}\index{twistor!dual twistor}
twistor} $W_i=(\sigma_\ald,\mu^\alpha)$ read as
\begin{equation}
\sigma^\ald\ =\ -\di x^{\alpha\ald}\mu_\alpha~.
\end{equation}
Such a twistor again corresponds to a two-dimensional totally null\index{twistor}
subspace of $M^c$, the so called {\em $\beta$-planes}. The\index{b-plane@$\beta$-plane}
parameterization is exactly the same as the one given in
\eqref{alphaplanes}. Note, however, that the r{\^o}le of
$\lambda_\ald$ and $\mu^\alpha$ have been exchanged, i.e.\ in this
case, $\lambda^\ald$ parameterizes the $\beta$-plane.\index{b-plane@$\beta$-plane}

\paragraph{Totally null hyperplanes.} Two $\alpha$- or
two $\beta$-planes either coincide or intersect in a single point.\index{b-plane@$\beta$-plane}
An $\alpha$- and a $\beta$-plane are either disjoint or intersect
in a line, which is null. The latter observation will be used in
the definition of ambitwistor spaces in section\index{ambitwistor space}\index{twistor}\index{twistor!ambitwistor}\index{twistor!space}
\ref{ssAmbitwistorSpace}. Furthermore, the correspondence between
points in $\PP\TT$ and planes in $M^c$ breaks down in the real
case.

\subsection{Penrose transform}\label{ssPTrafo}\index{Penrose transform}

The Penrose transform gives contour integral formul\ae{} for\index{Penrose transform}
mapping certain functions on the twistor space to solutions of the\index{twistor}\index{twistor!space}
massless field equations for particles with arbitrary helicity.\index{helicity}
For our considerations, we can restrict ourselves to the subspace
of $\CPP^3$, for which $\lambda_\ed\neq 0$ and switch for
simplicity to the inhomogeneous coordinates\index{homogeneous coordinates}\index{inhomogeneous coordinates}
$\lambda_\ald:=(1,\lambda)^T$.

\paragraph{Elementary states.}\label{pElStates}\index{elementary states}
A useful class of functions on twistor space are the so-called\index{twistor}\index{twistor!space}
{\em elementary states}. Let us again denote a twistor by\index{elementary states}
$(Z^i)=(\omega^\alpha,\lambda_\ald)$. Then an elementary state is
given by
\begin{equation}
f(Z)\ =\ \frac{(C_i Z^i)^c(D_iZ^i)^d}{(A_iZ^i)^{a+1}(B_iZ^i)^{b+1}}~,
\end{equation}
where $A_i,B_i,C_i,D_i$ are linearly independent and
$a,b,c,d\in\NN$. The Penrose transform will relate such an\index{Penrose transform}
elementary state to a field with helicity\index{helicity}
\begin{equation}
h=\tfrac{1}{2}(a+b-c-d)~,
\end{equation}
satisfying its massless equations of motion.

\paragraph{Negative helicity.} Consider the contour integral\index{helicity}
\begin{equation}\label{VII.21}
\phi_{\ald_1\ldots\ald_{2h}}(x)\ =\ \tfrac{1}{2\pi\di}\oint_\CC\dd
\lambda^\ald~ \lambda_\ald \lambda_{\ald_1} \ldots
\lambda_{\ald_{2h}} f(Z)~,
\end{equation}
where $\CC\cong S^1$ is a contour in $\CPP^1$, which is the
equator $|\lambda_\ed/\lambda_{\zd} |=1$ or a suitable deformation
thereof, if $f$ should become singulary on $\CC$. The contour
integral can only yield a finite result if the integrand is of
homogeneity zero and thus $f$ has to be a section of $\CO(-2h-2)$.
For these $f$, $\phi$ is a possibly non-trivial solution to the
massless field equations
\begin{equation}
\dpar^{\alpha\ald_1}\phi_{\ald_1\ldots\ald_{2h}}\ =\ 0
\end{equation}
for a field of helicity $-h$. This can be easily seen by\index{helicity}
substituting the incidence relations \eqref{lightraysincidence}\index{incidence relation}
into the primary spinor part of the twistor $Z=(\di\index{Spinor}\index{twistor}
x^{\alpha\ald}\lambda_\ald,\lambda_\ald)$ and taking the
derivative into the integral.

\paragraph{Zero helicity.} For the case of zero helicity, we can\index{helicity}
generalize the above formalism. We employ the same contour
integral as in \eqref{VII.21}, restricted to $h=0$:
\begin{equation}
\phi(x)\ =\ \tfrac{1}{2\pi\di}\oint_\CC\dd \lambda^\ald~ \lambda_\ald
f(Z)~,
\end{equation}
where $\phi$ is now a solution to the scalar field equation
\begin{equation}
\square
\phi\ =\ \tfrac{1}{2}\dpar^{\alpha\ald}\dpar_{\alpha\ald}\phi\ =\ 0~,
\end{equation}
the massless Klein-Gordon equation. Again, this fact is readily\index{Klein-Gordon equation}
seen by pulling the derivatives into the integral and for
non-vanishing $\phi$, $f$ is a section of $\CO(-2)$.

\paragraph{Positive helicity.} For positive helicities, we have to\index{helicity}
adapt our contour integral in the following way:
\begin{equation}
\phi_{\alpha_1\ldots\alpha_{2h}}(x)\ =\ \tfrac{1}{2\pi\di}\oint_\CC\dd
\lambda^\ald~ \lambda_\ald \der{\omega^{\alpha_1}
}\ldots\der{\omega^{\alpha_{2h}} } f(Z)~,
\end{equation}
which will give rise to the massless equations of motion for
helicity $h$\index{helicity}
\begin{equation}
\eps^{\alpha\alpha_1}\dpar_{\alpha\ald}\phi_{\alpha_1\ldots\alpha_{2h}}\ =\ 0
\end{equation}
and for nontrivial $\phi$, $f$ must be a section of $\CO(2h-2)$.
This result requires slightly more effort to be verified, but the
calculation is nevertheless straightforward.

\paragraph{Further remarks.} Altogether, we saw how to construct
solutions to massless field equations for fields with helicity $h$\index{helicity}
using functions on twistor space, which transform as sections of\index{twistor}\index{twistor!space}
$\CO(2h-2)$. One can prove that the (Abelian) \v{C}ech cohomology\index{Cech cohomology@\v{C}ech cohomology}
group $H^1(\CO(1)\oplus\CO(1),\CO(2h-2))$ is isomorphic to the
sheaf of solutions to the massless equations of motion for\index{sheaf}
particles with helicity $h$. Note that our convention for $h$\index{helicity}
differs to another very common one used e.g.\ in \cite{Wardbook}
by a sign.

Furthermore, this construction is reminiscent of the ``evaporation
of equations of motion'' for particles in the twistor approach and\index{twistor}
we will come across a generalization of this construction to Lie
algebra valued fields in section \ref{sSolutionGT}.\index{Lie algebra}

\section{Integrability}\index{integrability}

The final goal of this chapter is to construct the Penrose-Ward
transform for various field theories, which relates classical\index{Penrose-Ward transform}
solutions to some equations of motion to geometric data on a
twistor space. This transform is on the one hand founded on the\index{twistor}\index{twistor!space}
equivalence of the \v{C}ech and Dolbeault descriptions of
(topologically trivial) holomorphic vector bundles, on the other\index{holomorphic!vector bundle}
hand, its explicit construction needs the notion of linear systems\index{linear system}
and the corresponding compatibility conditions. Therefore, let us\index{compatibility conditions}
briefly comment on the property called (classical) {\em
integrability}, which is the framework for these entities.\index{integrability}

\subsection{The notion of
integrability}\label{ssNotionIntegrability}\index{integrability}

\paragraph{Basic Idea.} Although there seems to be no general
definition of integrability, quite generally speaking one can\index{integrability}
state that an integrable set of equations is an exactly solvable\index{integrable}
set of equations. Further strong hints that a set of equations may
be integrable are the existence of many conserved quantities and a\index{integrable}
description in terms of algebraic geometry. Until today, there is
no comprehensive algorithm to test integrability.\index{integrability}

\paragraph{Example.} To illustrate the above remarks, let us
briefly discuss an example given e.g.\ in \cite{Hitchin:1999at}:
The center-of-mass motion of a rigid body. Let $\omega$ be the\index{body}\index{rigid}
angular velocity and $I_i$ the principal moments of inertia. The
equations of motion then take the form\footnote{These equations
are the equations for a spinning top and they are related to the
Nahm equations cf.\ section \ref{ssNahm}.}\index{Nahm equations}
\begin{equation}
I_i\dot{\omega}_i\ =\ \eps_{ijk}(I_j-I_k)\omega_j\omega_k~.
\end{equation}
One can rescale these equations to the simpler form
\begin{equation}
\dot{u}_1\ =\ u_2u_3~,~~~\dot{u}_2\ =\ u_3u_1\eand\dot{u}_3\ =\ u_1u_2~.
\end{equation}

Let us now trace the above mentioned properties of integrable\index{integrable}
systems of equations. First, we have the conserved quantities
\begin{equation}
A\ =\ u_1^2-u_2^2\eand B\ =\ u_1^2-u_3^2~,
\end{equation}
since $\dot{A}=\dot{B}=0$ by virtue of the equations of motion.
Second, we find an elliptic curve by putting $y=\dot{u}_1$ and
$x=u_1$ and substituting the conserved quantities in the first
equation of motion $\dot{u}_1=u_2u_3$:
\begin{equation}
y^2\ =\ (x^2-A)(x^2-B)~.
\end{equation}
Thus, algebraic geometry is indeed present in our example.
Eventually, one can give explicit solutions, since the above
algebraic equation can be recast into the standard form
\begin{equation}
y^2\ =\ 4x^3-g_2x-g_3~,
\end{equation}
which is solved by the Weierstrass $\wp$-function with $x=\wp(u)$
and $y=\wp'(u)$. The solution is then given by $\dd t=\dd
\wp/\wp'$.

\paragraph{Ward conjecture.} This conjecture by Richard Ward\index{Ward conjecture}
\cite{Ward:1985gz} states that all the integrable equations in 1+1\index{integrable}
and 2+1 dimensions can be obtained from (anti-)self-dual
Yang-Mills theory in four dimensions by dimensional reduction. On\index{Yang-Mills theory}\index{dimensional reduction}\index{self-dual Yang-Mills theory}
commutative spaces, this conjecture can be regarded as confirmed
\cite{Masonbook}.

\subsection{Integrability of linear\index{integrability}
systems}\label{ssIntegrabilityLinearSystems}

\paragraph{A simple example.} In our subsequent discussion, we will
have to deal with a linear system of equations, which states that\index{linear system}
some $\sGL(n,\FC)$-valued function $\psi$ is covariantly constant
in several directions, e.g.\
\begin{equation}\label{linsys1}
\begin{aligned}
\nabla_1 \psi\ =\ 0\eand \nabla_2 \psi\ =\ 0~.
\end{aligned}
\end{equation}
Since this system is overdetermined, there can only exist a
solution if a certain condition obtained by cross-differentiating
is fulfilled:
\begin{equation}
\nabla_1\nabla_2 \psi-\nabla_2\nabla_1 \psi\ =:\ F_{12} \psi\ =\ 0~.
\end{equation}
Now a sufficient and necessary condition for this to hold is
$F_{12}=0$ as $\psi$ is invertible. This condition is called the
compatibility condition of the linear system \eqref{linsys1}.\index{linear system}
However, this system is too trivial to be interesting. The
solutions of $F_{12}=0$ are pure gauge and thus essentially\index{gauge!pure gauge}
trivial.

\paragraph{Self-Dual Yang-Mills theory.}\label{pSDYM} The obvious step to make\index{Yang-Mills theory}\index{self-dual Yang-Mills theory}
the compatibility conditions non-trivial is to introduce a\index{compatibility conditions}
so-called {\em spectral parameter} $\lambda\in\FC$ and consider\index{spectral parameter}
the linear system\index{linear system}
\begin{equation}\label{VII.34}
(\nabla_1-\lambda \nabla_3)\psi\ =\ 0\eand(\nabla_2+\lambda
\nabla_4)\psi\ =\ 0~.
\end{equation}
This linear system has a non-trivial solution if and only if\index{linear system}
\begin{equation}
[\nabla_1-\lambda \nabla_3,\nabla_2+\lambda\nabla_4]\ =\ 0~,
\end{equation}
or, written in terms of components of a Taylor expansion in the
spectral parameter if\index{spectral parameter}
\begin{equation}\label{VII.36}
F_{12}\ =\ F_{14}-F_{32}\ =\ F_{34}\ =\ 0~.
\end{equation}
The latter equations are, in a suitable basis, the self-dual
Yang-Mills equations, cf.\ section \ref{sssSDYM}, and
\eqref{VII.34} is exactly the linear system we will encounter\index{linear system}
later on in the twistor formulation. Note that the system\index{twistor}
\eqref{VII.36} is underdetermined (three equations for four
components) due to gauge invariance.

\paragraph{Further examples.} All the constraint
equations we encountered so far, i.e.\ the $\CN=1,\ldots,4$\index{constraint equations}
supersymmetrically extended self-dual Yang-Mills equations
\eqref{constraintN4SDYM} and also the full $\CN=3,4$ super
Yang-Mills theory \eqref{constraintN4SYM} can be obtained from\index{Yang-Mills theory}
linear systems. After introducing the simplifying spinorial\index{Spinor}\index{linear system}
notation $\lambda^{\ald}=(\lambda,-1)^T$ and
$\mu^{\alpha}=(\mu,-1)^T$, they read as
\begin{equation}
\begin{aligned}
\lambda^\ald(\dpar_{\alpha\ald}+\CA_{\alpha\ald})\psi&\ =\ 0~,\\
\lambda^\ald\left(\der{\eta^\ald_i}+\CA_\ald^i\right)\psi&\ =\ 0
\end{aligned}
\end{equation}
for the supersymmetric self-dual Yang-Mills equations and
\begin{equation}
\begin{aligned}
\mu^\alpha\lambda^\ald(\dpar_{\alpha\ald}+
\CA_{\alpha\ald})\psi&\ =\ 0~,\\
\lambda^\ald(D^i_{\ald}+\CA^i_{\ald})\psi&\ =\ 0~,\\
\mu^\alpha(D_{\alpha i}+\CA_{\alpha i})\psi&\ =\ 0~.
\end{aligned}
\end{equation}
in the case of the $\CN=3,4$ super Yang-Mills equations. Here, $i$
runs from $1$ to $\CN$. Note that in the latter case, really all
the conditions obtained from cross-differentiating are fulfilled
if the constraint equations \eqref{constraintN4SYM} hold.\index{constraint equations}

\section{Twistor spaces and the Penrose-Ward\index{twistor}\index{twistor!space}
transform}\label{sTwistorSpaces}

After this brief review of the basic ideas in twistor geometry and\index{twistor}
integrable systems let us now be more explicit in the spaces and\index{integrable}
and the conventions we will use. We will introduce several twistor\index{twistor}
correspondences between certain superspaces and modified and\index{super!space}
extended twistor spaces upon which we will construct Penrose-Ward\index{twistor}\index{twistor!space}
transforms, relating solutions to geometric data encoded in
holomorphic bundles.

\subsection{The twistor space}\label{ssTwistorSpace}\index{twistor}\index{twistor!space}

\paragraph{Definition.} We start from the complex projective
space $\CPP^3$ (the {\em twistor space}) with homogeneous\index{complex!projective space}\index{twistor}\index{twistor!space}
coordinates $(\omega^\alpha,\lambda_\ald)$ subject to the
equivalence relation
$(\omega^\alpha,\lambda_\ald)\sim(t\omega^\alpha,t\lambda_\ald)$
for all $t\in\FC^\times$, $\alpha=1,2$ and $\ald=\dot{1},\dot{2}$.
As we saw above, this is the space of light rays in complexified,\index{light ray}
compactified Minkowski space. To neglect the light cone at
infinity, we demand that $\lambda_\ald$ parameterizes a $\CPP^1$,
i.e.\ $\lambda_\ald\neq(0,0)^T$. Recall that the set which we
exclude by this condition is the Riemann sphere $\CPP^1$ and the\index{Riemann sphere}
resulting space $\CPP^3\backslash \CPP^1$ will be denoted by
$\CP^3$. This space can be covered by two patches $\CU_+$
($\lambda_{\dot{1}}\neq 0$) and $\CU_-$ ($\lambda_{\dot{2}}\neq
0$) with coordinates\footnote{A more extensive discussion of the
relation between these inhomogeneous coordinates and the\index{homogeneous coordinates}\index{inhomogeneous coordinates}
homogeneous coordinates is found in appendix \ref{AppDictionary}}
\begin{equation}\label{coords}
\begin{aligned}
z_+^\alpha&\ =\ \frac{\omega^\alpha}{\lambda_{\dot{1}}}~,~~~
z_+^3\ =\ \frac{\lambda_{\dot{2}}}{\lambda_{\dot{1}}}\ =:\ \lambda_+~~~
\mbox{on}~~\CU_+~,\\
z_-^\alpha&\ =\ \frac{\omega^\alpha}{\lambda_{\dot{2}}}~,~~~
z_-^3\ =\ \frac{\lambda_{\dot{1}}}{\lambda_{\dot{2}}}\ =:\ \lambda_-~~~
\mbox{on}~~\CU_-~,
\end{aligned}
\end{equation}
related by
\begin{equation}\label{coordstrafo}
z_+^\alpha\ =\ z_+^3z_-^\alpha~~~\mbox{ and }~~~z_+^3\ =\ \frac{1}{z_-^3}
\end{equation}
on the overlap $\CU_+\cap\CU_-$. Due to \eqref{coordstrafo}, the
space $\CP^3$ coincides with the total space of the rank two
holomorphic vector bundle\index{holomorphic!vector bundle}
\begin{equation}\label{bundle}
\CP^3\ =\ \CO(1)\oplus\CO(1)\ \rightarrow\  \CPP^1~,
\end{equation}
where the base manifold $\CPP^1$ is covered by the two patches\index{manifold}
$U_\pm:=\CU_\pm\cap \CPP^1$.

\paragraph{Moduli space of sections.} Holomorphic sections of the\index{moduli space}
complex vector bundle \eqref{bundle} are rational curves $\CPP^1_x\index{complex!vector bundle}
\embd \CP^3$ defined by the equations
\begin{equation}\label{sections}
z_+^\alpha\ =\ x^{\alpha \dot{1}}+\lambda_+ x^{\alpha \dot{2}}~~
\mbox{for}~~\lambda_+\in U_+~~~\mbox{ and }~~~
z_-^\alpha\ =\ \lambda_-x^{\alpha\dot{1}}+x^{\alpha\dot{2}}~~
\mbox{for}~~\lambda_-\in U_-
\end{equation}
and parameterized by moduli $x=(x^{\alpha\ald})\in\FC^4$. After
introducing the spinorial notation\index{Spinor}
\begin{equation}\label{lambda1}
\left(\lambda_\ald^+\right)\ :=\ \left(\begin{array}{c}1\\\lambda_+
\end{array}\right)~~~\mbox{and}~~~
\left(\lambda_\ald^-\right)\ :=\ \left(\begin{array}{c}\lambda_-\\1
\end{array}\right)~,
\end{equation}
we can rewrite \eqref{sections} as the {\em incidence relations}\index{incidence relation}
\begin{equation}\label{sections2}
z_\pm^\alpha\ =\ x^{\alpha\dot{\alpha}}\lambda_\ald^\pm~.
\end{equation}
Note the familiarity of these equations with the incidence
relations \eqref{lightraysincidence}. The meaning of\index{incidence relation}
\eqref{sections2} becomes most evident, when writing down a double
fibration:\index{double fibration}\index{fibration}
\begin{equation}\label{dblfibration}
\begin{aligned}
\begin{picture}(50,40)
\put(0.0,0.0){\makebox(0,0)[c]{$\CP^3$}}
\put(64.0,0.0){\makebox(0,0)[c]{$\FC^4$}}
\put(34.0,33.0){\makebox(0,0)[c]{$\CF^5$}}
\put(7.0,18.0){\makebox(0,0)[c]{$\pi_2$}}
\put(55.0,18.0){\makebox(0,0)[c]{$\pi_1$}}
\put(25.0,25.0){\vector(-1,-1){18}}
\put(37.0,25.0){\vector(1,-1){18}}
\end{picture}
\end{aligned}
\end{equation}
where $\CF^5:=\FC^4\times\CPP^1$, $\pi_1$ is the trivial
projection $\pi_1(x^{\alpha\ald},\lambda_\ald)=x^{\alpha\ald}$ and
$\pi_2$ is given by \eqref{sections2}. We thus obtain a twistor\index{twistor}
correspondence between points and subspaces of either spaces
$\FC^4$ and $\CP^3$:
\begin{equation}
\begin{aligned}
\left\{\,\mbox{projective lines $\CPP^1_x$ in $\CP^{3}$}\right\}&\
\longleftrightarrow\
\left\{\,\mbox{points $x$ in $\FC^4$}\right\}~,\\
\left\{\,\mbox{points $p$ in $\CP^3$}\right\}&\
\longleftrightarrow\ \left\{\,\mbox{null ($\alpha$-)planes
$\FC_p^2$ in $\FC^4$}\right\}~.
\end{aligned}
\end{equation}
While the first correspondence is rather evident, the second one
deserves a brief remark. Suppose $\hat{x}^{\alpha\ald}$ is a
solution to the incidence relations \eqref{sections2} for a fixed\index{incidence relation}
point $p\in \CP^{3}$. Then the set of all solutions is given by
\begin{equation}\label{superplanes}
\{x^{\alpha\ald}\}~~~\mbox{with}~~~ x^{\alpha\ald}\ =\
\hat{x}^{\alpha\ald}+\mu^\alpha\lambda^\ald_\pm~,
\end{equation}
where $\mu^\alpha$ is an arbitrary commuting two-spinor and we use\index{Spinor}
our standard convention of
$\lambda^\ald_\pm:=\eps^{\ald\bed}\lambda_\bed^\pm$ with
$\eps^{\ed\zd}=-\eps^{\zd\ed}=1$. One can choose to work on any
patch containing $p$. The sets defined in \eqref{superplanes} are
then called {\em null} or {\em $\alpha$-planes}.\index{a-plane@$\alpha$-plane}

The correspondence space $\CF^5$ is a complex five-dimensional
manifold, which is covered by the two patches\index{manifold}
$\tilde{\CU}_\pm=\pi_2^{-1}(\CU_\pm)$.

\paragraph{Vector fields.} On the complex manifold\index{complex!manifold}\index{manifold}
$\CP^3$, there is the natural basis $(\dpar/\dpar
\bz_\pm^\alpha,\dpar/\dpar \bz^3_\pm)$ of antiholomorphic vector
fields, which are related via\index{holomorphic!vector fields}
\begin{equation}
\der{\bz^\alpha_+}\ =\ \bz_-^3\der{\bz^\alpha_-}\eand
\der{\bz_+^3}\ =\
-(\bz_-^3)^2\der{\bz_-^3}-\bz_-^3\bz_-^\alpha\der{\bz_-^\alpha}~.
\end{equation}
on the intersection $\CU_+\cap\CU_-$. The leaves to the fibration\index{fibration}
$\pi_2$ in \eqref{dblfibration} are spanned by the vector fields
\begin{equation}\label{Vbars}
\bV_\alpha^\pm\ =\ \lambda^\ald_\pm \dpar_{\alpha\ald}~,
\end{equation}
which obviously annihilate the coordinates \eqref{sections2} on
$\CP^3$. The tangent spaces to the leaves of the fibration $\pi_2$\index{fibration}
are evidently of dimension $2$.

\paragraph{Real structure.}\label{pRealStructureP3} Recall that a\index{real structure}
real structure on a complex manifold $M$ is defined as an\index{complex!manifold}\index{manifold}
antiholomorphic involution $\tau:M \rightarrow M$. On the twistor\index{involution}\index{twistor}
space $\CP^3$, we can introduce three anti-linear transformations
of commuting two-spinors:\index{Spinor}\index{two-spinors}
\begin{subequations}
\begin{align}\label{trafo1}
(\omega^\alpha)\ \mapsto\ \tau_\eps(\omega^\alpha)\ =\ &\left(\begin{array}{cc}
0 & \eps\\1 & 0\end{array}\right)\left(\begin{array}{c}
\bar{\omega}^1\\\bar{\omega}^2\end{array}\right)\ =\ \left(\begin{array}{c}
\eps\bar{\omega}^2\\\bar{\omega}^1\end{array}\right)\ =:\ (\hat{\omega}^\alpha)~,
\\[0.2cm]\label{trafo3}
&\tau_0(\omega^\alpha)\ =\ (\bar{\omega}^\alpha)~,
\end{align}
\end{subequations}
where $\eps=\pm 1$. In particular, this definition implies
$(\hat{\omega}_\alpha):=\tau(\omega_\alpha)$ and
$(\hat{\lambda}^\ald):=\tau(\lambda^\ald)$, i.e.\ indices are
raised and lowered before $\tau$ is applied. For later reference,
let us give explicitly all the possible variants of the two-spinor\index{Spinor}
$\lambda_\ald^\pm$ given in \eqref{lambda1}:
\begin{equation}\label{alllambdas}
\begin{aligned}
(\lambda_+^\ald)&\ :=\ \left(\begin{array}{c} \lambda_+\\-1
\end{array}\right)~,&(\hl^+_\ald)&\ :=\ \left(\begin{array}{c} \eps\bl_+\\1
\end{array}\right)~,&(\hl_+^\ald)&\ :=\ \left(\begin{array}{c}
-\eps\\ \bl_+
\end{array}\right)~,\\
(\lambda_-^\ald)&\ :=\ \left(\begin{array}{c} 1\\ -\lambda_-
\end{array}\right)~,&
(\hl^-_\ald)&\ :=\ \left(\begin{array}{c} \eps\\\bl_-
\end{array}\right)~,&
(\hl_-^\ald)&\ :=\ \left(\begin{array}{c} -\eps\bl_-\\1
\end{array}\right)~.
\end{aligned}
\end{equation}

Furthermore, the transformations \eqref{trafo1}-\eqref{trafo3}
define {\em three} real structures on $\CP^3$ which in the\index{real structure}
coordinates \eqref{coords} are given by the formul\ae
\begin{subequations}
\begin{equation}\label{three}
\tau_\eps(z_+^1,z_+^2,z^3_+)\ =\ \left(\frac{\bz_+^2}{\bz^3_+},
\frac{\eps\bz_+^1}{\bz^3_+},\frac{\eps}{\bz^3_+}\right)~,~~~
\tau_\eps(z_-^1,z_-^2,z^3_-)\ =\ \left(\frac{\eps\bz_-^2}{\bz^3_-}
,\frac{\bz_-^1}{\bz^3_-},\frac{\eps}{\bz^3_-}\right)~,
\end{equation}
\begin{equation}
\tau_0(z_\pm^1,z_\pm^2,z^3_\pm)\ =\ (\bz_\pm^1,\bz^2_\pm,\bz^3_\pm)~.
\end{equation}
\end{subequations}

\paragraph{The dual twistor space.} For the dual twistor space,\index{twistor}\index{twistor!dual twistor}\index{twistor!space}
one starts again from the complex projective space $\CPP^3$, this\index{complex!projective space}
time parameterized by the two-spinors $(\sigma^\ald,\mu_\alpha)$.\index{Spinor}\index{two-spinors}
By demanding that $\mu_\alpha\neq (0,0)^T$, we again get a rank
two holomorphic vector bundle over the Riemann sphere:\index{Riemann sphere}\index{holomorphic!vector bundle}
\begin{equation}\label{dualbundle}
\CP^3_*\ =\ \CO(1)\oplus\CO(1)\ \rightarrow\  \CPP^1_*~.
\end{equation}
One should stress that the word ``dual'' refers to the
transformation property of the spinors and {\em not} to the dual\index{Spinor}
line bundles $\CO(-1)$ of the holomorphic line bundles $\CO(1)$\index{holomorphic!line bundle}
contained in the twistor space. This is why we denote these spaces\index{twistor}\index{twistor!space}
with a $*$ instead of a $\vee$. The dual twistor space $\CP^3_*$\index{twistor!dual twistor}
is covered again by two patches $\CU^*_\pm$ on which we have the
inhomogeneous coordinates\index{homogeneous coordinates}\index{inhomogeneous coordinates}
\begin{equation}
(u^\ald_\pm,\mu_\pm)\ewith u^\ald_+\ =\ \mu_+u^\ald_-\eand
\mu_+\ =\ \frac{1}{\mu_-}~.
\end{equation}
Sections of the bundle \eqref{dualbundle} are therefore
parameterized according to
\begin{equation}\label{dualincidence}
u^\ald_\pm\ =\ x^{\alpha\ald}\mu^\pm_\alpha\ewith\left(\mu_\alpha^+\right)
:=\left(\begin{array}{c}1\\\mu_+
\end{array}\right)\eand
\left(\mu_\alpha^-\right)\ :=\ \left(\begin{array}{c}\mu_-\\1
\end{array}\right)~,
\end{equation}
and one has again a double fibration analogously to\index{double fibration}\index{fibration}
\eqref{dblfibration}. The null planes in $\FC^4$ corresponding to
points in the dual twistor space via the above incidence relations\index{incidence relation}\index{twistor}\index{twistor!dual twistor}\index{twistor!space}
\eqref{dualincidence} are now called {\em $\beta$-planes}.\index{b-plane@$\beta$-plane}

Note that in most of the literature on twistor spaces, our dual\index{twistor}\index{twistor!space}
twistor space is called twistor space and vice versa. This is
related to focusing on anti-self-dual Yang-Mills theory instead of\index{Yang-Mills theory}\index{anti-self-dual}\index{self-dual Yang-Mills theory}
the self-dual one, as we do.

\paragraph{Real twistor space.}\label{pKleinianCase} It is\index{twistor}\index{twistor!real twistor space}\index{twistor!space}
obvious that the involution $\tau_{-1}$ has no fixed points but\index{involution}
does leave invariant projective lines joining $p$ and
$\tau_{-1}(p)$ for any $p\in\CP^3$. On the other hand, the
involutions $\tau_1$ and $\tau_0$ have fixed points which form a\index{involution}
three-dimensional real manifold\index{manifold}
\begin{equation}\label{realtwistor}
\CT^3\ =\ \FR P^3\backslash\FR P^1
\end{equation}
fibred over $S^1\cong\FR P^1\subset \CPP^1$. The space
$\CT^3\subset\CP^3$ is called {\em real twistor space}. For the\index{twistor}\index{twistor!real twistor space}\index{twistor!space}
real structure $\tau_1$, this space is described by the\index{real structure}
coordinates $(z_\pm^1,\,\de^{\di\chi}\,\bz_\pm^1,\,\de^{\di\chi})$
with $0\leq\chi<2\pi$, and for the real structure $\tau_0$, the\index{real structure}
coordinates $(z_\pm^1,\,z_\pm^2,\,\lambda_\pm)$ are real. These
two descriptions are equivalent. In the following we shall
concentrate mostly on the real structures $\tau_{\pm 1}$ since\index{real structure}
they give rise to unified formul\ae{}.

\paragraph{Metric on the moduli space of real curves.} The real\index{moduli space}
structures introduced above naturally induce real structures on\index{real structure}
the moduli space of curves, $\FC^4$:\index{moduli space}
\begin{equation}
\begin{aligned}
\tau_\eps(x^{1\ed})\ =\ \bar{x}^{2\zd}~,~~~ \tau_\eps(x^{1\zd})\
=\ \eps \bar{x}^{2\ed}\ewith\eps\ =\ \pm 1~,\\
\tau_0(x^{\alpha\ald})\ =\ \bar{x}^{\alpha\ald}~.\hspace{2.5cm}
\end{aligned}
\end{equation}
Demanding that $\tau(x^{\alpha\ald})=x^{\alpha\ald}$ restricts the
moduli space $\FC^4$ to $\FC^2\cong \FR^4$, and we can extract\index{moduli space}
four real coordinates via
\begin{equation}\label{2.13}
\bar{x}^{2\zd}\ =\ x^{1\ed}\ =:\ -(\eps x^4+\di x^3)\eand
x^{2\ed}\ =\ \eps\bar{x}^{1\zd}\ =:\ -\eps(x^2-\di x^1)~.
\end{equation}
Furthermore, the real moduli space comes naturally with the metric\index{moduli space}\index{metric}
given by
\begin{equation}\label{realmetric}
\dd s^2\ =\ \det(\dd x^{\alpha\ald})\ =\ g_{\mu\nu}\dd x^\mu\dd x^\nu
\end{equation}
with $g=(g_{\mu\nu})=\diag(+1,+1,+1,+1)$ for the involution\index{involution}
$\tau_{-1}$ on $\CP^3$ and $g=\diag(-1,-1,$ $+1,+1)$ for
$\tau_{1}$ (and $\tau_0$). Thus, the moduli space of real rational\index{moduli space}
curves of degree one in $\CP^3$ is the Euclidean space\footnote{In
our notation, $\FR^{p,q}=(\FR^{p+q},g)$ is the space $\FR^{p+q}$
with the metric\index{metric}
$g=\diag(\underbrace{-1,\ldots ,-1}_{q},\underbrace{+1,\ldots ,+1}_{p})$.}
$\FR^{4,0}$ or the Kleinian space $\FR^{2,2}$. It is not possible
to introduce a Minkowski metric on the moduli space of real\index{moduli space}\index{metric}
sections of the twistor space $\CP^3$. However, this will change\index{twistor}\index{twistor!space}
when we consider the ambitwistor space in section\index{ambitwistor space}\index{twistor!ambitwistor}
\ref{ssAmbitwistorSpace}.

\paragraph{Diffeomorphisms in the real case.}\label{pRealDiffeos}
It is important to note that the diagram
\begin{equation}\label{2.17}
\begin{aligned}
\begin{picture}(50,40)
\put(0.0,0.0){\makebox(0,0)[c]{$\CP^3_\eps$}}
\put(64.0,0.0){\makebox(0,0)[c]{$\FR^4$}}
\put(34.0,33.0){\makebox(0,0)[c]{$\FR^4\times \CPP^1$}}
\put(7.0,18.0){\makebox(0,0)[c]{$\pi_2$}}
\put(55.0,18.0){\makebox(0,0)[c]{$\pi_1$}}
\put(25.0,25.0){\vector(-1,-1){18}}
\put(37.0,25.0){\vector(1,-1){18}}
\end{picture}
\end{aligned}
\end{equation}
describes quite different situations in the Euclidean ($\eps=-1$)
and the Kleinian ($\eps=+1$) case. For $\eps=-1$, the map $\pi_2$
is a diffeomorphism,
\begin{equation}
\CP^3_{-1}\ \cong\  \FR^{4,0}\times \CPP^1~,
\end{equation}
and the double fibration \eqref{2.17} is simplified to the\index{double fibration}\index{fibration}
non-holomorphic fibration
\begin{equation}\label{2.19}\index{fibration}
\CP^3_{-1}\ \rightarrow\  \FR^{4,0}
\end{equation}
where $3$ stands for complex and $4$ for real dimensions. More
explicitly, this diffeomorphism reads
\begin{equation}\label{eq:2.19}
\begin{aligned}
x^{1\ed}\ =\ \frac{z^1_++z_+^3\bz^2_+}{1+z_+^3\bz_+^3}\ =\
\frac{\bz_-^3z_-^1+\bz_-^2}{1+z_-^3\bz_-^3}~,~~~ &x^{1\zd}\ =\
-\frac{\bz^2_+-\bz_+^3z^1_+}{1+z_+^3\bz_+^3}\ =\
-\frac{z_-^3\bz^2_--z_-^1}{1+z_-^3\bz_-^3}~,\\\lambda_\pm\ =\
&z_\pm^3~,
\end{aligned}
\end{equation}
and the patches $\CU_\pm$ are diffeomorphic to the patches\index{diffeomorphic}
$\tilde{\CU}_\pm$. Correspondingly, we can choose either
coordinates $(z^\alpha_\pm,z_\pm^3:=\lambda_\pm)$ or
$(x^{\alpha\ald},\lambda_\pm)$ on $\CP^3_{-1}$ and consider this
space as a complex $3$-dimensional or real $6$-dimensional
manifold. Note, however, that the spaces $\CP^3_{-1}$ and\index{manifold}
$\FR^{4,0}\times\CPP^1$ are not biholomorphic.\index{biholomorphic}

In the case of Kleinian signature $($$+$$+$$-$$-$$)$, we have a\index{Kleinian signature}
local isomorphisms\index{morphisms!isomorphism}
\begin{equation}
\sSO(2,2)\ \simeq\  \sSpin(2,2)\ \simeq\  \sSL(2,\FR)\times
\sSL(2,\FR)\ \simeq\ \sSU(1,1)\times \sSU(1,1)~,
\end{equation}
and under the action of the group $\sSU(1,1)$, the Riemann sphere\index{Riemann sphere}
$\CPP^1$ of projective spinors decomposes into the disjoint union\index{Spinor}
$\CPP^1=H^2_+\cup S^1\cup H_-^2=H^2\cup S^1$ of three orbits.
Here, $H^2=H_+^2\cup H_-^2$ is the two-sheeted hyperboloid and
$H_\pm^2=\{\lambda_\pm\in U_\pm||\lambda_\pm|<1\}\cong
\sSU(1,1)/\sU(1)$ are open discs. This induces a decomposition of
the correspondence space into
\begin{equation}
\FR^{4}\times \CPP^1\ =\ \FR^{4}\times H_+^2\cup \FR^{4}\times
S^1\cup \FR^{4}\times H_-^2\ =\ \FR^{4}\times H^2\cup
\FR^{4}\times S^1
\end{equation}
as well as a decomposition of the twistor space\index{twistor}\index{twistor!space}
\begin{equation}
\CP^{3}\ =\ \CP_+^{3}\cup \CP_0\cup\CP_-^{3}\ =:\
\tilde{\CP}^{3}\cup\CP_0~,
\end{equation}
where $\CP_\pm^{3}:=\CP^{3}|_{H^2_\pm}$ are restrictions of the
holomorphic vector bundle \eqref{bundle} to bundles over\index{holomorphic!vector bundle}
$H^2_\pm$. The space $\CP_0:=\CP^{3}|_{S^1}$ is the real
$5$-dimensional common boundary of the spaces $\CP^{3}_\pm$. There
is a real-analytic bijection between $\FR^{4}\times
H_\pm^2\cong\FC^2\times H_\pm^2$ and $\tilde{\CP}_\pm^{3}$, which
reads explicitly as
\begin{equation}
\begin{aligned}
x^{1\ed}\ =\ \frac{z^1_+-z_+^3\bz^2_+}{1-z_+^3\bz_+^3}\ =\
\frac{-\bz_-^3z_-^1+\bz_-^2}{1-z_-^3\bz_-^3}~,~~~ &x^{1\zd}\ =\
\frac{\bz^2_+-\bz_+^3z^1_+}{1-z_+^3\bz_+^3}\ =\
-\frac{z_-^3\bz^2_--z_-^1}{1-z_-^3\bz_-^3}~,\\\lambda_\pm\ =\
&z_\pm^3~.
\end{aligned}
\end{equation}
To indicate which spaces we are working with, we will sometimes
use the notation $\CP^{3}_{\eps}$ and imply
$\CP^{3}_{-1}:=\CP^{3}$ and $\CP^{3}_{+1}:=\tilde{\CP}^{3}\subset
\CP^{3}$. The situation arising for the real structure $\tau_0$\index{real structure}
can -- in principle -- be dealt with analogously.

\paragraph{Vector fields on $\CP^3_\eps$.} On $\CP_\eps^{3}$,
there is the following relationship between vector fields of type
(0,1) in the coordinates $(z_\pm^\alpha,z_\pm^3)$ and vector
fields \eqref{Vbars} in the coordinates
$(x^{\alpha\ald},\lambda_\pm)$:
\begin{equation}\label{VectorIdentity}
\begin{aligned}
\der{\bz_\pm^1}&\ =\ -\gamma_\pm\lambda_\pm^\ald\der{x^{2\ald}} \
=:\ -\gamma_\pm\bar{V}_2^\pm~,~~~& \der{\bz_\pm^2}&\ =\
\gamma_\pm\lambda_\pm^\ald\der{x^{1\ald}} \ =:\
-\eps\gamma_\pm\bar{V}_1^\pm~,\\\der{\bz_+^3}&\ =\
\der{\bl_+}+\eps\gamma_+ x^{\alpha\dot{1}}\bar{V}_\alpha^+~,~~~&
\der{\bz_-^3}&\ =\ \der{\bl_-}+\gamma_-
x^{\alpha\dot{2}}\bar{V}_\alpha^-~,
\end{aligned}
\end{equation}
where we introduced the factors
\begin{equation}\label{gammas}
\gamma_+\ =\ \frac{1}{1-\eps\lambda_+\bl_+}\ =\
\frac{1}{\hl^\ald_+\lambda^+_\ald}\eand \gamma_-\ =\
-\eps\frac{1}{1-\eps\lambda_-\bl_-}\ =\ \frac{1}{\hl^\ald_-
\lambda^-_\ald}~.
\end{equation}

\paragraph{The real twistor space $\CT^3$.} The set of fixed points\index{twistor}\index{twistor!real twistor space}\index{twistor!space}
under the involution $\tau_1$\footnote{Although $\tau_1$ was\index{involution}
defined on $\CP^{3}$, it induces an involution on $\CF^{5}$ which
we will denote by the same symbol in the following.} of the spaces
contained in the double fibration \eqref{dblfibration} form real\index{double fibration}\index{fibration}
subsets $\CT^{3}\subset\CP^{3}$, $\FR^{2,2}\subset\FC^{4}$ and
$\FR^{2,2}\times S^1\subset\CF^{5}$. Recall that the space $\CT^3$
is diffeomorphic to the space $\RPS^3\backslash\RPS^1$ (cf.\index{diffeomorphic}
\eqref{realtwistor}) fibred over $S^1\cong\RPS^1\subset\CPP^1$.
Thus, we obtain the real double fibration\index{double fibration}\index{fibration}
\begin{equation}\label{C5}
\begin{aligned}
\begin{picture}(50,50)
\put(0.0,0.0){\makebox(0,0)[c]{$\CT^{3}$}}
\put(74.0,0.0){\makebox(0,0)[c]{$\FR^{2,2}$}}
\put(38.0,37.0){\makebox(0,0)[c]{$\FR^{2,2}\times S^1$}}
\put(7.0,20.0){\makebox(0,0)[c]{$\pi_2$}}
\put(65.0,20.0){\makebox(0,0)[c]{$\pi_1$}}
\put(25.0,27.0){\vector(-1,-1){18}}
\put(47.0,27.0){\vector(1,-1){18}}
\end{picture}
\end{aligned}
\end{equation}
Here, $\pi_1$ is again the trivial projection and $\pi_2$ is given
by equations \eqref{sections2} with $|\lambda_\pm|=1$.

The tangent spaces to the two-dimensional leaves of the fibration\index{fibration}
$\pi_2$ in \eqref{C5} are spanned by the vector fields
\begin{equation}\label{C7}
\bv^+_\alpha\ :=\ \lambda_+^\ald\der{x^{\alpha\ald}}\ewith
\bv^+_2\ =\ -\lambda_+ \bv^+_1~,
\end{equation}
where $|\lambda_+|=1$. Equivalently, one could also use the vector
fields
\begin{equation}\label{C6.2}
\bv^-_\alpha\ :=\ \lambda_-^\ald\der{x^{\alpha\ald}}\ =\ \lambda_-
\bv^+_\alpha\ewith\lambda_-\ =\ \frac{1}{\lambda_+}\ =\ \bl_+~.
\end{equation}
The vector fields \eqref{C7} and \eqref{C6.2} are the restrictions
of the vector fields $\bar{V}^\pm_\alpha$ from \eqref{Vbars} to
$|\lambda_\pm|=1$.

\paragraph{Forms.} The forms $\bE^a_\pm$ with $a=1,2,3$ dual to the above vector
fields are given by the formul\ae{}
\begin{equation}\label{DefFormsBos}
\bE^\alpha_\pm\ =\ -\gamma_\pm\hl_\ald^\pm\dd x^{\alpha\ald}~,~~~
\bE^3_\pm\ =\ \dd \bl_\pm~.
\end{equation}

\paragraph{Flag manifolds.} There is a nice interpretation of the\index{flag manifold}\index{manifold}
double fibrations \eqref{dblfibration} and its dual version in\index{double fibration}\index{fibration}
terms of flag manifolds (see \ref{pFlagManifolds} of section\index{flag manifold}\index{manifold}
\ref{ssManifolds}). For this, however, we have to focus back on
the full complexified compactified twistor space $\CPP^3$. Upon\index{twistor}\index{twistor!space}
fixing the full space the flags will live in to be $\FC^4$, we can
establish the following double fibration:\index{double fibration}\index{fibration}
\begin{equation}\label{flgdblfibration1}
\begin{aligned}
\begin{picture}(75,40)
\put(0.0,0.0){\makebox(0,0)[c]{$F_{1,4}$}}
\put(64.0,0.0){\makebox(0,0)[c]{$F_{2,4}$}}
\put(32.0,33.0){\makebox(0,0)[c]{$F_{12,4}$}}
\put(7.0,18.0){\makebox(0,0)[c]{$\pi_2$}}
\put(55.0,18.0){\makebox(0,0)[c]{$\pi_1$}}
\put(25.0,25.0){\vector(-1,-1){18}}
\put(37.0,25.0){\vector(1,-1){18}}
\end{picture}
\end{aligned}
\end{equation}
Let $(L_1,L_2)$ be an element of $F_{12,4}$, i.e.\ $\dim_\FC
L_1=1$, $\dim_\FC L_2=2$ and $L_1\subset L_2$. Thus $F_{12,4}$
fibres over $F_{2,4}$ with $\CPP^1$ as a typical fibre, which
parameterizes the freedom to choose a complex one-dimensional
subspace in a complex two-dimensional vector space. The
projections are defined as $\pi_2(L_1,L_2)=L_1$ and
$\pi_1(L_1,L_2)=L_2$. The full connection to \eqref{dblfibration}\index{connection}
becomes obvious, when we note that
$F_{1,4}=\CPP^3=\CP^3\cup\CPP^1$ and that $F_{2,4}=G_{2,4}(\FC)$
is the complexified and compactified version of $\FR^4$. The
advantage of the formulation in terms of flag manifolds is related\index{flag manifold}\index{manifold}
to the fact that the projections are immediately clear: one has to
shorten the flags to suit the structure of the flags of the base
space.

The compactified version of the ``dual'' fibration is\index{fibration}
\begin{equation}\label{flgdblfibration2}
\begin{aligned}
\begin{picture}(80,40)
\put(0.0,0.0){\makebox(0,0)[c]{$F_{3,4}$}}
\put(64.0,0.0){\makebox(0,0)[c]{$F_{2,4}$}}
\put(32.0,33.0){\makebox(0,0)[c]{$F_{23,4}$}}
\put(7.0,18.0){\makebox(0,0)[c]{$\pi_2$}}
\put(55.0,18.0){\makebox(0,0)[c]{$\pi_1$}}
\put(25.0,25.0){\vector(-1,-1){18}}
\put(37.0,25.0){\vector(1,-1){18}}
\end{picture}
\end{aligned}
\end{equation}
where $F_{3,4}$ is the space of hyperplanes in $\FC^4$. This space
is naturally dual to the space of lines, as every hyperplane is
fixed by a vector orthogonal to the elements of the hyperplane.
Therefore, we have $F_{3,4}=F_{1,4}^*=\CPP_*^3\supset \CP^3_*$.

\subsection{The Penrose-Ward transform}\label{ssPenroseWardTrafo}\index{Penrose-Ward transform}

The Penrose-Ward transform gives a relation between solutions to\index{Penrose-Ward transform}
the self-dual Yang-Mills equations on $\FC^4$ and a topologically
trivial, holomorphic vector bundle $E$ over the twistor space\index{holomorphic!vector bundle}\index{twistor}\index{twistor!space}
$\CP^3$, which becomes holomorphically trivial upon restriction to
holomorphic submanifolds $\CPP^1\subset \CP^3$. We will first
discuss the complex case, and end this section with a remark on
the simplifications in the real setting.

\paragraph{The holomorphic bundle over $\CP^3$.} We start our
considerations from a rank $n$ holomorphic vector bundle $E$ over\index{holomorphic!vector bundle}
the twistor space $\CP^3$. We assume that $E$ is topologically\index{twistor}\index{twistor!space}
trivial, i.e.\ one can split its transition function $f_{+-}$\index{transition function}
according to $f_{+-}=\psi_+^{-1}\psi_-$, where $\psi_\pm$ are
smooth functions on the patches $\CU_\pm$. For the Penrose-Ward
transform to work, we have to demand additionally that $E$ becomes\index{Penrose-Ward transform}
holomorphically trivial, when we restrict it to an arbitrary
section of the vector bundle $\CP^3\rightarrow \CPP^1$. We will
comment in more detail on this condition in \ref{pHolTriv}.

\paragraph{Pull-back to the correspondence space.} The pull-back
bundle $\pi_2^*E$ over the correspondence space $\FC^4\times
\CPP^1$ has a transition function $\pi_2^*f_{+-}$, which due to\index{transition function}
its origin as a pull-back satisfies the equations
\begin{equation}\label{holcond}
\bV_\alpha^\pm (\pi_2^*f_{+-})\ =\ 0~.
\end{equation}
Furthermore, due to the holomorphic triviality of $E$ on any
$\CPP^1_x\embd \CP^3$, we can split the transition function\index{transition function}
according to
\begin{equation}\label{splitting}
\pi_2^*f_{+-}(x^{\alpha\ald}\lambda_\ald^\pm,\lambda_\ald^\pm)\ =\ \psi_+^{-1}(x^{\alpha\ald},\lambda_+)
\psi_-(x^{\alpha\ald},\lambda_-)~,
\end{equation}
where $\psi_\pm$ are holomorphic, matrix-group-valued functions in
the coordinates $(x^{\alpha\ald},\lambda_\pm)$ of the
correspondence space. This is easily seen by pulling back the
restrictions of $E$ to the $\CPP^1_x$ for each $x^{\alpha\ald}$
separately. This guarantees a splitting holomorphic in
$\lambda_\pm$ parameterizing the $\CPP^1_x$. Since the embedding
of the $\CPP^1_x$ in the twistor space $\CP^3$ is holomorphically\index{twistor}\index{twistor!space}
described by the moduli $x^{\alpha\ald}$, the splitting is
furthermore holomorphic in the latter coordinates.

\paragraph{Construction of a gauge potential.} On the
correspondence space, we obtain from \eqref{holcond} together with
\eqref{splitting} the equation
\begin{equation}\label{gluing}
\psi_+\bV^+_\alpha \psi_+^{-1}\ =\ \psi_-\bV_\alpha^+\psi_-^{-1}
\end{equation}
over $\CUt_+\cap\CUt_-$. One can expand $\psi_+$, $\psi_+^{-1}$
and $\psi_-$, $\psi_-^{-1}$ as power series in $\lambda_+$ and
$\lambda_-=\lambda_+^{-1}$, respectively. Upon substituting the
expansions into equations \eqref{gluing}, one sees that both sides
in \eqref{gluing} must be linear in $\lambda_+$; this is a
generalized Liouville theorem. One can introduce Lie algebra\index{Lie algebra}\index{Theorem!Liouville}
valued fields $A_\alpha$ whose dependence on $\lambda_\pm$ is made
explicit in the formul\ae
\begin{equation}\label{contractedA}
A^+_\alpha\ :=\  \lambda^\ald_+ \,
A_{\alpha\ald}\ =\ \lambda_+^\ald\,\psi_+\,\dpar_{\alpha\ald}\,\psi_+^{-1}\ =\ 
\lambda_+^\ald\,\psi_-\,\dpar_{\alpha\ald}\,\psi_-^{-1}~.
\end{equation}
The matrix-valued functions $A_{\alpha\ald}(x)$ can be identified
with the components of a gauge potential $A_{\alpha\ald}\dd
x^{\alpha\ald}+A_{\CPP^1}$ on the correspondence space
$\FC^4\times \CPP^1$ with $A_{\CPP^1}=0$: the component
$A_{\bl_+}$, vanishes as
\begin{equation}\label{holo3}
A_{\bl_+}\ =\ \psi_+\dpar^{\phantom{\pm}}_{\bl_+}\psi_+^{-1}\ =\ 0~.
\end{equation}

\paragraph{Linear system and the SDYM equations.} The equations\index{linear system}
\eqref{gluing} can be recast into a linear system
\begin{equation}\label{linsysSDYM}\index{linear system}
(\bV_\alpha^++A_\alpha^+)\psi_+\ =\ 0~,
\end{equation}
with similar equations for $\psi_-$. We encountered this linear
system already in section \ref{ssIntegrabilityLinearSystems},\index{linear system}
\ref{pSDYM}, and we briefly recall that the compatibility
conditions of this linear system are\index{compatibility conditions}\index{linear system}
\begin{equation}\label{compconds}
[\bar{V}_\alpha^++A_\alpha^+\,,\bar{V}_\beta^++A_\beta^+]\ =\ 
\lambda_+^\ald\lambda_+^\bed
[\dpar_{\alpha\ald}+A_{\alpha\ald}\,,\dpar_{\beta\bed}+A_{\beta\bed}]\ =:\ 
\lambda_+^\ald\lambda_+^\bed F_{\alpha\ald,\beta\bed}\ =\ 0~.
\end{equation}
{}To be satisfied for all $(\lambda_+^\ald)$, this equation has to
vanish to all orders in $\lambda_+$ separately, from which we
obtain the self-dual Yang-Mills (SDYM) equations
\begin{equation}
F_{1\dot{1},2\dot{1}}\ =\ 0~,~~~F_{1\dot{2},2\dot{2}}\ =\ 0~,~~~
F_{1\dot{1},2\dot{2}}+F_{1\dot{2},2\dot{1}}\ =\ 0
\end{equation}
for a gauge potential $(A_{\alpha\ald})$. Recall that in the
spinorial notation\index{Spinor}
$F_{\alpha\ald,\beta\bed}=\eps_{\alpha\beta}f_{\ald\bed}+\
\eps_{\ald\bed}f_{\alpha\beta}$ the SDYM equations $F=*F$ are
rewritten as
\begin{equation}
f_{\ald\bed}\ =\ 0~,
\end{equation}
i.e.\ the part of $F_{\alpha\ald\beta\bed}$ symmetric in the
indices $\ald\bed$ vanishes.

\paragraph{Holomorphic triviality.}\label{pHolTriv} Let us comment on the condition
of holomorphic triviality of $E$ upon reduction on subspaces
$\CPP^1_x$ in slightly more detail. For this, recall that every
rank $n$ holomorphic vector bundle over $\CPP^1$ is\index{holomorphic!vector bundle}
(holomorphically) equivalent to a direct sum of line bundles
\begin{equation}\label{6}
\CO(i_1)\oplus\ldots \oplus\CO(i_n)~,
\end{equation}
i.e., it is uniquely determined by a set of integers
$(i_1,\ldots,i_n)$. Furthermore, the sum of the $i_k$ is a
topological invariant and each of the $i_k$ is a holomorphic
invariant up to permutation.

Now consider a rank $n$ holomorphic vector bundle $E$ over the\index{holomorphic!vector bundle}
twistor space $\CP^{3}$. The set of equivalence classes of such\index{twistor}\index{twistor!space}
vector bundles $E$ which become holomorphically equivalent to the\index{holomorphically equivalent}
bundle \eqref{6} when restricted to any projective line $\CPP^1_x$
in $\CP^3$ will be denoted by $\CM(i_1,\ldots ,i_n)$. The moduli space\index{moduli space}
of holomorphic vector bundles on $\CP^3$ contains then all of the\index{holomorphic!vector bundle}
above moduli spaces:\index{moduli space}
\begin{equation}\label{8}
\CM\ \supset\ \bigcup_{i_1,\ldots ,i_n}\CM(i_1,\ldots ,i_n)~.
\end{equation}
Furthermore, $\CM$ contains also those holomorphic vector bundles\index{holomorphic!vector bundle}
whose restrictions to different $\CPP^1_x\embd \CP^3$ are not
holomorphically equivalent.\index{holomorphically equivalent}

The focus of interest in the literature, including this thesis, is
in general the moduli subspace
\begin{equation}\label{7}
\CM(0,\ldots,0)
\end{equation}
which is clearly a true subset of \eqref{8} and bijective to the
moduli space of solutions to the SDYM equations in four\index{moduli space}
dimensions. Although a generalization of Ward's construction to
the cases $\CM(i_1,\ldots ,i_n)$ for arbitrary $(i_1,\ldots ,i_n)$ were
studied in the literature e.g.\ by Leiterer in
\cite{Leiterer:1983aa}, a thorough geometric interpretation was
only given for some special cases of $(i_1,\ldots ,i_n)$. It seems
that the situation has not been yet completely clarified. From
\eqref{8} together with the Leiterer examples, it is, however,
quite evident that the moduli space $\CM(0,\ldots ,0)$ is not a dense\index{moduli space}
subset of the moduli space $\CM$. Furthermore, the statement is
irrelevant in the most important recent application of the
Penrose-Ward-transform in twistor string theory, cf.\ section\index{string theory}\index{twistor}\index{twistor!string theory}
\ref{ssTST}: a perturbative expansion in the vicinity of the
vacuum solution, which corresponds to a trivial transition
function $f_{+-}=\unit_n$. There, the necessary property of\index{transition function}
holomorphic triviality after restriction to any $\CPP^1_{x}\embd
\CP^{3}$ follows immediately from Kodaira's theorem.\index{Theorem!Kodaira}

\paragraph{Holomorphic Chern-Simons equations.} One can also
obtain a linear system directly on the twistor space by using the\index{linear system}\index{twistor}\index{twistor!space}
splitting of the transition function $f_{+-}$ of $E$ into smooth\index{transition function}
functions via
\begin{equation}
f_{+-}\ =\ \hpsi^{-1}_+\hpsi_-~.
\end{equation}
Such a splitting exists, as $E$ was assumed to be topologically
trivial. Note furthermore that $f_{+-}$ is the transition function\index{transition function}
of a {\em holomorphic} vector bundle and therefore satisfies
\begin{equation}\label{holf+-}
\der{\bz^a_\pm} f_{+-}\ =\ 0\ewith a\ =\ 1,\ldots,3~.
\end{equation}
Similarly to the case of the components $A_{\alpha\ald}$
introduced before, we find here a gauge potential
\begin{equation}
\hat{A}^{0,1}\ =\ \hpsi_+ \dparb \hpsi_+^{-1}\ =\ \hpsi_- \dparb
\hpsi_-^{-1}~,
\end{equation}
on $\CU_+\cap\CU_-$, which can be extended to a gauge potential on
the full twistor space $\CP^3=\CU_+\cup\CU_-$. Note that here, we\index{twistor}\index{twistor!space}
choose not to work with components, but directly with the
resulting Lie algebra valued $(0,1)$-form $A^{0,1}$. Again, one\index{Lie algebra}
can cast equations \eqref{holf+-} into a linear system\index{linear system}
\begin{equation}\label{linsyshCS}
(\dparb+\hat{A}^{0,1})\hat{\psi}_\pm\ =\ 0~,
\end{equation}
which has the holomorphic Chern-Simons equations
\begin{equation}\label{hCSeqn}
\dparb \hat{A}^{0,1}+\hat{A}^{0,1}\wedge \hat{A}^{0,1}\ =\ 0
\end{equation}
as its compatibility conditions. This aspect of the Penrose-Ward\index{compatibility conditions}
transform will become particularly important when dealing with the
supertwistor space $\CP^{3|4}$ in section\index{twistor}\index{twistor!space}
\ref{ssSuperextensionTwistorSpace}. There, we will be able to give
an action for holomorphic Chern-Simons theory due to the existence\index{Chern-Simons theory}
of a holomorphic volume form on $\CP^{3|4}$.\index{holomorphic!volume form}

One has to stress at this point that a solution to the hCS
equations \eqref{hCSeqn} corresponds to an arbitrary holomorphic
vector bundle over the twistor space $\CP^3$, which does not\index{holomorphic!vector bundle}\index{twistor}\index{twistor!space}
necessarily satisfy the additional condition of holomorphic
triviality on all the $\CPP^1_x\embd \CP^3$. In the following, we
will always imply the restriction to the appropriate subset of
solutions to \eqref{hCSeqn} when speaking about general solutions
to the hCS equations \eqref{hCSeqn} on $\CP^3$. As mentioned in
the previous paragraph, this restriction is irrelevant for
perturbative studies. Furthermore, it corresponds to those gauge
potentials, for which the component $\CAh_{\bl_\pm}$ can be gauged
away, as we discuss in the following paragraph.

\paragraph{Gauge equivalent linear system.} Recall that the\index{linear system}
trivializations defined by formula \eqref{splitting} correspond to\index{trivialization}
holomorphic triviality of the bundle $E|_{\CPP^1_x}$ for any
$\CPP^1_x\embd\CP^3$. Similarly, we may consider restrictions of
$E$ to fibres $\FC^2_\lambda$ of the fibration $\CP^3\rightarrow\index{fibration}
\CPP^1$. All these restrictions are holomorphically trivial due to
the contractibility of $\FC^2_\lambda$ for any $\lambda\in
\CPP^1$. Therefore there exist regular matrix-valued functions
$\breve{\psi}_\pm(z_\pm^\alpha,\lambda_\pm,\bl_\pm)$ depending
holomorphically on $z_\pm^\alpha$ (and non-holomorphically on
$\lambda_\pm$) such that
\begin{equation}\label{trafofunc2}
f_{+-}\ =\ \hat{\psi}_+^{-1}\hat{\psi}_-\ =\ (\breve{\psi}_+)^{-1}\breve{\psi}_-~,
\end{equation}
and
$\varphi:=\psi_+(\breve{\psi}_+)^{-1}=\psi_-(\breve{\psi}_-)^{-1}$
defines a gauge transformation\index{gauge!transformation}
\begin{equation}\label{gaugetrafo}
\left(\der{\bz^\alpha_\pm}\lrcorner \hat{A}^{0,1}\cb
\hat{A}_{\bl_+}\hspace{-4pt}\ =\ 0\right)~\stackrel{\varphi}{\longmapsto}~
\left(\der{\bz^\alpha_\pm}\lrcorner \breve{A}^{0,1}=0\cb
\breve{A}_{\bl_+}\right)
\end{equation}
to a special trivialization in which only $\breve{A}_{\bl_+}\neq\index{trivialization}
0$ and $\der{\bz^\alpha_\pm}\lrcorner\breve{A}^{0,1}=0$.

\paragraph{The Euclidean case.} Let us now consider the
Penrose-Ward transform in the Euclidean case. The important point\index{Penrose-Ward transform}
here is that the spaces $\CP^3_{-1}$ and $\FR^4\times \CPP^1$
become diffeomorphic, and thus the double fibration\index{diffeomorphic}\index{double fibration}\index{fibration}
\eqref{dblfibration} reduces to a single fibration
\begin{equation}
\CP^3\ \rightarrow\  \FR^4~.
\end{equation}
Therefore, we have the identification \eqref{VectorIdentity}
between the vector fields $\bV^\pm_\alpha$ and $\der{\bz^\alpha}$.
This implies furthermore that the linear system \eqref{linsysSDYM}\index{linear system}
and \eqref{linsyshCS} become (gauge) equivalent. We have for the
two splittings
\begin{equation}
f_{+-}\ =\ \psi_+^{-1}\psi_-\ =\ \hpsi_+^{-1}\hpsi_-\ewith
\hpsi_+\ =\ \varphi^{-1}\psi_\pm~,
\end{equation}
where $\varphi$ is a globally defined, regular matrix-valued
function on $\CP^3$. Decomposing the gauge potential
$\hat{A}^{0,1}$ into the components
$\hat{A}^\pm_\alpha:=\bV^\pm_\alpha\lrcorner \hat{A}^{0,1}$, we
have
\begin{equation}
\begin{aligned}
\hat{A}^+_\alpha&\ :=\ \hat{\psi}_+\bar{V}_\alpha^+\hat{\psi}_+^{-1}\ =\ 
\hat{\psi}_-\bar{V}_\alpha^+\hat{\psi}_-^{-1}&&\ =\ 
\varphi^{-1}(\psi_\pm\bar{V}_\alpha^+\psi_\pm^{-1})\varphi+
\varphi^{-1}\bar{V}_\alpha^+\varphi\\ &&&\ =\ 
\varphi^{-1}A_\alpha^+\varphi+\varphi^{-1}\bar{V}_\alpha^+\varphi~,\\
\hat{A}_{\bl_+}&\ :=\ \hat{\psi}_+\dpar^{\phantom{\pm}}_{\bl_+}\hat{\psi}_+^{-1}\ =\ 
\hat{\psi}_-\dpar^{\phantom{\pm}}_{\bl_+}\hat{\psi}_-^{-1}&&\ =\ 
\varphi^{-1}\dpar^{\phantom{\pm}}_{\bl_+}\varphi~,
\end{aligned}
\end{equation}
from which we indeed realize that $\varphi$ plays the r{\^o}le of a
gauge transformation.\index{gauge!transformation}

\paragraph{Vector bundles in the Kleinian case.} Consider a
real-analytic function $f_{+-}^\tau:\CT^{3}\rightarrow\sGL(n,\FC)$
on the twistor space $\CT^3$ which can be understood as an\index{twistor}\index{twistor!space}
isomorphism $f_{+-}^\tau:E_-^\tau\rightarrow E_+^\tau$ between two\index{morphisms!isomorphism}
trivial complex vector bundles $E^\tau_\pm\rightarrow\CT^{3}$. We\index{complex!vector bundle}
assume that $f_{+-}^\tau$ satisfies the reality condition
\begin{equation}\label{D1}
\left(f_{+-}^\tau\left(z_+^\alpha,\lambda_+\right)\right)^\dagger
\ =\ f_{+-}^\tau(z_+^\alpha,\lambda_+)~.
\end{equation}
Such a function $f_{+-}^\tau$ can be extend holomorphically into a
neighborhood $\CU$ of $\CT^{3}$ in $\CP^{3}$, such that the
extension $f_{+-}$ of $f_{+-}^\tau$ satisfies the reality
condition
\begin{equation}\label{DD3}
\left(f_{+-}\left(\tau_1(z_+^\alpha,\lambda_+)\right)\right)^\dagger
\ =\ f_{+-}(z_+^\alpha,\lambda_+)~,
\end{equation}
generalizing equation \eqref{D1}. The function $f_{+-}$ is
holomorphic on $\CU=\CU_+\cap\CU_-$ and can be identified with a
transition function of a holomorphic vector bundle $E$ over\index{holomorphic!vector bundle}\index{transition function}
$\CP^{3}=\CU_+\cup\CU_-$ which glues together two trivial bundles
$E_+=\CU_+\times\FC^n$ and $E_-=\CU_-\times \FC^n$. Obviously, the
two trivial vector bundles $E^\tau_\pm\rightarrow \CT^{3}$ are
restrictions of the trivial bundles $E_\pm\rightarrow \CU_\pm$ to
$\CT^{3}$.

The assumption that $E$ becomes holomorphically trivial upon
reduction to a subset $\CPP^1_x\subset \CP^3$ implied a splitting
of the transition function $f_{+-}$,\index{transition function}
\begin{equation}\label{DD5}
f_{+-}\ =\ \psi_+^{-1}\psi_-~,
\end{equation}
into regular matrix-valued functions $\psi_+$ and $\psi_-$ defined
on $\CU_+=\CP_+^{3}\cup\CU$ and $\CU_-=\CP_-^{3}\cup\CU$ and
holomorphic in $\lambda_+\in H_+^2$ and $\lambda_-\in H_-^2$,
respectively. Note that the condition \eqref{DD3} is satisfied if
\begin{equation}\label{DDD4}
\psi_+^{-1}(\tau_1(x^{\alpha\ald},\lambda_+))\ =\ 
\psi_-^\dagger(x^{\alpha\ald},\lambda_-)~.
\end{equation}
Restricting \eqref{DD5} to $S^1_{x}\embd\CPP^1_{x}$, we obtain
\begin{equation}\label{DDD2}
f_{+-}^\tau\ =\ (\psi_+^\tau)^{-1}\psi_-^\tau~~~\mbox{with}~~~
(\psi_+^\tau)^{-1}\ =\ (\psi_-^\tau)^\dagger~,
\end{equation}
where the $\psi_\pm^\tau$ are restrictions to $\FR^4\times S^1$ of
the matrix-valued functions $\psi_\pm$ given by \eqref{DD5} and
\eqref{DDD4}. Thus the initial twistor data consist of a\index{twistor}
real-analytic function\footnote{One could also consider the
extension $f_{+-}$ and the splitting \eqref{DDD2} even if
$f^\tau_{+-}$ is not analytic, but in this case the solutions to
the super SDYM equations can be singular. Such solutions are not
related to holomorphic bundles.} $f_{+-}^\tau$ on $\CT^{3}$
satisfying \eqref{D1} together with a splitting \eqref{DDD2}, from
which we construct a holomorphic vector bundle $E$ over $\CP^{3}$\index{holomorphic!vector bundle}
with a transition function $f_{+-}$ which is a holomorphic\index{transition function}
extension of $f_{+-}^\tau$ to $\CU\supset\CT^{3}$. In other words,
the space of real twistor data is the moduli space of holomorphic\index{moduli space}\index{twistor}
vector bundles $E\rightarrow\CP^{3}$ with transition functions\index{transition function}
satisfying the reality conditions \eqref{DD3}.

\paragraph{The linear system on $\CT^3$.}\index{linear system}
In the purely real setting, one considers a real-analytic
$\sGL(n,\FC)$-valued function $f^\tau_{+-}$ on $\CT^{3}$
satisfying the Hermiticity condition \eqref{D1} in the context of
the real double fibration \eqref{C5}. Since the pull-back of\index{double fibration}\index{fibration}
$f_{+-}^\tau$ to $\FR^{4}\times S^1$ has to be constant along the
fibres of $\pi_2$, we obtain the constraint equations\index{constraint equations}
$\bv^+_\alpha f^\tau_{+-}=0$ or equivalently $\bv^-_\alpha
f^\tau_{+-}=0$ with the vector fields $\bv^\pm_\alpha$ defined in
\eqref{C7}. Using the splitting \eqref{DDD2} of $f^\tau_{+-}$ on
fibres $S^1_{x}$ of the projection $\pi_1$ in \eqref{C5} and
substituting $f^\tau_{+-}=(\psi^\tau_+)^{-1}\psi^\tau_-$ into the
above constraint equations, we obtain the linear systems\index{constraint equations}\index{linear system}
\begin{equation}\label{T3linsys}
(\bv^+_\alpha+\CA^+_\alpha)\psi^\tau_+\ =\ 0~,~~~\mbox{or}~~~(\bv^-_\alpha+\CA^-_\alpha)\psi^\tau_-\ =\ 0~.
\end{equation}

Here, $\CA_\pm=(\CA_\alpha^\pm)$ are relative connections on the\index{connection}
bundles $E^\tau_\pm$. {}From \eqref{T3linsys}, one can find
$\psi^\tau_\pm$ for any given $\CA^\pm_\alpha$ and vice versa,
i.e.\ find $\CA^\pm_\alpha$ for given $\psi^\tau_\pm$ by the
formul\ae{}
\begin{equation}\label{D8s}
\begin{aligned}
\CA_\alpha^+&\ =\ \psi_+^\tau
\bv_\alpha^+(\psi_+^\tau)^{-1}\ =\ \psi_-^\tau
\bv_\alpha^+(\psi_-^\tau)^{-1}~,\\\CA_\alpha^-&\ =\ \psi_+^\tau
\bv_\alpha^-(\psi_+^\tau)^{-1}\ =\ \psi_-^\tau
\bv_\alpha^-(\psi_-^\tau)^{-1}~.
\end{aligned}
\end{equation}
The compatibility conditions of the linear systems\index{compatibility conditions}\index{linear system}
\eqref{T3linsys} read
\begin{equation}\label{D5}
\bv^\pm_\alpha
\CA^\pm_\beta-\bv^\pm_\beta\CA^\pm_\alpha+[\CA^\pm_\alpha,\CA^\pm_\beta]\ =\ 0~.
\end{equation}
Geometrically, these equations imply flatness of the curvature of\index{curvature}
the relative connections $\CA_\pm$ on the bundles $E^\tau_\pm$\index{connection}
defined along the real two-dimensional fibres of the projection
$\pi_2$ in \eqref{C5}.

Recall that $\psi^\tau_+$ and $\psi^\tau_-$ extend holomorphically
in $\lambda_+$ and $\lambda_-$ to $H_+^2$ and $H_-^2$,
respectively, and therefore we obtain from \eqref{D8s} that
$\CA^\pm_\alpha=\lambda_\pm^\ald\CA_{\alpha\ald}$, where
$\CA_{\alpha\ald}$ does not depend on $\lambda_\pm$. Then the
compatibility conditions \eqref{D5} of the linear systems\index{compatibility conditions}\index{linear system}
\eqref{T3linsys} reduce to the equations \eqref{compconds}. It was
demonstrated above that for $\eps=+1$, these equations are
equivalent to the field equations of SDYM theory on $\FR^{2,2}$.
Thus, there are bijections between the moduli spaces of solutions\index{moduli space}
to equations \eqref{D5}, the field equations of SDYM theory on
$\FR^{2,2}$ and the moduli space of $\tau_1$-real holomorphic\index{moduli space}
vector bundles $E$ over $\CP^{3}$.

\paragraph{Extension to $\tilde{\CP}^3$.} Consider now the extension
of the linear systems \eqref{T3linsys} to open domains\index{linear system}
$\CU_\pm=\CP^{3}_\pm\cup\,\CU\supset\CT^{3}$,
\begin{equation}
(\bar{V}_\alpha^\pm+\CA_\alpha^\pm)\psi_\pm\ =\ 0\eand
\dpar_{\bl_\pm}\psi_\pm\ =\ 0~,\label{DDD10}
\end{equation}
where here, the $\bar{V}_\alpha^\pm$ are vector fields of type
$(0,1)$ on $\CU^s_\pm:=\CU_\pm\backslash(\FR^4\times S^1)$ as
given in \eqref{Vbars} and \eqref{VectorIdentity}. These vector
fields annihilate $f_{+-}$ and from this fact and the splitting
\eqref{DD5}, one can also derive equations \eqref{DDD10}. Recall
that due to the existence of a diffeomorphism between the spaces
$\FR^{4}\times H^2$ and $\tilde{\CP}^{3}$ which is described in
\ref{pRealDiffeos}, the double fibration \eqref{2.17} simplifies\index{double fibration}\index{fibration}
to the nonholomorphic fibration
\begin{equation}\label{nonKleinian}\index{fibration}
\CP^3_{+1}\ \rightarrow\  \FR^4~.
\end{equation}
Moreover, since the restrictions of the bundle $E\rightarrow
\CP^3_{+1}$ to the two-dimensional leaves of the fibration\index{fibration}
\eqref{nonKleinian} are trivial, there exist regular matrix-valued
functions $\hat{\psi}_\pm$ on $\CU_\pm^s$ such that
\begin{equation}\label{DDD11}
f_{+-}\ =\ \hat{\psi}_+^{-1}\hat{\psi}_-
\end{equation}
on $\CU^s=\CU\backslash(\FR^{4}\times S^1)$. Additionally, we can
impose the reality condition
\begin{equation}\label{Drealcond}
\hat{\psi}_+^{-1}\left(x^{\alpha\ald},\frac{1}{\bl_+}\right)\ =\ 
\hat{\psi}_-^\dagger(x^{\alpha\ald},\lambda_-)
\end{equation}
on $\hat{\psi}_\pm$. Although $\CU^s$ consists of two disconnected
pieces, the functions $\hat{\psi}_\pm$ are not independent on each
piece because of the condition \eqref{Drealcond}, which also
guarantees \eqref{DD3} on $\CU^s$. The functions $\hat{\psi}_\pm$
and their inverses are ill-defined on $\FR^4\times
S^1\cong\CP_0^3$ since the restriction of $\pi_2$ to $\FR^4\times
S^1$ is a noninvertible projection onto $\CT^{3}$, see
\ref{pRealDiffeos}. Equating \eqref{DD3} and \eqref{DDD11}, one
sees that the singularities of $\hat{\psi}_\pm$ on $\FR^{4}\times
S^1$ split off, i.e.
\begin{equation}
\hat{\psi}_\pm\ =\ \varphi^{-1}\psi_\pm~,
\end{equation}
in a matrix-valued function $\varphi^{-1}$ which disappears from
\begin{equation}\label{DDD15}
f_{+-}\ =\ \hat{\psi}_+^{-1}\hat{\psi}_-\ =\ (\psi_+^{-1}\varphi)(\varphi^{-1}\psi_-)\ =\ \psi_+^{-1}\psi_-~.
\end{equation}
Therefore $f_{+-}$ is a nonsingular holomorphic matrix-valued
function on all of $\CU$.

{}From \eqref{DDD11}-\eqref{DDD15} it follows that on
$\tilde{\CP}^{3}$, we have a well-defined gauge transformation\index{gauge!transformation}
generated by $\varphi$ and one can introduce gauge potentials
$\hat{\CA}_+^{0,1}$ and $\hat{\CA}_-^{0,1}$ which are defined on
$\CU_+^s$ and $\CU_-^s$, respectively, but not on $\FR^{4}\times
S^1$. By construction,
$\hat{\CA}^{0,1}=(\hat{\CA}^{0,1}_+,\hat{\CA}^{0,1}_-)$ satisfies
the hCS equations \eqref{hCSeqn} on
$\tilde{\CP}^{3}=\CP_+^{3}\cup\CP_-^{3}$, which are equivalent to
the SDYM equations on $\FR^{2,2}$. Conversely, having a solution
$\hat{\CA}^{0,1}$ of the hCS field equations on the space
$\tilde{\CP}^{3}$, one can find regular matrix-valued functions
$\hat{\psi}_+$ on $\CU^s_+$ and $\hat{\psi}_-$ on $\CU_-^s$ which
satisfy the reality condition \eqref{Drealcond}. These functions
define a further function
$f^s_{+-}=\hat{\psi}_+^{-1}\hat{\psi}_-:\CU^s\rightarrow
\sGL(n,\FC)$ which can be completed to a holomorphic function
$f_{+-}:\CU\rightarrow \sGL(n,\FC)$ due to \eqref{DDD15}. The
latter one can be identified with a transition function of a\index{transition function}
holomorphic vector bundle $E$ over the full twistor space\index{holomorphic!vector bundle}\index{twistor}\index{twistor!space}
$\CP^{3}$. The restriction of $f_{+-}$ to $\CT^{3}$ is a
real-analytic function $f^\tau_{+-}$ which is {\em not
constrained} by any differential equation. Thus, in the case
$\eps=+1$ (and also for the real structure $\tau_0$), one can\index{real structure}
either consider two trivial complex vector bundles $E^\tau_\pm$\index{complex!vector bundle}
defined over the space $\CT^{3}$ together with an isomorphism\index{morphisms!isomorphism}
$f_{+-}^\tau:E_-^\tau\rightarrow E_+^\tau$ or a single complex
vector bundle $E$ over the space $\CP^{3}$. However, the\index{complex!vector bundle}
appropriate hCS theory which has the same moduli space as the\index{moduli space}
moduli space of these bundles is defined on $\tilde{\CP}^{3}$.
Moreover, real Chern-Simons theory on $\CT^{3}$ has no moduli,\index{Chern-Simons theory}
since its solutions correspond to flat bundles over $\CT^{3}$ with
constant transition functions\footnote{Note that these transition\index{transition function}
functions are in no way related to the transition functions
$f_{+-}$ of the bundles $E$ over $\CP^{3}$ or to the functions
$f_{+-}^\tau$ defined on the whole of $\CT^{3}$.} defined on the
intersections of appropriate patches covering $\CT^{3}$.

{}To sum up, there is a bijection between the moduli spaces of\index{moduli space}
solutions to equations \eqref{D5} and to the hCS field equations
on the space $\tilde{\CP}^{3}$ since both moduli spaces are\index{moduli space}
bijective to the moduli space of holomorphic vector bundles over\index{holomorphic!vector bundle}
$\CP^{3}$. In fact, whether one uses the real supertwistor space\index{twistor}\index{twistor!space}
$\CT^{3}$, or works with its complexification $\CP^{3}$, is partly\index{complexification}
a matter of taste. However, the complex approach is more
geometrical and more natural from the point of view of an action
principle and the topological B-model. For example, equations\index{topological!B-model}
\eqref{D5} cannot be transformed by a gauge transformation to a\index{gauge!transformation}
set of differential equations on $\CT^{3}$ as it was possible on
$\tilde{\CP}^{3}$ in the complex case. This is due to the fact
that the transition function $f_{+-}$, which was used as a link\index{transition function}
between the two sets of equations in the complex case does not
satisfy any differential equation after restriction to $\CT^{3}$.
{}From this we see that we cannot expect any action principle on
$\CT^{3}$ to yield equations equivalent to \eqref{D5} as we had in
the complex case. For these reasons, we will mostly choose to use
the complex approach in the following.

\paragraph{Reality of the gauge potential.} After imposing a reality
condition on the spaces $\CP^3$ and $\FC^4$, we have to do so for
the vector bundle and the objects it comes with, as well. Note
that $A_{\alpha\ald}\dd x^{\alpha\ald}$ will take values in the
algebra of anti-Hermitian $n\times n$ matrices if $\psi_\pm$
satisfies the following condition\footnote{Here, $\dagger$ means
Hermitian conjugation.}:
\begin{equation}
\psi_+^{-1}(x,\lambda_+)\ =\ \left(\psi_-\left(x,\frac{\eps}{\bl_-}\right)
\right)^\dagger~.
\end{equation}
The anti-Hermitian gauge potential components can be calculated
from \eqref{contractedA} to be
\begin{equation}\label{Acompsah}
A_{1\dot{2}}\ =\ \psi_+\dpar_{1\dot{2}}\psi_+^{-1}|^{\phantom{\pm}}_{\lambda_+=0}\ =\ 
-\eps A_{2\dot{1}}^\dagger~,~~~
A_{2\dot{2}}\ =\ \psi_+\dpar_{2\dot{2}}\psi_+^{-1}|^{\phantom{\pm}}_{\lambda_+=0}\ =\ 
-A_{1\dot{1}}^\dagger~.
\end{equation}

\paragraph{Explicit Penrose-Ward transform.} One can make the\index{Penrose-Ward transform}
Penrose-Ward transform more explicit. From the formula
\eqref{contractedA}, one obtains directly
\begin{equation}\label{pwtrafo1}
A_{\alpha\dot{1}}\ =\ -\oint_{S^1}\frac{\dd
\lambda_+}{2\pi\di}\frac{A_\alpha^+}{\lambda_+^2}~~~\mbox{and}~~~
A_{\alpha\dot{2}}\ =\ \oint_{S^1}\frac{\dd
\lambda_+}{2\pi\di}\frac{A_\alpha^+}{\lambda_+}~,
\end{equation}
where the contour $S^1=\{\lambda_+\in\CPP^1:|\lambda_+|=r<1\}$
encircles $\lambda_+=0$. Using \eqref{contractedA}, one can easily
show the equivalence of \eqref{pwtrafo1} to \eqref{Acompsah}. The
formul\ae{} \eqref{pwtrafo1} define the Penrose-Ward transform\index{Penrose-Ward transform}
\begin{equation}
\CP\mathcal{W}:(A^+_\alpha,A_{\bl_+}\hspace{-4pt}=0)\ \mapsto\ (A_{\alpha\ald})~,
\end{equation}
which together with a preceding gauge transformation\index{gauge!transformation}
\begin{equation}
(\hat{A}^+_\alpha,\hat{A}_{\bl_+})\stackrel{\varphi}{\longmapsto}(A^+_\alpha,A_{\bl_+}\hspace{-4pt}=0)
\end{equation}
maps solutions $(\hat{A}_\alpha^+,\hat{A}_{\bl_+})$ of the field
equations of hCS theory on $\CP^3$ to solutions $(A_{\alpha\ald})$
of the SDYM equations on $\FR^4$. Conversely, any solution
$(A_{\alpha\ald})$ of the SDYM equations corresponds to a solution
$(\hat{A}_\alpha^+,\hat{A}_{\bl_+})$ of the field equations of hCS
theory on $\CP^3$ which directly defines the inverse Penrose-Ward
transform $\CP\mathcal{W}^{-1}$. Note that gauge\index{Penrose-Ward transform}
transformations\footnote{Let us stress that there are two gauge
transformations for gauge potentials on two different spaces\index{gauge transformations}\index{gauge!transformation}
present in the discussion.} of
$(\hat{A}^+_\alpha,\hat{A}_{\bl_+})$ on $\CP^3$ and
$(A_{\alpha\ald})$ on $\FR^4$ do not change the transition
function $f_{+-}$ of the holomorphic bundle $E\rightarrow\CP^3$.\index{transition function}
Therefore, we have altogether a one-to-one correspondence between
equivalence classes of topologically trivial holomorphic vector
bundles over $\CP^3$ which become holomorphically trivial upon\index{holomorphic!vector bundle}
reduction to any $\CPP^1_x\subset \CP^3$ and gauge equivalence
classes of solutions to the field equations of hCS theory on
$\CP^3$ and the SDYM equations on $\FR^4$.

\paragraph{Anti-self-dual gauge fields.} The discussion of\index{anti-self-dual}
anti-self-dual gauge fields follows precisely the lines of the
discussion of the self-dual case. The first difference noteworthy
is that now the tangent spaces to the leaves of the fibration\index{fibration}
$\pi_2$ in the dual case are spanned by the vector fields
$\bV_\ald^\pm:=\mu^\alpha_\pm\dpar_{\alpha\ald}$ The definition of
the gauge potential $(A_{\alpha\ald})$ is then (cf.\
\eqref{contractedA})
\begin{equation}
A^+_\ald\ :=\  \mu^\alpha_+ \,
A_{\alpha\ald}\ =\ \mu^\alpha_+\,\psi_+\,\dpar_{\alpha\ald}\,\psi_+^{-1}\ =\ 
\mu^\alpha_+\,\psi_-\,\dpar_{\alpha\ald}\,\psi_-^{-1}~,
\end{equation}
which gives rise to the linear system\index{linear system}
\begin{equation}
(\bV_\ald^++A_\ald^+)\psi_+\ =\ 0~.
\end{equation}
The corresponding compatibility conditions are easily found to be\index{compatibility conditions}
\begin{equation}
[\bar{V}_\ald^++A_\ald^+\,,\bar{V}_\bed^++A_\bed^+]\ =\ 
\mu_+^\alpha\mu_+^\beta
[\dpar_{\alpha\ald}+A_{\alpha\ald}\,,\dpar_{\beta\bed}+A_{\beta\bed}]\ =:\ 
\mu_+^\alpha\mu_+^\beta F_{\alpha\ald,\beta\bed}\ =\ 0~,
\end{equation}
and these equations are equivalent to the anti-self-dual\index{anti-self-dual}
Yang-Mills equations $f_{\alpha\beta}=0$.

\paragraph{Example.} To close this section, let us consider an
explicit example for a Penrose-Ward transform, which will yield an\index{Penrose-Ward transform}
$\sSU(2)$ instanton. We start from a rank two holomorphic vector\index{instanton}
bundle over the real twistor space $\CP^3_{-1}$ given by the\index{twistor}\index{twistor!real twistor space}\index{twistor!space}
transition function\index{transition function}
\begin{equation}
f_{+-}\ =\ \left(\begin{array}{cc} \rho & \lambda_+^{-1} \\
-\lambda_+ & 0
\end{array}\right)
\end{equation}
(a special case of the Atiyah-Ward ansatz \cite{Atiyah:1977pw})\index{Atiyah-Ward ansatz}
together with the splitting
\begin{equation}
f_{+-}\ =\ \psi_+^{-1}\psi_-\ =\ 
\left(\begin{array}{cc} \phi+\rho_+ & -\lambda^{-1}_+\rho_+ \\
-\lambda_+ & 1
\end{array}\right)\frac{1}{\sqrt{\phi}}\frac{1}{\sqrt{\phi}}
\left(\begin{array}{cc} \phi+\rho_- & \lambda^{-1}_+ \\
\lambda\rho_- & 1
\end{array}\right)~.
\end{equation}
Here, we decomposed the function $\rho$ in its Laurent series
\begin{equation}
\rho\ =\ \sum_{n=-\infty}^\infty \rho_n\lambda^n\ =\ \rho_-+\phi+\rho_+
\end{equation}
and $\rho_\pm$ and $\phi$ denote the components holomorphic on
$U_\pm$ and $\CPP^1$, respectively. The gauge potential
$A^\pm_\alpha=\psi_+\bV_\alpha^\pm\psi_+^{-1}$ is then easily
calculated and the four-dimensional components $A_{\alpha\ald}$
are reconstructed via the formul\ae{} \eqref{pwtrafo1}.
Eventually, this calculation yields the result
\begin{equation}
A_\mu\ =\ \tfrac{1}{2\di}\bar{\eta}^a_{\mu\nu}\sigma^a(
\phi^{\frac{1}{2}}\dpar_\nu\phi^{-\frac{1}{2}}-\phi^{-\frac{1}{2}}\dpar_\nu\phi^{\frac{1}{2}})
+\tfrac{1}{2}\unit_2(
\phi^{\frac{1}{2}}\dpar_\mu\phi^{-\frac{1}{2}}+\phi^{-\frac{1}{2}}\dpar_\mu\phi^{\frac{1}{2}})~.
\end{equation}

\subsection{The ambitwistor space}\label{ssAmbitwistorSpace}\index{ambitwistor space}\index{twistor}\index{twistor!ambitwistor}\index{twistor!space}

\paragraph{Motivation.} The idea leading naturally to a
twistor space of Yang-Mills theory is to ``glue together'' both\index{Yang-Mills theory}\index{twistor}\index{twistor!space}
the self-dual and the anti-self-dual subsectors to the full\index{anti-self-dual}
theory. To achieve this, we will need two copies of the twistor\index{twistor}
space, one understood as dual to the other one, and glue them
together to the {\em ambitwistor space}. Roughly speaking, this\index{ambitwistor space}\index{twistor}\index{twistor!ambitwistor}\index{twistor!space}
gluing amounts to restricting to the diagonal in the two moduli
spaces. From this, we can already anticipate a strange property of\index{moduli space}
this space: The intersection of the $\alpha$- and $\beta$-planes\index{b-plane@$\beta$-plane}
corresponding to points in the two twistor spaces will be null\index{twistor}\index{twistor!space}
lines, but integrability along null lines is trivial. Therefore,\index{integrability}
we will have to consider infinitesimal neighborhoods of our new\index{infinitesimal neighborhood}
twistor space inside the product of the two original twistor\index{twistor}\index{twistor!space}
spaces, and this is the origin of the name {\em ambitwistor\index{twistor!ambitwistor}
space}. Eventually, this feature will find a natural
interpretation in terms of Gra{\ss}mann variables, when we will turn\index{Gra{\ss}mann variable}
to the superambitwistor space in section\index{ambitwistor space}\index{twistor}\index{twistor!ambitwistor}\index{twistor!space}
\ref{ssSuperambitwistorSpace}. This aspect verifies incidentally\index{incident}
the interpretation of Gra{\ss}mann directions of a supermanifold as an\index{super!manifold}
infinitesimal ``cloud of space'' around its body.\index{body}

\paragraph{The quadric $\CL^{5}$.} Consider the product of a twistor space\index{quadric}\index{twistor}\index{twistor!space}
$\CP^3$ with homogeneous coordinates\index{homogeneous coordinates}
$(\omega^\alpha,\lambda_\ald)$ and inhomogeneous coordinates\index{inhomogeneous coordinates}
$(z^\alpha_\pm,z^3_\pm=\lambda_\pm)$ on the two patches $\CU_\pm$
as introduced in section \ref{ssTwistorSpace} and an analogous
dual copy $\CP^3_*$ with homogeneous coordinates\index{homogeneous coordinates}
$(\sigma^\ald,\mu_\alpha)$ and inhomogeneous coordinates\index{inhomogeneous coordinates}
$(u^\ald_\pm,u^\drd_\pm=\mu_\pm)$ on the two patches $\CU^*_\pm$.
The space $\CP^3\times \CP^3_*$ is now naturally described by the
homogeneous coordinates\index{homogeneous coordinates}
$(\omega^\alpha,\lambda_\ald;\sigma^\ald,\mu_\alpha)$.
Furthermore, it is covered by the four patches
\begin{equation}
\CU_{(1)}\ :=\ \CU_+\times\CU^*_+~,~~~ \CU_{(2)}\ :=\
\CU_-\times\CU^*_+~,~~~ \CU_{(3)}\ :=\ \CU_+\times\CU^*_-~,~~~
\CU_{(4)}\ :=\ \CU_-\times\CU^*_-~,
\end{equation}
on which we have the evident inhomogeneous coordinates\index{homogeneous coordinates}\index{inhomogeneous coordinates}
$(z^\alpha_{(a)},z^3_{(a)};u^\ald_{(a)}, u^{\dot{3}}_{(a)})$. We
can consider $\CP^3\times \CP^3_*$ as a rank $4$ vector bundle
over the space $\CPP^1\times \CPP^1_*$. The global sections of
this bundle are parameterized by elements of $\FC^{4}\times
\FC^{4}_*$ in the following way:
\begin{equation}\label{ambisections}
z^\alpha_{(a)}\ =\ x^{\alpha\ald}\lambda_\ald^{(a)}~;~~~
u^\ald_{(a)}\ =\ x_*^{\alpha\ald}\mu_\alpha^{(a)}~.
\end{equation}
The {\em quadric} $\CL^5$ is now the algebraic variety in\index{algebraic variety}\index{quadric}
$\CP^3\times \CP^3_*$ defined by the equation
\begin{equation}\label{quadrichomo}
\omega^\alpha\mu_\alpha-\sigma^\ald\lambda_\ald\ =\ 0~.
\end{equation}
Instead of \eqref{quadrichomo}, we could have also demanded that
\begin{equation}\label{quadric2}
\kappa_{(a)}\ :=\
z^\alpha_{(a)}\mu_\alpha^{(a)}-u_{(a)}^\ald\lambda_\ald^{(a)}\
\stackrel{!}{=}\ 0~,
\end{equation}
on every $\CU_{(a)}$. These conditions -- or equally well
\eqref{quadrichomo} -- are indeed the appropriate ``gluing
conditions'' for obtaining a twistor space useful in the\index{twistor}\index{twistor!space}
description of Yang-Mills theory, as we will see. In the\index{Yang-Mills theory}
following, we will denote the restrictions of the patches
$\CU_{(a)}$ to $\CL^5$ by $\bar{\CU}_{(a)}:=\CU_{(a)}\cap\CL^5$.

\paragraph{Double fibration.} Because of the quadric condition\index{double fibration}\index{fibration}\index{quadric}
\eqref{quadric2}, the moduli $x^{\alpha\ald}$ and
$x_*^{\alpha\ald}$ are not independent on $\CL^{5}$, but one
rather has the relation
\begin{equation}
x^{\alpha\ald}\ =\ x_*^{\alpha\ald}~,
\end{equation}
which indeed amounts to taking the diagonal in the moduli space\index{moduli space}
$\FC^4\times \FC^4_*$. This will also become explicit in the
discussion in \ref{pPWtrafo}. With this identification, we can
establish the following double fibration using equations\index{double fibration}\index{fibration}
\eqref{ambisections}:
\begin{equation}\label{ambidblfibration2b}
\begin{aligned}
\begin{picture}(80,40)
\put(0.0,0.0){\makebox(0,0)[c]{$\CL^{5}$}}
\put(64.0,0.0){\makebox(0,0)[c]{$\FC^{4}$}}
\put(32.0,33.0){\makebox(0,0)[c]{$\CF^{6}$}}
\put(7.0,18.0){\makebox(0,0)[c]{$\pi_2$}}
\put(55.0,18.0){\makebox(0,0)[c]{$\pi_1$}}
\put(25.0,25.0){\vector(-1,-1){18}}
\put(37.0,25.0){\vector(1,-1){18}}
\end{picture}
\end{aligned}
\end{equation}
where $\CF^{6}\cong \FC^{4}\times\CPP^1\times\CPP^1_*$ and $\pi_1$
is the trivial projection. Here, we have the correspondences
\begin{equation}
\begin{aligned}
\left\{\,\mbox{subspaces $(\CPP^1\times \CPP_*^1)_{x}$ in
$\CL^{5}$} \right\}&\ \longleftrightarrow\
\left\{\,\mbox{points $x$ in $\FC^{4}$}\right\}~, \\
\left\{\,\mbox{points $p$ in $\CL^{5}$}\right\}&\
\longleftrightarrow\ \left\{\,\mbox{null lines in
$\FC^{4}$}\right\}~.
\end{aligned}
\end{equation}
The above-mentioned null lines are intersections of
$\alpha$-planes and the dual $\beta$-planes, as is evident from\index{a-plane@$\alpha$-plane}\index{b-plane@$\beta$-plane}
recalling the situation for both the twistor and the dual twistor\index{twistor}\index{twistor!dual twistor}
space. Given a solution $(\hat{x}^{\alpha\ald})$ to the incidence
relations \eqref{ambisections} for a fixed point $p$ in $\CL^{5}$,\index{incidence relation}
the set of points on such a null line takes the form
\begin{equation*}
\{(x^{\alpha\ald})\}~~\mbox{with}~~ x^{\alpha\ald}\ =\
\hat{x}^{\alpha\ald}+t\mu^\alpha_{(a)}\lambda^\ald_{(a)}~,
\end{equation*}
where $t$ is a complex parameter on the null line. The coordinates
$\lambda^\ald_{(a)}$ and $\mu^\alpha_{(a)}$ can be chosen from
arbitrary patches on which they are both well-defined.

\paragraph{Vector fields} The space $\CF^{6}$ is covered by four patches
$\tilde{\CU}_{(a)}:=\pi_2^{-1}(\bar{\CU}_{(a)})$ and the tangent
spaces to the one-dimensional leaves of the fibration\index{fibration}
$\pi_2:~\CF^{6}\rightarrow\CL^{5}$ in \eqref{ambidblfibration2b}
are spanned by the holomorphic vector field
\begin{equation}
W^{(a)}\ :=\ \mu_{(a)}^\alpha\lambda_{(a)}^\ald\dpar_{\alpha\ald}~.
\end{equation}

\paragraph{Flag manifolds.} As for the previously discussed twistor\index{flag manifold}\index{twistor}\index{manifold}
spaces, there is a description of the double fibration\index{double fibration}\index{fibration}
\eqref{ambidblfibration2b} in the compactified case in terms of
flag manifolds. The ambient space of the flags is again $\FC^4$,\index{flag manifold}\index{manifold}
and the double fibration reads\index{double fibration}\index{fibration}
\begin{equation}\label{flgdblfibration3}
\begin{picture}(80,40)
\put(0.0,0.0){\makebox(0,0)[c]{$F_{13,4}$}}
\put(64.0,0.0){\makebox(0,0)[c]{$F_{2,4}$}}
\put(32.0,33.0){\makebox(0,0)[c]{$F_{123,4}$}}
\put(7.0,18.0){\makebox(0,0)[c]{$\pi_2$}}
\put(55.0,18.0){\makebox(0,0)[c]{$\pi_1$}}
\put(25.0,25.0){\vector(-1,-1){18}}
\put(37.0,25.0){\vector(1,-1){18}}
\end{picture}
\end{equation}
where $F_{2,4}=G_{2,4}(\FC)$ is again the complexified and
compactified version of $\FR^{3,1}$. The flag manifold $F_{13,4}$\index{flag manifold}\index{manifold}
is topologically the zero locus of a quadric in\index{quadric}
$\CPP^3\times\CPP^3_*$. For further details and the super
generalization, see e.g.~\cite{Wardbook,Howe:1995md}.

\paragraph{Real structure $\tau_1$.} The Kleinian signature $(2,2)$ is\index{Kleinian signature}\index{real structure}
related to anti-linear transformations\footnote{We will not
consider the map $\tau_0$ here.} $\tau_1$ of spinors defined\index{Spinor}
before. Recall that
\begin{equation}
\tau_1\left(\begin{matrix}\omega^1 & \lambda_\ed & \sigma^\ed &
\mu_1 \\\omega^2 & \lambda_\zd & \sigma^\zd & \mu_2
\end{matrix}\right){=}\left(\begin{matrix}\bar{\omega}^2 & \bl_\zd & \bar{\sigma}^\zd & \bar{\mu}_2 \\
\bar{\omega}^1 & \bl_\ed & \bar{\sigma}^\ed & \bar{\mu}_1
\end{matrix}\right)~,
\end{equation}
and obviously $\tau_1^2=1$. Correspondingly for $(\lambda_\pm$,
$\mu_\pm)\in\CPP^1\times\CPP^1_*$, we have
\begin{equation}
\tau_1(\lambda_+)\ =\ \frac{1}{\bar{\lambda}_+}\ =\ \bl_-~,~~~
\tau_1(\mu_+)\ =\ \frac{1}{\bar{\mu}_+}\ =\ \bar{\mu}_-~
\end{equation}
with stable points
\begin{equation}
\{\lambda,\,\mu\,\in\CPP^1\times\CPP^1_*:\lambda\bl=1,\mu\bar{\mu}\ =\ 1\}\ =\ S^1\times
S^1_*\subset\CPP^1\times\CPP^1_*
\end{equation}
and parameterizing a torus $S^1\times S^1_*$. For the coordinates
$(x^{\alpha\ald})$, we have again
\begin{equation}
\tau_1\left(\begin{array}{cc}x^{1\dot{1}}&x^{1\dot{2}}\\
x^{2\dot{1}}&x^{2\dot{2}}
\end{array}\right)\ =\ \left(\begin{array}{cc}
\bar{x}^{2\dot{2}}&\bar{x}^{2\dot{1}}\\
\bar{x}^{1\dot{2}}&\bar{x}^{1\dot{1}}
\end{array}\right)
\end{equation}
and the real subspace $\FR^4$ of $\FC^4$ invariant under the
involution $\tau_1$ is defined by the equations\index{involution}
\begin{equation}\label{realcoord}
\bar{x}^{2\dot{2}}\ =\ x^{1\dot{1}}\ =:\ -(x^4+\di x^3)~~~\mbox{and}~~~
x^{2\dot{1}}\ =\ \bar{x}^{1\dot{2}}\ =:\ -(x^2-\di x^1)
\end{equation}
with a metric $\dd s^2=\det(\dd x^{\alpha\ald})$ of signature\index{metric}
$(2,2)$.

\paragraph{A $\tau_1$-real twistor diagram.}\index{twistor}
Imposing conditions \eqref{realcoord}, we obtain the real space
$\FR^{2,2}$ as a fixed point set of the involution\index{involution}
$\tau_1:\,\FC^{4}\rightarrow\FC^{4}.$ Analogously, for the twistor\index{twistor}
space $\CPP^{3}$ and its open subset $\CP^{3}$, we obtain real
subspaces $\RPS^{3}$ and $\CT^{3}$ (cf. \ref{pKleinianCase}).
Accordingly, a real form of the space $\CF^{6}$ is
$\CF^{6}:=\FR^{2,2}\times S^1\times S^1_*$, and we have a real
quadric $\CL^{5}\subset\CT^{3}\times\CT^{3}_*$ as the subset of\index{quadric}
fixed points of the involution\footnote{Again, we use the same\index{involution}
symbol $\tau_1$ for maps defined on different spaces.}
$\tau_1:\,\CL^{5}\rightarrow \CL^{5}$. This quadric is defined by\index{quadric}
equations \eqref{ambisections}-\eqref{quadric2} with the
$x^{\alpha\ald}$ satisfying \eqref{realcoord} and
$\lambda_+=\de^{\di\chi_1}=\lambda_-^{-1}$,
$\mu_+=\de^{\di\chi_2}= \mu_-^{-1}$, $0\leq\chi_1,\,\chi_2< 2\pi$.
Altogether, we obtain a real form
\begin{equation}\label{ambdblfibration3}
\begin{picture}(50,40)
\put(0.0,0.0){\makebox(0,0)[c]{$\CL^{5}$}}
\put(64.0,0.0){\makebox(0,0)[c]{$\FR^{2,2}$}}
\put(35.0,37.0){\makebox(0,0)[c]{$\CF^{6}$}}
\put(7.0,20.0){\makebox(0,0)[c]{$\pi_2$}}
\put(55.0,20.0){\makebox(0,0)[c]{$\pi_1$}}
\put(25.0,27.0){\vector(-1,-1){18}}
\put(37.0,27.0){\vector(1,-1){18}}
\end{picture}\vspace{2mm}
\end{equation}
of the double fibration \eqref{ambidblfibration2b}, where all the\index{double fibration}\index{fibration}
dimensions labelling the spaces are now {\em real} dimensions.

\paragraph{The Minkowskian involution $\tau_M$.} Let\index{involution}
us consider the manifold $\CP^{3}\times\CP^{3}_*$ with homogeneous\index{manifold}
coordinates $(\omega^\alpha,\lambda_\ald;\sigma^\ald,\mu_\alpha)$.
The antiholomorphic involution\index{involution}
\begin{equation}\label{Minvolution}
\tau_M:\,\CP^{3}\times\CP^{3}_*\ \rightarrow\ \CP^{3}\times\CP^{3}_*
\end{equation}
gives rise to Minkowski signature on the moduli space of sections.\index{moduli space}
It is defined as the map (see e.g.\ \cite{Maninbook})
\begin{equation}\label{Mtau}
\tau_M(\omega^\alpha,\lambda_\ald;\sigma^\ald,\mu_\alpha)\ =\ 
(-\overline{\sigma^\ald},\overline{\mu_\alpha};-\overline{\omega^\alpha},
\overline{\lambda_\ald})
\end{equation}
interchanging $\alpha$-planes and $\beta$-planes. One sees from\index{a-plane@$\alpha$-plane}\index{b-plane@$\beta$-plane}
\eqref{Mtau} that the real slice in the space
$\CP^{3}\times\CP^{3}_*$ is defined by the equation\footnote{Here,
$\alpha$ and $\ald$ denote {\em the same} number.}
\begin{equation}\label{Mreal1}
\sigma^\ald\ =\ -\overline{\omega^\alpha}~,~~~\mu_\alpha\ =\ \overline{\lambda_\ald}~.
\end{equation}
Finally, for coordinates $(x^{\alpha\ald})\in \FC^4$, we have
\begin{equation}
\tau_M(x^{\alpha\bed})\ =\ -\bar{x}^{\beta\ald}~,
\end{equation}
and the Minkowskian real slice $\FR^{3,1}\subset\FC^4$ is
parameterized by coordinates
\begin{equation}
\left(\begin{matrix}x^{1\dot{1}}&x^{1\dot{2}}\\
x^{2\dot{1}}&x^{2\dot{2}}\end{matrix}\right)^\dagger\ =\ 
-\left(\begin{matrix}x^{1\dot{1}}&x^{1\dot{2}}\\
x^{2\dot{1}}&x^{2\dot{2}}\end{matrix}\right)
\end{equation}
\begin{equation}\label{Mreal4}
\begin{aligned}
x^{1\dot{1}}&\ =\ -\di x^0-\di x^3~,~~~&x^{1\dot{2}}&\ =\ -\di x^1-x^2~,~\\
x^{2\dot{1}}&\ =\ -\di x^1+x^2~,~~~&x^{2\dot{2}}&\ =\ -\di x^0+\di
x^3~,
\end{aligned}
\end{equation}
with $(x^0,x^1,x^2,x^3)\in \FR^{3,1}$ and as in
\eqref{realmetric}, we define
\begin{equation}
\dd s^2\ =\ \det(\dd
x^{\alpha\ald})~\Rightarrow~g\ =\ \diag(-1,+1,+1,+1)~.
\end{equation}
One can also introduce coordinates
\begin{equation}
\tilde{x}^{\alpha\ald}\ =\ \di x^{\alpha\ald}~,
\end{equation}
yielding a metric with signature $(1,3)$. Recall that the\index{metric}
involution $\tau_M$ interchanges $\alpha$-planes and\index{a-plane@$\alpha$-plane}\index{involution}
$\beta$-planes and therefore exchanges opposite helicity states.\index{b-plane@$\beta$-plane}\index{helicity}
It might be identified with a $\RZ_2$-symmetry discussed recently
in the context of mirror symmetry~\cite{Aganagic:2004yh} and\index{mirror symmetry}
parity invariance~\cite{Witten:2004cp}.\index{parity}

\paragraph{A $\tau_M$-real twistor diagram.} Recall\index{twistor}
that $(\lambda_\ald)$ and $(\mu_\alpha)$ are homogeneous
coordinates on two Riemann spheres and the involution $\tau_M$\index{Riemann sphere}\index{homogeneous coordinates}\index{involution}
maps these spheres one onto another. Moreover, fixed points of the
map $\tau_M:\CPP^1\times\CPP^1_*\rightarrow \CPP^1\times\CPP^1_*$
form the Riemann sphere\index{Riemann sphere}
\begin{equation}
\CPP^1\ =\ \diag(\CPP^1\times\overline{\CPP}^1)\ ,
\end{equation}
where $\overline{\CPP}^1 (=\CPP^1_*)$ denotes the Riemann sphere\index{Riemann sphere}
$\CPP^1$ with the opposite complex structure. Therefore, a real\index{complex!structure}
slice in the space $\CF^{6}=\FC^{4}\times\CPP^1\times\CPP^1_*$
introduced in \eqref{ambidblfibration2b} and characterized as the
fixed point set of the involution $\tau_M$ is the space\index{involution}
\begin{equation}
\CF^{6}_{\tau_M}\ :=\ \FR^{3,1}\times\CPP^1
\end{equation}
of real dimension $6$.

The fixed point set of the involution \eqref{Minvolution} is the\index{involution}
diagonal in the space $\CP^{3}\times\bar{\CP}^{3}$, which can be
identified with the complex twistor space $\CP^{3}$ of real\index{twistor}\index{twistor!space}
dimension $6$. This involution also picks out a real quadric\index{involution}\index{quadric}
$\CL^{5}$ defined by equations \eqref{quadric2} and the reality
conditions \eqref{Mreal1}-\eqref{Mreal4}. Thus, we obtain a real
version of the double fibration \eqref{ambidblfibration2b},\index{double fibration}\index{fibration}
\begin{equation}\label{ambidblfibration3}
\begin{picture}(80,40)
\put(0.0,0.0){\makebox(0,0)[c]{$\CL^{5}$}}
\put(64.0,0.0){\makebox(0,0)[c]{$\FR^{3,1}$}}
\put(35.0,37.0){\makebox(0,0)[c]{$\CF^{6}$}}
\put(7.0,20.0){\makebox(0,0)[c]{$\pi_2$}}
\put(55.0,20.0){\makebox(0,0)[c]{$\pi_1$}}
\put(25.0,27.0){\vector(-1,-1){18}}
\put(37.0,27.0){\vector(1,-1){18}}
\end{picture}
\end{equation}
The dimensions of all spaces in this diagram are again real.

\paragraph{Yang-Mills equations from self-duality equations.}\label{pPWtrafo}
Consider a vector bundle $E$ over the space $\FC^4\times \FC^4$
with coordinates $r^{\alpha\ald}$ and $s^{\alpha\ald}$. On $E$, we
assume a gauge potential $A=A^r_{\alpha\ald}\dd
r^{\alpha\ald}+A^s_{\beta\bed}\dd s^{\beta\bed}$. Furthermore, we
introduce the coordinates
\begin{equation}
x^{\alpha\ald}\ =\
\tfrac{1}{2}(r^{\alpha\ald}+s^{\alpha\ald})\eand k^{\alpha\ald}\
=\ \tfrac{1}{2}(r^{\alpha\ald}-s^{\alpha\ald})
\end{equation}
on the base of $E$. We claim that the Yang-Mills equations
$\nabla^{\alpha\ald} F_{\alpha\ald\beta\bed}=0$ are then
equivalent to
\begin{equation}\label{condYM4}
\begin{aligned}
{}[\nabla^r_{\alpha\ald},\nabla^r_{\beta\bed}]&\ =\
\ast[\nabla^r_{\alpha\ald},\nabla^r_{\beta\bed}]+\CO(k^2)~,\\
{}[\nabla^s_{\alpha\ald},\nabla^s_{\beta\bed}]&\ =\
-\ast[\nabla^s_{\alpha\ald},\nabla^s_{\beta\bed}]+\CO(k^2)~,\\
{}[\nabla^r_{\alpha\ald},\nabla^s_{\beta\bed}]&\ =\ \CO(k^2)~,
\end{aligned}
\end{equation}
where we define\footnote{One could also insert an $\di$ into this
definition but on $\FC^4$, this is not natural.} $\ast
F^{r,s}_{\alpha\ald\beta\bed}:=\frac{1}{2}
\eps^{r,s}_{\alpha\ald\beta\bed\gamma\gad\delta\ded}
F_{r,s}^{\gamma\gad\delta\ded}$
separately on each $\FC^4$.

To understand this statement, note that equations \eqref{condYM4}
are equivalent to
\begin{equation}\label{condYM4simp}
\begin{aligned}
{}[\nabla^x_{\alpha\ald},\nabla^x_{\beta\bed}]&\ =\
[\nabla^k_{\alpha\ald},\nabla^k_{\beta\bed}]+\CO(k^2)~,\\
{}[\nabla^k_{\alpha\ald},\nabla^x_{\beta\bed}]&\ =\
\ast[\nabla^k_{\alpha\ald},\nabla^k_{\beta\bed}]+\CO(k^2)~,
\end{aligned}
\end{equation}
which is easily seen by performing the coordinate change from
$(r,s)$ to $(x,k$). These equations are solved by the expansion
\cite{Witten:1978xx,Isenberg:1978kk}
\begin{equation}\label{gaugepotYM4}
\begin{aligned}
A^k_{\alpha\ald}&\ =\
-\tfrac{1}{2}F^{x,0}_{\alpha\ald\beta\bed}k^{\beta\bed}-
\tfrac{1}{3}k^{\gamma\gad}\nabla^{x,0}_{\gamma\gad}(\ast
F^{x,0}_{\alpha\ald\beta\bed})k^{\beta\bed}~,\\
A^x_{\alpha\ald}&\ =\ A^{x,0}_{\alpha\ald} -\ast
F^{x,0}_{\alpha\ald\beta\bed}k^{\beta\bed}-
\tfrac{1}{2}k^{\gamma\gad}\nabla^{x,0}_{\gamma\gad}(
F^{x,0}_{\alpha\ald\beta\bed})k^{\beta\bed}~,
\end{aligned}
\end{equation}
if and only if
$\nabla_{x,0}^{\alpha\ald}F^{x,0}_{\alpha\ald\beta\bed}=0$ is
satisfied. Here, a superscript $0$ always denotes an object
evaluated at $k^{\alpha\ald}=0$. Thus we saw that a solution to
the Yang-Mills equations corresponds to a solution to equations
\eqref{condYM4} on $\FC^4\times \FC^4$.

\paragraph{Third order neighborhoods.} As
discussed before, the self-dual and anti-self-dual field strengths\index{anti-self-dual}\index{field strength}
solving the first and second equation of \eqref{condYM4} can be
mapped to certain holomorphic vector bundles over $\CP^3$ and\index{holomorphic!vector bundle}
$\CP^3_*$, respectively. On the other hand, the potentials given
in \eqref{gaugepotYM4} are now defined on a second order
infinitesimal neighborhood\footnote{not a thickening} of the\index{infinitesimal neighborhood}\index{thickening}
diagonal in $\FC^4\times\FC^4$ for which $\CO(k^3)=0$. In the
twistor description, this potential corresponds to a transition\index{twistor}
function $f_{+-}\sim\psi_+^{-1}\psi_-$, where the \v{C}ech
0-cochain $\{\psi_\pm\}$ is a solution to the equations
\begin{equation}
\begin{aligned}
\lambda^\ald_\pm\left(\der{r^{\alpha\ald}}+A^r_{\alpha\ald}\right)\psi_\pm&\
=\ \CO(k^4)~,\\
\mu^\alpha_\pm\left(\der{s^{\alpha\ald}}+A^s_{\alpha\ald}\right)\psi_\pm&\
=\ \CO(k^4)~.
\end{aligned}
\end{equation}
Roughly speaking, since the gauge potentials are defined to order
$k^2$ and since $\der{r^{\alpha\ald}}$ and $\der{s^{\alpha\ald}}$
contain derivatives with respect to $k$, the above equations can
indeed be rendered exact to order $k^3$. The exact definition of
the transition function is given by\index{transition function}
\begin{equation}
f_{+-,i}\ :=\
\sum_{j=0}^i\psi_{+,j}^{-1}\psi^{\phantom{-1}}_{-,i-j}~,
\end{equation}
where the additional indices label the order in $k$. On the
twistor space side, a third order neighborhood in $k$ corresponds\index{twistor}\index{twistor!space}
to a third order thickening in\index{thickening}
\begin{equation}
\kappa_{(a)}\ :=\
z^\alpha_{(a)}\mu_\alpha^{(a)}-u^\ald_{(a)}\lambda_\ald^{(a)}~.
\end{equation}

Altogether, we see that a solution to the Yang-Mills equations
corresponds to a topologically trivial holomorphic vector bundle\index{holomorphic!vector bundle}
over a third order thickening of $\CL^5$ in $\CP^3\times \CP^3_*$,\index{thickening}
which becomes holomorphically trivial, when restricted to any
$\CPP^1\times \CPP^1_*\embd\CL^5$.

\section{Supertwistor spaces}\label{ssupertwistorspaces}\index{twistor}\index{twistor!space}

So far we encountered two twistor spaces: the twistor space\index{twistor}\index{twistor!space}
$\CP^3$, which is an open subset of $\CPP^3$ and the ambitwistor\index{twistor!ambitwistor}
space, which is a third order thickening of the quadric $\CL^5$ in\index{quadric}\index{thickening}
$\CP^3\times \CP^3_*$. In this section, we discuss the extension
of the former by Gra{\ss}mann-odd directions \cite{Ferber:1977qx}.
Further extensions and the extension of the ambitwistor space will\index{ambitwistor space}\index{twistor}\index{twistor!ambitwistor}\index{twistor!space}
be discussed in subsequent sections.

\subsection{The superextension of the twistor\index{twistor}
space}\label{ssSuperextensionTwistorSpace}

\paragraph{Complex projective superspaces.} A super extension of\index{super!space}
the twistor space $\CPP^3$ is the supermanifold $\CPP^{3|\CN}$\index{super!manifold}\index{twistor}\index{twistor!space}
with homogeneous coordinates $(\omega^\alpha,\lambda_\ald,\eta_i)$\index{homogeneous coordinates}
subject to the identification
$(\omega^\alpha,\lambda_\ald,\eta_i)\sim
(t\,\omega^\alpha,t\,\lambda_\ald,t\,\eta_i)$ for any nonzero
complex scalar $t$. Here, $(\omega^\alpha,\lambda_\ald)$ are again
homogeneous coordinates on $\CPP^3$ and $\eta_i$ with\index{homogeneous coordinates}
$i=1,\ldots,\CN$ are Gra{\ss}mann variables. Interestingly, this\index{Gra{\ss}mann variable}
supertwistor space is a Calabi-Yau supermanifold in the case\index{Calabi-Yau}\index{Calabi-Yau supermanifold}\index{super!manifold}\index{twistor}\index{twistor!space}
$\CN=4$ and one may consider the topological B-model introduced in\index{topological!B-model}
section \ref{ssTopologicalBModel} with this space as target space\index{target space}
\cite{Witten:2003nn}.

\paragraph{Supertwistor spaces.} Let us now neglect the\index{twistor}\index{twistor!space}
super light cone at infinity similarly to the discussion in
section \ref{ssTwistorSpace}. That is, we consider analogously to
the space $\CP^3=\CPP^3\backslash\CPP^1=\CO(1)\oplus\CO(1)$ its
super extension $\CP^{3|\CN}$ covered by two patches,
$\CP^{3|\CN}=\CPP^{3|\CN}\backslash\CPP^{1|\CN}=
\hat{\CU}_+\cup\hat{\CU}_-$, with even coordinates \eqref{coords}
and odd coordinates
\begin{equation}\label{ferm1}
\eta_i^+\ =\ \frac{\eta_i}{\lambda_{\dot{1}}}~~\mbox{on}~~\hat{\CU}_+
~~~\mbox{and}~~~\eta_i^-\ =\ \frac{\eta_i}{\lambda_{\dot{2}}}~~
\mbox{on}~~\hat{\CU}_-
\end{equation}
related by
\begin{equation}\label{ferm2}
\eta^+_i\ =\ z_+^3\eta^-_i
\end{equation}
on $\hat{\CU}_+\cap\hat{\CU}_-$. We see from \eqref{ferm1} and
\eqref{ferm2} that the fermionic coordinates are sections of
$\Pi\CO(1)$. The supermanifold $\CP^{3|\CN}$ is fibred over\index{PO(n)@$\Pi\CO(1)$}\index{super!manifold}
$\CPP^{1|0}$,
\begin{equation}\label{superbundle}
\CP^{3|\CN}\ \rightarrow\ \CPP^{1|0}~,
\end{equation}
with superspaces $\FC^{2|\CN}_\lambda$ as fibres over\index{super!space}
$\lambda\in\CPP^{1|0}$. We also have a second fibration\index{fibration}
\begin{equation}\label{superbundle2}
\CP^{3|\CN}\ \rightarrow\ \CPP^{1|\CN}
\end{equation}
with $\FC_\lambda^{2|0}$ as fibres.

\paragraph{Global sections and their moduli.}\label{pGlobalSecs}
The global holomorphic sections of the bundle \eqref{superbundle}
are rational curves $\CPP^1_{x_R,\eta}\embd\CP^{3|\CN}$
parameterized by moduli
$(x_R,\eta)=(x_R^{\alpha\ald},\eta_i^\ald)\in\FC^{4|2\CN}_R$
according to
\begin{equation}
\label{supercoords}
\begin{aligned}
z_+^\alpha&\ =\ x^{\alpha\ald}_R\,\lambda_\ald^+\,,~~~
\eta_i^+\ =\ \eta_i^\ald\lambda_\ald^+~~~\mbox{for}~~~
(\lambda_\ald^+)\ =\ (1,\lambda_+)^T~,~~~\lambda_+\in
U_+~,\\
z_-^\alpha&\ =\ x^{\alpha\ald}_R\,\lambda_\ald^-\,,~~~
\eta_i^-\ =\ \eta_i^\ald\lambda_\ald^-~~~\mbox{for}~~~
(\lambda_\ald^-)\ =\ (\lambda_-,1)^T~,~~~\lambda_-\in U_-~.
\end{aligned}
\end{equation}
Here, the space $\FC^{4|2\CN}_R$ is indeed the anti-chiral
superspace. Equations \eqref{supercoords} define again a\index{chiral!superspace}\index{super!space}
supertwistor correspondence via the double fibration\index{double fibration}\index{fibration}\index{twistor}\index{twistor!correspondence}
\begin{equation}\label{superdblfibration}
\begin{aligned}
\begin{picture}(50,40)
\put(0.0,0.0){\makebox(0,0)[c]{$\CP^{3|\CN}$}}
\put(74.0,0.0){\makebox(0,0)[c]{$\FC_R^{4|2\CN}$}}
\put(42.0,37.0){\makebox(0,0)[c]{$\CF_R^{5|2\CN}$}}
\put(7.0,20.0){\makebox(0,0)[c]{$\pi_2$}}
\put(65.0,20.0){\makebox(0,0)[c]{$\pi_1$}}
\put(25.0,27.0){\vector(-1,-1){18}}
\put(47.0,27.0){\vector(1,-1){18}}
\end{picture}
\end{aligned}
\end{equation}
where $\CF^{5|2\CN}\cong\FC^{4|2\CN}_R\times \CPP^1$ and the
projections are defined as
\begin{equation}
\pi_1(x^{\alpha\ald}_R,\eta_i^\ald,\lambda_\ald^\pm)\ :=\
(x^{\alpha\ald}_R,\eta_i^\ald)\eand
\pi_2(x^{\alpha\ald}_R,\eta_i^\ald,\lambda_\ald^\pm)\ :=\
(x^{\alpha\ald}_R\lambda_\ald^\pm,\lambda_\pm,\eta_i^\ald\lambda^\pm_\ald)~.
\end{equation}
The supertwistor correspondence now reads explicitly\index{twistor}\index{twistor!correspondence}
\begin{equation*}
\begin{aligned}
\left\{\,\mbox{projective lines $\CPP^1_{x,\eta}$ in
$\CP^{3|\CN}$}\right\}&\ \longleftrightarrow\
\left\{\,\mbox{points $(x,\eta)$ in $\FC^{4|2\CN}$}\right\}~,\\
\left\{\,\mbox{points $p$ in $\CP^{3|\CN}$}\right\}&\
\longleftrightarrow\ \left\{\,\mbox{null ($\alpha_R$-)superplanes
$\FC_p^{2|2\CN}$ in $\FC^{4|2\CN}$}\right\}~.
\end{aligned}
\end{equation*}
Given a solution $(\hat{x}^{\alpha\ald},\hat{\eta}_i^\ald)$ to the
incidence relations \eqref{supercoords} for a fixed point $p\in\index{incidence relation}
\CP^{3|\CN}$, the set of all solutions is given by
\begin{equation}\label{ssuperplanes}
\{(x^{\alpha\ald},\eta_i^\ald)\}~~~\mbox{with}~~~ x^{\alpha\ald}\
=\ \hat{x}^{\alpha\ald}+\mu^\alpha\lambda^\ald_\pm\eand
\eta_i^\ald\ =\ \hat{\eta}_i^\ald+\eps_i \lambda_\pm^\ald~,
\end{equation}
where $\mu^\alpha$ is an arbitrary commuting two-spinor and\index{Spinor}
$\eps_i$ is an arbitrary vector with Gra{\ss}\-mann-odd entries. The
sets defined in \eqref{ssuperplanes} are then called {\em null} or
{\em $\alpha_R$-superplanes}, and they are of superdimension
$2|\CN$.

\paragraph{Global sections of a different kind.} In the previous
paragraph, we discussed sections of the bundle
\eqref{superbundle}, which is naturally related to the discussion
of the bosonic twistor space before. One can, however, also\index{twistor}\index{twistor!space}
discuss sections of the bundle \eqref{superbundle2}, which will
give rise to a relation to the dual supertwistor space\index{twistor}\index{twistor!space}
$\CP^{3|\CN}_*$ and its moduli space.\index{moduli space}

The global holomorphic sections of the bundle \eqref{superbundle2}
are spaces $\CPP^{1|\CN}_{x_L,\theta}\embd\CP^{3|\CN}$ defined by
the equations
\begin{equation}\label{supercoordsleft}
z^\alpha_\pm\ =\ x_L^{\alpha\ald}\lambda_\ald^\pm-2\theta^{\alpha
i}\eta^\pm_i~~~\mbox{with}~~(\lambda_\ald^\pm,\eta^\pm_i)\in
\hat{\CU}_\pm\cap\CPP^{1|\CN}
\end{equation}
and parameterized by the moduli
$(x_L,\theta)=(x^{\alpha\ald}_L,\theta^{\alpha
i})\in\FC_L^{4|2\CN}$. Note that the moduli space is the chiral\index{moduli space}
superspace $\FC_L^{4|2\CN}$, contrary to the anti-chiral\index{super!space}
superspace $\FC_R^{4|2\CN}$, which arose as the moduli space of\index{moduli space}
global sections of the bundle \eqref{superbundle}. Equations
\eqref{supercoordsleft} define another supertwistor\index{twistor}
correspondence,
\begin{equation}\label{superdblfibration1.5}
\begin{aligned}
\begin{picture}(50,40)
\put(0.0,0.0){\makebox(0,0)[c]{$\CP^{3|\CN}$}}
\put(74.0,0.0){\makebox(0,0)[c]{$\FC_L^{4|2\CN}$}}
\put(42.0,37.0){\makebox(0,0)[c]{$\CF_L^{5|3\CN}$}}
\put(7.0,20.0){\makebox(0,0)[c]{$\pi_2$}}
\put(65.0,20.0){\makebox(0,0)[c]{$\pi_1$}}
\put(25.0,27.0){\vector(-1,-1){18}}
\put(47.0,27.0){\vector(1,-1){18}}
\end{picture}
\end{aligned}
\end{equation}
where $\CF_L^{5|3\CN}:=\FC_L^{4|2\CN}\times\CPP^{1|\CN}$. The
twistor correspondence here reads\index{twistor}\index{twistor!correspondence}
\begin{equation*}
\begin{aligned}
\left\{\,\mbox{superspheres $\CPP^{1|\CN}_{x_L,\eta}$ in
$\CP^{3|\CN}$}\right\}&\ \longleftrightarrow\
\left\{\,\mbox{points $(x_L,\theta)$ in $\FC^{4|2\CN}_L$}\right\}~,\\
\left\{\,\mbox{points $p$ in $\CP^{3|\CN}$}\right\}&\
\longleftrightarrow\ \left\{\,\mbox{null ($\alpha_L$-)superplanes
$\FC_p^{2|2\CN}$ in $\FC_L^{4|2\CN}$}\right\}~.
\end{aligned}
\end{equation*}

\paragraph{Relation between the moduli spaces.} From \eqref{supercoords}\index{moduli space}
and \eqref{supercoordsleft} we can deduce that
\begin{equation}\label{coords4}
x_R^{\alpha\ald}\ =\ x^{\alpha\ald}-\theta^{\alpha i}\eta^\ald_i\eand
x_L^{\alpha\ald}\ =\ x^{\alpha\ald}+\theta^{\alpha i}\eta^\ald_i~,
\end{equation}
where $(x^{\alpha\ald})\in\FC^{4|0}$ are ``symmetric''
(non-chiral) bosonic coordinates. Substituting the first equation
of \eqref{coords4} into \eqref{supercoords}, we obtain the
equations
\begin{equation}
z_\pm^\alpha\ =\ x^{\alpha\ald}\lambda^\pm_\ald-\theta^{\alpha
i}\eta_i^\ald\lambda_\ald^\pm~,~~~\eta_i^\pm\ =\ \eta_i^\ald\lambda_\ald^\pm
\end{equation}
defining degree one curves $\CPP^1_{x,\theta,\eta}\embd
\CP^{3|\CN}$ which are evidently parameterized by moduli
$(x^{\alpha\ald},\theta^{\alpha i},\eta^\ald_i)\in\FC^{4|4\CN}$.
Therefore we obtain a third double fibration\index{double fibration}\index{fibration}
\begin{equation}\label{superdblfibration2}
\begin{aligned}
\begin{picture}(50,40)
\put(0.0,0.0){\makebox(0,0)[c]{$\CP^{3|\CN}$}}
\put(74.0,0.0){\makebox(0,0)[c]{$\FC^{4|4\CN}$}}
\put(42.0,37.0){\makebox(0,0)[c]{$\CF^{5|4\CN}$}}
\put(7.0,20.0){\makebox(0,0)[c]{$\pi_2$}}
\put(65.0,20.0){\makebox(0,0)[c]{$\pi_1$}}
\put(25.0,27.0){\vector(-1,-1){18}}
\put(47.0,27.0){\vector(1,-1){18}}
\end{picture}
\end{aligned}
\end{equation}
with coordinates
\begin{subequations}
\begin{align}
(x^{\alpha\ald}\cb\lambda_\ald^\pm\cb\theta^{\alpha
i}\cb\eta_i^\ald)~~~&\mbox{on}~~\CF^{5|4\CN}\ :=\ \FC^{4|4\CN}
\times\CPP^1~,\\\label{coordsc1} (x^{\alpha\ald}\cb\theta^{\alpha
i}\cb\eta^\ald_i)~~~&\mbox{on}~~\FC^{4|4\CN}~,\\
z_\pm^\alpha~,~~~\lambda_\ald^\pm~,~~~\label{coords6}
\eta_i^\pm~~~&\mbox{on}~~\CP^{3|\CN}~.
\end{align}
\end{subequations}
The definition of the projection $\pi_1$ is obvious and $\pi_2$ is
defined by \eqref{supercoords} and \eqref{coords4}.

The double fibration \eqref{superdblfibration2} generalizes both\index{double fibration}\index{fibration}
\eqref{superdblfibration} and \eqref{superdblfibration1.5} and
defines the following twistor correspondence:\index{twistor}\index{twistor!correspondence}
\begin{equation*}
\begin{aligned}
\left\{\,\mbox{projective line $\CPP^{1}_{x,\theta,\eta}$ in
$\CP^{3|\CN}$}\right\}&\ \longleftrightarrow\
\left\{\,\mbox{points $(x,\theta,\eta)$ in $\FC^{4|4\CN}$}\right\}~,\\
\left\{\,\mbox{points $p$ in $\CP^{3|\CN}$}\right\}&\
\longleftrightarrow\ \left\{\,\mbox{null ($\alpha$-)superplanes
$\FC_p^{2|3\CN}$ in $\FC^{4|4\CN}$}\right\}~.
\end{aligned}
\end{equation*}

\paragraph{Vector fields.} Note that one can project
from $\CF^{5|4\CN}$ onto $\CP^{3|\CN}$ in two steps: first from
$\CF^{5|4\CN}$ onto $\CF^{5|2\CN}_R$, which is given in
coordinates by
\begin{equation}\label{projection1}
(x^{\alpha\ald}\cb\lambda_\ald^\pm\cb\theta^{\alpha
i}\cb\eta^\ald_i) \ \rightarrow\ 
(x_R^{\alpha\ald}\cb\lambda_\ald^\pm\cb\eta_i^\ald)
\end{equation}
with the $x_R^{\alpha\ald}$ from \eqref{coords4}, and then from
$\CF_R^{5|2\CN}$ onto $\CP^{3|\CN}$, which is given in coordinates
by
\begin{equation}\label{projection2}
(x_R^{\alpha\ald}\cb\lambda_\ald^\pm\cb\eta_i^\ald)\ \rightarrow\ 
(x_R^{\alpha\ald}\lambda_\ald^\pm\cb\lambda_\ald^\pm\cb\eta_i^\ald
\lambda_\ald^\pm)~.
\end{equation}
The tangent spaces to the $(0|2\CN)$-dimensional leaves of the
fibration \eqref{projection1} are span\-ned by the vector fields\index{fibration}
\begin{equation}\label{vectorfields0}
D_{\alpha i}\ =\ \der{\theta^{\alpha
i}}+\eta_i^\ald\der{x^{\alpha\ald}}\ =:\  \dpar_{\alpha
i}+\eta_i^\ald\dpar_{\alpha\ald}
\end{equation}
on $\FC^{4|4\CN}\subset\CF^{5|4\CN}$. The coordinates
$x_R^{\alpha\ald}$, $\lambda_\ald^\pm$ and $\eta^\ald_i$ belong to
the kernel of these vector fields, which are also tangent to the
fibres of the projection $\FC^{4|4\CN}\rightarrow \FC^{4|2\CN}_R$
onto the anti-chiral superspace. The tangent spaces to the\index{chiral!superspace}\index{super!space}
$(2|\CN)$-dimensional leaves of the projection \eqref{projection2}
are spanned by the vector fields\footnote{For the definition of
$\lambda_\pm^\ald$, see section \ref{alllambdas}.}
\begin{subequations}
\begin{align}\label{vectorfields1}
\bar{V}_\alpha^\pm&\ =\ \lambda_\pm^\ald\dpar^R_{\alpha\ald}~,\\
\label{vectorfields2} \bar{V}_\pm^i&\ =\ \lambda_\pm^\ald\,
\dpar_\ald^i~~~\mbox{with}~~ \dpar_\ald^i\ :=\ \der{\eta^\ald_i}~,
\end{align}
\end{subequations}
where
$\dpar^R_{\alpha\ald}=\der{x_R^{\alpha\ald}}=\der{x^{\alpha\ald}}$.

\paragraph{Dual supertwistor space.} The dual supertwistor space\index{twistor}\index{twistor!space}
$\CP^{3|\CN}_*$ is obtained from the complex projective space\index{complex!projective space}
$\CPP^{3|\CN}_*$ with homogeneous coordinates\index{homogeneous coordinates}
$(\sigma^\ald,\mu_\alpha,\theta^i)$ by demanding that
$\mu_\alpha\neq (0,0)^T$. Thus, the space
$\CP_*^{3|\CN}=\CPP^{3|\CN}_*\backslash\CPP^{1|\CN}_*$ is covered
by the two patches $\hat{\CV}_\pm$ with the inhomogeneous
coordinates\index{homogeneous coordinates}\index{inhomogeneous coordinates}
\begin{eqnarray}\label{ip1}
&&u_+^\ald=\frac{\sigma^\ald}{\mu_1}~,~~
u_+^{\dot3}=\mu_+=\frac{\mu_2}{\mu_1}~~ \mbox{and} ~~
\theta_+^i=\frac{\theta^i}{\mu_1}~~ \mbox{on} ~~ \hat{\CV}_+~,\\
\label{ip2} &&w_-^\ald=\frac{\sigma^\ald}{\mu_2}~,~~
u_-^{\dot3}=\mu_-=\frac{\mu_1}{\mu_2}~~ \mbox{and} ~~
\theta_-^i=\frac{\theta^i}{\mu_2}~~ \mbox{on} ~~ \hat{\CV}_-~,\\
\label{ip3} &&u_+^\ald=\mu_+u_-^\ald~,~~~\mu_+=\mu_-^{-1}~, ~~~
\theta_+^i=\mu_+\theta^i_-~~~\mbox{on}~~~\hat{\CV}_+\cap\hat{\CV}_-~.
\end{eqnarray}
Sections of the bundle $\CP^{3|\CN}_*\rightarrow \CPP^{1|0}_*$
(degree one holomorphic curves
$\CPP^1_{x_L,\theta}\embd\CP_*^{3|\CN}$) are defined by the
equations
\begin{equation}\label{ssections}
w_\pm^\ald\ =\ x_L^{\alpha\ald}\mu_\alpha^\pm~,~~~
\theta_\pm^i\ =\ \theta^{\alpha i}\mu_\alpha^\pm~~~
\mbox{with}~~~(\mu_\alpha^+)\ =\ \left(\begin{array}{c}1\\\mu_+
\end{array}\right)~,~~~(\mu_\alpha^-)\ =\ \left(\begin{array}{c}\mu_-\\1
\end{array}\right)~,
\end{equation}
and parameterized by moduli $(x^{\alpha\ald}_L,\theta^{\alpha
i})\in\FC_L^{4|2\CN}$. Note that similarly to the supertwistor\index{twistor}
case, one can consider furthermore sections of the bundle
$\CP^{3|\CN}_*\rightarrow \CPP^{1|\CN}_*$.

Equations \eqref{ssections} give again rise to a double fibration\index{double fibration}\index{fibration}
\begin{equation}\label{superdblfibration4}
\begin{picture}(50,50)
\put(0.0,0.0){\makebox(0,0)[c]{$\CP_*^{3|\CN}$}}
\put(74.0,0.0){\makebox(0,0)[c]{$\FC_L^{4|2\CN}$}}
\put(42.0,37.0){\makebox(0,0)[c]{$\CF_{*}^{5|2\CN}$}}
\put(7.0,20.0){\makebox(0,0)[c]{$\pi_2$}}
\put(65.0,20.0){\makebox(0,0)[c]{$\pi_1$}}
\put(25.0,27.0){\vector(-1,-1){18}}
\put(47.0,27.0){\vector(1,-1){18}}
\end{picture}
\end{equation}
and the tangent spaces of the $(2|\CN)$-dimensional leaves of the
projection $\pi_2:\CF_{*}^{5|2\CN}\rightarrow\CP_*^{3|\CN}$ from
\eqref{superdblfibration4} are spanned by the vector fields
$\bV^\pm_\alpha=\lambda^\ald_\pm\dpar^L_{\alpha\ald}$ and
$\bV_i^\pm=\mu^\alpha\der{\theta^{i\alpha}}$.

\paragraph{Real structure.} Three real structures\index{real structure}
$\tau_{\pm 1},\tau_0$ can be imposed similarly to the bosonic
case. We will focus on the two real structures $\tau_\eps$ and\index{real structure}
define additionally to \eqref{three}
\begin{equation}\label{2.15}
\tau_1(\eta^\pm_i)\ =\
\left(\frac{\bar{\eta}^\pm_i}{\bz^3_\pm}\right)~,~~~
\tau_{-1}(\eta^\pm_1,\eta^\pm_2,\eta^\pm_3,\eta^\pm_4)\ =\
\left(\frac{\mp\bar{\eta}^\pm_2}{\bz^3_\pm},\frac{\pm\bar{\eta}^\pm_1}{\bz^3_\pm},
\frac{\mp\bar{\eta}^\pm_4}{\bz^3_\pm},\frac{\pm\bar{\eta}^\pm_3}{\bz^3_\pm}\right)~.
\end{equation}
For $\eps=+1$, one can truncate the involution $\tau_\eps$ to the\index{involution}
cases $\CN<4$, which is in the Euclidean case only possible for
$\CN=2$, see also the discussion in \ref{ssSpinors},
\ref{peuclideancase}. The corresponding reality conditions for the
fermionic coordinates on the correspondence and moduli spaces are\index{moduli space}
also found in section \ref{ssSpinors}. As before, we will denote
the real supertwistor spaces by $\CP^{3|4}_\eps$.\index{twistor}\index{twistor!space}

\paragraph{Identification of vector fields.} On $\CP_\eps^{3|4}$,
there is the following relationship between vector fields of type
(0,1) in the coordinates $(z_\pm^\alpha,z_\pm^3,\eta_i^\pm)$ and
vector fields in the coordinates
$(x^{\alpha\ald}_R,\lambda_\pm,\eta_i^\ald)$:
\begin{equation}\label{eq:2.21}
\begin{aligned}
\der{\bz_\pm^1}&\ =\ -\gamma_\pm\lambda_\pm^\ald\der{x^{2\ald}_R}
\ =:\ -\gamma_\pm\bar{V}_2^\pm~,~~~& \der{\bz_\pm^2}&\ =\
\gamma_\pm\lambda_\pm^\ald\der{x^{1\ald}_R} \ =:\
-\eps\gamma_\pm\bar{V}_1^\pm~,\\\der{\bz_+^3}&\ =\
\der{\bl_+}+\eps\gamma_+
x^{\alpha\dot{1}}_R\bar{V}_\alpha^++\eps\gamma_+\eta_i^\ed\bV^i_+~,~~~&
\der{\bz_-^3}&\ =\ \der{\bl_-}+\gamma_-
x^{\alpha\dot{2}}_R\bar{V}_\alpha^-+\gamma_-\eta_i^\zd\bV^i_-~.
\end{aligned}
\end{equation}
In the Kleinian case, one obtains additionally for the fermionic
vector fields
\begin{equation}
\der{\etab^\pm_i}\ =\ -\gamma_\pm \bV^i_\pm~,
\end{equation}
while in the Euclidean case, we have
\begin{equation}\label{eq:2.37}
\begin{aligned}
\der{\etab_1^\pm}&\ =\
\gamma_\pm\lambda^\ald_\pm\der{\eta_2^\ald}\ =:\
\gamma_\pm\bV_\pm^2~,& \der{\etab_2^\pm}&\ =\
-\gamma_\pm\lambda^\ald_\pm\der{\eta_1^\ald}\ =:\
-\gamma_\pm\bV_\pm^1~,\\ \der{\etab_3^\pm}&\ =\
\gamma_\pm\lambda^\ald_\pm\der{\eta_4^\ald}\ =:\
\gamma_\pm\bV_\pm^4~,& \der{\etab_4^\pm}&\ =\
-\gamma_\pm\lambda^\ald_\pm\der{\eta_3^\ald}\ =:\
-\gamma_\pm\bV_\pm^3~.
\end{aligned}
\end{equation}

\paragraph{Forms.} It will also be useful to introduce
$(0,1)$-forms $\bE^a_\pm$ and $\bE_i^\pm$ which are dual to
$\bV_a^\pm$ and $\bV^i_\pm$, respectively, i.e.\
\begin{equation}
\bV_a^\pm \lrcorner\bE^b_\pm\ =\ \delta_a^b\eand
\bV^i_\pm\lrcorner \bE_j^\pm\ =\ \delta_j^i~.
\end{equation}
Here, $\lrcorner$ denotes the interior product of vector fields\index{interior product}
with differential forms. Explicitly, the dual $(0,1)$-forms are
given by the formul\ae{}
\begin{equation}\label{DefForms}
\bE^\alpha_\pm\ =\ -\gamma_\pm\hl_\ald^\pm\dd x^{\alpha\ald}~,~~~
\bE^3_\pm\ =\ \dd \bl_\pm\eand\bE_i^\pm\ =\
-\gamma_\pm\hl_\ald^\pm\dd \eta^\ald_i~.
\end{equation}

In the case $\CN=4$, one can furthermore introduce the (nowhere
vanishing) holomorphic volume form $\Omega$, which is locally\index{holomorphic!volume form}
given as
\begin{equation}\label{Omega}
\Omega_\pm\ :=\ \Omega|_{\CU_\pm}\ :=\ \pm\dd \lambda_\pm\wedge\dd
z^1_\pm\wedge \dd z^2_\pm \dd \eta^\pm_1\ldots\dd\eta^\pm_4\ =:\
\pm\dd \lambda_\pm\wedge\dd z^1_\pm\wedge \dd
z^2_\pm~\Omega^\eta_\pm
\end{equation}
on $\CP^{3|4}$, independently of the real structure. The existence\index{real structure}
of this volume element implies that the Berezinian line bundle is\index{Berezinian}
trivial and consequently $\CP^{3|4}$ is a Calabi-Yau supermanifold\index{Calabi-Yau}\index{Calabi-Yau supermanifold}\index{super!manifold}
\cite{Witten:2003nn}, see also section \ref{ssCYsupermanifolds},
\ref{psCYex}. Note, however, that $\Omega$ is not a differential
form because its fermionic part transforms as a product of
Gra{\ss}mann-odd vector fields, i.e.\ with the inverse of the
Jacobian. Such forms are called integral forms.

\paragraph{Comment on the notation.} Instead of the shorthand
notation $\CP^{3|\CN}$, we will sometimes write
$(\CP^3,\CO_{[\CN]})$ in the following, which makes the extension
of the structure sheaf of $\CP^3$ explicit. The sheaf\index{sheaf}\index{structure sheaf}
$\CO_{[\CN]}$ is locally the tensor product of the structure sheaf
of $\CP^3$ and a Gra{\ss}mann algebra of $\CN$ generators.\index{Gra{\ss}mann algebra}\index{sheaf}\index{structure sheaf}

\subsection{The Penrose-Ward transform for $\CP^{3|\CN}$}\index{Penrose-Ward transform}

Similarly to the bosonic case, one can built a Penrose-Ward
transform between certain holomorphic vector bundles over the\index{Penrose-Ward transform}\index{holomorphic!vector bundle}
supertwistor space $\CP^{3|\CN}$ and solutions to the\index{twistor}\index{twistor!space}
supersymmetric self-dual Yang-Mills equations\footnote{For a
deformation of the supertwistor geometry yielding chiral mass\index{twistor}
terms, see \cite{Chiou:2005pu}.} on $\FC^4$
\cite{Semikhatov:1982ig,Volovich:1983ii,Volovich:1984kr,Volovich:1983aa,Tafel:1985qk}.

\paragraph{Holomorphic bundles over $\CP^{3|\CN}$.} In analogy to
the purely bosonic discussion, let us consider a topologically
trivial holomorphic vector bundle $\CE$ over the supertwistor\index{holomorphic!vector bundle}\index{twistor}
space $\CP^{3|\CN}$, which becomes holomorphically trivial, when
restricted to any subset $\CPP^1_{x,\eta}\subset \CP^{3|\CN}$.
Note that the vector bundle $\CE$ has ordinary, bosonic fibres and
thus is {\em not} a supervector bundle. Since the underlying base
manifold is a supermanifold, the sections of $\CE$ are, however,\index{super!manifold}\index{manifold}
vector-valued superfunctions. As usual, the bundle $\CE\rightarrow
\CP^{3|\CN}$ is defined by a holomorphic transition function\index{transition function}
$f_{+-}$ which can be split according to
\begin{equation}\label{supersplithCS}
f_{+-}\ =\ \hpsi_+^{-1}\hpsi_-~,
\end{equation}
where $\hpsi_\pm$ are smooth $\sGL(n,\FC)$-valued functions on the
patches $\hat{\CU}_\pm$ covering $\CP^{3|\CN}$.

\paragraph{Holomorphic Chern-Simons equations.} The splitting
\eqref{supersplithCS} together with the holomorphy of the
transition function\index{transition function}
\begin{equation}
\der{\bz^1_\pm}f_{+-}\ =\ \der{\bz^2_\pm}f_{+-}\ =\ \der{\bz^3_\pm}f_{+-}\ =\ 0
\end{equation}
leads to the equations
\begin{equation}
\hpsi_+\dparb\hpsi_+^{-1}\ =\ \hpsi_-\dparb\hpsi_-^{-1}~.
\end{equation}
Completely analogously to the purely bosonic case, we introduce a
gauge potential $\CA$ by
\begin{equation}
\CAh^{0,1}_\pm\ =\ \hpsi_\pm\dparb\hpsi_\pm^{-1}~,
\end{equation}
which fits into the linear system\index{linear system}
\begin{equation}\label{hCSlinsys}
(\dparb+\CAh^{0,1})\hpsi_\pm\ =\ 0~.
\end{equation}
The compatibility conditions of this linear system are again the\index{compatibility conditions}\index{linear system}
holomorphic Chern-Simons equations of motion
\begin{equation}\label{extrahCS}
\dparb \CAh+\CAh\wedge\CAh\ =\ 0~,
\end{equation}
and thus $\CAh^{0,1}$ gives rise to a holomorphic structure on\index{holomorphic!structure}
$\CP^{3|\CN}$.

In the following, we will always assume that we are working in a
gauge for which
\begin{equation}\label{gaugeholineta}
\der{\etab^i_\pm}\hpsi_\pm\ =\ 0 \ \Leftrightarrow \
\der{\etab_i^\pm}\lrcorner \CAh^{0,1}\ =\ 0~,
\end{equation}
for $a=1,2,3$.

\paragraph{Action for hCS theory.} In the case $\CN=4$, the
supertwistor space $\CP^{3|\CN}$ is a Calabi-Yau supermanifold,\index{Calabi-Yau}\index{Calabi-Yau supermanifold}\index{super!manifold}\index{twistor}\index{twistor!space}
and thus comes with the holomorphic volume form $\Omega$ defined\index{holomorphic!volume form}
in \eqref{Omega}. One can therefore introduce an action functional
\begin{equation}\label{actionhCS}
S_{\mathrm{hCS}}\ :=\
\int_{\CCP^{3|\CN}_\eps}\Omega\wedge\tr_\CG\left(\CAh^{0,1}\wedge\dparb\CAh^{0,1}+
\tfrac{2}{3}\CAh^{0,1}\wedge\CAh^{0,1}\wedge\CAh^{0,1}\right)~,
\end{equation}
where $\CCP^{3|\CN}_\eps$ is the subspace of $\CP^{3|\CN}_\eps$
for which\footnote{This restriction to a chiral subspace was
proposed in \cite{Witten:2003nn} and is related to self-duality.
It is not a contradiction to $\eta_i^\pm\neq 0$, but merely a
restriction of all functions on $\CP^{3|\CN}_\eps$ to be
holomorphic in the $\eta_i^\pm$.} $\bar{\eta}_i^\pm=0$
\cite{Witten:2003nn}. Note that the condition
\eqref{gaugeholineta}, which we introduced in the previous
paragraph, is necessary for \eqref{actionhCS} to be meaningful.

\paragraph{Pull-back of $\CE$ to the correspondence space.} The
pull-back of $\CE$ along $\pi_2$ is the bundle $\pi^*_2\CE$ with
transition function $\pi^*_2$ satisfying the equations\index{transition function}
\begin{equation}
\bV_\alpha^\pm(\pi_2^*f_{+-})\ =\ \bV^i_\pm(\pi_2^*f_{+-})\ =\ 0~.
\end{equation}
These equations, together with the splitting
\begin{equation}
\pi_2^*f_{+-}\ =\ \psi_+^{-1}\psi_-
\end{equation}
of the transition function into group-valued holomorphic functions\index{transition function}
$\psi_\pm$ on $\pi_2^{-1}(\CU_\pm)$, allow for the introduction of
matrix-valued components of a new gauge potential,
\begin{subequations}
\begin{align}
\CA_\alpha^+&\ :=\ \bar{V}_\alpha^+\lrcorner\,\CA&&\ =\ 
\psi_+\bar{V}_\alpha^+\psi_+^{-1}&&\ =\ \psi_-\bar{V}^+_\alpha\psi_-^{-1}&&\ =\ 
\lambda_+^\ald\CA_{\alpha\ald} (x_R,\eta)~,\\
\label{A3cpt} \CA_{\bl_+}&\ :=\ \dpar_{\bl_+}\lrcorner\,\CA&&\ =\ 
\psi_+\dpar_{\bl_+}\psi_+^{-1}&&\ =\ \psi_-\dpar_{\bl_+}\psi_-^{-1}&&\ =\ 0~,\\
\CA^i_+&\ :=\ \bar{V}^i_+\lrcorner\,\CA&&\ =\ 
\psi_+\bV_+^i\psi_+^{-1}&&\ =\ \psi_-
\dparb_+^i\psi_-^{-1}&&\ =\ \lambda_+^\ald\CA_\ald^i(x_R,\eta)~.
\end{align}
\end{subequations}

\paragraph{Linear system and super SDYM equations.} The gauge potential\index{linear system}
defined above fits into the linear system
\begin{subequations}\label{slinsys}\index{linear system}
\begin{eqnarray}\label{slinsys1}
(\bar{V}_\alpha^++\CA_\alpha^+)\psi_+&=&0~,\\
\dpar_{\bl_+}\psi_+&=&0~,\\\label{slinsys4}
(\bar{V}_+^i+\CA_+^i)\psi_+&=&0
\end{eqnarray}
\end{subequations}
of differential equations, whose compatibility conditions read\index{compatibility conditions}
\begin{equation}
\begin{aligned}
{}[\nabla_{\alpha\ald},\nabla_{\beta\bed}]+[\nabla_{\alpha\bed},
\nabla_{\beta\ald}]\ =\ 0~,~~~
[\nabla_{\ald}^i,\nabla_{\beta\bed}]+[\nabla_{\bed}^i,\nabla_{\beta\ald}]\ =\ 0~,\\
\{\nabla_\ald^i,\nabla_\bed^j\}+\{\nabla_\bed^i,\nabla_\ald^j\}\ =\ 0~.\hspace{2.5cm}
\end{aligned}
\label{compcon}
\end{equation}
Here, we have introduced covariant derivatives\index{covariant derivative}
\begin{equation}
\nabla_{\alpha\ald}\ :=\ \dpar_{\alpha\ald}^R
+\CA_{\alpha\ald}~~~\mbox{and}~~~
\nabla_\ald^i\ :=\ \dpar_\ald^i+\CA_\ald^i~.
\end{equation}
The equations \eqref{compcon} are the constraint equations for\index{constraint equations}
$\CN$-extended super SDYM theory.

\paragraph{Comments on the real case.} Although there is a
diffeomorphism between the correspondence space $\CF^{5|2\CN}_R$
and $\CP^{3|\CN}_\eps$ (up to the subtleties arising in the
Kleinian case $\eps=+1$), the linear systems \eqref{slinsys} and\index{linear system}
\eqref{hCSlinsys} do not coincide here. Instead, we have
\begin{equation}
\begin{aligned}
(\bar{V}_\alpha^++\CA_\alpha^+)\psi_+&\ =\ 0~,&&&(\bar{V}_\alpha^++\CAh_\alpha^+)\hpsi_+&\ =\ 0~,\\
\dpar_{\bl_+}\psi_+&\ =\ 0~,&&&(\dpar_{\bl_+}+\CAh_{\bl_+})\hpsi_+&\ =\ 0~,\\
(\bar{V}_+^i+\CA_+^i)\psi_+&\ =\ 0~,&&&\bar{V}_+^i\hpsi_+&\ =\ 0~,
\end{aligned}
\end{equation}
where the left-hand side is again \eqref{slinsys} and the
right-hand side is \eqref{hCSlinsys}, written in components
$\CAh_\alpha^+:=\bar{V}_\alpha^+\lrcorner \CAh^{0,1}$ and
$\CAh^i_+:=\bar{V}^i_+\lrcorner \CAh^{0,1}=0$. Thus, we can
schematically write for the gauge transformations between the\index{gauge transformations}\index{gauge!transformation}
trivializations $\psi_\pm$ and $\hpsi_\pm$\index{trivialization}
\begin{equation}
(\CAh_\alpha^\pm\neq 0,\CAh_{\bl_\pm}\neq 0,\CAh^i_\pm=0)\
\stackrel{\varphi}{\longrightarrow}\ (\CA_\alpha^\pm\neq
0,\CA_{\bl_\pm}\ =\  0,\CA^i_\pm\ \neq\  0)~.
\end{equation}
The main difference between the two gauges is that one can write
down an action for the one with $\CAh^{0,1}$ in the case $\CN=4$,
while this is never possible for the other gauge potential.

\paragraph{Super hCS theory.} In the following, we will discuss
holomorphic Chern-Simons theory using the components\index{Chern-Simons theory}
$\CAh_\alpha^+:=\bar{V}_\alpha^+\lrcorner \CAh^{0,1}$ introduced
above. The action \eqref{actionhCS} is rewritten as
\begin{equation}\label{actionhCScpt}
S_{\mathrm{hCS}}\ :=\ \int_{\CCP^{3|\CN}_\eps}\dd \lambda
\wedge\dd \bl \wedge\dd z^1\wedge \dd z^2\wedge \bar{E}^1\wedge
\bar{E}^2~\Omega^\eta \tr \eps^{abc}\left(\CA_a V_b\CA_c+
\tfrac{2}{3}\CA_a\CA_b\CA_c\right)~.
\end{equation}
Recall that we assumed in \eqref{actionhCS} that $\CA^i_\pm=0$.
The corresponding equations of motion read then e.g.\ on
$\hat{\CU}_+$ as
\begin{subequations}\label{shCS}
\begin{align}\label{shCS1}
\bar{V}_\alpha^+\hat{\CA}_\beta^+-
\bar{V}_\beta^+\hat{\CA}_\alpha^++
[\hat{\CA}_\alpha^+,\hat{\CA}_\beta^+]&\ =\ 0~,\\
\label{shCS2} \dpar_{\bl_+}\hat{\CA}_\alpha^+-
\bar{V}_\alpha^+\hat{\CA}_{\bl_+}+
[\hat{\CA}_{\bl_+},\hat{\CA}_\alpha^+]&\ =\ 0
\end{align}
\end{subequations}
and very similarly on $\hat{\CU}_-$. Here, $\hat{\CA}_\alpha^+$
and $\hat{\CA}_{\bl_+}$ are functions of
$(x_R^{\alpha\ald},\lambda_+,\bl_+,\eta^+_i)$. These equations are
equivalent to the equations of self-dual $\CN$-extended SYM theory
on $\FR^4$. As already mentioned, the most interesting case is
$\CN{=}4$ since the supertwistor space $\CP^{3|4}$ is a Calabi-Yau\index{Calabi-Yau}\index{twistor}\index{twistor!space}
supermanifold and one can derive equations \eqref{extrahCS} or\index{super!manifold}
\eqref{shCS1}, \eqref{shCS2} from the manifestly Lorentz invariant
action \eqref{actionhCS} \cite{Witten:2003nn, Sokatchev:1995nj}.
For this reason, we mostly concentrate on the equivalence with
self-dual SYM for the case $\CN{=}4$.

\paragraph{Field expansion for super hCS theory.} Recall that
$\hat{\CA}_\alpha$ and $\hat{\CA}_\bl$ are sections of the bundles
$\CO(1)\otimes\FC^2$ and $\bar{\CO}(-2)$ over $\CPP^1$ since the
vector fields $\bV_\alpha$ and $\dpar_{\bl_\pm}$ take values in
$\CO(1)$ and the holomorphic cotangent bundle of $\CPP^1$ is
$\CO(-2)$. Together with the fact that the $\eta^+_i$s take values
in the bundle $\Pi\CO(1)$, this fixes the dependence of\index{PO(n)@$\Pi\CO(1)$}
$\hat{\CA}^\pm_\alpha$ and $\hat{\CA}_{\bl_\pm}$ on
$\lambda^\ald_\pm$ and $\hat{\lambda}^\ald_\pm$. In the case
$\CN{=}4$, this dependence can be brought to the form
\begin{subequations}\label{expA}
\begin{align}\label{expAa}
\hat{\CA}_\alpha^+&\ =\ \lambda_+^\ald\,
A_{\alpha\ald}(x_R)+\eta_i^+\chi^i_\alpha(x_R)+
\gamma_+\,\tfrac{1}{2!}\,\eta^+_i\eta^+_j\,\hat{\lambda}^\ald_+\,
\phi_{\alpha \ald}^{ij}(x_R)+\\
\nonumber
&+\gamma_+^2\,\,\tfrac{1}{3!}\,\eta^+_i\eta^+_j\eta^+_k\,\hat{\lambda}_+^\ald\,
\hat{\lambda}_+^\bed\,
\tilde{\chi}^{ijk}_{\alpha\ald\bed}(x_R)+\gamma_+^3\,\tfrac{1}{4!}\,
\eta^+_i\eta^+_j\eta^+_k\eta^+_l\,
\hat{\lambda}_+^\ald\,\hat{\lambda}_+^\bed\,\hat{\lambda}_+^{\dot{\gamma}}\,
G^{ijkl}_{\alpha\ald\bed\dot{\gamma}}(x_R)~,\\
\label{expAl}
\hat{\CA}_{\bl_+}&\ =\ \gamma_+^2\,\tfrac{1}{2!}\,\eta^+_i\eta^+_j\,\phi^{ij}(x_R)+
\gamma_+^3\,\tfrac{1}{3!}\,\eta^+_i\eta^+_j\eta^+_k\,\hat{\lambda}_+^\ald\,
\tilde{\chi}^{ijk}_{\ald}
(x_R)+\\
\nonumber
&+\gamma_+^4\,\tfrac{1}{4!}\,\eta^+_i\eta^+_j\eta^+_k\eta^+_l\,
\hat{\lambda}_+^\ald\,\hat{\lambda}_+^\bed
G^{ijkl}_{\ald\bed}(x_R)~,
\end{align}
\end{subequations}
and there are similar expressions for $\hat{\CA}_\alpha^-,
\hat{\CA}_{\bl_-}$. Here,
$(A_{\alpha\ald}\cb\chi_\alpha^i\cb\phi^{ij}\cb\tilde{\chi}_{\ald
i})$ is the ordinary field content of $\CN{=}4$ super Yang-Mills
theory and the field $G_{\ald\bed}$ is the auxiliary field arising\index{Yang-Mills theory}
in the $\CN{=}4$ self-dual case, as discussed in section
\ref{sssSDYM}. It follows from \eqref{shCS2}-\eqref{expAl}
that\footnote{Here, $(\cdot)$ denotes again symmetrization, i.e.\
$(\ald\bed)=\ald\bed+\bed\ald$.}
\begin{equation}\label{expA2}
\phi^{ij}_{\alpha\ald}\ =\ -\nabla_{\alpha\ald}\phi^{ij}\ ,\quad
\tilde{\chi}^{ijk}_{\alpha\ald\bed}\ =\ -\tfrac{1}{4}\nabla_{\alpha(\ald}
\tilde{\chi}^{ijk}_{\bed)}\quad\mbox{and}\quad
G^{ijkl}_{\alpha\ald\bed\dot{\gamma}}\ =\ -\tfrac{1}{18}\nabla_{\alpha(\ald}
G^{ijkl}_{\bed\dot{\gamma})}\ ,
\end{equation}
i.e.\ these fields do not contain additional degrees of freedom.
The expansion \eqref{expAa}, \eqref{expAl} together with the field
equations \eqref{shCS1}, \eqref{shCS2} reproduces exactly the
super SDYM equations \eqref{SDYMeom}.

\paragraph{The cases $\CN<4$.} Since the $\eta^+_i$s are
Gra\ss mann variables and thus nilpotent, the expansion
\eqref{expA} for $\CN<4$ will only have terms up to order $\CN$ in
the $\eta^+_i$s. This exactly reduces the expansion to the
appropriate field content for $\CN$-extended super SDYM theory:
\begin{equation}
\begin{aligned}
\CN\ =\ 0 &~~~  A_{\alpha\ald}\\
\CN\ =\ 1 &~~~  A_{\alpha\ald},~~\chi^i_\alpha~~~\mbox{with}~~i\ =\ 1\\
\CN\ =\ 2 &~~~
A_{\alpha\ald},~~\chi^i_\alpha,~~\phi^{[ij]}~~~\mbox{with}~~i,j\ =\ 1,2\\
\CN\ =\ 3 &~~~
A_{\alpha\ald},~~\chi^i_\alpha,~~\phi^{[ij]},~~\chi^{[ijk]}_\ald~~~
\mbox{with}~~i,j,\ldots \ =\ 1,2,3\\
\CN\ =\ 4 &~~~
A_{\alpha\ald},~~\chi^i_\alpha,~~\phi^{[ij]},~~\chi^{[ijk]}_\ald,~~
G_{\ald\bed}^{[ijkl]}~~~\mbox{with}~~i,j,\ldots \ =\ 1,\ldots ,4~.
\end{aligned}
\end{equation}
One should note that the antisymmetrization $[\cdot]$ leads to a
different number of fields depending on the range of $i$. For
example, in the case $\CN{=}2$, there is only one real scalar
$\phi^{12}$, while for $\CN{=}4$ there exist six real scalars.
Inserting such a truncated expansion for $\CN{<}4$ into the field
equations \eqref{shCS1} and \eqref{shCS2}, we obtain the first
$\CN{+}1$ equations of \eqref{SDYMeom}, which is the appropriate
set of equations for $\CN{<}4$ super SDYM theory.

One should stress, however, that this expansion can only be
written down in the real case due to the identification of the
vector fields on $\CP^{3|\CN}$ with those along the projection
$\pi_2$. This is in contrast to the superfield expansion of the
gauge potentials participating in the constraint equations for\index{constraint equations}
$\CN$-extended supersymmetric SDYM theory, which also holds in the
complex case.

\paragraph{The Penrose-Ward transform for dual supertwistors.} The\index{Penrose-Ward transform}\index{twistor}
discussion of the Penrose-Ward transform over the dual
supertwistor space $\CP^{3|\CN}_*$ is completely analogous to the\index{twistor}\index{twistor!space}
above discussion, so we refrain from going in any detail. To make
the transition to dual twistor space, one simply has to replace\index{twistor}\index{twistor!dual twistor}\index{twistor!space}
everywhere all $\lambda$ and $\eta$ by $\mu$ and $\theta$,
respectively, as well as dualize the spinor indices\index{Spinor}
$\alpha\rightarrow \ald$, $\ald\rightarrow \alpha$ etc.\ and
change the upper R-symmetry indices to lower ones and vice versa.

\paragraph{\v{C}ech cohomology over supermanifolds.} Note that in\index{Cech cohomology@\v{C}ech cohomology}\index{super!manifold}
performing the Penrose-Ward transform, we have heavily relied on\index{Penrose-Ward transform}
both the \v{C}ech and the Dolbeault description of holomorphic
vector bundles. Recall that if the patches $\CU_a$ of the covering\index{holomorphic!vector bundle}
$\frU$ are Stein manifolds, one can show that the first \v{C}ech\index{Stein manifold}\index{manifold}
cohomology sets are independent of the covering $\frU$ and depend
only on the manifold $M$, e.g.\ $H^1(\frU,\frS)=H^1(M,\frS)$.\index{manifold}
Since the covering of the body of $\CP^{3|\CN}$ is obviously\index{body}
unaffected by the extension to an infinitesimal
neighborhood,\footnote{An infinitesimal neighborhood cannot be\index{infinitesimal neighborhood}
covered partially.} we can assume that $H^1$ is also independent
of the covering for supermanifolds.\index{super!manifold}

\paragraph{Summary.} We have described a one-to-one
correspondence between gauge equivalence classes of solutions to
the $\CN$-extended SDYM equations on $({\FR^4},g)$ with
$g=\diag(-\eps,-\eps,+1,+1)$ and equivalence classes of
holomorphic vector bundles $\CE$ over the supertwistor space\index{holomorphic!vector bundle}\index{twistor}\index{twistor!space}
$\CP^{3|\CN}$ such that the bundles $\CE$ are holomorphically
trivial on each projective line $\CPP^1_{x_R,\eta}$ in
$\CP^{3|\CN}$. In other words, there is a bijection between the
moduli spaces of hCS theory on $\CP^{3|\CN}$ and the one of\index{moduli space}
self-dual $\CN$-extended SDYM theory on $(\FR^4,g)$. It is assumed
that appropriate reality conditions are imposed. The Penrose-Ward
transform and its inverse are defined by the formul\ae{}\index{Penrose-Ward transform}
\eqref{expA}. In fact, these formul\ae{} relate solutions of the
equations of motion of hCS theory on $\CP^{3|\CN}$ to those of
self-dual $\CN$-extended SYM theory on $(\FR^4,g)$. One can also
write integral formul\ae{} of type \eqref{pwtrafo1} but we refrain
from doing this.

\section{Penrose-Ward transform using exotic supermanifolds}\index{Penrose-Ward transform}\index{exotic!supermanifold}\index{super!manifold}
\label{spwexotic}

\subsection{Motivation for considering exotic supermanifolds}\index{exotic!supermanifold}\index{super!manifold}

The Calabi-Yau property, i.e.\ vanishing of the first Chern class\index{Calabi-Yau}\index{Chern class}\index{first Chern class}
or equivalently the existence of a globally well-defined
holomorphic volume form, is essential for defining the B-model on\index{holomorphic!volume form}
a certain space. Consider the space $\CP^{3|4}$ as introduced in
the last section. Since the volume element $\Omega$ which is
locally given by $\Omega_\pm:=\pm\dd z_\pm^1\wedge\dd
z^2_\pm\wedge \dd \lambda_\pm\dd \eta_1^\pm\ldots \dd \eta_4^\pm$ is
globally defined and holomorphic, $\CP^{3|4}$ is a Calabi-Yau\index{Calabi-Yau}
supermanifold. Other spaces which have a twistorial\index{super!manifold}\index{twistor}
$\CO(1)\oplus\CO(1)$ body and are still Calabi-Yau supermanifolds\index{Calabi-Yau}\index{Calabi-Yau supermanifold}\index{body}
are, e.g., the weighted projective spaces\footnote{In fact, one\index{weighted projective spaces}
rather considers their open subspaces
$W\CPP^{3|2}(1,1,1,1|p,q)\backslash W\CPP^{1|2}(1,1|p,q)$.}
$W\CPP^{3|2}(1,1,1,1|p,q)$ with $(p,q)$ equal to $(1,3)$, $(2,2)$
and $(4,0)$ as considered in \cite{Popov:2004nk}. The topological
B-model on these manifolds was shown to be equivalent to $\CN=4$\index{topological!B-model}\index{manifold}
SDYM theory with a truncated field content. Additionally in the
cases $(2,2)$ and $(4,0)$, the parities of some fields are
changed.

An obvious idea to obtain even more Calabi-Yau supermanifolds\index{Calabi-Yau}\index{Calabi-Yau supermanifold}\index{super!manifold}
directly from $\CP^{3|4}$ is to combine several fermionic
variables into a single one,\footnote{A similar situation has been
considered in \cite{Lechtenfeld:2004cc}, where all the fermionic
variables where combined into a single even nilpotent one.} e.g.\
to consider coordinates $(\zeta_1:=\eta_1,$
$\zeta_2:=\eta_2\eta_3\eta_4)$. In an analogous situation for
bosonic variables, one could always at least locally find
additional coordinates complementing the reduced set to a set
describing the full space. Fixing the complementing coordinates to
certain values then means that one considers a subvariety of the
full space. However, as there is no inverse of Gra{\ss}mann variables,\index{Gra{\ss}mann variable}
the situation here is different. Instead of taking a subspace, we
rather restrict the algebra of functions (and similarly the set of
differential operators) by demanding a certain dependence on the
Gra{\ss}mann variables. One can indeed find complementing sets of\index{Gra{\ss}mann variable}
functions to restore the full algebra of functions on $\CP^{3|4}$.
Underlining the argument that we do not consider a subspace of
$\CP^{3|4}$ is the observation that we still have to integrate
over the full space $\CP^{3|4}$: $\int \dd \zeta_1\dd\zeta_2=\int
\dd \eta_1\ldots \dd \eta_4$. This picture has a slight similarity to
the definition of the body of a supermanifold as given in\index{body}\index{super!manifold}
\cite{DeWitt:1992cy,Cartier:2002zp}.

Possible inequivalent groupings of the Gra{\ss}mann coordinates of
$\CP^{3|4}$ are the previously given example
$(\zeta_1:=\eta_1,\zeta_2:=\eta_2\eta_3\eta_4)$ as well as
$(\zeta_1=\eta_1,\zeta_2=\eta_2,\zeta_3=\eta_3\eta_4)$,
$(\zeta_1=\eta_1\eta_2,$ $\zeta_2=\eta_3\eta_4)$, and
$(\zeta_1=\eta_1\eta_2\eta_3\eta_4)$. They correspond to exotic
supermanifolds of dimension $(3\oplus 0|2)$, $(3\oplus 1|2)$,\index{exotic!supermanifold}\index{super!manifold}
$(3\oplus 2|0)$, and $(3\oplus 1|0)$, respectively. Considering
hCS theory on them, one finds that the first one is equivalent to
the case $W\CPP^{3|2}(1,1,1,1|1,3)$ which was already discussed in
\cite{Popov:2004nk}. The case $(3\oplus 2|0)$ will be similar to
the case $W\CPP^{3|2}(1,1,1,1|2,2)$, but with a field content of
partially different parity. The case $(3\oplus 1|2)$ is a mixture\index{parity}
easily derived from combining the full case $\CP^{3|4}$ with the
case $(3\oplus 2|0)$. We restrict ourselves in the following to
the cases $(3\oplus 2|0)$ and $(3\oplus 1|0)$.

Instead of considering independent twistor correspondences between\index{twistor}\index{twistor!correspondence}
fattened complex manifolds and the moduli space of relative\index{complex!manifold}\index{fattened complex manifold}\index{moduli space}\index{manifold}
deformations of the embedded $\CPP^1$, we will focus on {\em
reductions} of the correspondence between $\CP^{3|4}$ and
$\FC^{4|8}$. This formulation allows for a more direct
identification of the remaining subsectors of $\CN=4$ self-dual
Yang-Mills theory and can in a sense be understood as a fermionic\index{Yang-Mills theory}\index{self-dual Yang-Mills theory}
dimensional reduction.\index{dimensional reduction}

\subsection{The twistor space $\SA$}\index{twistor}\index{twistor!space}

\paragraph{Definition of $\SA$.} The starting point of our discussion is
the supertwistor space $\CP^{3|4}=(\CP^3,\CO_{[4]})$. Consider the\index{twistor}\index{twistor!space}
differential operators
\begin{equation}
\CD^{i1}_\pm\ :=\ \eta_1^\pm\eta_2^\pm\der{\eta_i^\pm}\eand
\CD^{i2}_\pm\ :=\ \eta_3^\pm\eta_4^\pm\der{\eta_{i+2}^\pm}~~~\mbox{for}~~i\ =\ 1,2~~,
\end{equation}
which are maps $\CO_{[4]}\rightarrow\CO_{[4]}$. The space
$\CP^{3}$ together with the structure sheaf\index{sheaf}\index{structure sheaf}
\begin{equation}
\CO_{(1,2)}\ :=\ \bigcap_{i,j=1,2}\ker
\CD^{ij}_+\ =\ \bigcap_{i,j=1,2}\ker \CD^{ij}_-~,
\end{equation}
which is a reduction of $\CO_{[4]}$, is the fattened complex
manifold $\SA$, covered by two patches $\hat{\CU}_+$ and\index{complex!manifold}\index{fattened complex manifold}\index{manifold}
$\hat{\CU}_-$ and described by local coordinates
$(z^\alpha_\pm,\,\lambda_\pm,\,e^\pm_1:=\eta^\pm_1\eta^\pm_2,\,
e^\pm_2:=\eta^\pm_3\eta^\pm_4)$. The two even nilpotent
coordinates $e^\pm_i$ are each sections of the line bundle
$\CO(2)$ with the identification $(e^\pm_i)^2\sim 0$.

\paragraph{Derivatives on $\SA$.}\label{pderivativesSA}
As pointed out before, the coordinates $e^\pm_i$ do not allow for
a complementing set of coordinates, and therefore it is not
possible to use Leibniz calculus in the transition from the
$\eta$-coordinates on $(\CP^{3},\CO_{[4]})$ to the $e$-coordinates
on $(\CP^{3},\CO_{(1,2)})$. Instead, from the observation that
\begin{equation}
\begin{aligned}\label{derid}
\eta_2^\pm\der{e^\pm_1}\ =\ \left.\der{\eta^\pm_1}\right|_{\CO_{(1,2)}}~,~~~
\eta_1^\pm\der{e^\pm_1}\ =\ -\left.\der{\eta^\pm_2}\right|_{\CO_{(1,2)}}~,\\
\eta_4^\pm\der{e^\pm_2}\ =\ \left.\der{\eta^\pm_3}\right|_{\CO_{(1,2)}}~,~~~
\eta_3^\pm\der{e^\pm_2}\ =\ -\left.\der{\eta^\pm_4}\right|_{\CO_{(1,2)}}~,
\end{aligned}
\end{equation}
one directly obtains the following identities on
$(\CP^{3},\CO_{(1,2)})$:
\begin{equation}\label{deridb}
\der{e^\pm_1}\ =\ \der{\eta^\pm_2}\der{\eta^\pm_1}\eand
\der{e^\pm_2}\ =\ \der{\eta^\pm_4}\der{\eta^\pm_3}~.
\end{equation}
Equations \eqref{derid} are easily derived by considering an
arbitrary section $f$ of $\CO_{(1,2)}$:
\begin{equation}
f\ =\ a^0+a^1 e_1+a^2 e_2+a^{12}
e_1e_2\ =\ a^0+a^1\eta_1\eta_2+a^2\eta_3\eta_4+a^{12}\eta_1\eta_2\eta_3\eta_4~,
\end{equation}
where we suppressed the $\pm$ labels for convenience. Acting,
e.g., by $\der{\eta_1}$ on $f$, we see that this equals an action
of $\eta_2\der{e_1}$. It is then also obvious that we can make the
formal identification \eqref{deridb} on $(\CP^{3},\CO_{(1,2)})$.
Still, a few more comments on \eqref{deridb} are in order. These
differential operators clearly map $\CO_{(1,2)}\rightarrow
\CO_{(1,2)}$ and fulfill
\begin{equation}\label{derdel}
\der{e^\pm_i}\,e^\pm_j\ =\ \delta^i_j~.
\end{equation}
Note, however, that they do not quite satisfy the Leibniz rule,
e.g.:
\begin{equation}
1=\der{e^\pm_1}\,e^\pm_1\ =\ \der{e^\pm_1}\,(\eta^\pm_1\eta^\pm_2)
\ \neq\ 
\left(\der{e^\pm_1}\,\eta^\pm_1\right)\eta^\pm_2+\eta^\pm_1\left(\der{e^\pm_1}\,\eta^\pm_2\right)\ =\ 0~.
\end{equation}
This does not affect the fattened complex manifold $\SA$ at all,\index{complex!manifold}\index{fattened complex manifold}\index{manifold}
but it imposes an obvious constraint on the formal manipulation of
expressions involving the $e$-coordinates rewritten in terms of
the $\eta$-coordinates.

For the cotangent space, we have the identification $\dd
e^\pm_1=\dd \eta^\pm_2\dd \eta^\pm_1$ and $\dd e^\pm_2=\dd
\eta^\pm_4\dd \eta^\pm_3$ and similarly to above, one has to take
care in formal manipulations, as integration is equivalent to
differentiation.

\paragraph{Moduli space of sections.}\index{moduli space}
As discussed in section \ref{ssupertwistorspaces},
\ref{pGlobalSecs}, holomorphic sections of the bundle
$\CP^{3|4}\rightarrow \CPP^1$ are described by moduli which are
elements of the space $\FC^{4|8}=(\FC^4,\CO_{[8]})$. After the
above reduction, holomorphic sections of the bundle
$\SA\rightarrow\CPP^1$ are defined by the equations
\begin{equation}\label{secsA}
z^\ald_\pm\ =\ x^{\alpha\ald}\lambda_\ald^\pm\eand
e^\pm_i\ =\ e^{\ald\bed}_i\lambda^\pm_\ald\lambda^\pm_\bed~.
\end{equation}
While the Gra{\ss}mann algebra of the coordinates $\eta_k^\pm$ of\index{Gra{\ss}mann algebra}
$\CP^{3|4}$ immediately imposed a Gra{\ss}mann algebra on the moduli
$\eta_k^\ald\in\FC^{0|8}$, the situation here is more subtle. We
have\footnote{The brackets $(\cdot)$ and $[\cdot]$ denote
symmetrization and antisymmetrization, respectively, of the
enclosed indices with appropriate weight.}
$e_1^{(\ald\bed)}=\eta_1^{(\ald}\eta_2^{\bed)}$ and from this, we
already note that $(e_1^{\dot{1}\dot{2}})^2\neq0$ but only
$(e_1^{\dot{1}\dot{2}})^3=0$. Thus, the moduli space is a\index{moduli space}
fattening of order 1 in $e_1^{\dot{1}\dot{1}}$ and
$e_1^{\dot{2}\dot{2}}$, but a fattening of order 2 in
$e_1^{\dot{1}\dot{2}}$ which analogously holds for
$e_2^{\ald\bed}$. Furthermore, we have the additional identities
\begin{equation}\label{id1}
e_i^{\dot{1}\dot{2}}e_i^{\dot{1}\dot{2}}\ =\ -\tfrac{1}{2}
e_i^{\dot{1}\dot{1}}e_i^{\dot{2}\dot{2}}~~~\mbox{and}~~~
e_i^{\dot{1}\dot{2}}e_i^{\dot{2}\dot{2}}\ =\ e_i^{\dot{1}\dot{2}}e_i^{\dot{1}\dot{1}}\ =\ 0~.
\end{equation}
Additional conditions which appear when working with fattened
complex manifolds are not unusual and similar problems were\index{complex!manifold}\index{fattened complex manifold}\index{manifold}
encountered, e.g., in the discussion of fattened ambitwistor\index{twistor}\index{twistor!ambitwistor}
spaces in \cite{Eastwood:1987}.

More formally, one can introduce the differential operators
\begin{align}
&\CD^{1c}\ =\ (\eta_1^\ald\dpar^1_\ald-\eta_2^\ald\dpar^2_\ald)~,~~~
\CD^{2c}\ =\ (\eta_3^\ald\dpar^3_\ald-\eta_4^\ald\dpar^4_\ald)~,\\
&\CD^{1s}\ =\ (\dpar^2_{\dot{1}}\dpar^1_{\dot{2}}-\dpar^2_{\dot{2}}\dpar^1_{\dot{1}})~,~~~
\CD^{2s}\ =\ (\dpar^4_{\dot{1}}\dpar^3_{\dot{2}}-\dpar^4_{\dot{2}}\dpar^3_{\dot{1}})~
\end{align}
which map $\CO_{[8]}\rightarrow\CO_{[8]}$, and consider the
overlap of kernels
\begin{equation}
\CO_{(1;2,6)}\ :=\ \bigcap_{i=1,2}\left(\ker(\CD^{ic})\cap\ker(\CD^{is})\right)~.
\end{equation}
The space $\FC^{4}$ together with the structure sheaf\index{sheaf}\index{structure sheaf}
$\CO_{(1;2,6)}$, which is a reduction of $\CO_{[8]}$, is exactly
the moduli space described above, i.e.\ a fattened complex\index{moduli space}
manifold $\FC^{4\oplus 6|0}$ on which the coordinates\index{manifold}
$e_i^{\ald\bed}$ satisfy the additional constrains \eqref{id1}.

\paragraph{The double fibration.} Altogether, we have the following\index{double fibration}\index{fibration}
reduction of the full double fibration \eqref{superdblfibration}
for $\CN=4$:
\begin{equation}\label{dblfibrationA}
\begin{aligned}
\begin{picture}(50,45)
\put(0.0,0.0){\makebox(0,0)[c]{$(\CP^{3},\CO_{[4]})$}}
\put(64.0,0.0){\makebox(0,0)[c]{$(\FC^{4},\CO_{[8]})$}}
\put(34.0,39.0){\makebox(0,0)[c]{$(\FC^{4}\times\CPP^1,\CO_{[8]}\otimes\CO_{\CPP^1})$}}
\put(7.0,21.0){\makebox(0,0)[c]{$\pi_2$}}
\put(56.0,21.0){\makebox(0,0)[c]{$\pi_1$}}
\put(25.0,28.0){\vector(-1,-1){18}}
\put(37.0,28.0){\vector(1,-1){18}}
\end{picture}
\end{aligned}
\hspace{2.3cm}\ \longrightarrow\ \hspace{2.1cm}
\begin{aligned}
\begin{picture}(70,45)
\put(0.0,0.0){\makebox(0,0)[c]{$(\CP^{3},\CO_{(1,2)})$}}
\put(79.0,0.0){\makebox(0,0)[c]{$(\FC^{4},\CO_{(1;2,6)})$}}
\put(42.0,39.0){\makebox(0,0)[c]{$(\FC^{4}\times\CPP^1,\CO_{(1;2,6)}\otimes\CO_{\CPP^1})$}}
\put(7.0,21.0){\makebox(0,0)[c]{$\pi_2$}}
\put(56.0,21.0){\makebox(0,0)[c]{$\pi_1$}}
\put(25.0,28.0){\vector(-1,-1){18}}
\put(37.0,28.0){\vector(1,-1){18}}
\end{picture}
\end{aligned}
\end{equation}
where $\CO_{\CPP^1}$ is the structure sheaf of the Riemann sphere\index{Riemann sphere}\index{sheaf}\index{structure sheaf}
$\CPP^1$. The tangent spaces along the leaves of the projection
$\pi_2$ are spanned by the vector fields
\begin{equation}
\begin{aligned}
&\bV_\alpha^\pm\ =\ \lambda_\pm^\ald\dpar_{\alpha\ald}~,~~~~~~
&&\bV_\alpha^\pm\ =\ \lambda_\pm^\ald\dpar_{\alpha\ald}~,\\
&\bV^k_\pm\ =\ \lambda^\ald_\pm\der{\eta_k^\ald}~,~~~~~~ &&\bV^i_{\bed
\pm}\ =\ \lambda^\ald_\pm\der{e_i^{(\ald\bed)}}
\end{aligned}
\end{equation}
in the left and right double fibration in \eqref{dblfibrationA},\index{double fibration}\index{fibration}
where $k=1,\ldots ,4$. Note that similarly to \eqref{derid}, we have
the identities
\begin{equation}\label{derid2}
\begin{aligned}
&\eta_2^\ald\der{e_1^{(\ald\bed)}}\ =\ \left.\der{\eta_1^\bed}\right|_{\CO_{(1;2,6)}}~,~~~~~~&&
\eta_1^\ald\der{e_1^{(\ald\bed)}}\ =\ -\left.\der{\eta_2^\bed}\right|_{\CO_{(1;2,6)}}~,\\
&\eta_4^\ald\der{e_2^{(\ald\bed)}}\ =\ \left.\der{\eta_3^\bed}\right|_{\CO_{(1;2,6)}}~,~~~~~~&&
\eta_3^\ald\der{e_2^{(\ald\bed)}}\ =\ -\left.\der{\eta_4^\bed}\right|_{\CO_{(1;2,6)}}~,
\end{aligned}
\end{equation}
and it follows, e.g., that
\begin{equation}
\left.\bV^1_\pm\right|_{\CO_{(1;2,6)}}\ =\ \eta_2^\ald
\bV^1_{\ald\pm}\eand
\left.\bV^2_\pm\right|_{\CO_{(1;2,6)}}\ =\ -\eta_1^\ald
\bV^1_{\ald\pm}~.
\end{equation}

\paragraph{Holomorphic Chern-Simons theory on $\SA$.} The topological\index{Chern-Simons theory}
B-model on $\SA=(\CP^{3},\CO_{(1,2)})$ is equivalent to hCS theory
on $\SA$ since a reduction of the structure sheaf does not affect\index{sheaf}\index{structure sheaf}
the arguments used for this equivalence in
\cite{Witten:1992fb,Witten:2003nn}. Consider a trivial rank $n$
complex vector bundle\footnote{Note that the components of\index{complex!vector bundle}
sections of ordinary vector bundles over a supermanifold are\index{super!manifold}
superfunctions. The same holds for the components of connections\index{connection}
and transition functions.} $\CE$ over $\SA$ with a connection\index{transition function}
$\CAh$. The action for hCS theory on this space reads
\begin{equation}
S=\int_{\CSA} \Omega^{3\oplus
2|0}\wedge\tr\left(\CAh^{0,1}\wedge\bar{\dpar}\CAh^{0,1}+
\tfrac{2}{3}\CAh^{0,1}\wedge\CAh^{0,1}\wedge\CAh^{0,1}\right)~,
\end{equation}
where $\CSA$ is the subspace of $\SA$ for which $\bar{e}^\pm_i=0$,
$\CAh^{0,1}$ is the (0,1)-part of $\CAh$ and $\Omega^{3\oplus
2|0}$ is the holomorphic volume form, e.g.\ $\Omega^{3\oplus\index{holomorphic!volume form}
2|0}_+=\dd z^1_+\wedge\dd z^2_+\wedge\dd \lambda_+\dd e_1^+\dd
e_2^+$. The equations of motion read $\dparb
\CAh^{0,1}+\CAh^{0,1}\wedge \CAh^{0,1}=0$ and solutions define a
holomorphic structure $\dparb_\CAh$ on $\CE$. Given such a\index{holomorphic!structure}
solution $\CAh^{0,1}$, one can locally write
$\CAh^{0,1}|_{\hat{\CU}_\pm}=\hpsi_\pm\dparb\hpsi^{-1}_\pm$ with
regular matrix-valued functions $\hpsi_\pm$ smooth on the patches
$\hat{\CU}_\pm$ and from the gluing condition
$\hpsi_+\dparb\hpsi^{-1}_+=\hpsi_-\dparb\hpsi^{-1}_-$ on the
overlap $\hat{\CU}_+\cap\hat{\CU}_-$, one obtains
$\dparb(\hpsi^{-1}_+\hpsi_-)=0$. Thus,
$f_{+-}:=\hpsi^{-1}_+\hpsi_-$ defines a transition function for a\index{transition function}
holomorphic vector bundle $\tilde{\CE}$, which is (smoothly)\index{holomorphic!vector bundle}
equivalent to $\CE$.

\paragraph{The linear system on the correspondence space.} Consider\index{linear system}
now the pull-back of the bundle $\tilde{\CE}$ along $\pi_2$ in
\eqref{dblfibrationA} to the space $\FC^{4}\times \CPP^1$, i.e.\
the holomorphic vector bundle $\pi_2^*\tilde{\CE}$ with transition\index{holomorphic!vector bundle}
function $\pi_2^* f_{+-}$ satisfying $\bV^\pm_\alpha
\left(\pi_2^*f_{+-}\right)=\bV^k_{\pm} \left(\pi_2^*f_{+-}\right)=0$.
Let us suppose that the vector bundle $\pi_2^*\tilde{\CE}$ becomes
holomorphically trivial\footnote{This assumption is crucial for
the Penrose-Ward transform and reduces the space of possible\index{Penrose-Ward transform}
$\CAh^{0,1}$ to an open subspace around $\CAh^{0,1}=0$.} when
restricted to sections $\CPP^1_{x,e}\embd\CP^{3|4}$. This implies
that there is a splitting $\pi^*_2 f_{+-}=\hpsi_+^{-1}\hpsi_-$,
where $\hpsi_\pm$ are group-valued functions which are holomorphic
in the moduli $(x^{\alpha\ald},\eta^{\ald}_k)$ and $\lambda_\pm$.
From the condition $\bV^\pm_\alpha
\left(\pi_2^*f_{+-}\right)=\bV_\pm^{k}
\left(\pi_2^*f_{+-}\right)=0$ we obtain, e.g.\ on $\hat{\CU}_+$
\begin{equation}\label{defA}
\begin{aligned}
\hpsi_+
\bV^+_\alpha\hpsi_+^{-1}\ =\ \hpsi_-\bV^+_\alpha\hpsi_-^{-1}\ =:\ \lambda^\ald_+
\CA_{\alpha\ald}&\ =:\ \CA^+_\alpha~,\\
\hpsi_+ \bV^{k}_+
\hpsi_+^{-1}\ =\ \hpsi_-\bV^{k}_+\hpsi_-^{-1}\ =:\ \lambda^\ald_+
\CA^k_{\ald}&\ =:\ \CA^{k}_{+}~,\\\hpsi_+ \dpar_{\bl_+}
\hpsi_+^{-1}\ =\ \hpsi_-\dpar_{\bl_+}\hpsi_-^{-1}&\ =:\ \CA_{\bl_+}\ =\ 0~,\\
\hpsi_+
\dpar_{\bar{x}^{\alpha\ald}}\hpsi_+^{-1}\ =\ \hpsi_-\dpar_{\bar{x}^{\alpha\ald}}\hpsi_-^{-1}
&\ =\ 0~.
\end{aligned}
\end{equation}
Considering the reduced structure sheaves, we can rewrite the
second line of \eqref{defA}, e.g.\ for $k=1$ as
\begin{equation}\label{defA2}
\eta_2^\bed\hpsi_+ \bV^{1}_{\bed+}
\hpsi_+^{-1}\ =\ \eta_2^\bed\hpsi_-\bV^{1}_{\bed+}\hpsi_-^{-1}\ =:\ 
\eta_2^\bed\lambda^\ald_+ \CA^1_{\ald\bed}~,
\end{equation}
which yields $\eta_2^\bed \CA^1_{\ald\bed}=\CA^1_{\ald}$. From
this equation (and similar ones for other values of $k$) and the
well-known superfield expansion of $\CA^k_{\ald}$ (see e.g.\
\cite{Devchand:1996gv}), one can now construct the superfield
expansion of $\CA^i_{\ald\bed}$ by dropping all the terms, which
are not in the kernel of the differential operators $\CD^{jc}$ and
$\CD^{js}$. This will give rise to a bosonic subsector of $\CN=4$
SDYM theory.

\paragraph{Compatibility conditions.} To be more explicit,\index{compatibility conditions}
we can also use \eqref{defA2} and introduce the covariant
derivative\index{covariant derivative}
$\nabla_{\alpha\ald}:=\dpar_{\alpha\ald}+[\CA_{\alpha\ald},\cdot]$
and the first order differential operator
$\nabla_{\ald\bed}^i:=\dpar^i_{\ald\bed}+[\CA^i_{\ald\bed},\cdot]$,
which allow us to rewrite the compatibility conditions of the\index{compatibility conditions}
linear system behind \eqref{defA}, \eqref{defA2} for the reduced\index{linear system}
structure sheaf as\index{sheaf}\index{structure sheaf}
\begin{equation}
\begin{aligned}
{}[\nabla_{\alpha\ald},\nabla_{\beta\bed}]+[\nabla_{\alpha\bed},
\nabla_{\beta\ald}]\ =\ 0~,~~~~~
\eta^\gad_m\left([\nabla_{\ald\gad}^i,\nabla_{\beta\bed}]+[\nabla_{\bed\gad}^i,\nabla_{\beta\ald}]\right)\ =\ 0~,
\\\eta^\gad_m\eta^\ded_n\left([\nabla_{\ald\gad}^i,\nabla_{\bed\ded}^j]+
[\nabla_{\bed\gad}^i,\nabla_{\ald\ded}^j]\right)\ =\ 0~,\hspace{2.5cm}
\label{compconA}
\end{aligned}
\end{equation}
where $m=2i-1,2i$ and $n=2j-1,2j$. Note that $\nabla_{\ald\bed}^i$
is no true covariant derivative, as $\dpar_{\ald\bed}^i$ and\index{covariant derivative}
$\CA^i_{\ald\bed}$ do not have the same symmetry properties in the
indices. Nevertheless, the differential operators
$\nabla_{\alpha\ald}$ and $\nabla_{\ald\bed}^i$ satisfy the
Bianchi identities on $(\FC^{4},\CO_{(1;2,6)})$.

\paragraph{Constraint equations.} By eliminating\index{constraint equations}
all $\lambda$-dependence, we have implicitly performed the
push-forward of $\CA$ along $\pi_1$ onto
$(\FC^{4},\CO_{(1;2,6)})$. Let us define further tensor
superfields, which could roughly be seen as extensions of the
supercurvature fields and which capture the solutions to the above\index{curvature}
equations:
\begin{equation}\label{supercurvature}
\begin{aligned}
{}[\nabla_{\alpha\ald},\nabla_{\beta\bed}]\ =:\ \eps_{\ald\bed}
\CF_{\alpha\beta}~,~~~
[\nabla^i_{\ald\gad},\nabla_{\beta\bed}]\ =:\ \eps_{\ald\bed}
\CF^i_{\beta\gad}~,\\
[\nabla^i_{\ald\gad},\nabla^j_{\bed\ded}]\ =:\ \eps_{\ald\bed}
\CF^{ij}_{\gad\ded}~,\hspace{2.5cm}
\end{aligned}
\end{equation}
where $\CF_{\alpha\beta}=\CF_{(\alpha\beta)}$ and
$\CF^{ij}_{\gad\ded}=\CF^{(ij)}_{(\gad\ded)}+\CF^{[ij]}_{[\gad\ded]}$.
Note, however, that we introduced too many of these components.
Considering the third equation in \eqref{compconA}, one notes that
for $i=j$, the terms symmetric in $\gad,\ded$ vanish trivially.
This means that the components $\CF^{ii}_{(\gad\ded)}$ are in fact
superfluous and we can ignore them in the following discussion.
The second and third equations in \eqref{supercurvature} can be
contracted with $\eps^{\ald\gad}$ and $\eps^{\bed\ded}$,
respectively, which yields
\begin{equation}\label{cordef}
-2\nabla_{\beta\bed}
\CA^i_{[\dot{1}\dot{2}]}\ =\ \CF^i_{\beta\bed}\eand
-2\nabla_{\ald\gad}^i\CA^j_{[\dot{1}\dot{2}]}\ =\ \CF^{ij}_{\gad\ald}~.
\end{equation}
Furthermore, using Bianchi identities, one obtains immediately the
following equations:
\begin{equation}\label{eom}
\nabla^{\alpha\bed}\CF^i_{\alpha\gad}\ =\ 0\eand
\nabla_{\alpha\ald}\CF^{ij}_{\bed\gad}\ =\ \nabla^{i}_{\ald\bed}\CF_{\alpha\gad}^{j}~.
\end{equation}
Due to self-duality, the first equation is in fact equivalent to
$\nabla^{\alpha\bed}\nabla_{\alpha\bed}\CA^i_{[\dot{1}\dot{2}]}=0$,
as is easily seen by performing all the spinor index sums. From\index{Spinor}
the second equation, one obtains the field equation
$\nabla_{\alpha}{}^\bed\CF^{(12)}_{\bed\gad}=-
2[\CA^{(1}_{[\dot{1}\dot{2}]},\nabla_{\alpha\gad}\CA^{2)}_{[\dot{1}\dot{2}]}]$
after contracting with $\eps^{\ald\bed}$.

\paragraph{The superfield expansion.} To analyze the actual field
content of this theory, we choose {\em transverse gauge} as in
section \ref{sssSDYM}, \ref{pconstraintSDYM}, i.e.\ we demand
\begin{equation}
\eta_k^\ald\CA^k_\ald\ =\ 0~.
\end{equation}
Recall that this choice reduces the group of gauge transformations\index{gauge transformations}\index{gauge!transformation}
to ordinary, group-valued functions on the body of $\FC^{4|8}$. By\index{body}
using the identities $\eta_2^\bed \CA^1_{\ald\bed}=\CA^1_{\ald}$
etc., one sees that the above transverse gauge is equivalent to
the transverse gauge for the reduced structure sheaf:\index{sheaf}\index{structure sheaf}
\begin{equation}
e_i^{\ald\bed}\CA_{\ald\bed}^i\ =\ \eta^{(\ald}_1\eta^{\bed)}_2\CA^1_{\ald\bed}+
\eta^{(\ald}_3\eta^{\bed)}_4\CA^2_{\ald\bed}\ =\ 0~.
\end{equation}

In the expansion in the $e$s, the lowest components of
$\CF_{\alpha\beta}$, $\CA^i_{[\dot{1}\dot{2}]}$ and
$\CF^{(12)}_{(\ald\bed)}$ are the self-dual field strength\index{field strength}
$f_{\alpha\beta}$, two complex scalars $\phi^i$ and the auxiliary
field $G_{\ald\bed}$, respectively. The two scalars $\phi^i$ can
be seen as remainders of the six scalars contained in the $\CN=4$
SDYM multiplet, which will become even clearer in the real case.
The remaining components $\CA^i_{(\ald\bed)}$ vanish to zeroth
order in the $e$s due to the choice of transverse gauge. The field
$\CF^i_{\alpha\ald}$ does not contain any new physical degrees of
freedom, as seen from the first equation in \eqref{cordef}, but it
is a composite field. The same holds for $\CF^{[12]}_{[\gad\ald]}$
as easily seen by contracting the second equation in
\eqref{cordef} by $\eps^{\gad\ald}$:
$\CF^{[12]}_{[\dot{1}\dot{2}]}=-2[\CA^1_{[\dot{1}\dot{2}]},\CA^2_{[\dot{1}\dot{2}]}]$.

\paragraph{Equations of motion.} The superfield equations of motion
\eqref{eom} are in fact equivalent to the equations
\begin{equation}\label{eomr}
f_{\ald\bed}\ =\ 0~,~~~\square
\phi^i\ =\ 0~,~~~\eps^{\gad\ald}\nabla_{\alpha\ald}G_{\gad\ded}+
2[\phi^{(1},\nabla_{\alpha\ded}\phi^{2)}]\ =\ 0~.
\end{equation}
To lowest order in the $e$s, the equations obviously match. Higher
orders in the $e$s can be verified by defining the Euler operator\index{Euler operator}
(cf.\ section \ref{sssSDYM}, \ref{pconstraintSDYM})
$\CD:=e_i^{\ald\bed}\nabla_{(\ald\bed)}^i=e_i^{\ald\bed}\dpar_{(\ald\bed)}^i$
and applying $\CD$ on the superfields and equations of motion
which then turn out to be satisfied if the equations \eqref{eomr}
are fulfilled.

\subsection{The twistor space $\SB$}\index{twistor}\index{twistor!space}

\paragraph{Definition of $\SB$.} The discussion for $\SB$ follows
the same lines as for $\SA$ and is even simpler. Consider again
the supertwistor space $\CP^{3|4}=(\CP^3,\CO_{[4]})$. This time,\index{twistor}\index{twistor!space}
let us introduce the following differential operators:
\begin{equation}
\tilde{\CD}_\pm^{kl}\ :=\ \eta_k^\pm\der{\eta_l^\pm}~~~\mbox{for}~~k,l\ =\ 1,\ldots ,4~,
\end{equation}
which are maps $\CO_{[4]}\rightarrow\CO_{[4]}$. The space
$\CP^{3}$ together with the extended structure sheaf\footnote{The\index{sheaf}\index{structure sheaf}
same reduction can be obtained by imposing integral constraints
\cite{Lechtenfeld:2004cc}.}
\begin{equation}
\CO_{(1,1)}\ :=\ \bigcap_{k\neq l}\ker
\tilde{\CD}^{kl}_+\ =\ \bigcap_{k\neq l}\ker \tilde{\CD}^{kl}_-~,
\end{equation}
which is a reduction of $\CO_{[4]}$, is an order one thickening of\index{thickening}
$\CP^3$, which we denote by $\SB$. This manifold can be covered by\index{manifold}
two patches $\hat{\CU}_+$ and $\hat{\CU}_-$ on which we define the
coordinates
$(z^\alpha_\pm,\,\lambda_\pm,\,e^\pm:=\eta^\pm_1\eta^\pm_2\eta^\pm_3\eta^\pm_4)$.
The even nilpotent coordinate $e^\pm$ is a section of the line
bundle $\CO(4)$ with the identification $(e^\pm)^2\sim 0$.

\paragraph{Derivatives on $\SB$.} Similarly to the case $\SA$, we
have the following identities:
\begin{equation}
\begin{aligned}
&\eta^\pm_2\eta_3^\pm\eta_4^\pm\der{e^\pm}\ =\ \left.\der{\eta_1^\pm}\right|_{\CO_{(1,1)}}~,~~~
&\eta^\pm_1\eta_3^\pm\eta_4^\pm\der{e^\pm}\ =\ -\left.\der{\eta_2^\pm}\right|_{\CO_{(1,1)}}~,\\
&\eta^\pm_1\eta_2^\pm\eta_4^\pm\der{e^\pm}\ =\ \left.\der{\eta_3^\pm}\right|_{\CO_{(1,1)}}~,~~~
&\eta^\pm_1\eta_2^\pm\eta_3^\pm\der{e^\pm}\ =\ -\left.\der{\eta_4^\pm}\right|_{\CO_{(1,1)}}~
\end{aligned}
\end{equation}
which lead to the formal identifications
\begin{equation}
\der{e^\pm}\ =\ \der{\eta^\pm_4}\der{\eta^\pm_3}\der{\eta^\pm_2}\der{\eta^\pm_1}\eand
\dd
e^\pm\ =\ \dd{\eta^\pm_4}\dd{\eta^\pm_3}\dd{\eta^\pm_2}\dd{\eta^\pm_1}\ =\ \Omega^\eta_\pm~,
\end{equation}
but again with a restriction of the Leibniz rule in formal
manipulations of expressions written in the $\eta$-coordinates as
discussed in \ref{pderivativesSA}.

\paragraph{Moduli space and double fibration.}\index{double fibration}\index{fibration}\index{moduli space}
The holomorphic sections of the bundle $\SB\rightarrow \CPP^1$ are
defined by the equations
\begin{equation}\label{secsB}
z^\ald_\pm\ =\ x^{\alpha\ald}\lambda_\ald^\pm\eand
e^\pm\ =\ e^{(\ald\bed\gad\ded)}\lambda^\pm_\ald\lambda^\pm_\bed\lambda^\pm_\gad\lambda^\pm_\ded~.
\end{equation}
From the obvious identification
$e^{(\ald\bed\gad\ded)}=\eta_1^{(\ald}\eta_2^\bed\eta_3^\gad\eta_4^{\ded)}$
we see that a product
$e^{(\ald\bed\gad\ded)}e^{(\dot{\mu}\dot{\nu}\dot{\rho}\dot{\sigma})}$
will vanish, unless the number of indices equal to $\dot{1}$ is
the same as the number of indices equal to $\dot{2}$. In this
case, we have additionally the identity
\begin{equation}
\sum_{p}(-1)^{n_p} e^{p_1}e^{p_2}\ =\ 0~,
\end{equation}
where $p$ is a permutation of
$\dot{1}\dot{1}\dot{1}\dot{1}\dot{2}\dot{2}\dot{2}\dot{2}$, $p_1$
and $p_2$ are the first and second four indices of $p$,
respectively, and $n_p$ is the number of exchanges of a $\dot{1}$
and a $\dot{2}$ between $p_1$ and $p_2$, e.g.\
$n_{\dot{1}\dot{1}\dot{1}\dot{2}\dot{1}\dot{2}\dot{2}\dot{2}}=1$.

The more formal treatment is much simpler. We introduce the
differential operators
\begin{equation}
\begin{aligned}
&\tilde{\CD}^{klc}\ =\ \left(\eta^\ald_l\dpar_\ald^l-\eta^\ald_k\dpar^k_\ald\right)~~~\mbox{without
summation over $k$ and $l$}~,\\
&\tilde{\CD}^{kls}\ =\ \left(\dpar_{\dot{1}}^k\dpar_{\dot{2}}^l-\dpar_{\dot{2}}^k\dpar_{\dot{1}}^l\right)~,
\end{aligned}
\end{equation}
which map $\CO_{[8]}\rightarrow\CO_{[8]}$. Then the space
$\FC^{4}$ with the extended structure sheaf $\CO_{(1;2,5)}$\index{sheaf}\index{structure sheaf}
obtained by reducing $\CO_{[8]}$ to the overlap of kernels
\begin{equation}
\CO_{(1;2,5)}\ :=\ \bigcap_{k\neq l}
\left(\ker\tilde{\CD}^{klc}\cap\ker\tilde{\CD}^{kls}\right)
\end{equation}
is the moduli space described above. Thus, we have the following\index{moduli space}
reduction of the full double fibration \eqref{superdblfibration}\index{double fibration}\index{fibration}
for $\CN=4$:
\begin{equation}\label{dblfibrationB}
\begin{aligned}
\begin{picture}(50,45)
\put(0.0,0.0){\makebox(0,0)[c]{$(\CP^{3},\CO_{[4]})$}}
\put(64.0,0.0){\makebox(0,0)[c]{$(\FC^{4},\CO_{[8]})$}}
\put(34.0,39.0){\makebox(0,0)[c]{$(\FC^{4}\times\CPP^1,\CO_{[8]}\otimes\CO_{\CPP^1})$}}
\put(7.0,21.0){\makebox(0,0)[c]{$\pi_2$}}
\put(56.0,21.0){\makebox(0,0)[c]{$\pi_1$}}
\put(25.0,28.0){\vector(-1,-1){18}}
\put(37.0,28.0){\vector(1,-1){18}}
\end{picture}
\end{aligned}
\hspace{2.3cm}\ \longrightarrow\ \hspace{2.1cm}
\begin{aligned}
\begin{picture}(70,45)
\put(0.0,0.0){\makebox(0,0)[c]{$(\CP^{3},\CO_{(1,1)})$}}
\put(79.0,0.0){\makebox(0,0)[c]{$(\FC^{4},\CO_{(1;2,5)})$}}
\put(42.0,39.0){\makebox(0,0)[c]{$(\FC^{4}\times\CPP^1,\CO_{(1;2,5)}\otimes\CO_{\CPP^1})$}}
\put(7.0,21.0){\makebox(0,0)[c]{$\pi_2$}}
\put(56.0,21.0){\makebox(0,0)[c]{$\pi_1$}}
\put(25.0,28.0){\vector(-1,-1){18}}
\put(37.0,28.0){\vector(1,-1){18}}
\end{picture}
\end{aligned}
\end{equation}
where $\CO_{\CPP^1}$ is again the structure sheaf of the Riemann\index{sheaf}\index{structure sheaf}
sphere $\CPP^1$. The tangent spaces along the leaves of the
projection $\pi_2$ are spanned by the vector fields
\begin{equation}
\begin{aligned}
&\bV_\alpha^\pm\ =\ \lambda_\pm^\ald\dpar_{\alpha\ald}~,~~~
&&\bV_\alpha^\pm\ =\ \lambda_\pm^\ald\dpar_{\alpha\ald}~,\\
&\bV^k_\pm\ =\ \lambda^\ald_\pm\der{\eta_k^\ald}~,~~~
&&\bV^\pm_{\bed\gad\ded}\ =\ \lambda^{\ald}_\pm\dpar_{(\ald\bed\gad\ded)}
\end{aligned}
\end{equation}
in the left and right double fibration in \eqref{dblfibrationB},\index{double fibration}\index{fibration}
where $k=1,\ldots ,4$. The further identities
\begin{equation}
\begin{aligned}
&\eta_2^\bed\eta_3^\gad\eta_4^\ded\der{e^{(\ald\bed\gad\ded)}}\ =\ \left.\der{\eta_1^\ald}\right|_{\CO_{(1;2,5)}}~,~~~
&\eta_1^\bed\eta_3^\gad\eta_4^\ded\der{e^{(\ald\bed\gad\ded)}}\ =\ -\left.\der{\eta_2^\ald}\right|_{\CO_{(1;2,5)}}~,\\
&\eta_1^\bed\eta_2^\gad\eta_4^\ded\der{e^{(\ald\bed\gad\ded)}}\ =\ \left.\der{\eta_3^\ald}\right|_{\CO_{(1;2,5)}}~,~~~
&\eta_1^\bed\eta_2^\gad\eta_3^\ded\der{e^{(\ald\bed\gad\ded)}}\ =\ -\left.\der{\eta_4^\ald}\right|_{\CO_{(1;2,5)}}~
\end{aligned}
\end{equation}
are easily derived and from them it follows that e.g.\
\begin{equation}
\left.\bV^1_\pm\right|_{\CO_{(1;2,5)}}\ =\ \eta_2^\bed\eta_3^\gad\eta_4^\ded
\bV^\pm_{\bed\gad\ded}\eand
\left.\bV^2_\pm\right|_{\CO_{(1;2,5)}}\ =\ -\eta_1^\bed\eta_3^\gad\eta_4^\ded
\bV^\pm_{\bed\gad\ded}~.
\end{equation}

\paragraph{hCS theory on $\SB$ and linear system.} The topological\index{linear system}
B-model on $\SB$ is equivalent to hCS theory on $\SB$ and
introducing a trivial rank $n$ complex vector bundle $\CE$ over\index{complex!vector bundle}
$\SB$ with a connection $\CAh$, the action reads\index{connection}
\begin{equation}
S=\int_{\CSB} \Omega^{3\oplus
1|0}\wedge\tr\left(\CAh^{0,1}\wedge\bar{\dpar}\CAh^{0,1}+
\tfrac{2}{3}\CAh^{0,1}\wedge\CAh^{0,1}\wedge\CAh^{0,1}\right)~,
\end{equation}
with $\CSB$ being the chiral subspace for which $\bar{e}^\pm=0$
and $\CAh^{0,1}$ the $(0,1)$-part of $\CAh$. The holomorphic
volume form $\Omega^{3\oplus1|0}$ can be defined, e.g.\ on\index{holomorphic!volume form}
$\hat{\CU}_+$, as $\Omega^{3\oplus1|0}_+=\dd z^1_+\wedge\dd
z^2_+\wedge\dd \lambda_+\dd e^+$. Following exactly the same steps
as in the case $\SA$, we again obtain the equations
\begin{equation}\label{defB}
\begin{aligned}
\psi_+
\bV^+_\alpha\psi_+^{-1}\ =\ \psi_-\bV^+_\alpha\psi_-^{-1}\ =:\ \lambda^\ald_+
\CA_{\alpha\ald}&\ =:\ \CA^+_\alpha~,\\ \psi_+ \bV^{k}_+
\psi_+^{-1}\ =\ \psi_-\bV^{k}_+\psi_-^{-1}\ =:\ \lambda^\ald_+
\CA^k_{\ald}&\ =:\ \CA^{k}_{+}~,\\ \psi_+ \dpar_{\bl_+}
\psi_+^{-1}\ =\ \psi_-\dpar_{\bl_+}\psi_-^{-1}&\ =:\ \CA_{\bl_+}\ =\ 0~,\\
\psi_+
\dpar_{\bar{x}^{\alpha\ald}}\psi_+^{-1}\ =\ \psi_-\dpar_{\bar{x}^{\alpha\ald}}\psi_-^{-1}
&\ =\ 0~.
\end{aligned}
\end{equation}
and by considering the reduced structure sheaves, we can rewrite
the second line this time as
\begin{equation}\label{defB2}
\begin{aligned}
\eta_2^\bed\eta_3^\gad\eta_4^\ded\,\psi_+
\bV^+_{\bed\gad\ded}\psi_+^{-1}\ =\ 
\eta_2^\bed\eta_3^\gad\eta_4^\ded\,\psi_-\bV^+_{\bed\gad\ded}\psi_-^{-1}
&=:\eta_2^\bed\eta_3^\gad\eta_4^\ded\,\lambda^\ald_+
\CA_{\ald\bed\gad\ded}~,\\
&=:\eta_2^\bed\eta_3^\gad\eta_4^\ded\,\CA^+_{\bed\gad\ded}
\end{aligned}
\end{equation}
for $k=1$ which yields
$\eta_2^\bed\eta_3^\gad\eta_4^\ded\CA_{\ald\bed\gad\ded}=\CA^1_{\ald}$.
Similar formul\ae{} are obtained for the other values of $k$, with
which one can determine the superfield expansion of
$\CA_{\ald\bed\gad\ded}$ again from the superfield expansion of
$\CA^k_\ald$ by dropping the terms which are not in the kernel of
the differential operators $\tilde{\CD}^{klc}$ and
$\tilde{\CD}^{kls}$ for $k\neq l$.

\paragraph{Compatibility conditions.}\index{compatibility conditions}
Analogously to the case $\SA$, one can rewrite the linear system\index{linear system}
behind \eqref{defB}, \eqref{defB2} for the reduced structure
sheaf. For this, we define the covariant derivative\index{covariant derivative}\index{sheaf}\index{structure sheaf}
$\nabla_{\alpha\ald}:=\dpar_{\alpha\ald}+[\CA_{\alpha\ald},\cdot]$
and the first order differential operator
$\nabla_{\ald\bed\gad\ded}:=\dpar_{\ald\bed\gad\ded}+[\CA_{\ald\bed\gad\ded},\cdot]$.
Then we have
\begin{equation}\label{compconB}
\begin{aligned}
{}[\nabla_{\alpha\ald},\nabla_{\beta\bed}]+[\nabla_{\alpha\bed},
\nabla_{\beta\ald}]&\ =\ 0~,\\
\eta^{\dot{\nu}}_k\eta^{\dot{\rho}}_m\eta^{\dot{\sigma}}_n\left(
[\nabla_{\dot{\mu}\dot{\nu}\dot{\rho}\dot{\sigma}},\nabla_{\alpha\ald}]+
[\nabla_{\ald\dot{\nu}\dot{\rho}\dot{\sigma}},\nabla_{\alpha\dot{\mu}}]\right)&\ =\ 0~,
\\
\eta^\bed_r\eta^\gad_s\eta^\ded_t\eta^{\dot{\nu}}_k\eta^{\dot{\rho}}_m\eta^{\dot{\sigma}}_n
\left([\nabla_{\ald\bed\gad\ded},\nabla_{\dot{\mu}\dot{\nu}\dot{\rho}\dot{\sigma}}]+
[\nabla_{\dot{\mu}\bed\gad\ded},\nabla_{\ald\dot{\nu}\dot{\rho}\dot{\sigma}}]\right)
&\ =\ 0~,
\end{aligned}
\end{equation}
where $(rst)$ and $(kmn)$ are each a triple of pairwise different
integers between 1 and 4. Again, in these equations the
push-forward $\pi_{1*}\CA$ is already implied and solutions to
\eqref{compconB} are captured by the following extensions of the
supercurvature fields:\index{curvature}
\begin{equation}\label{supercurvatureB}
\begin{aligned}
{}[\nabla_{\alpha\ald},\nabla_{\beta\bed}]&\ =:\ \eps_{\ald\bed}
\CF_{\alpha\beta}~,\\\nonumber
[\nabla_{\dot{\mu}\dot{\nu}\dot{\rho}\dot{\sigma}},\nabla_{\alpha\ald}]&\ =:\ \eps_{\ald\dot{\mu}}
\CF_{\alpha\dot{\nu}\dot{\rho}\dot{\sigma}}~,\\
[\nabla_{\ald\bed\gad\ded},\nabla_{\dot{\mu}\dot{\nu}\dot{\rho}\dot{\sigma}}]&\ =:\ \eps_{\ald\dot{\mu}}
\CF_{\bed\gad\ded\dot{\nu}\dot{\rho}\dot{\sigma}}~,
\end{aligned}
\end{equation}
where $\CF_{\alpha\beta}=\CF_{(\alpha\beta)}$,
$\CF_{\alpha\dot{\nu}\dot{\rho}\dot{\sigma}}=\CF_{\alpha(\dot{\nu}\dot{\rho}\dot{\sigma})}$
and
$\CF_{\bed\gad\ded\dot{\nu}\dot{\rho}\dot{\sigma}}=\CF_{(\bed\gad\ded)(\dot{\nu}\dot{\rho}\dot{\sigma})}$
is symmetric under exchange of
$(\bed\gad\ded)\leftrightarrow(\dot{\nu}\dot{\rho}\dot{\sigma})$.
Consider now the third equation of \eqref{compconB}. Note that the
triples $(rst)$ and $(kmn)$ will have two numbers in common, while
exactly one is different. Without loss of generality, let $r\neq
k$, $s=m$ and $t=n$. Then one easily sees that the terms symmetric
in $\bed$, $\dot{\nu}$ vanish trivially. This means that the field
components $\CF_{\bed\gad\ded\dot{\nu}\dot{\rho}\dot{\sigma}}$
which are symmetric in $\bed$, $\dot{\nu}$ are again unconstrained
additional fields, which do not represent any of the fields in the
$\CN=4$ SDYM multiplet and we put them to zero, analogously to
$\CF^{ii}_{(\gad\ded)}$ in the case $\SA$.

\paragraph{Derivation of the superfield expansion.}
The second equation in \eqref{supercurvatureB} can be contracted
with $\eps^{\dot{\mu}\dot{\nu}}$ which yields
$2\nabla_{\alpha\ald}\CA_{[\dot{1}\dot{2}]\dot{\rho}\dot{\sigma}}=\CF_{\alpha\ald\dot{\rho}\dot{\sigma}}$
and further contracting this equation with $\eps^{\ald\dot{\rho}}$
we have
$\nabla_{\alpha}{}^\ald\CA_{[\dot{1}\dot{2}]\ald\dot{\sigma}}=0$.
After contracting the third equation with
$\eps^{\dot{\mu}\dot{\nu}}$, one obtains
\begin{equation}\label{Bkey}
-2\nabla_{\ald\bed\gad\ded}\CA_{[\dot{1}\dot{2}]\dot{\rho}\dot{\sigma}}\ =\ 
\CF_{\bed\gad\ded\ald\dot{\rho}\dot{\sigma}}~.
\end{equation}

The transversal gauge condition $\eta^\ald_k\CA^k_\ald=0$ is on
$\CO_{(1;2,5)}$ equivalent to the condition
\begin{equation}\label{gauge2}
e^{(\ald\bed\gad\ded)}\CA_{\ald\bed\gad\ded}\ =\ 0~,
\end{equation}
as expected analogously to the case $\SA$. To lowest order in
$e^{(\ald\bed\gad\ded)}$, $\CF_{\alpha\beta}$ can be identified
with the self-dual field strength $f_{\alpha\beta}$ and\index{field strength}
$\CA_{[\dot{1}\dot{2}]\ald\bed}$ with the auxiliary field
$G_{\ald\bed}$. The remaining components of
$\CF_{\bed\gad\ded\ald\dot{\rho}\dot{\sigma}}$, i.e.\ those
antisymmetric in $[\ald\bed]$, are composite fields and do not
contain any additional degrees of freedom which is easily seen by
considering equation \eqref{Bkey}.

\paragraph{Equations of motion.} Applying the Euler operator in\index{Euler operator}
transverse gauge
$\CD:=e^{(\ald\bed\gad\ded)}\nabla_{(\ald\bed\gad\ded)}=
e^{(\ald\bed\gad\ded)}\dpar_{(\ald\bed\gad\ded)}$, one can show
that the lowest order field equations are equivalent to the full
superfield equations of motion. Thus, \eqref{compconB} is
equivalent to
\begin{equation}
f_{\ald\bed}\ =\ 0\eand\nabla^{\alpha\ald}G_{\ald\bed}\ =\ 0~.
\end{equation}
Altogether, we found the compatibility condition for a linear
system encoding purely bosonic SDYM theory {\em including} the\index{linear system}
auxiliary field $G_{\ald\bed}$.

\subsection{Fattened real manifolds}\index{manifold}

The field content of hCS theory on $\SA$ and $\SB$ becomes even
more transparent after imposing a reality condition on these
spaces. One can directly derive appropriate real structures from\index{real structure}
the one on $\CP^{3|4}$, having in mind the picture of combining
the Gra{\ss}mann coordinates of $\CP^{3|4}$ to the even nilpotent
coordinates of $\SA$ and $\SB$. The real structure on $\CP^{3|4}$\index{real structure}
is discussed in detail in sections \ref{ssTwistorSpace} and
\ref{ssSuperextensionTwistorSpace}.

\paragraph{Real structures.} Recall the action of the\index{real structure}
two antilinear involutions $\tau_\eps$ with $\eps=\pm 1$ on the\index{antilinear involution}\index{involution}
coordinates $(z^1_\pm,z^2_\pm,z^3_\pm)$:
\begin{equation}\nonumber
\tau_\eps(z_+^1,z_+^2,z^3_+)\ =\ \left(\frac{\bz_+^2}{\bz^3_+},
\frac{\eps\bz_+^1}{\bz^3_+},\frac{\eps}{\bz^3_+}\right)\eand
\tau_\eps(z_-^1,z_-^2,z^3_-)\ =\ \left(\frac{\eps\bz_-^2}{\bz^3_-}
,\frac{\bz_-^1}{\bz^3_-},\frac{\eps}{\bz^3_-}\right)~.
\end{equation}
On $\SA$, we have additionally
\begin{equation}
\tau_\eps(e^1_+,e^2_+)\ =\ \left(\frac{\bar{e}^1_+}{(\bz^3_+)^2},\frac{\bar{e}^2_+}{(\bz^3_+)^2}\right)\eand
\tau_\eps(e^1_-,e^2_-)\ =\ \left(\frac{\bar{e}^1_-}{(\bz^3_-)^2},\frac{\bar{e}^2_-}{(\bz^3_-)^2}\right)~,
\end{equation}
and on $\SB$, it is
\begin{equation}
\tau_\eps(e_+)\ =\ \frac{\bar{e}_+}{(\bz^3_+)^4}\eand
\tau_\eps(e_-)\ =\ \frac{\bar{e}_-}{(\bz^3_-)^4}~.
\end{equation}
Recall that in the formulation of the twistor correspondence, the\index{twistor}\index{twistor!correspondence}
coordinates $z^3_\pm$ are usually kept complex for convenience
sake. We do the same while on all other coordinates, we impose the
condition $\tau_{\eps}(\cdot)=\cdot$. On the body of the moduli\index{body}
space, this will lead to a Euclidean metric $(+,+,+,+)$ for\index{metric}
$\eps=-1$ and a Kleinian metric $(+,+,-,-)$ for $\eps=+1$.

\paragraph{Real superfield expansion.} Recall that together with the
identification \eqref{VectorIdentity}
\begin{equation}
\der{\bar{z}^1_+}\ =\ \gamma_+ \bV_2^+\eand\der{\bar{z}^2_+}\ =\ \eps
\gamma_+ \bV_1^+~,
\end{equation}
we can rewrite the hCS equations of motion, e.g.\ on
$\hat{\CU}_+$, as
\begin{equation}\label{shCSb}
\begin{aligned}
\bV_\alpha^+\CA_\beta^+- \bV_\beta^+\CA_\alpha^++
[\CA_\alpha^+,\CA_\beta^+]&\ =\ &0~,\\
\dpar_{\bl_+}\CA_\alpha^+- \bV_\alpha^+\CA_{\bl_+}+
[\CA_{\bl_+},\CA_\alpha^+]&\ =\ &0~,
\end{aligned}
\end{equation}
where the components of the gauge potential are defined via the
contractions $\CA^\pm_\alpha:=\bV_\alpha^\pm \lrcorner\,
\CAh^{0,1}$, $\CA_{\bl_\pm}:=\dpar_{\bl_\pm} \lrcorner\,
\CAh^{0,1}$, and we assumed a gauge for which
$\CA^\pm_{i}:=\dpar_{\bar{e}_i^\pm} \lrcorner\, \CAh^{0,1}=0$, see
also \eqref{shCS}. On the space $\SA$ together with the field
expansion
\begin{subequations}\label{expAA}
\begin{eqnarray}\label{expAaA}
\CA_\alpha^+&=&\lambda_+^\ald\, A_{\alpha\ald}+
\gamma_+\,e^+_i\,\hat{\lambda}^\ald_+\, \phi_{\alpha \ald}^{i}+
\gamma_+^3\,e^+_1e^+_2\,\hat{\lambda}_+^\ald\hat{\lambda}_+^\bed\hat{\lambda}_+^{\dot{\gamma}}\,
G_{\alpha\ald\bed\dot{\gamma}}~,\\
\label{expAlA}
\CA_{\bl_+}&=&\gamma_+^2\,e^+_i\,\phi^i-2\eps\,\gamma_+^4\,e^+_1e^+_2\,
\hat{\lambda}_+^\ald\hat{\lambda}_+^\bed\, G_{\ald\bed}~,
\end{eqnarray}
\end{subequations}
the system of equations \eqref{shCSb} is equivalent to
\eqref{compconA}. Note that similarly to the expansion
\eqref{expA}, the expansion \eqref{expAA} is determined by the
geometry of $\SA$. Furthermore, one can identify
$\phi^i_{\alpha\ald}=-\frac{1}{2}\CF^i_{\alpha\ald}$ and
$G_{\alpha\ald\bed\gad}=\frac{1}{6}\nabla^{(1}_{\ald(\bed}\CF^{2)}_{\alpha\gad)}$.
On $\SB$, we can use
\begin{subequations}
\begin{eqnarray}\label{expAaB}
\CA_\alpha^+&=&\lambda_+^\ald\, A_{\alpha\ald}+
\gamma_+^3\,e^+\hat{\lambda}_+^\ald\hat{\lambda}_+^\bed\hat{\lambda}_+^{\dot{\gamma}}\,
G_{\alpha\ald\bed\dot{\gamma}}~,\\
\label{expAlB} \CA_{\bl_+}&=&\gamma_+^4\,e^+\,
\hat{\lambda}_+^\ald\hat{\lambda}_+^\bed\, G_{\ald\bed}~.
\end{eqnarray}
\end{subequations}
to have \eqref{shCSb} equivalent with \eqref{compconB} and
$G_{\alpha\ald\bed\gad}=\frac{1}{6}\CF_{\alpha\ald\bed\gad}$.

For compactness of the discussion, we refrain from explicitly
writing down all the reality conditions imposed on the component
fields and refer to section \ref{ssPenroseWardTrafo} for further
details.

\paragraph{Actions.} One can reconstruct two action functionals,
from which the equations of motion for the two cases arise. With
our field normalizations, they read
\begin{align}
S_{\SA}\ =\ &\int\dd^4 x~ \tr\left(G^{\ald\bed}f_{\ald\bed}-\phi^{(1}
\square
\phi^{2)}\right)~,\\
S_{\SB}\ =\ &\int\dd^4 x~ \tr\left(G^{\ald\bed}f_{\ald\bed}\right)~.
\end{align}
The action $S_{\SB}$ has first been proposed in
\cite{Chalmers:1996rq}.

\section{Penrose-Ward transform for mini-supertwistor\index{Penrose-Ward transform}\index{twistor}
spaces}\label{sPWMini}

It is well-known that the Bogomolny monopole equations are
obtained from the four-dimensional self-dual Yang-Mills equations
by the dimensional reduction $\FR^4\rightarrow \FR^3$ and that\index{dimensional reduction}
there is a twistor space, the so-called {\em mini-twistor space}\index{mini-twistor space}\index{twistor}\index{twistor!space}
\cite{Hitchin:1982gh} $\CP^2:=\CO(2)\rightarrow \CPP^1$, upon
which a Penrose-Ward transform for the dimensionally reduced\index{Penrose-Ward transform}
situation can be constructed. In this section, we will discuss the
corresponding superextension, the mini-supertwistor space\index{mini-supertwistor space}\index{twistor}\index{twistor!space}
\cite{Chiou:2005jn,Popov:2005uv}, which will lead to a
Penrose-Ward transform between certain holomorphic vector bundles\index{Penrose-Ward transform}\index{holomorphic!vector bundle}
and the supersymmetric Bogomolny monopole equations.

\subsection{The mini-supertwistor spaces}\index{mini-supertwistor space}\index{twistor}\index{twistor!space}

In the following, we will constrain the discussion for convenience
to the real case $\eps=-1$ with Euclidean signature. The Kleinian
signature will require some adjustments, similarly to the ones in\index{Kleinian signature}
the case of Kleinian twistor spaces discussed in section\index{twistor}\index{twistor!space}
\ref{ssTwistorSpace}.

\paragraph{Definition by dimensional reduction.} We start from the\index{dimensional reduction}
supertwistor space $\CP^{3|\CN}_{-1}$ with coordinates as defined\index{twistor}\index{twistor!space}
in section \ref{ssSuperextensionTwistorSpace}. Let $\CCG$ be the
Abelian group generated by the action of the vector field
$\CCT_2=\der{x^2}$. This group is the real part of the holomorphic
action of the complex group $\CCG_\FC\cong \FC$. In other words,
we have
\begin{equation}\label{eq:T4}
\begin{aligned}
\CCT_2&\ =\ \der{x^2}\ =\
\derr{z_+^a}{x^2}\der{z_+^a}+\derr{\bz_+^a}{x^1}\der{\bz_+^a}\\& \
=\
\left(-\der{z^2_+}+z_+^3\der{z^1_+}\right)+\left(-\der{\bz^2_+}+
\bz_+^3\der{\bz^1_+}\right) \ =:\ \CCT_+'+\bar{\CCT}_+'
\end{aligned}
\end{equation}
in the coordinates $(z_+^a,\eta_i^+)$ on $\hat{\CU}_+$, where
\begin{equation}\label{eq:T4compl}
\CCT_+'\ :=\ \CCT'|_{\hat{\CU}_+}\ =\
-\der{z^2_+}+z_+^3\der{z_+^1}
\end{equation}
is the holomorphic part of the vector field $\CCT_2$ on
$\hat{\CU}_+$. Similarly, we obtain
\begin{equation}
\CCT_2\ =\ \CCT'_-+\bar{\CCT}'_-~~~\mbox{with}~~~\CCT'_-\ :=\
\CCT'|_{\hat{\CU}_-}\ =\ -z^3_-\der{z_-^2}+\der{z^1_-}
\end{equation}
on $\hat{\CU}_-$ and $\CCT_+'=\CCT_-'$ on
$\hat{\CU}_+\cap\hat{\CU}_-$. Holomorphic functions $f$ on
$\CP^{3|4}_{-1}$ thus satisfy
\begin{equation}
\CCT_2f(z^a_\pm,\eta_i^\pm)\ =\ \CCT' f(z_\pm^a,\eta_i^\pm)
\end{equation}
and therefore $\CCT'$-invariant holomorphic functions on
$\CP^{3|4}_{-1}$ can be considered as ``free'' holomorphic
functions on a reduced space $\CP^{2|\CN}_{-1}\cong
\CP^{3|\CN}_{-1}/\CCG_\FC$ obtained as the quotient space of
$\CP^{3|4}_{-1}$ by the action of the complex Abelian group
$\CCG_\FC$ generated by $\CCT'$. For convenience, we will omit the
subscript $-1$ on twistor spaces in the remainder of this section.\index{twistor}\index{twistor!space}

\paragraph{Reduction diagram.} Let us summarize the effect of this
dimensional reductions on all the spaces involved in the double\index{dimensional reduction}
fibration by the following diagram:\index{fibration}
\begin{equation}\label{mrsuperdblfibration2}
\begin{aligned}
\begin{picture}(150,100)
\put(0.0,0.0){\makebox(0,0)[c]{$\CP^{2|\CN}$}}
\put(140.0,0.0){\makebox(0,0)[c]{$\FR^{3|2\CN}$}}
\put(0,90.0){\makebox(0,0)[c]{$\CP^{3|\CN}$}}
\put(30.0,88.0){\makebox(0,0)[c]{$\cong$}}
\put(73.0,90.0){\makebox(0,0)[c]{$\FR^{4|2\CN}\times S^2$}}
\put(102.0,90.0){\vector(1,0){20}}
\put(140,90.0){\makebox(0,0)[c]{$\FR^{4|2\CN}$}}
\put(73.0,50.0){\makebox(0,0)[c]{$\FR^{3|2\CN}\times S^2$}}
\put(26.0,27.0){\makebox(0,0)[c]{$\nu_2$}}
\put(110.0,27.0){\makebox(0,0)[c]{$\nu_1$}}
\put(59.0,37.0){\vector(-4,-3){40}}
\put(82.0,37.0){\vector(4,-3){40}}
\put(0.0,80.0){\vector(0,-1){70}}
\put(72.0,80.0){\vector(0,-1){20}}
\put(140.0,80.0){\vector(0,-1){70}}
\end{picture}
\end{aligned}
\end{equation}
Here, $\downarrow$ symbolizes projections generated by the action
of the groups $\CCG$ or $\CCG_\FC$ and $\nu_1$ is the canonical
projection. The projection $\nu_2$ will be described in the next
paragraphs.

\paragraph{Local coordinates.} The fibre coordinates on
$\CP^{2|\CN}$ are
\begin{equation}\label{eq:3.10}
\begin{aligned}
&w_+^1\ :=\ -\di(z_+^1+z_+^3z_+^2)~,~~~w_+^2\ :=\
z_+^3\eand\eta_i^+~~~\mbox{on}~~~\hat{\CU}_+~,\\&w_-^1\ :=\
-\di(z_-^2+z_-^3z_-^1)~,~~~w_-^2\ :=\
z_-^3\eand\eta_i^-~~~\mbox{on}~~~\hat{\CU}_-~,
\end{aligned}
\end{equation}
since $w_\pm^1$ is constant along the $\CCG_\FC$-orbits in
$\CP^{3|\CN}$ and thus descend to the patches
$\hat{\CV}_\pm:=\hat{\CU}_\pm\cap\CP^{2|\CN}$ covering the
mini-supertwistor space. On the overlap\index{mini-supertwistor space}\index{twistor}\index{twistor!space}
$\hat{\CV}_+\cap\hat{\CV}_-$, we have
\begin{equation}
w_+^1\ =\ \frac{1}{(w_-^2)^2}w_-^1~,~~~w_+^2\ =\
\frac{1}{w_-^2}\eand \eta_i^+\ =\ \frac{1}{w_-^2}\eta_i^-
\end{equation}
and thus the mini-supertwistor space coincides with the total\index{mini-supertwistor space}\index{twistor}\index{twistor!space}
space of a holomorphic vector bundle with typical fibre of\index{holomorphic!vector bundle}
dimension $1|\CN$:
\begin{equation}
\CO(2)\oplus\Pi\CO(1)\otimes\FC^\CN\ =\ \CP^{2|\CN}
\end{equation}
over the Riemann sphere $\CPP^1$. In the case $\CN=4$, this space\index{Riemann sphere}
is a Calabi-Yau supermanifold \cite{Chiou:2005jn} with a\index{Calabi-Yau}\index{Calabi-Yau supermanifold}\index{super!manifold}
holomorphic volume form\index{holomorphic!volume form}
\begin{equation}\label{eq:3.14}
\Omega|_{\hat{\CV}_\pm}\ :=\ \pm\dd w_\pm^1\wedge \dd w_\pm^2
\dd\eta_1^\pm\cdots\dd\eta_4^\pm~.
\end{equation}

As already mentioned, the body of the mini-supertwistor space\index{body}\index{mini-supertwistor space}\index{twistor}\index{twistor!space}
$\CP^{2|\CN}$ is the mini-twistor space \cite{Hitchin:1982gh}\index{mini-twistor space}
\begin{equation}
\CP^2\ \cong\ \CO(2)\ \cong\ T^{1,0}\CPP^1~,
\end{equation}
where $T^{1,0}\CPP^1$ denotes the holomorphic tangent bundle of
the Riemann sphere $\CPP^1$. Moreover, the space $\CP^{2|4}$ can\index{Riemann sphere}
be considered as an open subset of the weighted projective space
$W\CPP^{2|4}(2,1,1|1,1,1,1)$.

\paragraph{Real structure.} Clearly, a real structure $\tau_{-1}$ on\index{real structure}
$\CP^{2|\CN}$ is induced from the one on $\CP^{3|\CN}$. On the
local coordinates, $\tau_{-1}$ acts according to
\begin{equation}\label{eq:3.16}
\tau_{-1}(w_\pm^1,w_\pm^2,\eta_i^\pm)\ =\
\left(-\frac{\bar{w}_\pm^1}{(\bar{w}_\pm^2)^2},
-\frac{1}{\bar{w}_\pm^2},\pm\frac{1}{\bar{w}^2_\pm}
T_i{}^j\etab_j^\pm\right)~,
\end{equation}
where the matrix $T=(T_i{}^j)$ has already been defined in section
\ref{ssSpinors}, \ref{peuclideancase}.

\paragraph{Incidence relations.}  From \eqref{eq:3.16}, one sees\index{incidence relation}
that similarly to the case $\CP^{3|\CN}$, $\tau_{-1}$ has no fixed
points in $\CP^{2|\CN}$ but leaves invariant projective lines
$\CPP^1_{x,\eta}\embd\CP^{2|\CN}$ defined by the equations
\begin{equation}\label{eq:3.17}
\begin{aligned}
w_+^1&\ =\ y-2\lambda_+x^1-\lambda_+^2\bar{y}~,&\eta_i^+&\ =\
\eta_i^\ed+\lambda_+\eta_i^\zd&&\mbox{with}~~\lambda_+\ =\ w_+^2\
\in\
U_+~,\\
w_-^1&\ =\ \lambda_-^2y-2\lambda_-x^1-\bar{y}~,&\eta_i^-&\ =\
\lambda_-\eta_i^\ed+\eta_i^\zd&&\mbox{with}~~\lambda_-\ =\ w_-^2\
\in\ U_-~
\end{aligned}
\end{equation}
for fixed $(x,\eta)\in\FR^{3|2\CN}$. Here, $y=-(x^3+\di x^4)$,
$\bar{y}=-(x^3-\di x^4)$ and $x^1$ are coordinates on $\FR^3$.

We can use the coordinates $y^{\ald\bed}$ which were introduced in
\eqref{dimredcoordinates} with
\begin{equation}
y^{\ed\ed}\ =\-\bar{y}^{\zd\zd}\ =\ y\eand y^{\ed\zd}\ =\
\bar{y}^{\ed\zd}\ =\ -x^1~
\end{equation}
to rewrite \eqref{eq:3.17} concisely as
\begin{equation}\label{eq:3.18}
w_\pm^1\ =\
y^{\ald\bed}\lambda_\ald^\pm\lambda_\bed^\pm~,~~~w_\pm^2\ =\
\lambda_\pm\eand \eta_i^\pm\ =\ \eta_i^\ald\lambda_\ald^\pm~.
\end{equation}
In fact, the equations \eqref{eq:3.18} are the appropriate
incidence relations for the further discussion. They naturally\index{incidence relation}
imply the double fibration\index{double fibration}\index{fibration}
\begin{equation}\label{eq:3.19}
\begin{aligned}
\begin{picture}(50,40)
\put(0.0,0.0){\makebox(0,0)[c]{$\CP^{2|\CN}$}}
\put(64.0,0.0){\makebox(0,0)[c]{$\FR^{3|2\CN}$}}
\put(34.0,33.0){\makebox(0,0)[c]{$\CK^{5|2\CN}$}}
\put(7.0,18.0){\makebox(0,0)[c]{$\nu_2$}}
\put(55.0,18.0){\makebox(0,0)[c]{$\nu_1$}}
\put(25.0,25.0){\vector(-1,-1){18}}
\put(37.0,25.0){\vector(1,-1){18}}
\end{picture}
\end{aligned}
\end{equation}
where $\CK^{5|2\CN}\cong \FR^{3|2\CN}\times S^2$, $\nu_1$ is again
the canonical projection onto $\FR^{3|2\CN}$ and the projection
$\nu_2$ is defined by the formula
\begin{equation}\label{eq:3.20}
\nu_2(x^a,\lambda_\pm,\eta_i^\ald)\ =\
\nu_2(y^{\ald\bed},\lambda_\ald^\pm,\eta_i^\ald)\ =\
(w^1_\pm,w_\pm^2,\eta_i^\pm)~,
\end{equation}
where $a=1,2,3$. We thus have again the one-to-one correspondences
\begin{equation*}
\begin{aligned}
\{\,\mbox{$\tau_{-1}$-invariant projective lines $\CPP^1_{x,\eta}$
in $\CP^{2|\CN}$}\}\ \longleftrightarrow\ \{\,\mbox{points
$(x,\eta)$ in
$\FR^{3|2\CN}$}\}~,\\
\{\,\mbox{points $p$ in $\CP^{2|\CN}$}\}\ \longleftrightarrow\
\{\,\mbox{oriented $(1|0)$-dimensional lines $\ell_p$ in
$\FR^{3|8}$}\}~.~~~
\end{aligned}
\end{equation*}

\paragraph{Cauchy-Riemann structure on $\CK^{5|2\CN}$.} Although the\index{Cauchy-Riemann structure}
correspondence space $\CK^{5|2\CN}$ cannot be interpreted as a
complex manifold due to its dimensionality, one can consider it as\index{complex!manifold}\index{manifold}
a Cauchy-Riemann (CR) manifold, see \ref{ssCRstructures},
\ref{pCRmanifolds}. There are now several possible CR structures
on the body $\FR^3\times S^2$ of $\CK^{5|2\CN}$: One of them,\index{body}
which we denote by $\CCDb_0$, is generated by the vector fields
$\{\dpar_\by,\dpar_{\bl_\pm}\}$ and corresponds to the
identification $\CK_0^5:=(\FR^3\times S^2,\CCDb_0)\cong
\FR\times\FC\times \CPP^1$. Another one, denoted by $\CCDb$, is
spanned by the basis sections
$\{\dpar_{\bw_\pm^1},\dpar_{\bw^2_\pm}\}$ of the bundle
$T^c(\FR^3\times S^2)$. Note that $\CCDb$ is indeed a CR structure
as $\CCD\cap\CCDb=\{0\}$ and the distribution $\CCDb$ is
integrable: $[\dpar_{\bw^1_\pm},\dpar_{\bw^2_\pm}]=0$. Therefore,\index{integrable}
the pair $(\FR^3\times S^2,\CCDb)=:\CK^5$ is also a CR manifold.\index{manifold}
It is obvious that there is a diffeomorphism between the manifolds
$\CK^5$ and $\CK_0^5$, but this is not a CR diffeomorphism since
it does not respect the chosen CR structures. Note that a CR
five-manifold generalizing the above manifold $\CK^5$ can be\index{manifold}
constructed as a sphere bundle over an arbitrary three-manifold
with conformal metric \cite{LeBrun:1984}. Following\index{metric}
\cite{LeBrun:1984}, we shall call $\CK^5$ the {\em CR twistor\index{twistor}
space}.

Since the definition of CR structures naturally carries over to
the case of supermanifolds (see e.g.\ \cite{Howe:1995md}), we can\index{super!manifold}
straightforwardly define the CR supermanifold
\begin{equation}\label{eq:3.23}\index{super!manifold}
\CK^{5|2\CN}\ :=\  (\FR^{3|2\CN}\times S^2,\hat{\CCDb})\ewith
\hat{\CCDb}\ =\
\spn\left\{\der{\bw^1_\pm},\der{\bw^2_\pm},\der{\etab_i^\pm}\right\}~.
\end{equation}
A second interpretation of the space $\CK^{5|2\CN}$ as a CR
supermanifold is $\CK_0^{5|2\CN}\!:=\!(\FR^{3|2\CN}\times\index{super!manifold}
S^2,\hat{\CCDb}_0)$ with the distribution
$\hat{\CCDb}_0=\spn\{\der{\by},\der{\bl^{\phantom{\dagger}}_\pm},
\der{\etab_i^\ed}\}$. In both cases, the CR structures are of rank
$2|\CN$.

\paragraph{Coordinates on $\CK^{5|2\CN}$.} Up to now,
we have used the coordinates
$(y^{\ald\bed},\lambda_\ald^\pm,\hl_\ald^\pm,\eta_i^\ald)$ on the
two patches $\tilde{\CV}_\pm$ covering the superspace\index{super!space}
$\FR^{3|2\CN}\times S^2$. More convenient for the distribution
\eqref{eq:3.23} are, however, the coordinates \eqref{eq:3.17}
together with
\begin{equation}\label{eq:3.24}
\begin{aligned}
w_+^3&\ :=\ \tfrac{1}{1+\lambda_+\bl^{\phantom{\dagger}}_+}
\left[\bl_+y+(1-\lambda_+\bl_+)x^3+\lambda_+\by\right]
~~~\mbox{on}~~\tilde{\CV}_+~,\\
w_-^3&\ :=\ \tfrac{1}{1+\lambda_-\bl^{\phantom{\dagger}}_-}
\left[\lambda_-y+(\lambda_-\bl_--1)x^3+\bl_-\by\right]
~~~\mbox{on}~~\tilde{\CV}_-~.
\end{aligned}
\end{equation}
and we can write more concisely
\begin{equation}\label{eq:3.25}
w_\pm^1\ =\ y^{\ald\bed}\lambda_\ald^\pm\lambda_\bed^\pm ~,~~~
w_\pm^2\ =\ \lambda_\pm~,~~~ w_\pm^3\ =\ -\di\gamma_\pm
y^{\ald\bed}\lambda_\ald^\pm\hat{\lambda}^\pm_\bed\eand\eta_i^\pm\
=\ \eta_i^\ald\lambda_\ald^\pm~.
\end{equation}
Note that all the coordinates are complex except for $w_\pm^3$,
the latter being real.

\paragraph{Projection onto $\CP^{2|\CN}$.}\label{pProjectionP2}
The coordinates \eqref{eq:3.25} obviously imply that the
mini-supertwistor space $\CP^{2|\CN}$ is a complex\index{mini-supertwistor space}\index{twistor}\index{twistor!space}
subsupermanifold of the CR supermanifold $\CK^{5|2\CN}$, as they\index{super!manifold}
yield a projection
\begin{equation}\label{eq:3.26}
\nu_2\ :\ \CK^{5|2\CN}\ \rightarrow\ \CP^{2|\CN}~.
\end{equation}
The typical fibres of this projection are real one-dimensional
spaces $\ell\cong \FR$ parameterized by the coordinates $w_\pm^3$
and the pull-back of the real structure $\tau_{-1}$ on\index{real structure}
$\CP^{2|\CN}$ to $\CK^{5|2\CN}$ reverses the orientation of each
line $\ell$, since $\tau_{-1}(w_\pm^3)=-w_\pm^3$.

The geometry of the fibration \eqref{eq:3.26} becomes clearer when\index{fibration}
noting that the body $\CK^5$ of the supermanifold $\CK^{5|2\CN}$\index{body}\index{super!manifold}
can be seen as the sphere bundle
\begin{equation}
S(T\FR^3)\ =\ \{(x,u)\in T\FR^3\,|\,\delta_{ab}u^a u^b\ =\ 1\}\ \cong\
\FR^3\times S^2
\end{equation}
whose fibres at points $x\in\FR^3$ are spheres of unit vectors in
$T_x\FR^3$ \cite{Hitchin:1982gh}. Since this bundle is trivial,
its projection onto $\FR^3$ in \eqref{eq:3.19} is obviously
$\nu_1(x,u)=x$. Moreover, the complex two-dimensional mini-twistor\index{twistor}
space $\CP^2$ can be described as the space of all oriented lines
in $\FR^3$. That is, any such line $\ell$ is defined by a unit
vector $u^a$ in the direction of $\ell$ and a shortest vector
$v^a$ from the origin in $\FR^3$ to $\ell$, and one can easily
show \cite{Hitchin:1982gh} that
\begin{equation}
\CP^2\ =\ \{(v,u)\in
T\FR^3\,|\,\delta_{ab}u^av^b\ =\ 0\,,~\delta_{ab}u^au^b\ =\ 1\}\ \cong\
T^{1,0}\CPP^1\ \cong\ \CO(2)~.
\end{equation}

The fibres of the projection $\nu_2:\CK^5\rightarrow \CP^2$ are
the orbits of the action of the group $\CCG'\cong \FR$ on
$\FR^3\times S^2$ given by the formula
$(v^a,u^b)\mapsto(v^a+tu^a,u^b)$ for $t\in\FR$ and
\begin{equation}
\CP^2\ \cong\ \FR^3\times S^2/\CCG'~.
\end{equation}

Now $\CK^5$ is a (real) hypersurface in the twistor space $\CP^3$,\index{twistor}\index{twistor!space}
and $\CP^2$ is a complex two-dimensional submanifold of $\CK^5$.
Thus, we have
\begin{equation}
\CP^2\ \subset\ \CK^5\ \subset\ \CP^3\eand \CP^{2|\CN}\ \subset\
\CK^{5|2\CN}\ \subset\ \CP^{3|\CN}~.
\end{equation}

\paragraph{Vector fields on $\CK^{5|2\CN}$.} The vector fields on
$\CK^{5|2\CN}$ in the complex bosonic coordinates \eqref{eq:3.25}
are related to those in the coordinates
$(y,\by,x^3,\lambda_\pm,\bl_\pm)$ via the formul\ae{}
\begin{subequations}
\begin{align}
\begin{aligned}
\der{w_+^1}&\ =\
\gamma_+^2\left(\der{y}-\bl_+\der{x^3}-\bl_+^2\der{\by}\right)\
=:\ \gamma_+^2
W_1^+~,\\
\der{w_+^2}&\ =\
W_2^++2\gamma_+^2(x^3+\lambda_+\by)W_1^+-\gamma_+^2(\by-2\bl_+
x^3-\bl_+^2 y) W_3^+-\gamma_+\etab_i^\ed V_+^i~,\\
\der{w_+^3}&\ =\
2\gamma_+\left(\lambda_+\der{y}+\bl_+\der{\by}+\frac{1}{2}
(1-\lambda_+\bl_+)\der{x^3}\right)\ =:\ W_3^+~,
\end{aligned}
\intertext{as well as}
\begin{aligned}
\der{w_-^1}&\ =\ \gamma_-^2\left(\bl_-^2\der{y}-\bl_-\der{x^3}
-\der{\by}\right)\ =:\ \gamma_-^2W_1^-~,\\
\der{w_-^2}&\ =\
W_2^-+2\gamma_-^2(x^3-\lambda_-y)W_1^-+\gamma_-^2(\bl_-^2\by
-2\bl_-x^3-y)W_3^-+\gamma_-\etab^\zd_i
V_-^i~,\\
\der{w_-^3}&\ =\
2\gamma_-\left(\bl_-\der{y}+\lambda_-\der{\by}+\frac{1}{2}
(\lambda_-\bl_--1)\der{x^3}\right)\ =:\ W_3^-~,
\end{aligned}
\end{align}
\end{subequations}
where\footnote{Note that the vector field $W_3^\pm$ is real.}
$W_2^\pm:=\der{\lambda_\pm}$. With these identifications, we can
also use the vector fields $\bW_1^\pm$, $\bW_2^\pm$ and
$\bV_\pm^i$ to generate the CR structure $\hat{\CCDb}$. In the
simplifying spinorial notation we have furthermore\index{Spinor}
\begin{subequations}\label{eq:3.34all}
\begin{align}\label{eq:3.34}
\begin{aligned}
W_1^\pm\ =\
\hl^\ald_\pm\hl^\bed_\pm\dpar_{(\ald\bed)}~,~~~W_2^\pm\ =\
\dpar_{\lambda_\pm}~,~~~ W_3^\pm\ =\
2\gamma_\pm\hl^\ald_\pm\lambda^\bed_\pm\dpar_{(\ald\bed)}~,\\
V_\pm^i\ =\ -\hl_\pm^\ald T_j{}^i\der{\eta_j^\ald}~\hspace{4.0cm}~
\end{aligned}
 \intertext{as well
as} \label{eq:3.35}
\begin{aligned}
 \bW_1^\pm\ =\
-\lambda^\ald_\pm\lambda_\pm^\bed\dpar_{(\ald\bed)}~,~~~\bW_2^\pm\
=\ \dpar_{\bl_\pm}~,~~~ \bW_3^\pm\ =\ W_3^\pm\ =\
2\gamma_\pm\hl^\ald_\pm\lambda^\bed_\pm\dpar_{(\ald\bed)}~,\\
\bV_\pm^i\ =\ \lambda_\pm^\ald \der{\eta_i^\ald}~.\hspace{4.5cm}
\end{aligned}
\end{align}
\end{subequations}

\paragraph{Forms on $\CK^{5|2\CN}$.} The formul\ae{} for the
forms dual to the vector fields \eqref{eq:3.34} and
\eqref{eq:3.35} read
\begin{subequations}\label{eq:3.37}
\begin{align}
\begin{aligned}
\Theta_\pm^1\ :=\  \gamma^2_\pm\lambda_\ald^\pm\lambda_\bed^\pm\dd
y^{\ald\bed}~,~~~ \Theta_\pm^2\ :=\  \dd \lambda_\pm~,~~~
\Theta_\pm^3\ :=\  -\gamma_\pm\lambda_\ald^\pm\hl_\bed^\pm\dd
y^{\ald\bed}~,\\ E_i^\pm\ :=\ \gamma_\pm\lambda_\ald^\pm
T_i{}^j\dd \eta_j^\ald\hspace{4cm}
\end{aligned}
\intertext{and}
\begin{aligned}
\bTheta_\pm^1\ =\ -\gamma^2_\pm\hl_\ald^\pm\hl_\bed^\pm\dd
y^{\ald\bed}~,~~~\bTheta_\pm^2\ =\ \dd \bl_\pm~,~~~\bTheta_\pm^3\
=\ \Theta^3_\pm~,\hspace{1cm}\\\bar{E}_i^\pm\ =\
-\gamma_\pm\hl^\pm_\ald\dd \eta^\ald_i~,\hspace{4cm}
\end{aligned}
\end{align}
\end{subequations}
where $T_i{}^j$ has been given in section \ref{ssSpinors},
\ref{peuclideancase}. The exterior derivative on $\CK^{5|2\CN}$\index{exterior derivative}
accordingly given by
\begin{align}\nonumber
\dd|_{\tilde{\CV}_\pm}&\ =\ \dd w_\pm^1\der{w_\pm^1}+\dd
w_\pm^2\der{w_\pm^2}+\dd \bw_\pm^1\der{\bw_\pm^1}+\dd
\bw_\pm^2\der{\bw_\pm^2}+\dd
w_\pm^3\der{w_\pm^3}+\ldots\\
&\ =\ \Theta^1_\pm
W_1^\pm+\Theta_\pm^2W_2^\pm+\bTheta_\pm^1\bW_1^\pm+
\bTheta_\pm^2\bW_2^\pm+\Theta^3_\pm W_3^\pm +E_i^\pm
V^i_\pm+\bar{E}_i^\pm \bar{V}^i_\pm~,
\end{align}
where the dots stand for derivatives with respect to $\eta_i^\pm$
and $\etab_i^\pm$. Note again that $\Theta_\pm^3$ and $W_3^\pm$
are both real.\footnote{To homogenize the notation later on, we
shall also use $\bW_3^\pm$ and $\dpar_{\bw^\pm_3}$ instead of
$W_3^\pm$ and $\dpar_{w^\pm_3}$, respectively.}

\subsection{Partially holomorphic Chern-Simons theory}\index{Chern-Simons theory}

\paragraph{Outline.} In the following, we will discuss a
generalization of Chern-Simons theory on the correspondence space\index{Chern-Simons theory}
$\CK^{5|2\CN}$, which we call {\em partially holomorphic
Chern-Simons theory} or phCS theory for short. Roughly speaking,\index{Chern-Simons theory}
this theory is a mixture of Chern-Simons and holomorphic
Chern-Simons theory on the CR supertwistor space $\CK^{5|2\CN}$\index{Chern-Simons theory}\index{twistor}\index{twistor!space}
which has one real and two complex bosonic dimensions. Eventually,
we will find a one-to-one correspondence between the moduli space\index{moduli space}
of solutions to the equations of motion of phCS theory on
$\CK^{5|2\CN}$ and the moduli space of solutions to $\CN$-extended\index{moduli space}
supersymmetric Bogomolny equations on $\FR^3$, quite similar to\index{Bogomolny equations}
the correspondence between hCS theory on the supertwistor space\index{twistor}\index{twistor!space}
$\CP^{3|\CN}$ and $\CN$-extended supersymmetric SDYM theory in
four dimensions.

\paragraph{The integrable distribution $\CT$ on $\CK^{5|2\CN}$.}\index{integrable}\index{integrable distribution}
Combining the vector fields $\bW_1^\pm$, $\bW_2^\pm$, $\bV^i_\pm$
from the CR structure $\hat{\CCDb}$ with the vector field
$\bW_3^\pm$ yields an integrable distribution, which we denote by\index{integrable}\index{integrable distribution}
$\CT$. The distribution $\CT$ is integrable since we have
$[\bW_2^\pm,\bW_3^\pm]\ =\ \pm2 \gamma_\pm^2\bW_1^\pm$ and all
other commutators vanish. Also, $\CCV\ :=\ \CT\cap\bar{\CT}$ is a\index{commutators}
real one-dimensional and hence integrable. Note that $\CCV$ is\index{integrable}
spanned by the vector fields $\bW_3^\pm$ over the patches
$\tilde{\CV}_\pm\subset \CK^{5|2\CN}$. Furthermore, the
mini-supertwistor space $\CP^{2|\CN}$ is a subsupermanifold of\index{mini-supertwistor space}\index{super!manifold}\index{twistor}\index{twistor!space}
$\CK^{5|2\CN}$ transversal to the leaves of
$\CCV=\CT\cap\bar{\CT}$ and $\CT|_{\CP^{2|\CN}}=\hat{\CCDb}$.
Thus, we have an integrable distribution\index{integrable}\index{integrable distribution}
$\CT=\hat{\CCDb}\oplus\CCV$ on the CR supertwistor space\index{twistor}\index{twistor!space}
$\CK^{5|2\CN}$ and we will denote by $\CT_b$ its bosonic part
generated by the vector fields $\bW_1^\pm$, $\bW_2^\pm$ and
$\bW_3^\pm$.

\paragraph{Field equations of phCS theory.} Let $\CE$ be a
trivial rank $n$ complex vector bundle over $\CK^{5|2\CN}$ and\index{complex!vector bundle}
$\hat{\CA}_\CT$ a $\CT$-connection on $\CE$. We define the field\index{T-connection@$\CT$-connection}\index{connection}
equations of partial holomorphic Chern-Simons theory to be\index{Chern-Simons theory}
\begin{equation}\label{eq:4.16}
\dd_\CT\CAh_\CT+\CAh_\CT\wedge\CAh_\CT\ =\ 0~,
\end{equation}
In the nonholonomic basis $\{\bW_a^\pm, \bV_\pm^i\}$ of the
distribution $\CT$ over $\tilde{\CV}_\pm\subset\CK^{5|2\CN}$,
these equations read as
\begin{subequations}\label{eq:4.17}
\begin{align}\label{eq:4.17a}
\bW_1^\pm\CAh_2^\pm-\bW^\pm_2\CAh_1^\pm+[\CAh_1^\pm,\CAh_2^\pm]&\
=\ 0~,\\\label{eq:4.17b}
\bW_2^\pm\CAh_3^\pm-\bW^\pm_3\CAh_2^\pm+[\CAh_2^\pm,\CAh_3^\pm]\mp2
\gamma_\pm^2\CAh_1^\pm&\ =\ 0~,\\\label{eq:4.17c}
\bW_1^\pm\CAh_3^\pm-\bW^\pm_3\CAh_1^\pm+[\CAh_1^\pm,\CAh_3^\pm]&\
=\ 0~,
\end{align}
\end{subequations}
where the components $\CAh_a^\pm$ are defined via the contractions
$\CAh_a^\pm:=\bW^\pm_a\lrcorner \CAh_\CT$. Analogously to the case
of super holomorphic Chern-Simons theory on $\CP^{3|\CN}$, we\index{Chern-Simons theory}
assume that
\begin{equation}\label{eq:4.13}
\bV_\pm^i\lrcorner \CAh_\CT\ =\ 0\eand\bV_\pm^i(\CAh^\pm_a)\ =\
0~.
\end{equation}

\paragraph{Action functional.} When restricting to the case $\CN=4$,
we can write down an action functional for phCS theory. As it was
noted in \cite{Chiou:2005jn}, the $\CN=4$ mini-supertwistor space\index{mini-supertwistor space}\index{twistor}\index{twistor!space}
is a Calabi-Yau supermanifold and thus, there is a holomorphic\index{Calabi-Yau}\index{Calabi-Yau supermanifold}\index{super!manifold}
volume form $\Omega$ on $\CP^{2|4}$. Moreover, the pull-back
$\tilde{\Omega}$ of this form to $\CK^{5|8}$ is globally defined
and we obtain locally on the patches $\tilde{\CV}_\pm\subset
\CK^{5|8}$
\begin{equation}\label{eq:4.12}
\tilde{\Omega}|_{\tilde{\CV}_\pm}\ =\ \pm\dd w_\pm^1\wedge \dd
w_\pm^2\dd \eta_1^\pm\cdots\dd\eta_4^\pm~.
\end{equation}
Together with the assumptions in \eqref{eq:4.13}, we can write
down the CS-type action functional
\begin{equation}\label{eq:4.14}
S_{\mathrm{phCS}}\ =\
\int_{\CCK^{5|8}}\tilde{\Omega}\wedge\tr\left(\CAh_\CT\wedge
\dd_\CT\CAh_\CT+\tfrac{2}{3}
\CAh_\CT\wedge\CAh_\CT\wedge\CAh_\CT\right)~,
\end{equation}
where
\begin{equation}
\dd_\CT|_{\tilde{\CV}_\pm}\ =\ \dd \bw_\pm^a\der{\bw_\pm^a}+\dd
\etab_i^\pm\der{\etab_i^\pm}
\end{equation}
is the $\CT$-differential on $\CK^{5|8}$ and $\CCK^{5|8}$ is the\index{T-differential@$\CT$-differential}
chiral subspace of $\CK^{5|8}$ for which $\bar{\eta}^i=0$. The
action functional \eqref{eq:4.14} reproduces the phCS equations of
motion \eqref{eq:4.16}.

\paragraph{Supersymmetric Bogomolny equations.} The equations of\index{Bogomolny equations}
motion of the phCS theory defined above are equivalent to the
supersymmetric Bogomolny equations \eqref{eom-sB}. To show this,\index{Bogomolny equations}
we will give the explicit field expansion similar to \eqref{expA}
necessary to cast the equations \eqref{eq:4.17} into the form
\eqref{eom-sB}. As before in \eqref{expA}, we will only consider
the case $\CN=4$, from which all other cases $\CN<4$ can be
derived by truncation of the field expansion. First, note that due
to \eqref{eq:3.35}, we have
\begin{equation}
\bW_1^+\ =\ \lambda_+^2\bW_1^-~,~~~\bW_2^+\ =\
-\bl_+^{-2}\bW_2^-\eand \gamma_+^{-1}\bW_3^+\ =\
\lambda_+\bl_+\left(\gamma_-^{-1}\bW_3^-\right)
\end{equation}
and therefore $\CAh_1^\pm$, $\CAh_2^\pm$ and
$\gamma_\pm^{-1}\CAh_3^\pm$ take values in the bundles $\CO(2)$,
$\COb(-2)$ and $\CO(1)\otimes\COb(1)$, respectively. Together with
the definitions \eqref{eq:4.13} of $\CAh^\pm_a$ and
\eqref{eq:3.34all} of $\bW_a^\pm$ as well as the fact that the
$\eta_i^\pm$ are nilpotent and $\CO(1)$-valued, this determines
the dependence of $\CAh_a^\pm$ on $\eta_i^\pm$, $\lambda_\pm$ and
$\bl_\pm$ to be
\begin{equation}
\CAh_1^\pm\ =\ -\lambda_\pm^\ald \CB_\ald^\pm\eand \CAh_3^\pm\ =\
2\gamma_\pm\hl^\ald_\pm\CB_\ald^\pm
\end{equation}
with the abbreviations
\begin{subequations}\label{eq:4.19}
\begin{align}
&\begin{aligned} \CB_\ald^\pm\ :=\  &\lambda^\bed_\pm
B_{\ald\bed}+\di\eta_i^\pm\chi^i_\ald+\tfrac{1}{2!}
\gamma_\pm\eta_i^\pm\eta_j^\pm\hl^\bed_\pm\phi^{ij}_{\ald\bed}
+\tfrac{1}{3!}\gamma_\pm^2\eta_i^\pm\eta_j^\pm\eta_k^\pm
\hl^\bed_\pm\hl^\gad_\pm\tilde{\chi}^{ijk}_{\ald\bed\gad}\ +
\\&+\tfrac{1}{4!}\gamma_\pm^3\eta_i^\pm\eta_j^\pm\eta_k^\pm
\eta_l^\pm\hl_\pm^\bed\hl_\pm^\gad\hl_\pm^\ded
G^{ijkl}_{\ald\bed\gad\ded}
\end{aligned}\\\label{eq:4.19c}
&\begin{aligned} \CAh_2^\pm\ =\
\pm\left(\tfrac{1}{2!}\gamma_\pm^2\eta_i^\pm\eta_j^\pm\phi^{ij}+
\tfrac{1}{3!}\gamma_\pm^3\eta_i^\pm\eta_j^\pm\eta_k^\pm
\hl^\ald_\pm\tilde{\chi}_\ald^{ijk}\right.\\\left.+\tfrac{1}{4!}\gamma_\pm^4
\eta_i^\pm\eta_j^\pm\eta_k^\pm\eta_l^\pm \hl^\ald_\pm\hl^\bed_\pm
G^{ijkl}_{\ald\bed}\right)~.\hspace{0.5cm}
\end{aligned}
\end{align}
\end{subequations}
Note that in this expansion, all fields
$B_{\ald\bed},\chi_\ald^i,\ldots$ depend only on the coordinates
$(y^{\ald\bed})\in\FR^3$. Substituting \eqref{eq:4.19} into
\eqref{eq:4.17a} and \eqref{eq:4.17b}, we obtain the equations
\begin{equation}\label{eq:4.constr}
\begin{aligned}
\phi^{ij}_{\ald\bed}\ =\
-\left(\dpar_{(\ald\bed)}\phi^{ij}+[B_{\ald\bed},\phi^{ij}]
\right)~,~~~ \tilde{\chi}^{ijk}_{\ald(\bed\gad)}\ =\ -\tfrac{1}{2}
\left(\dpar_{(\ald(\bed)}\tilde{\chi}^{ijk}_{\gad)}+
[B_{\ald(\bed},\tilde{\chi}^{ijk}_{\gad)}]\right)~,\\
G^{ijkl}_{\ald(\bed\gad\ded)}\ =\
-\tfrac{1}{3}\left(\dpar_{(\ald(\bed)}G^{ijkl}_{\gad\ded)}+
[B_{\ald(\bed},G^{ijkl}_{\gad\ded)}]\right)\hspace{3.5cm}
\end{aligned}
\end{equation}
showing that $\phi_{\ald\bed}^{ij}$,
$\tilde{\chi}_{\ald\bed\gad}^{ijk}$ and
$G^{ijkl}_{\ald\bed\gad\ded}$ are composite fields, which do not
describe independent degrees of freedom. Furthermore, the field
$B_{\ald\bed}$ can be decomposed into its symmetric part, denoted
by $A_{\ald\bed}=A_{(\ald\bed)}$, and its antisymmetric part,
proportional to $\Phi$, such that
\begin{equation}\label{eq:4.B}
B_{\ald\bed}\ =\ A_{\ald\bed}-\tfrac{\di}{2}\eps_{\ald\bed}\Phi~.
\end{equation}
Defining additionally
\begin{equation}\label{eq:4.chi}
\tilde{\chi}_{i\ald}\ :=\
\tfrac{1}{3!}\eps_{ijkl}\tilde{\chi}_\ald^{jkl}\eand G_{\ald\bed}\
:=\  \tfrac{1}{4!}\eps_{ijkl}G^{ijkl}_{\ald\bed}~,
\end{equation}
we have thus recovered the field content of the $\CN=4$ super
Bogomolny equations together with the appropriate field equations\index{Bogomolny equations}
\eqref{eom-sB}. Up to a constant, the action functional of the
super Bogomolny model \eqref{action-sB} can be obtained from the\index{Bogomolny model}\index{super!Bogomolny model}
action functional of phCS theory \eqref{eq:4.14} by substituting
the above given expansion and integrating over the sphere
$\CPP^1_{x,\eta}\subset \CP^{2|4}$.

\paragraph{The linear systems.} To improve our\index{linear system}
understanding of the vector bundle $\CE$ over $\CK^{5|2\CN}$, let
us consider the linear system underlying the equations\index{linear system}
\eqref{eq:4.17}. This system reads
\begin{equation}\label{eq:4.39}
\begin{aligned}
(\bW_a^\pm+\CAh_a^\pm)\hpsi_\pm&\ =\ 0~,\\ \bV_\pm^i\hpsi_\pm&\ =\
0~,
\end{aligned}
\end{equation}
and has indeed \eqref{eq:4.17} as its compatibility conditions.\index{compatibility conditions}
Using the splitting $A_\CT|_{\tilde{\CV}_\pm}=
\psi_\pm\dd_\CT\psi_\pm^{-1}$, we can switch now to the \v{C}ech
description of an equivalent vector bundle $\tilde{\CE}$ with
transition function $f_{+-}=\psi_+^{-1}\psi_-$. Similarly to the\index{transition function}
description of the vector bundles involved in the Penrose-Ward
transform over the supertwistor space $\CP^{3|\CN}$, we can find a\index{Penrose-Ward transform}\index{twistor}\index{twistor!space}
gauge transformation generated by the globally defined group\index{gauge!transformation}
valued function $\varphi$, which acts by $\psi\mapsto
\hpsi_\pm=\varphi^{-1}\psi_\pm$ and leads to
\begin{equation}\label{eq:4.38}
\begin{aligned}
\CAh_1^\pm&\ \mapsto\  \CA_1^\pm\ =\ \varphi^{-1}\CAh_1^\pm
\varphi+\varphi^{-1}\bW_1^\pm
\varphi\ =\ \psi_\pm\bW_1^\pm\psi_\pm^{-1}~,\\
\CAh_2^\pm&\ \mapsto\  \CA_2^\pm\ =\ \varphi^{-1}\CAh_2^\pm
\varphi+\varphi^{-1}\bW_2^\pm
\varphi\ =\ \psi_\pm\bW_2^\pm\psi_\pm^{-1}\ =\ 0~,\\
\CAh^\pm_3&\ \mapsto\  \CA_3^\pm\ =\ \varphi^{-1}\CAh_3^\pm
\varphi+\varphi^{-1}\bW_3^\pm \varphi\ =\ \psi_\pm\bW_3^\pm
\psi_\pm^{-1}~,\\
0\ =\ \CAh_\pm^i\ :=\ \hpsi_\pm\bV^i_\pm\hpsi_\pm^{-1}&\ \mapsto\
\CA_\pm^i\ =\ \varphi^{-1}\bV^i_\pm \varphi\ =\
\psi_\pm\bV^i_\pm\psi_\pm^{-1}~,
\end{aligned}
\end{equation}
while $f_{+-}\ =\ \hpsi_+^{-1}\hpsi_-= \psi_+^{-1}\psi_-$ remains
invariant. In this new gauge, one generically has $\CA^i_\pm\neq
0$ and the new gauge potential fits into the following linear
system of differential equations:\index{linear system}
\begin{subequations}\label{eq:4.40}
\begin{align}\label{eq:4.40a}
(\bW_1^\pm+\CA_1^\pm)\psi_\pm&\ =\ 0~,\\\label{eq:4.40b}
 \bW_2^\pm\psi_\pm&\ =\ 0~,\\\label{eq:4.40c}
 (\bW_3^\pm+\CA_3^\pm)\psi_\pm&\ =\ 0~,\\
 (\bV_\pm^i+\CA_\pm^i)\psi_\pm&\ =\ 0~,
\end{align}
\end{subequations}
which is gauge equivalent to the system \eqref{eq:4.39}.

Due to the holomorphy of $\psi_\pm$ in $\lambda_\pm$ and the
condition $\CA_\CT^+=\CA_\CT^-$ on
$\tilde{\CV}_+\cap\tilde{\CV}_-$, the components $\CA_1^\pm$,
$\gamma_\pm^{-1}\CA_3^\pm$ and $\CA_\pm^i$ must take the form
\begin{equation}\label{eq:4.41}
\CA^\pm_1\ =\ -\lambda_\pm^\ald\lambda_\pm^\bed
\CB_{\ald\bed}~,~~~ \gamma_\pm^{-1}\CA_3^\pm\ =\
2\hl^\ald_\pm\lambda_\pm^\bed \CB_{\ald\bed}\eand \CA^i_\pm\ =\
\lambda^\ald_\pm\CAh^i_\ald~,
\end{equation}
with $\lambda$-independent superfields
$\CB_{\ald\bed}:=\CAh_{\ald\bed}-\frac{\di}{2}\eps_{\ald\bed}\Phi$
and $\CAh_\ald^i$. After introducing the first-order differential
operators $\nabla_{\ald\bed}:=\dpar_{(\ald\bed)}+\CB_{\ald\bed}$
and
$D_\ald^i=\der{\eta_i^\ald}+\CAh_\ald^i=:\dpar_\ald^i+\CAh_\ald^i$,
we can write the compatibility conditions of the linear system\index{compatibility conditions}\index{linear system}
\eqref{eq:4.40} as
\begin{equation}\label{eq:4.42}
\begin{aligned}
{}&[\nabla_{\ald\gad},\nabla_{\bed\ded}]+[\nabla_{\ald\ded},
\nabla_{\bed\gad}]\ =\ 0~,
~~~~[D^i_\ald,\nabla_{\bed\gad}]+[D_\gad^i,\nabla_{\bed\ald}]\ =\
0~,\\&\hspace{3cm} \{D_\ald^i,D_\bed^j\}+\{D_\bed^i,D_\ald^j\}\ =\
0~.
\end{aligned}
\end{equation}
These equations are the constraint equations of the super\index{constraint equations}
Bogomolny model, see also \eqref{mpconstraint}.\index{Bogomolny model}

\paragraph{The one-to-one correspondence.} Summarizing, we have
described a bijection between the moduli space\index{moduli space}
$\CM_{\mathrm{phCS}}$ of certain solutions to the field equations
\eqref{eq:4.17} of phCS theory and the moduli space\index{moduli space}
$\CM_{\mathrm{sB}}$ of solutions to the supersymmetric Bogomolny
equations \eqref{eom-sB}. The moduli spaces are obtained from the\index{Bogomolny equations}\index{moduli space}
respective solution spaces by factorizing with respect to the
action of the corresponding groups of gauge transformations.\index{gauge transformations}\index{gauge!transformation}

\subsection{Holomorphic BF theory}

So far, we defined a Chern-Simons type theory corresponding to the
super Bogomolny model, but this model was constructed on the\index{Bogomolny model}\index{super!Bogomolny model}
correspondence space $\CK^{5|2\CN}$, after interpreting it as a
partially holomorphic manifold. This is somewhat unusual, and one\index{manifold}
is naturally led to ask whether there is an equivalent model on
the mini-supertwistor space. In fact there is, and we will define\index{mini-supertwistor space}\index{twistor}\index{twistor!space}
it in this section. For simplicity, we will restrict all our
considerations from now on to the case $\CN=4$.

\paragraph{Holomorphic BF theory on $\CP^{2|4}$.} Consider the
mini-supertwistor space $\CP^{2|4}$ together with a topologically\index{mini-supertwistor space}\index{twistor}\index{twistor!space}
trivial holomorphic vector bundle $\CE$ of rank $n$ over $M$. Let\index{holomorphic!vector bundle}
$\CAh^{0,1}$ be the $(0,1)$-part of its connection one-form\index{connection}
$\CAh$, which we assume to satisfy the conditions
$\bV_\pm^i\lrcorner\CAh^{0,1}=0$ and
$\bV_\pm^i(\dpar_{\bw_\pm^{1,2}}\lrcorner\CAh^{0,1})=0$. Recall
that $\CP^{2|4}$ is a Calabi-Yau supermanifold and thus comes with\index{Calabi-Yau}\index{Calabi-Yau supermanifold}\index{super!manifold}
the holomorphic volume form $\Omega$ which is defined in\index{holomorphic!volume form}
\eqref{eq:3.14}. Hence, we can define a holomorphic BF (hBF) type
theory (cf.\ \cite{Popov:1999cq,Ivanova:2000xr,Baulieu:2004pv}) on
$\CP^{2|4}$ with the action
\begin{equation}\label{eq:5.1}
S_{\mathrm{hBF}}\ =\ \int_{\CCP^{2|4}}\Omega\wedge\tr\{B(\dparb
\CAh^{0,1}+\CAh^{0,1}\wedge\CAh^{0,1})\}\ =\
\int_{\CCP^{2|4}}\Omega\wedge \tr\{B \CF^{0,2}\}~,
\end{equation}
where $B$ is a scalar field in the adjoint representation of the\index{representation}
gauge group $\sGL(n,\FC)$, $\dparb$ is the antiholomorphic part of
the exterior derivative on $\CP^{2|4}$ and $\CF^{0,2}$ the $(0,2)$\index{exterior derivative}
part of the curvature two-form. The space $\CCP^{2|4}$ is the\index{curvature}
subsupermanifold of $\CP^{2|4}$ constrained\footnote{In string\index{super!manifold}
theory, one would regard $\CCP^{2|4}$ as the worldvolume of a
stack of $n$ not quite space-filling D3-branes.} by
$\etab_i^\pm=0$.

\paragraph{Equations of motion.} The corresponding equations of
motion of hBF theory are readily derived to be
\begin{subequations}\label{eq:5.2}
\begin{align}\label{eq:5.2a}
\dparb\CAh^{0,1}+\CAh^{0,1}\wedge\CAh^{0,1}&\ =\
0~,\\\label{eq:5.2b} \dparb B+[\CAh^{0,1},B]&\ =\ 0~.
\end{align}
\end{subequations}

Furthermore, both these equations as well as the Lagrangian in
\eqref{eq:5.1} can be obtained from the equations \eqref{eq:4.16}
and the Lagrangian in \eqref{eq:4.14}, respectively, by imposing
the condition $\dpar_{\bw_\pm^3}\CAh_{\bw_\pm^a}=0$ and
identifying
\begin{equation}
\CAh^{0,1}|_{\hat{\CV}_\pm}\ =\ \dd \bw_\pm^1\CAh_{\bw_\pm^1}+\dd
\bw_\pm^2\CAh_{\bw_\pm^2}\eand B^\pm\ :=\  B|_{\hat{\CV}_\pm}\ =\
\CAh_{\bw_\pm^3}~.
\end{equation}
On $\CP^{2|4}$, $\CAh_{\bw_\pm^3}$ behaves as a scalar and thus,
\eqref{eq:5.2} can be obtained from \eqref{eq:4.16} by demanding
invariance of all fields under the action of the group $\CCG'$.

\paragraph{Interpretation of the $B$-field.} By construction,
$B=\{B^\pm\}$ is a $\agl(n,\FC)$-valued function generating
trivial infinitesimal gauge transformations of $\CAh^{0,1}$ and\index{gauge transformations}\index{gauge!transformation}
therefore it does not contain any physical degrees of freedom. To
understand this statement, let us look at the infinitesimal level
of gauge transformations of $\CAh^{0,1}$, which take the form\index{gauge transformations}\index{gauge!transformation}
\begin{equation}
\delta\CA^{0,1}\ =\ \dparb B+[\CAh^{0,1},B]
\end{equation}
with $B\in H^0(\CP^{2|4},\sEnd E)$. Such a field $B$ solving
moreover \eqref{eq:5.2b} generates holomorphic transformations
such that $\delta \CAh^{0,1}\ =\ 0$. Their finite version is
\begin{equation}
\tilde{\CA}^{0,1}\ =\ \varphi\CAh^{0,1}\varphi^{-1}+\varphi\dparb
\varphi^{-1}\ =\ \CAh^{0,1}~,
\end{equation}
and for a solution $(\CAh^{0,1},B)$ to equations \eqref{eq:5.2} of
the form
\begin{equation}\label{eq:5.4}
\CA^{0,1}|_{\hat{\CV}_\pm}\ =\
\tilde{\psi}_\pm\dparb\tilde{\psi}_\pm^{-1}\eand B^\pm\ =\
\tilde{\psi}_\pm B^\pm_0\tilde{\psi}_\pm^{-1}~,
\end{equation}
such a $\varphi$ takes the form
\begin{equation}
\varphi_\pm\ =\
\tilde{\psi}_\pm\de^{B_0^\pm}\tilde{\psi}_\pm^{-1}~~~\mbox{with}~~~
\varphi_+\ =\ \varphi_-~~\mbox{on}~~\hat{\CV}_+\cap\hat{\CV}_-~.
\end{equation}

\paragraph{Full equivalences.} Altogether, we arrive at the
conclusion that the moduli space of solutions to hBF theory given\index{moduli space}
by the action \eqref{eq:5.1} is bijective to the moduli space of
solutions to the phCS-equations, and therefore we can sum up the
discussion up to now with the diagram
\begin{equation}
\begin{aligned}
\begin{picture}(180,70)(0,-5)
\put(0.0,0.0){\makebox(0,0)[c]{hBF theory on $\CP^{2|4}$}}
\put(160.0,14.0){\makebox(0,0)[c]{supersymmetric}}
\put(160.0,0.0){\makebox(0,0)[c]{ Bogomolny model on $\FR^3$}}\index{Bogomolny model}
\put(80.0,50.0){\makebox(0,0)[c]{phCS theory on $\CK^{5|8}$}}
\put(40.0,40.0){\vector(-1,-1){30}}
\put(10.0,10.0){\vector(1,1){30}}
\put(100.0,40.0){\vector(1,-1){18}}
\put(118.0,22.0){\vector(-1,1){18}} \put(56,0){\vector(1,0){37}}
\put(86,0){\vector(-1,0){30}}
\end{picture}
\end{aligned}
\end{equation}
describing equivalent theories defined on different spaces. Here,
it is again implied that the appropriate subsets of the solution
spaces to phCS and hBF theories are considered.

\section{Superambitwistors and mini-superambitwistors}\index{twistor}\index{twistor!ambitwistor}

\subsection{The superambitwistor\index{twistor}\index{twistor!ambitwistor}
space}\label{ssSuperambitwistorSpace}

Recall that in the construction of the ambitwistor space in\index{ambitwistor space}\index{twistor}\index{twistor!ambitwistor}\index{twistor!space}
section \ref{ssAmbitwistorSpace}, we ``glued together'' both the
self-dual and anti-self-dual subsectors of Yang-Mills theory to\index{Yang-Mills theory}\index{anti-self-dual}
obtain the full theory. It is now possible to define a
super-extension of this construction, which sheds more light on
the r{\^o}le played by the third order thickening.\index{thickening}

\paragraph{Definition.} For the definition of the superambitwistor\index{twistor}\index{twistor!ambitwistor}
space, we take a supertwistor space $\CP^{3|3}$ with coordinates\index{twistor!space}
$(z^\alpha_\pm,z^3_\pm,\eta_i^\pm)$ together with a ``dual''
copy\footnote{The word ``dual'' refers again to the spinor indices\index{Spinor}
and {\em not} to the line bundles underlying $\CP^{3|3}$.}
$\CP^{3|3}_*$ with coordinates
$(u^\ald_\pm,u^{\dot{3}}_\pm,\theta^i_\pm)$. The dual supertwistor\index{twistor}
space is considered as a holomorphic supervector bundle over the
Riemann sphere $\CPP^1_*$ covered by the patches $U_\pm^*$ with\index{Riemann sphere}
the standard local coordinates $\mu_\pm=u^{\dot{3}}_\pm$. For
convenience, we again introduce the spinorial notation\index{Spinor}
$(\mu_\alpha^+)=(1,\mu_+)^T$ and $(\mu_\alpha^-)=(\mu_-,1)^T$. The
two patches covering $\CP^{3|3}_*$ will be denoted by
$\CU_\pm^*:=\CP^{3|3}_*|_{U_\pm^*}$ and the product space
$\CP^{3|3}\times \CP^{3|3}_*$ of the two supertwistor spaces is\index{twistor}\index{twistor!space}
thus covered by the four patches
\begin{equation}
\CU_{(1)}\ :=\ \CU_+\times\CU^*_+~,~~~ \CU_{(2)}\ :=\
\CU_-\times\CU^*_+~,~~~ \CU_{(3)}\ :=\ \CU_+\times\CU^*_-~,~~~
\CU_{(4)}\ :=\ \CU_-\times\CU^*_-~,
\end{equation}
on which we have the coordinates
$(z^\alpha_{(a)},z^3_{(a)},\eta_i^{(a)};u^\ald_{(a)},
u^{\dot{3}}_{(a)},\theta^i_{(a)})$. This space is furthermore a
rank $4|6$ supervector bundle over the space $\CPP^1\times
\CPP^1_*$. The global sections of this bundle are parameterized by
elements of $\FC^{4|6}\times \FC^{4|6}_*$ in the following way:
\begin{equation}\label{sambisections}
z^\alpha_{(a)}\ =\
x_R^{\alpha\ald}\lambda_\ald^{(a)}~,~~~\eta^{(a)}_i\ =\
\eta_i^\ald\lambda_\ald^{(a)}~;~~~ u^\ald_{(a)}\ =\
x_L^{\alpha\ald}\mu_\alpha^{(a)}~,~~~ \theta^i_{(a)}\ =\
\theta^{\alpha i}\mu_\alpha^{(a)}~.
\end{equation}

The superambitwistor space is now the subspace $\CL^{5|6}\subset\index{ambitwistor space}\index{twistor}\index{twistor!ambitwistor}\index{twistor!space}
\CP^{3|3}\times \CP^{3|3}_*$ obtained from the {\em quadric\index{quadric}
condition} (the ``gluing condition'')
\begin{equation}\label{quadric2b}
\kappa_{(a)}\ :=\
z^\alpha_{(a)}\mu_\alpha^{(a)}-u_{(a)}^\ald\lambda_\ald^{(a)}+
2\theta_{(a)}^i\eta_i^{(a)}\ =\ 0~.
\end{equation}
In the following, we will denote the restrictions of $\CU_{(a)}$
to $\CL^{5|6}$ by $\hat{\CU}_{(a)}$.

\paragraph{Moduli space and the double fibration.}\index{double fibration}\index{fibration}\index{moduli space}
Due to the quadric condition \eqref{quadric2b}, the bosonic moduli\index{quadric}
are not independent on $\CL^{5|6}$, but one rather has the
relation
\begin{equation}\label{chiralcoords}
x_R^{\alpha\ald}\ =\ x^{\alpha\ald}-\theta^{\alpha
i}\eta_i^\ald~~~ \mbox{and}~~~x_L^{\alpha\ald}\ =\
x^{\alpha\ald}+\theta^{\alpha i}\eta^\ald_i~.
\end{equation}
The moduli $(x^{\alpha\ald}_R)$ and $(x^{\alpha\ald}_L)$ are
therefore indeed anti-chiral and chiral coordinates on the\index{chiral!coordinates}
(complex) superspace $\FC^{4|12}$ and with this identification,\index{super!space}
one can establish the following double fibration using equations\index{double fibration}\index{fibration}
\eqref{sambisections}:
\begin{equation}\label{ambidblfibration2}
\begin{aligned}
\begin{picture}(80,40)
\put(0.0,0.0){\makebox(0,0)[c]{$\CL^{5|6}$}}
\put(64.0,0.0){\makebox(0,0)[c]{$\FC^{4|12}$}}
\put(32.0,33.0){\makebox(0,0)[c]{$\CF^{6|12}$}}
\put(7.0,18.0){\makebox(0,0)[c]{$\pi_2$}}
\put(55.0,18.0){\makebox(0,0)[c]{$\pi_1$}}
\put(25.0,25.0){\vector(-1,-1){18}}
\put(37.0,25.0){\vector(1,-1){18}}
\end{picture}
\end{aligned}
\end{equation}
where $\CF^{6|12}\cong \FC^{4|12}\times\CPP^1\times\CPP^1_*$ and
$\pi_1$ is the trivial projection. Thus, one has the
correspondences
\begin{equation}
\begin{aligned}
\left\{\,\mbox{subspaces $(\CPP^1\times \CPP_*^1)_{x,\eta,\theta}$
in $\CL^{5|6}$} \right\}&\ \longleftrightarrow\
\left\{\,\mbox{points $(x,\eta,\theta)$ in $\FC^{4|12}$}\right\}~, \\
\left\{\,\mbox{points $p$ in $\CL^{5|6}$}\right\}&\
\longleftrightarrow\ \left\{\,\mbox{null superlines in
$\FC^{4|12}$}\right\}~.
\end{aligned}
\end{equation}
The above-mentioned null superlines are intersections of
$\alpha$-superplanes and dual $\beta$-super\-planes. Given a
solution
$(\hat{x}^{\alpha\ald},\hat{\eta}^\ald_i,\hat{\theta}^{\alpha i})$
to the incidence relations \eqref{sambisections} for a fixed point\index{incidence relation}
$p$ in $\CL^{5|6}$, the set of points on such a null superline
takes the form
\begin{equation*}
\{(x^{\alpha\ald},\eta^\ald_i,\theta^{\alpha i})\}~~\mbox{with}~~
x^{\alpha\ald}\ =\
\hat{x}^{\alpha\ald}+t\mu^\alpha_{(a)}\lambda^\ald_{(a)}~,~~
\eta^\ald_i\ =\ \hat{\eta}^\ald_i+\eps_i\lambda_{(a)}^\ald~,~~
\theta^{\alpha i}\ =\ \hat{\theta}^{\alpha
i}+\tilde{\eps}^i\mu^\alpha_{(a)}~.
\end{equation*}
Here, $t$ is an arbitrary complex number and $\eps_i$ and
$\tilde{\eps}^i$ are both three-vectors with Gra{\ss}\-mann-odd
components. The coordinates $\lambda^\ald_{(a)}$ and
$\mu^\alpha_{(a)}$ are chosen from arbitrary patches on which they
are both well-defined. Note that these null superlines are in fact
of dimension $1|6$.

\paragraph{Vector fields.} The space $\CF^{6|12}$ is covered by four patches
$\tilde{\CU}_{(a)}:=\pi_2^{-1}(\hat{\CU}_{(a)})$ and the tangent
spaces to the $1|6$-dimensional leaves of the fibration\index{fibration}
$\pi_2:~\CF^{6|12}\rightarrow\CL^{5|6}$ from
\eqref{ambidblfibration2} are spanned by the holomorphic vector
fields\index{holomorphic!vector fields}
\begin{align}\label{ambivec1}
W^{(a)}&\ :=\ \mu_{(a)}^\alpha\lambda_{(a)}^\ald\dpar_{\alpha\ald}~,~~~
D^i_{(a)}\ =\ \lambda_{(a)}^\ald D_\ald^i~~~\mbox{and}~~~
D^{(a)}_i\ =\ \mu_{(a)}^\alpha D_{\alpha i}~,
\end{align}
where $D_{\alpha i}$ and $D^i_\ald$ are the superderivatives
defined by
\begin{equation}
D_{\alpha i}\ :=\ \der{\theta^{\alpha
i}}+\eta_i^\ald\der{x^{\alpha\ald}}\eand D_\ald^i\ :=\
\der{\eta^\ald_i}+\theta^{\alpha i}\der{x^{\alpha\ald}}~.
\end{equation}

Recall that there is a one-to-one correspondence between
isomorphism classes of vector bundles and locally free sheaves and\index{morphisms!isomorphism}
therefore the superambitwistor space $\CL^{5|6}$ corresponds in a\index{ambitwistor space}\index{twistor}\index{twistor!ambitwistor}\index{twistor!space}
natural way to the sheaf $\CCL^{5|6}$ of holomorphic sections of\index{sheaf}
the bundle $\CL^{5|6}\rightarrow \CPP^1\times \CPP^1_*$.

\paragraph{$\CL^{5|6}$ as a Calabi-Yau supermanifold.} Just as the space\index{Calabi-Yau}\index{Calabi-Yau supermanifold}\index{super!manifold}
$\CP^{3|4}$, the superambitwistor space $\CL^{5|6}$ is a\index{ambitwistor space}\index{twistor}\index{twistor!ambitwistor}\index{twistor!space}
Calabi-Yau supermanifold. To prove this, note that it is\index{Calabi-Yau}\index{Calabi-Yau supermanifold}\index{super!manifold}
sufficient to show that the tangent bundle of the body $\CL^5$ of\index{body}
$\CL^{5|6}$ has first Chern number $6$, which is then cancelled by\index{Chern number}\index{first Chern number}
the contribution of $-6$ from the (unconstrained) fermionic
tangent directions. Consider therefore the map
\begin{equation}
\kappa:(z^\alpha_{(a)},\eta_i^{(a)},\lambda_\ald^{(a)},u^\ald_{(a)},\theta^i_{(a)},\mu_\alpha^{(a)})
\ \mapsto\  (\kappa_{(a)},\lambda_\ald^{(a)},\mu_\alpha^{(a)})~,
\end{equation}
where $\kappa_{(a)}$ has been defined in \eqref{quadric2b}. This
map is a vector bundle morphism and gives rise to the short exact
sequence\index{short exact sequence}
\begin{equation}\label{ambiexactseq}
0\ \longrightarrow\ \CL^5\ \longrightarrow\
\CO(1,0)\oplus\CO(1,0)\oplus\CO(0,1)\oplus\CO(0,1)\
\stackrel{\kappa}{\longrightarrow}\ \CO(1,1)\ \longrightarrow\ 0~,
\end{equation}
where $\CO(m,n)$ is a line bundle over the base
$\CPP^1\times\CPP^1_*$ having first Chern numbers $m$ and $n$ with\index{Chern number}\index{first Chern number}
respect to the two $\CPP^1$s in the base. The first and second
Chern classes of the bundles in this sequence are elements of\index{Chern class}
$H^2(\CPP^1\times\CPP^1,\RZ)\cong \RZ\times\RZ$ and
$H^4(\CPP^1\times\CPP^1,\RZ)\cong \RZ$, respectively. Let us
denote the elements of $H^2(\CPP^1\times \CPP^1,\RZ)$ by $i h_1+j
h_2$ and the elements of $H^4(\CPP^1\times \CPP^1,\RZ)$ by $k h_1
h_2$ with $i,j,k\in\RZ$. (That is, $h_1$, $h_2$ and $h_1h_2$ are
the generators of the respective cohomology groups.) Then the
short exact sequence \eqref{ambiexactseq} together with the\index{short exact sequence}
Whitney product formula yields\index{Whitney product formula}
\begin{equation}
(1+h_1)(1+h_1)(1+h_2)(1+h_2)\ =\ (1+\alpha_1 h_1+\alpha_2
h_2+\beta h_1 h_2)(1+ h_1+h_2)~,
\end{equation}
where $\alpha_1+\alpha_2$ and $\beta$ are the first and second
Chern numbers of $\CL^5$ considered as a holomorphic vector bundle\index{Chern number}\index{holomorphic!vector bundle}
over $\CPP^1\times \CPP^1_*$. It follows that $c_1=2$ (and
$c_2=4$), and taking into account the contribution of the tangent
space to the base\footnote{Recall that $T^{1,0}\CPP^1 \cong
\CO(2)$.} $\CPP^1\times \CPP^1_*$, we conclude that the tangent
space to $\CL^5$ has first Chern number 6.\index{Chern number}\index{first Chern number}

Since $\CL^{5|6}$ is a Calabi-Yau supermanifold, this space can be\index{Calabi-Yau}\index{Calabi-Yau supermanifold}\index{super!manifold}
used as a target space for the topological B-model. However, it is\index{target space}\index{topological!B-model}
still unclear what the corresponding gauge theory action will look
like. The most obvious guess would be some holomorphic BF-type
theory, see section \ref{ssrelFieldTheories}, \ref{pholBF}, with
$B$ a ``Lagrange multiplier (0,3)-form''.

\paragraph{Reality conditions on the superambitwistor space.}\index{ambitwistor space}\index{twistor}\index{twistor!ambitwistor}\index{twistor!space}
Recall that there is a real structure which leads to Kleinian\index{real structure}
signature on the body of the moduli space $\FR^{4|2\CN}$ of real\index{body}\index{moduli space}
holomorphic sections of the fibration $\pi_2$ in\index{fibration}
\eqref{superdblfibration}. Furthermore, if $\CN$ is even, one can
define a second real structure which yields Euclidean signature.\index{real structure}
Above, we saw that the superambitwistor space $\CL^{5|6}$\index{ambitwistor space}\index{twistor}\index{twistor!ambitwistor}\index{twistor!space}
originates from two copies of $\CP^{3|3}$ and therefore, we cannot
impose the Euclidean reality condition. However, besides the real
structure leading to a Kleinian signature, one can additionally\index{Kleinian signature}\index{real structure}
impose a reality condition for which we obtain a Minkowski metric\index{metric}
on the body of $\FR^{4|4\CN}$. In the following, we will focus on\index{body}
the latter.

Consider the anti-linear involution $\tau_M$ which acts on the\index{involution}
coordinates of $\CL^{5|6}$ according to
\begin{equation}\label{reality}
\tau_M(z^\alpha_\pm,\lambda_\ald^\pm,\eta_i^\pm;u^\ald,\mu_\alpha^\pm,\theta^i_\pm)\
:=\
\left(-\overline{u^\ald_\pm},\overline{\mu_\alpha^\pm},\overline{\theta^i_\pm};
-\overline{z^\alpha_\pm},\overline{\lambda_\ald^\pm},\overline{\eta_i^\pm}\right)~.
\end{equation}
Sections of the bundle $\CL^{5|6}\rightarrow \CPP^1\times
\CPP^1_*$ which are $\tau_M$-real are thus parameterized by moduli
satisfying
\begin{equation}
x^{\alpha\bed}\ =\ -\overline{x^{\bed\alpha}}\eand \eta_i^\ald\ =\
\overline{\theta^{\alpha i}}~.
\end{equation}
We can extract furthermore the contained real coordinates via the
identification
\begin{equation}\label{realcomponents}
\begin{aligned}
x^{1\dot{1}}&\ =\ -\di x^0-\di x^3~,~~~&x^{1\dot{2}}&\ =\ -\di x^1-x^2~,~\\
x^{2\dot{1}}&\ =\ -\di x^1+x^2~,~~~&x^{2\dot{2}}&\ =\ -\di x^0+\di
x^3~,
\end{aligned}
\end{equation}
and obtain a metric of signature $(3,1)$ on $\FR^4$ from $\dd\index{metric}
s^2:=\det(\dd x^{\alpha\ald})$. In this section, we will always
adopt this convention, even in the complexified Euclidean
situation.

\subsection{The Penrose-Ward transform on the\index{Penrose-Ward transform}
superambitwistor space}\label{ssPWSuperambitwistorSpace}\index{ambitwistor space}\index{twistor}\index{twistor!ambitwistor}\index{twistor!space}

\paragraph{The holomorphic vector bundle $\CE$.} Let $\CE$ be a\index{holomorphic!vector bundle}
topologically trivial holomorphic vector bundle of rank $n$ over
$\CL^{5|6}$ which becomes holomorphically trivial when restricted
to any subspace $(\CPP^1\times
\CPP^1)_{x,\eta,\theta}\embd\CL^{5|6}$. Due to the equivalence of
the \v{C}ech and the Dolbeault descriptions of holomorphic vector
bundles, we can describe $\CE$ either by holomorphic transition\index{holomorphic!vector bundle}
functions $\{f_{ab}\}$ or by a holomorphic structure\index{holomorphic!structure}
$\dparb_{\hat{\CA}}=\dparb+\hat{\CA}$: Starting from a transition
function $f_{ab}$, there is a splitting\index{transition function}
\begin{equation}
f_{ab}\ =\ \hpsi_a^{-1}\hpsi_b~,
\end{equation}
where the $\hpsi_a$ are smooth $\sGL(n,\FC)$-valued
functions\footnote{In fact, the collection $\{\hpsi_a\}$ forms a
\v{C}ech 0-cochain.} on $\CU_{(a)}$, since the bundle $\CE$ is
topologically trivial. This splitting allows us to switch to the
holomorphic structure $\dparb+\hat{\CA}$ with\index{holomorphic!structure}
$\hat{\CA}=\hpsi\dparb\hpsi^{-1}$, which describes a trivial
vector bundle $\hat{\CE}\cong \CE$. Note that the additional
condition of holomorphic triviality of $\CE$ on subspaces
$(\CPP^1\times \CPP^1)_{x,\eta,\theta}$ will restrict the explicit
form of $\hat{\CA}$.

\paragraph{Relation to $\CN=3$ SYM theory.}
Back at the bundle $\CE$, consider its pull-back $\pi_2^*\CE$ with
transition functions $\{\pi_2^*f_{ab}\}$, which are constant along\index{transition function}
the fibres of $\pi_2:\CF^{6|12}\rightarrow\CL^{5|6}$:
\begin{equation}\label{ambWcond1}
W^{(a)}\pi_2^*f_{ab}\ =\ D^i_{(a)}\pi_2^*f_{ab}\ =\
D_i^{(a)}\pi_2^*f_{ab}\ =\ 0~,
\end{equation}
The additional assumption of holomorphic triviality upon reduction
onto a subspace allows for a splitting
\begin{equation}\label{splittinghol}
\pi_2^*f_{ab}\ =\ \psi_a^{-1}\psi_b
\end{equation}
into $\sGL(n,\FC)$-valued functions $\{\psi_a\}$ which are
holomorphic on $\tilde{\CU}_{(a)}$: Evidently, there is such a
splitting holomorphic in the coordinates $\lambda_{(a)}$ and
$\mu_{(a)}$ on $(\CPP^1\times \CPP^1)_{x,\eta,\theta}$, since
$\CE$ becomes holomorphically trivial when restricted to these
spaces. Furthermore, these subspaces are holomorphically
parameterized by the moduli
$(x^{\alpha\ald},\eta^\ald_i,\theta^{\alpha i})$, and thus the
splitting \eqref{splittinghol} is holomorphic in all the
coordinates of $\CF^{6|12}$. Due to \eqref{ambWcond1}, we have on
the intersections $\tilde{\CU}_{(a)}\cap\tilde{\CU}_{(b)}$
\begin{subequations}
\begin{align}\label{ambpbgauge5}
\psi_a D^i_{(a)}\psi^{-1}_a\ =\ \psi_b D^i_{(a)}\psi^{-1}_b&\ =:\
\lambda^\ald_{(a)}\CA_\ald^i~,\\\label{ambpbgauge6} \psi_a
D_i^{(a)}\psi^{-1}_a\ =\ \psi_b D_i^{(a)}\psi^{-1}_b&\ =:\
\mu^\alpha_{(a)}\CA_{\alpha i}~,\\\label{ambpbgauge7} \psi_a
W^{(a)}\psi^{-1}_a\ =\ \psi_b W^{(a)}\psi^{-1}_b&\ =:\
\mu^\alpha_{(a)} \lambda^\ald_{(a)}\CA_{\alpha\ald}~,
\end{align}
\end{subequations}
where $\CA_\ald^i$, $\CA_{\alpha i}$ and $\CA_{\alpha\ald}$ are
independent of $\mu_{(a)}$ and $\lambda_{(a)}$. The introduced
components of the supergauge potential $\CA$ fit into the linear
system\index{linear system}
\begin{subequations}
\begin{align}\label{amblinsys21}
\mu_{(a)}^\alpha\lambda_{(a)}^\ald(\dpar_{\alpha\ald}+
\CA_{\alpha\ald})\psi_a&\ =\ 0~,\\
\lambda_{(a)}^\ald(D^i_{\ald}+\CA^i_{\ald})\psi_a&\ =\
0~,\\\label{amblinsys24} \mu_{(a)}^\alpha(D_{\alpha i}+\CA_{\alpha
i})\psi_a&\ =\ 0~,
\end{align}
\end{subequations}
whose compatibility conditions are\index{compatibility conditions}
\begin{equation}\label{ambcompcond1}
\begin{aligned}
\{\nabla_\ald^i,\nabla_\bed^j\}+\{\nabla_\bed^i,\nabla_\ald^j\}\ =\ 0~,~~~
\{\nabla_{\alpha i},\nabla_{\beta j}\}+\{\nabla_{\beta i},
\nabla_{\alpha j}\}\ =\ 0~,\\ \{\nabla_{\alpha
i},\nabla^j_{\ald}\}-2\delta_i^j\nabla_{\alpha\ald}\ =\ 0~.\hspace{3cm}
\end{aligned}
\end{equation}
Here, we used the obvious shorthand notations
$\nabla_\ald^i:=D_\ald^i+\CA_\ald^i$, $\nabla_{\alpha
i}:=D_{\alpha i}+\CA_{\alpha i}$, and
$\nabla_{\alpha\ald}=\dpar_{\alpha\ald}+\CA_{\alpha\ald}$.
Equations \eqref{ambcompcond1} are well known to be equivalent to
the equations of motion of $\CN=3$ SYM theory on\footnote{Note
that most of our considerations concern the complexified case.}
$\FC^4$ \cite{Witten:1978xx}, and therefore (up to a reality
condition) also to $\CN=4$ SYM theory on $\FC^4$.

We thus showed that there is a correspondence between certain
holomorphic structures on $\CL^{5|6}$, holomorphic vector bundles\index{holomorphic!structure}\index{holomorphic!vector bundle}
over $\CL^{5|6}$ which become holomorphically trivial when
restricted to certain subspaces and solutions to the $\CN=4$ SYM
equations on $\FC^4$. The redundancy in each set of objects is
modded out by considering gauge equivalence classes and
holomorphic equivalence classes of vector bundles, which renders
the above correspondences one-to-one.

\subsection{The mini-superambitwistor space\index{ambitwistor space}\index{mini-superambitwistor space}\index{twistor}\index{twistor!ambitwistor}\index{twistor!space}
$\CL^{4|6}$}\label{ssminisuperambi1}

In this section, we define and examine the mini-superambitwistor\index{twistor}\index{twistor!ambitwistor}
space $\CL^{4|6}$, which we will use to build a Penrose-Ward
transform leading to solutions to $\CN=8$ SYM theory in three\index{N=8 SYM theory@$\CN=8$ SYM theory}\index{Penrose-Ward transform}
dimensions. We will first give an abstract definition of
$\CL^{4|6}$ by a short exact sequence, and present more heuristic\index{short exact sequence}
ways of obtaining the mini-superambitwistor space later.\index{ambitwistor space}\index{mini-superambitwistor space}\index{twistor}\index{twistor!ambitwistor}\index{twistor!space}

\paragraph{Abstract definition of the mini-superambitwistor space.}\index{ambitwistor space}\index{mini-superambitwistor space}\index{twistor}\index{twistor!ambitwistor}\index{twistor!space}
The starting point is the product space $\CP^{2|3}\times
\CP^{2|3}_*$ of two copies of the $\CN=3$ mini-supertwistor space.\index{mini-supertwistor space}\index{twistor}\index{twistor!space}
In analogy to the space $\CP^{3|3}\times \CP^{3|3}_*$, we have
coordinates
\begin{equation}\label{coordminisuperambi}
\left(w^1_{(a)},~w^2_{(a)}\ =\ \lambda_{(a)},~\eta_i^{(a)};~
v^1_{(a)},~v^2_{(a)}\ =\ \mu_{(a)},~\theta^i_{(a)}\right)
\end{equation}
on the patches $\CV_{(a)}$ which are unions of $\CV_\pm$ and
$\CV_\pm^*$:
\begin{equation}
\CV_{(1)}\ :=\ \CV_+\times\CV^*_+~,~~~ \CV_{(2)}\ :=\
\CV_-\times\CV^*_+~,~~~ \CV_{(3)}\ :=\ \CV_+\times\CV^*_-~,~~~
\CV_{(4)}\ :=\ \CV_-\times\CV^*_-~.
\end{equation}
For convenience, let us introduce the subspace $\CPP^1_{\Delta}$
of the base of the fibration $\CP^{2|3}\times\index{fibration}
\CP^{2|3}_*\rightarrow \CPP^1\times \CPP^1_*$ as
\begin{equation}
\CPP^1_{\Delta}\ :=\ \diag(\CPP^1\times \CPP^1_*)\ =\
\{(\mu_\pm,\lambda_\pm)\in\CPP^1\times\CPP^1_*\,|\,\mu_\pm=\lambda_\pm\}~.
\end{equation}
Consider now the map $\xi:\CP^{2|3}\times \CP^{2|3}_*\rightarrow
\CO_{\CPP^1_{\Delta}}(2)$ which is defined by
\begin{equation}
\xi:
\left(w^1_{(a)},w^2_{(a)},\eta_i^{(a)};v^1_{(a)},v^2_{(a)},\theta^i_{(a)}\right)\ \mapsto\ 
\left\{\begin{array}{cl}
\left(w^1_\pm-v^1_\pm+2\theta^i_\pm\eta_i^\pm,\, w^2_\pm\right)
& \mbox{for } w^2_\pm\ =\ v^2_\pm \\
\left(0,\, w^2_{(a)}\right) & \mbox{else}
\end{array}\right.~,
\end{equation}
where $\CO_{\CPP^1_\Delta}(2)$ is the line bundle $\CO(2)$ over
$\CPP^1_\Delta$. In this definition, we used the fact that a point
for which $w^2_\pm=v^2_\pm $ holds, is located at least on one of
the patches $\CV_{(1)}$ and $\CV_{(4)}$. Note in particular that
the map $\xi$ is a morphism of vector bundles. Therefore, we can
define a space $\CL^{4|6}$ via the short exact sequence\index{short exact sequence}
\begin{equation}\label{superminiambiexactseq}
0\ \longrightarrow\ \CL^{4|6}\ \longrightarrow\ \CP^{2|3}\times
\CP^{2|3}_*\ \stackrel{\xi}{\longrightarrow}\
\CO_{\CPP^1_{\Delta}}(2)\ \longrightarrow\ 0~.
\end{equation}
We shall call this space the {\em mini-superambitwistor space} and\index{ambitwistor space}\index{mini-superambitwistor space}\index{twistor}\index{twistor!ambitwistor}\index{twistor!space}
denote the restrictions of the patches $\CV_{(a)}$ to $\CL^{4|6}$
by $\hat{\CV}_{(a)}$.

\paragraph{$\CL^{4|6}$ is not a vector bundle.} An important consequence
of this definition is that the sheaf $\CCL^{4|6}$ of holomorphic\index{sheaf}
sections of $\CL^{4|6}$ is {\em not} a locally free sheaf, because\index{locally free sheaf}
over any open neighborhood of $\CPP^1_{\Delta}$, it is impossible
to write $\CL^{4|6}$ as a direct sum of line bundles. This is
simply due to the fact that the stalks over $\CPP^1_{\Delta}$ are
isomorphic to the stalks of $\CCO_{\CPP^1_{\Delta}}(2)$, while the
stalks over $(\CPP^1\times\CPP^1_*)\backslash \CPP^1_{\Delta}$ are
isomorphic to the stalks of $\CCO_{\CPP^1\times \CPP^1_*}(2,2)$.

It immediately follows that the space $\CL^{4|6}$ is not a vector
bundle. However, one can easily see that $p:\CL^{4|6}\rightarrow
\CPP^1\times \CPP^1_*=:B$ is a fibration since the necessary\index{fibration}
homotopy lifting property is inherited from the one on\index{homotopy lifting property}
$\CL^{5|6}$. Given a commutative diagram
\begin{equation}
\begin{aligned}
\begin{picture}(50,50)
\put(0.0,40.0){\makebox(0,0)[c]{$X\times \{0\}$}}
\put(0.0,0.0){\makebox(0,0)[c]{$X\times [0,1]$}}
\put(65.0,40.0){\makebox(0,0)[c]{$\CL^{4|6}$}}
\put(65.0,0.0){\makebox(0,0)[c]{$B$}}
\put(27.5,0.0){\vector(1,0){25}} \put(27.5,40.0){\vector(1,0){25}}
\put(0.0,32.0){\vector(0,-1){20}}
\put(65.0,32.0){\vector(0,-1){20}}
\put(73.0,23.0){\makebox(0,0)[c]{$p$}}
\put(10.0,23.0){\makebox(0,0)[c]{$r$}}
\put(38,8.0){\makebox(0,0)[c]{$h_t$}}
\put(38,48.0){\makebox(0,0)[c]{$h$}}
\end{picture}
\end{aligned}
\end{equation}
the homotopy lifting property demands a map\index{homotopy lifting property}
$g:X\times[0,1]\rightarrow \CL^{4|6}$, which turns the commutative
square diagram into two commutative triangle diagrams. One can now
always lift the map $h$ to a map $\hat{h}:X\times\{0\}\rightarrow
\CL^{5|6}$ and since $\CL^{5|6}$ is a vector bundle over
$\CPP^1\times \CPP^1_*$ and thus a fibration, there is a map\index{fibration}
$\hat{g}:X\times [0,1]\rightarrow \CL^{5|6}$ which leads to two
commutative triangle diagrams in the square diagram
\begin{equation}
\begin{aligned}
\begin{picture}(50,50)
\put(0.0,40.0){\makebox(0,0)[c]{$X\times \{0\}$}}
\put(0.0,0.0){\makebox(0,0)[c]{$X\times [0,1]$}}
\put(65.0,40.0){\makebox(0,0)[c]{$\CL^{5|6}$}}
\put(65.0,0.0){\makebox(0,0)[c]{$B$}}
\put(27.5,0.0){\vector(1,0){25}} \put(27.5,40.0){\vector(1,0){25}}
\put(0.0,32.0){\vector(0,-1){20}}
\put(65.0,32.0){\vector(0,-1){20}}
\put(73.0,23.0){\makebox(0,0)[c]{$\hat{p}$}}
\put(10.0,23.0){\makebox(0,0)[c]{$\hat{r}$}}
\put(38,8.0){\makebox(0,0)[c]{$h_t$}}
\put(38,48.0){\makebox(0,0)[c]{$\hat{h}$}}
\end{picture}
\end{aligned}
\end{equation}
The function $g$ we are looking for is then constructed by
composition: $g=\pi\circ\hat{g}$, where $\pi$ is the natural
projection $\pi:\CL^{5|6}\rightarrow \CL^{4|6}$ with $p\circ
\pi=\hat{p}$.

The fact that the space $\CL^{4|6}$ is neither a supermanifold nor\index{super!manifold}
a supervector bundle over $B$ seems at first slightly disturbing.
However, once one is aware of this new aspect, it does not cause
any deep difficulties as far as the twistor correspondence and the\index{twistor}\index{twistor!correspondence}
Penrose-Ward transform are concerned.\index{Penrose-Ward transform}

\paragraph{The mini-superambitwistor space by dimensional\index{ambitwistor space}\index{mini-superambitwistor space}\index{twistor}\index{twistor!ambitwistor}\index{twistor!space}
reduction.} To obtain a clearer picture of the fibration\index{fibration}
$\CL^{4|6}$ and its sections, let us now consider the dimensional
reduction of the space $\CL^{5|6}$. We will first reduce the\index{dimensional reduction}
product space $\CP^{3|3}\times \CP^{3|3}_*$ and then impose the
appropriate reduced quadric condition. For the first step, we want\index{quadric}
to eliminate in both $\CP^{3|3}$ and $\CP^{3|3}_*$ the dependence
on the bosonic modulus $x^2$. Thus we should factorize by
\begin{equation}
\CCT_{(a)}\ =\ \left\{\begin{array}{l}
\der{z^2_+}-z_+^3\der{z^1_+}~~~\mbox{on}~~~\CU_{(1)}\\
z_-^3\der{z^2_-}-\der{z^1_-}~~~\mbox{on}~~~\CU_{(2)}\\
\der{z^2_+}-z_+^3\der{z^1_+}~~~\mbox{on}~~~\CU_{(3)}\\
z_-^3\der{z^2_-}-\der{z^1_-}~~~\mbox{on}~~~\CU_{(4)}
\end{array}\right.\eand
\CCT_{(a)}^*\ =\ \left\{\begin{array}{l}
\der{u^\zd_+}-u_+^\drd\der{u^\ed_+}~~~\mbox{on}~~~\CU_{(1)}\\
\der{u^\zd_+}-u_+^\drd\der{u^\ed_+}~~~\mbox{on}~~~\CU_{(2)}\\
u_-^\drd\der{u^\zd_-}-\der{u^\ed_-}~~~\mbox{on}~~~\CU_{(3)}\\
u_-^\drd\der{u^\zd_-}-\der{u^\ed_-}~~~\mbox{on}~~~\CU_{(4)}
\end{array}\right.~,
\end{equation}
which leads us to the orbit space
\begin{equation}
\CP^{2|3}\times \CP^{2|3}_*\ =\ (\CP^{3|3}/\CG)\times
(\CP^{3|3}_*/\CG^*)~,
\end{equation}
where $\CG$ and $\CG^*$ are the Abelian groups generated by $\CCT$
and $\CCT^*$, respectively. Recall that the coordinates we use on
this space have been defined in \eqref{coordminisuperambi}. The
global sections of the bundle $\CP^{2|4}\times
\CP^{2|4}_*\rightarrow \CPP^1\times \CPP^1_*$ are captured by the
parameterization
\begin{equation}\label{miniambisections}
w^1_{(a)}\ =\
y^{\ald\bed}\lambda_\ald^{(a)}\lambda_\bed^{(a)}~,~~~ v^1_{(a)}\
=\ y^{\ald\bed}_*\mu_\ald^{(a)}\mu_\bed^{(a)}~,~~~\theta^i_{(a)}\
=\ \theta^{\ald i}\mu_\ald^{(a)}~,~~~ \eta^{(a)}_i\ =\
\eta_i^\ald\lambda_\ald^{(a)}~,
\end{equation}
where we relabel the indices of $\mu_\alpha^{(a)}\rightarrow
\mu_\ald^{(a)}$ and the moduli $y^{\alpha\beta}_*\rightarrow
y^{\ald\bed}_*$, $\theta^{i \alpha}\rightarrow\theta^{i \ald}$,
since there is no distinction between left- and right-handed
spinors on $\FR^3$ or its complexification $\FC^3$.\index{Spinor}\index{complexification}

The next step is obviously to impose the quadric condition, gluing\index{quadric}
together the self-dual and anti-self-dual parts. Note that when\index{anti-self-dual}
acting with $\CCT$ and $\CCT^*$ on $\kappa_{(a)}$ as given in
\eqref{quadric2b}, we obtain
\begin{equation}\label{actiononkappa}
\begin{aligned}
\CCT_{(1)}\kappa_{(1)}&\ =\ \CCT_{(1)}^*\kappa_{(1)}\ =\
(\mu_+-\lambda_+)~,& \CCT_{(2)}\kappa_{(2)}&\ =\
\CCT_{(2)}^*\kappa_{(2)}\ =\ (\lambda_-\mu_+-1)~,\\
\CCT_{(3)}\kappa_{(3)}&\ =\ \CCT_{(3)}^*\kappa_{(3)}\ =\
(1-\lambda_+\mu_-)~,&\CCT_{(4)}\kappa_{(4)}&\ =\
\CCT_{(4)}^*\kappa_{(4)}\ =\ (\lambda_--\mu_-)~.
\end{aligned}
\end{equation}
This implies that the orbits generated by $\CCT$ and $\CCT^*$
become orthogonal to the orbits of $\der{\kappa}$ only at
$\mu_\pm=\lambda_\pm$. Therefore, it is sufficient to impose the
quadric condition $\kappa_{(a)}=0$ at $\mu_\pm=\lambda_\pm$, after\index{quadric}
which this condition will automatically be satisfied at the
remaining values of $\mu_\pm$ and $\lambda_\pm$. Altogether, we
are simply left with
\begin{equation}\label{quadricred}
\left.\left(w^1_{\pm}-v^1_{\pm}+
2\theta_{\pm}^i\eta_i^{\pm}\right)\right|_{\lambda_\pm=\mu_\pm} \ =\ \ 0~,
\end{equation}
and the subset of $\CP^{2|3}\times \CP^{2|3}_*$ which satisfies
this condition is obviously identical to the mini-superambitwistor\index{twistor}\index{twistor!ambitwistor}
space $\CL^{4|6}$ defined above.

The condition \eqref{quadricred} naturally fixes the
parameterization of global sections of the fibration $\CL^{4|6}$\index{fibration}
by giving a relation between the moduli used in
\eqref{miniambisections}. This relation is completely analogous to
\eqref{chiralcoords} and reads
\begin{equation}\label{chiralredcoords}
y^{\ald\bed}\ =\ y^{\ald\bed}_0-\theta^{(\ald i}\eta_i^{\bed)}~~~
\mbox{and}~~~y_*^{\ald\bed}\ =\ y_0^{\ald\bed}+\theta^{(\ald
i}\eta^{\bed)}_i~.
\end{equation}
We clearly see that this parameterization arises from
\eqref{chiralcoords} by dimensional reduction from\index{dimensional reduction}
$\FC^4\rightarrow \FC^3$. Thus indeed, imposing the condition
\eqref{quadricred} only at $\lambda_\pm=\mu_\pm$ is the
dimensionally reduced analogue of imposing the condition
\eqref{quadric2b} on $\CP^{3|3}\times \CP^{3|3}_*$.

\paragraph{Comments on further ways of constructing $\CL^{4|6}$.} Although
the construction presented above seems most natural, one can
imagine other approaches of defining the space $\CL^{4|6}$.
Completely evident is a second way, which uses the description of
$\CL^{5|6}$ in terms of coordinates on $\CF^{6|12}$. Here, one
factorizes the correspondence space $\CF^{6|12}$ by the groups
generated by the vector field $\CCT_2=\CCT_2^*$ and obtains the
correspondence space $\CK^{5|12}\cong \FC^{3|12}\times
\CPP^1\times \CPP^1_*$ together with equation
\eqref{chiralredcoords}. A subsequent projection $\pi_2$ from the
dimensionally reduced correspondence space $\CK^{5|12}$ then
yields the mini-superambitwistor space $\CL^{4|6}$ as defined\index{ambitwistor space}\index{mini-superambitwistor space}\index{twistor}\index{twistor!ambitwistor}\index{twistor!space}
above.

Furthermore, one can factorize $\CP^{3|3}\times \CP^{3|3}_*$ only
by $\CG$ to eliminate the dependence on one modulus. This will
lead to $\CP^{2|3}\times \CP^{3|3}_*$ and following the above
discussion of imposing the quadric condition on the appropriate\index{quadric}
subspace, one arrives again at \eqref{quadricred} and the space
$\CL^{4|6}$. Here, the quadric condition already implies the
remaining factorization of $\CP^{2|3}\times\CP^{3|3}_*$ by
$\CG^*$.

Eventually, one could anticipate the identification of moduli in
\eqref{chiralredcoords} and therefore want to factorize by the
group generated by the combination $\CCT+\CCT^*$. Acting with this
sum on $\kappa_{(a)}$ will produce the sum of the results given in
\eqref{actiononkappa}, and the subsequent discussion of the
quadric condition follows the one presented above.\index{quadric}

\paragraph{Double fibration.}\index{double fibration}\index{fibration}
Knowing the parameterization of global sections of the
mini-superambitwistor space fibred over $\CPP^1\times \CPP^1_*$ as\index{ambitwistor space}\index{mini-superambitwistor space}\index{twistor}\index{twistor!ambitwistor}\index{twistor!space}
defined in \eqref{chiralredcoords}, we can establish a double
fibration, similarly to all the other twistor spaces we\index{double fibration}\index{fibration}\index{twistor}\index{twistor!space}
encountered so far. Even more instructive is the following
diagram, in which the dimensional reduction of the involved spaces\index{dimensional reduction}
becomes evident:
\begin{equation}\label{ambdouble}
\begin{aligned}
\begin{picture}(100,85)
\put(0.0,0.0){\makebox(0,0)[c]{$\CL^{4|6}$}}
\put(0.0,52.0){\makebox(0,0)[c]{$\CL^{5|6}$}}
\put(96.0,0.0){\makebox(0,0)[c]{$\FC^{3|12}$}}
\put(96.0,52.0){\makebox(0,0)[c]{$\FC^{4|12}$}}
\put(51.0,33.0){\makebox(0,0)[c]{$\CK^{5|12}$}}
\put(51.0,85.0){\makebox(0,0)[c]{$\CF^{6|12}$}}
\put(37.5,25.0){\vector(-3,-2){25}}
\put(55.5,25.0){\vector(3,-2){25}}
\put(37.5,77.0){\vector(-3,-2){25}}
\put(55.5,77.0){\vector(3,-2){25}}
\put(0.0,45.0){\vector(0,-1){37}}
\put(90.0,45.0){\vector(0,-1){37}}
\put(45.0,78.0){\vector(0,-1){37}}
\put(24.0,78.0){\makebox(0,0)[c]{$\pi_2$}}
\put(72.0,78.0){\makebox(0,0)[c]{$\pi_1$}}
\put(24.0,26.0){\makebox(0,0)[c]{$\nu_2$}}
\put(72.0,26.0){\makebox(0,0)[c]{$\nu_1$}}
\end{picture}
\end{aligned}
\end{equation}
The upper half is just the double fibration for the quadric\index{double fibration}\index{fibration}\index{quadric}
\eqref{ambidblfibration2}, while the lower half corresponds to the
dimensionally reduced case. The reduction of $\FC^{4|12}$ to
$\FC^{3|12}$ is obviously done by factorizing with respect to the
group generated by $\CCT_2$. The same is true for the reduction of
$\CF^{6|12}\cong \FC^{4|12}\times\CPP^1\times\CPP^1_*$ to
$\CK^{5|12}\cong \FC^{3|12}\times\CPP^1\times\CPP^1_*$. The
reduction from $\CL^{5|6}$ to $\CL^{4|6}$ was given above and the
projection $\nu_2$ from $\CK^{5|12}$ onto $\CL^{4|6}$ is defined
by equations \eqref{miniambisections}. The four patches covering
$\CF^{6|12}$ will be denoted by
$\tilde{\CV}_{(a)}:=\nu_2^{-1}(\hat{\CV}_{(a)})$.

The double fibration defined by the projections $\nu_1$ and\index{double fibration}\index{fibration}
$\nu_2$ yields the following twistor correspondences:\index{twistor}\index{twistor!correspondence}
\begin{equation}
\begin{aligned}
\left\{\,\mbox{subspaces $(\CPP^1\times \CPP^1)_{y_0,\eta,\theta}$
in $\CL^{4|6}$} \right\}&\ \longleftrightarrow\
\left\{\,\mbox{points $(y_0,\eta,\theta)$ in $\FC^{3|12}$}\right\}~, \\
\left\{\,\mbox{generic points $p$ in $\CL^{4|6}$}\right\}&\
\longleftrightarrow\ \left\{\,\mbox{null superlines in
$\FC^{3|12}$}\right\}~,\\ \left\{\,\mbox{points $p$ in $\CL^{4|6}$
with $\lambda_\pm\ =\ \mu_\pm$}\right\}&\ \longleftrightarrow\
\left\{\,\mbox{superplanes in $\FC^{3|12}$}\right\}~.
\end{aligned}
\end{equation}
The null superlines and the superplanes in $\FC^{3|12}$ are
defined as the sets $\{(y^{\ald\bed},\eta_i^\ald,\theta^{\ald
i})\}$ with
\begin{equation*}
\begin{aligned}
y^{\ald\bed}&\ =\
\hat{y}^{\ald\bed}+t\lambda_{(a)}^{(\ald}\mu_{(a)}^{\bed)}~,~&\eta^\ald_i&\
=\ \hat{\eta}^\ald_i+\eps_i\lambda_{(a)}^\ald~,&\theta^{\ald i}\
=\
\hat{\theta}^{\ald i}+\tilde{\eps}^i\mu_{(a)}^\ald~,\\
y^{\ald\bed}&\ =\
\hat{y}^{\ald\bed}+\kappa^{(\ald}\lambda_{(a)}^{\bed)}~,&\eta^\ald_i&\
=\ \hat{\eta}^\ald_i+\eps_i\lambda_{(a)}^\ald~,&\theta^{\ald i}\
=\ \hat{\theta}^{\ald i}+\tilde{\eps}^i\lambda^\ald_{(a)}~,
\end{aligned}
\end{equation*}
where $t$, $\kappa^\ald$, $\eps_i$ and $\tilde{\eps}^i$ are an
arbitrary complex number, a complex commuting two-spinor and two\index{Spinor}
three-vectors with Gra{\ss}mann-odd components, respectively. Note
that in the last line, $\lambda^\ald_\pm=\mu_\pm^\ald$, and we
could also have written
\begin{equation*}
\{(y^{\ald\bed},\eta_i^\ald,\theta^{\ald i})\}~~~\mbox{with}~~~
y^{\ald\bed}\ =\
\hat{y}^{\ald\bed}+\kappa^{(\ald}\mu_{(a)}^{\bed)}~,~~~\eta^\ald_i\
=\ \hat{\eta}^\ald_i+\eps_i\mu_{(a)}^\ald~,~~~\theta^{\ald i}\ =\
\hat{\theta}^{\ald i}+\tilde{\eps}^i\mu_{(a)}^\ald~.
\end{equation*}

The vector fields spanning the tangent spaces to the leaves of the
fibration $\nu_2$ are for generic values of $\mu_\pm$ and\index{fibration}
$\lambda_\pm$ given by
\begin{equation}\label{vecfieldsminiambi}
\begin{aligned}
W^{(a)}&\ :=\ \mu_{(a)}^\ald\lambda_{(a)}^\bed\dpar_{(\ald\bed)}~,\\
\tilde{D}^i_{(a)}&\ :=\ \lambda_{(a)}^\bed\tilde{D}^i_\bed\ :=\
\lambda_{(a)}^\bed\left(\der{\eta_i^\bed}+\theta^{\ald
i}\dpar_{(\ald\bed)}\right)~,\\
D^{(a)}_i&\ :=\ \mu_{(a)}^\ald D_{\ald i}\ :=\
\mu_{(a)}^\ald\left(\der{\theta^{\ald
i}}+\eta_i^\bed\dpar_{(\ald\bed)}\right)~,
\end{aligned}
\end{equation}
where the derivatives $\dpar_{(\ald\bed)}$ have been defined in
\eqref{dimredcoordinates} and \eqref{dimredderivatives}. At
$\mu_\pm=\lambda_\pm$, however, the fibres of the fibration\index{fibration}
$\CL^{4|6}$ over $\CPP^1\times \CPP^1_*$ loose one bosonic
dimension. As the space $\CK^{5|12}$ is a manifold, this means\index{manifold}
that this dimension has to become tangent to the projection
$\nu_2$. In fact, one finds that over $\CPP^1_{\Delta}$, besides
the vector fields given in \eqref{vecfieldsminiambi}, also the
vector fields
\begin{equation}
\tilde{W}^\pm_\bed\ =\ \mu^\ald_\pm\dpar_{(\ald\bed)}\ =\
\lambda^\ald_\pm\dpar_{(\ald\bed)}
\end{equation}
annihilate the coordinates on $\CL^{4|6}$. Therefore, the leaves
to the projection $\nu_2:\CK^{5|12}\rightarrow \CL^{4|6}$ are of
dimension $2|6$ for $\mu_\pm= \lambda_\pm$ and of dimension $1|6$
everywhere else.

\paragraph{Real structure on $\CL^{4|6}$.} Quite evidently, a real\index{real structure}
structure on $\CL^{4|6}$ is inherited from the one on $\CL^{5|6}$,
and we obtain directly from \eqref{reality} the action of $\tau_M$
on $\CP^{2|4}\times \CP^{2|4}_*$, which is given by
\begin{equation}\label{realitymini}
\tau_M(w^1_\pm,\lambda_\ald^\pm,\eta_i^\pm;v^1_\pm,\mu_\ald^\pm,\theta^i_\pm)\
:=\
\left(-\overline{v^1_\pm},\overline{\mu_\ald^\pm},\overline{\theta^i_\pm};
-\overline{w^1_\pm},\overline{\lambda_\ald^\pm},\overline{\eta_i^\pm}\right)~.
\end{equation}
This action descends in an obvious manner to $\CL^{4|6}$, which
leads to a real structure on the moduli space $\FC^{3|12}$ via the\index{moduli space}\index{real structure}
double fibration \eqref{ambdouble}. Thus, we have as the resulting\index{double fibration}\index{fibration}
reality condition
\begin{equation}
y_0^{\ald\bed}\ =\ -\overline{y_0^{\bed\ald}}\eand \eta_i^\ald\ =\
\overline{\theta^{\ald i}}~,
\end{equation}
and the identification of the bosonic moduli $y^{\ald\bed}$ with
the coordinates on $\FR^3$ reads as
\begin{equation}\label{realcomponentsmini}
\begin{aligned}
y_0^{\ed\ed}\ =\ -\di x^0-\di x^3~,~~~y_0^{\ed\zd}\ =\
y_0^{\zd\ed}\ =\ -\di x^1~,~~~y_0^{\zd\zd}\ =\ -\di x^0+\di x^3~.
\end{aligned}
\end{equation}

The reality condition $\tau_M(\cdot)=\cdot$ is indeed fully
compatible with the condition \eqref{quadricred} which reduces
$\CP^{2|4}\times\CP^{2|4}_*$ to $\CL^{4|6}$. The base space
$\CPP^1\times \CPP^1_*$ of the fibration $\CL^{4|6}$ is reduced to\index{fibration}
a single sphere $S^2$ with real coordinates
$\frac{1}{2}(\lambda_\pm+\mu_\pm)=\frac{1}{2}(\lambda_\pm+\bl_\pm)$
and
$\frac{1}{2\di}(\lambda_\pm-\mu_\pm)=\frac{1}{2\di}(\lambda_\pm-\bl_\pm)$,
while the diagonal $\CPP^1_\Delta$ is reduced to a circle
$S^1_\Delta$ parameterized by the real coordinates
$\frac{1}{2}(\lambda_\pm+\bl_\pm)$. The $\tau_M$-real sections of
$\CL^{4|6}$ have to satisfy $w^1_{\pm}=\tau_M(w^1_\pm)=\bar{v}^1_\pm$.
Thus, the fibres of the fibration\index{fibration}
$\CL^{4|6}\rightarrow\CPP^1\times\CPP^1_*$, which are of complex
dimension $2|6$ over generic points in the base and complex
dimension $1|6$ over $\CPP^1_{\Delta}$, are reduced to fibres of
real dimension $2|6$ and $1|6$, respectively. In particular, note
that
$\theta^i_\pm\eta^\pm_i=\bar{\eta}^\pm_i\bar{\theta}^i_\pm=-\bar{\theta}^i_\pm\bar{\eta}_i^\pm$
is purely imaginary and therefore the quadric condition\index{quadric}
\eqref{quadricred} together with the real structure $\tau_M$\index{real structure}
implies that
$w^1_\pm=\bar{v}^1_\pm=\bar{w}^1_\pm+2\bar{\theta}^i_\pm\bar{\eta}^\pm_i$
for $\lambda_\pm=\mu_\pm=\bl_\pm$. Thus, the body $\z{w}^1_\pm$ of\index{body}
$w^1_\pm$ is purely real and we have $w^1_\pm\ =\ \z{w}^1_\pm
-\theta^i_\pm\eta_i^\pm$ and $v^1_\pm\ =\ \z{w}^1_\pm
+\theta^i_\pm\eta_i^\pm$ on the diagonal $S^1_{\Delta}$.

\paragraph{Interpretation of the involved real geometries.} For
the best-known twistor correspondences, i.e.\ the correspondence\index{twistor}\index{twistor!correspondence}
\eqref{dblfibration}, its dual and the correspondence
\eqref{ambidblfibration2b}, there is a nice description in terms
of flag manifolds, see e.g.\ the diagrams\index{flag manifold}\index{manifold}
\eqref{flgdblfibration1}, \eqref{flgdblfibration2} and
\eqref{flgdblfibration3} as well as the discussion in
\cite{Wardbook}. For the spaces involved in the twistor\index{twistor}
correspondences including mini-twistor spaces, one has a similarly\index{mini-twistor space}\index{twistor!space}
nice interpretation after restricting to the real situation. For
simplicity, we reduce our considerations to the
bodies\footnote{i.e.\ drop the fermionic directions} of the
involved geometries, as the extension to corresponding
supermanifolds is quite straightforward.\index{super!manifold}

Let us first discuss the double fibration for the mini-twistor\index{double fibration}\index{fibration}\index{twistor}
space, cf.\ \eqref{eq:3.19},
\begin{equation}\label{dblfibrationthreeself}
\begin{aligned}
\begin{picture}(50,40)
\put(0.0,0.0){\makebox(0,0)[c]{$\CP^{2|\CN}$}}
\put(64.0,0.0){\makebox(0,0)[c]{$\FC^{3|2\CN}$}}
\put(34.0,33.0){\makebox(0,0)[c]{$\CK^{5|2\CN}$}}
\put(7.0,18.0){\makebox(0,0)[c]{$\nu_2$}}
\put(55.0,18.0){\makebox(0,0)[c]{$\nu_1$}}
\put(25.0,25.0){\vector(-1,-1){18}}
\put(37.0,25.0){\vector(1,-1){18}}
\end{picture}
\end{aligned}
\end{equation}
and assume that we have imposed a suitable reality condition on
the sections of $\CP^{2|\CN}\rightarrow \CPP^1$, the details of
which are not important. We follow again the usual discussion of
the real case and leave the coordinates on the sphere complex.
\parpic[r]{\includegraphics[width=6.4cm,totalheight=4.5cm]{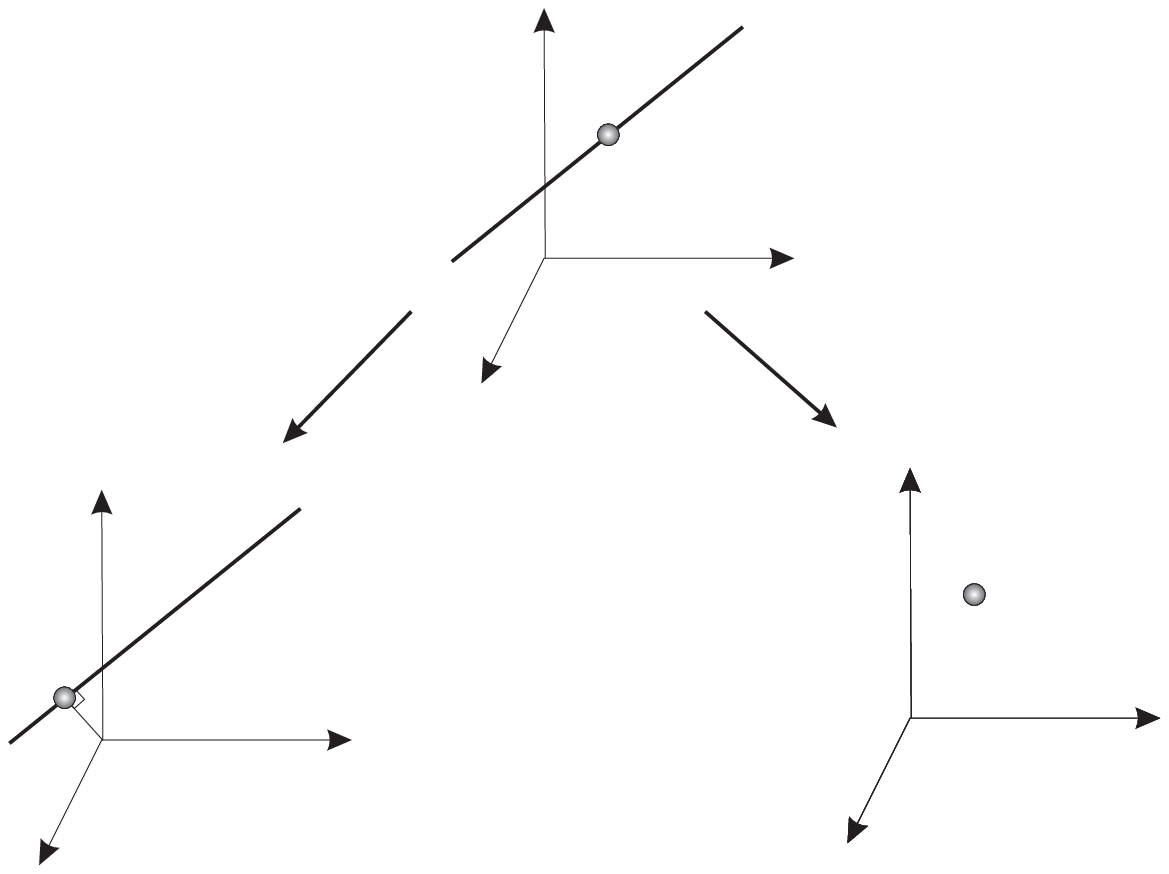}}
\noindent{}As correspondence space on top of the double fibration,\index{double fibration}\index{fibration}
we have thus the space $\FR^{3}\times S^2$, which we can
understand as the set of oriented lines\footnote{not only the ones
through the origin} in $\FR^3$ with one marked point. Clearly, the
point of such a line is given by an element of $\FR^3$, and the
direction of this line in $\FR^3$ is parameterized by a point on
$S^2$. The mini-twistor space $\CP^2\cong \CO(2)$ now is simply\index{mini-twistor space}\index{twistor}\index{twistor!space}
the space of all lines in $\FR^3$ \cite{Hitchin:1982gh}. Similarly
to the case of flag manifolds, the projections $\nu_1$ and $\nu_2$\index{flag manifold}\index{manifold}
in \eqref{dblfibrationthreeself} become therefore obvious. For
$\nu_1$, simply drop the line and keep the marked point. For
$\nu_2$, drop the marked point and keep the line. Equivalently, we
can understand $\nu_2$ as moving the marked point on the line to
its shortest possible distance from the origin. This leads to the
space $TS^2\cong \CO(2)$, where the $S^2$ parameterizes again the
direction of the line, which can subsequently be still moved
orthogonally to this direction, and this freedom is parameterized
by the tangent planes to $S^2$, which are isomorphic to $\FR^2$.
\parpic[l]{\includegraphics[width=6.4cm,totalheight=4.5cm]{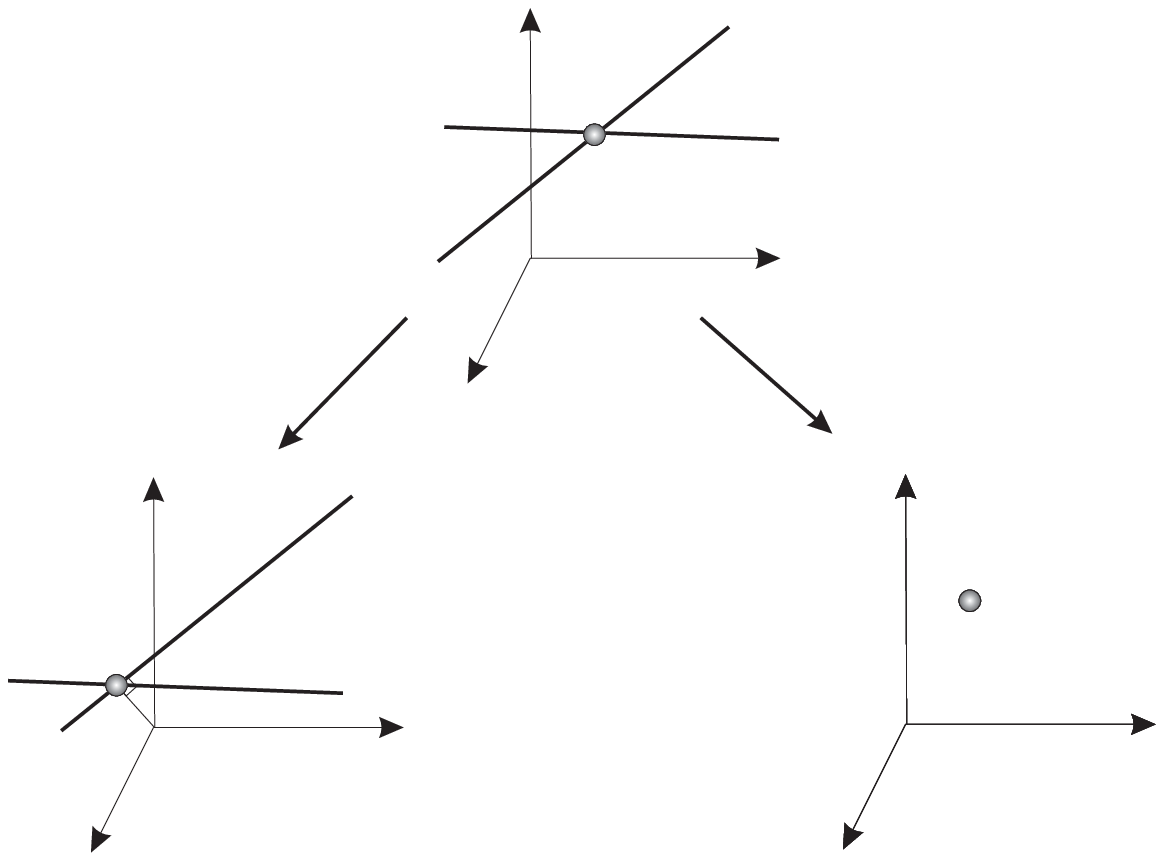}}
\noindent{}Now in the case of the fibration which is included in\index{fibration}
\eqref{ambdouble}, we impose the reality condition
\eqref{realitymini} on the fibre coordinates of $\CL^4$. In the
real case, the correspondence space $\CK^5$ becomes the space
$\FR^3\times S^2\times S^2$ and this is the space of two oriented
lines in $\FR^3$ intersecting in a point. More precisely, this is
the space of two oriented lines in $\FR^3$ each with one marked
point, for which the two marked points coincide. The projections
$\nu_1$ and $\nu_2$ in \eqref{ambdouble} are then interpreted as
follows. For $\nu_1$, simply drop the two lines and keep the
marked point. For $\nu_2$, fix one line and move the marked point
(the intersection point) together with the second line to its
shortest distance to the origin. Thus, the space $\CL^4$ is the
space of configurations in $\FR^3$, in which a line has a common
point with another line at its shortest distance to the origin.

Let us summarize all the above findings in the following table:
\begin{center}
\begin{tabular}[h]{|c|l|}
\hline Space & Relation to $\FR^3$\\\hline
\hline $\FR^3$ & marked points in $\FR^3$ $\phantom{\underbrace{\overbrace{i}}}$\\
\hline $\FR^3\times S^2$ &
oriented lines with a marked point in $\FR^3$ $\phantom{\underbrace{\overbrace{i}}}$\\
\hline \raisebox{-.2cm}[0pt][0pt]{$\CP^2\cong\CO(2)$} & oriented
lines in $\FR^3$ (with a marked point at shortest
dis-$\phantom{\overbrace{i}}$\\[-0.06cm]
& tance to the origin.)$\phantom{\underbrace{i}}$\\
\hline $\FR^3\times S^2\times S^2$ & two oriented lines with a
common
marked point in $\FR^3$ $\phantom{\underbrace{\overbrace{i}}}$\\
\hline \raisebox{-.2cm}[0pt][0pt]{$\CL^4$} & two oriented lines
with a
common marked point at shortest$\phantom{\overbrace{i}}$\\[-0.06cm]
&distance from one of the lines to the origin in
$\FR^3$$\phantom{\underbrace{i}}$\\\hline
\end{tabular}
\end{center}

\paragraph{Remarks concerning a topological B-model on $\CL^{4|6}$.} The\index{topological!B-model}
space $\CL^{4|6}$ is not well-suited as a target space for a\index{target space}
topological B-model since it is not a (Calabi-Yau) manifold.\index{Calabi-Yau}\index{topological!B-model}\index{manifold}
However, one clearly expects that it is possible to define an
analogous model since, if we assume that the conjecture in
\cite{Neitzke:2004pf} is correct, such a model should simply be
the mirror of the mini-twistor string theory considered in\index{string theory}\index{twistor}\index{twistor!string theory}
\cite{Chiou:2005jn}. This model would furthermore yield some
holomorphic Chern-Simons type equations of motion. The latter
equations would then define holomorphic $\CL^{4|6}$-bundles by an\index{L46-bundle@$\CL^{4|6}$-bundle}
analogue of a holomorphic structure. These bundles will be\index{holomorphic!structure}
introduced in section 4.3 and in our discussion, they substitute
the holomorphic vector bundles.\index{holomorphic!vector bundle}

Interestingly, the space $\CL^{4|6}$ has a property which comes
close to vanishing of a first Chern class. Recall that for any\index{Chern class}\index{first Chern class}
complex vector bundle, its Chern classes are Poincar{\'e} dual to the\index{complex!vector bundle}
degeneracy cycles of certain sets of sections (this is a
Gau{\ss}-Bonnet formula). More precisely, to calculate the first Chern
class of a rank $r$ vector bundle, one considers $r$ generic\index{Chern class}\index{first Chern class}
sections and arranges them into an $r\times r$ matrix $L$. The
degeneracy loci on the base space are then given by the zero locus
of $\det(L)$. Clearly, this calculation can be translated directly
to $\CL^{4|6}$.

We will now show that $\CL^{4|6}$ and $\CL^{5|6}$ have equivalent
degeneracy loci, i.e.\ they are equal up to a principal divisor,
which, if we were speaking of ordinary vector bundles, would not
affect the first Chern class. Our discussion simplifies\index{Chern class}\index{first Chern class}
considerably if we restrict our attention to the bodies of the two
supertwistor spaces and put all the fermionic coordinates to zero.\index{twistor}\index{twistor!space}
Instead of the ambitwistor spaces, it is also easier to consider\index{ambitwistor space}\index{twistor!ambitwistor}
the vector bundles $\CP^3\times \CP^3_*$ and $\CP^2\times\CP^2_*$
over $\CPP^1\times \CPP^1_*$, respectively, with the appropriately
restricted sets of sections. Furthermore, we will stick to our
inhomogeneous coordinates and perform the calculation only on the\index{homogeneous coordinates}\index{inhomogeneous coordinates}
patch $\CU_{(1)}$, but all this directly translates into
homogeneous, patch-independent coordinates. The matrices to be
considered are
\begin{equation*}
L_{\CL^5}\ =\ \left(\begin{array}{cccc} x_1^{1\ald}\lambda^+_\ald
& x_2^{1\ald}\lambda^+_\ald & x_3^{1\ald}\lambda^+_\ald &
x_4^{1\ald}\lambda^+_\ald \\
x_1^{2\ald}\lambda^+_\ald & x_2^{2\ald}\lambda^+_\ald &
x_3^{2\ald}\lambda^+_\ald &
x_4^{2\ald}\lambda^+_\ald \\
x_1^{\alpha 1}\mu^+_\alpha & x_2^{\alpha 1}\mu^+_\alpha  &
x_3^{\alpha 1}\mu^+_\alpha  &
x_4^{\alpha 1}\mu^+_\alpha \\
x_1^{\alpha 2}\mu^+_\alpha & x_2^{\alpha 2}\mu^+_\alpha  &
x_3^{\alpha 2}\mu^+_\alpha  & x_4^{\alpha 2}\mu^+_\alpha
\end{array}\right)~,~~~
L_{\CL^4}\ =\ \left(\begin{array}{cc}
y_1^{\ald\bed}\lambda^+_\ald\lambda^+_\bed &
y_2^{\ald\bed}\lambda^+_\ald\lambda^+_\bed \\
y_1^{\ald\bed}\mu^+_\ald\mu^+_\bed &
y_2^{\ald\bed}\mu^+_\ald\mu^+_\bed
\end{array}\right)~,
\end{equation*}
and one computes the degeneracy loci for generic moduli to be
given by the equations
\begin{equation}
(\lambda_+-\mu_+)^2\ =\
0\eand(\lambda_+-\mu_+)(\lambda_+-\varrho_+)\ =\ 0
\end{equation}
on the bases of $\CL^5$ and $\CL^4$, respectively. Here,
$\varrho_+$ is a rational function of $\mu_+$ and therefore it is
obvious that both degeneracy cycles are equivalent.

When dealing with degenerated twistor spaces, one usually retreats\index{twistor}\index{twistor!space}
to the correspondence space endowed with some additional symmetry
conditions \cite{Masonbook}. It is conceivable that a similar
procedure will help to define the topological B-model in our case.\index{topological!B-model}
Also, defining a suitable blow-up of $\CL^{4|6}$ over
$\CPP^1_{\Delta}$ could be the starting point for finding an
appropriate action.

\subsection{The Penrose-Ward transform using mini-ambitwistor spaces}\label{ssminisuperambi2}\index{Penrose-Ward transform}\index{ambitwistor space}\index{twistor}\index{twistor!ambitwistor}\index{twistor!space}

\paragraph{$\CL^{4|6}$-bundles.} Because the mini-superambitwistor\index{L46-bundle@$\CL^{4|6}$-bundle}\index{twistor}\index{twistor!ambitwistor}
space is only a fibration and not a manifold, there is no notion\index{fibration}\index{manifold}
of holomorphic vector bundles over $\CL^{4|6}$. However, our space\index{holomorphic!vector bundle}
is close enough to a manifold to translate all the necessary terms\index{manifold}
in a simple manner.

Let us fix the covering $\frU$ of the total space of the fibration\index{fibration}
$\CL^{4|6}$ to be given by the patches $\CV_{(a)}$ introduced
above. Furthermore, define $\frS$ to be the sheaf of smooth\index{sheaf}
$\sGL(n,\FC)$-valued functions on $\CL^{4|6}$ and $\frH$ to be its
subsheaf consisting of holomorphic $\sGL(n,\FC)$-valued functions\index{sheaf}\index{subsheaf}
on $\CL^{4|6}$, i.e.\ smooth and holomorphic functions which
depend only on the coordinates given in \eqref{miniambisections}
and $\lambda_{(a)},\mu_{(a)}$.

We define a {\em complex $\CL^{4|6}$-bundle} of rank $n$ by a\index{L46-bundle@$\CL^{4|6}$-bundle}
\v{C}ech 1-cocycle $\{f_{ab}\}\in  Z^1(\frU,\frS)$ on $\CL^{4|6}$
in full analogy with transition functions defining ordinary vector\index{transition function}
bundles, see section \ref{subseccohomology}. If the 1-cocycle is
an element of $Z^1(\frU,\frH)$, we speak of a {\em holomorphic
$\CL^{4|6}$-bundle}. Two $\CL^{4|6}$-bundles given by \v{C}ech\index{L46-bundle@$\CL^{4|6}$-bundle}
1-cocycles $\{f_{ab}\}$ and $\{f'_{ab}\}$ are called {\em\index{cocycles}
topologically equivalent} ({\em holomorphically equivalent}) if\index{holomorphically equivalent}
there is a \v{C}ech 0-cochain $\{\psi_{a}\}\in C^0(\frU,\frS)$ (a
\v{C}ech 0-cochain $\{\psi_{a}\}\in C^0(\frU,\frH)$) such that
$f_{ab}=\psi_a^{-1}f'_{ab}\psi_b$. An $\CL^{4|6}$-bundle is called\index{L46-bundle@$\CL^{4|6}$-bundle}
{\em trivial} ({\em holomorphically trivial}) if it is
topologically equivalent (holomorphically equivalent) to the\index{holomorphically equivalent}
trivial $\CL^{4|6}$-bundle given by $\{f_{ab}\}=\{\unit_{ab}\}$.\index{L46-bundle@$\CL^{4|6}$-bundle}

In the corresponding discussion of \v{C}ech cohomology on ordinary\index{Cech cohomology@\v{C}ech cohomology}
manifolds, one can achieve independence of the covering if the\index{manifold}
patches of the covering are all Stein manifolds. An analogous\index{Stein manifold}\index{manifold}
argument should be also applicable here, but for our purposes, it
is enough to restrict to the covering $\frU$.

Besides the \v{C}ech description, it is also possible to introduce
an equivalent Dolbeault description, which will, however, demand
an extended notion of Dolbeault cohomology classes.\index{Dolbeault cohomology}

\paragraph{The Penrose-Ward transform.} With the double fibration\index{Penrose-Ward transform}\index{double fibration}\index{fibration}
contained in \eqref{ambdouble}, it is not hard to establish the
corresponding Penrose-Ward transform, which is essentially a\index{Penrose-Ward transform}
dimensional reduction of the four-dimensional case presented in\index{dimensional reduction}
section 4.1.

On $\CL^{4|6}$, we start from a trivial rank $n$ holomorphic
$\CL^{4|6}$-bundle defined by a 1-cocycle $\{f_{ab}\}$ which\index{L46-bundle@$\CL^{4|6}$-bundle}
becomes a holomorphically trivial vector bundle upon restriction
to any subspace $(\CPP^1\times \CPP^1)_{y,\eta,\theta}\embd
\CL^{4|6}$. The pull-back of the $\CL^{4|6}$-bundle along $\nu_2$\index{L46-bundle@$\CL^{4|6}$-bundle}
is the vector bundle $\tilde{\CE}$ with transition functions\index{transition function}
$\{\nu_2^* f_{ab}\}$ satisfying by definition
\begin{equation}\label{ambWcond13d}
W^{(a)}\nu_2^*f_{ab}\ =\ \tilde{D}^i_{(a)}\nu_2^*f_{ab}\ =\
D_i^{(a)}\nu_2^*f_{ab}\ =\ 0~,
\end{equation}
at generic points of $\CL^{4|6}$ and for $\lambda_\pm=\mu_\pm$, we
have
\begin{equation}\label{ambWcond13Deltad}
\tilde{W}^{(a)}_\ald\nu_2^*f_{ab}\ =\
\tilde{D}^i_{(a)}\nu_2^*f_{ab}\ =\ D_i^{(a)}\nu_2^*f_{ab}\ =\ 0~.
\end{equation}
Restricting the bundle $\tilde{\CE}$ to a subspace $(\CPP^1\times
\CPP^1)_{y,\eta,\theta}\embd \CL^{4|6}\subset \CF^{5|12}$ yields a
splitting of the transition function $\nu_2^*f_{ab}$\index{transition function}
\begin{equation}
\nu_2^*f_{ab}\ =\ \psi_a^{-1}\psi_b~,
\end{equation}
where $\{\psi_a\}$ are again $\sGL(n,\FC)$-valued functions on
$\tilde{\CV}_{(a)}$ which are holomorphic. From this splitting
together with \eqref{ambWcond13d}, one obtains at generic points
of $\CL^{4|6}$ (we will discuss the situation over
$\CPP^1_{\Delta}$ shortly) that
\begin{equation}\label{ambpbgauge}
\begin{aligned}
\psi_a \tilde{D}^i_{(a)}\psi^{-1}_a\ =\ \psi_b
\tilde{D}^i_{(a)}\psi^{-1}_b&\ =:\
\lambda^\ald_{(a)}\tilde{\CA}_\ald^i~,\\
\psi_a D_i^{(a)}\psi^{-1}_a\ =\ \psi_b D_i^{(a)}\psi^{-1}_b&\ =:\
\mu^\ald_{(a)}\CA_{\ald i}~,\\\psi_a W^{(a)}\psi^{-1}_a\ =\ \psi_b
W^{(a)}\psi^{-1}_b&\ =:\ \mu^\ald_{(a)}
\lambda^\bed_{(a)}\CB_{\ald\bed}~,
\end{aligned}
\end{equation}
where $\CB_{\ald\bed}$ is a superfield which decomposes into a
gauge potential and a Higgs field $\Phi$:
\begin{equation}
\CB_{\ald\bed}\ :=\
\CA_{(\ald\bed)}+\tfrac{\di}{2}\eps_{\ald\bed}\Phi~.
\end{equation}
The zeroth order component in the superfield expansion of $\Phi$
will be the seventh real scalar joining the six scalars of $\CN=4$
SYM in four dimensions, which are the zeroth component of the
superfield $\Phi_{ij}$ defined in
\begin{equation}
\{D_{\ald i}+\CA_{\ald i},D_{\bed j}+\CA_{\bed j}\}\ =:\
-2\eps_{\ald\bed}\Phi_{ij}~.
\end{equation}
Thus, as mentioned above, the $\sSpin(7)$ R-symmetry group of
$\CN=8$ SYM theory in three dimensions will not be manifest in\index{N=8 SYM theory@$\CN=8$ SYM theory}
this description.

The equations \eqref{ambpbgauge} are equivalent to the linear
system\index{linear system}
\begin{equation}\label{amblinsys}
\begin{aligned}
\mu_{(a)}^\ald\lambda_{(a)}^\bed(\dpar_{(\ald\bed)}+
\CB_{\ald\bed})\psi_a&\ =\ 0~,\\
\lambda_{(a)}^\ald(\tilde{D}^i_{\ald}+\tilde{\CA}^i_{\ald})\psi_a&\
=\ 0~,\\ \mu_{(a)}^\ald(D_{\ald i}+\CA_{\ald i})\psi_a&\ =\ 0~.
\end{aligned}
\end{equation}
To discuss the corresponding compatibility conditions, we\index{compatibility conditions}
introduce the following differential operators:
\begin{equation}
\begin{aligned}
\tilde{\nabla}_\ald^i\ :=\
\tilde{D}_\ald^i+\tilde{\CA}_\ald^i~,~~~
\nabla_{\ald i}\ :=\ D_{\ald i}+\CA_{\ald i}~,\\
\nabla_{\ald\bed}\ :=\
\dpar_{(\ald\bed)}+\CB_{\ald\bed}~.\hspace{1.7cm}
\end{aligned}
\end{equation}
We thus arrive at
\begin{equation}\label{ambcompcond2}
\begin{aligned}
&\{\tilde{\nabla}_\ald^i,\tilde{\nabla}_\bed^j\}+\{\tilde{\nabla}_\bed^i,\tilde{\nabla}_\ald^j\}\
=\ 0~,~~~ \{\nabla_{\ald i},\nabla_{\bed j}\}+\{\nabla_{\bed i},
\nabla_{\ald j}\}\ =\ 0~,\\ &\hspace{3cm}\{\nabla_{\ald
i},\tilde{\nabla}^j_{\bed}\}-2\delta_i^j\nabla_{\ald\bed}\ =\ 0~,
\end{aligned}
\end{equation}
and one clearly sees that equations \eqref{ambcompcond2} are
indeed equations \eqref{ambcompcond1} after a dimensional
reduction $\FC^4\rightarrow \FC^3$ and defining $\Phi:=A_2$.\index{dimensional reduction}
(Recall that we are reducing the coordinates by $x^2$.) As it is
well known, the supersymmetry (and the R-symmetry) of $\CN=4$ SYM\index{super!symmetry}
theory are enlarged by this dimensional reduction and we therefore\index{dimensional reduction}
obtained indeed $\CN=8$ SYM theory on $\FC^3$.\index{N=8 SYM theory@$\CN=8$ SYM theory}

Let us now examine how the special case $\lambda_\pm=\mu_\pm$ fits
into the picture. One immediately notes that a transition function\index{transition function}
$\nu_2^* f_{ab}$, which satisfies \eqref{ambWcond13d} is of the
form
\begin{equation}
f_{ab}\ =\
f_{ab}(y^{\ald\bed}\lambda^{(a)}_{\ald}\lambda^{(a)}_{\bed},
y^{\ald\bed}\mu^{(a)}_{\ald}\mu^{(a)}_{\bed},\lambda^{(a)}_{\ald},\mu^{(a)}_{\ald})~,
\end{equation}
and thus the condition \eqref{ambWcond13Deltad} is obviously
fulfilled for $\lambda_\pm=\mu_\pm$. This implies in particular
that for $\lambda_\pm=\mu_\pm$, nothing peculiar happens, and it
suffices to consider the linear system \eqref{amblinsys}.\index{linear system}

Following the above analysis in a straightforward manner for
$\lambda_\pm=\mu_\pm$, one arrives at a linear system which\index{linear system}
contains singular operators on $\CPP^1_{\Delta}$ and the
compatibility conditions of this system cannot be pushed forward\index{compatibility conditions}
from the correspondence space $\CK^{5|12}$ down to $\FC^{3|12}$.
As mentioned above, we can ignore this point, as it will be
equivalent to considering the linear system \eqref{amblinsys} over\index{linear system}
$\CPP^1_{\Delta}$.

To sum up, we obtained a correspondence between holomorphic
$\CL^{4|6}$-bundles which become holomorphically trivial vector\index{L46-bundle@$\CL^{4|6}$-bundle}
bundles upon reduction to any subspace $(\CPP^1\times
\CPP^1)_{y,\eta,\theta}\embd\CL^{4|6}$ and solutions to the
three-dimensional $\CN=8$ SYM equations. As this correspondence
arises by a dimensional reduction of a correspondence which is\index{dimensional reduction}
one-to-one, it is rather evident that also in this case, we have a
bijection between both the holomorphic $\CL^{4|6}$-bundles and the\index{L46-bundle@$\CL^{4|6}$-bundle}
solutions after factorizing with respect to holomorphic
equivalence and gauge equivalence, respectively.

\paragraph{Yang-Mills-Higgs theory in three dimensions.} One can\index{Yang-Mills-Higgs theory}
translate the discussion of the ambitwistor space in\index{ambitwistor space}\index{twistor}\index{twistor!ambitwistor}\index{twistor!space}
\ref{ssAmbitwistorSpace} to the three-dimensional situation,
giving rise to a Penrose-Ward transform between holomorphic\index{Penrose-Ward transform}
$\CL^{4}$ bundles and the Yang-Mills-Higgs equations. First of
all, recall from section \ref{ssRelatedTheories}, \ref{pYMHiggs}
the appropriate Yang-Mills-Higgs equations obtained by dimensional
reduction are\index{dimensional reduction}
\begin{equation}\label{Yang-Mills-Higgs}
\nabla^{(\ald\bed)}F_{(\ald\bed)(\gad\ded)}\ =\
[\phi,\nabla_{(\gad\ded)}\phi]\eand\triangle \phi\ :=\
\nabla^{(\ald\bed)}\nabla_{(\ald\bed)}\phi\ =\ 0~,
\end{equation}
while the self-dual and anti-self-dual Yang-Mills equations\index{anti-self-dual}
correspond after the dimensional reduction to two Bogomolny\index{dimensional reduction}
equations which read
\begin{equation}
F_{(\ald\bed)(\gad\ded)}\ =\
\eps_{(\ald\bed)(\gad\ded)(\dot{\eps}\dot{\zeta})}\nabla^{(\dot{\eps}\dot{\zeta})}\phi\eand
F_{(\ald\bed)(\gad\ded)}\ =\
-\eps_{(\ald\bed)(\gad\ded)(\dot{\eps}\dot{\zeta})}\nabla^{(\dot{\eps}\dot{\zeta})}\phi~,
\end{equation}
respectively. Using the decomposition
$F_{(\ald\bed)(\gad\ded)}=\eps_{\ald\gad}f_{\bed\ded}+\eps_{\bed\ded}f_{\ald\gad}$,
the above two equations can be simplified to
\begin{equation}
f_{\ald\bed}\ =\ \tfrac{\di}{2}\nabla_{(\ald\bed)}\phi\eand
f_{\ald\bed}\ =\ -\tfrac{\di}{2}\nabla_{(\ald\bed)}\phi~.
\end{equation}
Analogously to the four-dimensional case, we start from a vector
bundle $E$ over the space $\FC^3\times \FC^3$ with coordinates
$p^{(\ald\bed)}$ and $q^{(\ald\bed)}$; additionally we introduce
the coordinates
\begin{equation}
y^{(\ald\bed)}\ =\
\tfrac{1}{2}(p^{(\ald\bed)}+q^{(\ald\bed)})\eand h^{(\ald\bed)}\
=\ \tfrac{1}{2}(p^{(\ald\bed)}-q^{(\ald\bed)})
\end{equation}
and a gauge potential
\begin{equation}
A\ =\ A^p_{(\ald\bed)}\dd p^{(\ald\bed)}+A^q_{(\ald\bed)}\dd
q^{(\ald\bed)}\ =\ A^y_{(\ald\bed)}\dd
y^{(\ald\bed)}+A^h_{(\ald\bed)}\dd h^{(\ald\bed)}
\end{equation}
on $E$. The differential operators we will consider in the
following are obtained from covariant derivatives by dimensional\index{covariant derivative}
reduction and take, e.g., the shape
\begin{equation}
\nabla_{\ald\bed}^y\ =\ \der{y^{(\ald\bed)}}+[A_{(\ald\bed)}^y+\tfrac{\di}{2}\eps_{\ald\bed}\phi^y,\,\cdot\,]~.
\end{equation}
We now claim that the Yang-Mills-Higgs equations
\eqref{Yang-Mills-Higgs} are equivalent to the equations
\begin{equation}\label{condYM3}
\begin{aligned}
{}[\nabla^p_{\ald\bed},\nabla^p_{\gad\ded}]&\ =\
\ast[\nabla^p_{\ald\bed},\nabla^p_{\gad\ded}]+\CO(h^2)~,\\
{}[\nabla^q_{\ald\bed},\nabla^q_{\gad\ded}]&\ =\
-\ast[\nabla^q_{\ald\bed},\nabla^q_{\gad\ded}]+\CO(h^2)~,\\
{}[\nabla^p_{\ald\bed},\nabla^q_{\gad\ded}]&\ =\ \CO(h^2)~,
\end{aligned}
\end{equation}
where we can use
$\ast[\nabla^{p,q}_{\ald\bed},\nabla^{p,q}_{\gad\ded}]=
\eps_{\ald\bed\gad\ded\dot{\eps}\dot{\zeta}}\nabla_{p,q}^{\dot{\eps}\dot{\zeta}}\phi^{p,q}$.
These equations can be simplified in the coordinates $(y,h)$ to
equations similar to \eqref{condYM4}, which are solved by the
field expansion
\begin{equation}\label{gaugepotYM3}
\begin{aligned}
A^h_{(\ald\bed)}&\ =\
-\tfrac{1}{2}F^{y,0}_{(\ald\bed)(\gad\ded)}h^{(\gad\ded)}-
\tfrac{1}{3}h^{(\gad\ded)}\nabla^{y,0}_{(\gad\ded)}
\eps_{(\ald\bed)(\dot{\eps}\dot{\zeta})(\dot{\sigma}\dot{\tau})}
(\nabla_{y,0}^{(\dot{\sigma}\dot{\tau})}\phi)h^{(\dot{\eps}\dot{\zeta})}~,\\
\phi^h&\ =\ \tfrac{1}{2}\nabla^{y,0}_{(\gad\ded)}\phi^{y,0}
h^{(\gad\ded)}+\tfrac{1}{6}h^{(\gad\ded)}\nabla^{y,0}_{(\gad\ded)}
\eps^{(\ald\bed)(\dot{\eps}\dot{\zeta})(\dot{\sigma}\dot{\tau})}
F^{y,0}_{(\dot{\eps}\dot{\zeta})(\dot{\sigma}\dot{\tau})}h_{(\ald\bed)}~,\\
A^y_{(\ald\bed)}&\ =\ A^{y,0}_{(\ald\bed)}
-\eps_{(\ald\bed)(\dot{\eps}\dot{\zeta})(\dot{\sigma}\dot{\tau})}
(\nabla_{y,0}^{(\dot{\sigma}\dot{\tau})}\phi^{y,0})h^{(\dot{\eps}\dot{\zeta})}-
\tfrac{1}{2}h^{(\gad\ded)}\nabla^{y,0}_{(\gad\ded)}(
F^{y,0}_{(\ald\bed)(\dot{\eps}\dot{\zeta})})h^{(\dot{\eps}\dot{\zeta})}~,\\
\phi^y&\ =\ \phi^{y,0}
+\tfrac{1}{2}\eps^{(\ald\bed)(\dot{\eps}\dot{\zeta})(\dot{\sigma}\dot{\tau})}
F^{y,0}_{(\dot{\eps}\dot{\zeta})(\dot{\sigma}\dot{\tau})}h_{(\ald\bed)}+
\tfrac{1}{2}h^{(\gad\ded)}\nabla^{y,0}_{(\gad\ded)}(\nabla_{(\ald\bed)}\phi^{y,0}
)h^{(\ald\bed)}~,
\end{aligned}
\end{equation}
if and only if the Yang-Mills-Higgs equations
\eqref{Yang-Mills-Higgs} are satisfied.

Thus, solutions to the Yang-Mills-Higgs equations
\eqref{Yang-Mills-Higgs} correspond to solutions to equations
\eqref{condYM3} on $\FC^3\times \FC^3$. Recall that solutions to
the first two equations of \eqref{condYM3} correspond in the
twistor description to holomorphic vector bundles over\index{holomorphic!vector bundle}\index{twistor}
$\CP^2\times \CP^2_*$. Furthermore, the expansion of the gauge
potential \eqref{gaugepotYM3} is an expansion in a second order
infinitesimal neighborhood of $\diag(\FC^3\times \FC^3)$. As we\index{infinitesimal neighborhood}
saw in the construction of the mini-superambitwistor space\index{ambitwistor space}\index{mini-superambitwistor space}\index{twistor}\index{twistor!ambitwistor}\index{twistor!space}
$\CL^{4|6}$, the diagonal for which $h^{(\ald\bed)}=0$ corresponds
to $\CL^4\subset \CP^2\times \CP^2_*$. The neighborhoods of this
diagonal will then correspond to {\em sub-thickenings} of $\CL^4$\index{sub-thickening}\index{thickening}
inside $\CP^2\times \CP^2_*$, i.e.\ for $\mu_\pm=\lambda_\pm$, we
have the additional nilpotent coordinate $\xi$. In other words,
the sub-thickening of $\CL^4$ in $\CP^2\times \CP^2_*$ is obtained\index{sub-thickening}\index{thickening}
by turning one of the fiber coordinates of $\CP^2\times \CP^2$
over $\CPP^1_{\Delta}$ into a nilpotent even coordinate (in a
suitable basis). Then we can finally state the following:

Gauge equivalence classes of solutions to the three-dimensional
Yang-Mills-Higgs equations are in one-to-one correspondence with
gauge equivalence classes of holomorphic $\CL^4$-bundles over a
third order sub-thickening of $\CL^4$, which become\index{sub-thickening}\index{thickening}
holomorphically trivial vector bundles when restricted to a
$\CPP^1\times \CPP^1$ holomorphically embedded into $\CL^4$.

\section{Solution generating techniques}\label{sSolutionGT}\index{solution generating techniques}

In this section, we will discuss solution generating techniques\index{solution generating techniques}
which are related to the twistorial description of field theories.\index{twistor}

The Atiyah-Drinfeld-Hitchin-Manin (ADHM) construction of
instantons \cite{Atiyah:1978ri} reduces the self-duality equations\index{instanton}
to a simple set of matrix equations. This construction has been
shown to be complete, i.e.\ all instanton solutions can be\index{instanton}
obtained by this algorithm. The original idea was to find an
instanton bundle over $\CP^3$ (a topologically trivial holomorphic\index{instanton}
vector bundle, which becomes holomorphically trivial upon
restriction to any $\CPP^1_x\subset \CP^3$) from a so-called
monad. Nevertheless, a very nice interpretation in terms of\index{monad}
D-brane configurations has been found later on\index{D-brane}
\cite{Witten:1995im,Douglas:1995bn,Douglas:1996uz}, see also
\cite{Dorey:2002ik,Tong:2005un}. Furthermore, supersymmetric
extensions of the ADHM construction have been proposed\index{ADHM construction}
\cite{Semikhatov:1982ig,Volovich:1983aa}.

The corresponding reduction to the three-dimensional Bogomolny
equations is given by the Nahm construction \cite{Nahm:1979yw}\index{Bogomolny equations}
with a D-brane interpretation developed in\index{D-brane}
\cite{Diaconescu:1996rk}. A corresponding superextension was
proposed in \cite{Lechtenfeld:2005xi}, and we will present this
extension in section \ref{ssNahm}.

We will present further solution generating techniques in section\index{solution generating techniques}
\ref{ssClassicalSolsMM}.

\subsection{The ADHM construction from monads}\index{ADHM construction}\index{monad}

In discussing the ADHM construction from monads, we follow\index{ADHM construction}\index{monad}
essentially the presentations in \cite{Wardbook} and
\cite{Feehan}. The technique of obtaining vector bundles from
monads stems originally from Horrocks \cite{Horrocks:1964}, see\index{monad}
also \cite{Atiyah:1979iu}.

\paragraph{Monads.} A {\em monad} $\frM$ over a manifold $M$ is a triple of\index{monad}\index{manifold}
vector bundles $A,B,C$ over $M$, which fits into the sequence of
vector bundles
\begin{equation}
A\stackrel{\alpha}{\longrightarrow}
B\stackrel{\beta}{\longrightarrow} C~,
\end{equation}
and thus the linear maps $\alpha$ and $\beta$ satisfy
$\beta\alpha=0$. The vector bundle $E=\mathrm{ker}
\beta/\mathrm{im} \alpha$ is called the {\em cohomology of the
monad}.\index{monad}

The rank and the total Chern class of the cohomology $E$ of the\index{Chern class}\index{total Chern class}
monad $\frM$ constructed above can be derived from the\index{monad}
corresponding data of the triple $A,B,C$ via the formul\ae{}
\begin{equation}
\begin{aligned}
\mathrm{rk}E&\ =\ \mathrm{rk}B-\mathrm{rk}A-\mathrm{rk}C~,\\
c(E)&\ =\ c(B)c(A)^{-1}c(C)^{-1}~.
\end{aligned}
\end{equation}

\paragraph{Annihilator.} The annihilator $U^0\subset V$ of a\index{annihilator}
subspace $U$ of a symplectic vector space\footnote{A symplectic\index{symplectic vector space}
vector space is a vector space equipped with a symplectic form\index{symplectic form}
$\omega$. That is, $\omega$ is a nondegenerate, skew-symmetric
bilinear form.} $V$ is given by those vectors $v\in V$, which
vanish upon pairing with any element of $U$ and applying the
symplectic form:\index{symplectic form}
\begin{equation}
U^0\ :=\ \left\{v\in V|\omega(v,u)=0\mbox{ for all }u\in U\right\}~.
\end{equation}

\paragraph{The instanton monad.} Let us now construct a monad\index{instanton}\index{monad}
$\frM$, which yields an instanton bundle as its cohomology. For
simplicity, we will restrict ourselves to the gauge group
$\sSU(2)$, but via embeddings, it is possible to generalize this
discussion to gauge groups $\sSU(n)$, $\sSO(n)$ and $\sSp(n)$.

Note that by introducing a symplectic form $\omega$ on a vector\index{symplectic form}
bundle $B$, $B$ can be identified with its own dual. Furthermore,
$A$ will be dual to $C$ and $\alpha$ to $\beta$. Thus, we can
reduce the data defining our monad $\frM$ to $A$, $(B,\omega)$ and\index{monad}
$\alpha$. For our construction, we choose $A$ to be $\CO^k(-1)$
over $\CPP^3$ and $B$ is the trivial bundle $\FC^{2k+2}\rightarrow
\CPP^3$.

It now remains to specify $\alpha$. For this, take two complex
vector spaces $V=\FC^{2k+2}$ and $W=\FC^k$ with a symplectic form\index{symplectic form}
$\omega$ on $V$, on both of which we have antilinear maps $\tau$,
with $\tau^2_W=\unit$ (a complex conjugation) and
$\tau^2_V=-\unit$ (induced from the real structure on $\CPP^3$),\index{real structure}
the latter being compatible with the symplectic form $\omega$:\index{symplectic form}
$\omega(\tau v_1,\tau v_2)=\overline{\omega(v_1,v_2)}$ and the
induced Hermitian form $h(v_1,v_2)=\omega(v_1,\tau v_2)$ for
$v_{1,2}\in V$ shall be positive definit. Additionally, we assume
a map
\begin{equation}\label{linmap}
\alpha:W\ \rightarrow\  V~~~\mbox{with}~~~\alpha\ =\ A_i
Z^i\ =\ A^\ald\lambda_\ald+A_\alpha\omega^\alpha~,
\end{equation}
where $(Z^i):=(\omega^\alpha,\lambda_\ald)$ will become the
homogeneous coordinates on the twistor space $\CPP^3$ and\index{homogeneous coordinates}\index{twistor}\index{twistor!space}
$A^\ald,A_\alpha$ are constant linear maps from $W$ to $V$. The
map $\alpha$ satisfies the compatibility condition $\tau \alpha(Z)
w=\alpha(\tau Z) \tau w$ with the maps $\tau$. Since $\alpha$ is
linear in $Z$, we can also see it as a homomorphism of vector
bundles $\alpha: W(-1)\rightarrow V\times \CPP^3$, where
$W(-1)=W\otimes \CO_{\CPP^3}(-1)$. From the map
$\alpha\dual:V\dual=V\rightarrow W\dual$, we obtain the monad\index{monad}
$\frM$
\begin{equation}
W(-1)\stackrel{\alpha}{\longrightarrow} V\times
\CPP^3\stackrel{\alpha\dual}{\longrightarrow} W\dual(1)~,
\end{equation}
where $W\dual(1)$ is the space $W\otimes (\CO_{\CPP^3}(-1))\dual$.

We impose now two additional conditions on the linear map
$\alpha$. First, the space $U_Z:=\alpha W$ is of dimension $k$ and
second, for all $Z\neq 0$, $U_Z$ is a subset of $U^0_Z$. The
latter condition is automatically satisfied for $k=1$. For $k>1$,
this amounts to the matrix equation $\alpha^T \omega \alpha=0$.
The instanton bundle over\footnote{Usually in this discussion, one\index{instanton}
considers the twistor space $\CPP^3$ of $S^4$, and imposes the\index{twistor}\index{twistor!space}
restrictions only for $Z^i\neq 0$.} $\CPP^3$ is then given by the
resulting cohomology
\begin{equation}
E_Z\ :=\ U_Z^0/U_Z
\end{equation}
of $\frM$. Since both $U_Z^0$ and $U_Z$ are independent of the
scaling, we have $E_Z=E_{tZ}$ and therefore the family of all
$E_Z$ is indeed a vector bundle $E$ over $\CPP^3$. In particular,
since $\dim U_Z=k$ and $\dim U_Z^0=k+2$, we have $\dim E_Z=2$,
which is the desired result for an $\sSU(2)$-instanton bundle. The\index{instanton}
symplectic form $\omega$ on $V$ induces a symplectic form $\omega$\index{symplectic form}
on $E_Z$, which renders the latter bundles structure group to
$\sSL(2,\FC)$.

One can verify that the bundle $E$ constructed in this manner is
in fact an instanton bundle \cite{Atiyah:1979iu}, and via the\index{instanton}
Penrose-Ward transform, one obtains the corresponding self-dual\index{Penrose-Ward transform}
gauge potential.

\paragraph{The picture over the moduli space.} Instead of\index{moduli space}
constructing the vector spaces $V$ and $W$ over the twistor space\index{twistor}\index{twistor!space}
$\CPP^3$ fibered over $S^4$, we can discuss them directly over the
space $S^4$. To this end, define $V$ and $W$ as before and choose
$\tau$ to be the complex conjugation on $W$. The symplectic form\index{symplectic form}
$\omega$ on $V$ is given by a skew-symmetric tri-band matrix of
dimension $(2k+2)\times (2k+2)$ with entries $\pm1$. The reality
condition $\tau \alpha(Z) w=\alpha(\tau Z) \tau w$ on the map
$\alpha$ can now be restated in the following way: Let us denote
the components of the matrix $B$ by $B^\ald_{i,j}$. Then for fixed
values of $m,n$, the $2\times 2$-matrix $B^\ald_{2m+\bed-1,n}$
should be a quaternion. Applying the same argument to $C$, we
arrive at a representation of the map $\alpha$ in terms of a\index{representation}
$(k+1)\times k$-dimensional matrix of quaternions
\begin{equation}\label{linmap2}
\Delta\ =\ A-C x~.
\end{equation}
The remaining condition that $\alpha(Z^i) W$ should be of
dimension $k$ for $Z^i\neq 0$ amounts to the fact that
$\bar{\Delta}(x)\Delta(x)$ is nonsingular and real for each $x$,
where $\bar{\Delta}$ is the conjugate transpose of $\Delta$. This
condition is equivalent to the so-called ADHM equations, which
will arise in the following section. One can easily
``supersymmetrize'' the above considerations, by considering a
supertwistor space $\CPP^{3|\CN}$ and adding appropriate linear\index{twistor}\index{twistor!space}
terms to \eqref{linmap} and \eqref{linmap2}.

\subsection{The ADHM construction in the context of D-branes}\label{ssADHM}\index{ADHM construction}\index{D-brane}

\paragraph{The D5-D9-brane system.}
As stated in the introduction to this section, the ADHM algorithm
for constructing instanton solutions has found a nice\index{instanton}
interpretation in the context of string theory. We start from a\index{string theory}
configuration of $k$ D5-branes bound to a stack of $n$ D9-branes,
which -- upon dimensional reduction -- will eventually yield a\index{dimensional reduction}
configuration of $k$ D(-1)-branes inside a stack of $n$ D3-branes.

\paragraph{D5-D5 strings.} From the perspective of the
D5-branes, the $\CN=2$ supersymmetry of type IIB superstring\index{super!symmetry}
theory is broken down to $\CN=(1,1)$ on the six-dimensional
worldvolume of the D5-brane, which is BPS. The fields in the
ten-dimensional Yang-Mills multiplet are rearranged into an
$\CN=2$ vector multiplet\index{vector multiplet}
$(\phi_a,A_{\alpha\ald},\chi_\alpha^i,\mub^\ald_i)$, where the
indices $i=1,\ldots,4$, $a=1,\ldots 6$ and $\alpha,\ald=1,2$ label
the representations of the Lorentz group $\sSO(5,1)\sim\sSU(4)$\index{Lorentz!group}\index{representation}
and the R-symmetry group $\sSO(4)\sim\sSU(2)_L\times \sSU(2)_R$,
respectively. Thus, $\phi$ and $A$ denote bosons, while $\chi$ and
$\mub$ refer to fermionic fields.

\parpic[r]{\includegraphics[width=5.7cm,totalheight=7.6cm]
{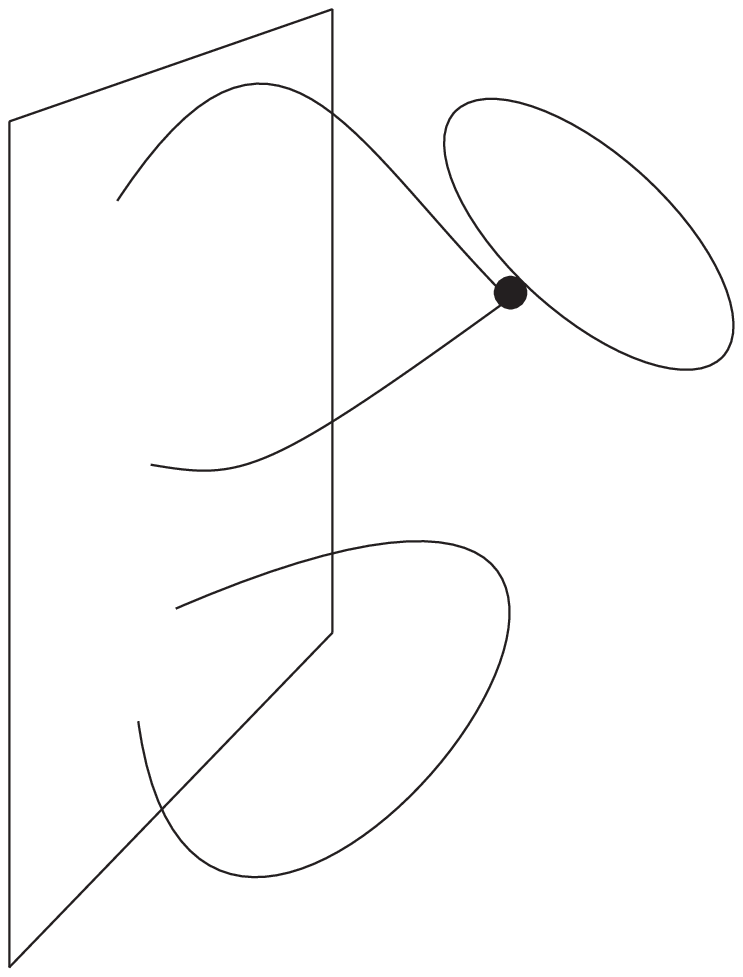}}

\noindent Note that the presence of the D$9$-branes will further
break supersymmetry down to $\CN=(0,1)$ and therefore the above\index{super!symmetry}
multiplet splits into the vector multiplet $(\phi_a,\mub^\ald_i)$\index{vector multiplet}
and the hypermultiplet $(A_{\alpha\ald},\chi_\alpha^i)$. In the\index{hypermultiplet}
following, we will discuss the field theory on the D5-branes in
the language of $\CN=(0,1)$ supersymmetry.\index{super!symmetry}

Let us now consider the vacuum moduli space of this theory which\index{moduli space}
is called the Higgs branch. This is the sector of the theory,
where the $D$-field, i.e.\ the auxiliary field for the $\CN=(0,1)$
vector multiplet, vanishes\footnote{This is often referred to as\index{vector multiplet}
the $D$-flatness condition.}. Therefore, we can restrict our
analysis in the following to a few terms of the action. From the
Yang-Mills part describing the vector multiplet, we have the\index{vector multiplet}
contribution $4\pi^2\alpha'{}^2\int
\dd^6x\tr_k\frac{1}{2}D^2_{\mu\nu}$, where we also introduce the
notation $D_{\mu\nu}=\tr_2(\vec{\sigma}\sigmab_{\mu\nu})\cdot
\vec{D}$. The hypermultiplet leads to an additional contribution\index{hypermultiplet}
of $\int \dd^6x\tr_k\di
\vec{D}\cdot\vec{\sigma}^\ald{}_\bed\bar{A}^{\alpha\bed}A_{\alpha\ald}$.
Note that we will use a bar instead of the dagger to simplify
notation. However, this bar must not be confused with complex
conjugation.

\paragraph{D5-D9 strings.} It remains to include the contributions
from open strings having one end on a D5-brane and the other one\index{open string}
on a D9-brane. These additional degrees of freedom are aware of
both branes and therefore form hypermultiplets under $\CN=(0,1)$\index{hypermultiplet}
supersymmetry. One of the hypermultiplets is in the\index{super!symmetry}
$(\bar{\mathbf{k}},\mathbf{n})$ representation of $\sU(k)\times\index{representation}
\sU(n)$, while the other one transforms as
$(\mathbf{k},\bar{\mathbf{n}})$. We denote them by
$(w_\ald,\psi^i)$ and $(\bar{w}^\ald,\bar{\psi}^i)$, where
$w_\ald$ and $\bar{w}^\ald$ and $\psi^i$ and $\bar{\psi}^i$ denote
four complex scalars and eight Weyl spinors, respectively. The\index{Spinor}
contribution to the $D$-terms is similar to the hypermultiplet\index{hypermultiplet}
considered above: $\int \dd^6x\tr_k\di
\vec{D}\cdot\vec{\sigma}^\ald{}_\bed\bar{w}^\bed w_\ald$.

\paragraph{The $D$-flatness condition.} Collecting all the
(algebraic) contributions of the $D$-field to the action and
varying them yields the equations of motion
\begin{equation}\label{Dflatness}
\alpha'{}^2\vec{D}\ =\
\frac{\di}{16\pi^2}\vec{\sigma}^\ald{}_\bed(\bar{w}^\bed
w_\ald+\bar{A}^{\alpha\bed}A_{\alpha\ald})~.
\end{equation}
After performing the dimensional reduction of the D5-brane to a\index{dimensional reduction}
D(-1)-brane, the condition that $\vec{D}$ vanishes is precisely
equivalent to the ADHM constraints.

\paragraph{The zero-dimensional Dirac operator.} Spelling out all
possible indices on our fields, we have $A_{\alpha\ald pq}$ and
$w_{up\ald}$, where $p,q=1,\ldots,k$ denote indices of the
representation $\mathbf{k}$ of the gauge group $\sU(k)$ while\index{representation}
$u=1,\ldots,n$ belongs to the $\mathbf{n}$ of $\sU(n)$. Let us
introduce the new combinations of indices $r=u\oplus
p\otimes\alpha=1,\ldots,n+2k$ together with the matrices
\begin{equation}\label{rearrangefields}
(a_{r q\ald})\ =\ \left(\begin{array}{c} w_{uq\ald} \\
A_{\alpha\ald pq}
\end{array}\right)~,~~~
(\bar{a}^{\ald r}_q)\ =\
\left(\bar{w}^\ald_{qu}~~A_{pq}^{\alpha\ald}
\right)\eand (b^\beta_{r q})\ =\ \left(\begin{array}{c} 0 \\
\delta_\alpha{}^\beta\delta_{pq}
\end{array}\right)~,
\end{equation}
which are of dimension $(n+2k)\times 2k$, $2k\times(n+2k)$ and
$(n+2k)\times 2k$, respectively. Now we are ready to define a
$(n+2k)\times 2k$ dimensional matrix, the zero-dimensional Dirac
operator of the ADHM construction, which reads\index{ADHM construction}
\begin{equation}
\Delta_{r p \ald}(x)\ =\ a_{r p \ald}+b^\alpha_{r p}
x_{\alpha\ald}~,
\end{equation}
and we put $\bar{\Delta}_p^{\ald r}:=(\Delta_{r p\ald})^*$.
Written in the new components \eqref{rearrangefields}, the ADHM
constraints amounting to the $D$-flatness condition read
$\vec{\sigma}^\ald{}_\bed(\bar{a}^\bed a_\ald)=0$, or, more
explicitly,
\begin{equation}\label{ADHMconstraints}
\bar{a}_\ald a_\bed+\bar{a}_\bed a_\ald\ =\ 0~,
\end{equation}
where we defined as usual
$\bar{a}_\ald=\eps_{\ald\bed}\bar{a}^\bed$. All further
conditions, which are sometimes also summarized under ADHM
constraints, are automatically satisfied due to our choice of
$b^\alpha_{r p}$ and the reality properties of our fields.

\paragraph{Construction of solutions.} The kernel of the
zero-dimensional Dirac operator is generally of dimension $n$, as
this is the difference between its numbers of rows and columns. It
is spanned by vectors, which can be arranged to a complex matrix
$U_{r u}$ which satisfies
\begin{equation}
\bar{\Delta}_p^{\ald r}U_{r u}\ =\ 0~.
\end{equation}
Upon demanding that the frame $U_{r u}$ is orthonormal, i.e.\ that
$\bar{U}^r_u U_{r v}=\delta_{uv}$, we can construct a self-dual
$\sSU(n)$-instanton configuration from\index{instanton}
\begin{equation}
(\CCA_{\alpha\ald})_{uv}\ =\ \bar{U}^r_u\dpar_{\alpha\ald} U_{r
v}~.
\end{equation}
Usually, one furthermore introduces the auxiliary matrix $f$ via
\begin{equation}
f\ =\ 2(\bar{w}^\ald
w_\ald+(A_{\alpha\ald}+x_{\alpha\ald}\otimes\unit_{k})^2)^{-1}~,
\end{equation}
which fits in the factorization condition $\bar{\Delta}_p^{\ald
r}\Delta_{r q\bed}=\delta^\ald_\bed(f^{-1})_{pq}$. Note that the
latter condition is equivalent to the ADHM constraints
\eqref{ADHMconstraints} arising from \eqref{Dflatness}. The matrix
$f$ allows for an easy computation of the field strength\index{field strength}
\begin{equation}\label{fieldstrength}
\CCF_{\mu\nu}\ =\ 4\bar{U}b\sigma_{\mu\nu}f\bar{b}U
\end{equation}
and the instanton number\index{instanton}\index{instanton number}
\begin{equation}
-\frac{1}{16\pi^2}\int \dd^4 x~ \tr_n \CCF^2_{\mu\nu}\ =\
\frac{1}{16\pi^2}\int \dd^4 x~ \square^2\tr_n\log f~.
\end{equation}
Note that the self-duality of $\CCF_{\mu\nu}$ in
\eqref{fieldstrength} is evident from the self-duality property of
$\sigma_{\mu\nu}$.

\subsection{Super ADHM construction and super\index{ADHM construction}
D-branes}\label{ssSADHM}\index{D-brane}

\paragraph{Superspace formulation of SYM theories.} First, recall from\index{super!space}
section \ref{sSYMtheories} that one can formulate the equations of
motion of $\CN=4$ super Yang-Mills theory and self-dual Yang-Mills\index{N=4 super Yang-Mills theory@$\CN=4$ super Yang-Mills theory}\index{Yang-Mills theory}
theory with arbitrary $\CN$ both in terms of ordinary fields on
$\FR^4$ (or its complexification $\FC^4$) and in terms of\index{complexification}
superfields on certain superspaces having $\FR^4$ as their body.\index{body}\index{super!space}
For $\CN=4$ SYM theory, the appropriate superspace is $\FR^{4|16}$
(or $\FC^{4|16}$), while for $\CN$-extended SDYM theory, one has
to use $\FR^{4|2\CN}$ (or $\FC^{4|2\CN}$). One can find an Euler
operator, which easily shows the equivalence of the superfield\index{Euler operator}
formulation with the formulation in terms of ordinary fields.

\paragraph{Superbrane system.} For the super ADHM construction,\index{ADHM construction}
let us consider $k$ D$5|8$-branes inside $n$ D$9|8$-branes. To
describe this scenario, it is only natural to extend the fields
arising from the strings in this configuration to superfields on
$\FC^{10|8}$ and the appropriate subspaces, respectively. In
particular, we will extend the fields $w_\ald$ and
$A_{\alpha\ald}$ entering into the $D$-flatness condition in the
purely bosonic setup to superfields living on $\FC^{6|8}$.
However, since supersymmetry is broken down to four copies of\index{super!symmetry}
$\CN=1$ due to the presence of the two stacks of D-branes, these\index{D-brane}
superfields can only be linear in the Gra{\ss}mann variables. From the\index{Gra{\ss}mann variable}
discussion in \cite{Harnad:1985bc}, we can then even state what
the superfield expansion should look like:
\begin{equation}\label{expansion}
w_\ald\ =\ \z{w}_\ald+\psi^i\eta_{i\ald}\eand A_{\alpha\ald}\ =\
\z{A}_{\alpha\ald}+\chi^i_\alpha\eta_{i\ald}~.
\end{equation}

\paragraph{Super ADHM equations.} The $D$-flatness condition we
arrive at after following the above discussion of the field
theories involved in the D-brane configurations reads again\index{D-brane}
\begin{equation}\label{Dflatness2}
\alpha'{}^2\vec{D}\ =\
\frac{\di}{16\pi^2}\vec{\sigma}^\ald{}_\bed(\bar{w}^\bed
w_\ald+\bar{A}^{\alpha\bed}A_{\alpha\ald})\ =\ 0~,
\end{equation}
but here, all the fields are true superfields. After performing
the dimensional reduction of the D$9|8$-D$5|8$-brane configuration\index{dimensional reduction}
to one containing D$3|8$- and D(-$1|8$)-branes, and arranging the
resulting field content according to \eqref{rearrangefields}, we
can construct the zero-dimensional super Dirac operator
\begin{equation}\label{superDirac}
\Delta_{r i \ald}\ =\ a_{r i \ald}+b^\alpha_{r i}x^R_{\alpha\ald}\
=\ \z{a}_{r i \ald}+b^\alpha_{r i}x^R_{\alpha\ald}+c_{r
i}^j\eta_{j \ald}~,
\end{equation}
where $(x_R^{\alpha\ald}, \eta_i^\ald)$ are coordinates on the
(anti-)chiral superspace $\FC^{4|8}$. That is, from the point of\index{chiral!superspace}\index{super!space}
view of the full superspace $\FC^{4|16}$ with coordinates
$(x^{\alpha\ald},\theta^{i\alpha},\eta_i^\ald)$, we have
$x_R^{\alpha\ald}=x^{\alpha\ald}+\theta^{i\alpha}\eta_i^\ald$. The
ADHM constraints are now turned into the super ADHM constraints,
which were discussed in \cite{Semikhatov:1982ig} for the first
time, see also \cite{Araki:2005jn} for a related recent
discussion.

Explicitly, these super constraints \eqref{ADHMconstraints} read
here
\begin{equation}
\z{\bar{a}}_\ald \z{a}_\bed+\z{\bar{a}}_\bed \z{a}_\ald\ =\ 0~,~~~
\z{\bar{a}}_\ald c_i-\bar{c}_i \z{a}_\ald\ =\ 0~,~~~ \bar{c}_i
c_j-\bar{c}_j c_i\ =\ 0~.
\end{equation}
The additional sign in the equations involving $c_i$ arises from
ordering and extracting the Gra{\ss}mann variables $\eta_i^\ald$ as\index{Gra{\ss}mann variable}
well as the definition $\overline{c_i\eta_i^\ald}=\eta_i^\ald
\bar{c}_i=-\bar{c}_i\eta_i^\ald$.

\paragraph{Construction of solutions.} As proven
in \cite{Semikhatov:1982ig,Volovich:1983ii}, this super ADHM
construction gives rise to solutions to the $\CN=4$\index{ADHM construction}
supersymmetrically extended self-dual Yang-Mills equations in the
form of the super gauge potentials
\begin{equation}\label{supersolutions}
\CCA_{\alpha\ald}\ =\ \bar{U}\dpar_{\alpha\ald} U\eand
\CCA^i_\ald\ =\ \bar{U} D^i_\ald U~,
\end{equation}
where $U$ and $\bar{U}$ are again zero modes of $\bar{\Delta}$ and
$\Delta$ and furthermore satisfy $\bar{U}U=\unit$. That is, the
super gauge potentials in \eqref{supersolutions} satisfy the
constraint equations of $\CN=4$ self-dual Yang-Mills theory\index{Yang-Mills theory}\index{constraint equations}\index{self-dual Yang-Mills theory}
\eqref{constraintN4SDYM}.

One might be tempted to generalize the Dirac operator in
\eqref{superDirac} to higher orders in the Gra{\ss}mann variables, but\index{Gra{\ss}mann variable}
this is unnatural both from the point of view of broken
supersymmetry due to the presence of D-branes and from the\index{D-brane}\index{super!symmetry}
construction of instanton bundles via monads (the original idea\index{instanton}\index{monad}
which gave rise to the ADHM construction). Besides this, higher\index{ADHM construction}
powers of Gra{\ss}mann variables will render the super ADHM equations\index{Gra{\ss}mann variable}
insufficient for producing solutions to the self-dual Yang-Mills
equations\footnote{In \cite{Dorey:1996bf}, a Dirac operator with
higher powers is mentioned, but it is not used to obtain solutions
in the way we do.}. Note also that this construction leads to
solutions of the $\CN=4$ SDYM equations, for which the Higgs
fields tend to zero as $x\rightarrow \infty$. Since the
Higgs-fields describe the motion of the D3-brane in the ambient
ten-dimensional space, this merely amounts to a choice of
coordinates: The axes of the remaining six directions go through
both ``ends'' of the stack of D3-branes at infinity. For a
discussion of the construction of solutions which do not tend to
zero but to a constant value $\sim\sigma^3$ see
\cite{Dorey:1996bf} and references therein.

The fact that solutions to the $\CN=4$ SDYM equations in general
do not satisfy the $\CN=4$ SYM equations does not spoil our
interpretation of such solutions as D(-$1|8$)-branes, since in our
picture, $\CN=4$ supersymmetry is broken down to four copies of\index{super!symmetry}
$\CN=1$ supersymmetry. Note furthermore that $\CN=4$ SYM theory
and $\CN=4$ SDYM theory can be seen as different weak coupling\index{weak coupling}
limits of {\em one} underlying field theory \cite{Witten:2003nn}.

\subsection{The D-brane interpretation of the Nahm\index{D-brane}
construction}\label{ssNahm}

Before presenting its super extension, let us briefly recollect
the ordinary Nahm construction \cite{Nahm:1979yw} starting from
its D-brane interpretation \cite{Diaconescu:1996rk} and\index{D-brane}
\cite{Hashimoto:2005yy}, see also \cite{Tong:2005un}. For
simplicity, we restrict ourselves to the case of
$\sSU(2)$-monopoles, but a generalization of our discussion to
gauge groups of higher rank is possible and rather
straightforward.

\parpic[r]{\includegraphics[width=8.7cm,totalheight=4.6cm]
{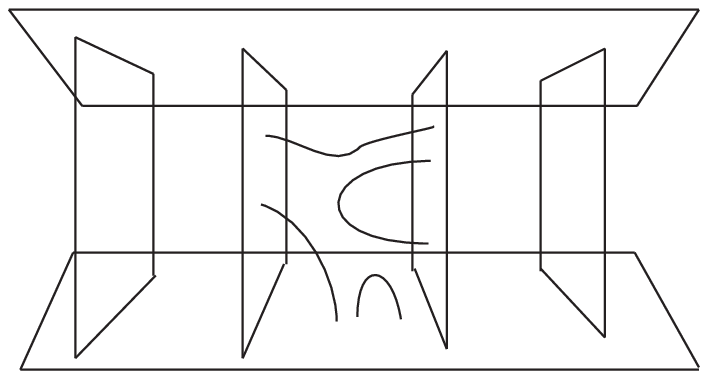}}

\paragraph{The D3-D1-brane system.} We start in ten-dimensional
type IIB superstring theory with a pair of D3-branes extended in\index{string theory}
the directions $1,2,3$ and located at $x^4=\pm 1$, $x^M=0$ for
$M>4$. Consider now a bound state of these D3-branes with $k$
D1-branes extending along the $x^4$-axis and ending on the
D3-branes. As in the case of the ADHM construction, we can look at\index{ADHM construction}
this configuration from two different points of view.

From the perspective of the D3-branes, the effective field theory
on their worldvolume is $\CN=4$ super Yang-Mills theory. The\index{N=4 super Yang-Mills theory@$\CN=4$ super Yang-Mills theory}\index{Yang-Mills theory}
D1-branes bound to the D3-branes and ending on them impose a BPS
condition, which amounts to the Bogomolny equations in three\index{Bogomolny equations}
dimensions
\begin{equation}\label{Bogomolny}
D_a\Phi\ =\ \tfrac{1}{2}\eps_{abc}F_{bc}~,
\end{equation}
where $a,b,c=1,2,3$. The end of the D1-branes act as magnetic
charges in the worldvolume of the D3-branes, and they can
therefore be understood as magnetic monopoles
\cite{Callan:1997kz}, whose field configuration $(\Phi,A^a)$
satisfy the Bogomolny equations. These monopoles are static\index{Bogomolny equations}
solutions of the underlying Born-Infeld action.\index{Born-Infeld action}

From the perspective of the D1-branes, the effective field theory
is first $\CN=(8,8)$ super Yang-Mills theory in two dimensions,\index{Yang-Mills theory}
but supersymmetry is broken by the presence of the two D3-branes\index{super!symmetry}
to $\CN=(4,4)$. As before, one can write down the corresponding
$D$-terms \cite{Tong:2005un} and impose a $D$-flatness condition:
\begin{equation}
D=\derr{X^a}{x^4}+[A_4,X^a]-\tfrac{1}{2}\eps_{abc}[X^b,X^c]+ R\ =\
0~,
\end{equation}
where the $X^a$ are the scalar fields corresponding to the
directions in which the D3-branes extend. The $R$-term is
proportional to $\delta(x^4\pm1)$ and allow for the D1-branes to
end on the D3-branes. They are related to the so-called Nahm
boundary conditions, which we do not discuss. The theory we thus
found is simply self-dual Yang-Mills theory, reduced to one\index{Yang-Mills theory}\index{self-dual Yang-Mills theory}
dimension.

\paragraph{Nahm equations.} By imposing {\em temporal gauge} $A_4=0$,\index{Nahm equations}\index{temporal gauge}
we arrive at the Nahm equations
\begin{equation}\label{Nahm}\index{Nahm equations}
\derr{X^a}{s}-\tfrac{1}{2}\eps_{abc}[X^b,X^c]\ =\
0~~~\mbox{for}~~~-1\ <\ s\ <\ 1~,
\end{equation}
where we substituted $s=x^4$. From solutions to these (integrable)\index{integrable}
equations, we can construct the one-dimensional Dirac operator
\begin{equation}\label{onedDirac}
\Delta^{\ald\bed}\ =\
(\unit_2)^{\ald\bed}\otimes\der{s}+\sigma^{(\ald\bed)}_a(x^a-X^a)~.
\end{equation}
The equations \eqref{Nahm} are analogously to the ADHM equations
the condition for $\bar{\Delta}\Delta$ to commute with the Pauli
matrices, or equivalently, to have an inverse $f$:\index{Pauli matrices}
\begin{equation}
\bar{\Delta}\Delta\ =\ \unit_2\otimes f^{-1}~.
\end{equation}

\paragraph{Construction of solutions.} The normalized zero
modes $U$ of the Dirac operator $\bar{\Delta}$ satisfying
\begin{equation}
\bar{\Delta}(s)U\ =\ 0~,~~~\int_{-1}^1 \dd s~ \bar{U}(s) U(s)\ =\
\unit
\end{equation}
then give rise to solutions to the Bogomolny equations\index{Bogomolny equations}
\eqref{Bogomolny} via the definitions
\begin{equation}
\varPhi(x,t)\ =\ \int_{-1}^1 \dd s~ \bar{U}(s) s U(s)\eand
\CCA_a(x,t)\ =\ \int_{-1}^1 \dd s~ \bar{U}(s) \dpar_a U(s)~.
\end{equation}
The verification of this statement is straightforward when using
the identity
\begin{equation}
U(s)\bar{U}(s')\ =\
\delta(s-s')-\overrightarrow{\Delta}(s)f(s,s')\overleftarrow{\Delta}(s')~.
\end{equation}
Note that all the fields considered above stem from D1-D1 strings.
The remaining D1-D3 strings are responsible for imposing the BPS
condition and the Nahm boundary conditions for the $X^a$ at $s=\pm
1$.

\paragraph{Super Nahm construction.} The superextension of the
Nahm construction is obtained analogously to the superextension of
the ADHM construction by extending the Dirac operator\index{ADHM construction}
\eqref{onedDirac} according to
\begin{equation}\label{onedDiracs}
\Delta^{\ald\bed}\ =\
(\unit_2)^{\ald\bed}\otimes\der{s}+\sigma^{(\ald\bed)}_a(x^a-X^a)+(\eta^{(\ald}_i\chi^{\bed)i})~.
\end{equation}
The fields $\chi^{\ald i}$ are Weyl-spinors and arise from the\index{Spinor}
D1-D1 strings. (More explicitly, consider a bound state of
D7-D5-branes, which dimensionally reduces to our D3-D1-brane
system. The spinor $\chi^{\ald i}$ is the spinor $\chi^i_\alpha$\index{Spinor}
we encountered before when discussing the $\CN=(0,1)$
hypermultiplet on the D5-brane.)\index{hypermultiplet}

\chapter{Matrix Models}

It is essentially three matrix models which received the most\index{matrix model}
attention from string theorists during the last years. First,
there is the Hermitian matrix model, which appeared in the early\index{Hermitian matrix model}\index{matrix model}
nineties in the context of two-dimensional gravity and $c=1$
non-critical string theory, see \cite{Ginsparg:1993is} and\index{string theory}
references therein. It experienced a renaissance in 2002 by the
work of Dijkgraaf and Vafa \cite{Dijkgraaf:2002fc}. Furthermore,
there are the two matrix models which are related to dimensional\index{matrix model}
reductions of ten-dimensional super Yang-Mills theory, the BFSS\index{Yang-Mills theory}
matrix model \cite{Banks:1996vh} and the IKKT matrix model\index{IKKT matrix model}\index{matrix model}
\cite{Ishibashi:1996xs}, see also \cite{Aoki:1998bq}. The latter
two are conjectured to yield non-perturbative and in particular
background independent definitions of M-theory and type IIB\index{M-theory}
superstring theory, respectively. The same aim underlies the work\index{string theory}
of Smolin \cite{Smolin:2000kc}, in which the simplest possible
matrix model, the cubic matrix model (CMM), was proposed as a\index{matrix model}
fundamental theory.

In this chapter, we will furthermore present the results of
\cite{Lechtenfeld:2005xi}, in which two pairs of matrix models\index{matrix model}
were constructed in the context of twistor string theory.\index{string theory}\index{twistor}\index{twistor!string theory}

\section{Matrix models obtained from SYM theory}\label{sSYMMM}\index{matrix model}

For the comparison with the twistor matrix models presented later,\index{matrix model}\index{twistor}
let us review some aspects of the BFSS and the IKKT matrix models.\index{IKKT matrix model}
The motivation for both these models was to find a
non-perturbative definition of string theory and M-theory,\index{M-theory}\index{string theory}
respectively.

\subsection{The BFSS matrix model}\label{ssBFSS}\index{BFSS matrix model}\index{matrix model}

In their famous paper \cite{Banks:1996vh}, Banks, Fishler, Shenkar
and Susskind conjectured that $\mathcal{M}$-theory in the infinite
momentum frame (IMF, see e.g.\ \cite{Bilal:1997fy}) is exactly\index{infinite momentum frame}
described by large $N$ supersymmetric matrix quantum mechanics.\index{matrix quantum mechanics}

\paragraph{Matrix quantum mechanics.} The Lagrangian for matrix\index{matrix quantum mechanics}
quantum mechanics with Minkowski time is given by
\begin{equation}
\CL\ =\ \tr\left(\tfrac{1}{2}\dot{\Phi}^2-U(\Phi)\right)~,
\end{equation}
where $\Phi$ is a Hermitian $N\times N$ matrix. This Lagrangian is
invariant under time-independent $\sSU(N)$ rotations. To calculate
further, it is useful to decompose $\Phi$ into eigenvalues and
angular degrees of freedom by using
$\Phi(t)=\Omega^\dagger(t)\Lambda(t)\Omega(t)$, where $\Lambda(t)$
is a diagonal matrix with the eigenvalues of $\Phi(t)$ as its
entries and $\Omega(t)\in \sSU(N)$. We can furthermore rewrite
$\tr
\dot{\Phi}^2=\tr\dot{\Lambda}^2+\tr[\Lambda,\dot{\Omega}\Omega^\dagger]^2$
and decompose $\dot{\Omega}\Omega^\dagger$ using symmetric,
antisymmetric and diagonal generators with coefficients
$\dot{\alpha}_{ij}$, $\dot{\beta}_{ij}$ and $\dot{\alpha}_i$,
respectively. After performing the trace, the Lagrangian reads as
\begin{equation}
\CL\ =\ \sum_i\left(\tfrac{1}{2}\dot{\lambda}^2_i+U(\lambda_i)\right)
+
\tfrac{1}{2}\sum_{i<j}(\lambda_i-\lambda_j)^2(\dot{\alpha}_{ij}^2
+ \dot{\beta}^2_{ij})~.
\end{equation}
The integration measure of the path integral is transformed to
$\CCD\Phi=\CCD\Omega \Delta^2(\Lambda)\Pi_i\dd \lambda_i$, where
$\Delta(\Lambda)=\prod_{i<j}(\lambda_i-\lambda_j)$ is the
so-called {\em Vandermonde determinant}. The corresponding\index{Vandermonde determinant}
Hamiltonian reads as
\begin{equation}
\CH\ =\ -\tfrac{1}{2\beta^2\Delta(\Lambda)}\sum_i
\frac{\dd^2}{\dd\lambda_i^2}\Delta(\Lambda) +\sum_iU(\lambda_i) +
\sum_{i<j}
\frac{\Pi_{ij}^2+\tilde{\Pi}_{ij}^2}{(\lambda_i-\lambda_j)^2}~,
\end{equation}
where $\Pi_{ij}$ and $\tilde{\Pi}_{ij}$ are the momenta conjugate
to $\alpha_{ij}$ and $\beta_{ij}$. For more details, see
\cite{Klebanov:1991qa}.

\paragraph{BFSS action.} The action of the BFSS model, describing
$N$ D0-branes, can be obtained by dimensional reduction of\index{dimensional reduction}
10-dimensional $\CN=1$ super Yang-Mills theory with gauge group\index{Yang-Mills theory}
$\sU(N)$ to $0+1$ dimensions in temporal gauge $A_0=0$:\index{temporal gauge}
\begin{equation}
S=\tfrac{1}{2g}\int \dd t~\left[\tr
  \dot{X}^i\dot{X}^i+2\theta^T\dot{\theta}+\frac{1}{2}\tr
      [X^i,X^j]^2-2\theta^T\gamma_i[\theta,X^i]\right].
\end{equation}
Here, the $X^i$ are nine Hermitian $N\times N$ matrices and
$\theta$ is a Majorana-Weyl spinor. Note that the bosonic part of\index{Majorana-Weyl spinor}\index{Spinor}
the BFSS Lagrangian is a matrix quantum mechanics Lagrangian.\index{matrix quantum mechanics}
Putting all the fermions to zero, one obtains the bosonic
equations of motion
\begin{equation}\label{BFFSeom}
\ddot{X}^i\ =\ -[[X^i,X^j],X^j]~.
\end{equation}

Restricting to the special class of classical (vacuum) solutions
which satisfy $[X^i,X^j]=0$, the matrices are simultaneously
diagonalizable and for gauge group $\sU(N)$ we can interpret such
solutions as a stack of $N$ D0-branes, whose positions in the
normal directions are given by the eigenvalues of the $X^i$.

The remaining classical solutions to \eqref{BFFSeom} do not
annihilate the positive-definite potential term and are thus no
vacuum solutions. They break supersymmetry, and in particular,\index{super!symmetry}
correspond to D0-branes, whose worldvolumes are smeared out in the
normal directions.

\paragraph{The BFSS model on a circle.} To describe D0-branes in a
spacetime, which has been compactified in one direction normal to
the worldvolume of the D0-branes, we consider infinitely many
copies of a D0-brane configuration and mod out the lattice
symmetry group afterwards. A good reference here is
\cite{Taylor:1997dy}.

To describe the copies of the D0-brane configuration, we extend
the $N\times N$-dimensional matrices $X^i$ to
$\infty\times\infty$-dimensional matrices $X^i_{nm}$ which are
divided into $N\times N$-dimen\-sional blocks, specified by the
indices $n,m\in \RZ$. We furthermore impose the condition
$X_{mn}=-X^\dagger_{nm}$ on the blocks. The new Lagrangian then
reads
\begin{equation}\label{Lag1}
\mathcal{L}\ =\ \frac{1}{2g}\left[\tr\dot{X}_{mn}^i\dot{X}_{nm}^i+
  \frac{1}{2}\tr(X^i_{mq}X^j_{qn}-X^j_{mq}X^i_{qn})(X^i_{nr}X^j_{rm}-
  X^j_{nr}X^i_{rm})\right].
\end{equation}

The periodicity condition from compactifying the $X^1$-direction
on a circle with radius $R$ translates into the following
conditions on the matrices $X^i_{mn}$:
\begin{eqnarray}
X^i_{mn}&=&X^i_{(m-1)(n-1)},~~~~ i>1\\
X^1_{mn}&=&X^1_{(m-1)(n-1)},~~~~ m\neq n\\
X^1_{mm}&=&X^1_{(m-1)(m-1)}+2\pi R\unit.
\end{eqnarray}
The first equation renders all blocks on diagonals equal for
$i>1$, the second equation does the same for some of the diagonals
of $X^1$. The third equation shifts subsequent blocks on the
principal diagonal for $i=1$ by an amount of $2\pi R$, the
circumference of the circle. Anti-Hermiticity of the $X^i_{mn}$
implies furthermore $(X^i_n)^\dagger=-X^i_{-n}$. We thus arrive at
\begin{equation}X^1\ =\ \left(
\begin{array}{ccccc}
\ddots & \vdots & \vdots & \vdots & \ddots  \\
\cdots & X^1_0-2\pi R\unit & X^1_1 & X^1_2 & \cdots \\
\cdots & X^1_{-1} &  X^1_0 & X^1_1 & \cdots\\
\cdots & X^1_{-2} & X^1_{-1} & X^1_0+2\pi R\unit & \cdots \\
\ddots & \vdots & \vdots & \vdots & \ddots  \\
\end{array}\right)
\end{equation}

Rewriting the Lagrangian (\ref{Lag1}) in terms of $X^i_n$ gives
the description of D0-branes moving in a compactified spacetime.
Expansions around the classical vacuum $[X^i,X^j]=0$ lead to the
expected mass terms proportional to the distance of the branes
plus the winding contribution $2\pi Rn$, similarly to our
discussion of T-duality in section \ref{ssTduality}.\index{T-duality}

Note that this matrix quantum mechanics is automatically a second\index{matrix quantum mechanics}
quantized formalism as it allows for an arbitrary number of
D0-branes. This analysis can also be easily generalized to more
than one compact dimension \cite{Taylor:1996ik}.

\paragraph{Reconstruction of spatial dimensions.}\label{pRecontructionDimension}
Let us consider the following correspondence:
\begin{equation}
\phi(\hat{x})\ =\ \sum_n\hat{\phi}_n\de^{\di n\hat{x}/\hat{R}}
\ \leftrightarrow\  \left(
\begin{array}{c}
\vdots\\
\hat{\phi}_{-1}\\
\hat{\phi}_{0}\\
\hat{\phi}_{1}\\
\vdots\\
\end{array}
\right)
\end{equation}
where $\hat{R}=\frac{1}{2\pi R}$. Then $X^1$ is a matrix
representation of the covariant derivative along the compactified\index{covariant derivative}\index{representation}
direction:
\begin{equation}
(\di\hat{\dpar}_1+A_1(\hat{x}))\phi(\hat{x})\ \leftrightarrow\ 
X^1\vec{\hat{\phi}}
\end{equation}
where $A^1(\hat{x})$ is a gauge potential whose Fourier modes are
identified with $X^1_n$:
\begin{equation}
A^1(\hat{x})\ =\ \sum_n A^1_n\de^{\di n \hat{x}/\hat{R}}\ =\ \sum_n
X^1_n\de^{\di
  n \hat{x}/\hat{R}}.
\end{equation}
Here, the derivative leads to the inhomogeneous terms $\sim 2\pi
n\unit$ and the gauge field gives rise to the remaining
components.

To return to the dual space, we can consider an analogous Fourier\index{dual space}
decomposition of $Y^i(\hat{x})=\sum_n X^i_n\de^{\di  n
\hat{x}/\hat{R}}$, with which we can rewrite the BFSS Lagrangian
as
\begin{equation*}
\CL\ =\ \int\frac{\dd x^1}{2\pi
R}\frac{1}{2g}\left[\tr\dot{Y}^i\dot{Y}^i+
  +\tr \dot{A}^1\dot{A}^1-\tr(\dpar_1
  Y^i-\di[A^1,Y^i])^2+\frac{1}{2}\tr [Y^i,Y^j]^2\right]~.
\end{equation*}
By integrating over $x^1$, we obtain again the Lagrangian
\eqref{Lag1}.

This result corresponds to T-duality in the underlying string\index{T-duality}
theory and describes $k$ D1-branes wrapped around a compact circle
of radius $R'=\frac{1}{2\pi R}$. Using this construction, we can
reduce the infinite-dimensional matrices of the model to finite
dimensional ones by introducing an additional integral.

Altogether, we have identified the degrees of freedom of a compact
$\sU(\infty)$ matrix model with the degrees of freedom of a\index{matrix model}
$\sU(N)$ gauge potential on a circle $\hat{S}^1$ with a radius
dual to $R$. In this manner, one can successively reconstruct all
spatial dimensions.

\subsection{The IKKT matrix model}\label{ssIKKTMM}\index{IKKT matrix model}\index{matrix model}

The IKKT model was proposed in \cite{Ishibashi:1996xs} by
Ishibashi, Kawai, Kitazawa and Tsuchiya. It is closely related to
the BFSS model, but while the latter is conjectured to give rise
to a description of M-theory, the former should capture aspects of\index{M-theory}
the type IIB superstring.

\paragraph{Poisson brackets.} A super Poisson structure has already been\index{Poisson brackets}\index{super!Poisson structure}
introduced in section \ref{ssSupergeneralities},
\ref{psuperPoisson}. Here, we want to be more explicit and
consider a two-dimensional Riemann surface $\Sigma$, i.e.\ the\index{Riemann surface}
worldsheet of a string. On $\Sigma$, we define\index{worldsheet}
\begin{equation}
\{X,Y\}\ :=\ \tfrac{1}{\sqrt{g}}\eps^{ab}\dpar_a X\dpar_b Y~,
\end{equation}
where $\sqrt{g}$ is the usual factor containing the determinant of
the worldsheet metric and $\eps^{ab}$ is the antisymmetric tensor\index{worldsheet}\index{metric}
in two dimensions.

\paragraph{Schild-type action.} Using the above Poisson brackets,\index{Poisson brackets}
we can write down the {\em Schild action}, which has been shown to\index{Schild action}
be equivalent\footnote{This equivalence is demonstrated by using
the equation for $\sqrt{g}$ in the Schild action.} to the usual\index{Schild action}
Green-Schwarz action \eqref{stringactionIIBNG} of the type IIB\index{Green-Schwarz action}
superstring in the Nambu-Goto form. It reads as\index{Nambu-Goto}
\begin{equation}\label{stringactionSchild}
S_{\mathrm{Schild}}\ =\ \int \dd^2\sigma~
\sqrt{g}\left(\alpha\left(\tfrac{1}{4}\{X^\mu,X^\nu\}^2-\tfrac{\di}{2}\bar{\psi}\Gamma^\mu
\{X^\mu,\psi\}\right)+\beta\right)~.
\end{equation}

\paragraph{$\CN=2$ supersymmetry.} The Schild action is invariant\index{Schild action}\index{super!symmetry}
under an $\CN=2$ (worldsheet) supersymmetry similarly to the\index{worldsheet}
Green-Schwarz action \eqref{stringactionIIBNG}. Here, the symmetry\index{Green-Schwarz action}
algebra is
\begin{equation}\label{susySchild}
\begin{aligned}
\delta^1\psi&\ =\ -\tfrac{1}{2}
\sigma_{\mu\nu}\Gamma^{\mu\nu}\eps~,~~~
&\delta^1X^\mu&\ =\ \di\bar{\eps}\Gamma^\mu\psi~,\\
\delta^2\psi&\ =\ \xi~,~~~&\delta^2 X^\mu&\ =\ 0~,
\end{aligned}
\end{equation}
where
\begin{equation}
\sigma_{\mu\nu}\ =\ \eps^{ab}\dpar_a X_\mu \dpar_b X\nu~.
\end{equation}

\paragraph{Quantization.} From the Schild action \eqref{stringactionSchild},\index{Schild action}\index{quantization}
one can construct a matrix model by the following (quantization)\index{matrix model}
prescription:
\begin{equation}\label{quantization}
\{\cdot,\cdot\}\ \rightarrow\  -\di[\cdot,\cdot]\eand \int
\dd^2\sigma~\sqrt{g}\ \rightarrow\  \tr~.
\end{equation}
This is consistent, and crucial properties of the trace like
\begin{equation}
\tr[X,Y]\ =\ 0\eand \tr(X[Y,Z])\ =\ \tr(Z[X,Y])
\end{equation}
are also fulfilled after performing the inverse transition to
\eqref{quantization}.

\paragraph{The matrix model.} Applying the rules\index{matrix model}
\eqref{quantization} to the Schild action\index{Schild action}
\eqref{stringactionSchild}, we arrive at
\begin{equation}
S_{\mathrm{IKKT}}\ =\ \alpha\left(-\frac{1}{4}\tr[A_\mu,A_\nu]^2-\frac{1}{2}\tr
\left(\bar{\psi}\gamma^\mu[A_\mu,\psi]\right)\right)+\beta\tr\unit~.
\end{equation}
Here, $A_\mu$ and $\psi$ are bosonic and fermionic anti-Hermitian
matrices, respectively. Except for the term containing $\beta$,
this action can also be obtained by dimensional reduction of\index{dimensional reduction}
ten-dimensional SYM theory to a point.

There is a remnant of the gauge symmetry, which is given by the
adjoint action of the gauge group on the matrices:
\begin{equation}
A_\mu\ \mapsto\  g A_\mu g^{-1}\eand \psi\ \mapsto\  g\psi g^{-1}~.
\end{equation}
The $\CN=2$ supersymmetry \eqref{susySchild} is directly\index{super!symmetry}
translated into
\begin{equation}
\begin{aligned}
\delta^1\psi&\ =\ \tfrac{\di}{2}[A_\mu,A_\nu]\Gamma^{\mu\nu}\eps~,~~~&\delta^1
A_\mu&\ =\ \di\bar{\eps}\Gamma_\mu\psi~,\\
\delta^2\psi&\ =\ \xi~,~~~&\delta^2 A_\mu&\ =\ 0~.
\end{aligned}
\end{equation}

\paragraph{Equations of motion.} The equations of motion
for the bosonic part for $\psi=0$ are
\begin{equation}
[A_\mu,[A_\mu,A_\nu]]\ =\ 0.
\end{equation}
One can interpret classical vacuum solutions, which are given by
matrices $A_\mu$ satisfying $[A_\mu,A_\nu]=0$, in terms of
D(-1)-branes, similar to the interpretation of such solutions in
the BFSS model in terms of D0-branes: The dimension of the
fundamental representation space of the gauge group underlying the\index{fundamental representation}\index{representation}
IKKT model (formally infinity) is the number of D(-1)-branes in
the picture and the diagonal entries of the matrices $A_\mu$,
after simultaneous diagonalization, correspond to their positions
in ten-dimensional spacetime.

\section{Further matrix models}\index{matrix model}

\subsection{Dijkgraaf-Vafa dualities and the Hermitian matrix model}\index{Hermitian matrix model}\index{matrix model}

\paragraph{Preliminary remarks.} Four years ago, Dijkgraaf and Vafa
showed in three papers
\cite{Dijkgraaf:2002fc,Dijkgraaf:2002vw,Dijkgraaf:2002dh} that one
can compute the effective superpotential in certain supersymmetric
gauge theories by performing purely perturbative calculations in
matrix models. While these computations were motivated by string\index{matrix model}
theory dualities, a proof was given in \cite{Dijkgraaf:2002dh}
using purely field theoretical methods.

\paragraph{Chain of dualities.} The chain of dualities used in
string theory to obtain a connection between supersymmetric gauge\index{connection}\index{string theory}
theories and Hermitian matrix models is presented e.g.\ in\index{Hermitian matrix model}\index{matrix model}
\cite{Dijkgraaf:2002dh}. One starts from a $\CN=2$ supersymmetric
gauge theory geometrically engineered within type IIB superstring
theory. By adding a tree-level superpotential, which translates\index{string theory}
into a deformation of the Calabi-Yau geometry used in the\index{Calabi-Yau}
geometric engineering, one breaks the supersymmetry of the gauge\index{geometric engineering}\index{super!symmetry}
theory further down to $\CN=1$. The engineered open string\index{open string}
description can now be related to a closed string description via\index{closed string}
a large $N$ duality, in which the Calabi-Yau geometry undergoes a\index{Calabi-Yau}
geometric transition (cf.\ the conifold transition in section\index{conifold}\index{conifold transition}
\ref{ssConifold}, \ref{pconifoldtrans}) and the D-branes are\index{D-brane}
turned into certain 3-form fluxes $H$. In this description, the
effective superpotential is just given by the integral $\int_X
H\wedge \Omega$, where $X$ is the new Calabi-Yau geometry.\index{Calabi-Yau}
Introducing a basis of homology cycles, one can rewrite this
integral in terms of a prepotential $\CF_0$ determined by the
Calabi-Yau geometry. It is furthermore known that the computation\index{Calabi-Yau}
of $\CF_0$ reduces to a calculation within closed topological
string theory, which in turn is connected to an open topological\index{string theory}\index{topological!string}
string theory via essentially the same large $N$ duality, we used
for switching between the open and closed ten-dimensional strings
of the geometric engineering picture above. The open topological\index{geometric engineering}
string reduced now not only to a holomorphic Chern-Simons theory,\index{Chern-Simons theory}
but under certain conditions, only zero-modes survive and we
arrive at a Hermitian matrix model, which is completely soluble.\index{Hermitian matrix model}\index{matrix model}

In the subsequently discussed variations of this gauge
theory/matrix model correspondence, several matrix models\index{matrix model}
appeared, which shall be briefly introduced in the following.

\paragraph{The Hermitian matrix model.} The partition function of\index{Hermitian matrix model}\index{matrix model}
the Hermitian matrix model is given by
\begin{equation}
Z=\frac{1}{\mathrm{vol}(G)}\int\dd \Phi~\de^{-\frac{1}{g_s}\tr
W(\Phi)}~,
\end{equation}
where $\Phi$ is a Hermitian $N\times N$ matrix, $W(\Phi)$ is a
polynomial in $\Phi$ and $\mathrm{vol}(G)$ is the volume of the
gauge group. One can switch to an eigenvalue formulation
analogously to the above discussed case of matrix quantum
mechanics. Here, the partition functions reads\index{matrix quantum mechanics}
\begin{equation}
Z=\int\prod_{i=1}^N\dd
\lambda_i~\Delta(\lambda)^2\de^{-\frac{1}{g_s}\sum_{i=1}^N
  W(\lambda_i)}~,
\end{equation}
where $\Delta(\lambda)$ is the usual Vandermonde determinant. The\index{Vandermonde determinant}
effective action is found via exponentiating the contribution of
the Vandermonde determinant, which yields\index{Vandermonde determinant}
\begin{equation}
S_{\mathrm{eff}}\ =\ \frac{1}{g_s}\sum_{i=1}^N
W(\lambda_i)-2\sum_{i<j}\log(\lambda_i-\lambda_j)~.
\end{equation}
This action describes $N$ repelling eigenvalues moving in the
potential $W(\lambda)$ and the corresponding equations of motion
read as
\begin{equation}
\frac{\dpar S_{\mathrm{eff}}}{\dpar
  \lambda_i}\ =\ 0\ =\ \frac{1}{g_s}W'(\lambda_i)-2\sum_{j\neq
  i}\frac{1}{\lambda_i-\lambda_j}~.
\end{equation}

\paragraph{The unitary matrix model.} The partition function of\index{matrix model}\index{unitary matrix model}
the unitary matrix model is formally the same as the one of the
Hermitian matrix model,\index{Hermitian matrix model}
\begin{equation}
Z=\frac{1}{\mathrm{vol}(\sU(N))}\int_{\sU(N)}\dd
U~\de^{-\frac{1}{g_s}\tr W(U)}~,
\end{equation}
but now the matrices $U$ are unitary $N\times N$ matrices. The
eigenvalue formulation of the partition function reads now
\begin{equation}
Z=\int\prod_{k=1}^N\dd
\alpha_k~\prod_{i<j}\sin^2\left(\frac{\alpha_i-\alpha_j}{2}\right)
\de^{-\frac{1}{g_s}\sum_{i=1}^N W(\alpha_k)}~,
\end{equation}
which is due to the fact that the Vandermonde determinant became\index{Vandermonde determinant}
periodic, as we identified $U=\exp(\di\Phi)$ where $\Phi$ is a
Hermitian matrix. After regularizing the expression, one obtains
the sine-term.

The corresponding equations of motion of the effective action are
\begin{equation}
W'(\alpha_i)-2g_s\sum_{j\neq
i}\cot\left(\frac{\alpha_i-\alpha_j}{2}\right)\ =\ 0~,
\end{equation}
and this system describes a Dyson gas of eigenvalues moving on the
unit circle.

The so-called Gross-Witten model is the unitary matrix model with\index{Gross-Witten model}\index{matrix model}\index{unitary matrix model}
potential
\begin{equation}
W(U)\ =\ \frac{\eps}{2g_s}\tr(U+U^{-1})\ =\ \frac{\eps}{g_s}\cos(\Phi)~.
\end{equation}

\paragraph{The holomorphic matrix model.} To introduce the holomorphic\index{holomorphic matrix model}\index{matrix model}
matrix model, which is proposed as the true matrix model
underlying the Dijkgraaf-Vafa conjecture and a cure for several of
its problems in \cite{Lazaroiu:2003vh}, we need the following
definitions:

Let $\sMat(N,\FC)$ be the set of complex $N\times N$ matrices and
$p_M(\lambda)=\det(\lambda\mathbbm{1}-M)$ the characteristic
polynomial of the matrix $M$. Let $\mathcal{M}$ be the subset of
matrices in $\sMat(N,\FC)$ with distinct roots. This implies that
elements of $\mathcal{M}$ are diagonalizable. The set of
eigenvalues of a matrix $M\in\CM$ is called the spectrum of $M$:
$\sigma(M)=\{\lambda_1,\ldots ,\lambda_N\}$. Choose a submanifold
$\Gamma\subset\mathcal{M}$ with
$\dim_\FR(\Gamma)=\dim_\FC(\mathcal{M})=N^2$ and define
furthermore the symplectic form on $\mathrm{M}_N(\FC)$ as\index{symplectic form}
$w=\wedge_{i,j}\dd M_{ij}$, where the indices are taken in
lexicographic order to fix the sign.

Given a potential $W(t)=\sum_m t_m z^m$, $t_m\in\FC$, the
partition function for the holomorphic matrix model is defined as\index{holomorphic matrix model}\index{matrix model}
\begin{equation}
Z_N(\Gamma,t)\ :=\ \CN\int_\Gamma w\de^{-N\tr W(M)}~.
\end{equation}

For its eigenvalue representation, we choose an open curve\index{representation}
$\gamma:\FR\rightarrow\FC$ without self-intersections and a
corresponding subset of matrices
$\Gamma(\gamma):=\{M\in\mathcal{M}|\sigma(M)\subset \gamma\}$.
After integrating out a group volume, giving rise to a new
normalization constant $\CN'$ we arrive at
\begin{equation}
Z_N(\Gamma(\gamma),t)\ =\ \CN'\int_\gamma\dd \lambda_1\ldots \dd
\lambda_N~ \prod_{i\neq
  j}(\lambda_i-\lambda_j)\de^{-N\sum_{j=1}^N
  W(\lambda_i)}.
\end{equation}

\subsection{Cubic matrix models and Chern-Simons theory}\index{Chern-Simons theory}\index{matrix model}

\paragraph{Action and equations of motion.} In \cite{Smolin:2000kc,
Smolin:2000fr,Livine:2002vq}, the simplest nontrivial matrix model\index{matrix model}
has been examined and proposes as a fundamental theory. The field
content consists of a single field $A$, which is a matrix with
Lie-algebra valued entries. For convenience, we decompose $A$
according to $A=A_a\tau^a$, where $A_a\in\sMat(N,\FR)$ and the
$\tau^a$ are generators of some (super-)Lie algebra\footnote{For a\index{Lie algebra}
super Lie algebra, the commutator and the trace have to be\index{super!Lie algebra}
replaced by the supercommutator and the supertrace, respectively,\index{super!commutator}\index{super!trace}
in the following discussion.} $\frg$. Its action is given by:
\begin{equation}\label{CAction}
S[A]\ =\ \tra{\sMat(N,\FR)}
A_\alpha^\beta[A_\beta^\gamma,A_\gamma^\alpha]\ =\ \tra{\frg}
A_i^j[A_j^k,A_k^i]\ =\ \tfrac{1}{3}\phi^{abc}\tra{\sMat(N,\FR)}A_a[A_b,A_c]~,
\end{equation}
where we introduced the usual structure constants
$\phi^{abc}=\frac{3}{2}\tra{\mathfrak{g}}\tau^a[\tau^b,\tau^c]$ of
the Lie algebra $\frg$. The indices $\alpha,\beta,\gamma$ belong\index{Lie algebra}
to the representation of $\frg$, while the indices $i,j,k$ are\index{representation}
indices for matrices in $\sMat(N,\FR)$.

This action has two symmetries: one by adjoint action of elements
of the group generated by $\mathfrak{g}$, the other one by adjoint
action of the group $\sGL(N,\FR)$.

The equations of motion, defining the classical solutions, are
\begin{equation}\label{eomsmolin}
\phi^{abc}[X_b,X_c]\ =\ 0~.
\end{equation}

By choosing different backgrounds (i.e., compactifications etc.)
and different Lie algebras $\mathfrak{g}$ one can obtain several\index{Lie algebra}
standard matrix models as BFSS or IKKT, as we will briefly discuss\index{matrix model}
in \ref{pCubicBFSSIKKT}.

\paragraph{Example $\frg=\asu(2)$.} Let us consider the example
$\frg=\asu(2)$, which reproduces in the large matrix limit and
after triple toroidal compactification Chern-Simons field theory
on a 3-torus. The structure constants are given by $\phi^{abc}=\di
\eps^{abc}$. We expand around a classical solution
$X_i,i\in\{1,2,3\}$ of (\ref{eomsmolin}):
\begin{equation}
S_X[A]\ :=\ S[X+A]-S[A]\ =\ \di\eps^{abc}\left(\tr
A_a[X_b,A_c]+\frac{1}{3}\tr A_a[A_b,A_c]\right)~.
\end{equation}
Now we use the usual trick of toroidal compactification described
in section \ref{ssBFSS}, \ref{pRecontructionDimension} to
compactify this system on the three-torus. For this, the large $N$
limit is taken in a special manner. We split the remaining
$N\times N$ matrices in $M\times M$ blocks in such a way that
\begin{equation}
\tra{\sMat(N,\FR)}A^i_a\ =\ \sum_{k=-n}^n \tra{\sMat(M,\FR)}
\tilde{A}^i(k)
\end{equation}
and $T=(2n+1)l_{Pl}$ is constant in the limiting process,
$l_{Pl}\rightarrow 0$ being an arbitrary length scale. Next, we
identify $X_i$ with the covariant derivative on the dual space and\index{covariant derivative}\index{dual space}
construct new functions $a(x)$ as:
\begin{equation}
X_i\ =\ \hat{\dpar}_i+a_i,~~~~ a_i(x)\ :=\ \sum_{k=-\infty}^\infty
\de^{\di k
  \hat{x}/\hat{R}}\tilde{A}^i(k)~.
\end{equation}
In the dual picture, the infinite trace becomes a finite trace and
an integral over the torus, as discussed above. So we eventually
obtain:
\begin{equation}
S=\eps^{ijk}\int_{T^3}\frac{\dd^3 x}{(2\pi
R')^3}\tr\left(a_i\dpar_j a_k + \tfrac{4}{3} a_i[a_j,a_k]\right)
\end{equation}
This can be regarded as the action for $\sU(M)$ or $\sGL(M,\FR)$
Chern-Simons theory on a 3-torus.\index{Chern-Simons theory}

An interesting point about this example is that, after performing
BRST quantization \cite{Livine:2002vq}, the one-loop effective\index{quantization}
action contains the quadric interaction term of the IKKT model.\index{quadric}

\paragraph{Cubic matrix model and the IKKT and BFSS models.}\label{pCubicBFSSIKKT}\index{matrix model}
It was claimed in \cite{Azuma:2001dr} that the IKKT model is
naturally contained in the CMM \eqref{CAction} with algebra
$\frg=\mathfrak{osp}(1|32)$. Furthermore, the field content and
the $\CN=2$ SUSY of the IKKT model was identified with the field
content and the supersymmetries of the CMM. This identification is
two-to-one, i.e.\ the CMM contains twice the IKKT in distinct
``chiral'' sectors.

The CMM was also considered in \cite{Bagnoud:2002iz}, but in this
case, a connection to the BFSS model was found. Here, one starts\index{connection}
from an embedding of the $\sSO(10,1)$-Poincar{\'e} algebra in\index{Poincar{\'e} algebra}
$\mathfrak{osp}(1|32)$, and switches to the IMF leaving $\sSO(9)$
as a symmetry group. After integrating out some fields, the
effective action is identical to the BFSS action.

\section{Matrix models from twistor string\index{matrix model}\index{twistor}
theory}\label{sTwistorMM}

\subsection{Construction of the matrix models}\index{matrix model}

\paragraph{Preliminary remarks.}
In this section, we construct four different matrix models. We\index{matrix model}
start with dimensionally reducing $\CN=4$ SDYM theory to a point,
which yields the first matrix model. The matrices here are just\index{matrix model}
finite-dimensional matrices from the Lie algebra of the gauge\index{Lie algebra}
group $\sU(n)$. The second matrix model we consider results from a\index{matrix model}
dimensional reduction of hCS theory on $\CP^{3|4}_\eps$ to a\index{dimensional reduction}
subspace $\CP_\eps^{1|4}\subset\CP^{3|4}_\eps$. We obtain a form
of matrix quantum mechanics with a complex ``time''. This matrix\index{matrix quantum mechanics}
model is linked by a Penrose-Ward transform to the first matrix\index{Penrose-Ward transform}
model.

By considering again $\CN=4$ SDYM theory, but on noncommutative
spacetime, we obtain a third matrix model. Here, we have\index{matrix model}\index{noncommutative spacetime}
finite-dimensional matrices with operator entries which can be
realized as infinite-dimensional matrices acting on the tensor
product of the gauge algebra representation space and the Fock\index{representation}
space. The fourth and last matrix model is obtained by rendering\index{matrix model}
the fibre coordinates in the vector bundle
$\CP^{3|4}_\eps\rightarrow \CPP^{1|4}$ noncommutative. In the
operator formulation, this again yields a matrix model with\index{matrix model}
infinite-dimensional matrices and there is also a Penrose-Ward
transform which renders the two noncommutative matrix models\index{Penrose-Ward transform}\index{matrix model}
equivalent.

In a certain limit, in which the ranks of the gauge groups
$\sU(n)$ and $\sGL(n,\FC)$ of the SDYM and the hCS matrix model\index{matrix model}
tend to infinity, one expects them to become equivalent to the
respective matrix models obtained from noncommutativity.\index{matrix model}

Dimensional reductions of holomorphic Chern-Simons theory on\index{Chern-Simons theory}\index{dimensional reduction}
purely bosonic local Calabi-Yau manifolds have been studied\index{Calabi-Yau}\index{manifold}
recently in \cite{Bonelli:2005dc,Bonelli:2005gt}, where also
D-brane interpretations of the models were given.\index{D-brane}

\paragraph{Matrix model of $\CN=4$ SDYM theory.}\index{matrix model}
We start from the Lagrangian in the action \eqref{LagrangianSDYM}
of $\CN=4$ supersymmetric self-dual Yang-Mills theory in four\index{Yang-Mills theory}\index{self-dual Yang-Mills theory}
dimension with gauge group $\sU(n)$. One can dimensionally reduce
this theory to a point by assuming that all the fields are
independent of $x\in\FR^4$. This yields the matrix model action\index{matrix model}
\begin{equation}\label{SDYMactionMM}
\begin{aligned}
S_{\mathrm{SDYMMM}}\ =\
\tr\Bigg(\Bigg.&G^{\ald\bed}\left(-\tfrac{1}{2}\eps^{\alpha\beta}[A_{\alpha\ald},A_{\beta\bed}]\right)+
\tfrac{\eps}{2}\eps_{ijkl}\tilde{\chi}^{\ald ijk}
[A_{\alpha\ald},\chi^{\alpha l}]\\
&+\Bigg. \tfrac{\eps}{4}\eps_{ijkl}
\phi^{ij}[A_{\alpha\ald},[A^{\alpha\ald},\phi^{kl}]]+
\eps_{ijkl}\phi^{ij}\chi^{\alpha k}\chi^l_{\alpha}\Bigg)~,
\end{aligned}
\end{equation}
which is invariant under the adjoint action of the gauge group
$\sU(n)$ on all the fields. This symmetry is the remnant of gauge
invariance. The corresponding equations of motion read
\begin{equation}\label{SDYMeomMM}
\begin{aligned}
\eps^{\alpha\beta}[A_{\alpha\ald},A_{\beta\bed}]&\ =\ 0~,\\
[A_{\alpha\ald},\chi^{\alpha i}]&\ =\ 0~,\\
\tfrac{1}{2}[A^{\alpha\ald},[A_{\alpha\ald},\phi^{ij}]]&\ =\
-\tfrac{\eps}{2}\{\chi^{\alpha
i},\chi^j_{\alpha}\}~,\\
[A_{\alpha\ald},\tilde{\chi}^{\ald ijk}]&\ =\ +
2\eps\,[\phi^{ij},\chi^k_{\alpha},]~,\\
\eps^{\ald\dot{\gamma}}[A_{\alpha\ald},G^{ijkl}_{\dot{\gamma}
\dot{\delta}}]&\ =\
+\eps\{\chi^{i}_{\alpha},\tilde{\chi}^{jkl}_{\dot{\delta}}\}
-\eps\,[\phi^{ij},[A_{\alpha\dot{\delta}},\phi^{kl}]]~.
\end{aligned}
\end{equation}

These equations can certainly also be obtained by dimensionally
reducing equations \eqref{SDYMeom} to a point. On the other hand,
the equations of motion of $\CN=4$ SDYM theory are equivalent to
the constraint equations \eqref{constraintN4SDYM} which are\index{constraint equations}
defined on the superspace $\FR^{4|8}$. Therefore,\index{super!space}
\eqref{SDYMeomMM} are equivalent to the equations
\begin{equation}\label{SDYMconstrainsMM}
[\CAt_{\alpha\ald},\CAt_{\beta\bed}]\ =\ \eps_{\ald\bed}
\CF_{\alpha\beta}~,~~~ \nabla^i_{\ald}\CAt_{\beta\bed}\ =\
\eps_{\ald\bed} \CF^i_{\beta}~,~~~
\{\nabla^i_{\ald},\nabla^j_{\bed}\}\ =\ \eps_{\ald\bed} \CF^{ij}
\end{equation}
obtained from \eqref{constraintN4SDYM} by dimensional reduction to\index{dimensional reduction}
the supermanifold\footnote{Following the usual nomenclature of\index{super!manifold}
superlines and superplanes, this would be a ``superpoint''.}
$\FR^{0|8}$.

Recall that the IKKT matrix model \cite{Ishibashi:1996xs} can be\index{IKKT matrix model}\index{matrix model}
obtained by dimensionally reducing $\CN=1$ SYM theory in ten
dimensions or $\CN=4$ SYM in four dimensions to a point. In this
sense, the above matrix model is the self-dual analogue of the\index{matrix model}
IKKT matrix model.\index{IKKT matrix model}

\paragraph{Matrix model from hCS theory.}\index{matrix model}
So far, we have constructed a matrix model for $\CN=4$ SDYM
theory, the latter being defined on the space $(\FR^{4|8},g_\eps)$
with $\eps=-1$ corresponding to Euclidean signature and $\eps=+1$
corresponding to Kleinian signature of the metric on $\FR^4$. The\index{Kleinian signature}\index{metric}
next step is evidently to ask what theory corresponds to the
matrix model introduced above on the twistor space side.\index{matrix model}\index{twistor}\index{twistor!space}

Recall that for two signatures on $\FR^4$ we use the supertwistor\index{twistor}
spaces
\begin{equation}\label{3.3'}
\CP_\eps^{3|4}\ \cong\  \Sigma^1_\eps \times \FR^{4|8}
\end{equation}
where
\begin{equation}
\Sigma_{-1}^1\ :=\ \CPP^1\eand \Sigma^1_{+1}\ :=\ H^2
\end{equation}
and the two-sheeted hyperboloid $H^2$ is considered as a complex
space. As was discussed in section 3.1, the equations of motion
\eqref{SDYMeomMM} of the matrix model \eqref{SDYMactionMM} can be\index{matrix model}
obtained from the constraint equations \eqref{constraintN4SDYM} by\index{constraint equations}
reducing the space $\FR^{4|8}$ to the supermanifold $\FR^{0|8}$\index{super!manifold}
and expanding the superfields contained in
\eqref{SDYMconstrainsMM} in the Gra{\ss}mann variables $\eta_i^\ald$.\index{Gra{\ss}mann variable}
On the twistor space side, this reduction yields the orbit spaces\index{twistor}\index{twistor!space}
\begin{equation}\label{3.4'}
\Sigma^1_\eps\times \FR^{0|8}\ =\ \CP^{3|4}_\eps/\CCG~,
\end{equation}
where $\CCG$ is the group of translations generated by the bosonic
vector fields $\der{x^{\alpha\ald}}$. Equivalently, one can define
the spaces $\CP^{1|4}_\eps$ as the orbit spaces
\begin{equation}
\CP^{1|4}_\eps\ :=\ \CP^{3|4}_\eps/\CCG^{1,0}~,
\end{equation}
where $\CCG^{1,0}$ is the complex Abelian group generated by the
vector fields $\der{z^\alpha_\pm}$. These spaces with $\eps=\pm 1$
are covered by the two patches $U^\eps_\pm \cong \FC^{1|4}$ and
they are obviously diffeomorphic to the spaces \eqref{3.4'},\index{diffeomorphic} i.e.\
\begin{equation}\label{3.5'}
\CP_\eps^{1|4}\ \cong\  \Sigma_\eps^1\times \FR^{0|8}
\end{equation}
due to the diffeomorphism \eqref{3.3'}. In the coordinates
$(z^3_\pm,\eta_i^\pm)$ on $\CP_\eps^{1|4}$ and
$(\lambda_\pm,\eta_i^\ald)$ on $\Sigma^1_\eps\times \FR^{0|8}$,
the diffeomorphism is defined e.g.\ by the formul\ae{}
\begin{equation}\label{eq:2.35}
\begin{aligned}
\eta_1^\ed&\ =\
\frac{\eta_1^+-z_+^3\bar{\eta}_2^+}{1+z_+^3\bz_+^3}\ =\
\frac{\bz_-^3\eta_1^--\bar{\eta}_2^-}{1+z_-^3\bz_-^3}~,&
\eta_2^\ed&\ =\
\frac{\eta_2^++z_+^3\bar{\eta}_1^+}{1+z_+^3\bz_+^3}\ =\
\frac{\bz_-^3\eta_2^-+\bar{\eta}_1^-}{1+z_-^3\bz_-^3}~,\\
\eta_3^\ed&\ =\
\frac{\eta_3^+-z_+^3\bar{\eta}_4^+}{1+z_+^3\bz_+^3}\ =\
\frac{\bz_-^3\eta_3^--\bar{\eta}_4^-}{1+z_-^3\bz_-^3}~,&
\eta_4^\ed&\ =\
\frac{\eta_4^++z_+^3\bar{\eta}_3^+}{1+z_+^3\bz_+^3}\ =\
\frac{\bz_-^3\eta_4^-+\bar{\eta}_3^-}{1+z_-^3\bz_-^3}~,
\end{aligned}
\end{equation}
in the Euclidean case $\eps=-1$. Thus, we have a dimensionally
reduced twistor correspondence between the spaces $\CP_\eps^{1|4}$\index{twistor}\index{twistor!correspondence}
and $\FR^{0|8}$
\begin{equation}\label{3.13'}
\begin{aligned}
\begin{picture}(50,40)
\put(0.0,0.0){\makebox(0,0)[c]{$\CP^{1|4}_\eps$}}
\put(64.0,0.0){\makebox(0,0)[c]{$\FR^{0|8}$}}
\put(34.0,33.0){\makebox(0,0)[c]{$\FR^{0|8}\times \Sigma_\eps^1$}}
\put(7.0,18.0){\makebox(0,0)[c]{$\pi_2$}}
\put(55.0,18.0){\makebox(0,0)[c]{$\pi_1$}}
\put(25.0,25.0){\vector(-1,-1){18}}
\put(37.0,25.0){\vector(1,-1){18}}
\end{picture}
\end{aligned}
\end{equation}
where the map $\pi_2$ is the diffeomorphism \eqref{3.5'}. It
follows from \eqref{3.13'} that we have a correspondence between
points $\eta\in \FR^{0|8}$ and subspaces $\CPP^1_\eta$ of
$\CP^{1|4}_\eps$.

Holomorphic Chern-Simons theory on $\CP^{3|4}_\eps$ with the\index{Chern-Simons theory}
action \eqref{actionhCS} is defined by a gauge potential
$\CA^{0,1}$ taking values in the Lie algebra of $\sGL(n,\FC)$ and\index{Lie algebra}
constrained by the equations $\bV_\pm^i(\bV_a^\pm\lrcorner
\CA^{0,1})=0$, $\bV_\pm^i\lrcorner \CA^{0,1}=0$ for $a=1,2,3$.
After reduction to $\CP_\eps^{1|4}$, $\CA^{0,1}$ splits into a
gauge potential and two complex scalar fields taking values in the
normal bundle $\FC^2\otimes \CO(1)$ to the space\index{normal bundle}
$\CP_\eps^{1|4}\embd \CP_\eps^{3|4}$. In components, we have
\begin{equation}
\CA^{0,1}_{\Sigma\pm}\ =\ \dd \bl_\pm\CA_{\bl_\pm}\eand
\CA_\alpha^\pm\ \mapsto\ \CX_\alpha^\pm~~~\mbox{on}~~~U_\pm^\eps~,
\end{equation}
where both $\CX_\alpha^\pm$ and $\CA_{\bl_\pm}$ are Lie algebra\index{Lie algebra}
valued superfunctions on the subspaces $U_\pm^\eps$ of
$\CP^{1|4}_\eps$. The integral over the chiral subspace
$\CCP^{3|4}_\eps\subset\CP_\eps^{3|4}$ should be evidently
substituted by an integral over the chiral subspace
$\CCP^{1|4}_\eps\subset \CP^{1|4}_\eps$. This dimensional
reduction of the bosonic coordinates becomes even clearer with the\index{dimensional reduction}
help of the identity
\begin{equation}\label{formreduction} \dd \lambda_\pm\wedge \dd
\bl_\pm\wedge\dd z_\pm^1\wedge \dd z_\pm^2\wedge \bE_\pm^1\wedge
\bE_\pm^2\ =\ \dd \lambda_\pm\wedge \dd \bl_\pm\wedge\dd
x^{1\dot{1}}\wedge\dd x^{1\dot{2}}\wedge\dd x^{2\dot{1}}\wedge\dd
x^{2\dot{2}}~.
\end{equation}
Altogether, the dimensionally reduced action reads
\begin{equation}\label{actionhCSred}
S_{\mathrm{hCS,red}}\ :=\ \int_{\CCP^{1|4}_\eps} \omega \wedge
\tr\eps^{\alpha\beta}\CX_\alpha\left(\dparb\CX_\beta+
\left[\CA_\Sigma^{0,1},\CX_\beta\right]\right)~,
\end{equation}
where the form $\omega$ is a restriction of the form $\Omega$ from
\eqref{Omega} and has components
\begin{equation}\label{defomega}
\omega_\pm\ :=\ \Omega|_{U_\pm^\eps}\ =\ \pm\dd\lambda_\pm
\dd\eta_1^\pm\ldots \dd \eta^\pm_4
\end{equation}
and thus takes values in the bundle $\CO(-2)$. Note furthermore
that $\dparb$ here is the Dolbeault operator on $\Sigma_\eps^1$
and the integral in \eqref{actionhCSred} is well-defined since the
$\CX_\alpha$ take values in the bundles $\CO(1)$. The
corresponding equations of motion are given by
\begin{subequations}\label{shCSred}
\begin{eqnarray}\label{shCSred1}
[\CX_1,\CX_2]&=&0~,\\
\label{shCSred2} \dparb\CX_\alpha+
[\CA^{0,1}_\Sigma,\CX_\alpha]&=&0~.
\end{eqnarray}
\end{subequations}
The gauge symmetry is obviously reduced to the transformations
\begin{equation}
\CX_\alpha\ \mapsto\ \varphi^{-1}\CX_\alpha\varphi\eand
\CA^{0,1}_\Sigma\ \mapsto\
\varphi^{-1}\CA^{0,1}_\Sigma\varphi+\varphi^{-1}\dparb\varphi~,
\end{equation}
where $\varphi$ is a smooth $\sGL(n,\FC)$-valued function on
$\CP_\eps^{1|4}$. The matrix model given by \eqref{actionhCSred}\index{matrix model}
and the field equations \eqref{shCSred} can be understood as
matrix quantum mechanics with a complex ``time''\index{matrix quantum mechanics}
$\lambda\in\Sigma_\eps^1$.

Both the matrix models obtained by dimensional reductions of\index{dimensional reduction}\index{matrix model}
$\CN=4$ supersymmetric SDYM theory and hCS theory are
(classically) equivalent. This follows from the dimensional
reduction of the formul\ae{} \eqref{expA} defining the\index{dimensional reduction}
Penrose-Ward transform. The reduced superfield expansion is fixed\index{Penrose-Ward transform}
by the geometry of $\CP_\eps^{1|4}$ and reads explicitly as
\begin{subequations}\label{fieldexpansionred}
\begin{eqnarray}\label{expredAa}
\CX_\alpha^+&=&\lambda_+^\ald\,
A_{\alpha\ald}+\eta_i^+\chi^i_\alpha+
\gamma_+\,\tfrac{1}{2!}\,\eta^+_i\eta^+_j\,\hl^\ald_+\,
\phi_{\alpha \ald}^{ij}+\\
\nonumber
&&+\gamma_+^2\,\,\tfrac{1}{3!}\,\eta^+_i\eta^+_j\eta^+_k\,\hl_+^\ald\,
\hl_+^\bed\,
\tilde{\chi}^{ijk}_{\alpha\ald\bed}+\gamma_+^3\,\tfrac{1}{4!}\,
\eta^+_i\eta^+_j\eta^+_k\eta^+_l\,
\hl_+^\ald\,\hl_+^\bed\,\hl_+^{\dot{\gamma}}\,
G^{ijkl}_{\alpha\ald\bed\dot{\gamma}}~,\\
\label{expredAl}
\CA_{\bl_+}&=&\gamma_+^2\eta^+_i\eta^+_j\,\phi^{ij}-
\gamma_+^3\eta^+_i\eta^+_j\eta^+_k\,\hl_+^\ald\,
\tilde{\chi}^{ijk}_{\ald}
+\\
\nonumber &&+2\gamma_+^4\eta^+_i\eta^+_j\eta^+_k\eta^+_l\,
\hl_+^\ald\,\hl_+^\bed G^{ijkl}_{\ald\bed}~,
\end{eqnarray}
\end{subequations}
where all component fields are independent of $x\in\FR^4$. One can
substitute this expansion into the action \eqref{actionhCSred} and
after a subsequent integration over $\CP^{1|4}_\eps$, one obtains
the action \eqref{SDYMactionMM} up to a constant multiplier, which
is the volume\footnote{In the Kleinian case, this volume is
na\"ively infinite, but one can regularize it by utilizing a
suitable partition of unity.} of $\Sigma^1_\pm$.

\paragraph{Noncommutative star product in spinor notation.}\index{Spinor}
Noncommutative field theories have received much attention\index{noncommutative field theories}
recently, as they were found to arise in string theory in the\index{string theory}
presence of D-branes and a constant NS $B$-field background\index{D-brane}
\cite{Seiberg:1999vs,Douglas:2001ba,Szabo:2001kg}.

There are two completely equivalent ways of introducing a
noncommutative deformation of classical field theory: a
star-product formulation and an operator formalism. In the first\index{operator formalism}
approach, one simply deforms the ordinary product of classical
fields (or their components) to the noncommutative star product
which reads in spinor notation as\index{Spinor}
\begin{equation}
(f\star g)(x)\ :=\
f(x)\exp\left(\tfrac{\di}{2}\overleftarrow{\dpar_{\alpha\ald}}
\theta^{\alpha\ald\beta\bed}\overrightarrow{\dpar_{\beta\bed}}\right)g(x)
\end{equation}
with $\theta^{\alpha\ald\beta\bed}=-\theta^{\beta\bed\alpha\ald}$
and in particular
\begin{equation}
x^{\alpha\ald}\star x^{\beta\bed}-x^{\beta\bed}\star
x^{\alpha\ald}\ =\ \di \theta^{\alpha\ald\beta\bed}~.
\end{equation}
In the following, we restrict ourselves to the case of a self-dual
$(\kappa=1)$ or an anti-self-dual $(\kappa=-1)$ tensor\index{anti-self-dual}
$\theta^{\alpha\ald\beta\bed}$ and choose coordinates such that
\begin{equation}
\theta^{1\ed 2\zd}\ =\ -\theta^{2\zd 1\ed}\ =\ -2\di\kappa\eps
\theta\eand \theta^{1\zd 2\ed}\ =\ -\theta^{2\ed 1\zd}\ =\ 2\di
\eps\theta~.
\end{equation}

The formulation of noncommutative $\CN=4$ SDYM theory on
$(\FR^4_\theta,g_\eps)$ is now achieved by replacing all products
in the action \eqref{LagrangianSDYM} by star products. For
example, the noncommutative field strength will read\index{field strength}
\begin{equation}
F_{\alpha\ald\beta\bed}\ =\
\dpar_{\alpha\ald}A_{\beta\bed}-\dpar_{\beta\bed}A_{\alpha\ald}+
A_{\alpha\ald}\star A_{\beta\bed}-A_{\beta\bed}\star
A_{\alpha\ald}~.
\end{equation}

\paragraph{Operator formalism.} For the matrix reformulation of\index{operator formalism}
our model, it is necessary to switch to the operator formalism,
which trades the star product for operator-valued coordinates
$\hat{x}^{\alpha\ald}$ satisfying
\begin{equation}\label{Heisenbergalgebra}
[\hat{x}^{\alpha\ald},\hat{x}^{\beta\bed}]\ =\ \di
\theta^{\alpha\ald\beta\bed}~.
\end{equation}
This defines the noncommutative space $\FR^4_\theta$ and on this
space, derivatives are inner derivations of the Heisenberg algebra\index{Heisenberg algebra}
\eqref{Heisenbergalgebra}:
\begin{equation}
\begin{aligned}
\der{\hat{x}^{1\ed}} f&\ :=\
-\frac{1}{2\kappa\eps\theta}[\hat{x}^{2\zd},f]~,~~~
&\der{\hat{x}^{2\zd}} f&\ :=\
+\frac{1}{2\kappa\eps\theta}[\hat{x}^{1\ed},f]~,\\
\der{\hat{x}^{1\zd}} f&\ :=\
+\frac{1}{2\eps\theta}[\hat{x}^{2\ed},f]~,~~~
&\der{\hat{x}^{1\zd}} f&\ :=\
-\frac{1}{2\eps\theta}[\hat{x}^{1\zd},f]~.
\end{aligned}
\end{equation}
The obvious representation space for the algebra\index{representation}
\eqref{Heisenbergalgebra} is the two-oscillator Fock space $\CH$
which is created from a vacuum state $|0,0\rangle$. This vacuum
state is annihilated by the operators
\begin{equation}
\begin{aligned}
\hat{a}_1&\ =\
\di\left(\frac{1-\eps}{2}\hat{x}^{2\ed}+\frac{1+\eps}{2}\hat{x}^{1\zd}\right)~,\\
\hat{a}_2&\ =\
-\di\left(\frac{1-\kappa\eps}{2}\hat{x}^{2\zd}+\frac{1+\kappa\eps}{2}\hat{x}^{1\ed}\right)
\end{aligned}
\end{equation}
and all other states of $\CH$ are obtained by acting with the
corresponding creation operators on $|0,0\rangle$. Thus,
coordinates as well as fields are to be regarded as operators in
$\CH$.

Via the Moyal-Weyl map
\cite{Seiberg:1999vs,Douglas:2001ba,Szabo:2001kg}, any function
$\Phi(x)$ in the star-product formulation can be related to an
operator-valued function $\hat{\Phi}(\hat{x})$ acting in $\CH$.
This yields the operator equivalent of star multiplication and
integration
\begin{equation}
f\star g\ \mapsto\ \hat{f}\hat{g}\eand \int \dd^4 x~ f\ \mapsto\
(2\pi\theta)^2\tr_\CH \hat{f}~,
\end{equation}
where $\tr_\CH$ signifies the trace over the Fock space $\CH$.

We now have all the ingredients for defining noncommutative
$\CN=4$ super SDYM theory in the operator formalism. Starting\index{N=4 super SDYM theory@$\CN=4$ super SDYM theory}\index{operator formalism}
point is the analogue of the covariant derivatives which are given\index{covariant derivative}
by the formul\ae{}
\begin{equation}
\begin{aligned}\nonumber
\hat{X}_{1\dot{1}}&\ =\
-\frac{1}{2\kappa\eps\theta}\hat{x}^{2\dot{2}}\otimes
\unit_n+\hat{A}_{1\dot{1}}~,& \hat{X}_{2\dot{2}}&\ =\
\frac{1}{2\kappa\eps\theta}\hat{x}^{1\dot{1}}\otimes
\unit_n+\hat{A}_{2\dot{2}}~,\\
\hat{X}_{1\dot{2}}&\ =\
\frac{1}{2\eps\theta}\hat{x}^{2\dot{1}}\otimes
\unit_n+\hat{A}_{1\dot{2}}~,& \hat{X}_{2\dot{1}}&\ =\
-\frac{1}{2\eps\theta}\hat{x}^{1\dot{2}}\otimes
\unit_n+\hat{A}_{2\dot{1}}~.
\end{aligned}
\end{equation}
These operators act on the tensor product of the Fock space $\CH$
and the representation space of the Lie algebra of the gauge group\index{Lie algebra}\index{representation}
$\sU(n)$. The operator-valued field strength has then the form\index{field strength}
\begin{equation}
\hat{F}_{\alpha\ald \beta\bed}\ =\
[\hat{X}_{\alpha\ald},\hat{X}_{\beta\bed}]+\di\theta_{\alpha\ald\beta\bed}\otimes
\unit_n~,
\end{equation}
where the tensor $\theta_{\alpha\ald\beta\bed}$ has components
\begin{equation}
\theta_{1\dot{1}2\dot{2}}\ =\ -\theta_{2\dot{2}1\dot{1}}\ =\
\di\frac{\kappa\eps}{2\theta}~,~~~ \theta_{1\dot{2}2\dot{1}}\ =\
-\theta_{2\dot{1}1\dot{2}}\ =\ -\di\frac{\eps}{2\theta}~,
\end{equation}
Recall that noncommutativity restricts the set of allowed gauge
groups and we therefore had to choose to work with $\sU(n)$
instead of $\sSU(n)$.

The action of noncommutative SDYM theory on
$(\FR^4_\theta,g_\eps)$ reads
\begin{equation}\label{SDYMactionNC}
\begin{aligned}
S^{\CN=4}_{\mathrm{ncSDYM}}\ =\
\tr_\CH\tr\Bigg(\Bigg.-\tfrac{1}{2}&\eps^{\alpha\beta}\hat{G}^{\ald\bed}
\left([\hat{X}_{\alpha\ald},\hat{X}_{\beta\bed}]+\di\theta_{\alpha\ald\beta\bed}\otimes
\unit_n\right)\\+
\tfrac{\eps}{2}\eps_{ijkl}\tilde{\hat{\chi}}^{\ald ijk}
[\hat{X}_{\alpha\ald},\hat{\chi}^{\alpha l}]&+
\tfrac{\eps}{2}\eps_{ijkl}
\hat{\phi}^{ij}[\hat{X}_{\alpha\ald},[\hat{X}^{\alpha\ald},\hat{\phi}^{kl}]]+
\eps_{ijkl}\hat{\phi}^{ij}\hat{\chi}^{\alpha
k}\hat{\chi}^l_{\alpha}\Bigg.\Bigg)~.
\end{aligned}
\end{equation}
For $\kappa=+1$, the term containing
$\theta_{\alpha\ald\beta\bed}$ vanishes after performing the index
sums. Note furthermore that in the limit of $n\rightarrow \infty$
for the gauge group $\sU(n)$, one can render the ordinary $\CN=4$
SDYM matrix model \eqref{SDYMactionMM} equivalent to\index{matrix model}
noncommutative $\CN=4$ SDYM theory defined by the action
\eqref{SDYMactionNC}. This is based on the fact that there is an
isomorphism of spaces $\FC^\infty\cong \CH$ and $\FC^n\otimes\index{morphisms!isomorphism}
\CH$.

\paragraph{Noncommutative hCS theory.}
The natural question to ask at this point is whether one can
translate the Penrose-Ward transform completely into the\index{Penrose-Ward transform}
noncommutative situation and therefore obtain a holomorphic
Chern-Simons theory on a noncommutative supertwistor space. For\index{Chern-Simons theory}\index{twistor}\index{twistor!space}
the Penrose-Ward transform in the purely bosonic case, the answer\index{Penrose-Ward transform}
is positive (see e.g.\
\cite{Kapustin:2000ek,Takasaki:2000vs,Lechtenfeld:2001ie,Lechtenfeld:2003vv}).

In the supersymmetric case, by defining the correspondence space
as $(\FR^{4|8}_\theta,g_\eps)\times \Sigma_\eps^1$ with the
coordinate algebra \eqref{Heisenbergalgebra} and unchanged algebra
of Gra{\ss}mann coordinates, we arrive together with the incidence
relations in \eqref{supercoords} at noncommutative\index{incidence relation}
coordinates\footnote{Observe that the coordinates on
$\Sigma^1_\eps$ stay commutative.} on the twistor space\index{twistor}\index{twistor!space}
$\CP^{3|4}_{\eps,\theta}$ satisfying the relations
\begin{equation}\label{noncommz}
\begin{aligned}
{}[\hat{z}^1_\pm,\hat{z}^2_\pm]&\ =\ 2(\kappa-1)\eps
\lambda_\pm\theta~,&[\hat{\bar{z}}^1_\pm,\hat{\bar{z}}^2_\pm]&\ =\
-2(\kappa-1)\eps\bl_\pm \theta~,\\
{}[\hat{z}^1_+,\hat{\bar{z}}^1_+]&\ =\
2(\kappa\eps-\lambda_+\bl_+) \theta~,&
[\hat{z}^1_-,\hat{\bar{z}}^1_-]&\ =\ 2(\kappa\eps\lambda_-\bl_--1)
\theta~,\\
[\hat{z}^2_+,\hat{\bar{z}}^2_+]&\ =\
2(1-\eps\kappa\lambda_+\bl_+)\theta~,&[\hat{z}^2_-,\hat{\bar{z}}^2_-]&\
=\ 2(\lambda_-\bl_--\eps\kappa)\theta~,
\end{aligned}
\end{equation}
with all other commutators vanishing. Here, we clearly see the\index{commutators}
advantage of choosing a self-dual deformation tensor $\kappa=+1$:
the first line in \eqref{noncommz} becomes trivial. We will
restrict our considerations to this case\footnote{Recall, however,
that the singularities of the moduli space of self-dual solutions\index{moduli space}
are not resolved when choosing a self-dual deformation tensor.} in
the following.

Thus, we see that the coordinates $z^\alpha$ and $\bz^\alpha$ are
turned into sections $\hat{z}^\alpha$ and $\hat{\bar{z}}^\alpha$
of the bundle $\CO(1)$ which are functions on $\CP^{1|4}_\eps$ and
take values in the space of operators acting on the Fock space
$\CH$. The derivatives along the bosonic fibres of the fibration\index{fibration}
$\CP^{3|4}_\eps\rightarrow \CP^{1|4}_\eps$ are turned into inner
derivatives of the algebra \eqref{noncommz}:
\begin{equation}
\der{\hat{\bz}^1_\pm}f\ =\ \frac{\eps}{2\theta}
\gamma_\pm\,[\hat{z}^1_\pm,f]~,~~~ \der{\hat{\bz}^2_\pm}f\ =\
\frac{1}{2\theta}\gamma_\pm\,[\hat{z}^2_\pm,f]~.
\end{equation}
Together with the identities \eqref{eq:2.21}, we can furthermore
derive
\begin{equation}\label{Vdef}
\hat{\bV}_1^\pm f\ =\
-\frac{\eps}{2\theta}\,[\hat{z}^2_\pm,f]\eand \hat{\bV}_2^\pm f\
=\ -\frac{\eps}{2\theta}\,[\hat{z}^1_\pm,f]~.
\end{equation}

The formul\ae{} \eqref{Vdef} allow us to define the
noncommutatively deformed version of the hCS action
\eqref{actionhCScpt}:
\begin{align}\label{actionhCSdef1}
S_{\mathrm{nchCS}}\ :=\ & \int_{\CCP^{1|4}_\eps}\omega\wedge
\tr_\CH \otimes\tr \left\{
\left(\hat{\CA}_2\dparb\hat{\CA}_1-\hat{\CA}_1\dparb\hat{\CA}_2\right)+2\hat{\CA}^{0,1}_\Sigma\left[\hat{\CA}_1,\hat{\CA}_2\right]-\right.
~,
\\\nonumber&\left.-\frac{\eps}{2\theta}\left(\hat{\CA}_1\left[\hat{z}^1,\hat{\CA}^{0,1}_\Sigma\right]-
\hat{\CA}^{0,1}_\Sigma\left[\hat{z}^1,\hat{\CA}_1\right]+\hat{\CA}^{0,1}_\Sigma\left[\hat{z}^2,\hat{\CA}_2\right]-
\hat{\CA}_2\left[\hat{z}^2,\hat{\CA}^{0,1}_\Sigma\right]\right)\right\}~,
\end{align}
where $\CCP^{1|4}_\eps$ is again the chiral subspace of
$\CP^{1|4}_\eps$ for which $\etab^i_\pm=0$, $\omega$ is the form
defined in \eqref{defomega} and $\tr_\CH$ and $\tr$ denote the
traces over the Fock space $\CH$ and the representation space of\index{representation}
$\agl(n,\FC)$, respectively. The hats indicate that the components
of the gauge potential $\hat{\CA}^{0,1}$ are now operators with
values in the Lie algebra $\agl(n,\FC)$.\index{Lie algebra}

We can simplify the above action by introducing the operators
\begin{equation}
\hat{\CX}^1_\pm\ =\ -\frac{\eps}{2\theta}\hat{z}^2_\pm\otimes
\unit_n+\hat{\CA}_1^\pm\eand \hat{\CX}^2_\pm\ =\
-\frac{\eps}{2\theta}\hat{z}^1_\pm\otimes \unit_n+\hat{\CA}_2^\pm~
\end{equation}
which yields
\begin{equation}\label{actionhCSdef2}
S_{\mathrm{nchCS}}\ =\
\int_{\CCP^{1|4}_\eps}\omega\wedge\tr_\CH\otimes
\tr\eps^{\alpha\beta}\hat{\CX}_\alpha\left(\dparb\hat{\CX}_\beta+
[\hat{\CA}^{0,1}_\Sigma,\hat{\CX}_\beta]\right)~,
\end{equation}
where the $\hat{\CX}_\alpha$ take values in the bundle $\CO(1)$,
so that the above integral is indeed well defined. Note that in
the matrix model \eqref{actionhCSred}, we considered matrices\index{matrix model}
taking values in the Lie algebra $\agl(n,\FC)$, while the fields\index{Lie algebra}
$\hat{\CX}_\alpha$ and $\hat{\CA}^{0,1}_\Sigma$ in the model
\eqref{actionhCSdef2} take values in $\agl(n,\FC)\otimes
\sEnd(\CH)$ and can be represented by infinite-dimensional
matrices.

\paragraph{String field theory.}\index{string field theory}
The form of the matrix model action given by \eqref{actionhCSdef2}\index{matrix model}
is identical to an action recently given as a cubic string field
theory for open $\CN=2$ strings \cite{Lechtenfeld:2004cc}. Let us\index{string field theory}
comment on that point in more detail.

First of all, recall cubic string field theory action from section\index{string field theory}
\ref{sssft} which reads as
\begin{equation}
S=\tfrac{1}{2}\int \left(\CA\star
Q\CA+\tfrac{2}{3}\CA\star\CA\star\CA\right)~.
\end{equation}
To qualify as a string field, $\CA$ is a functional of the
embedding map $\Phi$ from the string parameter space to the string
target space. For the case at hand, we take\index{target space}
\begin{equation}
\Phi:[0,\pi]\times G\ \rightarrow\  \CP^{3|4}_\eps~,
\end{equation}
where $\sigma\in[0,\pi]$ parameterizes the open string and $G\ni\index{open string}
v$ provides the appropriate set of Gra{\ss}mann variables on the\index{Gra{\ss}mann variable}
worldsheet. Expanding $\Phi(\sigma,v)=\phi(\sigma)+v\psi(\sigma)$,\index{worldsheet}
this map embeds the $\CN=2$ spinning string into supertwistor\index{twistor}
space. Next, we recollect
$\phi=(z^\alpha=x^{\alpha\ald}\lambda_\ald,\eta_i=\eta_i^\ald\lambda_\ald,\lambda,\bl)$
and allow the string to vibrate only in the $z^\alpha$-directions
but keep the $G$-even zero modes of $(\eta_i,\lambda,\bl)$, so
that the string field depends on
$\{z^{\alpha}(\sigma),\eta_i,\lambda,\bl;\psi^{\alpha\ddot{\alpha}}(\sigma)\}$
only \cite{Lechtenfeld:2004cc}. Note that with $\psi$ and $\eta$,
we have two types of fermionic fields present, since we are
implicitly working in the doubly supersymmetric description of
superstrings \cite{Sorokin:1999jx}, which we will briefly discuss
in section~4.2. Therefore, the two fermionic fields are linked via
a superembedding condition. We employ a suitable BRST operator
$Q=\bar{D}+\dparb$, where
$\bar{D}=\psi^{\alpha\ddot{1}}\lambda^\ald\dpar_{\sigma}x_{\alpha\ald}\in\CO(1)$
and $\dparb\in\CO(0)$ are type $(0,1)$ vector fields on the fibres
and the base of $\CP^{3|4}_\eps$, respectively, and split the
string field accordingly, $\CA=\CA_{\bar{D}}+\CA_{\dparb}$. With a
holomorphic integration measure on $\CP^{3|4}_\eps$, the
Chern-Simons action \eqref{actionhCS} projects to
\cite{Lechtenfeld:2004cc}
\begin{equation}\label{actionSFT}
S\ =\ \int_{\CCP^{1|4}_\eps} \omega \wedge\langle \tr
(\CA_{\bar{D}}\star\dparb\CA_{\bar{D}}+2\CA_{\bar{D}}\star
\bar{D}\CA_{\dparb}+
2\CA_{\dparb}\star\CA_{\bar{D}}\star\CA_{\bar{D}}\rangle~.
\end{equation}
Note that the string fields $\CA_{\bar{D}}$ and $\CA_{\dparb}$ are
fermionic, i.e.\ they behave in the action as if they were forms
multiplied with the wedge product. Furthermore, the
above-mentioned $\RZ$-grading of all the ingredients of this
action has to be adjusted appropriately. Giving an expansion in
$\eta_i$ for these string fields similar to the one in
\eqref{expAa} and \eqref{expAl}, one recovers the super string
field theory proposed by Berkovits and Siegel\index{string field theory}
\cite{Berkovits:1997pq}. Its zero modes describe self-dual $\CN=4$
SDYM theory.

By identifying $Q+\CA_{Q}$ with $\hat{\CX}$, $\dparb$ with
$\dparb_{\bl}$ and $\CA_\dparb$ with $\hat{\CA}^{0,1}_\Sigma$ and
adjusting the $\RZ_2$-grading of the fields, one obtains the\index{Z2-grading@$\RZ_2$-grading}
action\footnote{Note that $\dparb Q+Q\dparb=0$.}
\eqref{actionhCSdef2} from \eqref{actionSFT}. Therefore, we can
e.g.\ translate solution generating techniques which are at hand\index{solution generating techniques}
for our matrix model immediately to the string field theory\index{matrix model}\index{string field theory}
\eqref{actionSFT}.

\subsection{Classical solutions to the noncommutative matrix
model}\label{ssClassicalSolsMM}\index{matrix model}

Simple classical solutions of $\CN=4$ noncommutative SDYM theory
can be obtained by considering only the helicity $\pm 1$ part of\index{helicity}
the field content and putting all other fields to
zero.\footnote{This situation can also be obtained by taking
holomorphic Chern-Simons theory on a thickening of ordinary\index{Chern-Simons theory}\index{thickening}
twistor space as discussed in \cite{Saemann:2004tt} as a starting\index{twistor}\index{twistor!space}
point for constructing the matrix model.} Solutions with all the\index{matrix model}
fields being nontrivial can then be recovered by either acting
with supersymmetry transformations on the previously obtained seed\index{super!symmetry}
solutions or by a supersymmetric extension of the dressing method
\cite{Belavin:1978pa}. After considering the simple example of
Abelian instantons, we will present two solution generating\index{instanton}
techniques inspired by twistor methods similar to the ones\index{twistor}
presented in \cite{Popov:2005uv}.

\paragraph{Abelian instantons.} We start from the simplest case\index{instanton}
and consider Abelian instantons\footnote{Instantons on
noncommutative spaces were first discussed in
\cite{Nekrasov:1998ss}, cf.\ also \cite{Ihl:2002kz}, appendix B.},
i.e.\ let us choose the gauge group to be $\sU(1)$. Contrary to
commutative spaces, this gauge group allows for instantons in the\index{instanton}
noncommutative setting. In some cases, problematic singularities
in the moduli space of instantons are resolved by\index{instanton}\index{moduli space}
noncommutativity, but this will not be the case for self-dual
instantons on a space with self-dual deformation.\index{instanton}

The equations of motion for helicity $\pm1$ derived from\index{helicity}
\eqref{actionhCSdef2} read for this choice
\begin{equation}\label{eomhCSdef}
\eps^{\alpha\beta}[\hat{X}_{\alpha\ald},\hat{X}_{\beta\bed}]\ =\
0~\eand [\hat{X}_{\alpha\ald},\hat{G}^{\ald\gad}]\ =\ 0~.
\end{equation}
The first equation can be reduced to components and is then
spelled out as
\begin{equation}\label{eomcomp}
\begin{aligned}
{}[\hat{X}_{1\dot{1}},\hat{X}_{2\dot{1}}]\ =\ [\hat{X}_{1\dot{2}},\hat{X}_{2\dot{2}}]&\ =\ 0~,\\
{}[\hat{X}_{1\dot{1}},\hat{X}_{2\dot{2}}]-[\hat{X}_{2\dot{1}},\hat{X}_{1\dot{2}}]&\
=\ 0~.
\end{aligned}
\end{equation}
A trivial solution of these equations is the following:
\begin{equation}\label{seedsol}
\hat{X}_{1\dot{1}}\ =\ \di \hat{x}^{2\dot{2}}~,~~~
\hat{X}_{1\dot{2}}\ =\ \di
\hat{x}^{2\dot{1}}~,~~~\hat{X}_{2\dot{2}}\ =\ \di
\hat{x}^{1\dot{1}}~,~~~ \hat{X}_{2\dot{1}}\ =\ \di
\hat{x}^{1\dot{2}}~,
\end{equation}
where the factors of $\di$ were included to have anti-Hermitian
gauge potentials satisfying
\begin{equation}
\hat{X}_{1\dot{2}}\ =\ \hat{X}^\dagger_{2\dot{1}}\eand
\hat{X}_{2\dot{2}}\ =\ -\hat{X}^\dagger_{1\dot{1}}~,
\end{equation}
cf.\ equations \eqref{Acompsah}. These solutions can now be used
as seed solutions for the dressing method. In the two-oscillator
Fock space $\CH$, we introduce the projector on the vacuum
\begin{equation}
P_0\ =\ |0,0\rangle\langle 0,0|~,
\end{equation}
and two shift operators $S$ and $S^\dagger$ satisfying
\begin{equation}
S^\dagger S\ =\ \unit~,~~~S S^\dagger\ =\ \unit-P_0~,~~~ P_0 S\ =\
S^\dagger P_0\ =\ 0~.
\end{equation}
Then a new solution of \eqref{eomcomp} is given by
\begin{equation}
\tilde{\hat{X}}_{\alpha\ald}\ =\ S \hat{X}_{\alpha\ald}
S^\dagger~,
\end{equation}
where $\hat{X}_{\alpha\ald}$ is given in \eqref{seedsol}. In fact,
one can show that this method generates all possible solutions if
we allow for a more general projector $P_0$.

To obtain a seed solution to the second equation in
\eqref{eomhCSdef}, we consider this equation in components:
\begin{equation}\label{eomcompG}
\begin{aligned}\nonumber
[\hat{x}^{2\dot{2}},\hat{G}^{\dot{1}\dot{1}}]+[\hat{x}^{2\dot{1}},\hat{G}^{\dot{2}\dot{1}}]&\
=\ 0~,
&[\hat{x}^{1\dot{2}},\hat{G}^{\dot{1}\dot{1}}]+[\hat{x}^{1\dot{1}},\hat{G}^{\dot{2}\dot{1}}]&\ =\ 0~,\\
[\hat{x}^{1\dot{2}},\hat{G}^{\dot{1}\dot{2}}]+[\hat{x}^{1\dot{1}},\hat{G}^{\dot{2}\dot{2}}]&\
=\ 0~,
&[\hat{x}^{2\dot{2}},\hat{G}^{\dot{1}\dot{2}}]+[\hat{x}^{2\dot{1}},\hat{G}^{\dot{2}\dot{2}}]&\
=\ 0~.
\end{aligned}
\end{equation}
Additionally, when constructing a solution, we have to guarantee
$\hat{G}^{\ald\bed}=\hat{G}^{\bed\ald}$. One easily observes that
the problem of finding solutions to \eqref{eomcompG} decomposes
into finding solutions to the left two equations and solutions to
the right two equations and we choose the trivial solutions
\begin{equation}
\begin{aligned}
\hat{G}^{\ed\dot{1}}&\ =\ \hat{x}^{1\dot{1}}+c_1
\hat{x}^{2\dot{2}}~,&
\hat{G}^{\ed\dot{1}}&\ =\ \hat{x}^{2\dot{1}}+c_3 \hat{x}^{1\dot{2}}~,\\
\hat{G}^{\ed\dot{2}}\ =\ \hat{G}^{\zd\dot{1}}&\ =\
\hat{x}^{1\dot{2}}+\hat{x}^{2\dot{1}}~,&
\hat{G}^{\ed\dot{2}}\ =\ \hat{G}^{\zd\dot{1}}&\ =\ \hat{x}^{1\dot{1}}+\hat{x}^{2\dot{2}}~,\\
\hat{G}^{\zd\dot{2}}&\ =\ \hat{x}^{2\dot{2}}+c_2
\hat{x}^{1\dot{1}}~,& \hat{G}^{\ed\dot{1}}&\ =\
\hat{x}^{1\dot{2}}+c_4 \hat{x}^{2\dot{1}}~,
\end{aligned}
\end{equation}
respectively, where the $c_i$ are arbitrary complex constants. An
appropriate reality condition on the field $\hat{G}^{\ald\bed}$
demands, however, that $c_1^*=-c_2$ and $c_3=c_4^*$. Note that
both solutions can be linearly combined.

With the help of the shift operators $S$ and $S^\dagger$, one can
construct the ``dressed'' solution
\begin{equation}
\tilde{\hat{G}}^{\ald\bed}\ =\ S \hat{G}^{\ald\bed} S^\dagger~.
\end{equation}

\paragraph{Supersymmetric ADHM construction.} Although the solutions\index{ADHM construction}
obtained above by ``educated guessing'' can be made more
interesting by supersymmetry transformations, the procedure to\index{super!symmetry}
obtain further such solutions would be rather cumbersome.
Therefore, we will turn our attention in the following to more
sophisticated solution generating techniques.\index{solution generating techniques}

Besides a supersymmetric extension of the dressing method
\cite{Belavin:1978pa}, the most obvious candidate is the
supersymmetric extension of the ADHM construction, see section\index{ADHM construction}
\ref{ssSADHM}. The self-dual gauge potential obtained from this
construction will have components which are superfunctions, and
the component expansion is the one given in
\eqref{fieldexpansiondevchand}.

Although this method seems to be nicely suited for constructing
solutions, it proves to be rather difficult to restrict to a
nontrivial field content for the helicity $\pm1$ fields. An\index{helicity}
extension of well-known bosonic solutions, as e.g.\ the 't Hooft
instanton, to solutions including a nontrivial field\index{instanton}
$G^{\ald\bed}$ can be done in principle but requires an
unreasonable effort in calculations. Therefore we discard this
ansatz.

\paragraph{Infinitesimal deformations.}\label{secinfdefo} A nicer
and more interesting way of systematically obtaining
solutions for the auxiliary field $\hat{G}^{\ald\bed}$ is found by
observing that the equations of motion for $\hat{G}^{\ald\bed}$
with vanishing spinors and scalar fields coincide with the\index{Spinor}
linearized equations of motion\footnote{The linearized equations
of motion are obtained from the noncommutative version of
\eqref{SDYMeom} by considering a finite gauge potential while all
other fields in the supermultiplet are only infinitesimal.} for\index{supermultiplet}
$\delta\hat{G}^{\ald\bed}$:
\begin{equation}
\nabla_{\alpha\ald}\delta\hat{G}^{\ald\bed}\ =\
\nabla_{\alpha\ald}\hat{G}^{\ald\bed}\ =\ 0~.
\end{equation}
Thus, it is sufficient for our purposes to find solutions for
$\delta\hat{G}^{\ald\bed}$ and render them subsequently finite.
This can be easily achieved by considering perturbations of the
representant of the \v{C}ech cohomology class corresponding to a\index{Cech cohomology@\v{C}ech cohomology}
given holomorphic structure $\dparb+\CA^{0,1}$ as we will show in\index{holomorphic!structure}
the following.

For this, let us start from a gauge potential $\CA^{0,1}$ and
perform the gauge transformation \eqref{gaugetrafo}. The result is\index{gauge!transformation}
a gauge potential with components $(\CA^\pm_\alpha,\CA^i_\pm)$,
which satisfy the constraint equation of supersymmetric self-dual
Yang-Mills theory \eqref{constraintN4SDYM}. Let us recall that\index{Yang-Mills theory}\index{self-dual Yang-Mills theory}
these constraint equation are the compatibility conditions of the\index{compatibility conditions}
linear system\index{linear system}
\begin{equation}
(\bV^+_\alpha+\CA^+_\alpha)\psi_\pm\ =\
0~,~~~\dpar_{\bl_\pm}\psi_\pm\ =\ 0~,~~~
(\bV^i_++\CA^i_+)\psi_\pm\ =\ 0~,
\end{equation}
on $\CUt_+$ and a similar one on $\CUt_-$. Therefore, we can write
\begin{equation}
\begin{aligned}
\CA^+_\alpha&\ =\ \psi_\pm \bV^+_\alpha\psi_\pm^{-1}~,&
\CA_{\bl_+}&\ =\ \psi_\pm\dpar_{\bl_+}\psi^{-1}_\pm\ =\ 0~,&\CA^i_+&\ =\ \psi_\pm \bV^i_+\psi^{-1}_\pm~,\\
\CA^-_\alpha&\ =\ \psi_\pm
\bV^-_\alpha\psi_\pm^{-1}~,&\CA_{\bl_-}&\ =\ \psi_\pm\dpar_{\bl_-}\psi^{-1}_\pm\ =\ 0~,&\CA^i_-&\
=\ \psi_\pm \bV^i_-\psi^{-1}_\pm~,
\end{aligned}
\end{equation}
for some regular matrix valued functions $\psi_\pm$. On the
overlap $\CU_+\cap\CU_-$, we have the gluing condition
$\psi_+\bar{D}^\pm_I\psi_+^{-1}=\psi_-\bar{D}^\pm_I\psi_-^{-1}$,
where $\bar{D}^\pm_I=(\bV^\pm_\alpha,\dpar_{\bl_\pm},\bV^i_\pm)$.
This condition is equivalent to $\bar{D}_I(\psi_+^{-1}\psi_-)=0$.
Thus, we obtained an element $f_{+-}:=\psi_+^{-1}\psi_-$ of the
first \v{C}ech cohomology set, which contains the same information\index{Cech cohomology@\v{C}ech cohomology}
as the original gauge potential $\CA^{0,1}$, which was an element
of a Dolbeault cohomology group\footnote{One should stress that\index{Dolbeault cohomology}
the actual transition from the Dolbeault- to the \v{C}ech
description can be done directly without using the gauge
transformation \eqref{gaugetrafo}, which we included only for\index{gauge!transformation}
convenience sake.}. The function $f_{+-}$ is the transition
function of a holomorphic vector bundle over $\sts_\eps$, and,\index{holomorphic!vector bundle}\index{transition function}
when obtained from an anti-Hermitian gauge potential, this
function satisfies the reality condition
\begin{equation}\label{fpmrelcond}
f_{+-}(x,\lambda_+,\eta)\ \ =\  \
\left(f_{+-}(\tau(x,\lambda_+,\eta))\right)^\dagger~.
\end{equation}

An infinitesimal deformation\footnote{See also
\cite{Wolf:2004hp,Popov:2005uv}.} of $f_{+-}\mapsto f_{+-}+\delta
f_{+-}$ leads to infinitesimal deformations of the functions
$\psi_\pm$:
\begin{align}\nonumber
f_{+-}+\delta
f_{+-}&\ =\ (\psi_++\delta\psi_+)^{-1}(\psi_-+\delta\psi_-)\ =\ (\psi_+^{-1}-\psi_+^{-1}\delta\psi_+\psi_+^{-1})(\psi_-+\delta\psi_-)~\\
\Rightarrow~\delta f_{+-}&\ =\ f_{+-}\psi_-^{-1}\delta \psi_-
-\psi_+^{-1}\delta\psi_+ f_{+-}~.
\end{align}
It is convenient to introduce the auxiliary function
$\varphi_{+-}:=\psi_+(\delta f_{+-})\psi_-^{-1}$, which may be
written as the difference $\varphi_{+-}=\phi_+-\phi_-$ of two
regular matrix valued functions $\phi_\pm$, holomorphic in
$\lambda_\pm$. Note, however, that there is a freedom of assigning
sections constant in $\lambda_\pm$ to either $\phi_+$ or $\phi_-$,
which corresponds to a (partial) choice of the gauge. We can now
write $\delta \psi_\pm=-\phi_\pm\psi_\pm$ and from the functions
$\phi_\pm$, we can reconstruct the variation of the components of
the gauge potential:
\begin{subequations}
\begin{align}\label{defpot1}
\delta \CA_\alpha^+&\ =\ \delta \psi_\pm
\bV_\alpha^+\psi_\pm^{-1}+\psi_\pm
\bV_\alpha^+\delta(\psi_\pm^{-1})\ =\ \nabla_\alpha^+\phi_\pm~,\\
\delta \CA^i_+&\ =\ \delta \psi_\pm \bV^i_+\psi_\pm^{-1}+\psi_\pm
\bV^i_+\delta(\psi_\pm^{-1})\ =\
\nabla^i_+\phi_\pm~,\label{defpot2}
\end{align}
\end{subequations}
where we wrote $\nabla^+_\alpha=\bV^+_\alpha+[\CA^+_\alpha,\cdot]$
and $\nabla^i_+=\bV^i_++[\CA^i_+,\cdot]$ for the covariant
derivatives.\index{covariant derivative}

\paragraph{The ansatz and transverse gauge.}
As an ansatz to endow any bosonic self-dual gauge potential
$A_{\alpha\ald}$ with a nontrivial auxiliary field $G^{\ald\bed}$,
we choose
\begin{equation}\label{deformationsansatz}
\delta f_{+-}\ =\ \eta_1\eta_2\eta_3\eta_4 R(x,\lambda)\ =\ 
\eta_1\eta_2\eta_3\eta_4 [K,f_{+-}]~,
\end{equation}
where $K\in\agl(n,\FC)$. Using the commutator is important to
guarantee that the splitting $f_{+-}=\psi^{-1}_+\psi_-$ persists
after the deformation. Equations \eqref{defpot1} and
\eqref{defpot2} will moreover lead to additional components in the
$\eta$-expansion of the gauge potentials, which match the
appearances of $G^{\ald\bed}$ in \eqref{fieldexpansiondevchand},
i.e.\ this ansatz will give rise to terms of order $\eta^4$ in
$\CA_{\alpha}^\pm$ and of order $\eta^3$ in $\CA^i_\pm$. Note at
this point that although it produces a finite field, this
deformation is infinitesimal, as it leads to nilpotent expressions
for $\delta f_{+-}$, $\delta\psi_\pm$ and $\delta
(\psi_\pm^{-1})$.

To recover the physical field content, we have to obtain the
potentials $\CA_{\alpha\ald}$ and $\CA^i_\ald$ (which are
extracted from $\CA_\alpha$ and $\CA^i$) in transverse gauge.
Therefore, we carefully have to choose the splitting
$\varphi_{+-}=\phi_+-\phi_-$. First note that different splittings
are given by
\begin{equation}
\varphi_{+-}\ =\ \phi_+-\phi_-\ =\
(\phi_++\phi_0)-(\phi_-+\phi_0)\ =:\
\tilde{\phi}_+-\tilde{\phi}_-~,
\end{equation}
and our task is to find a suitable function $\phi_0$. With our
choice of $R$, i.e.\ with $R$ being homogeneous of order four in
the $\eta_i^\ald$, and starting from $\nabla^i_+=\bV^i_+$ we can
simply use
\begin{equation}
\phi_0\ :=\ -\tfrac{1}{4}\eta^\ald_i\CA^i_\ald~.
\end{equation}
Its additive contribution to the new potential
$\tilde{\CA}^i_\ald$ is $\dpar^i_\ald\phi_0$ (as $\phi_0$ is
independent of $\lambda_\pm$) and therefore\footnote{Note that
$\CD:=\eta^\ald_i\dpar_\ald^i$ satisfies $\CD\CD=h\CD$, when
acting on functions which are homogeneous in the $\eta^\ald_i$ of
degree $h$.}
\begin{equation}
\eta^\ald_i\tilde{\CA}^i_\ald\ =\
\eta^\ald_i\CA^i_\ald+\eta^\ald_i\dpar^i_\ald\phi_0 \ =\
\eta^\ald_i\CA^i_\ald-\eta^\ald_i\CA^i_\ald\ =\ 0~.
\end{equation}
This procedure gives obviously rise to additional terms of order
$\eta^4$ in $\tilde{\CA}_{\alpha}^\pm$ and of order $\eta^3$ in
$\tilde{\CA}^i_\pm$ only. Since our gauge potential is now in
transverse gauge, we know for sure that it is of the form
\eqref{fieldexpansiondevchand}, and thus one can consistently
extract the physical field content by comparing the expansions. In
particular, it is not necessary to worry about higher orders in
$\eta$ than the ones considered here, as the field content is
already completely defined at third order in $\eta$, cf.\
\eqref{fieldexpansiondevchand}. With our ansatz, we therefore are
certain to obtain a solution of the form $(A_{\alpha\ald}\neq
0,\chi^{\alpha i}=0,\phi^{ij}=0,\tilde{\chi}^{\ald
ijk}=0,G^{\ald\bed}\neq 0)$.

Note that the method described above directly translates to the
noncommutative setting.

\paragraph{Dressed Penrose-Ward transform.}\label{pDressedPW} Recalling the dressing\index{Penrose-Ward transform}
method, one is led to finding a solution to the equation
$\dpar_{\alpha\ald}G^{\ald\bed}=0$ by a Penrose
transform\footnote{See also the discussion of the Penrose\index{Penrose transform}
transform in section \ref{ssPTrafo}.} and inserting the dressing
factors accounting for a non-Abelian gauge group appropriately.
Thus, let us start (in the commutative setting) from the gauge
group $\sU(N)$ and choose a gauge potential $A_{\alpha\ald}=0$ as
a seed solution for the dressing method. The equation of motion
for the auxiliary field $G^{\ald\bed}$ then reduces to
$\nabla_{\alpha\ald}G^{\ald\bed}=\dpar_{\alpha\ald}G^{\ald\bed}$,
and the solutions to this equation are given by
\begin{equation}\label{trivPWtrafo}
G^{\ald\bed}\ =\ \oint_\gamma \frac{\dd
\lambda_+}{2\pi\di}\lambda_+^\ald\lambda_+^\bed f(Z^q)~,
\end{equation}
where $f(Z^q)$ is a Lie algebra valued function, which is of\index{Lie algebra}
homogeneity $-4$ in the twistor coordinates\index{twistor}
$(Z^q)=(x^{\alpha\ald}\lambda_\ald,\lambda_\ald)$, cf.\ the
discussion of elementary states in section \ref{ssPTrafo},\index{elementary states}
\ref{pElStates}. Furthermore, $f(Z^q)$ has two distinct poles
(without counting the multiplicities). The curve $\gamma$ is
chosen in such a way that it separates the poles from each other.
That this is in fact a solution of the given equation is easily
seen by pulling the derivative $\dpar_{\alpha\ald}$ into the
integral and noting that $\lambda_+^\ald\dpar_{\alpha\ald}
z^\beta=\lambda_+^\ald\lambda^+_\ald\delta_\alpha^\beta=0$.

When dressing the seed solution of $A_{\alpha\ald}$ to a
nontrivial solution
$\lambda^\ald_+\tilde{A}_{\alpha\ald}=\psi_+\bV^+_\alpha\psi_+^{-1}=\psi_-\bV^+_\alpha\psi_-^{-1}$,
we have to adapt equation \eqref{trivPWtrafo} in the following
way:
\begin{equation}\label{nontrivPWtrafo}
G^{\ald\bed}\ =\ \oint_\gamma \frac{\dd
\lambda_+}{2\pi\di}\lambda_+^\ald\lambda_+^\bed
\psi_+f(Z^q)\psi_+^{-1}~.
\end{equation}
Again, this is a solution since differentiating under the integral
yields
\begin{align}\nonumber
\dpar_{\alpha\ald}G^{\ald\bed}&\ =\ \oint_\gamma \frac{\dd
\lambda_+}{2\pi\di}\lambda_+^\bed
(\bV^+_\alpha\psi_+)f(Z^q)\psi_+^{-1}+\psi_+f(Z^q)(\bV^+_\alpha\psi_+^{-1})\\\nonumber
&\ =\ \oint_\gamma \frac{\dd
\lambda_+}{2\pi\di}\lambda_+^\ald\lambda_+^\bed
[f(Z^q),A_{\alpha\ald}]\\
&\ =\ [G^{\ald\bed},A_{\alpha\ald}]~.
\end{align}
From the above calculation, one also sees that one is free to
choose any combination of $\psi_\pm$ and $\psi_\pm^{-1}$ around
the function $f(Z^q)$.

\paragraph{Noncommutative example.}
Let us now turn to the noncommutative case, and complement the
noncommutative BPST instanton with a solution for the auxiliary\index{instanton}
field. We start from the noncommutative BPST instanton on
Euclidean spacetime ($\eps=-1$) as described e.g.\ in\footnote{For
doing the calculations in the following, we used the computer
algebra software Mathematica together with the ``Operator Linear
Algebra Package'' by Liu Zhao and the ``Grassmann'' package by
Matthew Headrick.} \cite{Horvath:2002bj}. The gauge potential
reads
\begin{equation}
\begin{aligned}
\hat{A}_{1\dot{1}}&\ =\ \left(\begin{array}{cc}
-\left(\frac{[-1]}{[0]}
-1\right)\frac{\hat{x}^{2\dot{2}}}{2\theta}
& \frac{\di}{[0][-1]}\\
0 &
-\left(\frac{[2]}{[1]}-1\right)\frac{\hat{x}^{2\dot{2}}}{2\theta}
\end{array}\right)~,&\hat{A}_{2\dot{2}}&\ =\ -\hat{A}_{1\dot{1}}^\dagger~,
\\
\hat{A}_{2\dot{1}}&\ =\ \left(\begin{array}{cc}
\frac{\hat{x}^{1\dot{2}}}{2\theta} \left(\frac{[-1]}{[0]}
-1\right) & 0\\\label{BPSTgaugepotential}
\frac{\hat{x}^{2\dot{2}}}{[0][-1]} &
\frac{\hat{x}^{1\dot{2}}}{2\theta}\left(\frac{[2]}{[1]}-1\right)
\end{array}\right)~,&
\hat{A}_{1\dot{2}}&\ =\ \hat{A}_{2\dot{1}}^\dagger~,~~~
\end{aligned}
\end{equation}
where we introduced the shorthand notation
$[n]:=\sqrt{\hat{r}^2+\Lambda^2+2n\theta}$ with
$\hat{r}^2=\hat{x}^{1\dot{1}}\hat{x}^{2\dot{2}}-\hat{x}^{2\dot{1}}\hat{x}^{1\dot{2}}$.
The parameter $\Lambda$ can be identified with the size of the
instanton. Moreover, we have the useful identities\index{instanton}
\begin{equation}
\begin{aligned}
\hat{x}^{1\dot{1}}[n]&\ =\ [n-1]
\hat{x}^{1\dot{1}}~,&\hat{x}^{1\dot{2}}[n]&\ =\ [n-1]\hat{x}^{1\dot{2}}~,\\
\hat{x}^{2\dot{1}}[n]&\ =\ [n+1]
\hat{x}^{2\dot{1}}~,&\hat{x}^{2\dot{2}}[n]&\ =\ [n+1]\hat{x}^{2\dot{2}}~
\end{aligned}
\end{equation}
which can be verified by interpreting $\hat{r}^2$ as the number
operator $\hat{N}$ plus a constant term, and then have it act on
an arbitrary state, decomposed into eigenstates of $\hat{N}$.

The corresponding matrix-valued functions $\hat{\psi}_+$,
$\hat{\psi}_-^{-1}$ and $\hat{f}_{+-}$ read {\small
\begin{align*}
\hat{\psi}_+&\ =\ -\frac{1}{\Lambda}\left(\begin{array}{cc}[-1] & 0 \\
0 & [+1]
\end{array}\right)
\left(\begin{array}{cc} -\hat{x}^{1\dot{2}}\hat{x}^{2\dot{1}}
+\Lambda^2+\lambda_+\hat{x}^{1\dot{2}}\hat{x}^{2\dot{2}} &
-\hat{x}^{1\dot{2}}\hat{x}^{1\dot{1}}+\lambda_+(\hat{x}^{1\dot{2}})^2
\\
\hat{x}^{2\dot{1}}\hat{x}^{2\dot{2}}-\lambda_+(\hat{x}^{2\dot{2}})^2
& \hat{x}^{2\dot{2}}\hat{x}^{1\dot{1}}
+\Lambda^2-\lambda_+\hat{x}^{1\dot{2}}\hat{x}^{2\dot{2}}
\end{array}\right)~,\\\nonumber
\hat{\psi}^{-1}_-&\ =\ -\frac{1}{\Lambda}\left(\begin{array}{cc}[-1] & 0 \\
0 & [+1]
\end{array}\right)
\left(\begin{array}{cc} -\hat{x}^{1\dot{2}}\hat{x}^{2\dot{1}}
+\Lambda^2+\frac{\hat{x}^{1\dot{1}}\hat{x}^{2\dot{1}}}{\lambda_+}
&
-\hat{x}^{1\dot{1}}\hat{x}^{1\dot{2}}+\frac{(\hat{x}^{1\dot{1}})^2}{\lambda_+}
\\
\hat{x}^{2\dot{2}}\hat{x}^{2\dot{1}}-\frac{(\hat{x}^{2\dot{1}})^2}{\lambda_+}
& \hat{x}^{2\dot{2}}\hat{x}^{1\dot{1}}
+\Lambda^2-\frac{\hat{x}^{1\dot{1}}\hat{x}^{2\dot{1}}}{\lambda_+}
\end{array}\right)~,\\\label{BPSTpsis}
\hat{f}_{+-}&\ =\ \frac{1}{\Lambda^2}\left(\begin{array}{cc} [1]^2-
\lambda_+\hat{x}^{1\dot{2}}\hat{x}^{2\dot{2}}
-\frac{\hat{x}^{2\dot{1}}\hat{x}^{1\dot{1}}}{\lambda_+}+
2\hat{x}^{2\dot{1}}\hat{x}^{1\dot{2}} &
-\frac{(\hat{x}^{1\dot{1}})^2}{\lambda_+}-\lambda_+(\hat{x}^{1\dot{2}})^2
+2\hat{x}^{1\dot{1}}\hat{x}^{1\dot{2}}\\
\frac{(\hat{x}^{2\dot{1}})^2}{\lambda_+}+\lambda_+(\hat{x}^{2\dot{2}})^2
-2\hat{x}^{2\dot{1}}\hat{x}^{2\dot{2}} &
[-1]^2-2\hat{x}^{1\dot{1}}\hat{x}^{2\dot{2}}+
\lambda_+\hat{x}^{1\dot{2}}\hat{x}^{2\dot{2}}+
\frac{\hat{x}^{2\dot{1}}\hat{x}^{1\dot{1}}}{\lambda_+}
\end{array}\right)~.
\end{align*}}
Additionally we use
\begin{equation}
\hat{\psi}_+^{-1}\ =\ \left(\begin{array}{cc}
\Lambda^2+\hat{x}^{1\dot{1}}\hat{x}^{2\dot{2}}-\lambda_+
\hat{x}^{2\dot{2}}\hat{x}^{1\dot{2}} &
-\lambda_+(\hat{x}^{1\dot{2}})^2+\hat{x}^{1\dot{2}}\hat{x}^{1\dot{1}}
\\
\lambda_+(\hat{x}^{2\dot{2}})^2-\hat{x}^{2\dot{1}}\hat{x}^{2\dot{2}}
& \Lambda^2-\hat{x}^{2\dot{1}}\hat{x}^{1\dot{2}}+\lambda
\hat{x}^{2\dot{2}}\hat{x}^{1\dot{2}}
\end{array}\right)~.
\end{equation}
As an ansatz for $f(Z^q)$, we choose the simple form
\begin{equation}
\hat{f}(\hat{z}^1_+,\hat{z}^2_+,\lambda_+)\ =\
\frac{\hat{z}^1_+}{\lambda_+^2(1+\lambda_+)}\sigma^3~.
\end{equation}
Note that it is not possible, to have the noncommutative
coordinates $\hat{z}^\alpha$ appear in a holomorphic way in the
denominator. This is due to the fact that the $\hat{z}^\alpha$ are
operators on a Fock space, for which only infinite-dimensional
representations exists. Therefore, the inverse of holomorphic\index{representation}
functions of $\hat{z}^\alpha$ does not exist in general.

The singularities at $\lambda_+=0$ and $\lambda_+=-1$ are
separated by a circle $\gamma$ of radius $r<1$ around
$\lambda_+=0$. Thus, the equations \eqref{trivPWtrafo} reduce to
\begin{equation}
\begin{aligned}
\hat{G}^{\dot{1}\dot{1}}&\ =\
\langle\hat{f}(Z^q)\rangle_{-3}~,~~~&
\hat{G}^{\dot{1}\dot{2}}&\ =\ -\langle\hat{f}(Z^q)\rangle_{-2}~,\\
\hat{G}^{\dot{2}\dot{1}}&\ =\
-\langle\hat{f}(Z^q)\rangle_{-2}~,~~~& \hat{G}^{\dot{2}\dot{2}}&\
=\ \langle\hat{f}(Z^q)\rangle_{-1}~,
\end{aligned}
\end{equation}
where $\langle\cdot\rangle_{n}$ denotes the $n$th coefficient in a
Laurent series. The undressed field $\hat{G}_0^{\ald\bed}$ is then
easily obtained and has as the only non-vanishing components
$\hat{G}^{\ald\bed}_0=\hat{G}^{\ald\bed3}_0\sigma^3$ with
\begin{equation}
\hat{G}_0^{\dot{1}\dot{2}3}\ =\ \hat{x}^{1\dot{1}}\eand
\hat{G}_0^{\dot{2}\dot{2}3}\ =\
(\hat{x}^{1\dot{1}}-\hat{x}^{1\dot{2}})~.
\end{equation}
The ``dressed'' solution on the other hand are determined by
equation \eqref{nontrivPWtrafo}, which reduces to
\begin{equation}
\begin{aligned}
\hat{G}^{\dot{1}\dot{1}}&\ =\
\langle\hat{\psi}_+\hat{f}(Z^q)\hat{\psi}_+^{-1}\rangle_{-3}~,~~~&
\hat{G}^{\dot{1}\dot{2}}&\ =\ -\langle\hat{\psi}_+\hat{f}(Z^q)\hat{\psi}_+^{-1}\rangle_{-2}~,\\
\hat{G}^{\dot{2}\dot{1}}&\ =\
-\langle\hat{\psi}_+\hat{f}(Z^q)\hat{\psi}_+^{-1}\rangle_{-2}~,~~~&
\hat{G}^{\dot{2}\dot{2}}&\ =\
\langle\hat{\psi}_+\hat{f}(Z^q)\hat{\psi}_+^{-1}\rangle_{-1}~.
\end{aligned}
\end{equation}
Performing the calculation and decomposing the field according to
$\hat{G}^{\ald\bed}=\hat{G}^{\ald\bed a}\sigma^a$, we arrive at
the following expressions: {
\fontsize{9}{9}%
\selectfont
\begin{align}\nonumber
\hat{G}^{\dot{1}\dot{1}k}\ =\ &0\\\nonumber
\hat{G}^{\dot{1}\dot{2}0}\ =\ &-\theta(
      3  \Lambda^2 \hat{x}^{1\dot{1}} + 2  (\hat{x}^{1\dot{1}})^2\hat{x}^{2\dot{2}}   -
        4  \hat{x}^{2\dot{1}}\hat{x}^{1\dot{1}}\hat{x}^{1\dot{2}}
        )\\\nonumber
\hat{G}^{\dot{1}\dot{2}1}\ =\ &\Lambda^2 \theta \hat{x}^{2\dot{1}} +
(\Lambda^2 - 2 \theta)
        (\hat{x}^{1\dot{1}})^2\hat{x}^{1\dot{2}}   + (\Lambda^2 + 4  \theta)
        \hat{x}^{2\dot{1}}\hat{x}^{1\dot{1}}\hat{x}^{2\dot{2}}   -
      \hat{x}^{2\dot{1}}(\hat{x}^{1\dot{1}})^2(\hat{x}^{1\dot{2}})^2   +
      \hat{x}^{2\dot{1}}(\hat{x}^{1\dot{1}})^2(\hat{x}^{2\dot{2}})^2
      \\\nonumber
\hat{G}^{\dot{1}\dot{2}2}\ =\ &\di(\Lambda^2 \theta \hat{x}^{2\dot{1}}
- (\Lambda^2 - 2 \theta)
        (\hat{x}^{1\dot{1}})^2\hat{x}^{1\dot{2}}   + (\Lambda^2 + 4  \theta)
        \hat{x}^{2\dot{1}}\hat{x}^{1\dot{1}}\hat{x}^{2\dot{2}}   +
      \hat{x}^{2\dot{1}}(\hat{x}^{1\dot{1}})^2(\hat{x}^{1\dot{2}})^2   +
      \hat{x}^{2\dot{1}}(\hat{x}^{1\dot{1}})^2(\hat{x}^{2\dot{2}})^2
      )\\\nonumber
\hat{G}^{\dot{1}\dot{2}3}\ =\ &\Lambda^2 (\Lambda^2 + \theta)
\hat{x}^{1\dot{1}} + (\Lambda^2 - 2 \theta)
        (\hat{x}^{1\dot{1}})^2\hat{x}^{2\dot{2}}   - (\Lambda^2 + 4  \theta)
        \hat{x}^{2\dot{1}}\hat{x}^{1\dot{1}}\hat{x}^{1\dot{2}}   -
      2  \hat{x}^{2\dot{1}}(\hat{x}^{1\dot{1}})^2\hat{x}^{1\dot{2}}\hat{x}^{2\dot{2}}
      \\\nonumber
\hat{G}^{\dot{2}\dot{2}0}\ =\ &\theta(4  \theta  \hat{x}^{1\dot{2}} -
\Lambda^2(3 \hat{x}^{1\dot{1}} + \hat{x}^{1\dot{2}}) -
      2  (\hat{x}^{1\dot{1}})^2\hat{x}^{2\dot{2}}   +
      4  \hat{x}^{1\dot{1}}\hat{x}^{1\dot{2}}\hat{x}^{2\dot{2}}   +
      4  \hat{x}^{2\dot{1}}\hat{x}^{1\dot{1}}\hat{x}^{1\dot{2}}  )\\\nonumber
\hat{G}^{\dot{2}\dot{2}1}\ =\ &\theta(-8 \theta  \hat{x}^{2\dot{2}} +
\Lambda^2(\hat{x}^{2\dot{1}} + \hat{x}^{2\dot{2}})) +
    (\hat{x}^{1\dot{1}})^2(\hat{x}^{2\dot{2}})^3   -
    (\hat{x}^{1\dot{1}})^2(\hat{x}^{1\dot{2}})^2\hat{x}^{2\dot{2}}   -\\\nonumber&
    2  \theta(\hat{x}^{1\dot{1}}(\hat{x}^{1\dot{2}})^2   - \hat{x}^{1\dot{1}}(\hat{x}^{2\dot{2}})^2   +
          (\hat{x}^{1\dot{1}})^2\hat{x}^{1\dot{2}}   + (2  \di)
            \hat{x}^{2\dot{2}}(\hat{x}^{1\dot{1}} - \hat{x}^{2\dot{2}})   -
          2  \hat{x}^{2\dot{1}}\hat{x}^{1\dot{1}}\hat{x}^{2\dot{2}}  ) -\\\nonumber&
    \hat{x}^{2\dot{1}}(\hat{x}^{1\dot{1}})^2(\hat{x}^{1\dot{2}})^2   +
    \hat{x}^{2\dot{1}}(\hat{x}^{1\dot{1}})^2(\hat{x}^{2\dot{2}})^2   + \Lambda^2(
      \hat{x}^{1\dot{1}}(\hat{x}^{2\dot{2}})^2   + (\hat{x}^{1\dot{1}})^2\hat{x}^{1\dot{2}}   +
        \hat{x}^{2\dot{1}}\hat{x}^{1\dot{1}}\hat{x}^{2\dot{2}}   -
        \hat{x}^{2\dot{1}}\hat{x}^{1\dot{2}}\hat{x}^{2\dot{2}}  )\\\nonumber
\hat{G}^{\dot{2}\dot{2}2}\ =\ &(\di) (\theta(-8 \theta
\hat{x}^{2\dot{2}} + \Lambda^2(\hat{x}^{2\dot{1}} +
\hat{x}^{2\dot{2}})) +
      (\hat{x}^{1\dot{1}})^2(\hat{x}^{2\dot{2}})^3   +
      (\hat{x}^{1\dot{1}})^2(\hat{x}^{1\dot{2}})^2\hat{x}^{2\dot{2}}   +
      2  \theta(\hat{x}^{1\dot{1}}(\hat{x}^{1\dot{2}})^2   +\\\nonumber& \hat{x}^{1\dot{1}}(\hat{x}^{2\dot{2}})^2   +
            (\hat{x}^{1\dot{1}})^2\hat{x}^{1\dot{2}}   - (2  \di)
              \hat{x}^{2\dot{2}}(\hat{x}^{1\dot{1}} - \hat{x}^{2\dot{2}})   +
            2  \hat{x}^{2\dot{1}}\hat{x}^{1\dot{1}}\hat{x}^{2\dot{2}}  ) +\\\nonumber&
      \hat{x}^{2\dot{1}}(\hat{x}^{1\dot{1}})^2(\hat{x}^{1\dot{2}})^2   +
      \hat{x}^{2\dot{1}}(\hat{x}^{1\dot{1}})^2(\hat{x}^{2\dot{2}})^2   + \Lambda^2(
        \hat{x}^{1\dot{1}}(\hat{x}^{2\dot{2}})^2   - (\hat{x}^{1\dot{1}})^2\hat{x}^{1\dot{2}}   +
          \hat{x}^{2\dot{1}}\hat{x}^{1\dot{1}}\hat{x}^{2\dot{2}}   -
          \hat{x}^{2\dot{1}}\hat{x}^{1\dot{2}}\hat{x}^{2\dot{2}}  ))\\\nonumber
\hat{G}^{\dot{2}\dot{2}3}\ =\ &\Lambda^4(\hat{x}^{1\dot{1}} -
\hat{x}^{1\dot{2}}) -
    4  \theta^2 \hat{x}^{1\dot{2}} + \Lambda^2 \theta(\hat{x}^{1\dot{1}} + \hat{x}^{1\dot{2}}) + (\Lambda^2 - 2  \theta)
      (\hat{x}^{1\dot{1}})^2\hat{x}^{2\dot{2}}   + \Lambda^2
      \hat{x}^{2\dot{1}}(\hat{x}^{1\dot{2}})^2   - \\\nonumber&(\Lambda^2 + 8  \theta)
      \hat{x}^{1\dot{1}}\hat{x}^{1\dot{2}}\hat{x}^{2\dot{2}}   - (\Lambda^2 + 4
      \theta)
      \hat{x}^{2\dot{1}}\hat{x}^{1\dot{1}}\hat{x}^{1\dot{2}}   -
    2 ((\hat{x}^{1\dot{1}})^2\hat{x}^{1\dot{2}}(\hat{x}^{2\dot{2}})^2   +
        \hat{x}^{2\dot{1}}(\hat{x}^{1\dot{1}})^2\hat{x}^{1\dot{2}}\hat{x}^{2\dot{2}}
        )~.
\end{align}}

\paragraph{Reality conditions for $\hat{G}^{\ald\bed}$.} Solutions
constructed from the two algorithms we discussed do not
necessarily satisfy any reality condition. Imposing a reality
condition a priori which would guarantee a real field
$\hat{G}^{\ald\bed}$ as an outcome of the constructions always
complicates the calculations. Instead, it is possible to adjust
the fields after all calculations are performed.

The easiest way to obtain the conditions which one should impose
on $\hat{G}^{\ald\bed}$ is the observation that this field appears
in the form $\hat{G}^{\ald\bed}\hat{f}_{\ald\bed}$ in the
Lagrangian. Thus the reality condition on $\hat{G}^{\ald\bed}$ has
to be the same as for $\hat{f}_{\ald\bed}$ to ensure that the
Lagrangian is either purely real or purely imaginary. From
$\hat{A}_{1\dot{1}}=-\hat{A}^\dagger_{2\dot{2}}$ and
$\hat{A}_{1\dot{2}}=\hat{A}^\dagger_{2\dot{1}}$ one concludes that
$\hat{f}_{\dot{1}\dot{1}}=-\hat{f}_{\dot{2}\dot{2}}^\dagger$ and
$\hat{f}_{\dot{1}\dot{2}}=\hat{f}_{\dot{2}\dot{1}}^\dagger$.
Therefore, we impose the following conditions:
\begin{equation}
\hat{G}^{\dot{1}\dot{1}}\ =\
-\left(\hat{G}^{\dot{2}\dot{2}}\right)^\dagger \eand
\hat{G}^{\dot{1}\dot{2}}\ =\
\left(\hat{G}^{\dot{2}\dot{1}}\right)^\dagger~,
\end{equation}
or, extracting an anti-Hermitian generator $\sigma^a$ of the gauge
group:
\begin{equation}
\hat{G}^{\dot{1}\dot{1}a}\ =\
\left(\hat{G}^{\dot{2}\dot{2}a}\right)^\dagger \eand
\hat{G}^{\dot{1}\dot{2}a}\ =\
-\left(\hat{G}^{\dot{2}\dot{1}a}\right)^\dagger~.
\end{equation}
Given a complex solution $\hat{G}^{\ald\bed}$, we can now
construct a real solution $\hat{G}^{\ald\bed}_r$ by
\begin{equation}
\begin{aligned}
\hat{G}_r^{\dot{1}\dot{1}a}&\ :=\
\tfrac{1}{2}\left(\hat{G}^{\dot{1}\dot{1}a}+
(\hat{G}^{\dot{2}\dot{2}a})^\dagger\right)~,&
\hat{G}_r^{\dot{1}\dot{2}a}\ :=\
\tfrac{1}{2}\left(\hat{G}^{\dot{1}\dot{2}a}-
(\hat{G}^{\dot{1}\dot{2}a})^\dagger\right)~,\\
\hat{G}_r^{\dot{2}\dot{2}a}&\ :=\
\tfrac{1}{2}\left(\hat{G}^{\dot{2}\dot{2}a}+
(\hat{G}^{\dot{1}\dot{1}a})^\dagger\right)~,
\end{aligned}
\end{equation}
which satisfies by construction the equations of motion, as one
easily checks by plugging these linear combinations into the
equations of motion.

\subsection{String theory perspective}\index{string theory}

\paragraph{Holomorphically embedded submanifolds and their normal bundles.}\index{normal bundle}
Recall that the equations
\begin{equation}\label{2.6}
z^\alpha_\pm\ =\ x^{\alpha\ald}\lambda^\pm_\ald\eand\eta_i^\pm\ =\ \eta_i^\ald\lambda_\ald^\pm
\end{equation}
describe a holomorphic embedding of the space $\CPP^1$ into the
supertwistor space $\sts$. That is, for fixed moduli\index{twistor}\index{twistor!space}
$x^{\alpha\ald}$ and $\eta^\ald_i$, equations \eqref{2.6} yield a
projective line $\CPP^1_{x,\eta}$ inside the supertwistor space.\index{twistor}\index{twistor!space}
The normal bundle to any $\CPP_{x,\eta}^1\embd \CP^{3|4}$ is\index{normal bundle}
$\CN^{2|4}=\FC^2\otimes\CO(1)\oplus\FC^4\otimes\Pi\CO(1)$ and we
have
\begin{equation}
h^0(\CPP_{x,\eta}^1,\CN^{2|4})\ =\ \dim_\FC
H^0(\CPP^1_{x,\eta},\CN^{2|4})\ \ =\ 4|8~.
\end{equation}
Furthermore, there are no obstructions to the deformation of the
$\CPP^{1|0}_{x,\eta}$ inside $\CP^{3|4}$ since
$h^1(\CPP^1_{x,\eta},\CN^{2|4})=0|0$.

On the other hand, one can fix only the even moduli
$x^{\alpha\ald}$ and consider a holomorphic embedding
$\CPP^{1|4}_x\embd\sts$ defined by the equations
\begin{equation}\label{moduli2}
z^\alpha_\pm\ =\ x^{\alpha\ald}\lambda_\ald^\pm~.
\end{equation}
Recall that the normal bundle to $\CPP^{1|0}_x\embd \CP^{3|0}$ is\index{normal bundle}
the rank two vector bundle $\CO(1)\oplus\CO(1)$. In the supercase,
the formal definition of the normal bundle by the short exact\index{normal bundle}
sequence
\begin{equation}\label{normalbundle14}
0\ \rightarrow\ T\CPP^{1|4}\ \rightarrow\ T\CP^{3|4}|_{\CPP^{1|4}}
\ \rightarrow\ \CN^{2|0}\ \rightarrow\ 0
\end{equation}
yields that $\CN^{2|0}=T\CP^{3|4}|_{\CPP^{1|4}}/T\CPP^{1|4}$ is a
rank two holomorphic vector bundle over $\CPP^{1|4}$ which is (in\index{holomorphic!vector bundle}
the real case) locally spanned by the vector fields $\gamma_\pm
V^\pm_\alpha$, where $V^\pm_\alpha$ is the complex conjugate of
$\bV^\pm_\alpha$. A global section of $\CN^{2|0}$ over
$\CU_\pm\cap \CPP^{1|4}$ is of the form
$s_\pm=T_\pm^\alpha\gamma_\pm V_\alpha^\pm$. Obviously, the
transformation of the components $T^\alpha_\pm$ from patch to
patch is given by $T^\alpha_+=\lambda_+T^\alpha_-$, i.e.\
$\CN^{2|0}=\CO(1)\oplus\CO(1)$.

\paragraph{Topological D-branes and the matrix models.}\index{D-brane}\index{matrix model}
The interpretation of the matrix model \eqref{actionhCSred} is now
rather straightforward. For gauge group $\sGL(n,\FC)$, it
describes a stack of $n$ almost space-filling D$(1|4)$-branes,
whose fermionic dimensions only extend in the holomorphic
directions of the target space $\CP^{3|4}_{\eps}$. These D-branes\index{D-brane}\index{target space}
furthermore wrap a $\CPP^{1|4}_{x}\embd \CP^{3|4}_{\eps}$.

We can use the expansion
$\CX_\alpha=\CX_\alpha^0+\CX_\alpha^i\eta_i+\CX_\alpha^{ij}\eta_i\eta_j+\ldots$
on any patch of $\CPP^{1|4}$ to examine the equations of motion
\eqref{shCS1} more closely:
\begin{equation}
\begin{aligned}
{}[\CX_1^0,\CX_2^0]&\ =\ 0~,\\
{}[\CX_1^i,\CX_2^0]+[\CX_1^0,\CX_2^i]&\ =\ 0~,\\
\{\CX_1^i,\CX_2^j\}-\{\CX_1^j,\CX_2^i\}+[\CX_1^{ij},\CX_2^0]+[\CX_1^0,\CX_2^{ij}]&\ =\ 0~,\\
\ldots
\end{aligned}
\end{equation}
Clearly, the bodies $\CX_\alpha^0$ of the Higgs fields can be
diagonalized simultaneously, and the diagonal entries describe the
position of the D$(1|4)$-brane in the normal directions of the
ambient space $\CP^{3|4}_\eps$. In the fermionic directions, this
commutation condition is relaxed and thus, the D-branes can be\index{D-brane}
smeared out in these directions even in the classical case.

\paragraph{Interpretation within $\CN=2$ string theory.}\index{N=2 string theory@$\CN=2$ string theory}\index{string theory}
Recall from section \ref{ssN2string} that the critical $\CN=2$
string has a four-dimensional target space and its open string\index{open string}\index{target space}
effective field theory is self-dual Yang-Mills theory (or its\index{Yang-Mills theory}\index{self-dual Yang-Mills theory}
noncommutative deformation \cite{Lechtenfeld:2000nm} in the
presence of a $B$-field). It has been argued \cite{Siegel:1992za}
that, after extending the $\CN=2$ string effective action in a
natural way to recover Lorentz invariance, the effective field\index{Lorentz!invariance}
theory becomes the full $\CN=4$ supersymmetrically extended SDYM
theory, and we will adopt this point of view in the following.

D-branes within critical $\CN=2$ string theory were already\index{D-brane}\index{N=2 string theory@$\CN=2$ string theory}\index{string theory}
discussed in \ref{ssFurtherDBranes}, \ref{pN2DBranes}; recall from
there that the low-energy effective action on such a D-brane is\index{D-brane}
SDYM theory dimensionally reduced to its worldvolume. Thus, we
have a first interpretation of our matrix model\index{matrix model}
\eqref{SDYMactionMM} in terms of a stack of $n$ D0- or
D$(0|8)$-branes in $\CN=2$ string theory, and the topological\index{N=2 string theory@$\CN=2$ string theory}\index{string theory}
D$(1|4)$-brane is the equivalent configuration in B-type
topological string theory.\index{string theory}\index{topological!string}

As usual, turning on a $B$-field background will give rise to
noncommutative deformations of the ambient space, and therefore
the matrix model \eqref{actionhCSdef2} describes a stack of $n$\index{matrix model}
D4-branes in $\CN=2$ string theory within such a background.\index{N=2 string theory@$\CN=2$ string theory}\index{string theory}

The moduli superspaces $\FR^{4|8}_\theta$ and $\FR^{0|8}$ for both\index{super!space}
the noncommutative and the ordinary matrix model can therefore be\index{matrix model}
seen as {\em chiral} D$(4|8)$- and D$(0|8)$-branes, respectively,
with $\CN=4$ self-dual Yang-Mills theory as the appropriate\index{Yang-Mills theory}\index{self-dual Yang-Mills theory}
(chiral) low energy effective field theory.

\subsection{SDYM matrix model and super ADHM construction}\index{ADHM construction}\index{matrix model}

\paragraph{The pure matrix model.} While a solution to the\index{matrix model}
$\CN=4$ SDYM equations with gauge group $\sU(n)$ and second Chern
number $c_2=k$ describes a bound state of $k$\linebreak\index{Chern number}
D(-$1|8$)-branes with $n$ D$(3|8)$-branes at low energies, the
SDYM matrix model obtained by a dimensional reduction of this\index{dimensional reduction}\index{matrix model}
situation describes a bound state between $k+n$ D(-$1|8$)-branes.
This implies that there is only one type of strings, i.e.\ those
having both ends on the D(-$1|8$)-branes. In the ADHM
construction, one can simply account for this fact by eliminating\index{ADHM construction}
the field content which arose previously from the open strings\index{open string}
having one endpoint on a D(-$1|8$)-brane and the other one on a
D($3|8$)-brane. That is, we put $w_{uq\ald}$ and $\psi^i$ to zero.

In fact, the remaining ADHM constraints read
\begin{equation}
\vec{\sigma}^\ald{}_\bed (\bar{A}^{\alpha\bed}A_{\alpha\ald})\ =\
0~,
\end{equation}
and one can use the reality conditions together with the
definition of the ordinary sigma matrices to show that these
equations are equivalent to
\begin{equation}
\eps^{\alpha\beta}[A_{\alpha\ald},A_{\beta\bed}]\ =\ 0~,
\end{equation}
which are simply the matrix-SDYM equations \eqref{SDYMeomMM} with
fields with more than one R-symmetry index put to zero. The
missing fermionic equations follow from this equation via the
expansion \eqref{expansion}, the latter being determined by the
expansion of superfields for both the full and the self-dual super
Yang-Mills theories. Thus, we recovered the equations of motion of
the matrix model \eqref{SDYMactionMM} in the ADHM construction as\index{ADHM construction}\index{matrix model}
expected from the interpretation via D-branes.\index{D-brane}

\paragraph{Extension of the matrix model.} It is now conceivable\index{matrix model}
that the D3-D(-1)-brane\footnote{For simplicity, let us suppress
the fermionic dimensions of the D-branes in the following.} system\index{D-brane}
explaining the ADHM construction can be carried over to the\index{ADHM construction}
supertwistor space $\CP^{3|4}$. That is we take a D1-D5-brane\index{twistor}\index{twistor!space}
system and analyze it either via open D5-D5 strings with
excitations corresponding in the holomorphic Chern-Simons theory\index{Chern-Simons theory}
to gauge configurations with non-trivial second Chern character or\index{Chern character}
by looking at the D1-D1 and the D1-D5 strings. The latter point of
view will give rise to a holomorphic Chern-Simons analogue of the
ADHM configuration, as we will show in the following.

The action for the D1-D1 strings is evidently our hCS matrix model\index{matrix model}
\eqref{actionhCSred}. To incorporate the D1-D5 strings, we can use
an action proposed by Witten in \cite{Witten:2003nn}\footnote{In
fact, he uses this action to complement the hCS theory in such a
way that it will give rise to full Yang-Mills theory on the moduli\index{Yang-Mills theory}
space. For this, he changes the parity of the fields $\alpha$ and\index{parity}
$\beta$ to be fermionic.}
\begin{equation}\label{extendedaction}
\int \dd\lambda \Omega^\eta~ \tr(\beta \dparb \alpha +\beta
\CA^{0,1}_\Sigma \alpha)~,
\end{equation}
where the fields $\alpha$ and $\beta$ take values in the line
bundles $\CO(1)$ and they transform in the fundamental and
antifundamental representation of the gauge group $\sGL(n,\FC)$,\index{fundamental representation}\index{representation}
respectively.

The equations of motion of the total matrix model which is the sum\index{matrix model}
of \eqref{actionhCSred} and \eqref{extendedaction} are then
modified to
\begin{equation}\label{exteoms}
\begin{aligned}
\dparb\CX_\alpha+[\CA^{0,1}_\Sigma,\CX_\alpha]&\ =\ 0~,\\
[\CX_1,\CX_2]+\alpha\beta&\ =\ 0~,\\
\dparb\alpha+\CA^{0,1}_\Sigma\alpha\ =\ 0\eand
\dparb\beta&+\beta\CA^{0,1}_\Sigma\ =\ 0~.
\end{aligned}
\end{equation}
Similarly to the Higgs fields $\CX_\alpha$ and the gauge potential
$\CA^{0,1}_\Sigma$, we can give a general field expansion for
$\beta$ and $\alpha=\bar{\beta}$:
\begin{equation}\label{abexpansion}
\begin{aligned}
\beta_+\ =\ & \lambda_+^\ald
w_\ald+\psi^i\eta^+_i+\gamma_+\tfrac{1}{2!}\eta^+_i\eta^+_j\hl^\ald_+\rho^{ij}_\ald+
\gamma^2_+\tfrac{1}{3!}\eta^+_i\eta^+_j\eta^+_k\hl^\ald_+\hl^\bed_+\sigma_{\ald\bed}^{ijk}\\&+\gamma^3_+
\tfrac{1}{4!}\eta^+_i\eta^+_j\eta^+_k\eta^+_l\hl^\ald_+\hl^\bed_+\hl^\gad_+\tau_{\ald\bed\gad}^{ijkl}~,\\
\alpha_+\ =\ &
\lambda^\ald_+\eps_{\ald\bed}\bar{w}^\bed_++\bar{\psi}^i\eta^+_i+\ldots~.
\end{aligned}
\end{equation}
Applying the equations of motion, one learns that the fields
beyond linear order in the Gra{\ss}mann variables are composite\index{Gra{\ss}mann variable}
fields:
\begin{equation}
\rho^{ij}_\ald\ =\ w_\ald\phi^{ij}~,~~~ \sigma_{\ald\bed}^{ijk}\
=\ \tfrac{1}{2}w_{(\ald}\tilde{\chi}_{\bed)}^{ijk}\eand
\tau_{\ald\bed\gad}^{ijkl}\ =\
\tfrac{1}{3}w_{(\ald}G_{\bed\gad)}^{ijkl}~.
\end{equation}

We intentionally denoted the zeroth order components of $\alpha$
and $\beta$ by $\lambda_\ald\bar{w}^\ald$ and $\lambda^\ald
w_\ald$, respectively, since this expansion together with the
field equations \eqref{exteoms} are indeed the (super) ADHM
equations
\begin{equation}
\vec{\sigma}^\ald{}_\bed(\bar{w}^\bed
w_\ald+A^{\alpha\ald}A_{\alpha\bed})\ =\ 0~,
\end{equation}
which are equivalent to the condition that
$\bar{\Delta}\Delta=\unit_2\otimes f^{-1}$. Recall, however, that
for superfields with components beyond linear order in the
Gra{\ss}mann fields, the super ADHM equations do {\em not} yield
solutions to the supersymmetric self-dual Yang-Mills equations.
Therefore, we additionally have to put these fields to zero in the
Higgs fields $\CX_\alpha$ and the gauge potential
$\CA^{0,1}_\Sigma$ (which automatically does the same for the
fields $\alpha$ and $\beta$).

This procedure seems at first slightly ad-hoc, but again it
becomes quite natural, when recalling that for the ADHM D-brane\index{D-brane}
configuration, supersymmetry is broken from $\CN=4$ to four times\index{super!symmetry}
$\CN=1$. Furthermore, the fields which are put to zero give rise
to the potential terms in the action, and thus, we can regard
putting these fields to zero as an additional ``$D$-flatness
condition'' arising on the topological string side.\index{topological!string}

With this additional constraint, our matrix model\index{matrix model}
\eqref{actionhCSred} together with the extension
\eqref{extendedaction} is equivalent to the ADHM equations and
therefore it is in the same sense dual to holomorphic Chern-Simons
theory on the full supertwistor space $\CP^{3|4}$, in which the\index{Chern-Simons theory}\index{twistor}\index{twistor!space}
ADHM construction is dual to SDYM theory.\index{ADHM construction}

Summarizing, the D3-D(-1)-brane system can be mapped via an
extended Penrose-Ward-transform to a D5-D1-brane system in
topological string theory. The arising super SDYM theory on the\index{string theory}\index{topological!string}
D3-brane corresponds to hCS theory on the D5-brane, while the
matrix model describing the effective action on the D(-1)-brane\index{matrix model}
corresponds to our hCS matrix model on a topological D1-brane. The
additional D3-D(-1) strings completing the picture from the
perspective of the D(-1)-brane can be directly translated into
additional D5-D1 strings on the topological side. The ADHM
equations can furthermore be obtained from an extension of the hCS
matrix model on the topological D1-brane with a restriction on the\index{matrix model}
field content.

\paragraph{D-branes in a nontrivial $B$-field background.} Except\index{D-brane}
for the remarks on the $\CN=2$ string, we have not yet discussed
the matrix model which we obtained from deforming the moduli space\index{matrix model}\index{moduli space}
$\FR^{4|8}$ to a noncommutative spacetime.\index{noncommutative spacetime}

In general, noncommutativity is interpreted as the presence of a
Kalb-Ramond $B$-field background in string theory. Thus, solutions\index{Kalb-Ramond}\index{Ramond}\index{string theory}
to the noncommutative SDYM theory \eqref{SDYMactionNC} on
$\FR^{4|8}_{\theta}$ are D(-$1|8$)-branes bound to a stack of
space-filling D$(3|8)$-branes in the presence of a $B$-field
background. This distinguishes the commutative from the
noncommutative matrix model: The noncommutative matrix model is\index{matrix model}
now dual to the ADHM equations, instead of being embedded like the
commutative one.

The matrix model on holomorphic Chern-Simons theory describes\index{Chern-Simons theory}\index{matrix model}
analogously a topological almost space-filling D$(5|4)$-brane in
the background of a $B$-field. Note that a noncommutative
deformation of the target space $\CP^{3|4}_\eps$ does not yield\index{target space}
any inconsistencies in the context of the topological B-model.\index{topological!B-model}
Such deformations have been studied e.g.\ in
\cite{Kapustin:2003sg} and \cite{Iqbal:2003ds}, see also
\cite{Kulaxizi:2004pa}.

On the one hand, we found two pairs of matrix models, which are\index{matrix model}
dual to each other (as the ADHM equations are dual to the SDYM
equations). On the other hand, we expect both pairs to be directly
equivalent to one another in a certain limit, in which the rank of
the gauge group of the commutative matrix model tends to infinity.\index{matrix model}
The implications of this observation might reveal some further
interesting features.

\subsection{Dimensional reductions related to the Nahm equations}\index{Nahm equations}\index{dimensional reduction}

After the discussion of the ADHM construction in the previous\index{ADHM construction}
section, one is led to try to also translate the D-brane\index{D-brane}
interpretation of the Nahm construction to some topological
B-model on a Calabi-Yau supermanifold. This is in fact possible,\index{Calabi-Yau}\index{Calabi-Yau supermanifold}\index{super!manifold}\index{topological!B-model}
but since the D-brane configuration is somewhat more involved, we\index{D-brane}
will refrain from presenting many details. In the subsequent
discussion, we strongly rely on results from \cite{Popov:2005uv}
presented in section \ref{sPWMini}, where further details
complementing our rather condensed presentation can be found. In
this section, we will constrain our considerations to real
structures yielding Euclidean signature, i.e.\ $\eps=-1$.\index{real structure}

\paragraph{The superspaces $\CQ^{3|4}$ and $\hat{\CQ}^{3|4}$.}\index{super!space}
We want to consider a holomorphic Chern-Simons theory which\index{Chern-Simons theory}
describes magnetic mono\-poles and their superextensions. For
this, we start from the holomorphic vector bundle\index{holomorphic!vector bundle}
\begin{equation}
\CQ^{3|4}\ =\ \CO(2)\oplus\CO(0)\oplus \FC^4\otimes \Pi\CO(1)
\end{equation}
of rank $2|4$ over the Riemann sphere $\CPP^1$. This bundle is\index{Riemann sphere}
covered by two patches $\tilde{\CV}_\pm$ on which we have the
coordinates $\lambda_\pm=w_2^\pm$ on the base space and
$w_1^\pm,w_3^\pm$ in the bosonic fibres. On the overlap
$\tilde{\CV}_+\cap\tilde{\CV}_-$, we have thus\footnote{The
labelling of coordinates is chosen to become as consistent as
possible with \cite{Popov:2005uv}.}
\begin{equation}\label{coordsQ56}
w^1_+\ =\ (w^2_+)^2 w^1_-~,~~~ w^2_+\ =\ \frac{1}{w^2_-}~,~~~
w^3_+\ =\ w^3_-~.
\end{equation}
The coordinates on the fermionic fibres of $\CQ^{3|4}$ are the
same as the ones on $\CP^{3|4}$, i.e.\ we have $\eta_i^\pm$ with
$i=1,\ldots 4$, satisfying $\eta_i^+=\lambda_+\eta_i^-$ on
$\tilde{\CV}_+\cap\tilde{\CV}_-$. From the Chern classes of the\index{Chern class}
involved line bundles, we clearly see that $\CQ^{3|4}$ is a
Calabi-Yau supermanifold.\index{Calabi-Yau}\index{Calabi-Yau supermanifold}\index{super!manifold}

Note that holomorphic sections of the vector bundle $\CQ^{3|4}$
are parameterized by elements $(y^{(\ald\bed)},y^4,\eta^\ald_i)$
of the moduli space $\FC^{4|8}$ according to\index{moduli space}
\begin{equation}
w^1_\pm\ =\
y^{\ald\bed}\lambda_\ald^\pm\lambda_\bed^\pm~,~~~w^3_\pm\ =\
y^4~,~~~ \eta_i^\pm\ =\ \eta_i^\ald\lambda_\ald^\pm\ewith
\lambda_\pm\ =\ w^2_\pm~.
\end{equation}

Let us now deform and restrict the sections of $\CQ^{3|4}$ by
identifying the modulus $y^4$ with
$-\gamma_\pm\lambda^\pm_\ald\hl^\pm_\bed y^{\ald\bed}$, where the
coordinates $\hl_\ald$ were defined in \eqref{alllambdas}. We
still have $w^3_+=w^3_-$ on the overlap
$\tilde{\CV}_+\cap\tilde{\CV}_-$, but $w^3$ no longer describes a
section of a holomorphic line bundle. It is rather a section of a\index{holomorphic!line bundle}
smooth line bundle, which we denote by $\hat{\CO}(0)$. This
deformation moreover reduces the moduli space from $\FC^{4|8}$ to\index{moduli space}
$\FC^{3|8}$. We will denote the resulting total bundle by
$\hat{\CQ}^{3|4}$.

\paragraph{Field theories and dimensional reductions.}\index{dimensional reduction}
First, we impose a reality condition on $\hat{\CQ}^{3|4}$ which is
(for the bosonic coordinates) given by
\begin{equation}
\tau(w_\pm^1,w_\pm^2)\ =\
\left(-\frac{\bar{w}_\pm^1}{(\bar{w}_\pm^2)^2},
-\frac{1}{\bar{w}_\pm^2}\right) \eand\tau(w^3_\pm)\ =\
\bar{w}^3_\pm~,
\end{equation}
cf.\ \eqref{eq:3.16}, and keep as usual the coordinate $w_\pm^2$
on the base $\CPP^1$ complex. Then $w^1_\pm$ remains complex, but
$w^3_\pm$ becomes real. In the identification with the real moduli
$(x^1,x^3,x^4)\in\FR^3$, we find that
\begin{equation}
y^{\ed\ed}\ =\ -(x^3+\di x^4)\ =\ -\bar{y}^{\zd\zd}\eand w^3_\pm\
=\ x^1\ =\ -y^{\ed\zd}~.
\end{equation}
Thus, the space $\hat{\CQ}^{3|4}$ reduces to a Cauchy-Riemann (CR)
manifold\footnote{Roughly speaking, a CR manifold is a complex\index{manifold}
manifold with additional real directions.}, which we label by
$\hat{\CQ}^{3|4}_{-1}=\CK^{5|8}$. This space has been extensively
studied in \cite{Popov:2005uv}, and it was found there that a
partial holomorphic Chern-Simons theory obtained from a certain\index{Chern-Simons theory}
natural integrable distribution on $\CK^{5|8}$ is equivalent to\index{integrable}\index{integrable distribution}
the supersymmetric Bogomolny model on $\FR^3$. Furthermore, it is\index{Bogomolny model}
evident that the complexification of this partial holomorphic\index{complexification}
Chern-Simons theory is holomorphic Chern-Simons theory on our\index{Chern-Simons theory}
space $\hat{\CQ}^{3|4}$. This theory describes holomorphic
structures $\dparb_\CA$ on a vector bundle $\CE$ over\index{holomorphic!structure}
$\hat{\CQ}^{3|4}$, i.e.\ a gauge potential $\CA^{0,1}$ satisfying
$\dparb\CA^{0,1}+\CA^{0,1}\wedge \CA^{0,1}=0$.

There are now three possibilities for (bosonic) dimensional
reductions\index{dimensional reduction}
\begin{equation}
\hat{\CQ}^{3|4}\ =\ \CO(2)\oplus\CO(0)\oplus \FC^4\otimes
\Pi\CO(1) \ \rightarrow\ \left\{\begin{array}{l}
\CP^{2|4}\ :=\ \CO(2)\oplus \FC^4\otimes \Pi\CO(1) \\
\hat{\CQ}^{2|4}\ :=\ \CO(0)\oplus \FC^4\otimes \Pi\CO(1) \\
\CPP^{1|4}\ :=\ \FC^4\otimes \Pi\CO(1)
\end{array}\right.~,
\end{equation}
which we want to discuss in the following.

The dimensional reduction of the holomorphic Chern-Simons theory\index{Chern-Simons theory}\index{dimensional reduction}
to the space $\CP^{2|4}$ has been studied in \cite{Popov:2005uv}.
It yields a holomorphic BF-theory, see section\index{holomorphic BF-theory}
\ref{ssrelFieldTheories}, \ref{pholBF}, where the scalar $B$-field
originates as the component $\der{\bar{w}^3_\pm}\lrcorner
\CA^{0,1}$ of the gauge potential $\CA^{0,1}$ on $\CE\rightarrow
\hat{\CQ}^{3|4}$. This theory is also equivalent to the
above-mentioned super Bogomolny model on $\FR^3$. It is\index{Bogomolny model}\index{super!Bogomolny model}
furthermore the effective theory on a topological D3-brane and --
via a Penrose-Ward transform -- can be mapped to static BPS gauge\index{Penrose-Ward transform}
configurations on a stack of D3-branes in type IIB superstring
theory. These gauge configurations have been shown to amount to\index{string theory}
BPS D1-branes being suspended between the D3-branes and extending
in their normal directions. Therefore, the holomorphic BF-theory\index{holomorphic BF-theory}
is the topological analogue of the D3-brane point of view of the
D3-D1-brane system.

From the above discussion, the field theory arising from the
reduction to $\hat{\CQ}^{2|4}$ is also evident. Note that
considering this space is equivalent to considering
$\hat{\CQ}^{3|4}$ with the additional restriction
$y^{\ed\ed}=y^{\zd\zd}=0$. Therefore, we reduced the super
Bogomolny model from $\FR^3$ to $\FR^1$, and we arrive at a\index{Bogomolny model}\index{super!Bogomolny model}
(partially) holomorphic BF-theory, which is equivalent to\index{holomorphic BF-theory}
self-dual Yang-Mills theory in one dimension. Since this theory\index{Yang-Mills theory}\index{self-dual Yang-Mills theory}
yields precisely the gauge-covariant Nahm equations, we conclude\index{Nahm equations}
that this is the D1-brane point of view of the D3-D1-brane system.

The last reduction proposed above is the one to $\CPP^{1|4}$. This
amounts to a reduction of the super Bogomolny model from $\FR^3$\index{Bogomolny model}\index{super!Bogomolny model}
to a point, i.e.\ SDYM theory in zero dimensions. Thus, we arrive
again at the matrix models \eqref{actionhCSred} and\index{matrix model}
\eqref{SDYMactionMM} discussed previously. It is interesting to
note that the matrix model cannot tell whether it originated from\index{matrix model}
the space $\CP^{3|4}$ or $\hat{\CQ}^{3|4}$.

\paragraph{The Nahm construction from topological D-branes.}\index{D-brane}
In the previous paragraph, we saw that both the physical D3-branes
and the physical D1-branes correspond to topological D3-branes
wrapping either the space $\CP^{2|4}\subset \hat{\CQ}^{3|4}$ or
$\hat{\CQ}^{2|4}\subset \hat{\CQ}^{3|4}$. The bound system of
D3-D1-branes therefore corresponds to a bound system of
D3-D3-branes in the topological picture. The two D3-branes are
separated by the same distance\footnote{In our presentation of the
Nahm construction, we chose this distance to be $1-(-1)=2$.} as
the physical ones in the normal direction $\CN_{\CP^{2|4}}\cong
\CO(2)$ in $\hat{\CQ}^{3|4}$. It is important to stress, however,
that since supersymmetry is broken twice by the D1- and the\index{super!symmetry}
D3-branes, in the topological picture, we have to put to zero all
fields except for $(A_a,\Phi,\chi^i_\ald)$.

It remains to clarify the r{\^o}le of the Nahm boundary conditions in
detail. In \cite{Diaconescu:1996rk}, this was done by considering
a D1-brane probe in a T-dualized configuration consisting of D7-
and D5-branes. This picture evidently cannot be translated into
twistor space. It would be interesting to see explicitly what the\index{twistor}\index{twistor!space}
boundary conditions correspond to in the topological setup.
Furthermore, it could be enlightening to study the topological
analogue of the Myers effect, which creates a funnel at the point
where the physical D1-branes end on the physical D3-branes.
Particularly the core of this ``bion'' might reveal interesting
features in the topological theory.

\paragraph{Summary of D-brane equivalences.} We gave an interpretation\index{D-brane}
of the matrix models derived from holomorphic Chern-Simons theory\index{Chern-Simons theory}\index{matrix model}
in terms of D-brane configurations within B-type topological\index{D-brane}
string theory. During this discussion, we established connections\index{connection}\index{string theory}
between topological branes and physical D-branes of type IIB\index{D-brane}
superstring theory, whose worldvolume theory had been reduced by\index{string theory}
an additional BPS condition due to the presence of a further
physical brane. Let us summarize the correspondences in the
following table:
\begin{equation*}
\begin{aligned}
\mbox{D$(5|4)$-branes in $\CP^{3|4}$} &\ \leftrightarrow\
\mbox{D$(3|8)$-branes in $\FR^{4|8}$}\\
\mbox{D$(3|4)$-branes wrapping $\CP^{2|4}$ in $\CP^{3|4}$ or
$\hat{\CQ}^{3|4}$} &\ \leftrightarrow\
\mbox{static D$(3|8)$-branes in $\FR^{4|8}$}\\
\mbox{D$(3|4)$-branes wrapping $\hat{\CQ}^{2|4}$ in
$\hat{\hat{\CQ}}^{3|4}$} &\ \leftrightarrow\
\mbox{static D$(1|8)$-branes in $\FR^{4|8}$}\\
\mbox{D$(1|4)$-branes in $\CP^{3|4}_\eps$} &\ \leftrightarrow\
\mbox{D(-$1|8$)-branes in $\FR^{4|8}$}~.
\end{aligned}
\end{equation*}
It should be stressed that the fermionic parts of all the branes
in $\CP^{3|4}_\eps$ and $\hat{\CQ}^{3|4}$ only extend into
holomorphic directions. It is straightforward to add to this list
the diagonal line bundle $\CD^{2|4}$, which is obtained from
$\CP^{3|4}$ by imposing the condition\footnote{or an appropriate
modification in the Euclidean case} $z^1_\pm=z^2_\pm$ on the local
sections
\begin{equation*}
\mbox{D$(3|4)$-branes wrapping $\CD^{2|4}$ in $\CP^{3|4}$}
\ \leftrightarrow\  \mbox{ D$(1|8)$-branes in $\FR^{4|8}$}~.
\end{equation*}

\chapter{Conclusions and Open Problems}

\section{Summary}

Let us briefly summarize the results presented in this thesis,
grouped according to the papers they were first published in.

\paragraph{Non-anticommutative deformations of superspaces.} One can define a\index{super!space}
non-anti\-com\-muta\-tive deformation of $\CN=4$ super Yang-Mills
theory by using the corresponding constraint equations. A\index{N=4 super Yang-Mills theory@$\CN=4$ super Yang-Mills theory}\index{Yang-Mills theory}\index{constraint equations}
Seiberg-Witten map can be motivated in that context.\index{Seiberg-Witten map}

Using Drinfeld twist techniques, one can make manifest the twisted\index{Drinfeld twist}\index{twist}
supersymmetry on non-anticommutative superspaces. This twisted\index{non-anticommutative superspace}\index{super!space}\index{super!symmetry}
supersymmetry can take over the r{\^o}le of ordinary supersymmetry in
the definition of chiral rings, supersymmetric Ward-Takahashi\index{chiral!ring}
identities and probably even non-renormalization theorems. These\index{non-renormalization theorems}\index{Theorem!non-renormalization}
constructions based on twisted supersymmetry may prove to be very\index{Twisted supersymmetry}\index{super!symmetry}
useful in explicit calculations within non-anticommutative field
theories, similarly to their cousins in ordinary supersymmetric\index{non-anticommutative field theories}
field theories.

\paragraph{Twistor string theory.} The concept of marrying twistor\index{string theory}\index{twistor}\index{twistor!string theory}
geometry and Calabi-Yau geometry using the supertwistor space\index{Calabi-Yau}\index{twistor!space}
$\CPP^{3|4}$ looks very promising. One can carry over the whole
Penrose-Ward transform to the case of supertwistors and describe\index{Penrose-Ward transform}\index{twistor}
solutions to the $\CN=4$ self-dual Yang-Mills equations in terms
of solutions to the holomorphic Chern-Simons equations on
$\CPP^{3|4}$.

Fattened complex manifolds (or exotic supermanifolds) arising\index{complex!manifold}\index{exotic!supermanifold}\index{fattened complex manifold}\index{super!manifold}\index{manifold}
naturally from the supertwistor space $\CPP^{3|4}$ can be used to\index{twistor}\index{twistor!space}
describe certain bosonic subsectors of $\CN=4$ self-dual
Yang-Mills theory. Furthermore, the concept of exotic Calabi-Yau\index{Calabi-Yau}\index{Yang-Mills theory}\index{self-dual Yang-Mills theory}
supermanifolds fits nicely into the framework of ordinary\index{super!manifold}
Calabi-Yau supermanifolds\index{Calabi-Yau}\index{Calabi-Yau supermanifold}

The mini-twistor space $\CO(2)$ can be supersymmetrically extended\index{mini-twistor space}\index{twistor}\index{twistor!space}
to a Calabi-Yau supermanifold. The topological B-model with this\index{Calabi-Yau}\index{Calabi-Yau supermanifold}\index{super!manifold}\index{topological!B-model}
space as its target space is equivalent to the supersymmetrically\index{target space}
extended Bogomolny equations in three dimensions.\index{Bogomolny equations}

A corresponding mini-superambitwistor space can be defined,\index{ambitwistor space}\index{mini-superambitwistor space}\index{twistor}\index{twistor!ambitwistor}\index{twistor!space}
although this space is neither the total space of a vector bundle
nor a manifold. Nevertheless, this space is still a fibration and\index{fibration}\index{manifold}
has all the necessary features for a Penrose-Ward transform\index{Penrose-Ward transform}
between certain generalized bundles over this fibration and\index{fibration}
solutions to the $\CN=8$ Yang-Mills-Higgs equations in three
dimensions.

Also matrix models can be consistently defined via dimensional\index{matrix model}
reduction of both the topological B-model on the supertwistor\index{topological!B-model}\index{twistor}
space $\CPP^{3|4}$ and its equivalent $\CN=4$ self-dual Yang-Mills
theory in four dimensions. One can interpret these matrix models\index{Yang-Mills theory}\index{matrix model}\index{self-dual Yang-Mills theory}
in terms of topological D-branes, (bound states of) D-branes in\index{D-brane}
type IIB superstring theory and D-branes within critical $\CN=2$\index{string theory}
string theory. Furthermore, one can extend the matrix models\index{matrix model}
obtained from the topological B-model to be equivalent to the ADHM\index{topological!B-model}
or even the Nahm equations. This extension allows furthermore for\index{Nahm equations}
establishing a Penrose-Ward transform between D-brane systems in\index{D-brane}\index{Penrose-Ward transform}
type IIB on the one side and topological B-branes on the other
side.

\section{Directions for future research}

In the derivation of the above results, several questions were
raised, which can be taken as starting point for quite interesting
future research.

\paragraph{Non-anticommutative deformations of superspaces.} In\index{super!space}
the definition of non-anticommutative $\CN=4$ super Yang-Mills
theory, it would be clearly interesting to see whether this\index{N=4 super Yang-Mills theory@$\CN=4$ super Yang-Mills theory}\index{Yang-Mills theory}
definition yields compatible results with the canonical definition
of non-anticommutative field theories by inserting star-products\index{non-anticommutative field theories}
into a superfield action. For this, one could either reduce the
amount of supersymmetry, or restrict the deformation tensor such\index{super!symmetry}
that it fits with the formulation of $\CN=4$ super Yang-Mills
theory in the language of $\CN=1$ superspace.\index{N=4 super Yang-Mills theory@$\CN=4$ super Yang-Mills theory}\index{Yang-Mills theory}\index{super!space}

Furthermore, it would be very interesting to substantiate the
definition of a Seiberg-Witten map in the non-anticommutative\index{Seiberg-Witten map}
setting. In \cite{Saemann:2004cf}, some arguments in favor of such
a map were given; however, these arguments are clearly not
sufficient.

It would also be illuminating to explore the connection of the\index{connection}
constraint equations and the underlying linear system of partial\index{constraint equations}\index{linear system}
differential equations. This system could subsequently be used to
generalize solution generating techniques (as e.g.\ the dressing\index{solution generating techniques}
and splitting methods) available for the corresponding linear
system in the undeformed case.\index{linear system}

Within the framework of Drinfeld twists, clearly the study of\index{Drinfeld twist}
twist-deformed superconformal invariance following the discussion
of twisted conformal invariance in \cite{Matlock:2005zn}, could
potentially yield further interesting results.

Moreover, our results on Drinfeld twists for non-anticommutative\index{Drinfeld twist}
superspaces may prove valuable for introducing a\index{super!space}
non-anticommutative deformation of supergravity. Building upon the
discussion presented in \cite{Aschieri:2005yw}, one could try to
construct a local version of the twisted supersymmetry.\index{Twisted supersymmetry}\index{super!symmetry}

Also, Seiberg's naturalness argument should be verified or at
least be motivated stronger in the non-anticommutative setting to
clarify the apparent inconsistency between the non-renormalization
theorem conjectured in \cite{Ihl:2005zd} and the further results\index{Theorem!non-renormalization}
for one-loop calculations in the literature.

\paragraph{Twistor string theory.} There remain essentially two open questions\index{string theory}\index{twistor}\index{twistor!string theory}
concerning the general supertwistor correspondence and its\index{twistor!correspondence}
application within topological string theory. First, it would be\index{string theory}\index{topological!string}
desirable to find an appropriate action functional for holomorphic
Chern-Simons theory on the superambitwistor space $\CL^{5|6}$. Up\index{Chern-Simons theory}\index{ambitwistor space}\index{twistor}\index{twistor!ambitwistor}\index{twistor!space}
to now, there have been two attempts in this direction
\cite{Movshev:2004ub,Mason:2005zm,Mason:2005kn}, but a more direct
construction would be desirable. Also, one could use the
Penrose-Ward transform built upon the supertwistor correspondence\index{Penrose-Ward transform}\index{twistor}\index{twistor!correspondence}
to establish or strengthen the long-sought relation between the
$\CN=2$ and $\CN=4$ topological strings.\index{topological!string}

The results on the supertwistor correspondence over exotic\index{twistor}\index{twistor!correspondence}
supermanifolds and in particular the results on the extension of\index{super!manifold}
Yau's theorem suggest that these space are quite natural to\index{Theorem!Yau}
consider as target spaces within topological string theory. One\index{string theory}\index{target space}\index{topological!string}
could try to establish mirror symmetry conjectures between certain\index{mirror symmetry}
exotic Calabi-Yau supermanifolds to enrich the set of examples of\index{Calabi-Yau}\index{Calabi-Yau supermanifold}\index{exotic!Calabi-Yau supermanifold}\index{super!manifold}
such a conjecture.

Partially holomorphic Chern-Simons theory leads naturally to\index{Chern-Simons theory}
studying twistor correspondences for further geometries, as shown\index{twistor}\index{twistor!correspondence}
also in the section discussing matrix models for the Nahm\index{matrix model}
equations. Other examples are certainly the Cauchy-Riemann
manifolds and the deformations of the mini-supertwistor space\index{Cauchy-Riemann manifold}\index{mini-supertwistor space}\index{twistor}\index{twistor!space}\index{manifold}
corresponding to turning on mass terms in the Bogomolny equations.\index{Bogomolny equations}
The most interesting question is certainly whether one can use
partially holomorphic geometry in the context of topological
M-theory \cite{Dijkgraaf:2004te}.\index{M-theory}

The construction of the mini-superambitwistor space $\CL^{4|6}$\index{ambitwistor space}\index{mini-superambitwistor space}\index{twistor}\index{twistor!ambitwistor}\index{twistor!space}
leads to a number of interesting questions. First of all, one
should find out, how to construct the topological B-model, which\index{topological!B-model}
has this space as a target space and its holomorphic\index{target space}
Chern-Simons-type equivalent theory. Second, one should
substantiate the mirror conjecture between the mini-supertwistor\index{twistor}
space $\CP^{2|4}$ and the mini-superambitwistor space $\CL^{5|6}$,\index{ambitwistor space}\index{mini-superambitwistor space}\index{twistor!ambitwistor}\index{twistor!space}
which naturally arises due to a similar conjecture between the
supertwistor space $\CP^{3|4}$ and the superambitwistor space\index{ambitwistor space}\index{twistor}\index{twistor!ambitwistor}\index{twistor!space}
$\CL^{5|6}$. As a third point, one might try to find the analogous
construction of an action for holomorphic Chern-Simons theory on\index{Chern-Simons theory}
the mini-superambitwistor space to the one proposed by Movshev\index{ambitwistor space}\index{mini-superambitwistor space}\index{twistor}\index{twistor!ambitwistor}\index{twistor!space}
\cite{Movshev:2004ub}. This might shed more light on the relevance
and usefulness of both the constructions of $\CL^{4|6}$ and the
action in \cite{Movshev:2004ub}.

In the area of matrix models within twistor string theory, one\index{matrix model}\index{string theory}\index{twistor}\index{twistor!string theory}
should first examine in more detail the topological D-brane\index{D-brane}
configuration yielding the Nahm equations. In particular, it is\index{Nahm equations}
desirable to obtain more results on the Myers effect and the core
of the ``bion'' in the topological setting. Second, one could
imagine to strengthen and extend the relations between D-branes in\index{D-brane}
type IIB superstring theory and the topological D-branes in the\index{string theory}
B-model. In the latter theory, the strong framework of derived
categories (see e.g.\ \cite{Aspinwall:2004jr}) might then be
carried over in some form to the full ten-dimensional string
theory. Eventually, it might also be interesting to look at the\index{string theory}
mirror of the presented configurations within the topological
A-model.\index{topological!A-model}

\appendix
\chapter{}
\pagestyle{fancyplain}
\lhead[\fancyplain{}{\bfseries\thepage}]{\fancyplain{}{\itshape\rightmark}}
\rhead[\fancyplain{}{\itshape
Appendices}]{\fancyplain{}{\bfseries\thepage}} \cfoot{}

\section{Further definitions}

In this appendix, we recall the notions of some elementary
mathematical objects. It turned out that the web page of
Wikipedia\footnote{{\tt http://en.wikipedia.org}} is a
surprisingly useful reference for looking up further mathematical
definitions.

\paragraph{Morphisms.} Given two groupoids $G_1,+_1$ and $G_2,+_2$, a map
$f:G_1\rightarrow G_2$ is called a {\em homomorphism} if it
satisfies
\begin{equation}
\forall a,b \in G_1: f(a+_1b)\ =\ f(a)+_2f(b)~.
\end{equation}
A bijective, injective, surjective homomorphism is called {\em
isomorphism}, {\em monomorphism}, {\em epimorphism}. If the\index{morphisms!epimorphism}\index{morphisms!isomorphism}\index{morphisms!monomorphism}
groupoids are identical: $(G_1,+_1)=(G_2,+_2)$ then the map is an
{\em automorphism}.\index{morphisms!automorphism}

\paragraph{Groups and representations.} A {\em\index{representation}
representation} of a group $\CCG$ is a homomorphism from $\CCG$ to
the space of linear transformations on a vector space. A {\em
faithful representation} is a representation which is furthermore\index{faithful representation}\index{representation}
isomorphic to $\CCG$. The {\em fundamental representation} of a\index{fundamental representation}
group is its lowest dimensional faithful representation. The {\em\index{faithful representation}
trivial representation} of $\CCG$ maps the whole group to the\index{trivial representation}
identity map $\unit$ on some vector space. A representation is
{\em irreducible} if it cannot be decomposed into block diagonal
form. All reducible representations can be built of the\index{representation}
irreducible representations.\index{irreducible representation}

Given an element $g$ of a group $G$, the {\em stabilizer subgroup}\index{stabilizer subgroup}
of $g$ (also called the {\em isotropy group} or {\em little
group}) is the subgroup $G_g$ of $G$ leaving $g$ invariant:
\begin{equation}
G_g\ :=\ \{x\in G|x\cdot g=g\}~.
\end{equation}

\paragraph{Short exact sequence.} A sequence of groups\index{short exact sequence}
\begin{equation}
\ldots A_i \ \stackrel{f_i}{\longrightarrow} \ A_{i+1} \
\stackrel{f_{i+1}}{\longrightarrow} \ A_{i+2}\ldots
\end{equation}
is called {\em exact} if the image of $f_i$ is equal to the kernel
of $f_{i+1}$. A {\em short exact sequence} is an exact sequence of\index{short exact sequence}
the form
\begin{equation}
0\ \longrightarrow\  A \ \stackrel{f}{\longrightarrow} \ B \
\stackrel{g}{\longrightarrow} \ C\ \longrightarrow\  0~.
\end{equation}
It follows that $f$ is a monomorphism and $g$ is an epimorphism.\index{morphisms!epimorphism}\index{morphisms!monomorphism}
Furthermore, if $A,B,C$ are vector bundles over a manifold $M$\index{manifold}
then $B=C\oplus A$.

\section{Conventions}

\paragraph{Metric conventions.} Our Minkowski metric follows the\index{metric}
``east coast convention'', i.e.\ it is mostly $+$.

\paragraph{Interior product.} For the interior product of a vector\index{interior product}
$V$ with a one-form $A$, we use the notation $V\lrcorner
A:=\langle V,A\rangle$. A second common notation for this product
is $\di_V A$.

\paragraph{Dual space.} Following Grothendieck, we denote the dual\index{dual space}
of a space $X$, i.e. the space of linear maps $X\rightarrow \FK$,
by $X^\vee$.

\paragraph{Commutators and parity.} We use square brackets $[\cdot]$ for the\index{commutators}\index{parity}
commutator, curly brackets $\{\cdot\}$ for the anticommutator and a
combination of both $\lsc\cdot\rsc$ for the graded or supercommutator:\index{super!commutator}
\begin{equation}
[A,B]\ :=\ A.B-B.A~,~~~\{A,B\}\ :=\ A.B+B.A~,~~~\lsc A,B
\rsc\ :=\ A.B-(-1)^{\tilde{A}\tilde{B}}B.A~,
\end{equation}
where $\tilde{A}\in\{0,1\}$ denotes the Gra{\ss}mann parity of $A$ and\index{parity}
$.$ denotes a product defined between $A$ and $B$.

\paragraph{Lie Algebra and gauge field conventions.} Almost all the fields\index{Lie algebra}
in this thesis live in the adjoint representation of a gauge\index{representation}
group: $(T^a)_{bc}=(f^a)_{bc}$. For this, we fix the trace
$\tr(T^aT^b)=-\delta^{ab}$ and choose the generators of the gauge
group to be anti-Hermitian: $[T^a,T^b]=f^{ab}{}_c T^c$ (note that
$f^a{}_{bc}=f^{ab}{}_c=f_{abc}=\ldots $). For the field strength, we\index{field strength}
use the definition $F_{\mu\nu}=\dpar_\mu A_\nu-\dpar_\nu
A_\mu+[A_\mu,A_\nu]$ and the covariant derivative of a field in\index{covariant derivative}
the adjoint representation is $D_\mu \psi=\dpar_\mu\index{representation}
\psi+[A_\mu,\psi]$. In terms of gauge components, this reads:
$F_{\mu\nu}^a=\dpar_\mu A_\nu-\dpar_\nu A_\mu+f^a{}_{bc}A_\mu^b
A_\nu^c$ and $(D_\mu
\lambda)^a=\dpar_\mu\lambda^a+f^a{}_{bc}A_\mu^b\lambda^b$.

\paragraph{SUSY conventions and identities.} We follow essentially
the conventions of Wess and Bagger \cite{Wess:1992cp}. Indices are
raised with the epsilon tensors according to
$\psi^\alpha=\eps^{\alpha\beta}\psi_\beta$,
$\psi_\alpha=\eps_{\alpha\beta}\psi^\beta$. We choose
$\eps_{12}=\eps_{\ed\zd} =-\eps^{12}=-\eps^{\ed\zd}=-1$ which
implies the relation
$\eps_{\alpha\beta}\eps^{\beta\gamma}=\delta_\alpha^\gamma$, and a
similar one for dotted indices. Short hand notations are:
$\psi\chi=\psi^\alpha\chi_\alpha=\chi^\alpha\psi_\alpha=\chi\psi$
and $\bar{\psi}\bar{\chi}=\bar{\psi}_\ald\bar{\chi}^\ald$.
Derivatives with respect to Gra{\ss}mann variables are defined in the\index{Gra{\ss}mann variable}
following manner:
$\overrightarrow{\dpar}_\alpha\theta^\beta=\delta^\beta_\alpha$
and
$\theta^\beta\overleftarrow{\dpar}_\alpha=-\delta^\beta_\alpha$,
together with the super Leibniz rule for differential operators\index{super!Leibniz rule}
$\dpar (fg)=(\dpar f)g+(-1)^{\tilde{\dpar}\tilde{f}}f(\dpar g)$,
where $\tilde{a}$ denotes again the Gra{\ss}mann parity of $a$, i.e.\index{parity}
$\tilde{a}=0$ for bosonic $a$ and $\tilde{a}=1$ for fermionic $a$.

\section{Dictionary: homogeneous $\leftrightarrow$ inhomogeneous
coordinates}\label{AppDictionary}\index{homogeneous coordinates}\index{inhomogeneous coordinates}

This appendix is a complement to the discussion of chapter
\ref{chTwistorGeometry} and provides information on how to switch
between homogeneous and inhomogeneous coordinates on twistor\index{homogeneous coordinates}\index{inhomogeneous coordinates}
space.

\paragraph{The Riemann sphere $\CPP^1$.} The sphere $S^2$ is
diffeomorphic to the complex projective space $\CPP^1$. Let us\index{complex!projective space}\index{diffeomorphic}
recall from section \ref{ssManifolds}, \ref{pGrassmannManifolds}
that this space can be para\-metri\-zed globally by complex
homogeneous coordinates $\lambda_{\dot{1}}$ and\index{homogeneous coordinates}
$\lambda_{\dot{2}}$ which are not simultaneously zero (in
projective spaces, the origin is excluded). Therefore the Riemann
sphere $\CPP^1$ can be covered by two coordinate patches\index{Riemann sphere}
\begin{equation}
U_+\ =\ \{\,[\lambda_{\dot{1}},\lambda_{\dot{2}}]\,|\,\lambda_{\dot{1}}\ \neq\
0\,\} ~~~\mbox{and}~~~
U_-\ =\ \{\,[\lambda_{\dot{1}},\lambda_{\dot{2}}]\,|\,\lambda_{\dot{2}}\ \neq\
0\,\}~,
\end{equation}
with coordinates
\begin{equation}
\lambda_+\ :=\ \frac{\lambda_{\dot{2}}}{\lambda_{\dot{1}}}~~~\mbox{on}~~
U_+~~~\mbox{and}~~~
\lambda_-\ :=\ \frac{\lambda_{\dot{1}}}{\lambda_{\dot{2}}}~~~\mbox{on}~~
U_-~.
\end{equation}
On the intersection $U_+\cap U_-$, we have
$\lambda_+=1/\lambda_-$.

\paragraph{Line bundles.} A global section of the holomorphic line
bundle $\CO(n)$ over $\CPP^1$ exists only for $n\geq 0$. Over\index{O(n)@$\CO(n)$}\index{holomorphic!line bundle}
$U_\pm$, it is represented by a polynomial $p^{(n)}_\pm$ of degree
$n$ in the coordinates $\lambda_\pm$ with $p^{(n)}_+=\lambda_+^n
p^{(n)}_-$ on $U_+\cap U_-$. The explicit expansion will look like
\begin{equation}
\begin{aligned}
p_+^{(n)}&\ =\  a_0+a_1\lambda_++a_2\lambda_+^2+\ldots +a_n\lambda_+^n~,\\
p_-^{(n)}&\ =\ a_0\lambda_-^n+\ldots +a_{n-2}\lambda_-^2+a_{n-1}\lambda_-+a_n~,
\end{aligned}
\end{equation}
and, multiplying the expansion in $\lambda_+$ by
$\lambda_{\dot{1}}^n$ (or the expansion in $\lambda_-$ by
$\lambda_{\dot{2}}^n$), one obtains a homogeneous polynomial of
degree $n$:
\begin{equation}
a_0\lambda_{\dot{1}}^n+a_1\lambda_{\dot{1}}^{n-1}\lambda_{\dot{2}}+\ldots
+a_{n-1}\lambda_{\dot{1}}\lambda_{\dot{2}}^{n-1}+a_n\lambda_{\dot{2}}^n~\ =:\ ~
Q^{\ald_1\ldots \ald_n}\lambda_{\ald_1}\ldots \lambda_{\ald_n}~.
\end{equation}

\paragraph{Gauge potentials.} Now let us consider the expansion
\eqref{expAa} and \eqref{expAl} of the super gauge potentials of
hCS theory on the supertwistor space. We get the following list of\index{twistor}\index{twistor!space}
objects:
\begin{eqnarray*}
\eta_i^+& \CO(1) & \eta_i=\lambda_{\dot{1}}\eta_i^+\\
\gamma_+& \CO(-1)\otimes\bar{\CO}(-1) & \gamma=
\frac{1}{\lambda_{\dot{1}}\bl_{\dot{1}}}\;
\gamma_+\left(=\frac{1}{\lambda^\ald\hat{\lambda}_\ald}\right)\\
\hat{\CA}^+_\alpha & \CO(1) &
\hat{\CA}_\alpha=\lambda_{\dot{1}}\;\hat{\CA}_\alpha^+\\
\hat{\CA}_{\bl_+} & \bar{\CO}(-2) &
\hat{\CA}_3=\frac{1}{\bl_{\dot{1}}\bl_{\dot{1}}}\;\hat{\CA}_{\bl_+}~.
\end{eqnarray*}
This implies the following expansions in homogeneous coordinates:\index{homogeneous coordinates}
\begin{subequations}
\begin{eqnarray}\label{hexpAa}
\hat{\CA}_\alpha&=&\lambda^\ald\,
A_{\alpha\ald}(x_R)+\eta_i\chi^i_\alpha(x_R)+
\gamma\,\tfrac{1}{2!}\,\eta_i\eta_j\,\hat{\lambda}^\ald\,\phi_{\alpha
\ald}^{ij}(x_R)+\\\nonumber
&&+\gamma^2\,\,\tfrac{1}{3!}\,\eta_i\eta_j\eta_k\,\hat{\lambda}^\ald\,
\hat{\lambda}^\bed\,
\tilde{\chi}^{ijk}_{\alpha\ald\bed}(x_R)+\gamma^3\,\tfrac{1}{4!}\,
\eta_i\eta_j\eta_k\eta_l\,
\hat{\lambda}^\ald\,\hat{\lambda}^\bed\,\hat{\lambda}^{\dot{\gamma}}\,
G^{ijkl}_{\alpha\ald\bed\dot{\gamma}}(x_R)~,\\\label{hexpAl}
\hat{\CA}_{3}&=&\gamma^2\,\tfrac{1}{2!}\,\eta_i\eta_j\,\phi^{ij}(x_R)+
\gamma^3\,\tfrac{1}{3!}\,\eta_i\eta_j\eta_k\,\hat{\lambda}^\ald\,
\tilde{\chi}^{ijk}_{\ald} (x_R)+\\\nonumber
&&+\gamma^4\,\tfrac{1}{4!}\,\eta_i\eta_j\eta_k\eta_l\,
\hat{\lambda}^\ald\,\hat{\lambda}^\bed G^{ijkl}_{\ald\bed}(x_R)~.
\end{eqnarray}
\end{subequations}

\paragraph{Equations of motion.} For rewriting the equations of motion
in terms of this gauge potential, we also need to rewrite the
vector fields \eqref{vectorfields1} and $\bV_3^\pm=\der{\bl_\pm}$
in homogeneous coordinates. The vector fields along the fibres are\index{homogeneous coordinates}
easily rewritten, analogously to the corresponding components of
the gauge potential. The vector field on the sphere can be
calculated by considering $\hat{\CA}_{\bl_+}\dd \bl_+= \hat{\CA}_3
\bar{\Theta}^3$. This implies
$\bar{\Theta}^3=\bl_{\dot{1}}\dd\bl_{\dot{2}}-\bl_{\dot{2}}\dd\bl_{\dot{1}}$,
which has a dual vector field $\bar{V}_3$ defined by
$\bar{V}_3\lrcorner\,\bar{\Theta}^3=1$. Altogether, we obtain the
basis
\begin{equation}
\bar{V}_\alpha\ =\ \lambda^\ald\der{x_R^{\alpha\ald}}~~~
\mbox{and}~~~\bar{V}_3\ =\ -\gamma\lambda^\ald\der{\hat{\lambda}^\ald}~.
\end{equation}
The field equations \eqref{shCS1} and \eqref{shCS2} now take the
form
\begin{equation}
\begin{aligned}
\bar{V}_\alpha \hat{\CA}_\beta-\bar{V}_\beta \hat{\CA}_\alpha
+[\hat{\CA}_\alpha,\hat{\CA}_\beta]&\ =\ 0~,\\
\bar{V}_3 \hat{\CA}_\alpha-\bar{V}_\alpha \hat{\CA}_3
+[\hat{\CA}_3,\hat{\CA}_\alpha]&\ =\ 0~,
\end{aligned}
\end{equation}
and yield the same equations \eqref{SDYMeom} for the physical
fields.

\newpage

\section{Map to (a part of) ``the jungle of TOE''}

\centerline{\includegraphics[width=16cm,height=23cm]{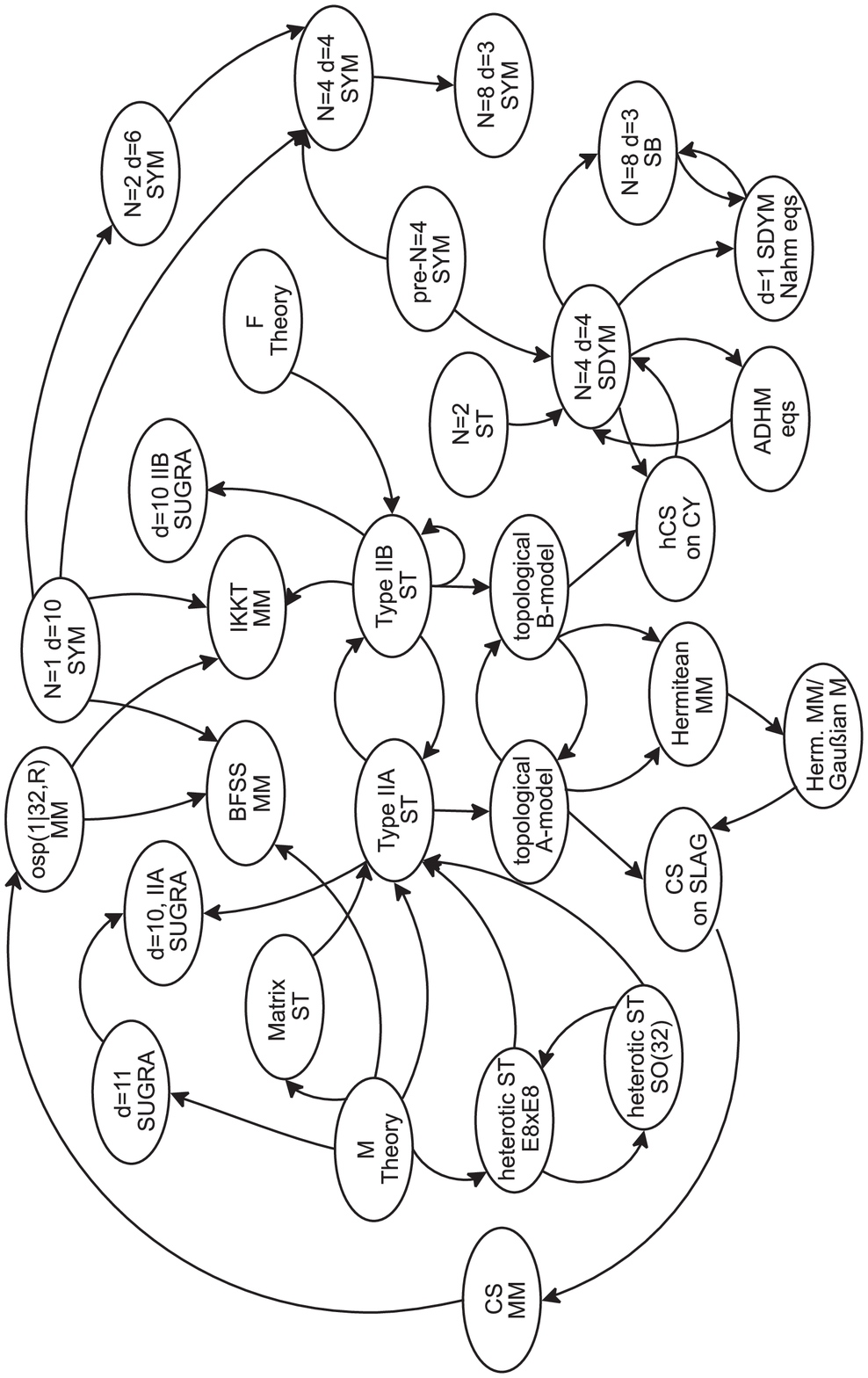}}

To put the different theories mentioned in this thesis into
context, the opposite map to a small part of ``the jungle of the
theory of everything'' might be helpful. The abbreviations used
are the following:
\begin{center}
\begin{tabular}{llllll}
CS & Chern-Simons theory & N & number of supersymmetries\\\index{Chern-Simons theory}
CY & Calabi-Yau manifold & SB & super Bogomolny model\\\index{Bogomolny model}\index{Calabi-Yau}\index{super!Bogomolny model}\index{manifold}
d & number of dimensions & SDYM & self-dual Yang-Mills theory\\\index{Yang-Mills theory}\index{self-dual Yang-Mills theory}
eqs & equations & SLAG & special Lagrangian manifold\\\index{manifold}
Gau{\ss}ian M & Gau{\ss}ian measure & ST & string theory\\\index{string theory}
hCS & holomorphic CS theory & SUGRA & supergravity\\
MM & matrix model & SYM & super Yang-Mills theory\\\index{Yang-Mills theory}\index{matrix model}
\end{tabular}
\end{center}
Furthermore, let us briefly comment on some of the arrows in the
map. Type IIA and type IIB superstring theory are linked by\index{string theory}
T-duality and the embedded topological models are related via\index{T-duality}
mirror symmetry. The arrow from type IIB to itself is so-called\index{mirror symmetry}
S-duality. The connection between holomorphic Chern-Simons theory\index{Chern-Simons theory}\index{connection}
and self-dual Yang-Mills theory is the Penrose-Ward transform\index{Penrose-Ward transform}\index{Yang-Mills theory}\index{self-dual Yang-Mills theory}
discussed extensively in this thesis. Most of the other links
correspond to dimensional reductions or to taking certain weak\index{dimensional reduction}
coupling limits.

\section{The quintic and the Robinson\index{quintic}
congruence}\label{sQuinticRobinson}

On the title page and on the back of this thesis there are two
pictures representing the essential geometries encountered in this
thesis: Calabi-Yau geometry and twistor geometry. The first is a\index{Calabi-Yau}\index{twistor}
cross-section of the quintic hypersurface, the second represents\index{quintic}
the Robinson congruence.\index{Robinson congruence}

\paragraph{The quintic.} To plot a cross-section through the\index{quintic}
quintic with Mathematica, we first mod out the projective symmetry
and arrive at
\begin{equation}
z^1+z^2+z^3+z^4+1\ =\ 0~,
\end{equation}
with $z^i\in\FC$. Furthermore, we assume constant values for $z^3$
and $z^4$ and put them to zero. The remaining real
four-dimensional object is then projected into three-dimensional
space by adding the real and imaginary parts of $z^2=u+\di v$ as a
function of $z^1=r+\di s$. The source code reads:

\begin{verbatim}
NSq[z_,n_]:=Abs[z]^(1/5)*Exp[I (Arg[z]+n*2*Pi)/5];

u[r_,s_,n_]:=2*Re[NSq[(1-(r+I s)^5),n]];
v[r_,s_,n_]:=2*Im[NSq[(1-(r+I s)^5),n]];

P=ParametricPlot3D[{{r,s,(u[r,s,1]+v[r,s,1])},
         {r,s,(u[r,s,2]+v[r,s,2])},{r,s,(u[r,s,3]+v[r,s,3])},
         {r,s,(u[r,s,4]+v[r,s,4])},{r,s,(u[r,s,5]+v[r,s,5])}},
         {r,-1,1},{s,-1,1}]
\end{verbatim}

In the notebook, we first defined a function $NSq(z,n)$ for the
$n$th of the five 5th roots of a complex number $z$. Then we
defined $z^2(z^1)$ for the five roots and plot all of them in a
single frame.

\paragraph{The Robinson congruence.} As discussed in section\index{Robinson congruence}
\ref{ssmotivation}, a null twistor corresponds to a geodesic in\index{twistor}
Minkowski space. A non-null twistor $Z^i$, on the other hand,
gives rise to a subspace of the dual twistor space $(\CPP^3)\dual$\index{twistor!dual twistor}\index{twistor!space}
via
\begin{equation}
S=\{W_i \in (\CPP^3)\dual | Z^iW_i\ =\ 0 \}~.
\end{equation}
The intersection of this space with the space of null twistors\index{twistor}
$\PP\TT_N$ is a three-dimensional space, which parameterizes a
spacetime filling family of geodesics in the compactified
spacetime $M$. By taking a time slice and projecting the tangent
vectors in $M$ onto this time slice, we recover Penrose's picture
of this Robinson congruence: A set of nested tori, with an axis in\index{Robinson congruence}
their middle as printed on the back of this thesis. The null axis
corresponds to a null twistor and the time evolution is a movement\index{twistor}
of the whole configuration along this axis, while the twisted
tangent vectors on the tori rotate. This observation originally
gave rise to the name twistor. For more details on this picture\index{twistor}
and its relation to the Hopf fibration, see \cite{Baird}.\index{fibration}

The Mathematica source code for generating the three-dimensional
projection reads:
\begin{verbatim}
r[theta_,omega_,phi_]:=2(-Tan[theta]Cos[omega+phi]+Sec[theta])/
                        (1+Tan[theta]Sin[omega+phi]Sin[omega+phi])

P[theta_]:=ParametricPlot3D[{r[theta,u,phi]Cos[phi],
        r[theta,u,phi]Sin[phi],-r[theta,u,phi]Tan[theta]Sin[u+phi]},
        {u,0,2 Pi},{phi,0,2 Pi}];

Show[{P[0.2],P[0.5],P[1.2]},
        PlotRange->{{-3.5,3.5},{-1,3.5},{-1,1}},
        ViewPoint->{0,-2.6,0.9}]
\end{verbatim}

\newpage
\pagestyle{fancyplain}
\renewcommand{\chaptermark}[1]{\markboth{#1}{}}
\renewcommand{\sectionmark}[1]{\markright{\thesection\ #1}}
\lhead[\fancyplain{}{\bfseries\thepage}]{\fancyplain{}{\itshape\rightmark}}
\rhead[\fancyplain{}{\itshape\leftmark}]{\fancyplain{}{\bfseries\thepage}}
\cfoot{}

\bibliographystyle{mydiss}

\bibliography{thesis}

\newpage
\pagestyle{plain} \addcontentsline{toc}{chapter}{Acknowledgements}
\begin{center}\LARGE\bfseries{\sc Acknowledgements}\end{center}

First of all, I would like to thank Prof.\ Olaf
Lechtenfeld for giving me the opportunity to do my PhD studies
here in Hannover, for always being ready to help with advice and
expertise, and for a pleasant collaboration on the paper
\cite{Lechtenfeld:2005xi}.

I am particularly grateful to my direct supervisor Dr.\ Alexander
Popov, who not only accompanied and guided essentially all of my
scientific steps towards this thesis and beyond in a thoughtful
and selfless way, but also came up with the patience to bear with
my problems in being as precise as possible. He taught me to apply
stricter standards to both my mathematical argumentation and my
scientific writing.

Many of the results presented in this thesis are the outcome of
joint work with several people, and I would like to express my
gratitude to my collaborators Matthias Ihl, Olaf Lechtenfeld,
Alexander Popov and Martin Wolf for fruitful teamworking.

Furthermore, learning and understanding physics and mathematics is
always incredibly simplified if one has the opportunity to discuss
one's thoughts and ideas with other people, who are willing to
share their knowledge and take the time to give helpful comments.
I therefore would like to thank Hendrik Adorf, Nikolas Akerblom,
Alexandra De Castro, Norbert Dragon, Michael Eastwood, Klaus
Hulek, Matthias Ihl, Alexander Kling, Petr Kulish, Olaf
Lechtenfeld, Guillaume Palacios, Stefan Petersen, Alexander Popov,
Emanuel Scheidegger, Albert Schwarz, Sebastian Uhlmann, Kirsten
Vogeler, Robert Wimmer, Martin Wolf and Boris Zupnik for numerous
discussions, helpful suggestions on how to improve the
publications I was involved in and partly also for proofreading
this thesis.

I am also very grateful to Prof.\ Holger Frahm, who kindly agreed
to being the second reviewer of this thesis.

Life clearly does not consist of science alone, and the people at
the ITP in Hannover were largely responsible for my time here
being as enjoyable as it was. Many thanks therefore to all of you:
Hendrik Adorf, Nikolas Akerblom, Cecilia Albertsson, Alexandra De
Castro, Henning Fehrmann, Michael Flohr, Matthias Ihl, Alexander
Kling, Michael Klawunn, Marco Krohn, Carsten Luckmann, Oleksiy
Maznytsya, Andreas Osterloh, Guillaume Palacios, Stefan Petersen,
Leonardo Quevedo, Sebastian Uhlmann, Kirsten Vogeler, Robert
Wimmer and Martin Wolf. In particular, I have to mention in this
list my office buddy Klaus Osterloh, who kept boredom out and
introduced the right amount of excitement into our office.

Finally, I would like to express my deepest gratitude to those
this thesis is dedicated to. There are many wonderful people who
taught and are still teaching me in all kinds of subjects; from
philosophy to science, from music to spirituality, from friendship
to love. I would not be appreciating the world's incredible beauty
in the way I do if it had not been for them.

\newpage
\pagestyle{plain} \addcontentsline{toc}{chapter}{Lebenslauf}
\begin{center}\LARGE\bfseries{\sc Lebenslauf}\end{center}
\vspace*{1cm} \hspace*{-1cm}\begin{tabular}{cl}
 {\it $~$23.4.1977} & geboren in Fulda\\[0.2cm]
 {\it 1983-1987} & Besuch der Bonifatius-Schule Fulda\\[0.2cm]
 {\it 1987-1988} & Besuch der Winfriedschule Fulda\\[0.2cm]
 {\it 1988-1992} & Besuch der Deutschen Schule Thessaloniki\\[0.2cm]
 {\it 1992-1996} & Besuch der Winfriedschule Fulda \\[0.2cm]
 {\it 1996} & Abitur\\[0.2cm]
 {\it 1996-1997} & Zivildienst in Fulda\\[0.2cm]
 {\it 1997-2000} & Studium der Mathematik und Physik an der Universit\"{a}t W\"{u}rzburg\\[0.2cm]
 {\it 1999} & Vordiplom in Mathematik und Physik\\[0.2cm]
 {\it 2000-2001} & Studium der Physik an der University of Texas at Austin\\[0.2cm]
 {\it 2001} & Master of Arts in Physik\\[0.2cm]
 {\it 1997-2002} & Stipendium der Studienstiftung des deutschen Volkes\\[0.2cm]
 {\it 2001-2002} & Besucher am Institute des Hautes
{\'E}tudes Scientifiques (IHES)\\
& und der Ecole Normale Sup{\'e}rieure (ENS), Paris\\[0.2cm]
 {\it 2002-2005} & Promotionsstudium der Physik an der Universit\"{a}t Hannover\\[0.2cm]
 {\it 2002-2005} & Stipendium des Graduiertenkollegs ``Quantenfeldtheoretische Methoden\\
 & in der Teilchenphysik, Gravitation, Statistischen Physik und Quantenoptik''\\
 & am Institut f\"{u}r theoretische Physik der Universit\"{a}t Hannover\\[0.2cm]
\end{tabular}

\thispagestyle{plain}

\pagestyle{fancyplain}
\lhead[\fancyplain{}{\bfseries\thepage}]{\fancyplain{}{\itshape
Index}} \rhead[\fancyplain{}{\itshape
Index}]{\fancyplain{}{\bfseries\thepage}}

\printindex


\newpage
\pagestyle{empty}{} \vspace*{1cm}


\newpage
\pagestyle{empty}{} \vspace*{3cm}

\centerline{\includegraphics[width=19cm,height=14cm]{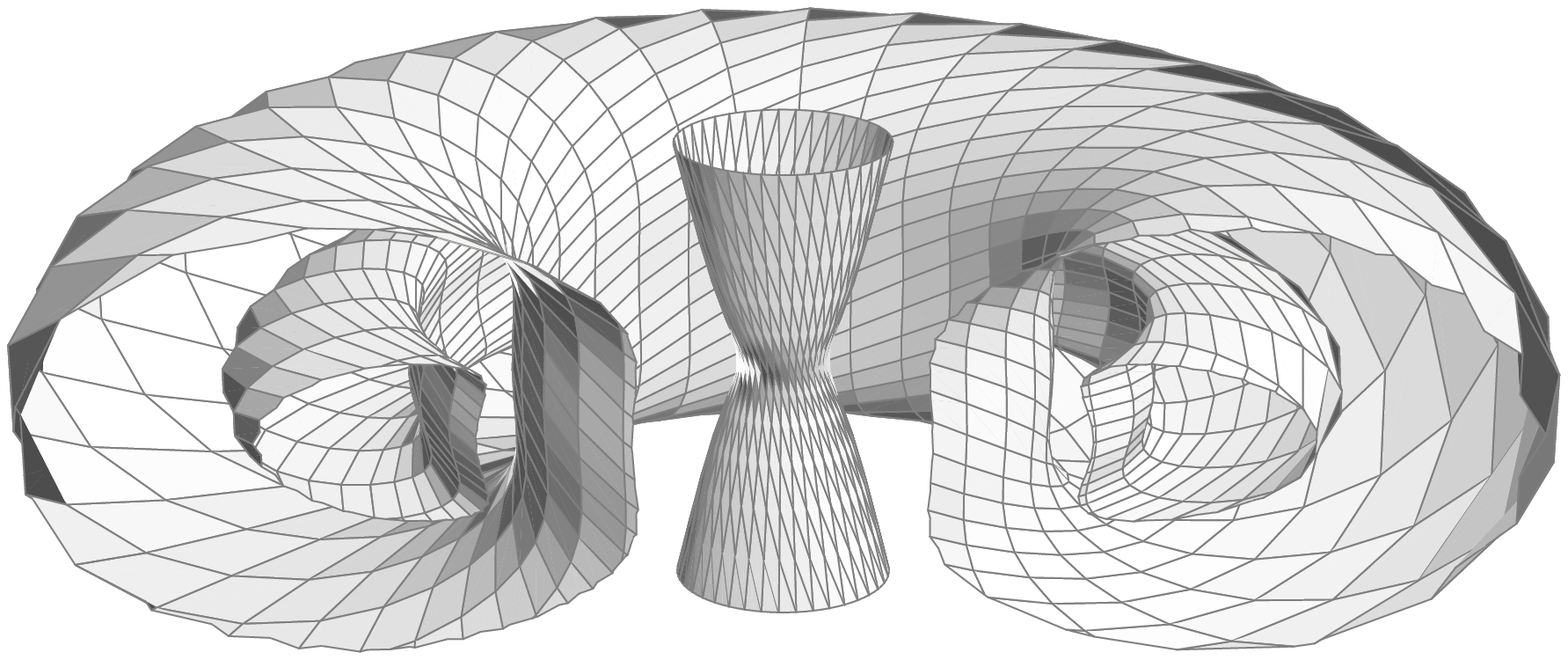}}
\end{document}